\documentclass[10pt]{report} 
\usepackage[margin=10pt,font=singlespacing,labelfont=bf,labelsep=period]{caption}
\usepackage[doublespacing]{setspace}
\usepackage{longtable}

\usepackage{amssymb}
\usepackage{amsmath}
\usepackage{latexsym}
\usepackage[utf8]{inputenc} 
\usepackage{mathrsfs}

\usepackage{geometry} 
\usepackage[pdftex]{graphicx}
\usepackage{epstopdf}
\usepackage{graphics,epsfig}

\usepackage{amsmath}
\usepackage{amssymb}
\usepackage{latexsym}
\usepackage{lscape}
\usepackage{rotating}
\usepackage{hyperref}
\usepackage{longtable}

\usepackage{booktabs} 
\usepackage{array}
\usepackage{paralist} 
\usepackage{verbatim}
\usepackage{subfig} 

\usepackage{fancyhdr} 
\usepackage{sectsty}
\allsectionsfont{\sffamily\mdseries\upshape} 

\usepackage[nottoc,notlof,notlot]{tocbibind} 
\usepackage[titles,subfigure]{tocloft}

\title{The Role of the Volume in Black Hole Thermodynamics}

\author{by\\
William John Victor Ballik \\
\\
Based on a thesis submitted to the Department of Physics, \\
Engineering Physics and Astronomy
\\
in conformity with the requirements for 
\\
the degree of Doctor of Philosophy \\
\\
\\
\\
Queen's University\\
\\
Kingston, ON, Canada
\\
\\
Thesis: February, 2025\\
This document: June, 2026
\\
\\
Copyright \copyright  William John Victor Ballik, February 2025, June 2026
\\
\\
Contact: ballik.william@queensu.ca}
\date{} 
\numberwithin{equation}{section}

\begin{document}
\pagenumbering{roman}

\section*{Introductory note} 

The following is based on my thesis, ``The Role of the Volume in Black Hole Thermodynamics,'' submitted to the Department of Physics, Engineering Physics and Astronomy in conformity with the requirements for the degree of Doctor of Philosophy, for Queen's University. The final copy of my thesis was submitted to Queen's University in February, 2025. 

This following document is somewhat edited from that thesis. Most of the changes are minor, including elaborating on certain points, fixing minor typos, streamlining derivations, modifying language, and so on. I also moved the content of Section \ref{AreaRevisited} to Appendix \ref{AreaCalculationAppendix}. The most significant changes are the deletion of one of the appendices in my original thesis and the addition of Chapter \ref{additionalNote}: Additional Note and the associated Appendix \ref{AdditionalNoteAppendix}, which are based on research I completed since the thesis was submitted. Finally, an effort has been made to add in more references to the literature than appeared in the thesis to better place my work in a broader context, especially when it comes to the integrability of Hamiltonian charges.

Despite the title, the focus of this thesis and document is primarily on black hole \emph{mechanics} for Kerr--anti-de Sitter black holes, and interpreting the role that the so-called geometric and thermodynamic volumes have from a black hole mechanics perspective. 

Comments are welcome.

\maketitle

\newcommand{\nc}{\newcommand}
\nc{\G}{\Gamma}
\nc{\de}{\delta}
\nc{\na}{\nabla}
\nc{\nn}{\nonumber \\}
\nc{\f}{\frac}
\nc{\ep}{\epsilon}
\nc{\bs}{\boldsymbol}
\nc{\La}{\Lambda}
\nc{\lie}{\mc L}
\nc{\om}{\omega}
\nc{\Th}{\Theta}
\nc{\Si}{\Sigma}
\nc{\z}{\zeta}
\nc{\mc}{\mathcal}
\nc{\hde}{\hat \de}
\nc{\la}{\lambda}
\nc{\pa}{\partial} 
\nc{\cd}{\cdot}
\nc{\cds}{\ldots}
\nc{\ka}{\kappa}
\nc{\Heq}{\overset{H}{=}}
\nc{\Om}{\Omega}
\nc{\tht}{\theta}
\nc{\sq}{\sqrt}
\nc{\ve}{\varepsilon}
\nc{\Ups}{\Upsilon}
\nc{\mr}{\mathring}
\nc{\vp}{\varphi}
\nc{\g}{\gamma}
\nc{\ti}{\tilde}
\nc{\ups}{\upsilon}
\nc{\bt}{\beta}
\nc{\al}{\alpha}

\begin{abstract}
\addcontentsline{toc}{chapter}{Abstract}

Gibbons et al.~\cite{GibbonsPerry} found the energy $\mc{E}$ of Kerr--anti-de Sitter black holes by integrating the first law of black hole thermodynamics, $\de\mc{E}= \sum_i\Om_i\de\mc{J}_i+T\de S$, with black hole angular momenta $\mc{J}_i$, angular velocity $\Om_i$, temperature $T$ and entropy $S$. They showed that $\mc{E}$ corresponds to the Ashtekar--Magnon--Das (AMD) energy, calculated in frame adapted to the Killing vector $\xi^a$ which is asymptotically timelike and hypersurface-orthogonal. In Cveti\v{c} et al.~\cite{Cvetic}, the first law was extended by interpreting $\mc{E}$ as an enthalpy and the cosmological constant $\tilde\La$ as being proportional to a pressure $P$ according to $P = -(D-2)\ti \La/16\pi$. The modified first law is $\de\mc{E}=\sum_i\Om_i\de\mc{J}_i+T\de S+V_{th}\de P$ with ``thermodynamic volume'' $V_{th}$. Due to scaling symmetry, the Smarr relation $(D-3)\mc{E}=(D-2)\left( \sum_i\Om_i\mc{J}_i+TS\right)-2PV_{th}$ is automatically satisfied. In a frame adapted to the Killing vector $\bt^a = (D-1)^{-1}\na_b\bs{h}^{ba}$ where $\bs h$ is the Principal Conformal Killing--Yano tensor, the corresponding AMD energy $\mc F$ and angular velocities $\om_i$ satisfy the Smarr relation $(D-3)\mc{F}=(D-2)\left(\sum_i\om_i\mc{J}_i+TS\right)-2PV_{geo}$ with ``geometric volume'' $V_{geo}$. 

I extend the work of Parikh~\cite{Parikh} to define the vector volume $\mc{V}_C$ of a $D$-dimensional stationary black hole to be equal to the rate of growth of the $D$-volume of the black hole along the flow of the stationarity Killing vector. I show that $V_{geo}=\mc{V}_C$.

These papers and my work suggest the following questions: why is it necessary to use a frame adapted to $\xi^a$ rather than $\bt^a$ to recover the first law? Why does $\mc V_C$ appear more naturally in the $\bt^a$ frame? Adapting Barnich and Comp\`ere \cite{BarnichCompere}, I define a $(D-2)$-form $\bs I_\chi$ associated with each Killing vector $\chi^a$. $\bs I_\chi$ makes use of the Kerr--Schild decomposition of Kerr--anti-de Sitter as an anti-de Sitter background plus perturbation. The integral of $\bs{I}_\chi$ over an arbitrary $(D-2)$-surface enclosing the black hole gives a conserved quantity $H^{\bs I}_\chi = \oint\bs{I}_\chi$, with $\mc{E}=H^{\bs I}_\xi$ and $\mc{F}=H^{\bs I}_\bt$. I show that the first law will be satisfied with quantities constructed from $\bs{I}_\chi$ if the background anti-de Sitter metric and the vector $\chi^a$ both have unvarying components. This is possible to construct for $\xi^a$ but not $\bt^a$, explaining why the first law works for $\mc{E}$ but not $\mc{F}$. I show that $\mc{V}_C$ appears in the $\bt$-associated Smarr relation due to simplifications related to $\bs{h}$.

\setcounter{page}{2}
\end{abstract}
\setcounter{page}{3}

\chapter*{Co-Authorship}
\addcontentsline{toc}{chapter}{Co-Authorship}

My supervisor, Dr.~Kayll Lake, co-authored the paper \cite{Ballik} which makes up Chapter \ref{PRDPaper}.

\chapter*{Acknowledgments}
\addcontentsline{toc}{chapter}{Acknowledgments}

Thanks to my supervisor Dr.~Kayll Lake for his ongoing, continued, longstanding support and championing of my work and his breadth and depth of knowledge of GR. Thanks to my friends in the GR group, Dr.~Majd Abdelqader, Dr.~Sheref Aboelhassan, Dr.~Dominic Rochfort, Matt McNish, and especially Dr.~Dmitri Lebedev, who has been on the same path as me forever and is one of my oldest friends. I think it might be a quirk of GR, but this group is some of the most original, interesting, brightest and best overall people I've ever met.

I want to thank Dr.~Don Page, Dr.~Greg Smith, Dr.~Joe Bramante and Dr.~Ken Clark for serving on my examining committee.

Thanks to the department, particularly Dr.~Aaron Vincent for his strong encouragement in getting me to finish writing, Dr.~Jordan Morelli for his friendship, Dr.~Larry Widrow for his obliging me with last minute letters of reference, Dr.~Rob Knobel, for giving me teaching opportunity and encouragement, and Graduate Chairs Larry, Aaron and Ken and Graduate Program Assistants Loanne Meldrum and Alla Kryachkova for all their work keeping my degree running smoothly.

I would like to thank Dr.~David Kubiz\v{n}ak for a helpful discussion at the Atlantic General Relativity Conference (2017), which affected the direction of my research, and Dr.~Don Page for helpful comments on reading the abstract of this thesis, which led to some rewriting of the abstract as well as the body of Chapter~\ref{PRDPaper}.

My research has been financially supported by the Ontario Graduate Scholarship, the Queen Elizabeth II Scholarship in Science Technology, the Queen's Graduate Entrance Tuition Award, Queen's Graduate Award, and the  Conference Travel Award, for which I am grateful. The paper which appears as Chapter \ref{PRDPaper} was supported in part by a grant to Dr.~Lake from the Natural Sciences and
Engineering Research Council of Canada. Portions of this work were completed using \emph{GRTensor} versions II and III \cite{GrTensor}. \emph{Maple} was in general invaluable for performing calculations.

Lastly, I want to thank and send my love to my friends and especially to my family.

\tableofcontents

\listoftables

\addcontentsline{toc}{chapter}{List of Tables}

\listoffigures

\addcontentsline{toc}{chapter}{List of Figures}

\chapter*{List of Abbreviations and Symbols}
\addcontentsline{toc}{chapter}{List of Abbreviations and Symbols}

 \begin{longtable}[c]{| c | c | c |}
 \caption{List of commonly used abbreviations and symbols. If applicable, an equation, section or bibliographic reference to where the abbreviation/symbol is introduced, defined, or near where it is first used in the text body. If there is no text reference the symbol or abbreviation appears in the Nomenclature section. \label{CommonlyUsedSymbols}}\\

 \hline
 \multicolumn{3}{| c |}{List of Abbreviations and Symbols}\\
 \hline
 Symbol & Meaning & Text Reference \\
 \hline
 \endfirsthead

 \hline
 \multicolumn{3}{|c|}{Continuation of Table \ref{CommonlyUsedSymbols}}\\
 \hline
 Abbr./Sym. & Meaning & Text Reference \\ 
 \hline
 \endhead

 \hline
 \endfoot

 \hline
 \multicolumn{3}{| c |}{End of Table \ref{CommonlyUsedSymbols}}\\
 \hline\hline
 \endlastfoot
 $\hat 0$ & Index $2n+1$ in odd dimensions for orthonormal basis & Sect.~\ref{GKNAdSSubsection} \\
 $a,b,c,\ldots$ & Tensor indices &  \\
 $A,B,C,\ldots$ & Orthonormal basis indices & \\
  $a_i$ & Kerr--AdS Rotation Parameters(s) ($a$ for $D = 4$) & Sect.~\ref{KAdSForms} \\ 
 $A$ & Black hole horizon area & \eqref{AGibbonsLu} \\
 $\bs A$ & Maxwell electromagnetic potential 1-form & \eqref{KerrNewmanAdSA} \\ 
 $A^{(j)}$ & Elementary symmetric polynomials in $x_\mu^2$ & \eqref{Aj},\eqref{Aj2} \\ 
 $A_\mu^{(j)}$ & Elementary symmetric polynomials in $x_\nu^2$ with $x_\mu^2$ omitted & \eqref{Amuj} \\
 $\mc A_N$ & Surface ``area'' ($N$-volume) of unit $N$-sphere & \eqref{AsubD} \\ 
 ABL & Alternate Boyer--Lindquist coordinates & \eqref{ABL} \\
 ADM & Arnowitt--Deser--Misner (conserved quantity) & Chap.~\ref{NoetherChapter}\\
 AdS & Anti-de Sitter spacetime &  \\ 
$\alpha, \bt, \gamma, \ldots$ & Values of $\mu, \nu, \ldots$ excluding $n$ in orthonormal basis & Sect.~\ref{GKNAdSSubsection} \\ 
 $\hat \alpha, \hat \bt, \hat \gamma, \ldots$ & Equal to $\alpha + n, \bt + n, \gamma + n, \ldots$ & Sect.~\ref{GKNAdSSubsection} \\
 AMD  & Ashtekar--Magnon--Das (conserved quantity) & \eqref{EJiQC} \\
 $\bs b$ & Potential one-form for PCKY tensor & \eqref{bpotentialdefinition} \\ 
 $\mc B$ & Black hole region & Chap.~\ref{PRDPaper} \\
 $B^\mu_{(j)}$ & Inverse of $A_\mu^{(j)}$ & \eqref{Bmuj} \\
 BC & Barnich and Comp\`ere (2005)  & Ref.~\cite{BarnichCompere}\\
 $\bt$ & Principal vector & \eqref{betadef} \nn 
 BL & Boyer--Lindquist coordinates & \eqref{BLmetric} \\
 BLKS & Boyer--Lindquist--Kerr--Schild form & \eqref{BLKS} \\
 $C$ & $\prod_{1\leq i < j \leq n-1+\ve} (a_i^2-a_j^2)$ & \eqref{Cdefinition} \\
 $c$ & $\prod_{j=1}^{n-1+\ve} a_j^2$  & \eqref{Xmuc} \\ 
 $C_{abcd}$ & Weyl tensor & \\ 
 $C^{(l)}_i$ & Elementary symmetric polynomials in $a_j^2$ excluding $j = 0$ & \eqref{Cil} \\ 
 $\hat C^{(l)}_i$ & Elementary symmetric polynomials in $a_j^2$ including $j = 0$ & \eqref{Chatil} \\ 
 $\bs C_{\chi;\gamma}$ & Differential form used by BC to calculate $H_\chi$ & \eqref{BC3.9} \\
 CGKP & Cveti\v{c}, Gibbons, Kubiz\v{n}\'ak, and  Pope (2011)  & Ref.~\cite{Cvetic}  \\
 $\chi$ & Generic Killing vector & Sect.~\ref{ImportantKilling} \\ 
 CLP & Chen, L\"u and Pope (2006) & Ref.~\cite{Chen} \\ 
 $D$ & Dimension of spacetime ($D \geq 4$) & \\
 $d$ & Exterior derivative or standard differential & \\ 
 $D^i_{(l)}$ & Inverse of $C^{(l)}_i$ & \eqref{Cil} \\
 $\hat D^i_{(l)}$ & Inverse of $\hat C^{(l)}_i$ & \eqref{hatDhatUps} \\ 
 $\de$ & Variation of some quantity & \\ 
 $\de^a_b, \de_{ij}$ etc.&Kronecker delta & \\ 
 $\Delta, \Delta_\tht$ & Functions appearing in four-dimensional KAdS metric &\eqref{KS4D},\eqref{BL4D} \\
 $\pa_a, \pa/\pa x^a$ & Partial derivative with respect to $x^a$ & \\
 $\pa \mc R$ & Boundary of region $\mc R$ & Chap.~\ref{PRDPaper} \\
 $\textrm{div}$ & Divergence & \\
 $ds^2$ & Line element, $ds^2 = g_{ab} dx^a dx^b$ & \\ 
 $d \bar s^2$ & Line element for background, $d\bar s^2 = \bar g_{ab} dx^a dx^b$ & \\ 
 $d S$ & $(D-2)$-surface area element & \\
 $d S_{ab}$ & $(D-2)$-surface directed area element  &\\
 $d \Si$ & Hypersurface $(D-1)$-volume element &\\
 $d \Si_a$ & Directed hypersurface $(D-1)$-volume element & \\ 
 $\mc E$ & Energy (or enthalpy) & Chap.~\ref{NoetherChapter}\\ 
 $e^A$ & One-form orthonormal basis & \eqref{emuehatmuehat0} \\ 
 $e_A$ & Vector orthonormal basis & \eqref{eAvectorbasis} \\
 EH & Einstein--Hilbert & \eqref{LEH} \\
 EM & Electromagnetic & \eqref{LEM} \\ 
 $\ve$ & $D \mod 2$ & \eqref{varepsilon} \\ 
 $\bs \ep$ & Levi-Civita symbol & \\ 
 $\eta_i$ & Azimuthal Killing vector (for $D=4$ only $\eta$) & Sect.~\ref{ImportantKilling} \\
 $\mc F$ & Conserved quantity associated with $\bt$  & \eqref{Fdefinition} \\
 $F$ & $k_r$ value & \eqref{GibbonsXiWF} \\
 $\bs F_{ab}$ & Electromagnetic field & \eqref{LEM} \\
 $\flat$ & Lowers a vector to a one-form using the metric & \\
 $g$ & Metric tensor determinant & \\ 
 $\bar g$ & Background metric tensor determinant & \\ 
 $g_{ab}$ & Metric tensor & \\ 
 $\bar g_{ab}$ & Background metric tensor, typically pure anti-de Sitter & \\ 
 $G_{ab}$ & Einstein tensor & \\ 
 $\Gamma_i$ & $\prod_{\nu=1}^n (a_i^2-x_\nu^2)$ & \eqref{Gammatildephi} \\
 $\G^a_{bc}$ & Christoffel symbol & \\ 
 GKAdS & Generalized Kerr--AdS & Sect.~\ref{GKAdSSection} \\
 GKNAdS & Generalized Kerr--NUT--AdS & Sect.~\ref{GKNAdSSection} \\
 GPP & Gibbons, Perry and Pope (2005)  & Ref.~\cite{GibbonsPerry}\\ 
 $H$ & Kerr--Schild scalar & \\ 
 $H$ & Black hole horizon & Sect.~\ref{BHMechanicsSection}\\ 
 $h_{ab}$ & Kerr--Schild perturbation & \\ 
 $\bs h_{ab}$ & Principal Conformal Killing--Yano Tensor & \eqref{PCKYdef} \\ 
 $H_\chi$ & Hamiltonian associated with vector $\chi$ & Sect.~\ref{HamiltonianNoetherSection} \\
 $H^{\bs I}_\chi$ & $\oint \bs I_\chi$ & \eqref{HIchi} \\
 $I_D$ & Action (for $D$-dimensional spacetime) & \eqref{PhiTID} \\
 $\bs I_\chi$ & Differential form for defining ``conserved quantities'' & Sect.~\ref{KSBC} \\ 
 $\mc J_i$ & Angular momentum (for $D=4$ only $\mc J$) & Chap.~\ref{NoetherChapter}\\ 
 $\bs J_\chi$ & Noether current form associated with $\chi$ & \eqref{Jchi} \\ 
 $k$ & Kerr--Schild null vector & \eqref{kmu} \\ 
 $\bs K^K_\chi$ & Komar differential form associated with $\chi$ & \eqref{dKxiKdefinition} \\
 $\bs k_\chi[\de \phi;\phi]$ & Differential form related to conserved quantities & \eqref{kchi} \\ 
 $\bs k^{EH}_\chi[\de g;g]$ & Value of $\bs k_\chi[\de g;g]$ for Einstein--Hilbert gravity & \eqref{kEHincludingdeltachi} \\
 KAdS & Kerr--anti-de Sitter & Chap.~\ref{GKAdSChapter} \\ 
 $\kappa$ & Black hole surface gravity & \eqref{horizonzetakappa} \\
 KBL & Katz--Bi\v{c}\'ak--Lynden-Bell (superpotential, etc.) & Sect.~\ref{KBLSection} \\
 KS & Kerr--Schild coordinates & \eqref{dbarsspheroidal} \\ 
 $l$ & Anti-de Sitter radius of curvature & \eqref{l2} \\ 
 $\bs L$ & Lagrangian differential form & \eqref{IisintL} \\
 $\ell^a$ & ``Outgoing'' null geodesic in Kerr--AdS spacetimes & \eqref{elldef} \\
 $\lie_\chi$ & Lie derivative with respect to $\chi$ & \\
 $\La$ & ``Cosmological constant'' & \\
 $\tilde \La$ & $2 \La/(D-2)$ & \\ 
 $m$ & Kerr--AdS mass parameter & \eqref{His2mbyU} \\
 $m_\mu, m_\mu^*, \tilde m_\mu, \tilde m_\mu^*$ & Null eigenvectors of PCKY tensor & Sect.~\ref{PCKYNull} \\ 
 $\mu, \nu, \ldots$ & Indices for orthonormal basis (between 1 and $n$) & Sect.~\ref{GKNAdSSubsection} \\
 $\hat \mu, \hat \nu, \ldots$ & Equal to $\mu+n, \nu+n, \ldots$ & Sect.~\ref{GKNAdSSubsection} \\
 $\mu_i$ & Latitude coordinates/direction cosines (spheroidal) & \eqref{summu} \\
 $\hat \mu_i$ & Latitude coordinates/direction cosines (spherical polar) & \eqref{dbarsspherical} \\
 $\mu(r)$ & ``Mass function'' in GKAdS spacetimes & Sect.~\ref{GKAdSSection} \\ 
 $n$ & $\lfloor D/2 \rfloor$ & \eqref{nDby2} \\ 
 $n^a$ & Auxiliary null vector on horizon & Sect.~\ref{BHMechanicsSection} \\
 $\na_a$ & Covariant derivative & \\ 
 $\nu_i$ & ``Latitude coordinate'' including radius & \eqref{nusum1} \\ 
 NUT & Newman--Unti--Tamburino spacetime & Sect.~\ref{GKNAdSSection} \\
 $\mc O$ & Order (in Taylor expansions) & \\ 
 $\Om_i$ & Angular velocity & \eqref{Omegaomega} \\
 $\om_i$ & Angular velocity (asymptotically rotating frame) & \eqref{Omegaomega} \\
 $\bs \om^\mu$ & 2-form $e^\mu \wedge e^{\hat \mu}$ & \eqref{omegamu} \\ 
 $\bs \om_\chi$ & Killing potential 2-form for $\chi$ & \eqref{omegachidef} \\ 
 $p$ & $r / r_+$ & Sect.~\ref{variationsinfourdimensions} \\ 
 $P$ & Pressure, $P = - (D-2) \tilde \La/16\pi$ & \eqref{PLambdaRelation} \\
 $P$ & $\prod_{1\leq \mu < \nu \leq n} (x_\mu^2-x_\nu^2)$ & \eqref{Pdefinition} \\ 
 $\tilde P$ & $\prod_{1 \leq \alpha < \beta \leq n-1} (y_\alpha^2 - y_\beta^2)$ & \eqref{tildePdef} \\ 
 PCKY tensor & Principal Conformal Killing--Yano tensor $\bs h$ & \eqref{PCKYdef} \\
 $\Phi$ & Thermodynamic potential & \eqref{SmarrGibbsDuhem} \\ 
 $\Phi_\alpha$ & Electromagnetic potential (generally on horizon) & \eqref{firstlawwithThetadLambda} \\
 $\phi$ & Generic fields (including metric tensor) under variation & \eqref{deltaL} \\
 $\phi_i$ & Azimuthal angles (KS coordinates) & \eqref{dbarsspheroidal} \\ 
 $\vp_i$ & Azimuthal angles (BL coordinates) & \eqref{phivarphi} \\ 
 $\hat \vp_i$ & Azimuthal angles (ABL coordinates) & \eqref{dphidphihat} \\ 
 $\Phi_i$ & Azimuthal angles (Eddington--Finkelstein-like coordinates) & \eqref{phiPhiv} \\ 
 $\breve \phi_i$ & Azimuthal angles (BLKS coordinates) & \eqref{brevephi} \\ 
 $\tilde \vp_i$ & Rescaled time--azimuthal coordinates & \eqref{Gammatildephi} \\
 $\psi$ & Azimuthal angle ($D=4$ Eddington--Finkelstein) & \eqref{kadsIngoing4D} \\ 
 $\psi_j$ & Time--azimuthal coordinates adapted to orthonormal basis & Sect.~\ref{GKNAdSSubsection} \\ 
 $\bs Q_\chi$ & Noether charge form associated with $\chi$ & \eqref{NoetherCurrentCharge} \\ 
 $Q$ & Kerr--Newman--AdS parameter related to electric charge & \eqref{KerrNewmanAdSA} \\ 
 $Q_C[\chi]$ & AMD charge associated with $\chi$ & \eqref{EJiQC} \\
 $\mc Q$ & Electric charge (Kerr--Newman--AdS) & \eqref{mcQdef} \\ 
 $\mc Q_\alpha$ & Electric charge & \eqref{firstlawwithThetadLambda} \\ 
 $Q_\mu$ & $X_\mu/U_\mu$ in Generalized Kerr--NUT--AdS spacetimes & \eqref{QmuUmuSdefinition} \\
 $r$ & Spheroidal radius & Sect.~\ref{KAdSForms} \\ 
 $r_+$ & Event horizon radius & Sect.~\ref{BHHorizonGKAdS} \\
 $R_{abcd}$ & Riemann tensor & \\
 $R_{ab}$ & Ricci tensor & \\ 
 $R$ & Ricci scalar & \\ 
 $\bs R_{AB}$ & Riemann curvature two-forms in orthonormal basis & App.~\ref{curvature} \\ 
 $\mc R_{AB}$ & Ricci tensor orthonormal basis components & \eqref{Riccicanonical} \\ 
 $\rho^2$ & $r^2+a^2\cos^2\tht$ (four dimensions) & \eqref{KS4D} \\
 $S$ & Entropy & Chap.~\ref{NoetherChapter}\\
 $S$ & Function appearing in $e^{\hat 0}$ in odd dimensional GKNAdS spacetimes & \eqref{QmuUmuSdefinition} \\ 
 $(s)$ & Solution space parameter (as superscript) & Sect.~\ref{BCsection} \\
 $\sharp$ & Raises a one-form to a vector using the metric & \\
 $\Sigma$ & Hypersurface & \\ 
 $t$ & Time parameter (KS coordinates) & \eqref{dbarsspheroidal} \\ 
 $T$ & Temperature & Chap.~\ref{NoetherChapter}\\
 $T_{ab}$ & Stress--energy tensor & \\
 $\tau$ & Time parameter (BL or ABL coordinates) & \eqref{ttau} \\ 
 $\tht$ & Latitude angle (four dimensions) & \eqref{KS4D} \\ 
 $\bs \Th[\de g;g]$ & Symplectic potential form &  \eqref{ThetaEH} \\ 
 $\bs \tht^{EH}$ & Differential form used to calculate $\bs I_\chi$ & \eqref{thetaEH} \\ 
 $\Th$ & Thermodynamic conjugate to $\tilde \La$ & \eqref{firstlawwithThetadLambda} \\ 
 $\Th'$ & Thermodynamic conjugate to $\tilde \La$ in asymptotically rotating frame & \eqref{SmarrwithF} \\ 
 $u$ & $\cos \tht$ (four dimensions) & Chap.~\ref{ExplicitGKAdSChapter}\\
 $U$ & Scalar function in denominator of Kerr--Schild $H$ & \eqref{Udefinition} \\
 $U_\mu$ & $\prod_{\nu \neq \mu} (x_\nu^2-x_\mu^2)$ & \eqref{QmuUmuSdefinition} \\
 $u,v,x_i,y_i,z$ & Pseudo-Cartesian coordinates & Sect.~\ref{pseudoCartesian} \nn 
 $\Ups_i$ & $\prod_{j=1,j\neq i}^{n-1+\ve} (a_j^2-a_i^2)$ & \eqref{Cil} \\ 
 $\hat \Ups_i$ & $\prod_{j=0,j\neq i}^{n-1+\ve} (a_j^2-a_i^2)$ & \eqref{hatDhatUps} \\
 $v$ & Ingoing Eddington--Finkelstein-like coordinate & \eqref{phiPhiv} \\ 
 $V$ & Function appearing in BL, ABL forms of KAdS metric & \eqref{GibbonsV} \\
 $v^{EH}_a$ & $\na^b \de g_{ab} - g^{bc} \na_a \de g_{bc}$ & \eqref{ThetaEH} \\
 $V^{EH}_a$ & Term related to $v^{EH}_a$ given by $(V^{EH})^b = \bar \na_a h^{ab}$ & \eqref{VEHdef} \\
 $\mc V_{v,\mc R}$ & Vector volume associated with vector $v$ and region $\mc R$ & Chap.~\ref{PRDPaper} \\
 $\mc V_N$ & $N$-volume of unit $N$-ball & \eqref{VN} \\ 
 $V_{geo}$ & Geometric volume & \eqref{ThetaprimeVgeo} \\ 
 $ V_{th}$ & Thermodynamic volume & \eqref{Vth} \\ 
 $W$ & $k_t$ value & \eqref{GibbonsXiWF} \\
 WAND & Weyl-Aligned Null Direction & \eqref{TypeD} \\
 $x^a$ & Coordinates ($a = 1, \ldots, D$) & \\ 
 $x_\mu$ & Coordinates $x_\alpha = y_\alpha, \alpha < n, x_n = ir$ & \eqref{xn},\eqref{xalpha} \\ 
 $X$ & $-X_n$ & \eqref{Xdefinition} \\ 
 $\xi$ & Asymptotically static Killing vector & \eqref{xietai} \\ 
 $\bar X$ & $X$ value for background AdS spacetime & \eqref{Xbar} \\ 
 $X_\mu$ & Functions of $x_\mu$ appearing in GKNAdS spacetimes & \eqref{Xmu} \\ 
 $\bar X_\mu$ & $X_\mu$ values for background AdS spacetime & \eqref{Xbaralpha},\eqref{Xbar} \\ 
 $\Xi_i$ & $1-a_i^2/l^2$ ($\Xi = 1-a^2/l^2$ if $D = 4$) & \eqref{GibbonsXiWF} \\ 
 $Y$ & Function of $y$, $Y(y) = (-1)^{1-\varepsilon} \f{(1-y^2/l^2)}{y^{2\varepsilon}} \prod_{j = 1}^{n-1+\ve} (a_j^2 - y^2)$ & \eqref{Yofy} \\
 $y$ & Spherical radius & \eqref{dbarsspherical} \\ 
 $y_\alpha$ & Jacobi transformed latitude coordinates & \eqref{Jacobi} \\ 
 $\z^a$ & Killing vector tangent to null generators of horizon & \eqref{zetabreakdown} \\ 
 $\cdot$ & Contraction & \\ 
 $\wedge$ & Wedge product & \\
 $*$ & Hodge dual & \\
 $\bullet$ & Complete contraction of differential forms & \\
 $\Box$ & d'Alembert operator $\na_a \na^a$ & \\
 $!$ & Factorial $(n! = n(n-1)!, 0!=1$) & \\
 $!!$ & Double factorial $(n!!=n(n-2)!!,0!!=(-1)!!=1$) & \\
 \end{longtable}

\chapter*{Nomenclature, Key Concepts and Key Formulas} \label{Nomenclature}
\addcontentsline{toc}{chapter}{Nomenclature}

This thesis is concerned with General Relativity. I use geometric units ($G=c=\hbar = k_B = 1$) throughout, except where otherwise indicated. My general conventions follow Wald \cite{Wald84}, with the exception of Chapter \ref{PRDPaper}, which is the published paper \cite{Ballik} and has somewhat different notation. 

I consider spacetime of dimension $D \geq 4$. I generally assume Lorentzian signature, or $(- + \ldots +)$. The coordinates are denoted $x^a$, with $a = 1, \ldots, D$. I use Roman characters for these coordinate and tensor indices (Greek in Chapter \ref{PRDPaper}). 

Because I will be dealing with higher dimensions, and, in particular, area and volume calculations in higher-dimensions, I will use factorials: $n! = n\cdot (n-1) \ldots 1$ for a positive integer $n$, with $0! = 1$ by convention. 

I will also use the double factorial $n!!$ which is defined for integers $n$ as
\begin{align} 
    n!! &= n \cdot (n-2)!!, n \geq 1 \nn 
    0!! &= (-1)!! = 1. \label{doublefactorial}
\end{align}
Note that $n!! \neq (n!)!$ in general.

When doing Taylor expansions I use $\mc O$ to represent the order, e.g.~$\exp x = 1 + x + \mc O(x^2)$. 

The line element $ds^2$ can be written in terms of the metric tensor $g_{ab}$ as
\begin{align}
    ds^2 = g_{ab} dx^a dx^b.
\end{align}

A vector $v$ can either be represented by its components, either contravariant $v^a$ or covariant $v_a$, or ``directly,'' using the interpretation of $v$ 
as a directional derivative on a scalar function $f$ on a manifold. In the latter case we can expand out $v$ as
\begin{align}
    v &= v^a \f{\pa}{\pa x^a},
\end{align}
so that $v(f) = v^a \f{\pa f}{\pa x^a}$. Partial derivatives can also be written
\begin{align}
    \pa_a &\equiv \f{\pa}{\pa x^a}.
\end{align}
A covariant vector is also called a one-form. One-forms can be expressed in terms of the basis one-forms $dx^a$, which doubles as the exterior derivative of the coordinates $x^a$:
\begin{align}
    \bs b &= \bs b_a dx^a. 
\end{align}

The use of $x^a$ for coordinates is complicated by the convention, beginning in Chapter \ref{GKAdSChapter} and continuing, of using $x_\mu$ to denote a \emph{subset} of the coordinates for a specific set of spacetimes. In this case the index $\mu$ is not really a coordinate index, and does not run from 1 to $D$. This convention is one in use in the literature (see \cite{Chen,KrtousKubiznak,Hamamoto} and others) but may cause some confusion. 

The Kronecker delta $\de^a_b$ is defined so that $\de^a_b = 1$ if $a = b$ and 0 otherwise. The contravariant metric $g^{ab}$ satisfies 
\begin{align}
    g^{ab} g_{bc} &= \de^a_c.
\end{align}
The metric can thus be used to both ``raise'' and ``lower'' tensors, via the covariant and contravariant forms of the metric. 

When not using indices, I sometimes use musical isomorphisms to show how vectors and one-forms can be related by the metric (``raising'' and ``lowering''):
\begin{align}
    V^\flat &= V_a dx^a \nn 
    \bs b^\sharp &= \bs b^a \pa_a.
\end{align}

A general tensor can be represented by its components in the standard way.

The metric-compatible covariant derivative operator is represented by $\na_a$, satisfying $\na_a g_{bc} = 0$. For a vector $v$, 
\begin{align}
    \na_a v^b &= \pa_a v^b + \G^b_{ac} v^c,
\end{align}
where 
\begin{align}
    \G^a_{bc} &= \f12 g^{ad} ( \pa_b g_{dc} + \pa_c g_{bd} - \pa_d g_{bc})
\end{align}
are the \emph{Christoffel symbols}. Covariant derivatives of other tensors can be found in the standard way, using the fact that $\na_a f = \pa_a f$ for scalars $f$ and the product rule (that $\na_c (A_a B^b) = (\na_c A_a) B^b + A_a \na_c B^b$ and so on). 

The covariant derivative in the direction of a vector $v$ is given by
\begin{align}
    \na_v \equiv v^a \na_a.
\end{align}

A curve in spacetime is given by $x^a = x^a(\la)$ for some $\la$. Its tangent vector at each point is given by $v^a = d x^a/d\la$. If the curve is geodesic, then $v^a$ satisfies
\begin{align}
    \na_v v \propto v,
\end{align}
or
\begin{align}
    v^b \na_b v^a \propto v^a.
\end{align}
If 
\begin{align}
    v^b \na_b v^a = 0,
\end{align}
$v$ is said to be \emph{affinely parametrized}.

The Riemann tensor ${R^a}_{bcd}$ is defined so that
\begin{align}
    \na_b \na_a A^m - \na_a \na_b A^m = - {R^m}_{n a b} A^n,
\end{align}
with the Ricci tensor $R_{ab}$ defined as
\begin{align}
    R_{a b} &= {R^c}_{acb}
\end{align}
and Ricci scalar $R$ by
\begin{align}
    R &= R_{ab} g^{ab}.
\end{align}
The Einstein tensor $G_{ab}$ is defined by
\begin{align}
    G_{ab} = R_{ab} - \f12 R g_{ab}.
\end{align}
Vacuum is taken to be a spacetime satisfying
\begin{align}
    G_{ab} + \La g_{ab} &= 0, \label{GplusLag}
\end{align}
where $\La$ is a constant (``the cosmological constant''), including in higher dimensions. I do not consider other theories of gravity (such as Lovelock gravity). Outside vacuum, 
\begin{align}
    G_{ab} + \La g_{ab} = 8\pi T_{ab}
\end{align}
where $T_{ab}$ is the stress--energy tensor.

\eqref{GplusLag} implies that the Ricci tensor satisfies
\begin{align}
    R_{ab} &= \f{2}{D-2} \La g_{ab}.
\end{align}
If $\La < 0$, it is convenient to define a radius of curvature $l$ which is constant for the spacetime, satisfying
\begin{align}
    R_{ab} &= -\f{D-1}{l^2} g_{ab}.
\end{align}
I take $l$ to be positive. Flat space, which is to say $\La = 0$, is recovered in the $l \to \infty$ limit.

Another convention in use in \cite{Cvetic} is to use $\La$ to represent the constant of proportionality between the Ricci tensor and metric tensor for vacuum solutions. Because I am using $\La$ for its appearance in \eqref{GplusLag}, I will instead use $\tilde \La$ for the constant satisfying
\begin{align}
    R_{ab} = \tilde \La g_{ab},
\end{align}
in vacuum. This satisfies
\begin{align}
    \tilde \La &\equiv \f{2 \La}{D-2}.
\end{align}
Of course if $D = 4$, $\tilde \La = \La$. $l$ is given by 
\begin{align}
    \tilde \La &= -\f{D-1}{l^2} \nn 
    \La &= - \f{(D-1)(D-2)}{2l^2}.
\end{align}
Spacetimes for which $\La < 0$, including spacetimes which are not in vacuum, are referred to as being anti-de Sitter or AdS spacetimes. Pure AdS is what I use to describe the anti-de Sitter spacetime itself, which is described in more detail in Chapter \ref{GKAdSChapter}.

The Weyl tensor $C_{abcd}$ is the ``trace-free'' part of the Riemann tensor, given by
\begin{align}
    C_{abcd} = R_{abcd} - \f{2}{D-2} \left( g_{a[c} R_{d]b} - g_{b[c} R_{d]a}\right) + \f{2}{(D-1)(D-2)} R g_{a[c} g_{d]b}. \label{Weyldefinition}
\end{align}

At times I make use of an orthonormal basis. The basis of one-forms is written $e^A$, with capital Roman letters. The convention then is that $e^A$ has covariant components in a coordinate basis given by $e^A_a$. The associated vector basis is given by $e_A$, which has contravariant components $e_A^a$. The orthonormality condition means that 
\begin{align}
    g^{ab} e^A_a e^B_b &= \de^{AB},
\end{align}
where here $\de^{AB}$ is again the Kronecker delta, either equal to 1 if $A=B$ and 0 otherwise. The right-hand side is $\de^{AB}$ rather than $\eta^{AB}$ because while the spacetime is Lorentzian in signature, the standard orthonormal basis is of Euclidean signature, with one of the one-forms being imaginary-valued. Similarly,
\begin{align}
    g_{ab} e_A^a e_B^b &= \de_{AB},
\end{align}
with $\de_{AB}$ equal to 1 if $A=B$ and 0 otherwise. The orthonormality also implies that 
\begin{align}
    g^{ab} e^A_a = e_A^b.
\end{align}
Note that for a Lorentzian-signature metric, at least one of the basis one-forms must be complex-valued (not real-valued).

I make extensive use of differential forms. A good review of some key results of Wald and others regarding first law in the symplectic formalism is by Rossi in \cite{Rossi} and some of my conventions are chosen to match those. I use boldface for differential forms, except for, in some cases, one-forms.

To define my conventions, first I explain what I mean by symmetrization and antisymmetrization. 

Let $S_p$ ($p$ a positive integer) be the set of distinct permutations of the integers $1, \ldots, p$, and let $\sigma \in S_p$ be an element of the set. Then $\sigma = \{ \sigma(1), \ldots, \sigma(p)\}$, where $\sigma(1), \ldots ,\sigma(p)$ are distinct elements of $\{1, \ldots , p\}$. Let an elementary permutation be one for which two elements are swapped. Let an even permutation of $\{1, \ldots, p\}$ be one which can be reached through an even number of elementary permutations and an odd permutation be one which can be reached through an odd number of steps. Let $(-1)^{|\sigma|} = +1$ if $\sigma$ is an even permutation of $\{1, \ldots, p\}$ and $-1$ if $\sigma$ is an odd permutation. 

Square brackets represent antisymmetrization, where
\begin{align}
    \omega_{[a_1 \ldots a_p]} &= \frac{1}{p!} \sum_{\sigma \in S_p}(-1)^{|\sigma|} \omega_{\sigma(a_1) \ldots \sigma(a_p)}.
\end{align}
Parentheses represent symmetrization, where
\begin{align}
    h_{(a_1 \ldots a_p)} &= \frac{1}{p!} \sum_{\sigma \in S_p} h_{\sigma(a_1) \ldots \sigma(a_p)}.
\end{align}

A purely antisymmetric covariant tensor can be written as a differential form. The conventions used by Wald in \cite{Wald84} for differential forms are (letting $\bs \omega$ be a $p$-form)
\begin{align}
    \bs \om &= \bs \om_{a_1 \ldots a_p} dx^{a_1} \ldots dx^{a_p} \nn 
    \bs \omega_{a_1 \ldots a_p} &= \bs \omega_{[a_1 \ldots a_p]}
\end{align}
The exterior derivative $d$ of a $p$-form $\bs \om$ is a $(p+1)$-form with components
\begin{align}
    (d \bs \omega)_{b a_1 \ldots a_p} &= (p+1) \partial_{[b} \bs \omega_{a_1 \ldots a_p]}. \label{domega}
\end{align}
Note that the derivative $\partial_b$ can be replaced with a covariant derivative associated with a torsion-free connection, such as the Levi-Civita connection, since the connection terms will vanish under the antisymmetrization, so that 
\begin{align*}
    (d \bs \om)_{b a_1 \ldots a_p} &= (p+1) \na_{[b} \bs \om_{a_1 \ldots a_p]}.
\end{align*}
In particular, if $v$ is a vector and $v^\flat$ is its associated one-form,
\begin{align}
    (d v^\flat)_{a b} &= 2 \pa_{[a} v_{b]}.
\end{align}

A form which is the exterior derivative of another form is called exact, as in $\bs \om = d \bs \psi$. A closed form is one whose exterior derivative is zero, as in $\bs \om$ such that $d \bs \om = 0$. All exact forms are closed. 

The Lie derivative with respect to a vector field $v$ is represented by $\lie_v$. In particular, when acting on a scalar $f$,
\begin{align}
    \lie_v f &= v(f) = v^a \pa_a f,
\end{align}
and when acting on another vector $u^a$ the result is
\begin{align}
    \lie_v u^a &= v^b \pa_b u^a - u^b \pa_b v^a.
\end{align}
These properties along with the product derivative rule $\lie_v (A_a B^b) = (\lie_v A_a) B^b + A_a \lie_v B^b$ give the Lie derivatives of other tensors. 

For differential forms $\bs \om$, the Lie derivative satisfies the Cartan identity,
\begin{align}
    \lie_v \bs \om &= v \cdot d \bs \om + d (v \cdot \bs \om).  \label{Cartan}
\end{align}

A Killing vector $\chi$ is a vector for which the Lie derivative of the metric $g_{ab}$ vanishes,
\begin{align}
    \lie_\chi g_{ab} &= 0.
\end{align}
This is equivalent to the statement
\begin{align}
    \na_{(a} \chi_{b)} = 0.
\end{align}

Let $\Box$ be the d'Alembert operator 
\begin{align}
    \Box &\equiv \na_a \na^a.
\end{align}
When acting on a Killing vector $\chi^a$,
\begin{align}
    \Box \chi^a &= \na_b \na^b \chi^a \nn 
    &= -R^a_b \chi^b. \label{Boxchi}
\end{align}

The wedge product $\wedge$ of a $p$-form $\bs \omega$ and a $q$-form $\bs \mu$ is 
\begin{align}
    (\bs \omega \wedge \bs \mu)_{a_1 \ldots a_p b_1 \ldots b_q} &= \frac{(p+q)!}{p!q!} \bs \omega_{[a_1 \ldots a_p} \bs \mu_{b_1 \ldots b_q]}.
\end{align}

$\bs \ep$ is the volume form associated with the metric $g_{ab}$. In coordinates $(x^1, \ldots, x^D)$ it is given by
\begin{align}
    \bs \epsilon &= \pm \sqrt{-g} dx^1 \wedge \ldots \wedge dx^D,
\end{align}
with
\begin{align}
    \bs \epsilon_{1 \ldots D} &= \pm \sqrt{-g}.
\end{align}
Here $g$ is the determinant of the metric. Lorentzian signature metrics, with the ``mostly pluses'' convention, have $g < 0$ when the coordinates are real. $\bs \ep$ is defined up to choice of orientation. 

Let $e_{a_1 \ldots a_D}$ be the symbol which has $e_{a_1 \ldots a_D} = +1 \: (-1)$ if $\{a_1,\ldots, a_D\}$ is an even (odd) permutation of $\{1, \ldots, D\}$ and zero otherwise. Then 
\begin{align}
    \bs \ep_{a_1 \ldots a_D} =  \pm\sqrt{-g} e_{a_1 \ldots a_D}
\end{align}

The $\pm$ sign is to account for the choice of orientation; generally we make the assumption of the positive sign. $\bs \ep$ is a tensor.

$\bs \ep^{a_1 \ldots a_D}$ is also fully antisymmetric, with
\begin{align}
    \bs \ep^{1 \ldots D} =  \mp \f{1}{\sqrt{-g}}.
\end{align}

A vector dotted with a differential form means contraction in the first index, e.g.
\begin{align}
    (v \cdot \bs \epsilon)_{a b c} &= v^d \bs \ep_{d a b c},
\end{align}
and is equivalent to the interior derivative $i_v$: $v \cdot \bs \ep = \iota_v \bs \epsilon$. A dot will also be used to represent the inner product of two vectors. If $u,v$ are two vectors,
\begin{align}
    u \cdot v &= g_{a b} u^a v^b.
\end{align}
Occasionally I will use the convention that a vector squared will be given by its inner product with itself, as in
\begin{align}
    v^2 &\equiv v \cdot v = g_{ab} v^a v^b.
\end{align}
I will call $v^2$ the norm of $v$. A vector is timelike if its norm is negative, spacelike if its norm is positive, and null if its norm is zero.

Let $\bullet$ represent the complete contraction of two $p$-forms, defined as
\begin{align}
    \bs \omega \bullet \bs \psi &= \bs \omega_{a_1 ... a_p} \bs \psi^{a_1 \ldots a_p},
\end{align}
using the convention of \cite{KrtousKubiznak}.

A spacetime with a Killing vector which is timelike (at least for much of the spacetime) is called stationary, with the timelike Killing vector referred to as the stationarity vector. If the stationarity vector is also orthogonal to some hypersurface---that is, a $(D-1)$-dimensional embedded submanifold---then the spacetime and the vector are called static. 

For AdS (not necessarily pure AdS) spacetimes, I refer to a Killing vector as ``asymptotically static'' in the following case. There is a spherical coordinate $y$ or spheroidal radius coordinate $r$ (explained in more detail in Chapter \ref{GKAdSChapter}), and I will describe the large-$y$ or large-$r$ values of the spacetime as the asymptotic region. If there is a Killing vector $\xi$ and hypersurface $\Si$ with normal $n_a$ such that, in the large-$y$ or large-$r$ limit, $\xi^a/\sqrt{-\xi_b\xi^b} = n^a$, then I consider the spacetime to be asymptotically static. I will use $\xi$ for this vector, and it will turn out to be particularly important. 

\section*{Kerr--Schild Decomposition} 

A \emph{Kerr--Schild spacetime} is one in which the metric $g_{ab}$ can be decomposed into a background spacetime metric $\bar g_{ab}$ and a perturbation $h_{ab}$, as in
\begin{align}
    g_{ab} = \bar g_{ab} + h_{ab}.
\end{align}
Here $h_{ab}$ can be written as a scalar function $H$ multiplied by a null vector $k_a$ as
\begin{align}
    h_{ab} = H k_a k_b.
\end{align}
Some authors use Kerr--Schild to mean exclusively the case where the background metric is Minkowski space. By ``Kerr--Schild spacetime'' what I mean is the ``Generalized Kerr--Schild'' spacetime of Taub \cite{Taub81}, which allows the background to be arbitrary (not necessarily Minkowski). Usually I will have anti-de Sitter (or Minkowski) be the background spacetime. Specifically I will usually denote the background spacetime by $\bar g_{ab}$. I will use bars to denote quantities which are calculated in the background spacetime, such as $\bar \na_a$ meaning the background covariant derivative, making use of the background metric Christoffel symbols $\bar \G^a_{bc}$. Let $\bar g^{ab}$ be the contravariant form of $\bar g_{ab}$, satisfying $\bar g^{ab} \bar g_{ac} = \de^a_c$. 

I also always use Kerr--Schild metrics in which the Kerr--Schild vector $k^a$ is an affinely parametrized geodesic. If $k^a$ is an affinely parametrized geodesic with respect to the background spacetime, it is also one with respect to the full spacetime (as shown in \cite{Taub81}):
\begin{align}
    k^a \bar \na_a k^b = k^a \na_a k^b = 0. \label{geodesicinfullandbackground}
\end{align}

Note that $k^a$ can be raised and lowered using either metric;
\begin{align}
    k^a &= g^{ab} k_b = \bar g^{ab} k_b \nn 
    k_a &= g_{ab} k^b = \bar g_{ab} k^b.
\end{align}

The metric determinants for the full and background spacetimes, $g$ and $\bar g$ respectively, are equal.

\section*{Conventions for Integration of Differential Forms}

To integrate a $p$-dimensional form $\bs \om$ over a $p$-dimensional surface $\Sigma$, choose coordinates $y^1, \ldots , y^p$ on $\Sigma$ and then pull back the form $\bs \om$ to the coordinates intrinsic to $\Sigma$ so that it can be written as $\bs \omega|_\Si = f dy^1 \wedge \cds \wedge dy^p$. Then the integral is
\begin{align}
    \int_\Sigma \bs \omega = \int f dy^1 \ldots dy^p,
\end{align}
so that the wedge products in the differential form are removed in order to perform the integration. In particular, the integral of the volume element $\bs \ep$ over some region $D$-dimensional region $\mc R$ is
\begin{align}
    \int_{\mc R} \bs \ep = \int_{\mc R} \sqrt{-g} dx^1 \cds dx^D.
\end{align}

The convention for orientation when integrating differential forms is the one for which the generalized Gauss--Stokes theorem holds (I refer to it as Gauss' law or Stokes' law). Given a $p$-dimensional region $\Sigma$ with boundary $\partial\Sigma$, for a $p$-form $\bs \omega$, the Gauss--Stokes law is that
\begin{align}
    \int_\Sigma \bs \omega = \int_{\partial \Sigma} d \bs \omega, \label{StokesTheorem}
\end{align}
using the induced orientation from $\Sigma$ onto $\partial \Sigma$. 

In addition to integrating differential forms directly, it is also useful to  write the quantity $d\Sigma_a$ as the directed surface element associated with a $(D-1)$-dimensional hypersurface $\Sigma$. This satisfies
\begin{align}
    (V \cdot \bs \ep)|_\Sigma = V^a d \Sigma_a \label{VdotdSigma}
\end{align}
for an arbitrary vector $V$, where $|_\Sigma$ means the restriction to $\Sigma$. 

For timelike or spacelike hypersurfaces $\Sigma$, 
\begin{align}
    d \Sigma_a = \pm n_a d \Sigma, \label{dSigmaa}
\end{align}
where $d\Sigma$ is the volume element intrinsic to $\Sigma$ and $n_a$ is the unit normal satisfying $n_a n^a = \pm 1$. By convention, $n^a$ is chosen to be future-oriented where possible and $n^a d \Sigma_a = + d \Sigma$, so that the sign in \eqref{dSigmaa} matches that of the sign of $n_a n^a$. 

In coordinates intrinsic to the hypersurface $\Sigma$, $y^i = (y^1, \ldots, y^{D-1})$ and given the induced metric ${}^{(D-1)}g_{ij}$, the volume element $d \Sigma$ is
\begin{align}
    d \Sigma &= \pm\sqrt{|{}^{(D-1)}g|} dy^1  \ldots  d y^{D-1},
\end{align}
where the sign $\pm$ depends on orientation and $|{}^{(D-1)}g|$ is the absolute value of the determinant of ${}^{(D-1)}g_{ij}$. The induced metric satisfies
\begin{align}
    ds^2|_\Sigma &= {}^{(D-1)}g_{i j} dy^i dy^j.
\end{align}

An integral can then be written as
\begin{align}
    \int_\Sigma V \cdot \bs \ep &= \oint_\Sigma V^a d\Sigma_a.
\end{align}

On a null hypersurface, the normal one-form $k_a$ is not normalizable and $d \Sigma = 0$, so that \eqref{dSigmaa} is not meaningful, though \eqref{VdotdSigma} still holds. 

Along similar lines, for a $(D-2)$-dimensional surface $S$ we define (also following Poisson) the directed surface element $d S_{a b}$ in the following way. Let $j^{a b}$ be an antisymmetric tensor. Then let
\begin{align}
    \bs J_{a_1 \ldots a_{D-2}} &= j^{b c} \bs \ep_{b c a_1 \ldots a_{D-2}}.
\end{align}
We will then write
\begin{align}
    \bs J = j^{ab} d S_{ab},
\end{align}
so that
\begin{align}
    \int_S \bs J = \int_S j^{a b} d S_{a b}.
\end{align}

I usually use $\oint$ when integrating a differential form over a closed manifold, which is to say, one whose boundary is empty. 

Following Poisson \cite{Poisson}, $dS_{ab}$ can be written in terms of a binormal. The surface $S$ will in general have two orthogonal normal directions. If one, say $n_a$, is timelike and one, say $r_a$, is spacelike, both normalized $(n_an^a = -1$ and $r_a r^a = +1$), $dS_{ab}$ can be written as
\begin{align}
    dS_{ab} &= -2 n_{[a} r_{b]} d S,
\end{align}
where $dS$ is the ($D-2$)-volume element intrinsic to the surface. If $S$ has coordinates $\tht^A = \tht^1, \ldots, \tht^{D-2}$, with induced metric $ds^2|_S = \sigma_{A B} d \tht^A d \tht^B$, then
\begin{align}
    dS &= \sqrt{\sigma} d \tht^1 \ldots d \tht^{D-2}.
\end{align}
We also note,
\begin{align}
    dS &= \f12 n^{[a}r^{b]} dS_{a b}.
\end{align}
Alternatively, if $S$ has two null normal directions, say $\z^a$ and $n^a$ satisfying $\z^an_a = -1$ (by convention), the binormal can be written as
\begin{align}
    dS_{ab} &= 2 \z_{[a}n_{b]} d S.
\end{align}
(The sign is defined up to a choice of orientation.) 

The Gauss--Stokes theorem relating an integral over a $(D-1)$-surface $\Sigma$ to that over its boundary $\partial \Sigma$ in terms of $d\Sigma_a, dS_{ab}$ is, for an antisymmetric tensor field $\bs B^{a b}$,
\begin{align}
    \int_\Sigma \na_b \bs B^{a b} d \Sigma_a &= \f12 \oint_{\pa \Si} \bs B^{ab} dS_{ab}. \label{StokesLawCodim1}
\end{align}

\section*{Hodge Dual, Divergence and Komar Forms}

Let the Hodge dual, $*$, act on a $p$-form $\bs \om$ and return a $(D-p)$-form $* \bs \om$ according to
\begin{align}
    (*\bs \om)_{a_{p+1} \ldots a_D} &\equiv \bs \om^{a_1 \ldots a_p} \bs \ep_{a_1 \ldots a_D}. 
\end{align}
Using this convention, $d *\bs \om$ is a $(D-p+1)$-form given by
\begin{align}
    d * \bs \om &= (-1)^{p-1} p * \textrm{div} \bs \om, \label{dstaromega}
\end{align}
where $\textrm{div} \bs \om$ is the divergence of $\bs \om$, also a differential form, given by
\begin{align}
    (\textrm{div} \bs \om)_{b_1 \ldots b_{p-1}} &= \na^a \bs \om_{a b_1 \ldots b_{p-1}}. \label{divomega}
\end{align}
For a derivation of the above (useful because there are so many conventions in use in the literature that it is important to keep track of what the solutions are in these ones), see Appendix \ref{HodgeAppendix}. 

Put this way, letting $\bs B$ be a 2-form, the left-hand side of \eqref{StokesLawCodim1} is  $\f12 \int_\Si d * \bs B$ and the right-hand side is $\f12 \oint_{\pa \Si} * \bs B$ confirming that \eqref{StokesLawCodim1} is a special case of \eqref{StokesTheorem}.

The following formulas are commonly used. For a vector $v$, its divergence $\na_a v^a$ can be computed by
\begin{align}
    \na_a v^a &= \f{1}{\sqrt{-g}} \pa_a \left( \sqrt{-g} v^a\right).
\end{align}
For an antisymmetric 2-form $\bs J^{ab}$,
\begin{align}
    \na_a \bs J^{ab} = \f{1}{\sqrt{-g}} \pa_a \left(\sqrt{-g} \bs J^{ab}\right). \label{twoformdivergence}
\end{align}
These are special cases of \eqref{dstaromega}. 

Using (see e.g.~\cite{KrtousKubiznak})
\begin{align}
    \bs \ep^{a_1 ... a_r c_{r+1} ... c_D} \bs \ep_{b_1 ... b_r c_{r+1} ... c_D} &= r! (D-r)! \de^{[a_1}_{b_1} \ldots \de^{a_r]}_{b_r},
\end{align}
for a $p$-form $\bs \om$, $**\bs \om$ is a $p$-form proportional to $\bs \om$, equal to
\begin{align}
    * * \bs \om &= (-1)^{p(D-p)+1} p! (D-p)! \bs \om. \label{starstaromega}
\end{align}
This allows us to define the Hodge dual inverse of $\bs \om$ as
\begin{align}
    *^{-1} \bs \om &= (-1)^{p(D-p)+1} \f{1}{p! (D-p)!} * \bs \om,
\end{align}
so that
\begin{align}
    * *^{-1} \bs \om = *^{-1} * \bs \om = \bs \om.
\end{align}

Let $V$ be a vector and $\bs \om$ be a $p$-form. $V \cdot * \bs \om$ evaluates to
\begin{align}
    (V \cdot * \bs \om)_{a_{p+2} \ldots a_D} &= V^{a_{p+1}} (*\bs \om)_{a_{p+1} \ldots a_D} \nn 
    &= V^{a_{p+1}} \bs \om^{a_1 \ldots a_p} \bs \ep_{a_1 \ldots a_D} \nn 
    &= \bs \om^{[a_1 \ldots a_p} V^{a_{p+1}]} \bs \ep_{a_1 \ldots a_D} \nn 
    &= \f{p! 1!}{(p+1)!} (\bs \om \wedge V^\flat)^{a_1 \ldots a_{p+1}} \bs \ep_{a_1 \ldots a_D} \nn
    &= \f{1}{p+1} (* (\bs \om \wedge V^\flat))_{a_{p+2} \ldots a_D} \nn 
    V \cdot * \bs \om &= \f{1}{p+1} * (\bs \om \wedge V^\flat). \label{Vdotstaromega}
\end{align}

For Killing vectors $\chi$, I will define the Komar differential form by (following the notation of \cite{BarnichCompere,Compere}) 
\begin{align}
    (\bs K^K_\chi)_{a_1 \ldots a_{D-2}} &= \f1{16\pi} \bs \ep_{a_1 \ldots a_{D-2} b c} \na^b \chi^c.
\end{align}
Note that $\na^b \chi^c$ is antisymmetric because $\chi$ is a Killing vector. We then have,
\begin{align}
    (d \bs K^K_\chi)_{a_2 \ldots a_D} &= \f{2}{16\pi} \bs \ep_{b a_2 \ldots a_D} \na_c \na^b \chi^c \nn 
    &= - \f{1}{8\pi} \bs \ep_{b a_2 \ldots a_D} \na_c \na^c \chi^b \nn 
    &=  \f{1}{8\pi} \bs \ep_{b a_2 \ldots a_D} R^b_c \chi^c \label{dKKxigeneral}
\end{align}
using \eqref{Boxchi} on the last line. In AdS vacuum,
\begin{align}
    (d \bs K^K_\chi)_{a_2 \ldots a_D} &= -\f{D-1}{8\pi l^2} \bs \ep_{b a_2 \ldots a_D} \chi^b \nn 
    d \bs K^K_\chi &= -\f{D-1}{8\pi l^2} * \chi^\flat. \label{dKKxi}
\end{align}
During some of this thesis I will be comparing different spacetimes. When there is an ambiguity about which spacetime metric is being used to evaluate the Komar integral, I will specify the metric in use in square brackets, such as $\bs K^K_\chi[g]$. 

\chapter{Introduction and Key Literature Review}
\pagenumbering{arabic}

\emph{As stated on the opening page of the document, the following document is closely based on my thesis, submitted February, 2025. I have edited it in the following ways:
\begin{itemize}
    \item I made some minor typographic corrections and some simplifications of notation and some mildly cleaned up arguments, derivations, etc. I moved some material from the main body to an appendix. I added some more references to the literature and more comments contextualizing my work in the literature. I added a few comments to possible future work.
    \item I removed one appendix chapter and modified some text referring to it.
    \item I added Chapter \ref{additionalNote}, as well as the associated Appendix \ref{AdditionalNoteAppendix}, based on an observation I made after the thesis was submitted.   
    \item I moved the bulk of Section \ref{AreaRevisited} to Appendix \ref{AreaCalculationAppendix}.
    \item I added subsections \ref{noteonkerrdesitter} and \ref{betajsection}.
\end{itemize}}

In \cite{BCH}, Bardeen, Carter and Hawking developed a set of laws of black hole mechanics, in which the black hole surface area was shown to be nondecreasing, and it was argued that it was analogous to the black hole entropy. This analogy was strengthened to an equality (up to a constant multiple) due to the existence of Hawking radiation \cite{HawkingRad}. This raises the question of why it is the area, and not the volume, which plays such a central role in black hole thermodynamics.

There was some interest in the subject of black hole volumes in the early 2000's. Whereas the black hole surface area is unambiguous, at least for a stationary black hole, the black hole volume is more ambiguous. In a typical four-dimensional setting, it would at first glance seem reasonable to define the black hole volume as the three-dimensional volume of some spatial hypersurface extending below the horizon down to the singularity. This depends very strongly on the choice of hypersurface. One idea from Christodoulou and Rovelli \cite{ChristodoulouRovelli} is to find the three-dimensional volume of a maximal hypersurface; subsequent papers (by Christodoulou and others) continued this idea (e.g.~\cite{ChristodoulouDeLorenzo}). This is not the approach that we took. 

In contrast to the three-dimensional volume, the four-volume (four-dimensional volume) below the black hole horizon is invariant and does not depend on a choice of hypersurface, but is formally infinite. The approach Dr.~Lake and I decided on was to use the growth rate of the four-volume---or, in higher-dimensions, the growth rate of the $D$-volume---of the black hole, leading to the definition of the vector volume.

\section{The Vector Volume}

In \cite{BallikLake10}, Dr.~Lake and I proposed taking the growth rate of the black hole four-volume with respect to an affine parameter of the null generators of a stationary black hole, showing that the four-volume grows logarithmically with the affine parameter for degenerate black holes. After posting on the arXiv we were informed of the work of M.~K.~Parikh \cite{Parikh} who had made a similar calculation previously. I generalized this idea to defining a vector volume as a rate of growth of the four-volume of a black hole along the flow of the vector field. This work was published as \cite{Ballik} and is included as Chapter \ref{PRDPaper} in this thesis. 

A brief summary of the results of \cite{Ballik} is the following. In $D$-dimensional spacetime, let $v^a$ be a divergence-free vector $(\na_a v^a = 0)$. Let $\mc R$ be a ($D$-dimensional) region, with boundary $\pa \mc R$ to which $v^a$ is \emph{tangent}. Then we define the \emph{vector volume} associated with $v$ and $\mc R$ to be
\begin{align}
    \mc V_{v,\mc R} &= \int_\Sigma v \cdot \bs \ep,
\end{align}
where $\Sigma$ is any hypersurface which ``spans'' $\mc R$, in the sense that $\partial \Sigma$ is entirely contained within $\pa \mc R$. Gauss' law tells us that, up to orientation, this quantity is independent of the actual hypersurface $\Sigma$. In particular, if $\xi$ is a Killing vector, $\na_a \xi^a = 0$ so the vector volume (or Killing volume in this case) can be used.

Other important features:
\begin{itemize}
    \item It is convenient to rewrite the above as $\int_\Sigma v^a d \Sigma_a$.
    \item Killing volumes associated with azimuthal symmetry Killing vectors---if the region is azimuthally symmetric---are zero, because a choice of $\Sigma$ can be made such that $\eta^a d\Sigma_a = 0$. This also means that if $v$ is a non-azimuthal symmetric vector and $\eta$ is an azimuthal symmetry vector, $\mc V_{v+\Omega \eta,\mc R} = \mc V_{v,\mc R}$.
    \item In a Kerr--Schild spacetime, the metric determinants for the background and full metric are the same, and therefore the vector volumes are the same.
    \item In Kerr--Schild spacetimes where the backgrounds are Minkowski, anti-de Sitter or de Sitter, there is a specific sense in which the vector volumes can be identified as a Euclidean volume---that is to say, the spatial volume of a region of Euclidean space matching the spatial region of the black hole in the background spacetime. 
\end{itemize}

We call the \emph{canonical black hole volume} $\mc V_C$ the vector volume of a black hole associated with the canonically normalized stationarity Killing vector (details in Section \ref{canon}). Our main application to black hole mechanics/thermodynamics is for Kerr--anti-de Sitter spacetimes. This application is related to the work of Gibbons, Perry and Pope (2005), hereafter GPP \cite{GibbonsPerry} and Cveti\v{c}, Gibbons, Kubiz\v{n}\'ak and Pope (2001), hereafter CGKP \cite{Cvetic}. 

\section{Brief Review of Two Key Papers in the Mechanics of AdS Black Holes}

\subsection{Background: First Law and Smarr Relations} 

As is well-known, the four-dimensional Kerr black hole, with energy $\mc E$ (which is equal to the Kerr mass parameter $m$), angular momentum $\mc J$, angular velocity $\Omega$, horizon area $A$ and surface gravity $\kappa$, obeys the Smarr relation
\begin{align}
    \f12 \mc E = \Omega \mc J + \frac{\kappa A}{8 \pi} \label{Smarr1}
\end{align}
It was shown by Bardeen, Carter and Hawking \cite{BCH} that in transforming between two nearby states, the Kerr black hole obeys the relationship
\begin{align}
    d \mc E = \Omega d \mc J + \frac{\kappa}{8 \pi} d A.
\end{align}
This was referred to as the \emph{first law of black hole mechanics}, or simply \emph{first law}, and an analogy with the first law of thermodynamics became further solidified when later work (e.g.~\cite{HawkingRad}) fixed the temperature associated with a black hole as being $T = \kappa/2\pi$ and the entropy as being $S = A/4$ (in units with $\hbar = G = c = k_B = 1$). We will generally refer to relations that relate quantities such as the energy, area, and angular momentum of a black hole in ``integral form'' such as \eqref{Smarr1} as Smarr relations and relations which relate these quantities to each other in \emph{differential form} as ``first law'' relations. The first law and Smarr relation can be extended to other cases, such as the Kerr--Newman black hole which also has electric charge. 

The first law and Smarr relation are related and follow from a scaling symmetry. In four dimensions (again in geometric units, $\hbar = G = c = k_B = 1$), $\mc E$ has dimensions of length, $S$ and $\mc J$ have dimensions of length squared, and $T$ and $\Omega$ have dimensions of inverse length. Further, $\mc E$ must be a homogeneous function of $T$ and $S$. This means that with one solution $\tilde {\mc E}, \tilde {\mc J}, \tilde \Omega, \tilde T,\tilde S$ for the first law, another solution $\mc E = \la \tilde {\mc E}, \mc J = \la^2 \tilde {\mc J}, \Om = \la^{-1} \tilde \Om, S = \la^2 \tilde S, T = \la^{-1} \tilde T$ also exists. The second then satisfies
\begin{align}
    d \mc E &= d (\la \tilde {\mc E}) \nn 
    &= \la d \tilde{\mc E} + \tilde {\mc E} d  la \nn 
    &= \la d \tilde { \mc E} + \mc E d \ln \la
\end{align}
Similarly, $d \mc J = \la^2 \tilde {\mc J} + 2 \la \tilde{\mc  J} d \la = \la^2 \tilde {\mc J} + 2 \mc J d \ln \la$, and similarly for $dS$. We then have, applying the first law,
\begin{align}
    d\mc E - \Om d \mc J - T d S &= \la (d \tilde {\mc E} - \tilde \Om d \tilde {\mc J} - \tilde T d \tilde S) + (\mc E - 2 \Om \mc J - 2 T S) d \ln \lambda \nn
    0 &= (\mc E - 2 \Om \mc J - 2 T S) d \ln \la,
\end{align}
applying the first law for both $d \mc E$ and $d \tilde{\mc  E}$. Under an arbitrary rescaling, then, the Smarr relation is recovered from the first law.

Kerr--anti-de Sitter black holes, given in arbitrary dimension in \cite{GibbonsLu,GibbonsLu2}, are solutions to Einstein's equations with
\begin{align}
    G_{ab} + \La g_{ab} &= 0, \label{EinsteinEquation}
\end{align}
where $\La$ (``cosmological constant'') is a negative constant, $g_{ab}$ is the metric, and $G_{ab}$ is the Einstein tensor. In my notation, $\La$ appears as in \eqref{EinsteinEquation}. A related constant, which I will call $\tilde \La$, is given by
\begin{align}
    R_{ab} &= \tilde \La g_{ab},
\end{align}
where $R_{ab}$ is the Ricci tensor. For solutions to \eqref{EinsteinEquation}, $\tilde \La$ is constant, equal to
\begin{align}
    \tilde \La &= \f{2 \La}{D-2}.
\end{align}
Note that when $D = 4$, $\La = \tilde \La$. Another useful quantity is the radius of curvature $l$ for an asymptotically anti-de Sitter spacetime, given by
\begin{align}
    \tilde \La &= -\f{D-1}{l^2}
\end{align}
where $D$ is the spacetime dimension.

I will now summarize the papers of GPP and CGKP to describe how my work relates to those papers. 

\subsection{Gibbons, Perry and Pope (2005)}

GPP \cite{GibbonsPerry} surveyed the literature regarding definitions for key quantities for Kerr--anti-de Sitter black holes. Kerr--anti-de Sitter black holes have an event horizon with an associated area $A$ and surface gravity $\kappa$. The surface gravity, horizon area, and angular momenta were generally universally accepted, but there were several incompatible expressions for energy and angular velocity. Using the usual identifications of temperature $T$ and entropy $S$ with  $\kappa$ and $A$ according to $T = \kappa/2\pi, S = A/4$ (in units where $G = \hbar = 1$), GPP proposed that any valid definitions for energy $\mc E$, angular momenta $\mc J_i$ and corresponding $\Omega_i$ (in higher dimensions; in four dimensions, only one $\mc J$ and $\Omega$ need be present) would satisfy the first law,
\begin{align}
    d \mc E = \sum_i \Omega_i d \mc J_i + T d S. \label{firstlaw1}
\end{align}

The $d$ here represents a variation between two ``nearby solutions,'' comparing two Kerr--AdS black holes with parameters $(\mc E,\mc J_i,\Om_i,T,S)$ and $(\mc E+d\mc E,\mc J_i+d \mc J_i,\Om_i+d\Om_i,T+dT,S+dS)$. Here, $\mc J_i$ are defined by the Komar integrals associated with the rotational Killing vectors, $A$ found by straightforward calculation of the area of a spatial section of the horizon and $\kappa$ by using the properties of the Killing vector tangent to the null generators of the horizon. 

For the $\Omega_i$, GPP emphasize the importance of choosing the angular velocity relative to a frame which is nonrotating at infinity. As I will discuss, this frame is adapted to a Killing vector $\xi^a$ which is asymptotically hypersurface orthogonal. 
Using these values for the $\Omega_i$, the first law can be integrated according to \eqref{firstlaw1}, so that, choosing $\mc E = 0$ in pure anti-de Sitter space, $\mc E$ can be fixed exactly. 

Also in wide use is a specific frame which is rotating at infinity. Letting the angular velocities in this frame be (in my notation) $\omega_i$, then not only is the value of $\mc E$ not recovered, but $\sum_i \omega_i d \mc J_i + T d S$ is not even a total differential, 
so that no energy could be defined that way satisfying a variation law. 

Having their expressions for energy and angular velocity, they compare their results to the literature and find agreement in some cases and disagreement in others. They clear up some of the ambiguity in the literature by pointing out that in several cases the energies in the literature were calculated (through various methods) in the frame which is asymptotically rotating. 

GPP further calculate the conserved energy in higher dimensions using the conformal techniques of Ashtekar, Magnon and Das (AMD). To do so, they find that the energy $\mc E$ obtained by integrating the first law is consistent with the conformal charge associated with a timelike Killing vector which is nonrotating at infinity, which I will call $\xi$. They find the conformal charge associated with a different Killing vector, which I will call $\bt$, which is rotating at infinity and is related to $\xi$ by 
\begin{align}
    \bt^a &= \xi^a + \sum_i \f{a_i}{l^2} \eta_i^a,
\end{align}
where $\eta_i^a$ are azimuthal symmetry Killing vectors and $a_i$ are the rotation parameters. I will denote the conformal charge (``energy'') associated with $\bt$ $\mc F$, and note that $\mc F-\sum_i \om_i \mc J_i = \mc E - \sum_i \Om_i \mc J_i$. For lower dimensions, $\mc F$ was consistent with several definitions of energy in the literature, which do not satisfy a first-law relationship. It was also shown shortly after their paper first appeared in pre-print form that the energy could also be found by the superpotential of Katz, Bi\v{c}\'ak and Lynden-Bell (KBL) in \cite{DeruelleKatz05} (see Section \ref{KBLSection}).

GPP also gave the Quantum Statistical Relation, 
\begin{align}
    \mc E - T S - \sum_i \Omega_i \mc J_i = T I_D,
\end{align}
where $I_D$ is the Euclidean action of the $D$-dimensional Kerr--anti-de Sitter metric.  

GPP pointed out that the scaling argument cannot be applied straightforwardly to relate the first law to a Smarr relation in this case because there is another parameter, $\Lambda$, which has dimensions of inverse square length, which does not appear (directly) in the first law relation. This means that $\mc E$ is a function not only of $\mc J_i$ and $S$ but also of $\Lambda$.

Several important papers followed. Of particular importance of this work is CGKP. 

\subsection{Cveti\v{c}, Gibbons, Kubiz\v{n}\'ak and Pope (2011)}

CGKP \cite{Cvetic} developed thermodynamics for AdS black holes by allowing $\tilde \Lambda$ to vary in the first law. The basic argument is to use the same definitions of energy and angular momentum as developed in GPP, and then to allow $\tilde \Lambda$ to vary in the Kerr--anti-de Sitter (and other cases) equations. This paper builds on an argument by Kastor \cite{KastorEtal:2009} that the ``energy'' $\mc E$ associated with a black hole in anti-de Sitter space should be taken to correspond to enthalpy, rather than energy, and to satisfy (in the absence of rotation) a law like
\begin{align}
    d \mc E = T d S + V d P,
\end{align}
where $P$ is a pressure term, corresponding (up to a numerical factor) to the cosmological constant $\tilde \Lambda$, and $V$ is a volume term conjugate to this pressure. Incorporating rotation as well as charges $\mc Q_\alpha$ and associated potentials $\Phi_\alpha$, CGKP~write the first law as
\begin{align}
    d \mc E = \sum_i \Om_i d \mc J_i + T d S + \sum_\alpha \Phi_\alpha d \mc Q_\alpha + \Th d \tilde \La.
\end{align}
Here, $\mc E$ is taken as a function of $\mc J_i, S, \mc Q_\alpha$ and $\tilde \La$, with, again, the definitions of $\mc E, \mc J_i$ and $S$ all taken to correspond to the ones found by GPP. Then the terms $\Om_i, T$ and, significantly, $\Th$ are found by varying $\mc E$ with respect to these parameters:
\begin{align}
    \Om_i &= \left.\frac{\pa \mc E}{\pa \mc J_i}\right|_{\mc Q_\alpha, T, \tilde \La} \nn 
    T &= \left. \frac{\pa \mc E}{\pa S}\right|_{\mc J_i,\tilde \La} \nn 
    \Phi_\alpha &= \left. \frac{\pa \mc E}{\pa \mc Q_\alpha}\right|_{\mc J_i,T,\tilde \La} \nn 
    \Th &= \left. \frac{\pa \mc E}{\pa \tilde \La}\right|_{\mc J_i,\mc Q_\alpha,T}.
\end{align}
The first law of GPP is then a special case where $\tilde \La$ remains fixed, and so that (in the absence of charge) the $\Om_i$ and $T$ necessarily agree with those in GPP.

CGKP also give a Smarr relation associated with this class of black holes, which takes the form (in arbitrary dimension)
\begin{align}
    (D-3) \mc E = (D-2) \left(T S + \sum_i \Om_i \mc J_i\right) + (D-3) \sum_\alpha \Phi_\alpha \mc Q_\alpha - 2 \Th \tilde \La.
\end{align}
They point out that this Smarr relation can be derived from the first law by a scaling relation, since now $E$ is a homogeneous function of $(\mc J_i,\mc  Q_\alpha, T, \tilde \La)$, all of which are allowed to vary, and here (in higher dimensions), if $\hbar = 1$ is relaxed so that there is a length scale, $\mc E$ and $\mc Q_\alpha$ have dimensions of length to the $(D-3)$ power, $A = 4 \hbar S$ and $\mc J_i$ have dimensions length to the $(D-2)$ power, and $\tilde \La$ still has dimensions of inverse square length.

To further the identification between pressure and $\tilde \La$, they identify $\tilde \La$ with pressure $P$ and the parameter $\Th$ as a ``thermodynamic volume'' (or ``effective volume'') $V_{th}$ by
\begin{align}
    P &= - \frac{D-2}{16\pi} \tilde \La \nn 
    V_{th} &= - \f{16\pi}{D-2} \Th.
\end{align}
This thermodynamic volume is compared to what CGKP~define to be the geometric volume $V_{geo}$, defined by
\begin{align}
    V_{geo} &= \int_\Sigma \sqrt{-g} d^{D-1} x
\end{align}
where $d^{D-1}x$ is the product of the coordinate differentials excluding $dt$ (where $t$ is the time parameter), $g$ is the metric determinant, and $\Sigma$ the region between the singularity (at $r = 0$ in even dimensions, though a little more care is required in odd dimensions---a point I will revisit in Section \ref{ringsingularity}) and the event horizon. This volume, for the classes of black holes studied (mostly Kerr--AdS, but also a few others), is observed to satisfy the intriguing relation
\begin{align}
    V_{geo} &= \frac{r_+ A}{D-1}, \label{rArelationship}
\end{align}
where $A$ is the horizon area and $r_+$ is the value of the usual Boyer--Lindquist radius coordinate (Kerr--AdS) or aerial radius (spherical symmetry) on the horizon. 

In the spherically symmetric examples considered by \cite{Cvetic}, $V_{th} = V_{geo}$. In the case of rotation, they are found to differ by the following terms,
\begin{align}
    V_{geo} &= V_{th} - \f{8\pi}{(D-1)(D-2)} \sum_i a_i \mc J_i, \label{VgeominusVth}
\end{align}
where $a_i$ are the rotation parameters appearing in the Kerr--AdS metric.

If instead of the asymptotically nonrotating frame, the asymptotically rotating frame associated with $\bt$ is used, then the first law relation does not hold, as found by GPP, but the Smarr relation takes a different form in which the geometric volume appears in a natural way. In this frame, the Smarr relation reads
\begin{align}
    (D-3) \mc F &= (D-2) \left( \sum_i \om_i \mc J_i + TS\right) + (D-3) \sum_\alpha \Phi_\alpha \mc Q_\alpha - 2 P V_{geo}. \label{Fsmarr}
\end{align}
The geometric volume thus appears directly in the asymptotically rotating frame Smarr relation but not in the asymptotically nonrotating frame, which suggests that there is some significance to this asymptotically rotating frame despite the fact that the first law is not satisfied with it ($d \mc F \neq \sum_i \om_i d \mc J_i + T dS + V_{geo}d P$). 

The vector $\bt$, which in the asymptotically rotating Boyer--Lindquist coordinates is given by $\bt = \pa/\pa \tau$, is equal, up to a numerical factor, to the divergence of the \emph{Principal Conformal Killing--Yano} (PCKY) tensor $\bs h$ of the spacetime. 

CGKP assign to each Killing vector $\chi$ a Killing potential 2-form $\bs \omega_{\chi}$ satisfying $\chi^b = \na_a \bs \om_\chi^{ab}$. Following \cite{KastorEtal:2009}, they give a procedure for constructing the energy $\mc E$ and the Smarr relation as a result, making use of a combination of the Komar integral associated with the time-symmetry Killing vector on the black hole horizon as combined with the integral of that Killing vector's Killing potential. They note that there is an ambiguity in the choice of Killing potential, since it can only be defined up to a co-closed form. This ambiguity means that a gauge choice for the Killing potential must be made to give the correct value of $\mc E$, which must be determined through other means.  They provide a construction of a Killing potential using the PCKY tensor, a prescription which is essentially unique. Using this form of the Killing potential, they show that they can, using the Killing potential for $\bt$, recover $\mc F$ and its associated Smarr relation \eqref{Fsmarr}, and consequently the geometric volume $V_{geo}$, but not $\mc E$ and its associated Smarr relation. For $\mc E$, a Killing potential can be constructed, but a gauge choice must be made to fix the form of the Killing potential, which means that it is necessary to find $\mc E$ by other means. They conclude by noting ``This suggests that the other remarkable properties of the
geometric volume, such as the fact that it is given by the Euclidean space formula \eqref{rArelationship}, might be related to the existence of the hidden symmetries of the Kerr--AdS metrics,'' by which they mean the existence of the closed conformal Killing--Yano tensor $\bs h$. 

I pointed out in \cite{Ballik} that the geometric volume corresponds to my canonical black hole volume $\mc V_C = \mc V_{\xi,\mc B}$, where $\xi$ is the canonically normalized stationarity vector (in a sense explained in Section \ref{NormalizationCanon}) and $\mc B$ is the black hole region.

CGKP also state the existence of a ``reverse isoperimetric inequality'' relating the thermodynamic volume and the area satisfied by all black holes under consideration. I do not address this result in my work.

The results for the frames adapted to $\xi$ and $\bt$ are summarized in Table \ref{summarytable}.

   \begin{table} \centering
        \begin{tabular}{|c|c|c|c|c|}
\hline

\textbf{Killing Vector} $\chi$ & \textbf{Associated Energy} & \textbf{Associated Frame}  & \textbf{First Law?} & \textbf{Smarr} \\ \hline \hline
& & & & \\
$\xi$ & $\mc E$ & Asymptotically nonrotating & YES & $V_{th} \neq \mc V_C$ \\ & & & &\\ \hline
& & & &\\ 
$\bt$ & $\mc F$ & Asymptotically rotating & NO  & $V_{geo} = \mc V_C$ \\
& & & &\\ \hline
\end{tabular} \caption{Summary of GPP, CGKP results. The first law works for the asymptotically nonrotating frame adapted to $\xi$, but not for the rotating frame adapted to $\bt$. The Smarr relation gives $V_{th} \neq \mc V_C$ for the asymptotically nonrotating frame adapted to $\xi$, but gives $V_{geo} = \mc V_C$ for the asymptotically rotating frame adapted to $\bt$. } \label{summarytable}
    \end{table}

\section{``Open Questions'' and My Answers} \label{openquestions}

The GPP and CGKP papers suggest the following issues for study (see Table \ref{summarytable} for the first two points):  
\begin{itemize}
    \item Why is it necessary to use the asymptotically-nonrotating frame, rather than the asymptotically-rotating frame, in order to achieve a working first law relation? (Or: is there something more we can say on this point?)
    \item Why does the geometric volume, or the canonical black hole volume (which are identical), appear in the Smarr relation at all, and, in particular, why does it appear in the Smarr relation associated with the asymptotically rotating frame, rather than the asymptotically static one? 
    \item Why does the Killing potential construction resulting from the PCKY tensor produce $\mc F$ and its associated Smarr relation but not $\mc E$? 
    \item Why does the area--volume relationship \eqref{rArelationship} hold?
\end{itemize}

There have been various methods of addressing the first questions in the literature, some of which have overlap with the method I used, and which I address in Section \ref{clarifyingcomments} below. The main argument in these papers (such as \cite{HajianSheikh-Jabbari,Chrusciel,Blagojevic:2020edq,Jing:2017jxw}) comes down to the fact that the Hamiltonian associated with $\xi$ is integrable, but the one associated with $\bt$ is not. My approach fits well into this general line of argument, and is focused on using the Kerr--Schild form of the metric to provide an \emph{explanation} for the integrability of the Hamiltonian associated with $\xi$ and the non-integrability of the Hamiltonian associated with $\bt$. There are also various methods of addressing the Smarr relation in the literature, which I also discuss in brief in Section \ref{clarifyingcomments} as well as what is distinct about my method. Now I will describe my approach. 

In order to address these questions, I wanted a way to define the conserved quantities $\mc E$ and $\mc J_i$ such that they would satisfy the first-law relation and be possible to evaluate on the horizon directly. 
Most definitions of conserved thermodynamic quantities in GR are defined ``at infinity,'' and their values on the horizon are consequently found indirectly by equating them with their values at infinity. (The use of a Killing potential construction for $\mc E$ does give a definition that can be evaluated at the horizon, but the gauge freedom means that the gauge must be fixed by finding $\mc E$ from another method.)  I made use of a paper by Barnich and Comp\`ere \cite{BarnichCompere} (hereafter BC). Making use of the conformal methods of Wald, BC suggested a method for integrating through solution space from a background reference spacetime to a given solution, in order to derive a Smarr relation with quantities that, when vary, produce a first law. They applied it to the Kerr--AdS solution, comparing the Kerr--AdS solution to an AdS background, and recovered the conserved quantities of GPP. 

The covariant phase space methods of Wald (often referred to as the Iyer--Wald formalism from \cite{IyerWald94,IyerWald95}) allow for the definition of a conserved charge associated with a Killing vector $\chi$, equal to the Hamiltonian $H_\chi$ associated with $\chi$, which varies under change in the metric represented by $\de$ according to
\begin{align}
    \de H_\chi &= \oint_C \bs k_\chi[\de g;g], \label{deHchiIntro}
\end{align}
where $C$ is a $(D-2)$-dimensional surface and $\bs k_\chi[\de g;g]$ is a $(D-2)$-form depending on the Killing vector $\chi$, the metric $g_{ab}$, variation between metric solutions $\de g_{ab}$, and the specific theory under consideration. When the Einstein--Hilbert Lagrangian is used, the form is denoted $\bs k^{EH}_\chi[\de g;g]$. 

The important Killing vectors are $\eta_i$ (the azimuthal symmetry Killing vectors), $\xi$ (the asymptotically hypersurface orthogonal Killing vector),  $\bt$ (the Killing vector related to the PCKY tensor $\bs h$) and $\z$ (the Killing vector tangent to the null generators of the event horizon). The associated Hamiltonians $H_\chi$ are the energy $\mc E$ for $\xi$ and the negative of the angular momentum $-\mc J_i$ for $\eta_i$. We can also associated a ``conserved Hamiltonian'' to $\bt$, called $\mc F$, and to $\z$, which is, by linearity, $\mc E - \sum_i \Om_i \mc J_i, \mc F - \sum_i \om_i \mc J_i$, though it turns out, as pointed out by, e.g.,~\cite{HajianSheikh-Jabbari,Chrusciel,Blagojevic:2020edq,Jing:2017jxw}, that these are not true Hamiltonians due to the failure of integrability of \eqref{deHchiIntro} for these values of $\chi$; I will nevertheless list them as ``$H_\chi$'' values in the table. 

   \begin{table}
   \centering
        \begin{tabular}{|c|c|c|}
\hline

\textbf{Killing Vector} $\chi$ & \textbf{Meaning} & ``$H_\chi$'' \\ \hline \hline
& & \\
$\eta_i$ & Azimuthal Symmetry & $-\mc J_i$ (angular momentum) \\ & &\\ \hline
& & \\ 
$\xi$ & Stationarity (asymptotically  & $\mc E$ (energy in \\
& hypersurface orthogonal) & asympt.~static frame) \\ \hline
& & \\ 
$\bt$ & Principal vector/stationarity  & $\mc F$ (energy in \\ 
& (not asympt.~hyp.~orth.) & asympt.~rotating frame) \\ \hline
& & \\ 
$\z$ & Tangent to null generators of horizon & $\mc E - \sum_i \Om_i \mc J_i, \mc F - \sum_i \om_i \mc J_i$ \\  & &  \\
\hline

\end{tabular} \caption{Summary of Killing vectors, their meanings, and their associated Hamiltonians.} \label{summaryHtable}
    \end{table}

It is also shown by BC, based on Bardeen, Carter and Hawking \cite{BCH}, that, when comparing a black hole solution with a Killing horizon and a perturbed solution, if $\bar \z^a$ is the Killing vector tangent to the null generators of the horizon in the unperturbed spacetime and $H$ is a $(D-2)$-surface on the horizon, 
\begin{align}
    \oint_H \bs k^{EH}_{\bar \z}[\de g;g] = \f{\kappa \de A}{8\pi},
\end{align}
even when the black holes are not solutions to Einstein's equations. This allows for a proof of the first law of black hole mechanics, and also suggests that if it is possible to find some quantity with variation equal to $\bs k^{EH}_\chi[\de g;g]$, it could be used to construct conserved quantities which will automatically satisfy a first law variation.

My modification to the BC method is to make use of the Kerr--Schild form of the Kerr--AdS solutions, as given by \cite{GibbonsLu,GibbonsLu2} where the higher-dimensional Kerr--AdS (and Kerr--de Sitter) solutions were first identified. In a Kerr--Schild metric, a full spacetime $g_{ab}$ can be written as the sum of a background metric $\bar g_{ab}$ and a Kerr--Schild perturbation $h_{ab}$, of arbitrary size, but of the form
\begin{align}
    h_{ab} = H k_a k_b,
\end{align}
where $H$ is a scalar and $k^a$ is a null geodesic vector. In this case, the background spacetime is AdS. Then it is possible to define a $(D-2)$-form $\bs I_\chi$ associated with the Killing vector $\chi$ satisfying
\begin{align}
    \de \bs I_\chi &= \bs k^{EH}_\chi[\de g;g]
\end{align}
provided that the variation satisfies certain requirements. This quantity is equal to
\begin{align}
    \bs I_\chi &= -\bs K^K_\chi + \overline{\bs K^K_\chi} - \chi \cdot \bs \tht,
\end{align}
where $\bs K^K_\chi$ is the Komar integrand in the full spacetime associated with $\chi$,
\begin{align}
    (\bs K^K_\chi)_{a_1 \ldots a_{D-2}} &= \f{1}{16\pi} \na^c \chi^d \bs \ep_{c d a_1 \ldots a_{D-2}},
\end{align}
where $\bs \ep_{a_1 \ldots a_D}$ is the Levi-Civita tensor, $\overline{\bs K^K_\chi}$ is the equivalent Komar integral associated with the \emph{background} spacetime, and $\bs \tht$ is given by
\begin{align}
    \bs \tht_{a_1 \ldots a_{D-1}} &= \f{1}{16\pi} \na_b h^{a b} \bs \ep_{a a_1 \ldots a_{D-1}}.
\end{align}
The conserved energy and angular angular momenta can then be written as $\oint_C \bs I_\chi$ with $\chi$, respectively, the stationarity Killing vector and the negative of the azimuthal symmetry Killing vectors. I also show generalization to include an electric charge. 

I then show that the Komar integral associated with the \emph{background} spacetime, when evaluated on the black hole horizon, is proportional to $\La \mc V_{\xi,\mc B}$ and so is the reason for the appearance of the geometric volume in the Smarr relations. 

I use the properties of the PCKY tensor to show that the construction of the Killing potentials using this tensor naturally causes $\mc V_{\xi,\mc B}$ to appear and to explain the area--volume relation \ref{rArelationship}. The argument is summarized as follows. Because of the non-contribution of angular Killing vectors to the vector volume, the canonical volume $V_C = \int_\Si \xi^a d \Si_a = \int_\Si \bt^a d \Si_a$, where $\Si$ is the region between the black hole singularity and event horizon. Because $\bt^a = (D-1)^{-1} \na_b \bs h^{ba}$, the latter can be expressed through the Gauss--Stokes law as $-\f{1}{2(D-1)} \oint_{\pa \Si} \bs h^{ab} d S_{ab} = -\f{1}{2(D-1)} \oint_H \bs h^{ab} d S_{ab}$, where $dS_{ab}$ is the directed surface element and $H$ is the horizon, which follows from the vanishing of $\oint \bs h^{ab} d S_{ab}$ on a surface just enclosing the singularity. Then we can show that $dS_{ab} = 2 \z_{[a} n_{b]} d S$ where $\z_a$ is the Killing vector tangent to the null generators of the horizon, $n_a$ is an auxiliary null vector with $n_a \z^a = -1$, and $dS$ is the undirected area element. Because $\z_a$ is an eigenvector of $\bs h^{ab}$ with eigenvalue $r$, where $r$ is the Boyer--Lindquist radial coordinate, $\bs h^{ab} d S_{ab}$ works out to $-2r_+ dS$, where $dS$ is the area  on the horizon, and so integrates to $-2 r_+A$. This shows that the area--volume relation $\mc V_C = r_+A/(D-1)$ follows directly from the definition of the vector volume and the properties of the PCKY tensor, in a chain of reasoning summarized by \eqref{chainofreasoningforrplus}.

I show that the $(D-2)$-form $\bs I_\chi$, when integrated on the horizon (or anywhere) gives the values for the energy and angular momenta which match those in GPP and CGKP. To analyze the strengths and limitations of the use of $\bs I_\chi$, I examine a broader class of spacetimes, which I call the Generalized Kerr--AdS spacetimes, by replacing $m$ in the Kerr--AdS solutions with a function of $r$, $\mu(r)$, and check under what circumstances the resulting integrals of $\bs I_\chi$ on the horizon can be combined in a way to satisfy a first law relation.

I argue that the failure of first law relation when using an asymptotically rotating frame, in which the conserved energy is based on $\bt$ rather than on an asymptotically static vector, is not exclusively because of the use of an asymptotically rotating vector per se and more because, when the Kerr rotation parameters $a_i$ are used, $\bt$ points ``in different directions'' with respect to the AdS background, when the Kerr--AdS spacetime is varied. This becomes a problem because the relationship $\de \bs I_\chi = \bs k_\chi^{EH}[\de g;g]$ requires both that $\chi^a$ and the background metric $\bar g_{ab}$ do not change under the perturbation. Because $\bt^a$'s direction in the background metric depends on the rotation parameters $a_i$, it is not possible for varying $a_i$ to find coordinates in which \emph{both} $\bt^a$ and $\bar g_{ab}$ are unvarying, so that $\de \bs I_\bt \neq \bs k^{EH}_\bt [\de g;g]$. 

Finally I address the question of why the Smarr relation works with conserved quantities based around $\bt$ rather than the asymptotically static vector by considering, in detail, a number of cancellations in $\bs I_\bt$, showing eventually that the ``Komar difference'' and ``non-Komar'' parts of the integral of $\bs I_\bt$ are equal up to a factor of $D-3$. This makes use of several properties of $\bt$ that result from it being proportional to the divergence of the PCKY tensor. 

For interest I also give the general results for the integrals of $\bs I_\chi$ for various values of the Killing vector $\chi$, on surface of constant time and spheroidal radius $r$, when $m \to \mu(r)$. In doing so I highlight the differences between the value of the energy and angular momentum in the exact solutions to Einstein's equations and in the more general setting with $m \to \mu(r)$. 

\section{Clarifying Comments} \label{clarifyingcomments}

In the $\La$CDM model, it is believed that $\La$ is a small positive quantity (estimated at $(2.846 \pm 0.076) \times 10^{-122} m_{Pl}^2$ in natural units by the 2018 Planck Collaboration \cite{Miville-Deschenes}). It is nevertheless satisfying to consider anti-de Sitter spacetimes due to the absence of a cosmological horizon, as well as the AdS/CFT (Conformal Field Theory) correspondence (introduced by Maldacena in \cite{Maldacena}). 

Why study higher dimensions? Partly, I want to address the results for higher-dimensional Kerr--AdS black holes that appear in the papers I am mentioning, particularly GPP and CGKP. Higher dimensions of course occur in many interesting fields, such as string theory. Another important factor is that I think that studying higher dimensions can give a more detailed understanding of the four-dimensional case. To what extent is something the result of the dimension being 4, and to what extent does it appear to be universal? There are some results which simplify in four dimensions---compare, for example, the ratio of angular momentum to energy (or enthalpy), which is the rotational parameter $a$ in four dimensions but has a much more complicated form in higher dimensions. Understanding this may help avoid overstating the importance of some results which occur only in four dimensions. Many results hold equally well for all $D \geq 4$. The distinction may turn out to be important for understanding why some $D = 4$ results occur, even for a reader otherwise uninterested in higher dimensions. 

Since the CKGP paper, fifteen years ago, there has been an explosion of papers on the burgeoning field referred to as Black Hole Chemistry (see \cite{KubiznakMannTeo,Mann} for reviews), furthering the interpretation of $\mc E$ as an enthalpy for a black hole, not just in the setting of Kerr--AdS but in more general settings. My focus in this thesis was on the black hole mechanics/geometric side, and so is somewhat orthogonal to the fascinating direction taken in the literature. (See also: \cite{Wei:2021lmo}.) 

As stated, my approach in considering the conserved quantities was in based on a modification of the BC method. There have been a number of other approaches in the literature, many of which are related in some way to the BC method and to the covariant phase space methods of Wald. A full survey is beyond my scope, but I want to make a few observations of works that share some commonalities with mine. Works such as \cite{HajianSheikh-Jabbari,Chrusciel,Blagojevic:2020edq,Jing:2017jxw} consider under what conditions Hamiltonian conserved charges (either through the Hamiltonian formalism or through the covariant phase space formalism) are \emph{integrable}, identifying integrability as the determiner of whether a conserved charge can be defined in a way that (as \cite{Blagojevic:2020edq} puts it) ``makes sense.'' It is pointed out by \cite{HajianSheikh-Jabbari,Chrusciel,Blagojevic:2020edq,Jing:2017jxw} that, in my notation, a conserved charge associated with $\xi$ is integrable, but one associated with $\bt$ is not. This statement comes down to the notion that for the Hamiltonian associated with a given Killing vector $\chi$ to be well-defined, it must satisfy \eqref{deHchiIntro}. Additionally, the right-hand side (and thus the left) must be a total differential in $\de$ for $H_\chi$ to be well-defined.  It turns out that this is the case for $\chi = \xi, \eta$ but \emph{not} $\chi = \bt$, which shows that the conserved quantity associated with $\bt$ is not integrable. The fact that such a conserved charge is not integrable essentially guarantees the failure of the first-law when using a conserved quantity which one would naively associate with $\bt$ (that is, $\mc F$). (\cite{Chrusciel} uses a different argument, not using the Iyer--Wald phase space formalism, but the argument is equivalent.) To summarize: one can use $\xi$ but not $\bt$ to base the first-law on because $H_\xi$ is integrable but $H_\bt$ is not, and this is well-known. 

My approach is similar in that it is also gives an argument for the non-integrability associated with the Killing vector $\bt$. The contribution that I believe to be unique is that, by using the Kerr--Schild form, I show that the conserved quantity associated with $\xi$ is integrable because there exist coordinates in which both the Kerr--Schild background $\bar g_{ab}$ and the contravariant components $\xi^a$ remain constant even when varying between different Kerr--AdS solutions (provided that $\La$ is unvaried). This singles out the nonrotating frame for a specific reason that I do not believe has been identified. (There is something close in \cite{Blagojevic:2020edq}, who observe that the background geometry found by setting $m = 0$ into the metric for Kerr--AdS in a rotating frame is $a$-dependent. They note that Henneaux and Teitelboim \cite{HenneauxTeitelboim85} originally formulated their conserved charges in a frame which had the usual spherical-polar form of AdS as background, and then note in their paper that it is sufficient to go to a frame in which the time and azimuthal coordinates take the same form as in \cite{HenneauxTeitelboim85}, but not necessarily the radial or latitude coordinates. Implicit there is the notion that the background ``cannot change too much'' for the charge to be integrable, which is the finding I present here.) 

Along related lines, the Smarr relation, and, relatedly, the first-law including the variation in $\La$, has been explored from a number of angles, many of which overlap very closely with my work here. Several authors take the approach where they allow $\La$ to vary, and then modify some methods for calculating conserved charges to account for that, finding the thermodynamic volume this way. In doing so they derive expressions for the thermodynamic volume (and, consequently the Smarr relation). In comparing different Kerr--AdS solutions with the same $\La$, by varying, say, $\mc E$ and $\mc J$ in four dimensions, there is the justification that the two spacetimes are different solutions to the same theory, with $\La$ being fixed in the theory. Conceptually, varying $\La$ (or $\tilde \La$) is more complicated. As pointed out by, for instance, Henneaux and Teitelboim \cite{HenneauxTeitelboim85}, varying $\La$ directly in the action of a black hole leads to a poorly-posed variational problem. There are ways around this---Henneaux and Teitelboim, for instance, point out that a gauge field of dimension $D-1$ can be added to the action and can replicate the equations of motion associated with a nonzero $\La$. Further work considering $\La$ as being related to a type of charge associated with a gauge field and potential is pursued in \cite{Chernyavsky,HajianOzsahin}, which is intriguing but is not our approach. In these cases, the gauge of the field associated with $\La$ ``is fixed if one demands that the mass, angular momentum, and other charges be reproduced correctly by the covariant formulation of charges,'' \cite{HajianOzsahin} so that there is a similar ambiguity to the ambiguity in the gauge of the Killing potential $\bs \om_\chi$.   The difference between the the thermodynamic and geometric volumes is similarly pointed out in \cite{Chernyavsky} to be fixed by demanding that the entropy be integrable.

While the CGKP paper treats $\tilde \La$ as a dynamical variable, my approach is, \emph{for the most part}, to keep $\tilde \La$ fixed and to show how it enters into the Smarr relation. This is done in part to sidestep the issues that arise from varying $\tilde \La$. These two methods (keeping $\tilde \La$ fixed while checking the Smarr relation, or varying $\tilde \La$) are equivalent to an extent, in that the $\Th$ or $V_{th}$ which appears in the first law with a varying $\tilde \La$ is necessarily via the scaling argument the same $\Th$ or $V_{th}$ appearing in the Smarr relation. This being the case, I generally use the term ``energy'' rather than ``enthalpy'' to describe the conserved quantity associated with the asymptotically static Killing vector, and think of $\tilde \La$ and thus $\La$ as a true constant. Part of the reason for this is that the Kerr--Schild decomposition method works best when the Kerr--Schild background spacetime is fixed, and the Kerr--Schild breakdown of Kerr--AdS is of a background pure AdS spacetime with a given $\La$ plus a perturbation. Additionally, the covariant methods of Wald imply that $\mc E$ is the Hamiltonian associated with the asymptotically static Killing vector and thus the term energy is appropriate for this. My using the term energy to describe $\mc E$ is not meant to be a ruling out of the interpretation of enthalpy, particularly in the case of varying $\ti \La$, but to reflect the way my focus was on the first law with unvarying $\ti \La$ and on the Smarr relations, with varying $\ti \La$ coming only as a consequence. 

Other methods that address a variable $\La$ include the ``off-shell ADT (Abbott--Deser--Tekin)'' formalism \cite{Hyun:2017nkb} and the Extended or Modified Iyer--Wald formalism (e.g.~\cite{Gao,Campos25,Xiao,Xiao25,Couch}). \cite{Gao} gives a formula for the expected $\mc V$ which appears in the variation law, but notes that it must be modified to make charges integrable by what amounts to a gauge term. There are also several very recent papers \cite{Gao,Campos24,Campos25,Campos26} which examine conserved charges based on a vector defined by $\bt/\sqrt{\Xi}$ where $\Xi = 1-a^2/l^2$ (in four dimensions), where $a$ is the rotation parameter and $l$ is the AdS radius, and they demonstrate that when constructed from $\bt/\sqrt{\Xi}$, the first law is satisfied \emph{and} the thermodynamic and vector volume associated with $\bt/\sqrt{\Xi}$ coincide; \cite{Campos24,Campos25,Campos26} refers to this as the Alternative Thermodynamic Theory. That the thermodynamic and geometric volumes coincide is essentially a consequence of the vector volume's appearance in the Smarr relation associated with $\bt$. I do not consider charges constructed out of $\bt/\sqrt{\Xi}$, though it is an intriguing avenue of study.

\cite{Xiao} points out relationships between the Komar integrals and the geometric volume, and identifies the difference between the thermodynamic volume and geometric volume as being related to the difference between the actual energy and the energy as calculated through the Komar integral, which is a similar argument to the one I lay out here. In \cite{Xiao25}, Xiao et al.~provide an explicit formula for the thermodynamic volume using the extended Iyer--Wald approach. Very recently, Campos et al.~\cite{Campos24,Campos25,Campos26} identify the vector volume as appearing naturally in the volume term which appears in the first-law (with varying $\La$) and in the Smarr relation, and relate it to Komar integrals. Another recent approach to the Smarr relation is in \cite{Golshani} in terms of Noether charges, reproducing the Smarr relation from Barnich and Comp\`ere \cite{BarnichCompere}.

What I believe to be novel in my work is that I identify a reason why the geometric volume appears in the Smarr relation associated with $\bt$ but not with $\xi$ by showing that the terms associated with the difference in the Komar integrals between the full and background spacetime for $\bt$ match up with an ``extra term'' which appears in $\bs I_\bt$ and thus in $\mc F = \oint \bs I_\bt$, due to simplifications related to the existence of the Principal Conformal Killing--Yano tensor.

While I have made an effort to reference works that have some overlap with mine, black hole thermodynamics for Kerr--anti-de Sitter is an active area of research and I may have made some oversights. These are not intentional, and comments are welcome on gaps in my description of the literature. 

\section{Thesis Organization} 

The thesis is organized as follows. The first few chapters introduce important background material and provide new definitions, and the final chapters apply them to address 
the ``open questions'' listed in Section \ref{openquestions}. 

Chapter \ref{PRDPaper}, originally published as \cite{Ballik}, introduces the vector volume, defined as a quantity which measures the rate of growth of $D$-dimensional volume in a $D$-dimensional region $\mc R$ along the flow of a vector field. I show that this can be written as $\int_\Si v^a d \Si_a$, where $\Si$ is a $(D-1)$-dimensional region spanning $\mc R$ and $v$ is the vector. I show that this quantity is independent of the particular choice of $\Si$ when $v^a$ is divergence-free and parallel to the boundary of $\mc R$. Since all Killing vectors are divergence-free, it is natural to use a Killing vector for the vector volume. I show that for axially symmetric regions the vector volume based on the azimuthal Killing vector vanishes identically. I show that my black hole vector volume, using the canonically normalized time Killing vector, is equal to CGKP's geometric volume. 

Chapter \ref{GKAdSChapter} reviews the Kerr--anti-de Sitter spacetimes in arbitrary dimension $D \geq 4$, states a natural generalization by letting the mass parameter $m$ vary with spheroidal radius $r$ as a function $\mu(r)$, and then goes on to list several important properties of these spacetimes. These spacetimes can be written in Kerr--Schild form, with a background anti-de Sitter metric and with Kerr--Schild vector $k$, and also admit the existence of a Principal Conformal Killing--Yano tensor $\bs h$. The divergence of $\bs h$ (up to a numerical constant) is given by the Killing vector $\bt$, which is called the principal vector of the spacetime. I list some important properties of these spacetimes which will be needed in our later analysis. Finally I write these metrics in what I call pseudo-Cartesian coordinates, which I will use for volume calculations later on.

Chapter \ref{NoetherChapter} reviews the papers by GPP and CGKP. I list a few questions that are raised by these papers which I would like to address. In order to discuss the conserved charges of energy and angular momentum, and their relationship to black hole mechanics, I review the covariant phase space formalism of Wald. Then I discuss a paper by Barnich and Comp\`ere (BC) \cite{BarnichCompere} which proposes a method for integrating through solution space to find conserved charges which automatically satisfy a black hole variation law, and which can be used to produce a Smarr relation. I then apply the BC method to the particular case of a Kerr--Schild spacetime and in doing so define a $(D-2)$-form $\bs I_\chi$ associated with each Killing vector $\chi$. I show that under certain circumstances, the integral of the variation of this quantity over a $(D-2)$-surface gives the variation in energy or angular momentum, or $\kappa \de A/8\pi$ (depending on the value of $\chi$). By showing how $\bs I_\chi$ involves the Komar integral of $\chi$ in the background spacetime and then relating this background spacetime Komar integral to the vector volume of the black hole, I show how the black hole vector volume comes into the conserved charges in a natural way. I list some further properties of the $\bs I_\chi$ differential form, including a modification in the presence of electric charge and the relationship of $\bs I_\chi$ to the stress--energy tensor. 

Chapter \ref{VolumeAreaChapter} describes some important results about volume and area. Here I introduce integrals which provide the higher-dimensional volumes and areas of unit balls or spheres. I calculate explicitly the area of the black hole event horizon for the Generalized Kerr--AdS spacetime in higher dimensions. I use the pseudo-Cartesian coordinates introduced in Chapter \ref{GKAdSChapter} to show that we can interpret $\mc V_{\xi,\mc B}$ as being the Euclidean volume of the ellipsoid which constitutes the black hole region in the spatial part of the pseudo-Cartesian coordinates. I show how the vector volume appears in the use of Euclidean methods in GPP and resolve an open question from the CGKP paper by showing why the vector volume of black holes enters naturally when CGKP's method for constructing a Killing potential is used. Finally I use the properties of the PCKY tensor $\bs h$ to explain the fact that the black hole vector volume is equal to the black hole surface area multiplied by the spheroidal radius at the horizon divided by $D-1$.

Chapter \ref{ExplicitGKAdSChapter} applies the results of Chapter \ref{NoetherChapter} by examining in detail $\bs I_\chi$. After calculating $\bs I_\chi$ for Kerr--AdS associated with the asymptotically static Killing vector, the azimuthal symmetry Killing vectors, and the asymptotically rotating vector $\bt$, I show that these results correspond to those found by GPP and CGKP. I calculate the solutions in four dimensions in the more general setting with $m \to \mu(r)$ and calculate their variation, checking under what circumstances a first law relationship is recovered. I check under what circumstances the formula $\oint_H \bs k_{\bar \z}[\de g;g] = \kappa \de A/8\pi$ can be used. 

I argue that some of the limitations of the use of $\de \bs I_\chi = \bs k_\chi^{EH}[\de g;g]$ come down to the fact that this statement holds exactly only if both the contravariant components of $\chi^a$ and the covariant components of the \emph{background} metric $\bar g_{ab}$ are unchanging under the perturbation, and show that if the rotational parameter $a$ varies that these do not hold, even if the background spacetime remains anti-de Sitter. I also point out that the statement $\oint_H \bs k^{EH}_{\bar \z}[\de g;g] = \kappa \de A/8\pi$ relies on the black hole horizon remaining at the same coordinates, which is not in general true. To compensate for these I use the Lie derivative along a vector corresponding to the difference between two sets of coordinates to relate the coordinates being varied to a modified set of coordinates where (respectively) the background metric or the location of the horizon are kept fixed. I use this to demonstrate how we can still use the variation law with the conserved quantities based on $\bs I_\chi$ even in certain cases where $\de \bs I_\chi \neq \bs k_\chi^{EH}[\de g;g]$, and use $\oint_H \bs k_{\bar \z}^{EH}[\de g;g] = \kappa \de A/8\pi$ even when the horizon radius changes. I argue that the failure of the first law when constructed using the asymptotically rotating frame is a consequence of the fact that we cannot find a set of coordinates where both the AdS background metric $\bar g_{ab}$ \emph{and} the vector $\bt^a$ are unvarying in a perturbation (if the rotational parameter $a$ varies). 

In arbitrary dimension I evaluate $\oint_C \bs I_\chi$ for various values of $\chi$ where $C$ is a $(D-2)$-surface of constant time coordinate $t$ and spheroidal radius $r$. I show how a number of the properties of $\bt$ introduced in Chapter \ref{GKAdSChapter} significantly simplify the expression for $\bs I_\bt$ and how, in vacuum, this causes the Komar-difference and non-Komar parts of $\oint_C \bs I_\bt$ to be equal up to a factor of $D-3$, thus addressing why the geometric volume arises naturally in the Smarr relationship based around $\bt$ rather than around the asymptotically static vector. I conclude the chapter by describing a method of using the stress--energy tensor of a time-varying spacetime to define conserved quantities in such a way that they automatically satisfy the first law. 

Chapter \ref{conclusionchapter} concludes and proposes future work. (An additional note, written since the submission of the thesis, appears as Chapter \ref{additionalNote}.) The bibliography and appendices follow.

\chapter{The Vector Volume of Black Holes: Paper} \label{PRDPaper}

The following was originally published as \cite{Ballik} by the author of this thesis along with Dr.~Kayll Lake. The formatting has been changed to present it as a chapter in a larger work. The notation/conventions are different from in the larger document: most notably, Greek letters are used for abstract indices (following \cite{Poisson}) whereas Roman characters are used in the rest of this thesis (following \cite{Wald84}). There are a few minor typographical corrections made to the text of the paper, and one footnote deleted, but otherwise the content is unchanged. A few a additional notes are included in brackets. 

Throughout this chapter we use Greek indices to go from 0 to $D-1$, where $D$ is the dimensionality of the spacetime; Roman indices usually go from $1$ to $D-1$ but should be clear in context.

An addendum concerning the vector volume associated with an azimuthal vector field, not appearing in the original paper and written at the time as the rest of the thesis, is included as Section \ref{addendum}.

\section{Abstract}

By examining the rate of growth of an invariant volume $\mathcal V$ of some spacetime region along a divergence-free vector field $v^\alpha$, we introduce the concept of a ``vector volume" $\mathcal{V}_v$. This volume can be defined in various equivalent ways.  For example, it can be given as $\mathrm d \mathcal V(\mu) / \mathrm d \mu$, where  $v^\alpha \partial_\alpha = \mathrm d / \mathrm d \mu$, and $\mu$ is a parameter distance  along the integral curve of $v$. Equivalently, it can be defined as $\int v^\alpha \mathrm d \Sigma_\alpha$, where $\mathrm d \Sigma_\alpha$ is the directed surface element.  We find that this volume is especially useful for the description of black holes, but it can be used in other contexts as well. Moreover, this volume has several properties of interest.  Among these is the fact that the vector volume is linear with respect to the the choice of vector $v^\alpha$.   As a result, for example, in stationary axially symmetric spacetimes with timelike Killing vectors $t^\alpha$ and axial symmetric Killing vectors $\phi^\alpha$, the vector volume of an axially symmetric region with respect to the vector $t^\alpha + \Omega \phi^\alpha$ is equal for any value of $\Omega$, a consequence of the additional result that $\phi^\alpha$ does not contribute to $\mathcal{V}_v$. Perhaps of most interest is the fact that in Kerr--Schild spacetimes the volume element for the full spacetime is equal to that of the background spacetime. We discuss different ways of using the vector volume to define volumes for black holes.  Finally, we relate our work to the recent wide-spread thermodynamically motivated study of the ``volumes" of black holes associated with nonzero values of the cosmological constant $\Lambda$.

\section{Introduction}

It is well known (e.g.~\cite{Poisson}) that, following the pioneering work of Bardeen, Carter and Hawking \cite{BCH}, the ``surface area'' of black holes ($\mathcal{A}$) is of fundamental interest.  It is non-decreasing through classical (non-quantum) processes and is thus generally associated with the black hole entropy.  This raises the naive question as to whether or not the ``volume" of black holes is important. This question has seen little interest until recently. With the surge in interest in the cosmological constant $\Lambda$, since $\Lambda$ is, naively, a ``pressure" term, one can certainly ask where the ``$PdV$" term went in the first law of black hole thermodynamics. There is now wide-spread interest in this ``$V$" term, e.g.~\cite{CaldarelliCognola, KastorEtal:2009, Dolan:2010, Dolan:2011a, Dolan:2011b, Dolan:2012, Cvetic, LarranagaCardenas:2012, LarranagaMojica:2012,Gibbons:2012, KubiznakMann:2012, GunasekaranEtal:2012, BelhajEtal:2012,  LuEtal:2012, SmailagicSpallucci:2012, HendiVahinidia:2012, DolanKastor}.  However, while the surface area is well-defined and invariant for a $D$-dimensional black hole, the volume intrinsic to a $(D-1)$-dimensional hypersurface depends on the choice of slicing, and the full $D$-dimensional volume of the black hole region is formally infinite.

The attempt to find a definition of volume which is both invariant and finite is the inspiration for this work in which we define a ``vector volume''. This can be thought of as the rate of change of the invariant four-volume of some region along a vector field.  The vector volume turns out to be a generalization of several other definitions of volume in use by other authors.  Moreover, the vector volume has many interesting properties which we examine here.  We show, for example, that in certain ``Kerr--Schild" cases, the vector volume reduces to the Euclidean volume of a subspace region. We also offer a new definition of the surface gravity as the ratio of two vector volumes.

The paper [chapter] is structured as follows: In Section \ref{review}, we review recent developments in which various authors working independently have used similar expressions for black hole volume.  In Section \ref{vecvolintrodef} we define the vector volume in several different ways which we show to be equivalent.  In Section \ref{generalproperties} we show some of the vector volume's interesting properties.  Section \ref{kerrschildtype} discusses the vector volume in Kerr--Schild-type metrics.  Section \ref{canon} discusses a ``canonical black hole volume'' based on the vector volume using the time-translation Killing vector in stationary black holes which is closely related to work from Parikh \cite{Parikh} and Cveti\v{c} et al. \cite{Cvetic}  Section \ref{ourvolume} discusses a ``null generator volume'' for stationary black holes, first defined in \cite{BallikLake10}, and how this can be used alongside the canonical black hole volume to define the surface gravity of black holes in a novel way.  Section \ref{otherauthors} further explores the connection between the vector volume and other works, concentrating on those of Cveti\v{c} et al.~\cite{Cvetic} and Hayward \cite{Hayward}.  Section \ref{PRDconclusion} summarizes our analysis.

We work largely in dimension $D = 4$ for clarity, though in some cases we keep the dimension $D$ of the spacetime general. Most of what follows is generalizable to higher dimensions though this generalization is not our principal concern here. [The thesis, as a whole, \emph{is} concerned with arbitrary-dimension $D$.]

\section{Review of Some Recent Developments} \label{review}

Here we review several independent recent volume definitions for black holes.  It turns out that these are all special cases of a more general vector volume which we present in this paper.  The work is from Parikh \cite{Parikh}, Cveti\v{c} et al.~\cite {Cvetic}, and Hayward (e.g.~\cite{Hayward}).  For the most part we use the notation used by the authors in their papers.  We note that the authors of this paper have independently developed a definition of volume which is in a similar vein to these \cite{BallikLake10}.

\subsection{Parikh Volume} \label{Parikhintro}

In 2006, Parikh \cite{Parikh} defined a volume for stationary black holes.  In his paper, Parikh begins by considering a $D$-dimensional spherically symmetric spacetime with a timelike Killing vector and a horizon, with line element 
\begin{equation}\label{metricone}
\mathrm d s^2 = -\alpha(r) \mathrm d t_s^2 + \frac{\mathrm d r^2}{\alpha(r)} + r^2 \mathrm d \Omega_{D-2}^2(\vec x)
\end{equation}
where $\mathrm d \Omega_{D-2}$ is the line element for the $(D-2)$-sphere, $r$ is the aerial radius, $\vec x$ is a vector representation of the coordinates on the $(D-2)$-sphere, $t_s$ is the static time coordinate and $\alpha(r)$ is some function of $r$.  The Killing vector $\partial_{t_s}$ is the Killing vector corresponding to staticity, which is timelike outside the horizon.  The horizon is at a radius $r = r_+$ for which $\alpha(r) = 0$.  At this value, the metric is non-regular, so Parikh introduces a new time coordinate $t$ defined by
\begin{equation}
t_s = \lambda t + f(r,\vec x).
\end{equation}
$\partial_t$ will be a Killing vector for any constant $\lambda$, but in order to preserve the time orientation and asymptotic normalization of the Killing vector, $\lambda$ is set to +1 in Parikh's definition, so that $\partial_t = \partial_{t_s}$.  The advantage to this new coordinate $t$ over $t_s$ is that for certain functions $f$ it is possible to obtain a slicing that extends through the horizon.

Parikh then notes that while the $(D-1)$-dimensional volume of the region $0 \leq r \leq r_+$ on the hypersurfaces of constant $t$ depends on the choice of function $f$, one can instead define a ``differential spacetime volume'' $\mathrm d \mathcal V_D$ which is invariant.  This differential spacetime volume is the $D$-volume of the region where $t'$ varies between $t$ and $t + \mathrm d t$:
\begin{equation}
\mathrm d \mathcal V_D = \int_t^{t + \mathrm d t} \mathrm d t' \int_0^{r_+} \mathrm d r \int d^{D-2}x \sqrt{- g_D}.
\end{equation}
Here, $g_D$ is the determinant of the full $D$-dimensional metric.  Since $\partial_t$ is a Killing vector, the metric $g_{\alpha \beta}$ is independent of $t$, and thus $t$ enters into $\mathrm d \mathcal V_D$ only through the multiplicative term $\mathrm d t$. The Parikh volume is defined as the ratio of this differential spacetime volume to $\mathrm d t$:
\begin{equation}
\mathcal V_{P} \equiv \frac{\mathrm d \mathcal V_D}{\mathrm d t} = \int \mathrm d^{D-1}x \sqrt{-g_D} \label{Parikhdef}
\end{equation}
where $\mathrm d^{D-1}x$ is the product of the differentials except for $\mathrm d t$.  Essentially, one uses $g_D$ instead of $g_{D-1}$ (the determinant of the metric of the $t = const.$ hypersurfaces), which makes the volume constant in time for all choices of Killing time, and invariant under the choice of stationary time slices.

Though he used static spherical symmetry as an example, Parikh notes that his volume definition can similarly be applied to any stationary black hole. In particular, Parikh notes that the volume for static, spherically symmetric black holes in 4 dimensions, as before with horizon at $r = r_+$, is
\begin{equation}
\mathcal V_{P} = \frac{4\pi}{3} r_+^3,
\end{equation}
which is of course simply the Euclidean volume for a sphere of radius $r_+$. The volume for the (four-dimensional) Kerr black hole is given by Parikh as
\begin{equation} \label{vpk}
\mathcal V_{P} = \frac{4\pi}{3} r_+(r_+^2 + a^2)
\end{equation}
where $r_+$ and $a$ have their usual meanings as the value of radius $r$ at the outer horizon and rotation parameter respectively.

Before continuing, we make a further note about the Parikh volume.  Using the well-known horizon area $\mathcal{A}=4 \pi (r_{+}^2+a^2)$ we can write (\ref{vpk}) in the form
\begin{equation} \label{vpa}
\mathcal V_{P} = \frac{r_{+} \mathcal{A}}{3},
\end{equation}
a result we revisit in future sections.

\subsection{Geometric Volume} \label{cveticgeometricvolume}

The laws of black hole thermodynamics in spherical symmetry with a nonzero cosmological constant term $\Lambda$, as well as the generalized Smarr formula, were considered by \cite{KastorEtal:2009}. \cite{Cvetic} extended this to include rotation and we follow \cite{Cvetic}. (See as well the references therein and subsequent work \cite{LarranagaCardenas:2012, LarranagaMojica:2012,Gibbons:2012, KubiznakMann:2012, GunasekaranEtal:2012, BelhajEtal:2012,  LuEtal:2012, SmailagicSpallucci:2012, HendiVahinidia:2012, DolanKastor}.)  The argument goes in essence as follows.  In black hole spacetimes with $\Lambda$ [actually the value satisfying $R_{ab} = \La g_{ab}$, and so $\tilde \La$ in the notation of the rest of this thesis], the black hole thermodynamic variation laws can be written in terms of a black hole enthalpy $E$, giving rise to a modified first law of thermodynamics,
\begin{equation}
\mathrm d E = T \mathrm d S + \sum_i \Omega_i \mathrm d J_i + \sum_\alpha \Phi_\alpha \mathrm d Q_\alpha + \Theta \mathrm d \Lambda,
\end{equation}
or, in non-differential Smarr--Gibbs--Duhem form,
\begin{equation}\label{sgd}
E = \frac{D-2}{D-3}\left(T S + \sum_i \Omega_i J_i\right) + \sum_\alpha \Phi_\alpha Q_\alpha - \frac{2}{D-3} \Theta \Lambda,
\end{equation}
where $T$ is the effective temperature of the black hole, $S$ is the entropy, $J_i$ are the components of the angular momenta, $\Omega_i$ are the corresponding angular velocities, $Q_\alpha$ are the conserved charges, $\Phi_\alpha$ are the potentials corresponding to those charges, and $\Theta$ is the conjugate to $\Lambda$.  Since $\Lambda$ can be interpreted as a pressure (up to a multiplicative constant), $\Theta$ is interpreted as being proportional to a volume for the black hole, by analogy with the classical thermodynamical first law for enthalpy $H$ in terms of temperature $T$, entropy $S$, pressure $P$, volume $V$ and non-$Pd V$ work $W$,
\begin{equation}
\mathrm d H = T \mathrm d S  - \delta W + V \mathrm d P.
\end{equation}
This yields a relationship between a ``thermodynamic" volume $\mathcal V_{th}$ and $\Theta$:
\begin{equation}
\mathcal V_{th} = - \frac{16 \pi \Theta}{D-2},
\end{equation}
where $D$ is the dimension of the spacetime. With spherical symmetry, $\mathcal V_{th}$ corresponds to the ``naive" geometrical volume
\begin{equation}
\mathcal V_{geo} = \int \mathrm d r \int \mathrm d \Omega \sqrt{- g_D} \label{Vgeo}
\end{equation}
where $r$ ranges over the black hole and $\mathrm d \Omega$ is the surface element on the unit $(D-2)$ sphere. For black holes with nonzero angular momentum, the thermodynamic and geometric volumes differ by \cite{Cvetic}
\begin{equation}\label{vthermo}
\mathcal V_{th} - \mathcal V_{geo} =  \frac{8 \pi}{(D-1)(D-2)} \sum_i a_i J_i,
\end{equation}
where $a_i$ are the rotational parameters for the black hole corresponding to the $J_i$.  The geometric volume satisfies the relation
\begin{equation}
\mathcal V_{geo} = \frac{r_+ \mathcal{A}}{D-1} \label{VrA1}
\end{equation}
for all black holes of the Kerr--Newman--de Sitter family, generalizing (\ref{vpa}). Again, we return to this below. Of central importance here is the fact that
\begin{equation}
\mathcal V_{geo} = \mathcal V_{P} \label{geop}.
\end{equation}
The geometrical and Parikh volumes are equivalent.

\subsection{Kodama Volume} \label{HaywardKodama}
In several papers (for example \cite{Hayward}, among others), Hayward defines a volume for dynamical black holes in terms of the Kodama vector, an analogue to the Killing vector in dynamical spacetimes.  He makes a similar development for cylindrical symmetry in \cite{Haywardcylinder}, but we will focus here on the spherical symmetry case as an example.

The line element for four-dimensional dynamic spherically symmetric spacetime can be written in the form
\begin{equation}
\mathrm d s^2 = g_{A B} \mathrm d x^A \mathrm d x^B + r^2 \mathrm d \Omega_2^2 \label{Haywardmetric}
\end{equation}
where there are two coordinates $x^A$ in addition to the coordinates $(\theta,\phi)$ within the 2-sphere metric $\mathrm d \Omega_2^2$.  Here, $r(x^A)$ is the aerial radius.

An analogue to the Killing vector in dynamic spherical symmetry is the Kodama vector, which we will label by $K^\alpha$.  An important property of the Kodama vector is that it becomes null on and only on the trapping horizon of a dynamic black hole, a property that is analogous to the Killing vector becoming null on and only on the Killing horizon of a stationary black hole.  The Kodama vector $K^\alpha$ is defined as the curl of the aerial radius,
\begin{equation}
K^\alpha = \epsilon^{\alpha \beta} \nabla_\beta r, \label{Kodamadef}
\end{equation}
where $\epsilon^{\alpha \beta}$ is the volume form associated with the 2-metric $g_{A B}$ from \eqref{Haywardmetric}, or
\begin{equation}
\epsilon^{\alpha \beta} = \epsilon^{A B}\delta^\alpha_A \delta^\beta_B
\end{equation}
where $\epsilon^{A B}$ is the Levi-Civita tensor for the two dimensions $x^A$ in \eqref{Haywardmetric}.  The Kodama vector agrees with the usual timelike Killing vector in stationary spherically symmetric spacetimes if $K^\alpha$ commutes with $\nabla^\alpha r$. In these cases the line element can be written as
\begin{equation}
\mathrm d s^2 = -\left( 1 - \frac{2 E(r,t)}{r}\right) \mathrm d t^2 + \left( 1 - \frac{2 E(r,t)}{r}\right)^{-1} \mathrm d r^2 +  r^2 \mathrm d \Omega_2^2 \label{Eofrmetric}
\end{equation}
where the Kodama vector is $t^\alpha = \delta^\alpha_t$, which is also a Killing vector if $\pa E/\pa t = 0$.  In general, if $E(r)$ is allowed to vary with $t$ in line elements of the form $\eqref{Eofrmetric}$ then the Kodama vector is $K = \partial_t$, though it is obviously only a Killing vector if the line element is $t$-independent.

The Kodama vector (in spherical symmetry only) has the property
\begin{equation}
\nabla_\alpha K^\alpha = 0,
\end{equation}
which, along with the Gauss theorem (see, for example, \cite{Poisson}), implies a conserved quantity Hayward defines to be the volume,
\begin{equation}
\mathcal V_{K} = \left|\int_\Sigma K^\alpha \mathrm d \Sigma_\alpha\right| \label{HaywardV}
\end{equation}
where $\Sigma$ is a spacelike hypersurface and $\mathrm d \Sigma_\alpha$ is the volume element of the surface times a future directed normal.~(The absolute value signs are to avoid having to deal with the sign of the result.)  If the horizon of a black hole is located by $r = r_+$, then the volume can be defined as \eqref{HaywardV} with $\Sigma$ being the  region $r \leq r_+$, with result $\mathcal V_{K}=4 \pi r_+^3 / 3$. In the spherically symmetric case and with the Kodama vector the usual timelike Killing vector then
\begin{equation}
\mathcal V_{K} = \mathcal V_P = \mathcal V_{geo} \label{kc}.
\end{equation}

\subsection{Null Generator Volume} \label{OurVolumeIntro}

\begin{figure}[ht]
\centering 
\epsfig{file=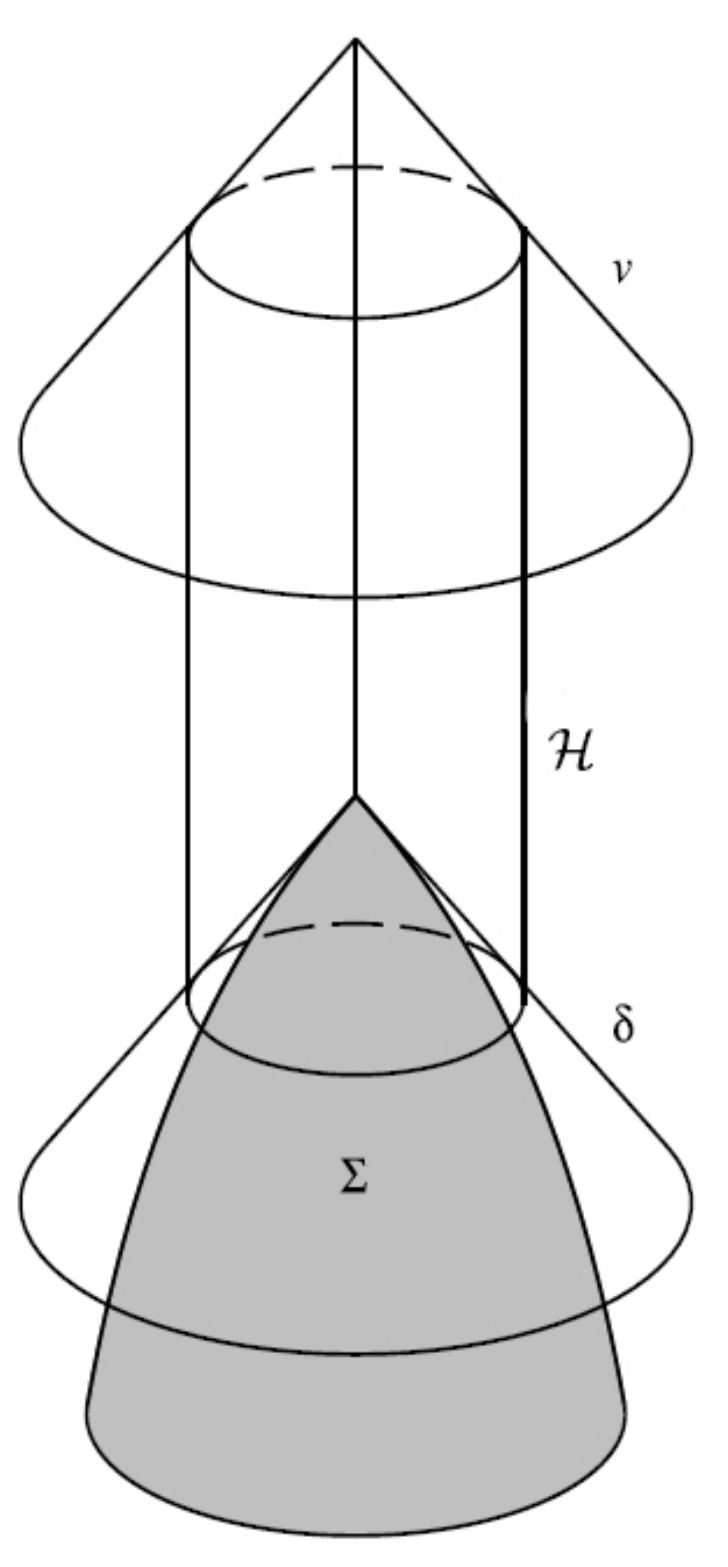,height=3.5in,width=2
in,angle=0}
\caption{The collapse of a timelike boundary surface $\Sigma$ that terminates at the central singularity simultaneously with the null cone $\delta$ and produces a black hole with horizon $\mathcal H$.  The null cone $v$ is any null cone to the future of $\delta$.  The invariant four-volume $\mathcal V$ calculated here is bounded by $\delta$ and $v$ and is to the interior of $\mathcal H$. [Clarification: the details of this are in \cite{BallikLake10}, but the idea is that for one of the calculations in \cite{BallikLake10}, as depicted here, we considered the formation of a Schwarzschild black hole by the collapse of a timelike boundary $\Si$. We considered the four-volume bounded by two sets of ingoing light cones, the first of which is labelled by $\de$, and intersects $\Si$ at $r = 0$, so that all ingoing light cones to the future of $\de$ terminate at the curvature singularity at $r=0$.]} \label{diagram}
\end{figure}

In \cite{BallikLake10}, we defined a volume rate for stationary non-degenerate black holes.  To define this rate, we considered a region of the black hole bounded by the event horizon and two distinct ingoing null cones, as shown schematically in Figure~\ref{diagram}. Then, by allowing the intersection of the ingoing null cone with the event horizon to vary, it was found that the four-volume of this region grows as $\mathcal V \propto \ln \lambda$, where $\lambda$ is the affine parameter on the null generators of the horizon and the constant of proportionality is the volume $\mathcal{V}_{P}$ divided by the surface gravity $\kappa$ of the black hole.  Thus we define the null generator volume as
\begin{equation}\label{firstkappa}
\mathcal V_{\mathcal{N}} \equiv \frac{\mathrm d \mathcal V(\lambda)}{\mathrm d \ln \lambda}=\frac{\mathcal{V}_{P}}{\kappa}.
\end{equation}

The easiest way to demonstrate this volume is to use ingoing Eddington--Finkelstein-like coordinates. For metrics of the form (\ref{metricone}) (dropping the superscript $s$ on $t$) define $\mathrm d V \equiv \mathrm d t + \mathrm d r / \alpha(r)$, from which the line element becomes
\begin{equation}
\mathrm d s^2 = - \alpha(r) \mathrm d V^2 + 2 \mathrm d V \mathrm d r + r^2 \mathrm d \Omega_{D-2}^2.
\end{equation}
Here, $V = const.$ labels sets of ingoing null geodesics.  The $D$-volume of the region between $V = V_0$ and a larger (arbitrary) value of $V$ is
\begin{eqnarray}
\mathcal V = \int_{V_0}^V \mathrm d V' \int_0^{r_+} \mathrm d r \int \mathrm d \Omega_{D-2} \sqrt{-g}\\ \nonumber = (V - V_0) \int \mathrm d^{D-1} x \sqrt{-g},
\end{eqnarray}
and so
\begin{equation}
\frac{\mathrm d \mathcal V}{\mathrm d V} = \int \mathrm d^{D-1} x \sqrt{-g}.
\end{equation}
Assuming asymptotic flatness ($\alpha(r) \to 1$ as $r \to \infty$), the relationship between $t$ and thus $V$ and an affine parameter $\lambda$ on the event horizon is given by \cite{Poisson}
\begin{equation}
\frac{\mathrm d V}{\mathrm d \lambda} = \frac{1}{\kappa \lambda}
\end{equation}
where $\kappa$ is the surface gravity of the black hole; in this case then $V = \ln \lambda / \kappa$ up to an additive constant, and we can write
\begin{equation}
\frac{\mathrm d \mathcal V}{\mathrm d \ln \lambda} = \kappa^{-1} \int \mathrm d^{D-1} x \sqrt{-g}
\end{equation}
as claimed.

This procedure can be generalized for other non-degenerate stationary metrics. For example, in Kerr--Newman, we can use ingoing coordinates in which the ingoing principal null geodesics have all coordinates constant except $r$.

\section{Vector Volume} \label{vecvolintrodef}

In this section we define a vector volume rate (which we also refer to as the ``vector volume") in any space or spacetime with respect to any divergence-free vector field $v^\alpha$ ($\nabla_\alpha v^\alpha = 0$). The set of divergence-free vectors of course includes all Killing vectors. The vector volume of a $D$-dimensional region $\mathcal R$ with respect to $v^\alpha$ is written as  $\mathcal V_{v}$. The vector $v^\alpha$ must satisfy $v^{\alpha}n_{\alpha}=0$ where $n_{\alpha}$ is normal to the boundary of $\mathcal R$. We will now introduce two definitions of $\mathcal V_{v}$.  Section \ref{differentiallike} defines it as a derivative of the $D$-volume of region $\mathcal R$ along the vector field $v^\alpha$.  Section \ref{haywardlike} defines it as an integral of the vector $v^\alpha$ over a hypersurface.  These two definitions are then shown to be equivalent.  A third definition, which uses perhaps less familiar terminology, is included as Appendix \ref{vve}. [See also Section \ref{addendum} for caveats.]

\subsection{Definition 1} \label{differentiallike}

\begin{figure}[ht]
\centering
\epsfig{file=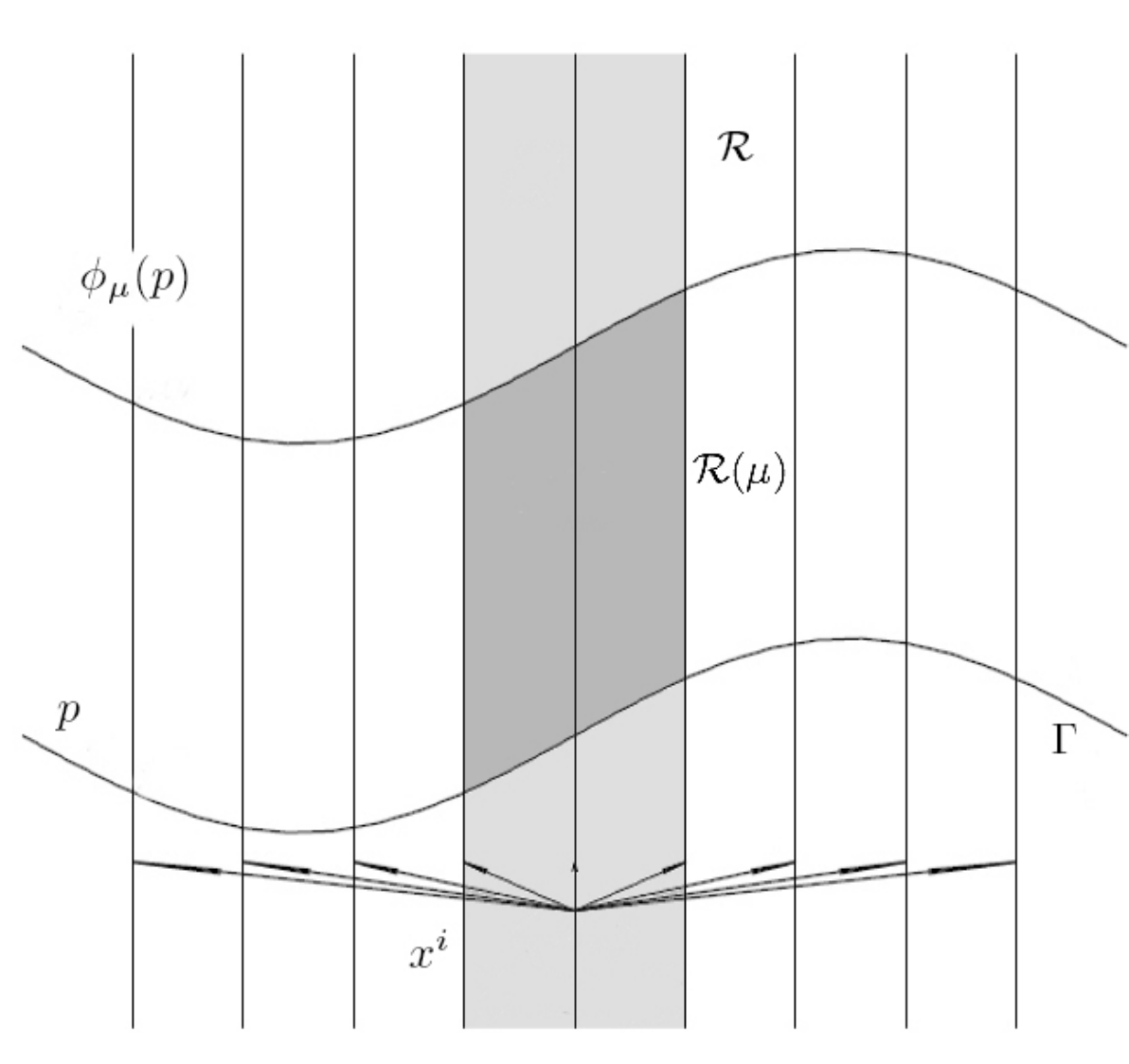,height=3in,width=3in,angle=0}
\caption{\label{diagram2} The vertical lines represent the congruence of integral curves of $v$.  The region $\mathcal R$ is shown in light grey and the specific region $\mathcal R(\mu)$ shown in darker grey.  The hypersurface $\Gamma$ is labelled and the other surface represents a movement a parameter distance $\mu$ along the integral curves. The coordinates $x^i=const$ are shown.}
\end{figure}

Define $\mathcal V(\mathcal R)$ as the $D$-dimensional volume of region $\mathcal R$. The essential point is to define a volume for which we use the derivative of the scalar volume along the vector field $v$:
\begin{equation}
v^\alpha \partial_\alpha ( \mathcal V(\mathcal R)).\label{meaningless}
\end{equation}
In order to provide a meaning to \eqref{meaningless} we must define $\mathcal V(\mathcal R)$ as a quantity which depends in some specific way on the coordinates. One method is to consider the congruence of integral curves of $v$.  Let $\phi_\mu(p)$ be the point lying at parameter distance $\mu$ along the integral curve of $v$ starting at point $p$.  Define $\Gamma$ as an arbitrary hypersurface of dimension $(D-1)$ which intersects every integral curve of $v$ exactly once.  Then $\Gamma \cap \mathcal R$ is a hypersurface region which lies entirely within $\mathcal R$ and intersects each integral curve within $\mathcal R$ exactly once.  Finally, define $\mathcal R(\mu)$ as the region which lies within $\mathcal R$, for which each point within $\mathcal R(\mu)$ can be expressed as
\begin{equation}
\phi_\nu(p), \;\;\;\; 0 \leq \nu \leq \mu, \;\;\;\; p \in  \Gamma \cap \mathcal R.
\end{equation}
In other words, $\mathcal R (\mu)$ is the subregion of $\mathcal R$ for which every point is at most a parameter distance of $\mu$ from $\Gamma$ along the integral curves of $v$. The situation is shown schematically in Figure \ref{diagram2}.

Since
\begin{equation}
 v^\alpha \partial_\alpha = \frac{\mathrm d}{\mathrm d \mu},
\end{equation}
from (\ref{meaningless}) we have
\begin{equation}
\mathcal V_{v} = \frac{\mathrm d}{\mathrm d \mu} \mathcal V(\mathcal R(\mu)). \label{derivativebased}
\end{equation}
Further, since $\mathcal V_{v}$ is, as we will show, a constant, this volume rate can be written as the ratio
\begin{equation}
\mathcal V_{v} = \frac{\mathcal V(\mathcal R(\mu))}{\mu}.
\end{equation}
These relations provide a simple interpretation of the vector volume: it is the growth rate of the $D$-dimensional volume $\mathcal R$ along the vector field $v$.  We show later in this section and in Section \ref{haywardlike} that $\mathcal V_{v}$ is independent of  $\mu$ and the particular choice of hypersurface $\Gamma$.

For ease of computation here we now use adapted coordinates $x^\alpha$ in which  $v^\alpha = \delta^\alpha_0$, and where $\Gamma$ is defined by $x^0 = 0$. The computation of $\mathcal V(\mathcal R(\mu))$ is now straightforward.  The region $\mathcal R(\mu)$ is simply equal to the subregion of $\mathcal R$ for which $0 \leq x^0 \leq \mu$. We can define this region as $\mathcal R(x^0)$ for $x^0 \geq 0$.  As a result, the vector volume can be written as

\begin{equation}
\mathcal V_{v} = \frac{\partial \mathcal V(\mathcal R(x^0))}{\partial x^0}.
\end{equation}

Since $\mathcal R$ is defined by the normal to its boundaries being perpendicular to $v^\alpha$, we can define $\mathcal R$ in these coordinates as $x^i \in \Sigma$ for some $(D-1)$-dimensional region $\Sigma$.  (This results in $n_\alpha$ having no $x^0$ component, confirming that $v^\alpha n_\alpha = 0$ as required.)  Then $\mathcal R(x^0)$ becomes the set of points for which $x^i \in \Sigma$ and the $x^0$ values which lie within the region vary between $0$ and the $x^0$ parameter within the expression.  $\mathcal V(\mathcal R(x^0))$ is given by the $D$-space integral,
\begin{eqnarray}
\mathcal V(\mathcal R(x^0)) = \int_0^{x^0} \mathrm d x^0 \int_\Sigma \sqrt{|g_D|} \mathrm d^{D-1} x   \\ \nonumber = x^0 \int_\Sigma \sqrt{|g_D|} \mathrm d^{D-1} x \label{VRx0}
\end{eqnarray}
where $g_D$ is the determinant of the metric $g_{\alpha \beta}$ and $\mathrm d^{D-1} x $ is the product of the differentials excepting $\mathrm d x^0$.  We note further that since $\nabla_\mu v^\mu = 0$, and using the relationship
\begin{equation}
\nabla_\mu A^\mu = |g_D|^{-\frac{1}{2}} \partial_\mu (|g_D|^{\frac{1}{2}} A^\mu)
\end{equation}
for an arbitrary vector field $A^\mu$, we find
\begin{equation}
\partial_\mu(|g_D|^{\frac{1}{2}} v^\mu) = \partial_0(|g_D|^{\frac{1}{2}}) = 0
\end{equation}
from which we find that the metric determinant is independent of coordinate $x^0$. [Note added, not in original paper: it is because $\pa_0 \sqrt{|g_D|} = 0$ and that $\Sigma$ does not depend on $x^0$ that we could take the $\int d x^0$ integral ``outside'' the $\Sigma$ integral in \eqref{VRx0}.]
This implies that the vector volume is expressible as
\begin{equation}
\mathcal V_{v} = \int_{x^i \in \Sigma} \sqrt{|g_D|} \mathrm d^{D-1} x, \label{gexpress}
\end{equation}
which is independent of $x^0$ and thus $\mu$. We also note that this is the same form as the definition for the Parikh volume \eqref{Parikhdef} and of the Cveti\v{c} et al.~geometric volume \eqref{Vgeo}.  Thus the vector volume is a generalization of the Parikh and geometric volumes. This connection will be further explored in Section \ref{canon}. We can show the invariance under the choice of slicing hypersurface $\Gamma$ here but it will be easier to do so in the following section.

\subsection{Definition 2} \label{haywardlike}

The second definition is similar to that used by Hayward to construct the Kodama volume as discussed in Section \ref{HaywardKodama} and given by (\ref{HaywardV}).  Recall there that $\Sigma$ slices the region $\mathcal B$ bounded by the black hole trapping horizon (whose normal is orthogonal to the Kodama vector [in the case where the spacetime is time-independent and so the trapping horizon is at constant $r$; in general $K^\alpha \pa_\alpha r = 0$]) and $\mathrm d \Sigma_\alpha$ is the directed surface element of $\Sigma$.  We note that the Kodama vector is, in certain situations, divergence-free.  We can generalize this definition to write the vector volume in a similar form. Replacing $K^\alpha$ with $v^\alpha$,  $\mathcal B$ with a region $\mathcal R$ whose normal is orthogonal to $v^\alpha$, and $\Sigma$ with $\Gamma \cap \mathcal R$, with $\Gamma$ as defined in the previous subsection, we have
\begin{equation}
\mathcal V_{v} = \int_{\Gamma \cap \mathcal R} v^\alpha \mathrm d \Sigma_\alpha \label{haywarddef}
\end{equation}
where here $\mathrm d \Sigma_\alpha$ is the directed surface element of $\Gamma$.  We note here that we can always choose the orientation of $\Gamma \cap \mathcal R$ such that the integral is positive.  We choose an adapted coordinate system wherein $v^\alpha = \delta^\alpha_0$, $\mathcal R$ is defined by $x^i \in \Sigma$, and $x^0 = 0$ corresponds to $\Gamma$. Then $\mathrm d \Sigma_\alpha$ is equal to $\delta^0_\alpha \sqrt{|g_D|} \mathrm d^{D-1} x$ where $\mathrm d^{D-1} x$ is the product of the differentials of $\mathrm d x^i$. The integral \eqref{haywarddef} becomes
\begin{equation}
\mathcal V_{v} = \int_{x^i \in \Sigma} \sqrt{|g_D|} \mathrm d^{D-1} x.
\end{equation}
This recovers \eqref{gexpress} and thus shows that \eqref{haywarddef} is equivalent to the definitions given in Section \ref{differentiallike}. For a parallel development using differential forms see Appendix \ref{vve}.

We now demonstrate that this volume is independent of the choice of particular hypersurface $\Gamma$.  To do this, we define a region $\mathcal Q$ which is bounded by two possible (nonintersecting) hypersurfaces $\Gamma$, say $\Gamma_1$ and $\Gamma_2$, as well as the boundaries of $\mathcal R$.  We now say that [the normal to] $\Gamma_2$ points ``outward'' and [the normal to] $\Gamma_1$ points ``inward,'' so that the closed integral of the vector field over the boundary of $\mathcal Q$ is given by
\begin{equation}
\oint_\mathcal{\partial Q}v^\alpha \mathrm d\Sigma_\alpha = \int_{\Gamma_2} v^\alpha \mathrm d \Sigma_\alpha - \int_{\Gamma_1} v^\alpha \mathrm d \Sigma_\alpha,
\end{equation}
where of course $v^\alpha \mathrm d \Sigma_\alpha$ is zero on $\partial \mathcal R$ since the normal is orthogonal to $v^\alpha$.  We note that the first term, by the divergence theorem, is equal to $\int_{\mathcal Q} \nabla_\alpha v^\alpha \mathrm d^{D} \mathcal V$ where $\mathrm d^{D} \mathcal V$ is the volume element $\sqrt{|g_D|} \mathrm d^{D} x$.  Rearranging, this implies
\begin{equation}
\int_{\Gamma_2} v^\alpha \mathrm d \Sigma_\alpha - \int_{\Gamma_1} v^\alpha \mathrm d \Sigma_\alpha = \int_\mathcal Q \nabla_\alpha v^\alpha \mathrm d^{D} \mathcal V.
\end{equation}
Since $\nabla_\alpha v^\alpha = 0$,
\begin{equation}\label{Gauss}
\int_{\Gamma_2} v^\alpha \mathrm d \Sigma_\alpha = \int_{\Gamma_1} v^\alpha \mathrm d \Sigma_\alpha,
\end{equation}
a standard application of Gauss' theorem.  In words, $v^{\alpha}$ is the ``flux" associated with the conservation of $\mathcal{V}_v$. However, note that for any vector field $u^\alpha$ which is not divergence free, an attempt to use the definition here for the vector volume will yield a result that is dependent on the choice of hypersurface $\Gamma$.

\section{Vector Volume -- General Properties} \label{generalproperties}

Since the covariant derivative and thus the divergence are linear functions, the divergence of any linear combination of two divergence-free vectors is zero.  As a result, if $v^\alpha$ and $w^\alpha$ are valid vectors to define a vector volume, so will a linear combination of those vectors.  In what follows in this section we examine how linear combinations of choices of vector affects the resulting vector volume.

\subsection{Constant Multiplication}

Let us first show that the vector volume of a region $\mathcal R$ with respect to a nonzero positive constant $C$ times a given vector field $v$ is that same constant $C$ times the vector volume of $\mathcal R$ with respect to $v$. That is,
\begin{equation}
\mathcal V_{Cv} = C \mathcal V_{v}.
\end{equation}
To show this we use \eqref{haywarddef}.  Since the integral curves for the vector fields $v$ and $Cv$ are the same, we can use the same hypersurface region $\Gamma$ to represent the hypersurface which lies within $\mathcal R$ and intersects the integral curve(s) exactly once.  Then we can write
\begin{eqnarray}
\mathcal V_{Cv} = \int_{\Gamma\cap\mathcal R} C v^\alpha \mathrm d \Sigma_\alpha = \\ \nonumber C \int_{\Gamma\cap\mathcal R} v^\alpha \mathrm d \Sigma_\alpha = C \mathcal V_{v}
\end{eqnarray}
as required. If $C$ is a negative constant, then the result will be $\mathcal V_{Cv} = |C| \mathcal V_{v}$ because the orientation of the hypersurface, and thus the sign of $\mathrm d \Sigma_\alpha$, is always chosen so that the volume rate is positive [if it is nonzero].  We cannot choose $C = 0$ because the vector volume with respect to a zero vector field is not defined.

\subsection{Two Vectors} \label{twovectorsSection}

Here we seek the relationship between the vector volumes of divergence-free vectors $v^\alpha$ and $w^\alpha$ and the vector volume of $v^\alpha + w^\alpha$.  We can only compare the vector volumes for $v^\alpha, w^\alpha$ and $v^\alpha + w^\alpha$ if they are volumes corresponding to a common  region $\mathcal R$.  This only occurs if the normal to $\mathcal R$ is perpendicular throughout to both $v^\alpha$ and $w^\alpha$. If we can find a case in which the boundary of $\mathcal R$ is parallel to both $v$ and $w$, and $\Gamma$ intersects the integral curves of both exactly once, while they are oriented the same direction, then the vector volume is a simple sum since
\begin{eqnarray}
\mathcal V_{v+w} = \int_{\Gamma\cap\mathcal R} (v^\alpha + w^\alpha) \mathrm d \Sigma_\alpha \\ \nonumber = \int_{\Gamma\cap\mathcal R} v^\alpha \mathrm d \Sigma_\alpha + \int_{\Gamma\cap\mathcal R} w^\alpha \mathrm d \Sigma_\alpha.
\end{eqnarray}

One possible situation in which the normal to $\partial \mathcal R$ is perpendicular to two appropriate vector fields $v^\alpha$ and $w^\alpha$ is the case in which $w^\alpha$ is a vector tangent to closed, cyclic curves. Assume that we can choose an adapted coordinate system where $v^\alpha = \delta^\alpha_0$, $w^\alpha = \delta^\alpha_1$, and orbits of $w$ of length $P$ are closed, such that the points $(x^0, x^1, x^A)$ and $(x^0, x^1 + P, x^A)$ are coincident.  The most obvious instance of this is in spaces or spacetimes with azimuthal symmetry wherein $w^\alpha = \phi^\alpha$ and $P = 2\pi$ for the usual azimuthal symmetry vector $\phi^\alpha$.  A region $\mathcal R$ whose boundary normal is perpendicular to both $v^\alpha$ and $w^\alpha$ must have the form $x^A \in \Psi$ for some $(D-2)$-dimensional region $\Psi$, where there are no boundaries on $x^0$ or $x^1$.  We note that the integral curves of $v^\alpha$ can be represented by $x^i = const.$, and the integral curves of $v^\alpha + H w^\alpha$, for nonzero constant $H$, can be written as $x^A = const.$, $x^1 = H x^0 + B$ for some $B$ constant along each integral curve.  We note that the choice of $\Gamma$ as the hypersurface region $x^0 = 0$, $x^A \in \Psi$ has the property that it intersects each integral curve of $v^\alpha$ and $v^\alpha + H w^\alpha$ exactly once, so it is a suitable choice for the hypersurface $\Gamma$.  We can then calculate $\mathcal V_{v + Hw}$:
\begin{equation}
\mathcal V_{v + H w} = \int_{\Gamma\cap\mathcal R} v^\alpha \mathrm d \Sigma_\alpha + H \int_{\Gamma\cap\mathcal R} w^\alpha \mathrm d \Sigma_\alpha.
\end{equation}
In our adapted coordinates, $\mathrm d \Sigma_\alpha = \sqrt{|g_D|} \delta^0_\alpha$, so that
\begin{equation}
\int_{\Gamma\cap\mathcal R} w^\alpha \mathrm d \Sigma_\alpha = \int_{x^A \in \Psi} \sqrt{|g_D|} \delta^\alpha_1 \delta^0_\alpha = 0, \label{cyclicvectorvolumevanishing}
\end{equation}
and so we find that in this particular situation
\begin{equation}
\mathcal V_{v + H w} = \mathcal V_{v}. \label{twovectors}
\end{equation}
We can summarize this by saying that if $v^\alpha$ is a divergence free vector, $w^\alpha$ is a ``cyclic'' divergence-free vector (as defined above), and $\mathcal R$ is a region whose boundary normal is perpendicular to both, then the vector volume $\mathcal V_{v + H w}$ with respect to the vector $v^\alpha + H w^\alpha$ is equal to the vector volume $\mathcal V_{v}$.  Combining this result with the result of the previous subsection, we have
\begin{equation}\label{cvhw}
\mathcal V_{C v + H w} = C \mathcal V_{v}.
\end{equation}
We make use of (\ref{cvhw}) below.

\section{Kerr--Schild Metrics} \label{kerrschildtype}

A generalized Kerr--Schild spacetime (e.g.~\cite{Sopuerta}) has the form
\begin{equation}
\mathrm d s^2 = g_{\alpha \beta} \mathrm d x^\alpha \mathrm d x^\beta,
\end{equation}
where
\begin{equation}
g_{\alpha \beta} = \bar g_{\alpha \beta} + 2 K k_\alpha k_\beta, \qquad g^{\alpha \beta} = \bar g^{\alpha \beta} - 2 K k^\alpha k^\beta,
\end{equation}
where $\bar g_{\alpha \beta}$ with inverse $\bar g^{\alpha \beta}$ is some ``background'' metric, often flat space, $K$ is a scalar function, and $k_\alpha$ is a vector which is null in both the background and full metric. Further, the components of $k_\alpha$ can be raised and lowered using either metric:
\begin{equation}
g^{\alpha \beta} k_\beta = \bar g^{\alpha \beta} k_\beta = k^\alpha, \qquad g_{\alpha \beta} k^\beta = \bar g_{\alpha \beta} k^\beta = k_\alpha.
\end{equation}
Examples of spacetimes which can be expressed as Kerr--Schild metrics with a Minkowski background are certain spherically symmetric spacetimes and the Kerr--Newman spacetime. Kerr--Newman--(anti) de Sitter is an example of a spacetime which can be written in Kerr--Schild form with a non-flat background (in this case, (anti) de Sitter).

The two spacetimes with metrics $g_{\alpha \beta}$ and $\bar g_{\alpha \beta}$ can be expressed in the same coordinates, and we can introduce a vector $v^\alpha$ which is well defined according to both backgrounds by setting its components to be equal.  Note that the covector associated with $v^\alpha$ is in general not the same when $v^\alpha$ is lowered by both metrics, since in general
\begin{equation}
v_\alpha \equiv g_{\alpha \beta} v^\beta = \bar g_{\alpha \beta} v^\beta + 2 K k_\alpha (k_\beta v^\beta).
\end{equation}

The Matrix Determinant Lemma (e.g.~\cite{Harville}) states that if $A$ is an invertible square matrix and $u$ and $v$ are column vectors, then
\begin{equation}
\text{det}(A + u v^T) = (1 + v^T A^{-1} u) \text{det}(A)
\end{equation}
where $v^T$ is the transpose of vector $v$ and $A^{-1}$ is the inverse of matrix $A$. Now consider the determinant of
\begin{equation}
\bar g_{\alpha \beta} + 2 K k_\alpha k_\beta.
\end{equation}
We can represent this sum as a square matrix with the $\alpha$ index changing along the rows and the $\beta$ index changing along the columns. In this case we can represent $\bar g_{\alpha \beta}$ by $A$, $\bar g^{\alpha \beta}$ by $A^{-1}$, $2 K k_\alpha$ by $v^T$ and $k_\beta$ by $u$.  It is easy to check that $v^T A^{-1} u = \bar g^{\alpha \beta} (2 K k_\alpha k_\beta)$ in this representation.  Thus we find
\begin{eqnarray}
g = \text{det}(g_{\alpha \beta}) = \text{det}(\bar g_{\alpha \beta} + 2 K k_\alpha k_\beta) = \\ \nonumber (1 + \bar g^{\alpha \beta} 2 K k_\alpha k_\beta) \text{det}(\bar g_{\alpha \beta}) = \bar g
\end{eqnarray}
where $g$ and $\bar g$ represent the determinants of $g_{\alpha \beta}$ and $\bar g_{\alpha \beta}$ respectively.  This relies on the fact that $k_\alpha$ is null with respect to $\bar g_{\alpha \beta}$.

The equality of the determinants of the full metric and background metric implies that the volume element for the full spacetime ($\mathcal{F}$) is equal to the volume element for the background spacetime ($\mathcal{B}$):
\begin{equation}
\mathrm d^{D} \mathcal V_{\mathcal{F}} = \sqrt{|g|} \mathrm d^{D} x = \sqrt{|\bar g|} \mathrm d^{D} x = \mathrm d^{D} \mathcal V_{\mathcal{B}}.
\end{equation}
This indicates that the $D$-volume of some region can be calculated using either the full metric or the background metric.  This is not in general true for $N$-volumes where $N < D$ within the spacetime (such as lengths when $D > 1$, areas when $D>2$ etc.),~since the sub-determinants are not in general equal for $g_{\alpha \beta}$ and $\bar g_{\alpha \beta}$.

We note now that Definition 1 of the vector volume requires only a vector $v^\alpha$ and the $D$-volume of some region $\mathcal R$ parameterized by $\mu$, which is a function of the coordinates.  Since $v^\alpha$, the coordinates, and the $D$-volume are equivalent in the full spacetime and in the background spacetime, we find that the vector volume with respect to $v^\alpha$ of a region $\mathcal R$ in a Kerr--Schild spacetime $g_{\alpha \beta}$ is identical to the vector volume with respect to $v^\alpha$ of the background spacetime.  We can also see that the vector volume is the same in both the full spacetime and background using expression \eqref{gexpress} and using the equality of the two determinants.

We note before continuing that a vector which is divergence-free according to $g_{\alpha \beta}$ will also be divergence-free according to $\bar g_{\alpha \beta}$:
\begin{eqnarray}
\left.\nabla_\alpha v^\alpha\right|_{\mathcal{F}} = \frac{1}{\sqrt{|g|}} \partial_\alpha (\sqrt{|g|} v^\alpha) = \\ \nonumber \frac{1}{\sqrt{|\bar g|}} \partial_\alpha (\sqrt{|\bar g|} v^\alpha) = \left.\nabla_\alpha v^\alpha\right|_{\mathcal{B}}.
\end{eqnarray}
This allows us to see that a vector which is valid for calculating the vector volume in the background spacetime will be valid for calculating the vector volume in the full spacetime.

The fact that the vector volume with respect to a full spacetime is equal to the vector volume calculated for its background spacetime is of particular importance and interest when the background spacetime is Minkowski space.  Let $\bar g_{\alpha \beta} = \eta_{\alpha \beta}$ and move for the moment into Lorentzian coordinates $(T,x^i)$ where the $x^i$ are $D-1$ spatial coordinates.  The metric can be expressed as
\begin{equation}
\eta_{\alpha \beta}\mathrm d x^\alpha \mathrm d x^\beta = - \mathrm d T^2 + \sum_i (\mathrm d x^i)^2.
\end{equation}
Now define $T^\alpha$ according to $T^\alpha \partial_\alpha \equiv \partial_T$.  Then if we have some region $\mathcal R$ defined by $x^i \in \Sigma$, where $\Sigma$ is some $(D-1)$-dimensional spatial region, the vector volume of $\mathcal R$ with respect to $T^\alpha$ in the background space is, from \eqref{gexpress},
\begin{equation}\label{euclid}
\mathcal V_{T} = \int_{x^i \in \Sigma} {|\eta|} \mathrm d^{D-1} x = \int_{x^i \in \Sigma} \mathrm d^{D-1} x \equiv \mathcal V_E,
\end{equation}
where $\mathcal V_{E}$ is the Euclidean volume of the spatial component of region $\mathcal R$ in the flat background spacetime.  Since the vector volume is equal regardless of whether we choose the flat background or the full metric, the vector volume with respect to $T^\alpha$ for the full spacetime $g_{\alpha \beta}$ will be equal to the Euclidean volume for the spatial part of the region, as calculated in the flat background.

\subsection{``Double'' Kerr--Schild Form} \label{doubleKS}

We can extend the argument given above further.  With the specific example of the Kerr-(anti) de Sitter metric, we note that it is possible to have a metric which can be expressed in Kerr--Schild form with a non-Minkowski background, which \emph{itself} can be expressed in Kerr--Schild form.  In other words, we let the full metric be expressed, as before, in the form
\begin{equation}
g_{\alpha \beta} = \bar g_{\alpha \beta} +2 K k_\alpha k_\beta, \qquad g^{\alpha \beta} = \bar g^{\alpha \beta} - 2 K k^\alpha k^\beta, \label{firstform}
\end{equation}
where, in the particular case of Kerr--(anti) de Sitter, $\bar g_{\alpha \beta}$ is the metric for (anti) de Sitter spacetime and $k_\alpha$ is a null vector with respect to both $g_{\alpha \beta}$ and $\bar g_{\alpha \beta}$.  From the previous section we have $g = \bar g$.  In the Kerr-(anti) de Sitter case, we can break this down even further by expressing the (anti) de Sitter metric in terms of the flatspace metric, say $h_{\alpha \beta}$, and a vector $l_\alpha$, which is ``null'' with respect to $h_{\alpha \beta}$ and $\bar g_{\alpha \beta}$, but not necessarily $g_{\alpha \beta}$.  The reason for this is that the (anti) de Sitter spacetime can itself be expressed in Kerr--Schild form. However, the ``null vector'' in this form will not necessarily be null in the global Kerr-(anti) de Sitter spacetime.  We can write this as
\begin{equation}
\bar g_{\alpha \beta} = h_{\alpha \beta} +2L l_\alpha l_\beta.
\end{equation}
Now we set $\bar g^{\alpha \beta}$ and $h^{\alpha \beta}$ as the inverses of $\bar g_{\alpha \beta}$ and $h_{\alpha \beta}$, and (for lack of a better name) set $\tilde l^\alpha = \bar g^{\alpha \beta} l_\beta = h^{\alpha \beta} l_\beta$ as the ``contravariant'' form of $l_\alpha$ within the background spacetime (which will not, in general, be equal to $l^\alpha = g^{\alpha \beta} l_\beta$ for the full spacetime).   Note that~$\bar g = h$, where $h = \text{det}(h_{\alpha \beta})$. We know also from \eqref{firstform} that $\bar g^{\alpha \beta}$ can be written as
\begin{equation}
\bar g^{\alpha \beta} = h^{\alpha \beta} - 2L \tilde l^\alpha \tilde l^\beta, \label{contra1}
\end{equation}
and so we can decompose the spacetime into a form
\begin{equation}
g_{\alpha \beta} = h_{\alpha \beta} + 2 L l_\alpha l_\beta + 2 K k_\alpha k_\beta,
\end{equation}
where $h_{\alpha \beta}$ is the metric tensor for flat space, $2 L l_\alpha l_\beta$ is a correction from Minkowski to (anti) de Sitter, and $2 K k_\alpha k_\beta$ is a correction from (anti) de Sitter to Kerr--(anti) de Sitter.  Since $g = \bar g$ and $\bar g = h$, we find that $g = h$. The determinant of the metric tensor, and thus the associated vector volume, is the same for the Kerr-(anti)de Sitter spacetime and the background-background metric. We can now retrace the argument given above and arrive back at (\ref{euclid}).

Note that the contravariant metric equation for this ``double Kerr--Schild'' form is slightly complicated because $\tilde l^\alpha = \bar g^{\alpha \beta} l_\beta \neq l^\alpha = g^{\alpha \beta}l_\beta$; in fact $\tilde l^\alpha = l^\alpha + 2 K k^\alpha k^\beta l_\beta$.  If we set $\gamma \equiv k^\alpha l_\alpha$, we can write the contravariant metric equation somewhat compactly as
\begin{equation}
g^{\alpha \beta} = h^{\alpha \beta} -  2L (l^\alpha + 2 K k^\alpha \gamma) (l^\beta + 2 K k^\beta \gamma) - 2 K k^\alpha k^\beta.
\end{equation}

The primary application of the vector volume is to stationary black holes, i.e.~black holes with a Killing vector corresponding to time translation which is timelike at large radius. [For asymptotically flat or asymptotically anti-de Sitter spacetimes, the time translation Killing vector is timelike at infinite radius. The de Sitter case is complicated by the existence of the cosmological horizon.]  We expand on two cases below. The first is the canonical black hole volume, defined and discussed in Section \ref{canon}, which is equivalent to the volume considered by Parikh as reviewed in Section \ref{Parikhintro} and the geometric volume defined by Cveti\v{c} et al.~as reviewed in Section \ref{cveticgeometricvolume} and is similar to Hayward's volume \cite{Hayward}.  The second is related to the volume introduced in Section \ref{OurVolumeIntro}. This is examined in Section \ref{ourvolume}.

\section{Canonical Black Hole Volume} \label{canon}

In this section we define and examine the ``canonical black hole volume'' for stationary spacetimes.  We define this as the vector volume of the region below the event horizon of a black hole, with respect to the canonical Killing vector which corresponds to stationarity.  We will label this volume $\mathcal V_\mathcal C$.  In general, for time coordinate $t$ we will write the corresponding Killing vector as $t^\alpha$, where $t^\alpha \partial_\alpha \equiv \partial_t$; instead of $t$ sometimes $T$ or $\tau$ or another coordinate may be used.

We note that the canonical black hole volume is equivalent to the volume that Parikh defines in his paper \cite{Parikh} as well as the geometric volume defined by Cveti\v{c} et al.~\cite{Cvetic}

To define the canonical black hole volume explicitly, assume that there is a stationary $D$-dimensional black hole which can be written in coordinates $(t,x^i)$ adapted to the Killing vector which corresponds to the black hole's stationarity, which takes the form $\partial_t$.  We also demand that $\partial_t$ be properly normalized, if possible, a condition which we elaborate on in Section~\ref{NormalizationCanon}.  Let the region below the horizon be $x^i \in \Sigma$, where $\Sigma$ is a $(D-1)$-dimensional region.  The line element can be written as
\begin{equation}
\mathrm d s^2 = g_{t t}(x^i) \mathrm d t^2 + 2 g_{t i}(x^i) \mathrm d t \mathrm d x^i + g_{i j}(x^i) \mathrm d x^i \mathrm d x^j,
\end{equation}
where, since these coordinates are adapted to the Killing vector $\partial_t$, the components of $g_{\alpha \beta}$ are independent of $t$.  If we let $g_D$ be the determinant of the metric $g_{\alpha \beta}$, then from \eqref{gexpress} we can write the canonical black hole volume as
\begin{equation}
\mathcal V_\mathcal C \equiv \mc V_t \equiv \int_{x^i \in \Sigma} \sqrt{|g_{(D)}|} \mathrm d^{D-1} x.
\end{equation} 
Compare to \eqref{Parikhdef} and \eqref{Vgeo}.

Remarkably, if $\phi^\alpha \partial_\alpha = \partial_\phi$ is a Killing vector corresponding to axial symmetry such that $\phi$ is a cyclic coordinate, it follows from \eqref{twovectors} that the vector volumes for the black hole calculated using Killing vectors $t^\alpha$ and $t^\alpha + \Omega \phi^\alpha$ are identical for any constant $\Omega$:
\begin{equation}
\mathcal V_\mathcal C = \mathcal V_t = \mathcal V_{t + \Omega \phi}. \label{canonwithomega}
\end{equation}
A consequence of this, for example, is that in stationary, axisymmetric black holes with some angular momentum, we can calculate the canonical black hole (BH) volume using either the stationarity Killing vector $\partial_t$ or the Killing vector $\xi = \partial_t + \Omega_H \partial_\phi$ which is tangent to the null generators of the horizon.

An important point remains. We have to ensure that the vector $\partial_{t}$ has the proper normalization.  In his paper, Parikh suggests fixing the asymptotic normalization of the Killing vector, but this becomes problematic in spacetimes with $\Lambda$.

\subsection{Killing Vector Normalization} \label{NormalizationCanon}

Consider first an asymptotically flat stationary black hole with axial symmetry.  How should we fix the asymptotic normalization of the Killing vector corresponding to stationarity?  Following Carter \cite{CarterLesHouches}, the normalization for the vector corresponding to stationary, say $T^\alpha$, is set by
\begin{equation}
-T^\alpha T_\alpha \to 1
\end{equation}
where the limit is taken at spatial infinity.  In spacetime with time symmetry but no spatial translational symmetry, this defines the Killing vector uniquely. (If a translational spacelike Killing vector exists, it is possible to create a new well-normalized timelike Killing vector by a linear combination of the time Killing vector with the spatial Killing vector, as in Minkowski space.)  Spacetimes with axial symmetry have a rotation Killing vector, say $m^\alpha$, which is zero on the rotation axis, whose normalization can be fixed by requiring
\begin{equation}
\frac{ \partial_\alpha X \partial^\alpha X}{4 X} \to 1
\end{equation}
where $X = m^\alpha m_\alpha$ and the limit is taken on the rotation axis.  This is what ensures the standard periodicity $2 \pi$ \cite{CarterLesHouches}.

As explained above, any Killing vector of the form $\tilde T^\alpha \equiv T^\alpha + \Omega m^\alpha$ will yield the same canonical volume. Now the magnitude squared of $\tilde T^\alpha$ will be everywhere equal to that of $T^\alpha$ only if $\Omega = 0$, since
\begin{equation}
\tilde T^\alpha \tilde T_\alpha = T^\alpha T_\alpha + 2 \Omega T^\alpha m_\alpha + \Omega^2 X.
\end{equation}
However, since $m^\alpha = 0$ on the rotation axis, any valid Killing vector $\tilde T^\alpha$ for calculating the volume will have
\begin{equation}
-\tilde T^\alpha \tilde T_\alpha \to -T^\alpha T_\alpha \to 1
\end{equation}
with the limit taken to spatial infinity on the rotation axis.  We can use this as the condition for the normalization of the Killing vector to produce the proper canonical BH volume in axisymmetric spacetimes which are asymptotically flat.

If the space is \emph{not} asymptotically flat, then it is more challenging to define the proper normalization.  In the case of static spherical symmetry, metrics of the form (\ref{metricone}) (with $t_s=t$), i.e.~ones wherein $g_{r r} g_{t t} = -1$, are important due to the vanishing radial null-null component of the Ricci tensor \cite{Jacobson}.  Such spacetimes include the Reissner--Nordstr\"om--(anti) de Sitter class. Now if we take $T^{\alpha}\partial_{\alpha} \equiv \partial_{t}$, the condition $g_{r r} g_{t t} = -1$ suggests that the normalization condition
\begin{equation}
-g_{r r} \tilde T^\alpha \tilde T_\alpha = 1.
\end{equation}
Here, $r$ is a uniquely defined coordinate since surfaces of $r = const.$ are spheres with surface area $4\pi r^2$ (in the 4 dimensional case). The spherically symmetric spacetimes I am interested in, the Reissner--Nordström--(anti-)de Sitter black holes, all have $g_{tt} g_{rr} = -1$.

This still leaves spacetimes which are not asymptotically flat and which do not possess spherical symmetry; an example here is Kerr--Newman--(anti-)de Sitter case, which has line element
\begin{equation}
\mathrm d s^2 = -\frac{\Delta_r}{\rho^2} \left( \mathrm d \tau - \frac{a \sin^2 \theta \mathrm d \varphi}{\Xi}\right)^2 + \frac{\Delta_\theta \sin^2\theta}{\rho^2} \left(a \mathrm d \tau - \frac{(r^2+a^2)\mathrm d \varphi}{\Xi}\right)^2 + \frac{\rho^2 \mathrm d r^2}{\Delta_r} + \frac{\rho^2 \mathrm d \theta^2}{\Delta_\theta}. \label{KdSLine}
\end{equation}
Here, $\Delta_r = \left(1 - \frac{\Lambda r^2}{3}\right)(r^2 + a^2) - 2 m r + Q^2$, $\Delta_\theta = 1 + \frac{\Lambda a^2}{3}\cos^2 \theta$, $\Xi = 1 + \frac{\Lambda a^2}{3}$, $\rho^2 = r^2 + a^2\cos^2\theta$, and $\Lambda$ is the cosmological constant, opposite in sign to that in \cite{CarterLesHouches}. As in \cite{CarterLesHouches} we assume here that $\Xi>0$ for $\Lambda<0$. The $r \to \infty$ limit of this metric, we note, is outside the cosmological horizon for $\Lambda>0$, and the metric does not reduce to flat space asymptotically.  Not only that, but $r$ is no longer a coordinate for which the surfaces $r = const.$ represent spheres of symmetry, but ellipsoids instead. Nevertheless, by analogy with the requirement that a well-normalized Killing vector $\tilde T^\alpha$ has $\tilde T^\alpha \tilde T_\alpha \to -1$ on the rotation axis in asymptotically flat spacetime with axial symmetry, and that $g_{r r} g_{t t} = -1$ is a proper normalization in spherical symmetry, we are led to the following question: What if we require, in the case of Kerr--Newman--(anti) de Sitter,
\begin{equation}
-g_{r r} \tilde T^\alpha \tilde T_\alpha \to 1
\end{equation}
where the limit is taken for large $r$ on the rotation axis?  Remarkably, in the Kerr--Newman--(anti) de Sitter metric, setting $\tilde T^\alpha \partial_\alpha = \partial_\tau + \Omega \partial_\varphi$ (for any constant $\Omega$) yields
\begin{equation}
-g_{r r} \tilde T^\alpha \tilde T_\alpha = 1
\end{equation}
\emph{everywhere} on the rotation axis (where $g_{r r}$ and the norm of $\tilde T^\alpha$ are defined). This saves us the difficulty of worrying about, for example, going beyond the cosmological horizon and seems to be a reasonable normalization requirement.

There is another potential method for choosing the ``correct'' time-symmetry Killing vector for metrics which can be expressed in Kerr--Schild or ``double Kerr--Schild'' form with a flat background.

\subsection{Kerr--Schild Forms} \label{PKS}

If we have a metric which can be expressed in Kerr--Schild form or ``double Kerr--Schild form'' wherein there is a Minkowski background, another sensible way to choose the Killing vector $T^\alpha$ corresponding to stationarity is to choose a Killing vector such that, if $\eta_{\alpha \beta}$ represents the flat background,
\begin{equation}
-\eta_{\alpha \beta} T^\alpha T^\beta = 1.
\end{equation}
This definition allows us to write the background spacetime in coordinates $(T,x^i)$, where $T^\alpha\partial_\alpha = \partial_T$.  In cases where the metric is axially symmetric, we can extend this definition to any vector $\tilde T^\alpha = T^\alpha + \Omega m^\alpha$ (where $m^\alpha$ is the Killing vector for axial symmetry) by requiring
\begin{equation}
-\eta_{\alpha \beta} \tilde T^\alpha \tilde T^\beta = 1 \text{ on the rotation axis}.
\end{equation}

The advantage here is that if there is a flat background $\eta_{\alpha \beta}$, the canonical black hole volume would then, from the results of Section \ref{kerrschildtype}, be the Euclidean volume for the spatial component of the black hole, as calculated in the flat background.  As we will show in the following subsection, in the cases of static spherical symmetry in the form (\ref{metricone}), Kerr--Newman black holes, and Kerr--Newman--(anti) de Sitter black holes, this normalization scheme, and the ones presented in the previous subsection, are consistent with each other.  Thus, in these cases, we can show that the canonical black hole volume is equal to the Euclidean volume of the spatial component of the black hole region.

\subsection{Examples}

We will work in four dimensions for these examples.

\subsubsection{Spherical Symmetry}

Consider the spherically symmetric metric (\ref{metricone}). If this is a black hole, there is some value $r_+$ for which $\alpha(r_+) = 0$ and $\alpha(r) > 0$ just outside it.  Thus the black hole region is defined by $0 \leq r \leq r_+$.  We note that $t^\alpha$ with $t^\alpha \partial_\alpha = \partial_t$ is a Killing vector for which $-g_{r r} (t^\alpha t_\alpha) = 1$, so $t^\alpha$ is a valid vector for calculating the canonical black hole volume:
\begin{equation}
\mathcal V_\mathcal C = \mathcal V_{t} =  \int_{r \leq r_+} \sqrt{|g_{(D)}|} \mathrm d^{D-1} x = \frac{4 \pi r_+^3}{3},
\end{equation}
which, as Parikh noted when he did his calculation for a spherically symmetric black hole, is the Euclidean volume of a 2-sphere of radius $r_+$.

We can show that this makes sense by rewriting (\ref{metricone}) in Kerr--Schild form by choosing a new coordinate system $(T,r,\theta,\phi)$ where
\begin{equation}
\mathrm d T = \mathrm d t + \frac{1 - \alpha}{\alpha} \mathrm d r
\end{equation}
and defining a null vector $k_\alpha$ by
\begin{equation}
k_\alpha  \mathrm d x^\alpha = \mathrm d T  + \mathrm d r.
\end{equation}
Then, setting $\eta_{\alpha \beta} \mathrm d x^\alpha \mathrm d x^\beta = - \mathrm d T^2 + \mathrm d r^2 + r^2 \mathrm d \Omega_2^2$, which is Minkowski space in spherical polar coordinates, we can write the metric as
\begin{equation}
g_{\alpha \beta} = \eta_{\alpha \beta} + (1 - \alpha(r) ) k_\alpha k_\beta,
\end{equation}
which is in Kerr--Schild form.  Further, $\eta_{\alpha \beta} t^\alpha t^\beta = -1$.  This confirms that the volume of a region with respect to $t^\alpha$ is the Euclidean volume of the spatial component of the region, as calculated in the flat background with metric tensor $\eta_{\alpha \beta}$. This provides a clear explanation for why the canonical black hole volume is the Euclidean volume for spherical black holes.

\subsubsection{Kerr--Newman}

The line element for the Kerr--Newman black hole, which is asymptotically flat, can be written in Boyer--Lindquist coordinates as
\begin{equation}
\mathrm d s^2 = \left(\frac{\mathrm d r^2}{\Delta}+ \mathrm d \theta^2\right)\rho^2 - \left(\mathrm d t - a \sin^2 \theta \mathrm d \phi\right)^2 \frac{\Delta}{\rho^2} + \left( (r^2 + a^2) \mathrm d\phi - a \mathrm d t \right)^2 \frac{\sin^2 \theta}{\rho^2}, \label{BL}
\end{equation}
where $\rho^2 = r^2 + a^2 \cos^2\theta$, $\Delta = r^2 - 2 M r + a^2 + Q^2$ with mass $M$, charge $Q$ and rotational parameter $a$.  In these coordinates, the outer horizon lies at $r= r_+$.  The vector $t^\alpha$ with $t^\alpha \partial_\alpha = \partial_t$ is Killing and $t_\alpha t^\alpha \to -1$ asymptotically, so it is a good vector for the calculation of the canonical black hole volume.  The canonical black hole volume then is
\begin{equation}
\mathcal V_\mathcal C = \mathcal V_{t} = \int_{r \leq r_+} \sqrt{|g_{(D)}|} \mathrm d^{D-1} x = \frac{4\pi}{3} r_+ (r_+^2 + a^2).
\end{equation}

While it is not immediately obvious that the expression on the right is equal to the Euclidean volume of the background metric in Kerr--Newman, this is also the case.  The Kerr--Schild form of the Kerr--Newman black hole in Lorentzian coordinates $(T,x,y,z)$ can be written as \cite{Poisson}
\begin{equation}
g_{\alpha \beta} = \eta_{\alpha \beta} + f k_\alpha k_\beta,
\end{equation}
where $f =  \frac{r^2}{r^4 + a^2 z^2}(2 M r - Q^2)$, $\eta_{\alpha \beta} = \mathrm {diag} (-1,1,1,1)$ and $k_\alpha$ is a null vector with
\begin{equation}
k_\alpha = \left(1, \frac{r x + a y}{r^2 + a^2}, \frac{r y - a x}{r^2 + a^2}, \frac{z}{r}\right).
\end{equation}
The coordinate $r$ is the usual Boyer--Lindquist $r$ defined in these Cartesian coordinates by
\begin{equation}
\frac{x^2 + y^2}{r^2 + a^2} + \frac{z^2}{r^2} = 1.
\end{equation}
Surfaces of constant $r$ are \emph{ellipsoids} in $(x,y,z)$, where the semi-major axes are $\sqrt{r^2+a^2}$ for the $x$ and $y$ directions and $r$ for the $z$-direction.

The relationship between $T$ and the coordinates $(t,r)$ is
\begin{equation}
\mathrm d T = \mathrm d t + \frac{2 M r - Q^2}{\Delta} \mathrm d r.
\end{equation}
We can confirm that $\partial_T = \partial_t$.  Thus, we confirm that $\eta_{\alpha \beta} t^\alpha t^\beta=-1$ and so the canonical black hole volume for the Kerr--Newman black hole is equal to the the Euclidean volume of the spatial component of the flat background of the Kerr--Newman black hole when written in Kerr--Schild coordinates.

Since the horizon is at $r = r_+$, the flatspace interpretation of the volume of the region $0 \leq r \leq r_+$ is not the volume of a sphere but the volume of an \emph{ellipsoid} with semi-major axes $\sqrt{r_+^2 + a^2}, \sqrt{r_+^2 + a^2}$, and $r$ (for $x,y$ and $z$ directions respectively).  The Euclidean volume of an ellipsoid with semi-major axes $u,v$ and $w$ is $4\pi u v w/3$, which in this case is $4\pi r_+(r_+^2+a^2)/3$, confirming that the canonical black hole volume of the Kerr--Newman black hole is in fact its Euclidean volume in this set of coordinates.

\subsubsection{Kerr--Newman--(anti) de Sitter} \label{KNAdSPaper}

The line element for Kerr--Newman--(anti) de Sitter is given in \eqref{KdSLine} and in these coordinates the metric determinant $g_{(4)}$ is
\begin{equation}
g_{(4)} = - \frac{\sin^2 \theta (r^2 + a^2 \cos^2\theta)^2}{\Xi^2}.
\end{equation}
The black hole region is given by $r \leq r_0$ where $r_0$ is the appropriate root to $\Delta_r = 0$. If we use the vector $\partial_\tau$ to calculate volume as discussed in Section \ref{NormalizationCanon}, then we obtain
\begin{equation}
\mathcal V_\mathcal C = \mathcal V_{\tau} =  \int_{r \leq r_0} \sqrt{|g_{(D)}|} \mathrm d^{D-1} x = \frac{4\pi}{3} \frac{r_0 (r_0^2 + a^2)}{\Xi}
\end{equation}
Again, the expression on the right is not immediately familiar, but as explained below this corresponds to a Euclidean volume expression.

Kerr--Newman--(anti) de Sitter, with line element of the form \eqref{KdSLine}, cannot be written in ``single'' Kerr--Schild form but can be written in ``double'' Kerr--Schild form.  Introducing new coordinates $t, \phi$ defined by
\begin{eqnarray}
\mathrm d t = \mathrm d \tau + \frac{2 m r - Q^2} {(1-\lambda r^2) \Delta_r }\mathrm d r, \\ \nonumber \qquad \mathrm d \phi = \mathrm d \varphi - \lambda a \mathrm d \tau + \frac{ (2 m r - Q^2) a r }{(r^2+a^2) \Delta_r} \mathrm d r,
\end{eqnarray}
the line element for the full spacetime overall can be written in terms of the background (anti) de Sitter metric by writing
\begin{equation}
\mathrm d s^2 = \mathrm d \bar s^2 + \frac{2 m r - Q^2}{\rho^2} (k_\alpha \mathrm d x^\alpha)^2,
\end{equation}
where
\begin{equation}
k_\alpha \mathrm d x^\alpha = \frac{\Delta_\theta \mathrm d t}{\Xi} + \frac{\rho^2 \mathrm d r}{(1 - \lambda r^2)(r^2+a^2)} - \frac{a \sin^2 \theta \mathrm d \phi}{\Xi},
\end{equation}
\begin{equation}
\mathrm d \bar s^2 = - \frac{\Delta_\theta ( 1 - \lambda r^2)}{\Xi} \mathrm d t^2 + \frac{\rho^2}{(1 - \lambda r^2)(r^2+a^2)} \mathrm d r^2 + \frac{\rho^2}{\Delta_\theta} \mathrm d \theta^2 + \frac{r^2+a^2}{\Xi} \sin^2 \theta \mathrm d \phi^2, \label{dbars}
\end{equation}
$\lambda = \Lambda/3$ and $\rho^2, \Delta_\theta, r$ and $a$ have the same meanings as in \eqref{KdSLine} \cite{GibbonsLu}.  It is not readily apparent that the expression for $\mathrm d \bar s^2$ is the line element for (anti) de Sitter spacetime, but by making the substitutions defined implicitly by \cite{AckayMatzner}
\begin{equation}
R^2 = \frac{r^2 \Delta_\theta + a^2 \sin^2\theta}{\Xi}, \qquad R \cos \Theta = r \cos\theta,
\end{equation}
we find that $\mathrm d \bar s^2$ becomes
\begin{equation}
\mathrm d \bar s^2 = -(1 - \lambda R^2) \mathrm d t^2 + \frac{\mathrm d R^2}{1 - \lambda R^2} + R^2 (\mathrm d \Theta^2 + \sin^2\Theta \mathrm d \phi)
\end{equation}
which is the familiar form [of (anti) de Sitter] with radial coordinate $R$ and angular coordinate $\Theta$.  By defining coordinate $T$ by
\begin{equation}
\mathrm d T  = \mathrm d t + \frac{\lambda R^2}{1 - \lambda R^2} \mathrm d R
\end{equation}
we can rewrite $\mathrm d \bar s^2$ as
\begin{equation}
\mathrm d \bar s^2 = \mathrm d s^2_{flat} + \lambda R^2 (l_\alpha \mathrm d x^\alpha)^2
\end{equation}
where $l_\alpha \mathrm d x^\alpha = \mathrm d T + \mathrm d R$ (which is a null vector in the (anti) de Sitter background and the Minkowski background) and $\mathrm d s^2_{flat} = -\mathrm d T^2 + \mathrm d R^2 + R^2 (\mathrm d \Theta^2 + \sin^2\Theta \mathrm d \phi^2)$ which is the familiar form for Minkowski space in spherical polar coordinates.  This means that the full Kerr--Newman--(anti) de Sitter metric becomes
\begin{equation}
g_{\alpha \beta} = \eta_{\alpha \beta} + \lambda R^2 l_\alpha l_\beta + \frac{2 m r - Q^2}{\rho^2} k_\alpha k_\beta
\end{equation}
where $\eta_{\alpha \beta}$ is the Minkowski metric, $k_\alpha$ is a null vector in the full Kerr--Newman--(anti) de Sitter spacetime and the (anti) de Sitter background, and $l_\alpha$ is a null vector in the (anti) de Sitter background and the flat background for the (anti) de Sitter background.

If we call the Killing vector used to define the canonical black hole volume $t^\alpha \partial_\alpha = \partial_\tau$ from before, we note that it is also equal to $t^\alpha \partial_\alpha = \partial_\tau = \partial_t = \partial_T$.  Thus we note immediately that $-\eta_{\alpha \beta} t^\alpha t^\beta = 1$ everywhere, which is the suggestion made in Section \ref{PKS} for dealing with spacetimes such as Kerr-(anti) de Sitter for which asymptotic properties are not well defined.  According to the results of Section \ref{doubleKS}, as well, the vector volume of a region in a ``double'' Kerr--Schild metric for which we use $\partial_T$ as the vector will yield the Euclidean spatial volume of that region in the flat background.  As a result, if we use $\partial_t = \partial_T$ for the vector to calculate the canonical black hole volume we will again calculate the Euclidean volume of the flat background.

In the case of Kerr--Newman--(anti) de Sitter, the black hole region is bounded by one of the solutions to $\Delta_r  = 0$; let us call the solution which gives the event horizon $r = r_0$.  The surface defined by $r = r_0$ is not a sphere in the Minkowski background, because the Minkowski background has $R$, not $r$, as a radial coordinate.  If we define Euclidean coordinates ($x,y,z$) by $x = R \sin \Theta \cos \phi, y = R \sin \Theta \sin\phi, z = R \cos \Theta$ in the usual way, we find a relationship between $(x,y,z)$ and $r$ as
\begin{equation}
\frac{x^2 + y^2}{\Xi^{-1} (r^2+a^2)} + \frac{z^2}{r^2} = 1.
\end{equation}
This implies that surfaces of constant $r$ are ellipsoids with two semi-major axes $\sqrt{ (r^2+a^2)/\Xi}$ and one semi-major axis $r$.  The Euclidean volume inside the region bounded by $r = r_0$ then would be $4\pi r_0 (r_0^2+a^2)/3\Xi$ as we discovered.  Once again the canonical black hole volume corresponds to the Euclidean volume of the black hole region.

\subsection{Connection to Area}

The invariant surface area of the horizon $\mathcal{A_H}$ (calculated in the usual way) of a spherically symmetric black hole with horizon radius $r_+$ is simply $4\pi r_+^2$.  The surface area of a Kerr--Newman black hole horizon is $4 \pi (r_+^2 +a^2)$ and the surface area of a Kerr--Newman--(anti) de Sitter black hole horizon is $4 \pi (r_+^2 + a^2) / \Xi$.  We note then that in all three cases the ratio of the canonical black hole volume $\mathcal V_C$ to the black hole surface area is:
\begin{equation}
\mathcal V_\mathcal C = \mathcal {V}_{geo} =\frac{r_+ \mathcal A_H}{3}, \label{volarearatio}
\end{equation}
which recovers \eqref{VrA1}.

In spherical symmetry with coordinates of a form like (\ref{metricone}), surfaces of constant $t, r$ have surface area $4 \pi r^2$ for all $r$.  However, in the Kerr--Newman (and Kerr--Newman--(anti) de Sitter) case the area only has the above form on the horizons; for example, in Kerr--Newman, taking a 2-surface $\Gamma$ defined by $r= R, t = const.$ gives a line element (from \eqref{BL})
\begin{equation}
\mathrm d s_\Gamma^2 = (R^2 + a^2 \cos^2\theta) \mathrm d \theta^2 + \frac{\left((R^2+a^2)^2 - \Delta(R) a^2 \sin^2\theta\right)\sin^2\theta}{R^2 + a^2 \cos^2\theta}\mathrm d \phi^2,
\end{equation}
from which the area is
\begin{equation}
\mathcal{A} = \int \sqrt{(R^2+a^2)^2 - \Delta(R) a^2 \sin^2\theta} \sin\theta \mathrm d \theta \mathrm d \phi.
\end{equation}
For simplicity, define
\begin{equation}
B \equiv \frac{\Delta a^2}{(R^2 + a^2)^2},
\end{equation}
which will be zero only on the horizons.  Then the area of a surface of constant $r,t$ can be written as
\begin{equation}
\mathcal{A} = \left\{
\begin{array}{rl}
2\pi(R^2 + a^2) \left( 1 + \left( |B|^{-\frac{1}{2}} + |B|^{\frac{1}{2}}\right)\arcsin\sqrt{\frac{B}{B-1}}\right) & \text{if } B < 0\\
\\
4\pi(R^2 + a^2) & \text{if } B = 0\\
\\
2\pi (R^2 + a^2) \left( 1 + \left( B^{-\frac 1 2} - B^{\frac 1 2}\right)\mathrm{arcsinh}\sqrt{\frac{B}{1-B}} \right) & \text{if } 0 < B \leq 1 \\
\\
2\pi(R^2 + a^2)\left( 1 + \left(B^{\frac{1}{2}} - B^{-\frac{1}{2}}\right)\left( \frac \pi 2 - \ln\left(1 - \frac{1}{\sqrt{B}} \right)\right)\right) & \text{if } B > 1
\end{array} \right.
\end{equation}
This shows that the relationship \eqref{volarearatio} is a property of the horizon alone, since $B=0$ if and only if $\Delta = 0$.

We note also that the area is calculated through very different means than the vector volume.  While the canonical black hole volume, for example, turns out to be equal to the Euclidean volume of the flat space background, the surface area of the horizon in Kerr--Newman and Kerr--Newman--(anti) de Sitter is not what is expected from the surface area of ellipsoids in Euclidean space.  In Kerr--Newman, wherein the Minkowski representation of $r = R, t = const.$ is an ellipsoid with one semi-axis $R$ and two semi-axes $\sqrt{R^2 + a^2}$, the Euclidean surface area of these 2-surfaces is
\begin{equation}
\mathcal{A}_{E} = \pi \left(2( R^2 + a^2) + \frac{R^2}{e}\ln\left(\frac{1 + e}{1 - e}\right)\right)
\end{equation}
where $e=a/\sqrt{R^2+a^2}$ is the ellipticity of the ellipsoid \cite{Weisstein}.

\section{Null Generator Volume and Surface Gravity} \label{ourvolume}

As stated in Section \ref{OurVolumeIntro}, in \cite{BallikLake10} we defined a black hole volume by
\begin{equation}
\mathcal V_{\mathcal{N}}= \frac{\mathrm d \mathcal V}{\mathrm d \ln \lambda} \nonumber
\end{equation}
where $\lambda$ is an affine parameter on the null generators on the horizon.  We will now show that in stationary, axially symmetric black holes this volume is equal to the vector volume of the black hole region with respect to the (unique) Killing vector $k^\alpha$ for which
\begin{equation}
k^\beta \nabla_\beta k^\alpha = k^\alpha
\end{equation}
on the horizon.

Spacetimes which are stationary and axially symmetric permit a Killing vector $\xi^\alpha$ which is tangent to the null generators of the horizon, for which (again, on the horizon only) \cite{Poisson}
\begin{equation}
\xi^\beta \nabla_\beta \xi^\alpha = \kappa \xi^\alpha \label{horizonequation}
\end{equation}
where $\kappa$ is the surface gravity of the horizon.  Generally speaking, $\xi^\alpha$ is written as
\begin{equation}
\xi^\alpha = t^\alpha + \Omega_H \phi^\alpha
\end{equation}
where $t^\alpha$ is the Killing vector corresponding to stationarity and $\phi^\alpha$ the Killing vector corresponding to axial symmetry, both of which have the properties explained in Section \ref{NormalizationCanon}, and $\Omega_H$ is the angular velocity which is a constant.

If $\xi^\alpha$ is multiplied by a nonzero constant $K$, then \eqref{horizonequation} becomes
\begin{equation}
 K\xi^\beta \nabla_\beta (K \xi^\alpha) = K^2 \kappa \xi^\alpha = K \kappa (K \xi^\alpha)
\end{equation}
which means that for the new vector $K \xi^\alpha$, we effectively have a new value for the constant of proportionality which was once held by $\kappa$.  If we define $K = 1/\kappa$ and set
\begin{equation}
k^\alpha \equiv \xi^\alpha/\kappa = (t^\alpha + \Omega_H \phi^\alpha)/\kappa, \label{kdefinition}
\end{equation}
then we have
\begin{equation}
k^\beta \nabla_\beta k^\alpha = k^\alpha. \label{unitycondition}
\end{equation}
This is the unique Killing vector which is both proportional to the null generators on the horizon and which satisfies \eqref{unitycondition} on the horizon.  (If the horizon is degenerate this vector will not exist since the Killing vectors tangent to the horizon have $\xi^\beta \nabla_\beta \xi^\alpha = 0$ and $\kappa = 0$ in this case.) 

Since $k^\alpha$ is a Killing vector, we can choose a system of adapted coordinates $(\bar k, x^i)$ where $k^\alpha \partial_\alpha = \partial_{\bar k}$.  Since $k^\alpha$ is tangent to the null generators on the horizon, on the horizon we have $x^i = const., \bar k = \bar k(\lambda)$, where $\lambda$ is an affine parameter.  The affine parameter can be found from \cite{Poisson}
\begin{equation}
\frac{\mathrm d \lambda}{\mathrm d \bar k} = \exp \left[\int^{\bar k } \mathrm d {\bar k'} \right] = \exp \bar k,
\end{equation}
which, rearranging, implies that up to a linear transformation,
\begin{equation}
\bar k = \ln \lambda.
\end{equation}

Writing $k^\alpha \partial_\alpha = \mathrm d / \mathrm d \ln \lambda$, from \eqref{derivativebased} we have
\begin{equation}
\mathcal V_{k} = \frac{\mathrm d \mathcal V_{\mathcal B}}{\mathrm d \ln \lambda}
\end{equation}
where $\mathcal V_{k}$ is the vector volume of the black hole region $\mathcal B$ with respect to the vector $k^\alpha$, and $\lambda$ is the affine parameter on the horizon.
The advantage of using this as the volume of stationary black holes is that the choice of vector $k$ relies only on local parameters, rather than, as with the canonical black hole volume, requiring asymptotic flatness or a Kerr--Schild form.

The relationship between the volume generated by $k^\alpha$ and the canonical black hole volume can be found by using \eqref{kdefinition}.
With the help of (\ref{cvhw}) we find that for a region $\mathcal{B}$
\begin{equation}
\mathcal V_{k} = \mathcal V_{\kappa^{-1} t + \kappa^{-1}\phi} = \kappa^{-1} \mathcal V_{t},
\end{equation}
in agreement with (\ref{firstkappa}).
This allows us to give a new definition for the surface gravity $\kappa$ in terms of the ratio of these volumes:
\begin{equation}
\kappa \equiv \frac{\mathcal V_{t}}{\mathcal V_{k}} = \frac{\mathcal V_\mathcal C}{\mathcal V_\mathcal N}, \label{newkappadefinition}
\end{equation}
where the second equality makes explicit that the surface gravity is the ratio of the canonical and null-generator volumes.  In the special case where the spacetime is a Kerr--Schild spacetime with a flat background and $\eta_{\alpha \beta} t^\alpha t^\beta = -1$, then $\mathcal V_{t} \equiv \mathcal V_{E}$, the Euclidean spatial volume, calculated in the background space. We then have
\begin{equation}
\kappa \equiv \frac{\mathcal V_{E}}{\mathcal V_{k}}.
\end{equation}

The usual physical meaning given to the surface gravity is, as explained in \cite{Poisson}, ``the force required of an observer at infinity to hold a particle (of unit mass) stationary at the event horizon.'' The interpretation given here, that the surface gravity is the ratio of the canonical black hole volume (and for Kerr--Schild spacetimes, the Euclidean spatial volume of the background spacetime) to the rate of change with respect to the logarithm of the affine parameter of the invariant four-volume of a black hole, is a local interpretation that would appear to be new.

An important question is, can we use this to gain further insights into black hole mechanics? The most obvious conclusion regards the third law of black hole mechanics, $\kappa \nrightarrow 0$. Since $\mathcal V_{t} > 0$ even in the degenerate case (consider, for example, the static spherically symmetric case) we see that the third law demands that the rate of growth $\mathcal V_{k}$ must remain finite.  In order to violate the third law we need $\mathcal V_{k} \to \infty$ and since $\mathrm d \mathcal V_{\mathcal{B}}/\mathrm d \lambda$ is finite, this requires $\lambda \to \infty$ in agreement with the formulation of Israel \cite{Israel86}. That is, in a sequence of quasi-static steps, the reduction of $\kappa$ to zero would take infinite advanced time.

\section{Connection to Other Authors} \label{otherauthors}

Here we further discuss the connection between the vector volume and some of the other works referenced in Section \ref{review}.

\subsection{Cveti\v{c} et al.}

Following the geometrical approach introduced in \cite{KastorEtal:2009}, and commenting here specifically on the detailed discussion in \cite{Cvetic}, it should be clear now that the geometric volume defined in \cite{Cvetic} as quoted in \eqref{Vgeo} is equal to the canonical black hole volume and thus is an instance of the use of the more general vector volume.  Additionally, they noted the relationship \eqref{VrA1} which is a higher-dimensional generalization of \eqref{volarearatio}. Of particular interest is the modified Smarr--Gibbs--Duhem relation (\ref{sgd}) and the thermodynamic volume relation (\ref{vthermo}). These results involve integrals over an infinite $(D-2)$-surface with modifications to remove the contribution from $\Lambda$ to the $E$ and $J$ integrals.  In particular, $\mathcal{V}_{th}$ is eventually derived as an integral over a 2-surface of the Killing potential for $\chi^\alpha$, the Killing vector proportional to the null generators on the horizon.

In order to demonstrate how the vector volume enters into (\ref{sgd}) in a somewhat natural way, we demonstrate a somewhat similar relation, modified in such a way as to focus on the vector volume.  The derivation here largely follows \cite{Poisson}.

For a hypersurface $\Sigma$ with boundary $S$, one form of Gauss' theorem relating to an antisymmetric tensor $B^{\alpha \beta}$ is
\begin{equation}
\int_\Sigma \nabla_\beta B^{\alpha \beta} \mathrm d \Sigma_\alpha = \frac{1}{2}\oint_S B^{\alpha \beta} \mathrm d S_{\alpha \beta}
\end{equation}
where $\mathrm d \Sigma_\alpha$ is the volume element and $\mathrm d S_{\alpha \beta}$ is the surface element.  For a Killing vector $\xi^\alpha$, $\nabla_\beta \nabla^\beta \xi^\alpha = -R^\alpha_\beta \xi^\beta$.  Additionally, $\nabla^\beta \xi^\alpha$ is an antisymmetric tensor.  As a result, for Killing vectors we can write Gauss' law as
\begin{equation}
\oint_S \nabla^\alpha \xi^\beta \mathrm d S_{\alpha \beta} = 2 \int_\Sigma R^\alpha_\beta \xi^\beta \mathrm d \Sigma_\alpha \label{Rtensor}
\end{equation}

Now let the hypersurface $\Sigma$ be a particular hypersurface spanning the black hole region. Its outer boundary is the horizon $H$.  Let its inner boundary be defined by $S'$, which we let be an arbitrarily small surface which encloses the singularity; in the Kerr--Newman--(anti) de Sitter class of spacetimes, we can define this by $t = const.,$ $r = \delta$ where we let $\delta \to 0$.  Now we use \eqref{Rtensor} in this case with our vector $k$ as defined in Section \ref{ourvolume}, noting that the integral over the boundary surface will consist of two parts---one for the horizon $H$ and one for the inner surface $S'$.
\begin{equation}
\int_\Sigma R^\alpha_\beta k^\beta \mathrm d \Sigma_\alpha = \frac{1}{2} \left(\oint_H \nabla^\alpha k^\beta \mathrm d S_{\alpha \beta} - \oint_{S'} \nabla^\alpha k^\beta \mathrm d S_{\alpha \beta}\right) \label{areavolume}
\end{equation}
For ease of representation, let $I_H = \oint_H \nabla^\alpha k^\beta \mathrm d S_{\alpha \beta}$ and $I_{S'} = \oint_{S'} \nabla^\alpha k^\beta \mathrm d S_{\alpha \beta}$.

We can now use Einstein's equations,
\begin{equation}
R^\alpha_\beta = 8\pi \left(T^\alpha_\beta - \frac{1}{2}T \delta^\alpha_\beta\right) + \Lambda \delta^\alpha_\beta
\end{equation}
to rewrite the left hand side of (\ref{areavolume}) as
\begin{equation}
8\pi \int_\Sigma T^\alpha_\beta k^\beta \mathrm d \Sigma_\alpha - 4\pi \int_\Sigma T k^\alpha \mathrm d \Sigma_\alpha + \Lambda \int_\Sigma k^\alpha \mathrm d \Sigma_\alpha,
\end{equation}
where of course $\int_\Sigma k^\alpha \mathrm d \Sigma_\alpha = \mathcal V_{k}$, volume of $\Sigma$, which, as $S'$ becomes smaller, approaches the volume of the black hole.  Let us define the integral over the stress-energy tensor as $I_T$,
\begin{equation}
I_T \equiv 4\pi \int_\Sigma \left(2T^\alpha_\beta k^\beta - T k^\alpha\right)\mathrm d \Sigma_\alpha.
\end{equation}
The left-hand side of \eqref{areavolume} is thus equal to $\Lambda \mathcal V_{k} + I_T$.  We note that in vacuum (or vacuum with $\Lambda$), $I_T$ will be identically zero.

We now find $I_H$.  As discussed in Section \ref{ourvolume}, on $H$ the Killing vector $k^\alpha$ has the property that it is tangent to the null generators, $k^\beta \nabla_\beta k^\alpha = k^\alpha$.  We can write $\mathrm d S_{\alpha \beta} = 2 k_{[\alpha} N_{\beta]} \mathrm d S$, where the square brackets denote anti-symmetrization and $N_\beta$ is an auxiliary null vector defined by $N^\alpha k_\alpha = -1$ and where $N^\alpha$ is orthogonal to the vectors $e^\alpha_A = \partial x^\alpha / \partial \theta^A$, where the $\theta^A$ represent coordinates on the horizon.  In this case, following a procedure similar to that given in \cite{Poisson}, we can define the area $\mathcal A$ in terms of $I_H$ as follows:
\begin{equation}
I_H = \oint_H \nabla^\alpha k^\beta (2 k_{[\alpha} N_{\beta]}) \mathrm d S = 2 \oint_H k^\alpha \nabla_\alpha k^\beta N_\beta \mathrm d S = 2 \oint_H k^\beta N_\beta \mathrm d S = -2\mathcal A,
\end{equation}
where in the last equality we used $k^\beta N_\beta = -1$ and that the integral of the surface element over the surface is $\mathcal A$.

To define $I_{S'}$, we first review the definitions of mass and angular momentum from the Komar formulae.  The black hole mass $M_H$ and angular momentum $J_H$ can be defined using Komar formulae as integrals over the horizon:
\begin{equation}
M_H = - \frac{1}{8\pi} \oint_H \nabla^\alpha t^\beta \mathrm d S_{\alpha \beta} \label{komarm}
\end{equation}
and
\begin{equation}
J_H = \frac{1}{16\pi} \oint_H \nabla^\alpha \phi^\beta \mathrm d S_{\alpha \beta}, \label{komarj}
\end{equation}
where $t^\alpha$ is the well-normalized time Killing vector and $\phi^\alpha$ is the well-normalized axial symmetry Killing vector.
These appear in the Smarr formula for stationary black holes along with the surface gravity $\kappa$ and the surface area of the horizon $A$,
\begin{equation}
M_H - 2 \Omega_H J_H = \frac{\kappa \mathcal A}{4\pi}.
\end{equation}
Along similar lines, then, define $M_{S'}$ and $J_{S'}$ by integrating over the limiting surface around the horizon, $S'$, instead of over the horizon:
\begin{equation}
M_{S'} = - \frac{1}{8\pi} \oint_{S'} \nabla^\alpha t^\beta \mathrm d S_{\alpha \beta} \label{komarms}
\end{equation}
\begin{equation}
J_{S'} = \frac{1}{16\pi} \oint_{S'} \nabla^\alpha \phi^\beta \mathrm d S_{\alpha \beta}. \label{komarj1}
\end{equation}
Since $k^\alpha = \kappa^{-1} (t^\alpha + \Omega_H \phi^\alpha)$, where $\kappa$ is the surface gravity and $\Omega_H$ the angular velocity of the black hole, we can now write
\begin{equation}
I_{S'} = \oint_{S'} \nabla^\alpha k^\beta \mathrm d S_{\alpha \beta} = \frac{-8 \pi M_{S'} + 16 \pi \Omega_H J_{S'}}{\kappa}.
\end{equation}

As a result, \eqref{areavolume} gives rise to a modified Smarr relation,
\begin{equation}
\Lambda \mathcal V_{k} + I_T = -\mathcal A +\frac{4\pi M_{S'}}{\kappa} - \frac{8 \pi \Omega_H J_{S'}}{\kappa}.
\end{equation}
If we now rewrite the volume term in terms of the volume $\mathcal V_{t} = \kappa \mathcal V_{k}$, this expression becomes
\begin{equation}
M_{S'} = \frac{\kappa \mathcal A}{4\pi} + 2 \Omega_H J_{S'} + \frac{\Lambda \mathcal V_{t}}{4 \pi} + \frac{\kappa I_T}{4 \pi}. \label{ourSmarr}
\end{equation}

To confirm that these definitions of mass and angular momentum might have meaning, we first check them in four-dimensional Schwarzschild--(anti) de Sitter space
($D=4$ and $\alpha(r) = 1 - 2 m /r - \Lambda r^2/3$ in (\ref{metricone})). We find that $M_{S'}$ approaches $m$ (as the $r = const.$ surface approaches $r = 0$), whereas $M_H = m - \Lambda r_+^3/3$ [originally erroneously written $m-\La r_+^2/3$], where $r_+$ is the value of $r$ on the horizon.  This helps give some weight to the definitions as presented.  In Kerr spacetime (line element \eqref{BL} with $Q = 0$), $M_H = M_{S'} = m$ and $J_H = J_{S'} = m a$.  In Kerr-(anti) de Sitter (line element (\ref{KdSLine}) with $Q=0$), we find a slightly different form,
\begin{equation}
M_{S'} = \frac{m\left(1 - \Lambda a^2 / 3\right)}{\Xi^2}, \qquad J_{S'} = \frac{m a}{\Xi^2}.
\end{equation}
Equation (\ref{ourSmarr}) is in fact very similar in form to (\ref{sgd}). In cases where the charge is nonzero, the Komar formulae integrated over $S'$ diverge and so some method to subtract out the charge contribution would need to be introduced.

[A more thorough discussion of Cveti\v{c} et al.'s~paper, including work done since the original publication of the content of this chapter, begins in Section \ref{CveticSection}.]

\subsection{Kodama Vector}

Since the Kodama vector from Section \ref{HaywardKodama} has a zero expansion in spherical symmetry, Hayward's Kodama volume \eqref{HaywardV},
wherein $K^\alpha$ is orthogonal to [the normal to] the boundary of $\Sigma$ [where $\Si$ is a region $0 \leq r \leq r_0$ for constant $r_0$], is clearly, in the case of spherical symmetry, a vector volume. If we write the metric for dynamic spherical symmetry in the form \eqref{Eofrmetric}, then $g_{r r} g_{t t} = -1$, so that the metric determinant is the same as that for Minkowski space in spherical polar coordinates, which explains why the Kodama volume for the region $0 \leq r \leq r_0$ is equal to the Euclidean volume of a sphere of $r=r_0$.  This implies that the Kodama volume is potentially a sensible generalization of the canonical black hole volume. 

\section{Conclusion} \label{PRDconclusion}

We have defined a vector volume in spacetime and have shown that this volume is a conserved, invariant quantity with several notable properties.  We defined a canonical black hole volume and showed that in Kerr--Schild metrics with a Minkowski background,  this volume corresponds to the Euclidean volume of the spatial component of the black hole region.  We have shown that the work of Parikh, Cveti\v{c} et al.~and Hayward involve the use of specific instances of the vector volume.  In addition to these, we proposed a null generator volume for non-degenerate stationary black holes. This volume has the advantage that it depends on neither the asymptotic properties of the spacetime nor the background metric in Kerr--Schild type spacetimes. Combining the canonical black hole volume and null generator volume, both of which are special cases of the vector volume, we have arrived at a new definition of the surface gravity.

\section{Paper Appendix: Vector volume element via differential forms} \label{vve}

This is a slightly more abstract version of the definition given in Section \ref{haywardlike}. We refer the reader to the readable account on differential forms by Israel \cite{Israel79}. The volume \eqref{haywarddef} can be expressed in terms of differential forms as follows.  We let $\epsilon$ be the Levi-Civita tensor, or volume $D$-form, such that the $D$-volume $\mathcal V$ of a generic region $\mathcal Q$ is
\begin{equation}
\mathcal V_\mathcal Q = \int_\mathcal Q \epsilon.
\end{equation}
We now define a  $(D-1)$-form ``vector volume element,'' $\delta \mathcal V_v$, by
\begin{equation}
\delta \mathcal V_v \equiv i_v \epsilon
\end{equation}
where $i_X \alpha$ is the $(n-1)$-form interior product of the $n$-form $\alpha$ with the vector $X$.  We then define the vector volume by integrating this $(D-1)$-form along the hypersurface region $\Gamma$ as defined in Section \ref{differentiallike}:

\begin{equation}
\mathcal V_{v} = \int_{\Gamma\cap\mathcal R} \delta \mathcal V_v = \int_{\Gamma\cap\mathcal R} i_v \epsilon.
\end{equation}

In adapted coordinates where $v^\alpha = \delta^\alpha_0$, $\epsilon_{\alpha \beta \gamma \ldots \mu} = \sqrt{|g_D|} [\alpha \: \beta \: \gamma \ldots \mu]$, where $ [\alpha \: \beta \: \gamma \ldots \mu]$ is equal to $1$ ($-1$) if $\alpha,\beta,\gamma\ldots \mu$ is an even (odd) permutation of $0,1,2\ldots D-1$ and zero otherwise.  This implies that $i_v \epsilon$ is a $(D-1)$-form with components
\begin{eqnarray}
(i_v \epsilon)_{\alpha \beta \ldots \mu} = v^\nu \epsilon_{\nu \alpha \beta \ldots \mu} = \epsilon_{0 \alpha \beta \ldots \mu} = \\ \nonumber \sqrt{|g_D|} [0 \: \alpha \: \beta \ldots \mu].
\end{eqnarray}

The definition of the integral of an $n$-form over an $n$-dimensional manifold is given by Wald \cite{Wald84} as follows.  If there exist coordinates $x^0, x^1, \ldots x^{n-1}$ on the manifold, then an $n$-form $\alpha$ defined on the manifold can be written in the form $\alpha = a \mathrm d x^0 \wedge \mathrm d x^1 \wedge \ldots \mathrm d x^{n-1}$, where the wedge symbol denotes a totally antisymmetric product and $a$ is a scalar.  The integral of $\alpha$ is then defined as
\begin{equation}
\int \alpha = \int a \mathrm d x^0 \mathrm d x^1 \ldots \mathrm d x^{n-1},
\end{equation}
or the integral of the scalar over the product of the differentials.

The manifold over which $i_v \epsilon$ is defined is the hypersurface $\Gamma$, and we can use coordinates $x^1, x^2, \ldots x^{D-1}$ for the $(D-1)$-dimensional hypersurface.  In these coordinates, we can write
\begin{equation}
i_v \epsilon = \sqrt{|g_D|} \: \mathrm d x^1 \wedge \mathrm d x^2 \ldots \mathrm d x^{D-1},
\end{equation}
so that
\begin{equation}
\int_{\Gamma\cap\mathcal R} \de \mathcal V_v = \int_{\Gamma\cap\mathcal R} i_v \epsilon = \int_{x^i \in \Sigma} \sqrt{|g_D|} \mathrm d^{D-1} x,
\end{equation}
and we recover \eqref{gexpress}.  $\mathcal V_v$, as the [integral of the] interior product of the vector field $v$ with the volume form $\epsilon$, has a very clear interpretation, which is the advantage of presentation given in this Appendix.

We can now show that the reason for the invariance of $\mathcal V_{v}$, under the choice of $\Gamma$, is because the contribution of the ``vector volume element'' from each individual integral curve of $v$ is the same regardless of its position along the curve.  To demonstrate this, we take the Lie derivative $\mathcal L_v$ of the vector volume element along the vector field $v$, in its formulation as a $(D-1)$-form $i_v \epsilon$. This gives
\begin{equation}
\mathcal L_v (\delta \mathcal V_v) = \mathcal L_v (i_v \epsilon) = \text{div}(v) \text{ } i_v \epsilon, \label{liev}
\end{equation}
where $\text{div}(v) = \nabla_\mu v^\mu$ is the divergence of $v$.  (We show that $\mathcal L_v(i_v \epsilon) = \text{div}(v) \, i_v\epsilon$ in Section \ref{lieproof}.)  This demonstrates that the vector volume element contribution from each integral curve is independent of position along the curve if and only if vector field be divergence-free. This definition emphasizes that the total vector field volume can be interpreted as the Riemann sum of contributions $\delta \mathcal V_v$ from each individual integral curve of $v$, and the result \eqref{liev} shows that $\delta \mathcal V_v$ is constant along each integral curve. This is the reason why $\mathcal V_{v}$ is independent of the choice of hypersurface.

\section{Paper Appendix: Lie Derivative Proof} \label{lieproof}
We wish to prove that
\begin{equation}\label{bb}
\mathcal L_v (i_v \epsilon) = \text{div}(v) i_v \epsilon
\end{equation}
as required by (\ref{liev}). We start with Cartan's identity which states that
\begin{equation}
\mathcal L_X \omega = \mathrm d (i_X \omega) + i_X \mathrm d \omega
\end{equation}
for vector $X$, differential form $\omega$, and exterior derivative $\mathrm d$. Note that $i_X^2 = 0$ and $\mathrm d^2 = 0$. We have
\begin{equation}
\mathcal L_v (i_v \epsilon) = \mathrm d(i_v (i_v \epsilon) )+ i_v \mathrm d (i_v \epsilon)
\end{equation}
with $\mathrm d (i_v (i_v \epsilon)) = \mathrm d (i_v^2 \epsilon) = 0$ since $i_v^2 \omega = 0$ for any differential form $\omega$.  Further, we can write $\mathrm d(i_v \epsilon) = \mathcal
L_v \epsilon - i_v \mathrm d \epsilon$ by applying Cartan's identity again.  The $i_v \mathrm d \epsilon$ term becomes zero when the interior product $i_v$ is taken with it.  As a result, we can write
\begin{equation}
\mathcal L_v (i_v \epsilon) = i_v \mathcal L_v \epsilon.
\end{equation}
Now from the definition of divergence we have $\mathcal L_v \epsilon = \text{div}(v) \epsilon$ and so
\begin{equation}
\mathcal L_v (i_v \epsilon) = i_v (\text{div}(v) \epsilon).
\end{equation}
Since the interior product is the contraction of a form with the vector field, the scalar $\text{div}(v)$ can be brought outside the interior product and we arrive at (\ref{bb}) as required.

\section{Addendum} \label{addendum}

\emph{The following addendum was written with the rest of this thesis, and not in the original paper:}

A clarification, inspired by comments from Dr.~Don Page on a previous version of this document's abstract, is the following. For simplicity consider the four-dimensional case, where the boundary of $\mc R$ is a three-dimensional hypersurface. The main scenario I was considering in this chapter is one where the boundary to $\mc R$ is topologically essentially an ``infinite cylinder,'' something like $R \times S^2$ where $R$ is the real line and $S^2$ the two-sphere. More generally, since $\mc R$ in adapted coordinates is chosen to have $-\infty < x^0 < \infty, x^i \in \Si$, $\pa \mc R$ is $-\infty < x^0 < \infty, x^i \in \pa \Si$, and so the topology is the product of the real line with the topology of $\pa \Si$. That is to say, assume that the vertical lines in Fig.~\ref{diagram2} extend infinitely.

\begin{figure}[ht]
\centering
\epsfig{file=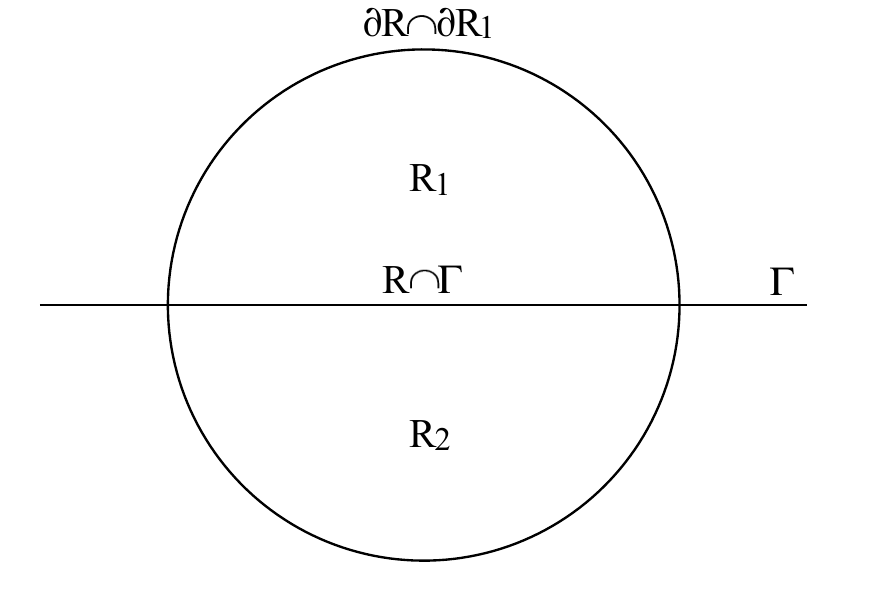,height=2in,width=3in,angle=0}
\caption{\label{sphereDiagram} The situation where the region $\mc R$ is something like a closed ball, with boundary $\pa \mc R$ corresponding to a 3-sphere (if $\mc R$ is four-dimensional). Here a hypersurface $\G$ cuts $\mc R$ into two regions, $\mc R_1$ and $\mc R_2$, shown. The boundary $\pa \mc R_1$ is divided into one region which is a common boundary to both $\mc R$ and $\mc R_1$, identified as $\pa \mc R \cap \pa \mc R_1$, and the intersection of $\mc R$ with $\G$, $\mc R \cap \G$.}
\end{figure}

If instead the boundary of $\pa \mc R$ has a topology like $S^3$, then the hypersurface $\G$ will divide the boundary of $\mc R$ into two regions with the topology of 3-balls, that is (topologically) the objects to the interior of the 2-sphere including the boundary. Let $\G$ divide $\mc R$ into $\mc R_1$ and $\mc R_2$. Then consider $\mc R_1$. Its boundary will be one part which shares its boundary with that of $\mc R$, which we can label $\pa \mc R_1 \cap \mc \pa \mc R$, and another part which is the intersection of $\G$ with $\mc R$. This situation is shown schematically in Fig.~\ref{sphereDiagram}. 

Then we have, choosing suitable orientation,
\begin{align}
    \int_{\mc R_1} \na_\alpha v^\alpha d^4 \mc V = \oint_{\pa \mc R_1 \cap \mc R} v^\alpha d\Sigma_\alpha - \oint_{\G \cap \mc R} v^\alpha d\Sigma_\alpha.
\end{align}
Since $\na_\alpha v^\alpha = 0$ and $\oint_{\pa \mc R_1} v^\alpha d \Sigma_\alpha = 0$, we conclude that in this case,
\begin{align}
    \oint_{\G \cap \mc R} v^\alpha d \Sigma_\alpha = 0,
\end{align}
so that the vector volume is a constant independent of the choice of $\G$ but that this constant is zero. 

\begin{figure}[ht]
\centering
\epsfig{file=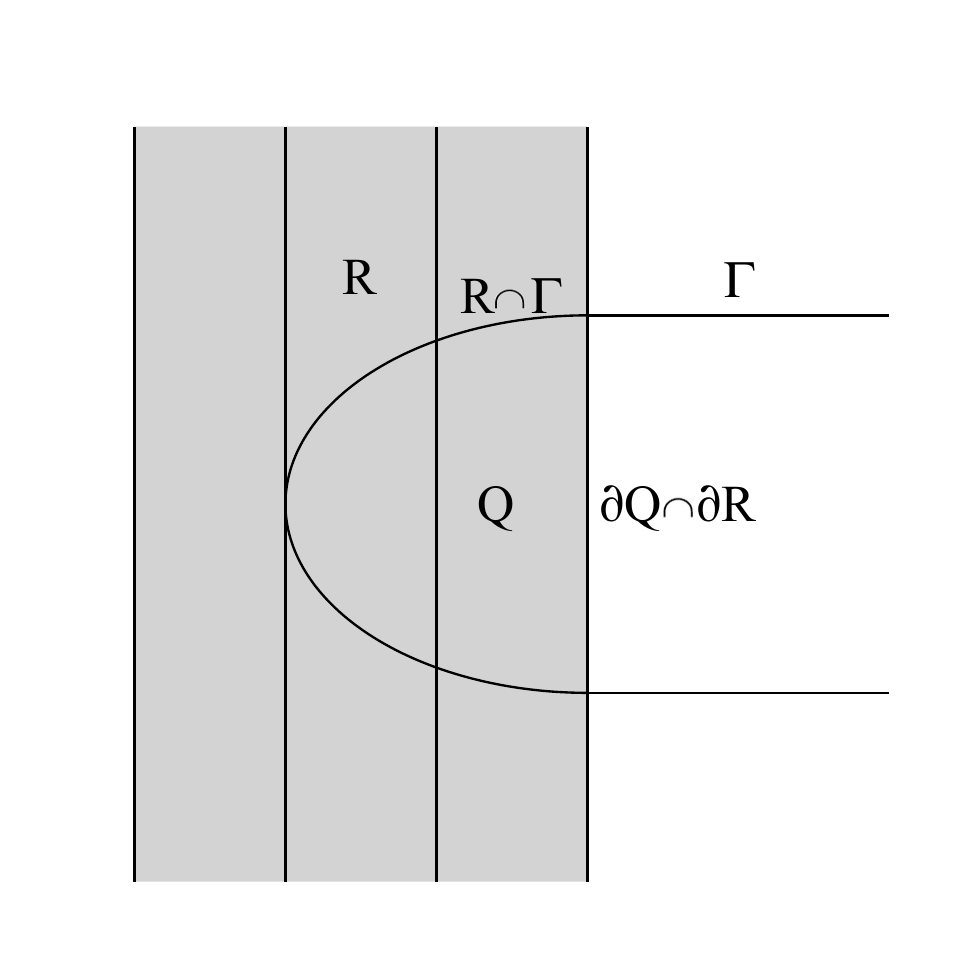,height=3in,width=3in,angle=0}
\caption{\label{sidecutoutdiagram} The shaded region is $\mc R$, and the vertical lines represent the integral curves  of vector $v$ within it. $\G$ here is a hypersurface which in this case ``cuts into the side of $\mc R$,'' and so does not intersect with each integral curve once (intersecting some twice and some zero times). The region $\mc Q$ is the region ``between $\G$ and $\pa \mc R$'' and the boundaries of $\mc Q$ are the portion of $\pa \mc Q$ lying within $\pa \mc R$, given by $\pa \mc Q \cap \pa \mc R$, and the portion of $\G$ lying within $\mc R$, given by $\G \cap \mc R$.}
\end{figure}

The fact that the vector volume associated with the ``infinite cylinder'' scenario usually considered is nonvanishing comes down to the fact that if $\pa \mc R$ starts as an ``infinite cylinder,'' $\Gamma$ will divide $\mc R$ into two  ``half-infinite cylinders,'' which do not ``close back up,'' so that to get a finite region for the Gauss' law integration requires another hypersurface $\Gamma_2$. By contrast, if $\Gamma$ is chosen in the ``infinite cylinder'' $\mc R$ case to ``take a chunk out of the side of $\mc R$,'' as shown in Fig.~\ref{sidecutoutdiagram}, such that it divides $\mc R$ into one region $\mc Q$ with a closed boundary consisting of one part that it shares with $\pa \mc R$ (given by $\pa \mc Q \cap \pa \mc R$) and another part which is $\mc R \cap \G$, then by the same argument as above, $\int_\mc Q \na_\alpha v^\alpha = 0$ and so $\oint_{\pa \mc Q} v^\alpha d \Si_\alpha = 0$, so that we must have $\int_{\G \cap \mc R} v^\alpha d \Sigma_\alpha = 0$. The way the chapter precluded this instance is the stipulation in Section \ref{differentiallike} that $\Gamma$ intersect each of the integral curves of $v^\alpha$ exactly once. In the scenario depicted in Fig.~\ref{sidecutoutdiagram}, some of the integral curves are intersected twice, and some not at all. 

I will thus make the following clarification. Let $\G$ divide $\pa \mc R$ into two components, which I will call $\pa \mc R_1$ and $\pa \mc R_2$, so that $\pa \mc R_1 \cup \pa \mc R_2 = \pa \mc R$, $\pa \mc R_1 \cap \pa \mc R_2 = \pa \mc R \cap \G$. If either $\pa \mc R_1$ or $\pa \mc R_2$ are \emph{contractible} (can be continuously deformed into a point), then $\int_{\G \cap \mc R} v^\alpha d \Sigma_\alpha = 0$. For a nonzero vector volume, it is required that neither $\pa \mc R_1$ nor $\pa \mc R_2$ is contractible.  

Another point that is important is that the construction of adapted coordinates in which $\sqrt{-g}$ is independent of $x^0$ and $v^\alpha = \de^\alpha_0$ implicitly assumed that, in these adapted coordinates, the metric determinant remained independent of $x^0$ throughout the portion of the manifold covered by this coordinate chart.

It might be that we are only interested in a subset of the full manifold, if we cannot find a set of adapted coordinates which cover the entire manifold at once. For example, in the Schwarzschild black hole, we may consider Eddington--Finkelstein coordinates 
\begin{align}
    ds^2 = - (1-2m/r) dv^2 + 2 dv dr + r^2 (d\tht^2 + \sin^2 \tht d \phi^2)
\end{align}
which are regular across the (future) horizon. Of course $0 \leq \tht \leq \pi$ and $0 \leq \phi < 2\pi$ with a periodic identification of the $\phi$ coordinate with the points $\phi = \phi_0$ and $\phi = \phi_0 + 2\pi$ identified with each other, as usual. The $v$ coordinate satisfies $-\infty < v < \infty$. Consider $\xi^\alpha = \de^\alpha_v$, a Killing vector. In these coordinates, defined for $-\infty < v < \infty$, the black hole region $\mc B$ defined by $r_0 \leq r \leq 2m$ with $-\infty < v < \infty$ (and with arbitrary $\tht, \phi$) has boundary given by $r=r_0$ and $r = 2m$; we can let $r_0$ become arbitrarily small (but positive). $\xi^\alpha$ is tangent to the outer and inner boundaries consisting of $r = r_0$ and $r = 2m$, and no other boundaries appear in the portion of the manifold covered by this coordinate chart. Integral curves of $\xi^\alpha$ are the curves of constant $(r,\tht,\phi)$ extending over all $v$. A region $\G$ which intersects each of the integral curves of $\xi^\alpha$ exactly once within the part of the manifold covered by this coordinate chart has the form $v = \Phi(r,\tht,\phi)$ (for some function $\Phi$). The vector volume associated with $\xi$ and $\mc B$ will be $4\pi( (2m)^3 - r_0^3)/3$ or $4\pi (2m)^3/3$ in the limit as $r_0 \to 0$. The statement that $\G$ intersects all the integral curves of $v^a$ once does not create any problems so far. If we include the whole manifold, however, say by using Kruskal--Szekeres coordinates, the black hole region contains the bifurcation surface where $\xi^\alpha$ vanishes, which will necessarily not appear in a set of adapted coordinates. Finding a hypersurface $\G$ which intersects each of the integral curves of $\xi^\alpha$ will not be possible in this case. 

A more precise statement then is the following: given a vector $v^\alpha$, assume that there is some portion of the spacetime in which $v^\alpha$ is divergence-free, and that we can then find a coordinate chart adapted to $v^\alpha$ such that $v^\alpha = \de^\alpha_0$ and such that, in these coordinates, $-\infty < x^0 < \infty$, with the other coordinates denoted $x^i$. Then let $\mc R$ be an infinite region which lies entirely within the portion of the manifold covered by this coordinate chart, corresponding to $x^i \in \Si, -\infty < x^0 < \infty$, for some $(D-1)$-dimensional region $\Si$. Then let $\G$ be a hypersurface given by $x^0 = \Phi(x^i)$, where $\Phi$ is a smooth function. Then $\G$ automatically intersects the integral curves of $v^\alpha$ \emph{appearing in the part of the manifold covered by this coordinate chart} exactly once, and $\int_{\G \cap \mc R} v^\alpha d \Si_\alpha = \int_{\Si} \sqrt{-g} d^{D-1}x$ is independent of the choice of $\G$. The statement is then tied to the specific adapted coordinate system and the global situation for the full manifold might be more complicated.

A final clarification is that when discussing the vector volume of black holes, I am generally for simplicity assuming stationary black holes with a global Killing vector, for which the adapted time coordinate associated with stationarity extends from $- \infty$ to $+ \infty$. The volume calculation in Section \ref{OurVolumeIntro} considers a scenario where a black hole is formed by the collapse of some timelike boundary, but specifically requires that the region over which the growth rate is calculated be ``after the black hole has already been formed.''

\chapter{The Generalized Kerr--anti-de Sitter Spacetime and its Properties} \label{GKAdSChapter}

In this chapter, I describe the Kerr--anti-de Sitter (Kerr--AdS) metrics (in arbitrary dimension) and define a generalized form of them by allowing the mass function to vary with radius function $r$. I introduce these metrics and show them in various forms in Section \ref{KAdSForms}.  I then relate these to the broader class of Generalized Kerr--NUT--AdS spacetimes and describe their symmetries, including existence of the Principal Conformal Killing-Yano Tensor $\bs h$ and its associated vector $\beta$, in Section \ref{GKNAdSSection}. (My terminology is different from Houri et al.'s~\cite{Houri08} meaning of Generalized Kerr--NUT--AdS---what I call Generalized Kerr--NUT--AdS is what Houri et al.~refer to as Kerr--NUT--AdS or off-shell Kerr--NUT--AdS; I apologize for any confusion.) I specify back to the Generalized Kerr--AdS spacetimes and make some notes on their Kerr--Schild null vector and black hole horizons in Section \ref{KadSKSForm}. In Section \ref{pseudoCartesian}, I introduce a set of pseudo-Cartesian coordinates for the spacetime, which I believe to be novel for Kerr--AdS. The Generalized Kerr--anti-de Sitter spacetimes will be the main focus throughout the rest of this thesis. 

\section{Forms of the Kerr--(anti-)de Sitter Metric} \label{KAdSForms}

The Kerr--(anti-)de Sitter metrics in arbitrary dimension were first given by Gibbons, L\"u, Page and Pope (2005) \cite{GibbonsLu,GibbonsLu2}, who gave (simultaneously) the Kerr--de Sitter and Kerr--anti-de Sitter solutions. I will focus on the Kerr--anti-de Sitter solutions in this work, but the basic forms of the solutions are identical, and can be written in a form which shows the solution for arbitrary $\La$ (including the Myers--Perry black holes, which are the higher-dimensional generalization of the $\La = 0$ Kerr solutions). These expressions were found by assuming a Kerr--Schild ansatz, in which a background anti-de Sitter spacetime, written in spheroidal coordinates, is modified by a term related to a null vector (as shown below). The method was to take the known solutions for dimension $D = 4, 5$, to generalize them in a natural way, and then to check the solutions. The solutions are then expressed in three forms: the Kerr--Schild form (KS), a Boyer--Lindquist form (BL) and an Alternative Boyer--Lindquist form (ABL). I will slightly modify notation as well as incorporate some notation from \cite{GibbonsPerry}. 

The solutions given in \cite{GibbonsLu,GibbonsLu2} and developed/clarified in subsequent papers are for dimension $D \geq 4$ only, with the case $D < 4$ not considered. 

In Kerr--anti-de Sitter, (as in pure anti-de Sitter), the Ricci tensor can be written as
\begin{align}
    R_{a b} &= -\frac{D-1}{l^2} g_{ab}, \label{RicciTensorEquation}
\end{align}
where we take $l$ to be a positive constant. The Kerr--de Sitter solution can be found by letting $\la = -l^{-2}$ in all the subsequent equations and allowing it to be positive, and the Myers--Perry solutions can be found by setting $\la = 0$. Note that since $G_{ab} + \La g_{ab} = 0$ and $G_{ab} = R_{ab} - \f12 R g_{ab}$, $R_{ab} = 2 (D-2)^{-1} \La g_{ab}$ and so $l^2$ satisfies
\begin{align}
    l^2 &= -\frac{(D-1)(D-2)}{2 \La}. \label{l2}
\end{align}
We take $l$ to be positive.

For a Kerr--Schild metric of the form $g_{ab} = \bar g_{ab} + h_{ab}$, where $h_{ab} = H k_a k_b$, where $k_a$ is null, the Kerr--Schild correction can be raised and lowered using either metric. The Ricci tensor for the full spacetime is related to that of the background spacetime according to \cite{Dereli}.
\begin{align}
    R^a_b &= \bar R^a_b - h^a_c \bar R^c_b + \f12 \left( \bar \na_c \bar \na_b h^{ac} + \bar \na^c \bar \na^a h_{bc} - \bar \na^c \bar \na_c h^a_b\right). \label{KSRicciRelationship}
\end{align}

In dimension $D$, let 
\begin{align}
    n &\equiv \left\lfloor \f{D}{2}\right\rfloor, \label{nDby2}
\end{align}
where $\lfloor x\rfloor$ is the floor function applied to $x$ (the largest integer less than or equal to $x$). Further, let
\begin{align}
    \varepsilon &\equiv D \bmod 2, \label{varepsilon}
\end{align}
so that $\varepsilon = 1$ if $D$ is odd and $0$ if $D$ is even. These conventions are chosen to match with \cite{KrtousKubiznak} and related papers. (\cite{KrtousKubiznak} also uses $k = n +\varepsilon$, but I am using $k$ for other quantities with enough frequency that I will just use $n+\varepsilon$ when needed. The original papers by Gibbons et al.~\cite{GibbonsLu,GibbonsLu2} uses $N = \lfloor(D-1)/2\rfloor = n-1+\ve$ and $\epsilon = 1-\varepsilon$.)

Introduce $n$ direction cosines $\mu_i$, subject to the constraint
\begin{align}
    \sum_{i = 1}^{n} \mu_i^2 = 1.\label{summu}
\end{align}
There will also be associated $n-1+\ve$ azimuthal angular coordinates $\phi_i$, a radial coordinate $r$, and the time coordinate $t$. 

Each azimuthal coordinate is canonically normalized and has associated period $2\pi$. When $D$ is odd, all the $\mu_i$ lie in the interval $0 \leq \mu_i \leq 1$. When $D$ is even, the $\mu_i$ lie in this interval except that $\mu_{n}$ is allowed to satisfy $-1 \leq \mu_{n} \leq 1$. Note that there are in total $D+1$ coordinates, together with the constraint \eqref{summu}. Since $\mu_n$ is also the latitude variable which is eliminated and so is not integrated over, there is an extra factor of 2 that must be inserted into the results of integrals over all the latitude variables, which corresponds to the fact that there are two solutions $\mu_n$ for each set of $\{\mu_1, \ldots, \mu_{n-1}\}$ (except for the trivial case when $\mu_n = 0$).

The Kerr--Schild form is 
\begin{align}
    d s^2 &= d \bar s^2 + H (k_a dx^a)^2,
\end{align}
or $g_{ab} = \bar g_{ab} + H k_a k_b$, where $ds^2$ is the full spacetime line element (in KS form) with $d \bar s^2$ the line element of pure anti-de Sitter in spheroidal coordinates. $H$ is a scalar function, which for Kerr--AdS can be decomposed into
\begin{align}
    H &= \f{2m}{U}, \label{His2mbyU}
\end{align}
where $m$ is a constant parameter, associated with but not necessarily equal to the mass of the black hole. (Later on, beginning in Section \ref{GKAdSSection}, I will generalize the Kerr--AdS solutions by replacing $m$ with a generic function $\mu(r)$ of $r$.) The values for $d\bar s^2, k_a$ and $U$ are given below. 

Let 
\begin{align}
    \Xi_i &= 1 - \frac{a_i^2}{l^2} \nn 
    W &= \sum_{i = 1}^{n} \frac{\mu_i^2}{\Xi_i} \nn
    F &= \frac{r^2}{1 + r^2 l^{-2}} \sum_{i = 1}^{n} \frac{\mu_i^2}{r^2+a_i^2}. \label{GibbonsXiWF}
\end{align}

The KS null vector can be expressed in covariant and contravariant form as
\begin{align}
    k_a dx^a &= W dt + F dr - \sum_{i = 1}^{n-1+\ve} \f{a_i \mu_i^2}{\Xi_i} d\phi_i \nn 
    k^a \pa_a &= - \f{1}{1+r^2l^{-2}} \f{\pa}{\pa t} + \f{\pa}{\pa r} - \sum_{i = 1}^{n-1+\ve} \f{a_i}{r^2+a_i^2} \f{\pa}{\pa \phi_i}. \label{kmu}
\end{align}
Note that this implies that $k$ is past-pointing. (I am following the convention given by \cite{GibbonsLu,GibbonsLu2} here.) Note that if we let $t = l \phi_0$ and $l = a_0$, then $k^a \partial_a$ reduces to $\partial_r - \sum_{i = 0}^{n-1+\ve} \frac{a_i}{r^2+a_i^2} \partial_{\phi_i}$. The scalar function $U$ is given by
\begin{align}
    U &= r^{1-\varepsilon} \sum_{i = 1}^{n} \frac{\mu_i^2}{r^2+a_i^2} \prod_{j = 1}^{n-1+\ve} (r^2+a_j^2). \label{Udefinition}
\end{align}

In even dimension, let us call $\phi_n \equiv 0$ and $a_n \equiv 0$, which allows for more compact expressions. The angular $\phi_i$ are nonzero for $i$ varying from 1 to $n-1+\ve$.  

Finally, the background metric is given by 
\begin{align}
    d\bar s^2 &= - W (1 + r^2l^{-2}) dt^2 + F dr^2 + \sum_{i = 1}^{n} \frac{r^2+a_i^2}{\Xi_i} (d\mu_i^2 + \mu_i^2 d\phi_i^2) \nn
    &\qquad - \frac{1}{W (r^2+l^2)} \left(\sum_{i = 1}^{n} \frac{(r^2+a_i^2)\mu_i d \mu_i}{\Xi_i}\right)^2. \label{dbarsspheroidal}
\end{align}
(Note again that, in even dimensions, the term $d\phi_{n} = 0$.)

The background metric is pure anti-de Sitter in unusual coordinates. The usual form of anti-de Sitter, using coordinate $y$ to represent the aerial radius, is 
\begin{align}
    d\bar s^2 &= -(1 + y^2l^{-2}) dt^2 + \frac{dy^2}{1 + y^2 l^{-2}} + y^2 \sum_{i = 1}^{n} ( d \hat \mu_i^2 + \hat \mu_i^2 d\phi_i^2), \label{dbarsspherical}
\end{align}
where the $\hat \mu_i$ are spherical polar latitude coordinates, subject to the constraint
\begin{align}
    \sum_{i = 1}^{n} \hat \mu_i^2 = 1.
\end{align}
The coordinate transformation required to bring the background metric \eqref{dbarsspheroidal} into the form \eqref{dbarsspherical} is
\begin{align}
    y^2 \hat \mu_i^2 &= \frac{(r^2+a_i^2) \mu_i^2}{\Xi_i} \label{spheroidaltransformation}
\end{align}
for $1 \leq i \leq n$. Note that $t$ and $\phi_i$ are unchanged, so that the $t$ and $\phi_i$ coordinates maintain their meanings; in particular, the Killing vectors $\pa/\pa t$ and $\pa /\pa\phi_i$ refer to the same spacetime symmetries in the two versions of the AdS background metric. 

The KS form for four-dimensional Kerr--anti-de Sitter is (using $\mu_1 = \sin \tht, \mu_2 = \cos \tht$)
\begin{align}
    ds^2 &= d\bar s^2 + \frac{2m}{U} (k_a dx^a)^2 \nn
    d \bar s^2 &= - \frac{(1+r^2l^{-2}) \Delta_\tht dt^2}{\Xi} + \frac{\rho^2 dr^2}{(1+r^2l^{-2}) (r^2+a^2)} + \frac{\rho^2 d\tht^2}{\Delta_\tht} + \frac{(r^2+a^2) \sin^2\tht d\phi^2}{\Xi} \nn 
    \Xi &= 1 -\frac{a^2}{l^2} \nn 
    \rho^2 &= r^2 + a^2 \cos^2\tht \nn 
    \Delta_\tht &= 1 - \frac{a^2 \cos^2\tht}{l^2} \nn 
    k_a dx^a &= \frac{\Delta_\tht d t}{\Xi} + \frac{\rho^2 dr}{(1+r^2l^{-2})(r^2+a^2)} - \frac{a \sin^2\tht d \phi}{\Xi} \nn 
    k^a \partial_a &= -\f{1}{1+r^2l^{-2}} \partial_t + \partial_r - \f{a}{r^2+a^2} \partial_\phi \nn 
    U &= \rho^2/r. \label{KS4D}
\end{align}

Gibbons et al.~were led to the higher-dimensional expressions using the Kerr--Schild forms of the Kerr--(anti-)de Sitter metrics in $D=4,5$ and making ``natural generalizations.'' The Ricci tensor equation \eqref{RicciTensorEquation} was then checked up to $D = 11$ using \eqref{KSRicciRelationship} using \emph{Mathematica}.

Gibbons et al.~also give a Boyer--Lindquist (BL) form of the metric, obtained by defining coordinates $\tau$ and $\varphi_i$ related to the KS coordinates by
\begin{align}
    dt &= d \tau + \frac{2 m dr}{(1 + r^2l^{-2})(V - 2m)} \label{ttau} \\
    d \phi_i &= d \varphi_i + a_i l^{-2} d \tau + \frac{2 m a_i dr}{(r^2+a_i^2)(V - 2m)} \label{phivarphi}
\end{align}
where
\begin{align}
    V &= \frac{U}{F} = \f{1}{r^{1+\varepsilon}} (1 + r^2l^{-2})\prod_{i = 1}^{n-1+\ve} (r^2+a_i^2), \label{GibbonsV}
\end{align}
which we note is a function of $r$ only (so we can write $V = V(r)$). 

Under this coordinate transformation the metric becomes
\begin{align}
    ds^2 &= -W(1+r^2l^{-2})d \tau^2 + \frac{U dr^2}{V-2m} + \frac{2m}{U} \left(d \tau - \sum_{i = 1}^{n-1+\ve} \frac{a_i\mu_i^2 d\varphi_i}{\Xi_i}\right)^2 + \nn
    &\qquad \sum_{i = 1}^{n} \frac{r^2+a_i^2}{\Xi_i} \left( d\mu_i^2 + \mu_i^2 (d \varphi_i - a_i l^{-2} d\tau)^2\right) - \frac{1}{W(r^2+l^2)} \left( \sum_{i = 1}^{n} \frac{(r^2+a_i^2) \mu_i d \mu_i}{\Xi_i}\right)^2. \label{BLmetric}
\end{align}

The four-dimensional solution was found by Carter \cite{CarterMetric}, and can be expressed in BL coordinates in the form given by \eqref{KdSLine}, as in \cite{GibbonsPerry}, or
\begin{align}
    ds^2 &= - \f{\Delta}{\rho^2} \left[ d \tau - \f{a}{\Xi} \sin^2\tht d \varphi\right]^2 + \f{\rho^2 dr^2}{\Delta} + \f{\rho^2 d \tht^2}{\Delta_\tht} + \f{\Delta_\tht \sin^2\tht}{\rho^2} \left[ a d\tau - \f{r^2+a^2}\Xi d \varphi\right]^2, \label{BL4D}
\end{align}
with $\rho^2, \Xi, \Delta_\tht$ as in \eqref{KS4D} and 
\begin{align}
    \Delta = (r^2+a^2)(1+r^2l^{-2}) - 2m r.
\end{align}

A third set of coordinates, which I will call the Alternative Boyer--Lindquist (ABL) coordinates, is obtained by performing the coordinate transformation \eqref{ttau} but, instead of using coordinates $\varphi_i$, using $\hat \varphi_i$, related to $\phi_i$ by
\begin{align}
    d \phi_i &= d \hat \varphi_i + \frac{2 m a_i dr}{(r^2+a_i^2)(V-2m)}. \label{dphidphihat}
\end{align}
Make the choices for the integration constants so that $\hat \varphi_i = \varphi_i + a_i l^{-2} \tau$.

The metric in this form is given by
\begin{align}
    ds^2 &= - W(1+r^2l^{-2}) d\tau^2 + \frac{2m}{U} \left( W d \tau - \sum_{i = 1}^{n-1+\ve} \frac{a_i \mu_i^2 d \hat \varphi_i}{\Xi_i}\right)^2 + \sum_{ i = 1}^{n-1+\ve} \frac{r^2+a_i^2}{\Xi_i} \mu_i^2 d \hat \varphi_i^2 \nn &\qquad + \frac{U dr^2}{V-2m} + \sum_{ i = 1}^{n} \frac{r^2+a_i^2}{\Xi_i} d\mu_i^2 - \frac{1}{W (r^2+l^2)} \left( \sum_{ i = 1}^{n}\frac{r^2+a_i^2}{\Xi_i} \mu_i d \mu_i\right)^2. \label{ABL}
\end{align}
The four-dimensional form is
\begin{align}
    ds^2 &= - \f{\Delta}{\Xi^2 \rho^2} \left[ \Delta_\tht d \tau - a \sin^2\tht d \hat \varphi\right]^2 + \f{\rho^2 dr^2}{\Delta} + \f{\rho^2 d\tht^2}{\Delta_\tht} + \f{\Delta_\tht \sin^2\tht}{\Xi^2 \rho^2} \left[ a \left(1 + \f{r^2}{l^2}\right) d \tau - (r^2+a^2) d \hat \varphi\right]^2. \label{ABL4D}
\end{align}

A crucial difference between the BL and ABL coordinates is that the ABL coordinates, which are also used as the starting point for Chen et al.~\cite{Chen}, are asymptotically static, and the BL coordinates are not. By ``asymptotically static'' I mean that in the BL coordinates,
\begin{align}
    g_{\tau \varphi_i} &= -\f{r^2}{l^2} \sum_{i = 1}^{n} \frac{a_i \mu_i^2}{\Xi_i} + \mc O(r)
\end{align}
whereas in the ABL coordinates,
\begin{align}
    \lim_{r\to \infty}g_{\tau \hat \varphi_i} &= 0. 
\end{align}
This difference turns out to be very important and will be discussed further.

The Kerr--Schild null vector in the ABL coordinates is
\begin{align}
    k_a dx^a &= \frac{U dr}{V-2m} + W d \tau - \sum_{i = 1}^{n-1+\ve} \frac{a_i \mu_i^2 d \hat \varphi_i}{\Xi_i} \nn 
    k^a \partial_a &= \frac{\pa}{\pa r} - \frac{V}{V-2m} \left( \f{1}{1+r^2l^{-2}} \frac{\pa}{\pa \tau} + \sum_{i = 1}^{n-1+\ve} \frac{a_i}{r^2+a_i^2} \frac{\pa}{\pa \hat \varphi_i}\right).
\end{align}
In four dimensions,
\begin{align}
    k_a dx^a &= \f{\rho^2 d r}{\Delta} + \f{\Delta_\tht d \tau}{\Xi}  - \f{a \sin^2\tht d \hat \vp}{\Xi} \nn 
    k^a \pa_a &= \f{\pa}{\pa r} + \f{\Delta+2mr}{\Delta} \left( \f{1}{1+r^2/l^2} \f{\pa}{\pa \tau} + \f{a}{r^2+a^2} \f{\pa}{\pa \hat \vp}\right). \label{k4DABL}
\end{align}

It is the ABL coordinates that are used in GPP \cite{GibbonsPerry} to do certain crucial calculations related to black hole thermodynamics; they give the (square root of the negative of the) determinant of the metric, after applying the constraint to eliminate $\mu_{n}$ by writing $\mu_{n} = \sqrt{1- \sum_{i = 1}^{n-1} \mu_i^2}$, as
\begin{align}
    \sqrt{-g} = \frac{r U \prod_{i=1}^{n-1+\ve} \mu_i}{\mu_{n} \prod_{j = 1}^{n} \Xi_j}. \label{Gibbonssqrtg}
\end{align}
This is also the correct expression for $\sqrt{-g}$ for the KS (in either the full or background spacetime) and BL coordinates, because the Jacobian for the coordinate transformations \eqref{ttau} together with \eqref{phivarphi} (with $r =r, \mu_i = \mu_i)$ as well as \eqref{phivarphi} by itself (with $r=r,\tau=\tau,\mu_i=\mu_i$) is 1. 

(This is easy to see if we write the Jacobian matrix for the transformations \eqref{ttau} and \eqref{phivarphi} as, excluding the $\mu_i$ since these are trivially preserved,
\begin{align}
    \begin{bmatrix} 
    \f{\partial r}{\partial r} & \f{\partial r}{\partial \tau} & \f{\partial r}{\partial \varphi_j} \\ 
    \f{\partial t}{\partial r} & \f{\partial t}{\partial \tau} & \f{\partial t}{\partial \varphi_j} \\
    \f{\partial \phi_i}{\partial r} & \f{\partial \phi_i}{\partial \tau} & \f{\partial \phi_i}{\partial \varphi_j}
    \end{bmatrix} &= \begin{bmatrix} 1 & 0 & 0 \\
    \f{2m}{(1+r^2l^{-2})(V-2m)} & 1 & 0 \nn \f{2ma_i}{(r^2+a-i^2)(V-2m)} & a_il^{-2} & \de_{ij}\end{bmatrix}
\end{align}
Since the Jacobian matrix is a lower triangular matrix, its determinant is just the product of its diagonal entries, which is 1. The same argument applies for coordinate transformations including \eqref{phivarphi}.)

It will also be convenient to have a form that is analogous to the Eddington--Finkelstein ingoing coordinates. These coordinates are useful for covering the interior of a black hole horizon, down to the curvature singularity, though they do not cover the entire manifold. This transformation leaves the $(r,\mu_i)$ coordinates unchanged but introduces new time--azimuthal coordinates $(v,\Phi_i)$, such that the Kerr--Schild vector becomes $k = \partial_r$. Consequently, along the ``ingoing'' null direction, corresponding to the KS vector, all coordinates except $r$ remain constant, with $r$ acting as an affine parameter. (Note that $k$ is past-directed.) These transformations are
\begin{align}
    \phi_i &= \Phi_i + \arctan(r/a_i) \nn 
    t &= v + l \arctan(r/l). \label{phiPhiv}
\end{align}
In these coordinates, 
\begin{align}
    k_a dx^a &= W d v - \sum_{i = 1}^{n-1+\ve} \frac{a_i \mu_i^2}{\Xi_i} d \Phi_i,
\end{align}
and the background metric becomes
\begin{align}
    d \bar s^2 &= -W (1 + r^2l^{-2}) d v^2 - 2 W dr d v + \sum_{i = 1}^{n}\left[ \frac{r^2+a_i^2}{\Xi_i} \left( d \mu_i^2 + \mu_i^2 d \Phi_i^2\right) + \f{2 a_i r \mu_i^2}{\Xi_i} dr d \Phi_i\right] \nn 
    &\qquad - \f{1}{W(r^2+l^2)} \left( \sum_{i = 1}^{n} \frac{(r^2+a_i^2)\mu_i d \mu_i}{\Xi_i}\right)^2.
\end{align}
We find that the metric now has no $g_{rr}$ terms, at the cost that we now have a greater number of cross terms ($drdv$, etc.). In four dimensions letting $u = \cos \tht$ and letting, for this metric specifically, $\psi$ match with the $\Phi$ above, this gives 
\begin{align}
    ds^2 &= - \frac{(l^2 - a^2 u^2)(l^2 + r^2)}{l^2 (l^2 - a^2)} d v^2 + 2 \frac{l^2 - a^2 u^2}{l^2 - a^2} d v d r - 2 \frac{l^2 a (1-u^2)}{l^2 - a^2} d r d \psi + \nonumber \\
    & \qquad \frac{l^2 (r^2 + a^2 u^2)}{ (1-u^2)(l^2 - a^2 u^2)} du^2 + \frac{l^2 (r^2+a^2)(1-u^2)}{l^2-a^2} d\psi^2 + \nonumber \\
    &\qquad \frac{2 m r}{r^2 + a^2 u^2} \left( -\frac{l^2 - a^2 u^2}{l^2 - a^2} d v + \frac{l^2 a (1-u^2)}{l^2 - a^2} d \psi\right)^2. \label{kadsIngoing4D}
\end{align}

We will also wish to have a Kerr--Schild form of the metric which is asymptotically non-static in a way that is analogous to the way in which the BL coordinates are asymptotically non-static. To do so we define coordinates say $\breve{\phi}_i$ by 
\begin{align} 
    \phi_i = \breve{\phi}_i + a_i l^{-2} t. \label{brevephi}
\end{align}
Then the null vector has the form
\begin{align}
    k_a dx^a &= dt + F dr - \sum_{i = 1}^{n-1+\ve} \frac{a_i \mu_i^2}{\Xi_i} d \breve \phi_i \nn 
    k^a \partial_a &= -\frac{1}{1+r^2/l^2} \f{\pa}{\pa t} + \f{\pa}{\pa r} - \sum_{i = 1}^{n-1+\ve} \f{a_i \Xi_i}{(r^2+a_i^2)(1+r^2/l^2)} \f{\pa}{\pa \breve{\phi}_i}.
\end{align}

The functional form of $H$ is unchanged, since the $r$ and $\mu_i$ coordinates are unchanged. The background metric becomes
\begin{align}
    d \bar s^2 &= - W(1+r^2l^{-2}) d t^2 + F dr^2 + \sum_{i = 1}^{n} \f{r^2+a_i^2}{\Xi_i} d \mu_i^2 - \f{1}{W(r^2+l^2)} \left(\sum_{i = 1}^{n} \f{(r^2+a_i^2)\mu_i d \mu_i}{\Xi_i}\right)^2\nn
    &\qquad + \sum_{i = 1}^{n-1+\ve} \f{(r^2+a_i^2) \mu_i^2}{\Xi_i} (d \breve \phi_i + a l^{-2} d t)^2. \label{BLKS}
\end{align}
We note that the background metric is equal to the BL metric with $m$ set to zero and with $\tau \to t, \varphi_i \to \breve \phi_i$.

In four dimensions,
\begin{align}
    d \bar s^2 &= - \frac{(1+r^2l^{-2}) \Delta_\tht dt^2}{\Xi} + \frac{\rho^2 dr^2}{(1+r^2l^{-2}) (r^2+a^2)} + \frac{\rho^2 d\tht^2}{\Delta_\tht} + \frac{(r^2+a^2) \sin^2\tht (d\breve \phi+a l^{-2} d t)^2}{\Xi} \nn
    k_a dx^a &= d t + \frac{\rho^2 dr}{(1+r^2l^{-2})(r^2+a^2)} - \frac{a \sin^2\tht d \breve \phi}{\Xi} \nn 
    k^a \partial_a &= -\f{1}{1+r^2l^{-2}} \partial_t + \partial_r - \f{a \Xi}{(r^2+a^2)(1+r^2l^{-2})} \partial_{\breve{\phi}} \nn 
    \phi &= \breve \phi + a l^{-2} t . \label{BLKS4D}
\end{align}
I call this the BLKS (Boyer--Lindquist--Kerr--Schild) form. I will not use this form often, but will use it in Section \ref{checkinghorizonareavariationrule}.

\subsection{Generalization to Include Mass as Function of Radius} \label{GKAdSSection}

There is an obvious generalization of the Kerr--anti-de Sitter metrics simply found by taking the $m$ function which appears and sending $m \to \mu(r)$ in all the above formulas, where $\mu(r)$ is an as yet-unspecified function of $r$. (The use of $\mu$ here is to choose a symbol which is related to but distinct from $m$, and it is not intended to be connected to the latitude variables, for which $\mu_i$ is used.) I will refer to this as Generalized Kerr--anti-de Sitter (GKAdS). Kerr--anti-de Sitter itself is recovered by simply setting $\mu(r) = m$. Note further that the same coordinate transformations apply between the KS, BL and ABL forms of the metric but with $m \to \mu(r)$ in \eqref{phivarphi} and \eqref{dphidphihat}. 

The Kerr--Newman--anti-de Sitter solution in four dimensions is given by $\mu(r) = m -  Q^2/2r$, where $Q$ is a constant associated with the electric monopole charge, as in \eqref{KdSLine} (which is shown in BL coordinates). In fact the general Kerr--Newman solution includes both electric and magnetic monopole components, in which case the $Q^2$ above is supplemented by a $P^2$, but I will only consider the electric case. The Kerr--Newman--AdS solution is equipped with an electromagnetic field potential one-form $\bs A$, which can be written in BL coordinates as \cite{CaldarelliCognola}
\begin{align}
    \bs A = - \f{Q r}{\rho^2} \left( d \tau - \f{a \sin^2 \tht}{\Xi} d \vp\right), \label{KerrNewmanAdSA}
\end{align}
up to a gauge transformation, $\bs A \to \bs A' = \bs A + d \lambda$ for some scalar $\lambda$. Another convenient form will be introduced in Section \ref{KSNullVector}. In ABL coordinates this is
\begin{align}
    \bs A &= -\f{Q r}{\Xi \rho^2} \left( \Delta_\tht d \tau - a \sin^2\tht d \hat \vp\right). \label{AABL}
\end{align}

Of the GKAdS spacetimes, only the Kerr--anti-de Sitter spacetimes are actually vacuum solutions. The only case I am considering of non-constant $\mu(r)$ of physical interest is Kerr--Newman--AdS. Nevertheless I think it is worth considering a more general class of spacetimes in order to tease out which properties are the result of a more general set of geometries conditions and which require Kerr--AdS (or, possibly, Kerr--Newman--AdS) specifically to hold. Some of this is out of general geometric interest, but the main goal is ultimately to understand the Kerr--AdS solution more clearly.

In Section \ref{GKNAdSSection}, I will discuss an even broader class of spacetimes, of which the GKAdS is a subset, the Generalized Kerr--NUT-Anti-de Sitter (GKNAdS: note the N) class of solutions. These possess a number of general properties which I will use for the GKAdS (note the lack of N) solutions. The GKNAdS solutions are referred to as the ``off-shell'' Kerr--NUT--AdS solutions in, e.g.,~\cite{Krtous16}.  

\subsection{Important Killing Vectors} \label{ImportantKilling}

At this point I will define the following Killing vectors, associated with the ABL coordinates. The vector associated with time symmetry will be called $\xi$ and the vectors associated with azimuthal symmetry $\eta_i$:
\begin{align}
    \xi &\equiv \f{\pa}{\pa \tau} \nn 
    \eta_i &\equiv \f{\pa}{\pa \hat \vp_i}. \label{xietai}
\end{align}
In four dimensions there is only one azimuthal coordinate so we can just write
\begin{align}
    \eta &\equiv \f{\pa}{\pa \hat \vp}.
\end{align}
Because of the way the ABL coordinates are related to the KS coordinates, we also have, in KS coordinates,
\begin{align}
    \xi &= \f{\pa}{\pa t} \nn 
    \eta_i &= \f{\pa}{\pa \phi_i},
\end{align}
with $\eta = \pa/\pa \phi$ in four dimensions. 

The $\eta_i$ are azimuthal symmetry Killing vectors. For the GKAdS spacetimes they are tangent to closed circles, which have period $2 \pi$. $\xi$ is the asymptotically-static Killing vector, in the sense that it is the Killing vector associated with the time coordinate which is, in the large-radius limit, orthogonal to the constant-$\tau$ or constant-$t$ surfaces. In particular,
\begin{align}
    \lim_{r\to\infty} \xi \cdot \eta_i = 0.
\end{align}

In the BL coordinates, we find
\begin{align}
    \eta_i &= \f{\pa}{\pa \vp_i},
\end{align}
but we have, again in BL coordinates specifically, that $\pa/\pa \tau$ is not equal to $\xi$, but is instead
\begin{align}
    \f{\pa}{\pa \tau} &= \xi + \sum_{i=1}^{n-1+\ve} \f{a_i}{l^2} \eta_i.
\end{align}
It will turn out that this quantity, $\xi + \sum_{i=1}^{n-1+\ve} a_il^{-2} \eta_i$, is important and closely related to the symmetries of the spacetime and I will call it $\beta$. I will define this in a different way after introducing the Generalized Kerr--NUT--AdS metric. Whereas $\xi$ is asymptotically static, $\beta$ is not (unless the $a_i$ are zero, in which case $\beta = \xi$). 

As a consequence, the ABL and KS coordinates represent an asymptotically static or asymptotically nonrotating frame, in that $\xi = \pa/\pa \tau$ (ABL) and $\xi = \pa/\pa t$ (KS) is the asymptotically static Killing vector. The BL coordinates represent an asymptotically non-static, rotating frame, in that $\beta = \pa/\pa \tau$ (BL) is not static, and is given by $\beta = \xi + \sum_i a_i l^{-2} \eta_i$, which is rotating relative to the asymptotically nonrotating/static frame, and $\beta \cdot \eta_i$ does not have a zero limit as $r \to \infty$. In fact, $\beta \cdot \eta_i$ diverges for large $r$. Again the exception is if $a_i = 0$ in which case $\beta = \xi$.

Later I will introduce the vector $\z^a$ which is the Killing vector tangent to the null generators of the horizon.

I will reserve the symbol $\chi$ for a generic Killing vector. 

\section{The Generalized Kerr--NUT--AdS Metrics, their Principal Conformal Killing--Yano Tensor, and its Associated Killing Vector} \label{GKNAdSSection}

A generalization of the Kerr--AdS solutions to include NUT charges was found by Chen, L\"u and Pope \cite{Chen} (hereafter CLP). This could then be generalized still further to include non-vacuum solutions which are mathematically related to the Kerr--NUT--AdS solutions. I will refer to this class of spacetimes, which includes Kerr--NUT--AdS (and, consequently, Kerr--AdS), as Generalized Kerr--NUT--AdS (GKNAdS). The GKNAdS solutions were then studied in several papers \cite{KrtousKubiznak, KrtousFrolov, Hamamoto, Houri} (and many others), which demonstrated the existence of a Closed Conformal Killing-Yano Tensor $\bs h$, which is very important for the symmetries of the spacetime. An excellent review article on the Closed Conformal Killing--Yano Tensor and its consequences for the Kerr--NUT--AdS spacetimes is \cite{FrolovReview}.

This general class of spacetimes has a number of important symmetries, which end up being important for the narrower class of spacetimes, the GKAdS spacetimes, which I am considering. I will start by introducing the class of spacetimes found by CLP, then show where the GKAdS solutions fit in. Doing this in arbitrary dimension requires some care, and this section contains a lot of bookkeeping to show the transformation between different coordinate systems. This is necessary, particularly for Chapter \ref{ExplicitGKAdSChapter}.

\subsection{Generalized Kerr--NUT--AdS Solution} \label{GKNAdSSubsection}

CLP transformed the known Kerr--AdS solutions from Gibbons et al.~\cite{GibbonsLu2} to a new set of coordinates by \emph{Jacobi transformations}. These transformations were applied to the latitude variables to transform from the $n$ constrained $\mu_i$ to $n-1$ unconstrained variables $y_\alpha$.

The Jacobi transformations used by CLP are based on the following. Given the metric of a $(D-2)$-sphere, in the form
\begin{align}
    ds^2_{S^{D-2}} &= \sum_{i=1}^{n} d\mu_i^2 + \sum_{i=1}^{n-1+\ve} \mu_i^2 d \hat \varphi_i^2, \label{Dm2spheremetric}
\end{align}
subject to $\sum_{i=1}^n \mu_i^2 = 1$, the metric can be diagonalized and written in terms of $n-1$ \emph{unconstrained} parameters $y_\alpha$ defined implicitly by
\begin{align}
    \mu_i^2 &= \frac{\prod_{\alpha = 1}^{n-1} (a_i^2 - y_\alpha^2)}{\prod_{j=1,j\neq i}^{n} (a_i^2 - a_j^2)}, \label{Jacobi}
\end{align}
where $a_i$ are a series of distinct constants. (The product in the denominator is, as should be clear, over all integers $1 \leq j \leq n$ except $j = i$, which is excluded. Note also that $a_n$ is included in the product in both odd and even dimensions, but that in even dimensions $a_n = 0$.) Under this transformation, the line element becomes
\begin{align}
    ds^2_{S^{D-2}} &= \sum_{\alpha = 1}^{n-1} \left( \frac{-y_\alpha^2{\prod}_{\beta =1,\beta\neq \alpha}^{n-1}(y_\alpha^2 - y_\beta^2)}{\prod_{j=1}^n(a_j^2-y_\alpha^2)}\right) dy_\alpha^2 + \sum_{i=1}^{n-1+\ve} \frac{\prod_{\alpha = 1}^{n-1} (a_i^2 - y_\alpha^2)}{{\prod}_{j=1,j\neq i}^{n} (a_i^2 - a_j^2)} d \hat \varphi_i^2.
\end{align}

CLP apply the Jacobi transformations \eqref{Jacobi} to the $\mu_i^2$ in the ABL form of the metric, using the rotational parameters $a_i$ for the $a_i$ constants which appear in this metric form, and then find that (``remarkably,'' as CLP state) the segment of the metric involving the latitude variables becomes diagonal. This is particularly interesting because while the metric on surfaces of $r,\tau$ constant are topologically spheres, they are still ``distorted'' spheres and do not have the metric \eqref{Dm2spheremetric}. Consequently, it is perhaps not initially clear why the Jacobi transformations would diagonalize the latitude part of the metric. I have some comments on why the Jacobi transform works as well as it does in Section \ref{JacobiTransformationsSection}. 

After applying the Jacobi transformation, CLP found that the $y_\alpha$ coordinates enter the metric onto a nearly parallel footing with the radial coordinate $r$, or more precisely to the combination $i r$ (where $i$ is the imaginary unit). Applying a Wick rotation by defining 
\begin{align}
    x_n &\equiv i r  \label{xn}
\end{align}
and also writing
\begin{align}
    x_\alpha \equiv y_\alpha \label{xalpha}
\end{align}
for $\alpha = 1, \ldots, n-1$ brings out the symmetry between the $ir$ and $y_\alpha$ coordinates more explicitly. For the most part, the $y_\alpha$ and $r$ are real, so that $x_\alpha$ are real but $x_n$ is imaginary. There are exceptions, particularly in the region below the black hole horizon, in which $r$ can take on imaginary values, which I will return to and which will become important in Section \ref{AreaVolumeRelationship}.

Let $\mu, \nu, \rho,\sigma,\tau$ be indices which are generally taken to vary from 1 to $n$, and will typically be used in association with the $x_\mu$ variables (in some way). $\alpha, \beta$ etc.~will be taken to vary between $1$ and $n-1$ when associated with the $x_\alpha$ (and usually in this case I will write $y_\alpha$ instead, to specify that $x_n = ir$ is not being considered).

Finally, a linear transformation is applied from the ABL time and the azimuthal coordinates $(\tau, \hat \varphi_i)$ to a new set of coordinates $\psi_j$ which also demonstrate the symmetries of the GKNAdS spacetime more clearly. The actual relationship between the $(\tau, \hat \vp_i)$ and $\psi_j$ coordinates will be discussed in Section \ref{timeazimuthal}.

I will now state the solutions to the GKNAdS spacetime, and \emph{then} specify which spacetimes of those are Kerr--NUT--AdS, and then which of those are Kerr--AdS. While extremely rich and interesting, the Kerr--NUT--AdS solutions are beyond my scope, but there are some general results of the GKNAdS class of metrics, particularly from \cite{Hamamoto}, which apply to Kerr--AdS, which I will use. 

The GKNAdS metric is most simply expressed by using an orthonormal basis, which I will refer to as the canonical orthonormal basis. My conventions follow most closely \cite{KrtousKubiznak} and \cite{Hamamoto}. I will use capital Roman letters to represent the basis indices, so that the orthonormal basis one-forms are $e^A$, $A = 1, \ldots, D$, with components $e^A_a$ (lowercase Roman indices still representing abstract indices). The orthonormal condition implies
\begin{align}
    g^{a b} e^A_a e^B_b &= \delta^{AB}.
\end{align}
The (also orthonormal) basis of vectors dual to the one-form basis is represented by $e_A$ (with components $e_A^a$):
\begin{align}
    e_A^a e^B_a &= \delta^B_A \nn 
    g_{a b} e_A^a e_B^b &= \delta_{AB}.
\end{align}

As before let $\mu$ (and $\nu, \rho,\sigma$) vary from $1$ to $n$. Following \cite{KrtousKubiznak} and other such papers, let $\hat \mu = n+\mu$ so that $\hat \mu$ can take on values from $n+1$ to $2n$. Finally in odd dimensions let $\hat 0 = D =2n+1$. 

We will use the $n$ coordinates $x_\mu$ (already described) and $n+\ve$ coordinates $\psi_j$, where $0 \leq j \leq n-1+\ve$ which we now introduce and are linear combinations of the time and azimuthal coordinates in a way I will show later. The $e^\mu$ will then be proportional to $dx_\mu$ one-forms and the $e^{\hat \mu}$ and $e^{\hat 0}$ (in odd dimension) will be a combination of the $d\psi_j$.

The metric can be expressed as
\begin{align}
    ds^2 &= \sum_{A=1}^D (e^A)^2 \nn 
    &= \sum_{\mu=1}^n \left( (e^\mu)^2 + (e^{\hat \mu})^2\right) + \varepsilon (e^{\hat 0})^2. \label{oneforms}
\end{align}

The one-forms are
\begin{align}
    e^\mu &= \frac{d x_\mu}{\sqrt{Q_\mu}} \nonumber \\
    e^{\hat \mu} &= \sqrt{Q_\mu} \sum_{j = 0}^{n-1} A_\mu^{(j)} d\psi_j \nonumber \\
    e^{\hat 0} &= \sqrt{S}\sum_{j = 0}^{n} A^{(j)} d \psi_j. \label{emuehatmuehat0}
\end{align}
Here,
\begin{align}
    Q_\mu &= \frac{X_\mu}{U_\mu} \nonumber \\
    U_\mu &= \prod_{\nu \neq \mu} (x_\nu^2 - x_\mu^2) \nonumber \\
    S &= \frac{-c}{A^{(n)}}. \label{QmuUmuSdefinition}
\end{align}
$X_\mu$ is a function of $x_\mu$ only (as yet arbitrary). $c$ is a constant. We assume that the $x_\mu$ are all distinct, so that none of the $U_\mu$ are vanishing. (The case where the $x_\mu$ are not distinct requires special care; see, e.g.~\cite{Houri}. We won't worry about this here.) \emph{A priori} we allow for the $x_\mu$ or $\psi_j$ to be complex, but will specify them later so that physical spacetimes with this basis will have real line elements. (Note that $S$ appears as $+c/A^{(n)}$ in \cite{Hamamoto} but $S = -c/A^{(n)}$ in \cite{KrtousKubiznak}; I am following \cite{KrtousKubiznak} here.) The $A^{(j)}$ are the elementary symmetric polynomials in the collection $\{x_1^2, \ldots, x_n^2\}$ of degree $j$,
\begin{align}
    A^{(j)} &= \sum_{\mu_i < ... < \mu_j} \prod_{i = 1}^j x^2_{\mu_i},  j > 0 \nn 
    A^{(0)} &= 1. \label{Aj}
\end{align}

For example,
\begin{align}
    A^{(0)} &= 1 \nn 
    A^{(1)} &= x_1^2 + \ldots + x_n^2 \nn 
    A^{(2)} &= x_1^2 x_2^2 + x_1^2 x_3^2 + \ldots + x_1^2 x_n^2 + x_2^2 x_3^2 + \ldots + x_2^2 x_n^2 + \ldots + x_{n-1}^2 x_n^2, \label{Aj2}
\end{align}
and so on, so that $A^{(j)}$ is the sum of all unique terms which are the products of $j$ distinct $x_\mu^2$. These can be generated by
\begin{align}
    \prod_{\mu = 1}^n (1 + \alpha x_\mu^2) &= \sum_{j = 0}^{n} A^{(j)} \alpha^j
\end{align}
where $\alpha$ is an arbitrary parameter. Similarly, the $A_\mu^{(j)}$ are the elementary symmetric polynomials with the $x_\mu$ term omitted,
\begin{align}
    A_\mu^{(j)} &= \sum_{\substack{\nu_1 < ... < \nu_j \\ \nu_i \neq \mu}} \prod_{i = 1}^j x^2_{\nu_i}, j > 0  \nonumber \\
    A_\mu^{(0)} &= 1 \nn
    \prod_{\nu \neq \mu} (1 + \alpha x_\nu^2) &= \sum_{j = 0}^{n-1} A_\mu^{(j)} \alpha^j. \label{Amuj}
\end{align}
$A_\mu^{(j)}$ can also be found by setting $x_\mu \to 0$ in $A^{(j)}$. 

Treating the $A_\mu^{(j)}$ as an $n \times n$ matrix, with $\mu$ varying from 1 to $n$ and $j$ from $0$ to $n-1$, there is an inverse matrix $B^\mu_{(j)}$ which acts as both a left and right inverse, given by 
\begin{align}
    B^\mu_{(j)} &= \f{(-x_\mu^2)^{n-1-j}}{U_\mu}, \label{Bmuj}
\end{align}
so that 
\begin{align}
    \sum_{j=0}^{n-1} A_\mu^{(j)} B^\nu_{(j)} &= \de^\nu_\mu, 1 \leq \mu,\nu \leq n \nn 
    \sum_{\mu = 1}^n A_\mu^{(l)} B^\mu_{(j)} &= \de^l_j, 0 \leq j,l \leq n-1.
\end{align}
This inverse relation (given in, for instance, \cite{KrtousKubiznak}) is an algebraic relation having to do with Vandermonde matrices \cite{Turner}; it is satisfied for any set of distinct values $x_\mu$, and I will use variants of this relationship later on. Note further that $B^\mu_{(j)}$ is defined for arbitrary $j$ (not just within the range $0 \leq j \leq n-1$), even though it is only properly in this range for which it acts as an inverse for $A_\mu^{(j)}$.

It is convenient to let 
\begin{align}
    \bs \om^\mu &= e^\mu \wedge e^{\hat \mu} \label{omegamu}
\end{align}
be a two-form which is associated with the $e^\mu, e^{\hat \mu}$ directions. The $\bs \om^\mu$ satisfy
\begin{align}
    \bs \om^\mu \bullet \bs \om^\nu &= 2 \delta^{\mu \nu}.
\end{align}

The associated vector basis is
\begin{align}
    e_\mu &= \sqrt{Q_\mu} \partial_{x_\mu} \nonumber \\
    e_{\hat \mu} &= \frac{1}{\sqrt{Q_\mu} U_\mu} \sum_{j = 0}^{n-1+\ve} (-x_\mu^2)^{n-1-j} \partial_{\psi_j} = \f{1}{\sqrt{ Q_\mu}} \sum_{j=0}^{n-1+\ve} B^\mu_{(j)} \pa_{\psi_j} \nonumber \\ 
    e_{\hat 0} &= (-c A^{(n)})^{-1/2} \partial_{\psi_n}, \label{eAvectorbasis}
\end{align}
satisfying
\begin{align}
    e_A^a &= e^A_b g^{ab}.
\end{align}

CLP showed that this class of spacetimes, for specific choices of functions $X_\mu$, include the higher-dimensional Kerr--NUT, Kerr--NUT--de Sitter and Kerr--NUT--AdS spacetimes. (The Taub--NUT spacetime was discovered by Taub \cite{Taub51} and then extended to a larger manifold by Newman, Tamburino and Unti \cite{NUT}; the Kerr--NUT solution is a generalization of that.) The Kerr--NUT--AdS spacetimes are vacuum ($T_{ab} = 0$) with negative $\La$ and reduce to the Kerr--AdS spacetimes if the NUT parameters are set to zero. If the NUT parameters are not set to zero, the spacetimes are not globally asymptotically anti-de Sitter, in that they have a different topology than anti-de Sitter asymptotically. See \cite{Rodriguez} for a recent treatment of the thermodynamics of Kerr--NUT--AdS black holes. I will only \emph{touch on} the Kerr--NUT--AdS solutions with nonzero NUT parameters and will in general in this thesis set NUT parameters to zero. If the vacuum condition is dropped then the Generalized Kerr--NUT--AdS spacetimes are recovered, but then $T_{ab} \neq 0$ in general.

The Riemann curvature for the CLP metrics were calculated by Hamamoto et al.~\cite{Hamamoto} in terms of the functions $X_\mu$. The explicit values are stated in Appendix \ref{curvature}. (I use the curvature forms to show in Appendix \ref{r2zeroinfourdimensions} that the curvature invariant $r_2 = S^a_b S^b_c S^c_a$ vanishes for all four-dimensional Kerr--NUT--AdS spacetimes, where $S_{ab}$ is the traceless Ricci tensor.) An important result is that the canonical frame components of the Ricci tensor, which I will call $\mc R_{A B} \equiv R_{ab} e_A^a e_B^b$, are diagonal (in both odd and even dimension), with $\mc R_{AB} = 0$ if $A\neq B$, and also have $\mc R_{\mu \mu} = \mc R_{\hat \mu \hat \mu}$. This is true for the whole class of GKNAdS spacetimes. Explicitly,
\begin{align}
    R_{a b} &= \sum_{A,B} \mc R_{A B} e^A_a e^B_b \nn
    &= \sum_{\mu = 1}^n \mc R_{\mu \mu} ( e^\mu_a e^\mu_b + e^{\hat \mu}_a e^{\hat \mu}_b) + \varepsilon \mc R_{\hat 0 \hat 0} e^{\hat 0}_a e^{\hat 0}_b, \label{Riccicanonical}
\end{align}
where $e^A_a$ (and so on) are the coordinate components of the orthonormal frame vector $e^A$. 

Having calculated the frame components, Hamamoto et al.~showed that the space for which the Ricci curvature tensor is given by $R_{a b} = \la g_{a b}$ is 
\begin{align}
    (-1)^{n-1} X_\mu &= \sum_{j = 0}^{n} c_{2j} x_\mu^{2j} + b_\mu x_\mu, \textrm{ (even dimension)} \nn 
    (-1)^{n-1} X_\mu &= \sum_{j = 1}^n c_{2j} x_\mu^{2j} + b_\mu + \f{(-1)^{n+1} c}{x_\mu^2} \textrm{ (odd dimension)}, \label{Xmu}
\end{align}
where $c_{2j}, c$ and $b_\mu$ are free parameters ($b_\mu$ are distinct for each $\mu$), except that $c_{2n}=l^{-2}$ to give AdS. The factor of $(-1)^{n-1}$ is to account for the difference between Hamomoto's conventions for $U_\mu$ and mine; similarly the factor multiplying $c$ is different by a sign due to a difference in how $S$ is defined. (The constant curvature space, which is pure Minkowski, anti-de Sitter or de Sitter, is recovered if $b_\mu = 0$ in even dimension, or $b_\mu = C$, a constant independent of $\mu$, in odd dimension.) If the spacetime is interpreted as vacuum with $\Lambda$, then $\la = 2\La/(D-1)(D-2)$. The solutions of \cite{Hamamoto}, calculated to arbitrary dimensions, match those for CLP \cite{Chen}, which were checked to satisfy $R_{a b} = (D-1)\la g_{ab}$ explicitly up to a finite number of dimensions. These solutions can be written, following the notation of \cite{KrtousKubiznak}, as 
\begin{align}
    X_\mu &= (-1)^{1-\varepsilon} \frac{1 + \lambda x_\mu^2}{x_\mu^{2\varepsilon}} \prod_{j = 1}^{n-1+\ve} (a_j^2 - x_\mu^2) + 2 M_\mu (-x_\mu)^{1 - \varepsilon}, \nn 
    c &= \prod_{j = 1}^{n-1+\ve} a_j^2, \label{Xmuc}
\end{align}
for all $x_\mu$, where $a_i$ are constants, which in the case of Kerr--AdS are will be related to the angular momenta. In the case of anti-de Sitter we can set $\lambda = -l^{-2}$ where $l$ is the associated radius of curvature of AdS. 

The $a_i$ represent rotation parameters, $M_\alpha$ for $\alpha = 1, \ldots, (n-1)$ represent NUT parameters, and $m = - i^{1 + \varepsilon} M_n$ is the mass parameter $m$. Kerr--AdS is recovered if $M_\alpha = 0$. 

It is worth writing the form of the Kerr--AdS metric in these coordinates somewhat more explicitly, since it is so central to this work. I will follow conventions also used by \cite{Chen}, where they let $X \equiv -X_n$. (As of yet I have not established the relationship between $U_n$ and the $U$ from \eqref{Udefinition}.) Explicitly for Kerr--AdS itself,
\begin{align}
    X &= -X_n = \begin{cases}
        \f{1+r^2/l^2}{r^2} \prod_{i = 1}^n (r^2+a_i^2) - 2m &\textrm{if }D\textrm{ is odd} \\
        (1+r^2/l^2)\prod_{i=1}^{n-1}(r^2+a_i^2) - 2mr &\textrm{if }D\textrm{ is even}.
    \end{cases} \label{Xdefinition} 
\end{align}
The other $X_\alpha$ functions are conveniently written by letting them be equal to $\bar X_\alpha$, with the bar to signify that they take on their ``background AdS'' values (with $M_\alpha = 0$). Define the function $Y(y)$ to be
\begin{align}
    Y(y) &\equiv (-1)^{1-\varepsilon} \f{(1-y^2/l^2)}{y^{2\varepsilon}} \prod_{j = 1}^{n-1+\ve} (a_j^2 - y^2). \label{Yofy}
\end{align}
Then
\begin{align}
    \bar X_\alpha &= Y(y_\alpha). \label{Xbaralpha}
\end{align}
In this case we have
\begin{align}
    ds^2 &= \sum_{\alpha=1}^{n-1}  \f{U_\alpha dy_\alpha^2}{ \bar X_\alpha} + \f{U_n dr^2}{X}  + \sum_{\alpha = 1}^{n-1} \f{\bar X_\alpha}{U_\alpha} \left(\sum_{j = 0}^{n-1} A_\alpha^{(j)} d \psi_j\right)^2 - \f{X}{U_n} \left( \sum_{j = 0}^{n-1} A_n^{(j)} d \psi_j\right)^2 + \varepsilon S \left(\sum_{j = 0}^n A^{(j)} d \psi_j\right)^2. \label{ds2KAdSpsicoords}
\end{align}
The horizon is located at the solution to $X = 0$, which in these coordinates is singular (since $g_{r r}$ diverges).

In this solution, in the part of the spacetime of physical interest, $U_\mu \geq 0$ for all $\mu$, $X_\alpha \geq 0$ and $X$ is positive outside the horizon. This means that, in the region outside the black hole horizon, all the one-forms are real except $e^{\hat n}$, which is imaginary (so that $i e^{\hat n}$ is a real unit timelike one-form).

Let us also define $\bar X$, the value for $X$ with $m = 0$, as
\begin{align}
    \bar X &= -Y(x_n) \nn 
    &= \f{1}{r^{2\ve}} (1+r^2/l^2) \prod_{j=1}^{n-1+\ve} (a_j^2+r^2) \nn 
    &= \f{1}{l^2 r^{2\ve}} \prod_{j=0}^{n-1+\ve} (r^2+a_j^2). \label{Xbar}
\end{align}
The AdS solution can be written as \eqref{ds2KAdSpsicoords} with $X \to \bar X$. We also note that $V$ from \eqref{GibbonsV} is
\begin{align}
    V &= \f{\bar X}{r^{1-\ve}}.
\end{align}
This also tells us that
\begin{align}
    V - 2m = \f{X}{r^{1-\ve}}, \label{Vm2mX}
\end{align}
in Kerr--AdS. For GKAdS (not GKNAdS), we have instead $V - 2\mu(r) = \f{X}{r^{1-\ve}}$.

For large $r$, $X$ is real. $U_n$ is also real. There is some ambiguity in taking the square root, $\sqrt{X_n}$; I will let $\sqrt{X_n} \equiv i \sqrt{X}$ and similarly $\sqrt{\bar X_n} \equiv i \sqrt{\bar X}$. With this choice we can rewrite $e^n, e^{\hat n}, e_n, e_{\hat n}$ as follows:
\begin{align}
    e^n &= \sqrt{\f{U_n}{X}} dr \nn 
    e^{\hat n} &= i \sqrt{\f{X}{U_n}} \sum_{j=0}^{n-1} A_n^{(j)} d \psi_j \nn 
    e_n &= \sqrt{\f{X}{U_n}} \f{\pa}{\pa r} \nn 
    e_{\hat n} &= -i\sqrt{\f{U_n}{X}} \sum_{j=0}^{n-1+\ve} B^n_{(j)} \f{\pa}{\pa \psi_j}.
\end{align}
Since the $A_n^{(j)}$ and $B^n_{(j)}$ are real, this helps show that $e^{\hat n}$ is an imaginary-valued one-form so that $i e^{\hat n}$ is a real-valued unit timelike one-form.

I verify explicitly some of the consequences of the Jacobi transformation, including the new forms of some of the metric components, in Appendix \ref{CalculationsAssociatedWithJacobiTransformations}.

\subsection{Limits on Jacobi Transformed Coordinates} \label{limitsonjacobitransformedcoordinates}

The limits on the Jacobi transformed coordinates $y_\alpha$ are those for which all the $\mu_i^2$ are real and nonnegative. Since the constraint $\sum_{i=1}^{n} \mu_i^2 = 1$ is automatically satisfied, as long as all the $\mu_i^2$ are real and nonnegative, all the $\mu_i^2$ will also be no greater than 1. 

The $a_i$ must all be distinct in order to use the Jacobi transformation. (See \cite{FrolovReview} for a discussion, which also includes the case with nonzero NUT parameters. Note that they use a different ordering than I am using here, with $a_1$ the smallest and $a_n$ the largest.) Without loss of generality we can choose an ordering
\begin{align}
    a_1 > ... > a_n \geq 0
\end{align}
(with $a_n = 0$ in even dimension). The requirement that the right hand side of \eqref{Jacobi} be nonnegative restricts the sign of $y_\alpha$. One possible choice is
\begin{align}
    a_1^2 \geq y_1^2 \geq a_2^2 \geq \ldots \geq y_{n-1}^2 \geq a_n^2.
\end{align}
Indeed, up to relabeling, this is the only sensible choice because we also require that the $y_\alpha$ be distinct (to avoid the vanishing of $U_\alpha$). In other words, for all $\alpha$ from 1 to $n-1$, $a_\alpha^2 \geq y_\alpha^2 \geq a_{\alpha+1}^2$.  In odd dimensions, the $\mu_i$ all vary from 0 to 1, so that the value of all $\mu_i$ is determined uniquely from $\mu_i^2$. In this case, we can similarly take all $y_\alpha$ to be positive and to vary between the positive and negative values. In even dimension, all $\mu_i$ except $\mu_n$ vary from 0 to 1. $\mu_n$ is allowed to vary between $-1$ and $1$, and, with $a_n = 0$, it will have the form
\begin{align}
    \mu_n^2 &= \f{\prod_{\alpha=1}^{n-1} y_\alpha^2}{\prod_{j=1}^{n-1} a_j^2},
\end{align}
so that we can demand
\begin{align}
    \mu_n &= \f{\prod_{\alpha=1}^{n-1} y_\alpha}{\prod_{j=1}^{n-1} a_j}.
\end{align}
Taking all $a_j$ (except $a_{n}$) to be positive, we can let all $y_\alpha$ except $y_{n-1}$ be strictly positive, and allow $a_{n-1} \geq y_{n-1} \geq -a_{n-1}$ so that $\mu_n$ is allowed to take on both positive and negative values. Thus in odd dimensions we have
\begin{align}
    a_1 \geq y_1 \geq a_2 \geq y_2 \geq \ldots \geq a_{n-1} \geq y_{n-1} \geq a_n > 0 \label{oddyalphalimits}
\end{align}
and in even dimensions
\begin{align}
    a_1 \geq y_1 \geq a_2 \geq y_2 \geq \ldots \geq a_{n-1} \geq y_{n-1} \geq -a_{n-1}. \label{evenyalphalimits}
\end{align}
(Again, my ordering of is different from in \cite{FrolovReview} and references therein, but the principle is the same.) 

\subsection{Metric Time--Azimuthal Components} \label{timeazimuthal}

We wish to convert back and forth between different forms of the Kerr--AdS metric. (The notation here is similar to, but distinct from, that appearing in \cite{FrolovReview} sect.~4.2.1.) To do so it will be useful to define quantities which are analogous to $A_\mu^{(j)}, B^\mu_{(j)}$ and $U_\mu$, but where the polynomials are constructed from the $(n+\ve)$ values $a_i^2$ ($i = 0, \ldots, n-1+\ve)$ (that is to say, including $a_0 = l$), rather than the $n$ values $x_\mu^2$. Let $\hat C_i^{(l)}$, analogous to $A_\mu^{(j)}$, be given by
\begin{align}
    \prod_{j=0,j \neq i}^{n-1+\ve} (1 + \alpha a_j^2) &\equiv \sum_{l = 0}^{n-1+\ve} \alpha^l \hat C^{(l)}_i \nn 
    \hat C_i^{(l)} &= \sum_{\substack{0 \leq j_1 < \ldots < j_l \leq n-1+\ve \\ j_k \neq i}} \prod_{k=1}^l a_{j_k}^2, \qquad l > 0 \nn 
    \hat C_i^{(0)} &= 1. \label{Chatil}
\end{align}
Then let $\hat D^i_{(l)}$ and $\hat \Ups_i$, analogous to $B^\mu_{(j)}$ and $U_\mu$, be
\begin{align}
    \hat D^i_{(l)} &\equiv \f{(-a_i^2)^{n-1+\ve-l}}{\hat \Ups_i} \nn 
    \hat \Ups_i &\equiv \prod_{j=0, j \neq i}^{n-1+\ve} (a_j^2-a_i^2). \label{hatDhatUps}
\end{align}
These are defined for $0 \leq i,l \leq n-1+\ve$. We then have the inverse relations
\begin{align}
    \sum_{i=0}^{n-1+\ve} \hat C_i^{(j)} \hat D^i_{(l)} &= \de^j_l \nn 
    \sum_{l=0}^{n-1+\ve} \hat C_i^{(l)} \hat D^j_{(l)} &= \de^j_i,
\end{align}
where $0 \leq i,j,l \leq n-1+\ve$. 

It is also convenient to define
\begin{align}
    \Gamma_i &\equiv \prod_{\nu = 1}^n (a_i^2-x_\nu^2), \qquad 0 \leq i \leq n-1+\ve \nn 
    \tilde \varphi_0 &\equiv -\f{\tau}{\prod_{i = 1}^{n-1+\ve} (l^2-a_i^2)} = -\f{\tau}{l^{2(n-1+\ve)}\prod_{i=1}^{n} \Xi_i} \nn 
    \tilde \varphi_i &\equiv (-1)^{n+\ve} \f{\hat \varphi_i}{a_i \Xi_i \Upsilon_i}, \qquad 1 \leq i \leq n-1+\ve, \label{Gammatildephi}
\end{align}
where $(\tau,\hat \varphi_i)$ are the time--azimuthal coordinates associated with the ABL metric. 

Because sometimes we are interested in the set $\{a_1, \ldots, a_{n-1+\ve}\}$ and sometimes we are interested in the set $\{a_0 = l, a_1, \ldots, a_{n-1+\ve}\}$, let $C_i^{(l)}, D^i_{(l)}, \Ups_i$ be the equivalents to $\hat C_i^{(l)}, \hat D^i_{(l)}$ and $\hat \Ups_i$ based on the set which excludes $a_0=l$, so that $1 \leq i \leq n-1+\ve$ and $0 \leq l \leq n-2+\ve$. (The hat terms include $l$ and the terms without a hat do not.) Then,
\begin{align}
    \prod_{j = 1, j \neq i}^{n-1+\ve}(1+\alpha a_j^2) &= \sum_{l = 0}^{n+\ve -2} \alpha^l C_i^{(l)} \nn 
    C_i^{(l)} &= \sum_{\substack{1 \leq j_1 < \ldots < j_l \leq n-1+\ve \\ j_k \neq i}} \prod_{k=1}^l a^2_{j_k}, \qquad 1 \leq l \leq n-2+\ve \nn 
    C_i^{(0)} &= 1 \nn 
    D^i_{(l)} &\equiv \f{(-a_i^2)^{n-2+\ve-l}}{\Ups_i} \nn 
    \Ups_i &\equiv \prod_{j=1, j \neq i}^{n-1+\ve} (a_j^2-a_i^2).  \label{Cil}
\end{align}
In four dimensions, $\Ups_1 = 1$. These quantities satisfy the useful relations
\begin{align}
    \hat C_i^{(k)} &= C_i^{(k)} + l^2 C_i^{(k-1)}, \qquad i>0,k>0 \nn 
    \hat \Ups_i &= l^2 \Xi_i \Ups_i,\qquad i > 0 \nn 
    \hat \Ups_0 &= (-l^2)^{n-1+\ve} \prod_{i=1}^n \Xi_i \nn 
    \sum_{i=1}^{n-1+\ve}  C_i^{(k)}  D^i_{(l)} &= \de^k_l \nn 
    \sum_{l = 0}^{n-2+\ve} C_i^{(l)} D^j_{(l)} &= \de_i^j. \label{hatnonthatrelationsforCetc}
\end{align}

Note that because $a_n = 0$ if $\ve = 0$, $\Xi_n$ in this case is automatically 1. This means that $\prod_{i=1}^{n-1+\ve} \Xi_i = \prod_{i=1}^{n} \Xi_i$, so I will favour the latter for simpler writing.

Let us also define
\begin{align}
    \hat \vp_0 &\equiv \tau/l.
\end{align}
Then the $\tilde \vp_i$ for $0 \leq i \leq n-1+\ve$ can be written compactly as
\begin{align}
    \tilde \vp_i &= \f{l^2 (-1)^{n+\ve} \hat \vp_i}{a_i \hat \Ups_i}.
\end{align}

Then the frame components for the time--azimuthal sector can also be written as (adapting \cite{Chen})
\begin{align}
    e^{\hat \mu} &= - \sqrt{Q_\mu} \sum_{i = 0}^{n-1+\ve} \f{a_i^{2\varepsilon} \Gamma_i d \tilde \varphi_i}{a_i^2 - x_\mu^2} \nn 
    e^{\hat 0} &= -\sqrt{S} \sum_{i = 0}^n \Gamma_i d \tilde \varphi_i.\label{ehatmualt}
\end{align}
The sign choices (such as $e^{\hat 0} = -\sqrt{S}\sum_{i = 0}^n \Gamma_i d \tilde \varphi_i$ instead of $+\sqrt S \sum_{i = 0}^n \Gamma_i d \tilde \varphi_i$) are made to have convenient relationships between the $d \psi_j$ and $d \tilde \phi_i$ coordinates; of course the frame one-forms only appear in the metric squared, so either choice is possible \emph{a priori}. These can be rewritten in terms of $\hat \vp_i$ directly as
\begin{align}
    e^{\hat \mu} &= \sq{Q_\mu} (-1)^{n-1+\ve} \sum_{i=0}^{n-1+\ve} \f{l^2 a_i^{2\ve-1} \G_i d \hat \vp_i}{\hat \Ups_i(a_i^2-x_\mu^2)} \nn 
    e^{\hat 0} &= \sqrt{S} (-1)^{n} \sum_{i=0}^{n} \f{l^2 \G_i d\hat \vp_i}{a_i \hat \Ups_i}. \label{ehatmuhatphi}
\end{align}

In the KS time and azimuthal coordinates (with $\phi_0 = t/l$) we have instead (recall $1 \leq \alpha \leq n-1$) (see Appendix Section \ref{canonicalbasisKScoordssection} for derivation)
\begin{align}
    e^{\hat \alpha} &= \sq{Q_\alpha} (-1)^{n-1+\ve} \sum_{i=0}^{n-1+\ve} \f{l^2 a_i^{2\ve-1}\G_i d \phi_i}{\hat \Ups_i (a_i^2-y_\alpha^2)} \nn 
    e^{\hat n} &= (-1)^{n-1+\ve} \sqrt{Q_n} \sum_{i=0}^{n-1+\ve} \f{l^2 a_i^{2\ve-1} \G_i}{\hat \Ups_i (a_i^2+r^2)} d \phi_i  - \f{\sqrt{Q_n}}{\bar Q_n} \f{2mr^{1-\ve} dr}{X_n} \nn
    e^{\hat 0} &= \sqrt{S} (-1)^n \sum_{i=0}^n \f{l^2 \G_i d\phi_i}{a_i \hat \Ups_i}. \label{canonicalbasisKScoords}
\end{align}
Here $\bar Q_n \equiv \bar X_n/U_n$. Thus we find that the functional forms of $e^{\hat \alpha}$ and $e^{\hat 0}$ are unchanged except that $\hat \vp_i \to \phi_i$, and that $e^{\hat n}$ involves both this change as well as an extra $dr$ term. This is for Kerr--AdS specifically; for the GKAdS we just send $m \to \mu(r)$. 

CLP \cite{Chen} state that the relationship between $\psi_j$ and $(\tau,\hat \varphi_i)$ can be recovered by comparing the expressions \eqref{ehatmualt} to those in \eqref{emuehatmuehat0} but they do not give them explicitly. I will do so here. (Again, the results are also those appearing in \cite{FrolovReview}, in different notation.) We have,
\begin{align}
    \f{\Gamma_i}{a_i^2 - x_\mu^2} &= \prod_{\nu \neq \mu} (a_i^2 - x_\nu^2) \nn 
    &= (-1)^{n-1} \sum_{j = 0}^{n-1} A_\mu^{(j)} (-a_i^2)^{n-1-j}, \label{Gammabya2mx2}
\end{align}
from which
\begin{align}
    \sum_{i = 0}^{n-1+\ve} \f{a_i^{2 \varepsilon} \Gamma_i d \tilde \varphi_i}{a_i^2 - x_\mu^2} &= \sum_{i = 0}^{n-1+\ve} a_i^{2\varepsilon} d \tilde \varphi_i \sum_{j = 0}^{n-1} (-1)^{n-1} A_\mu^{(j)} (-a_i^2)^{n-1-j} \nn 
    &= \sum_{j =0}^{n-1} A_\mu^{(j)} \left[ (-1)^{n-1} \sum_{i =0}^{n-1+\ve} (-1)^\varepsilon (-a_i^2)^{n-1+\varepsilon-j} d \tilde \varphi_i \right] \nn 
    &= \sum_{j =0}^{n-1} A_\mu^{(j)} \left[  (-1)^{n-1+\ve} \sum_{i = 0}^{n-1+\ve} (-a_i^2)^{n-1+\ve-j} d \tilde \varphi_i \right].
\end{align}
Similarly, 
\begin{align}
    \Gamma_i &= \prod_{\mu = 1}^n (a_i^2-x_\mu^2) \nn 
    &= (-1)^{n} \sum_{j = 0}^n A^{(j)} (-a_i^2)^{n-j} \nn 
    \sum_{i = 0}^n \Gamma_i d \tilde \varphi_i &= \sum_{j = 0}^n A^{(j)} \left[(-1)^n\sum_{i=0}^n  (-a_i^2)^{n-j} d \tilde \varphi_i\right].
\end{align}
Equating $e^{\hat \mu} = e^{\hat \mu}$ and $e^{\hat 0} = e^{\hat 0}$ from \eqref{emuehatmuehat0} and \eqref{ehatmualt},
\begin{align}
    d \psi_j &=- (-1)^{n-1+\ve} \sum_{i=0}^{n-1+\ve} (-a_i^2)^{n-1+\ve - j} d \tilde \vp_i.
\end{align}
Using the expression for $\tilde \vp_i$ in terms of $\hat \vp_i$ and using the $\hat C_i^{(j)}, \hat D^i_{(j)}$ inverse relations, we have
\begin{align}
    d \psi_j &= \sum_{i = 0}^{n-1+\ve}\hat D^i_{(j)}  \f{l^2 d \hat \varphi_i}{a_i} \nn 
    d \hat \varphi_i &= \f{a_i}{l^2} \sum_{j = 0}^{n-1+\ve} \hat C^{(j)}_i d \psi_j \nn 
    \f{\pa}{\pa \hat \vp_i} &= \sum_{j=0}^{n-1+\ve} \hat D^i_{(j)} \f{l^2}{a_i} \f{\pa}{\pa \psi_j} \nn
    \f{\pa}{\pa \psi_j} &= \sum_{i = 0}^{n-1+\ve} \hat C_i^{(j)} \f{a_i}{l^2} \f{\pa}{\pa \hat \varphi_i} \label{psivarphiequations}
\end{align}
In particular, $\pa/\pa \psi_0$ can be written
\begin{align}
    \f{\pa}{\pa \psi_0} &= \sum_{i = 0}^{n-1+\ve} \f{a_i}{l^2} \f{\pa}{\pa \hat \varphi_i} \nn 
    &= \f{\pa}{\pa \tau} + \sum_{i = 1}^{n-1+\ve} \f{a_i}{l^2} \f{\pa}{\pa \hat \varphi_i} \nn 
    &= \xi + \sum_{i = 1}^{n-1+\ve} \f{a_i}{l^2} \eta_i.
\end{align}
This is the vector found for $\pa/\pa \tau$ in BL coordinates as introduced in Section \ref{ImportantKilling}. It turns out this is equal to the $\beta$ vector which will be introduced in Section \ref{PCKYIntro}. 

Note that in \cite{FrolovReview}, it is pointed out that taking $\tau = \psi_0$ in the form \eqref{oneforms}, \eqref{emuehatmuehat0} of the metric brings the metric into a ``Carter-like form.''

The expressions for the canonical basis vectors in $(t,r,y_\alpha,\phi_i)$ coordinates are given in Appendix \ref{basisvectorsKS}.

The pure AdS metric in coordinates $(r,y_\alpha, t,\phi_i)$ can be found by setting $M_\mu = 0$, using the $\tilde \phi_i$ form of the basis one-forms and noting that $\tau = t, \hat \varphi_i = \phi_i$ if $M_\mu = 0$. It is written as
\begin{align}
    d \bar s^2 &= \sum_{\alpha=1}^{n-1}\left[ \f{U_\alpha dy_\alpha^2}{\bar X_\alpha} +\f{\bar X_\alpha}{U_\alpha} \left( \sum_{i = 0}^{n-1+\ve} \f{l^2 a_i^{2\varepsilon - 1} \Gamma_i d \phi_i}{(a_i^2-y_\mu^2)\hat \Ups_i} \right)^2\right] + \f{U_n dr^2}{\bar X}   - \f{\bar X}{U_n}\left( \sum_{i = 0}^{n-1+\ve} \f{l^2 a_i^{2\varepsilon - 1} \Gamma_i d \phi_i}{(a_i^2+r^2)\hat \Ups_i} \right)^2  \nn
    & \qquad + \varepsilon S \left(\sum_{i = 0}^{n-1+\ve} \f{l^2 \Gamma_i d \phi_i}{a_i \hat \Ups_i}\right)^2, \label{adsrytphi}
\end{align}
using $\phi_0 \equiv t/l$ and $a_0 \equiv l$ for convenience. 

While this form maintains the general structure associated with the natural orthonormal basis, the background AdS can also be written in a form which is diagonal in the coordinates $(t,r,y_\alpha, \phi_i)$. The easiest way to see it is to take the form of the AdS metric 
\eqref{dbarsspheroidal} for the $(t,\phi_i)$ portion of the metric and apply the Jacobi transformations \eqref{Jacobi} to $\mu_i$. The result is
\begin{align}
    d \bar s^2 &= \sum_{\alpha = 1}^{n-1} \f{U_\alpha dy_\alpha^2}{\bar X_\alpha} + \f{U_n dr^2}{\bar X} + \sum_{i = 0}^{n-1+\ve} \f{(-1)^{n+\ve} l^2 \Gamma_i}{a_i^{2(1-\ve)} \hat \Ups_i} d \phi_i^2 \nn 
    &= \sum_{\alpha=1}^{n-1} \f{U_\alpha dy_\alpha^2}{\bar X_\alpha} + \f{U_n dr^2}{\bar X} + \sum_{i=1}^{n-1+\ve} \f{\Gamma_i}{\Xi_i (-1)^{n+\ve} a_i^{2(1-\ve)} \Upsilon_i} d\phi_i^2 - \f{(1+r^2/l^2)\prod_{\alpha=1}^{n-1} (1-y_\alpha^2/l^2)}{\prod_{j=1}^n \Xi_j} dt^2. \label{backgroundds2rytphi}
\end{align}

The advantage of the coordinates $(\tau,\hat \varphi_i)$ (or $(\tau,\varphi_i), (t,\phi_i)$ etc.)~over $\psi_j$ coordinates is that they better demonstrate the global symmetries of the spacetime. In the presence of NUT parameters, things can get more complicated, but in the case of Kerr--AdS (or background AdS), or even a generalized Kerr--AdS where $X$ is a general function of $r$ but the other $X_\alpha = \bar X_\alpha = Y(y_\alpha)$, each of $\hat \varphi_i, \varphi_i, \phi_i$ are canonically normalized to have period $2\pi$, such that the points $(\tau, \hat \varphi_1, \ldots \hat \varphi_i, \ldots, \hat \varphi_{n-1+\ve},r,y_\alpha)$ and $(\tau, \hat \varphi_1, \ldots, \hat \varphi_i + 2\pi, \ldots, \hat \varphi_{n-1+\ve},r,y_\alpha)$ are identified (and similarly for $\varphi_i, \phi_i$). By contrast, the $\psi_j$'s periodicities are functions of the $a_i$ in a complicated way (and must basically be constructed from the periodicities of the $\hat \varphi_i$ coordinates). On the other hand, the $\psi_j$ coordinates simplify the metric expressions and as we will see they also represent important symmetries in the spacetime. $\beta = \partial/\partial \psi_0$ in particular will turn out to be very important. 

Note that since the $\psi_j$ are linearly related to $\tau, \hat \varphi_i$, all of the $\pa/\pa \psi_j$ are Killing vectors. (This is also apparent from the fact that the basis one-forms, and thus the metric, do not depend on $\psi_j$.)

Since \eqref{backgroundds2rytphi} gives a diagonal form of the metric, it is straightforward to calculate the metric determinant $\bar g$:
\begin{align}
    -\bar g &= -\bar g_{tt} \left(\prod_{i = 1}^{n-1+\ve} \bar g_{\phi_i \phi_i}\right) \bar g_{rr} \prod_{\alpha=1}^{n-1} \bar g_{y_\alpha y_\alpha}\nn 
    &= \f{(1+r^2/l^2) \prod_{\alpha = 1}^{n-1} (1-y_\alpha^2/l^2)}{\prod_{j=1}^n \Xi_j} \left(\prod_{i = 1}^{n-1+\ve} \f{\Gamma_i}{\Xi_i (-1)^{n-2+\ve} a_i^{2(1-\ve)} \Upsilon_i} \right) \f{U_n}{\bar X} \prod_{\alpha = 1}^{n-1} \f{U_\alpha}{\bar X_\alpha}.
\end{align}

The product $\prod_{\mu=1}^n U_\mu$ can be evaluated by noting that the binomial $x_\mu^2-x_\nu^2$ for each pair of distinct $\mu,\nu$ appears twice in the product, once with a positive sign and once with a negative sign. There are $n(n-1)/2$ such pairs. Letting $P$ be defined by
\begin{align}
    P &\equiv \prod_{1 \leq \mu < \nu \leq n} (x_\mu^2-x_\nu^2), \label{Pdefinition}
\end{align}
which I note includes $x_n^2 = -r^2$, then 
\begin{align}
    \prod_{\mu = 1}^n U_\mu &= (-1)^{n(n-1)/2} P^2.
\end{align}
Similarly, let $C$ be defined by
\begin{align}
    C &\equiv \prod_{1 \leq i < j \leq n-1+\ve} (a_i^2 - a_j^2). \label{Cdefinition}
\end{align}

We then have,
\begin{align}
    \prod_{i=1}^{n-1+\ve} (-1)^{n-2+\ve} \Upsilon_i &= (-1)^{(n-2+\ve)(n-1+\ve)}  (-1)^{(n-1+\ve)(n+\ve)/2} C^2 \nn 
    &= (-1)^{(n-1+\ve)(n-2+\ve)/2} C^2.
\end{align}
The term $(-1)^{(n-2+\ve)(n-1+\ve)} = 1$ because at least one of $n-2+\ve$ and $n-1+\ve$ must be even.

We then have
\begin{align}
    \f{\prod_{\mu=1}^{n} U_\mu}{\prod_{i=1}^{n-1+\ve} (-1)^{n-2+\ve} \Upsilon_i} &= (-1)^{n(n-1)/2 - (n-1+\ve)(n-2+\ve)/2} \f{P^2}{C^2} \nn 
    &= (-1)^{(1-\ve)(2n-2+\ve)/2} \f{P^2}{C^2}.
\end{align}
If $\ve = 1$, the above sign is 1. If $\ve = 0$, it is $(-1)^{n-1}$.

In even dimension, the $(1+r^2/l^2),(1-y_\alpha^2/l^2)$ and $\Gamma_i$ terms cancel the $\bar X, \bar X_\alpha$ terms up to a factor of $(-1)^{n-1}$, which cancels the above factor of $(-1)^{n-1}$. Consequently, in even dimensions, 
\begin{align}
    -\bar g &= \left( \f{P}{C \prod_{i=1}^{n-1} a_i \Xi_i}\right)^2.
\end{align}

In odd dimensions, the argument is similar and the result is 
\begin{align}
    -\bar g &= \left(\f{P r \prod_{\alpha = 1}^{n-1} y_\alpha}{C \prod_{j = 1}^n \Xi_j}\right)^2.
\end{align}

Combining the two expressions,
\begin{align}
    \sqrt{-\bar g} &= \f{P}{C \prod_{i = 1}^{n} \Xi_i} \f{(r \prod_{\alpha=1}^{n-1} y_\alpha)^\varepsilon}{(\prod_{j=1}^{n-1} a_j)^{1-\varepsilon}}. \label{sqrtgrytphi}
\end{align}
This is up to an overall sign, which depends on the choice of orientation.   

For interest we can also state the metric determinant in $(r,y_\alpha,\psi_j)$ coordinates, which is calculated in Appendix \ref{Determinantinxmupsijcoords} to be
\begin{align}
    \sqrt{-g} &= P \left( \sqrt c r \prod_{\alpha=1}^{n-1} y_\alpha\right)^\ve, 
\end{align}
again up to choice of orientation. This expression is simpler than the one for $(r,y_\alpha,\phi_i)$ coordinates, but it is easier to work with the $(t,r,y_\alpha,\phi_i)$ coordinates overall because the $0 \leq \phi_i \leq 2\pi$ periodicities (for $i \neq 0$) are clearer than the limits on $\psi_j$.

I will say a quick word on dimensions (in the sense of, ``dimensions of length''). As usual let $c = G = k_B = 1$, but for the moment relax the $\hbar = 1$ condition. Then the only dimension of interest is length, which I designate $L$. $ds^2$ has dimensions of length squared or $L^2$. That means that each $e^A$ has dimensions of $L^1$. $r, x_\mu, a_i, l$ all have dimensions of $L^1$, and $Q_\mu$ all have dimensions $L^0$. $X_\mu, \bar X_\mu, X, \bar X,$ and $U_\mu$ all have dimensions $L^{2(n-1)}$. $t, \tau$ have dimensions $L^1$ and $\phi_i, \hat \vp_i, \vp_i$ are dimensionless. The $A_\mu^{(j)}$ and $A^{(j)}$ have dimension $L^{2j}$ and $B^\mu_{(j)}$ have dimension $L^{-2j}$. Interestingly the $\psi_j$ are not dimensionless, but have dimension $L^{1-2j}$. 

\subsection{Principal Conformal Killing--Yano Tensor and Associated Vector} \label{PCKYIntro}

As shown in \cite{KrtousFrolov, Houri}, the existence of the basis given above is equivalent to the statement that there exists a Principal Conformal Killing--Yano (PCKY) tensor of the spacetime, a Closed Conformal KY tensor which has the form
\begin{align}
    \bs h &= \sum_{\mu = 1}^n x_\mu \bs \om^\mu, \label{PCKYdef}
\end{align}
or $\bs h_{ab} = \sum_{\mu = 1}^n 2 x_\mu e^\mu_{[a} e^{\hat \mu}_{b]}$. To distinguish from the Kerr--Schild correction $h_{ab}$, I will use a boldface for the PCKY tensor at all times. 

Associated with $\bs h$ is the vector which I call $\beta$ (and which is called $\xi$ in those papers, which I am reserving for the asymptotically static vector), which is given by the divergence of $\bs h$:
\begin{align}
    \beta^b &\equiv \frac{1}{D-1} \nabla_a \bs h^{a b} \label{betadef}
\end{align}
$\beta$ is a Killing vector which also satisfies
\begin{align}
    \mc L_\beta \bs h = 0,
\end{align}
where $\mc L_\beta$ is the Lie derivative with respect to the vector $\beta$. Both of these statements automatically follow from the existence of the PCKY tensor \cite{FrolovReview}. In fact, all the quantities $x_\mu, \psi_j, Q_\mu$ and $S$ can be generated entirely from the PCKY tensor and its eigenvalues and eigenvectors.

$\bs h$ can also be expressed in terms of a potential one-form $\bs b$ as (see e.g.~\cite{KrtousFrolov})
\begin{align}
    \bs h &= d \bs b \nonumber \\
    \bs b&= \frac{1}{2} \sum_{j = 0}^{n-1} A^{(j+1)} d\psi_j. \label{bpotentialdefinition}
\end{align}
The closed condition on $\bs h$ is that $d\bs h = 0$, which follows because $\bs h$ is exact ($\bs h = d\bs b$). The CKY (conformal Killing--Yano) condition is the statement that
\begin{align}
    \nabla_{(a} \bs h_{b) c} &= g_{ab} \beta_c - \beta_{(a} g_{b) c}.
\end{align}
Along with the closedness condition $d \bs h = 0$, some algebra gives (see e.g.~\cite{FrolovReview}) 
\begin{align}
    \na_a \bs h_{bc} &= \f{2}{3} (\na_{(a} \bs h_{b)c} - \na_{(a} \bs h_{c)b}) + \na_{[a} \bs h_{bc]} \nn 
    &= g_{ab} \bt_c - g_{ac} \bt_b.
\end{align} 

The $\beta$ vector in coordinates $(x_\mu, \psi_j)$ has the form (see, e.g.,~\cite{KrtousKubiznak})
\begin{align}
    \beta &= \frac{\partial}{\partial \psi_0},
\end{align}
which can be expanded as
\begin{align}
    \beta &= \sum_{\mu = 1}^n \sqrt{Q_\mu} e_{\hat \mu} + \varepsilon \sqrt{S} e_{\hat 0}. \label{betacanonicalbasis}
\end{align}
Both the metric and $\bs h$ are independent of $\psi_0$, which demonstrates $\lie_\beta g = \lie_\beta \bs h = 0$. Moreover, all the basis one-forms are independent of $\psi_0$ so that $\lie_\beta e^A = 0$. 

The coordinate expression for $\beta^\flat$ is
\begin{align}
    \beta^\flat &= \sum_{\mu = 1}^n \sqrt{Q_\mu} e^{\hat \mu} + \varepsilon \sqrt{S} e^{\hat 0} \nonumber \\
    &= \sum_{\mu =1}^n Q_\mu \sum_{j = 0}^{n-1} A_\mu^{(j)} d\psi_j + \varepsilon S \sum_{j = 0}^n A^{(j)} d \psi_j.
\end{align}

The quantity $\beta^2 \equiv \beta \cdot \beta$ can be written as
\begin{align}
    \beta^2 &\equiv \beta \cdot \beta \nn
    &= \sum_{\mu=1}^n Q_\mu + \varepsilon S. \label{beta2}
\end{align}
This quantity turns out to be important and will recur. 

Additionally, after some work (see Appendix Section \ref{dbetaSection}), we have
\begin{align}
    d \beta^\flat &= \sum_{\mu = 1}^n \frac{\partial \beta^2}{\partial x_\mu}  e^\mu \wedge e^{\hat \mu} = \sum_{\mu=1}^n \f{\pa \beta^2}{\pa x_\mu} \bs \omega^\mu. \label{dbetaflat}
\end{align}
To be clear, here and elsewhere, I mean
\begin{align}
    \f{\pa \beta^2}{\pa x_\mu} &= \f{\pa}{\pa x_\mu} \left( \beta^2\right).
\end{align}

Note how similar $d \beta^\flat$ is to $\bs h$ in terms of its structure (expansion in terms of the $\bs \omega^\mu$). 

To tease out the reason for this form, it is perhaps useful to show how $d \beta^\flat$ relates to $\bs h$ and the curvature tensors associated with the spacetime. $\na_a \bt_b$ is given in (e.g.)~\cite{FrolovReview} as 
\begin{align}
    (D-2) \na_a \bt_b &= - R_{ac} {\bs h^c}_b + \f12 \bs h_{cd} {R^{cd}}_{ab},
\end{align}
which (along with the fact that $\bt$ is a Killing vector and so has $\na_{(a}\bt_{b)}=0$) can be rewritten as 
\begin{align}
    \na_a \beta_b &= \na_{[a}\beta_{b]} = \f{1}{2(D-2)}\left[ C_{abcd} \bs h^{c d} + \f{2(D-4)}{D-2} R^d_{[a} \bs h_{b] d} -\f{2}{(D-1)(D-2)} R \bs h_{ab} \right]. \label{nabetacurvatureresult}
\end{align}
(See Appendix \ref{dbetacurvature} for more comments on this.) Interestingly, if $D = 4$, the Ricci tensor term drops out. Appendix \ref{dbetacurvature} also shows how the Riemann decomposition of $\na_a \beta_b$ and the symmetries of the Riemann and Ricci tensors imply that $\na_a \beta_b$ can be expanded in terms of the $\bs \om^\mu_{ab}$ as in \eqref{dbetaflat}.

Now consider pure AdS, so that  $ C_{abcd} = 0$, $ R^a_b = -(D-1)l^{-2} \de^a_b$ and $ R = - D(D-1)l^{-2}$. Then, 
\begin{align}
    {\na_a \beta_b} &= \f{1}{l^2} {\bs h_{a b} }\nn 
    {d \beta^\flat} &= \f{2}{ l^2} {\bs h}. \label{dbetaflatAdS}
\end{align}
This tells us that in pure AdS, $\beta$ is (up to constant multiple) both the divergence of $\bs h$ and also a potential for it.  
(The reason for the extra factor of two is because $(d \beta^\flat)_{ab} = 2 \na_{[a}\beta_{b]}.$) 

(In \cite{FrolovWeaklyCharged}, a weak electromagnetic field is constructed in a Kerr--NUT--(A)dS spacetime of the form
\begin{align}
    \bs F_{ab} = 2 q \left( \pa_{[a} \bt_{b]} + \f{2 \La}{(D-1)(D-2)} \bs h_{ab}\right),
\end{align}
and it is pointed out that, since $\bs h_{ab}$ has the same form in the background (A)dS as the full spacetime---a point I will return to in Section \ref{canonicalbasisbackground}, ``the operation of `upgrading' the Killing vector ansatz for an electromagnetic field can be interpreted as a subtraction from $2 \pa_{[a} \bt_{b]}$ a similar quantity, calculated in the corresponding (anti-)de Sitter background metric'' [slightly modifying notation from the quote]. I mention this in part because later we will encounter expressions of the form $\na^a \bt^b - \overline{\na^a \bt^b}$, where $\overline{\na^a \bt^b}$ represents the covariant derivative of $\bt$ as calculated in the background AdS. This term is equivalent to the term appearing in \cite{FrolovWeaklyCharged}.)

\subsection{Null Vectors} \label{PCKYNull}

Following \cite{KrtousFrolov}, let $m_\mu$ be the (complex) vector
\begin{align}
    m_\mu &\equiv \frac{i e_\mu + e_{\hat \mu}}{\sqrt 2}.
\end{align}
We note immediately that $m_\mu$ is null.

$m_\mu$ has the property
\begin{align}
    m_\mu \cdot \bs \omega^\nu &= \delta^\nu_\mu i x_\mu m_\mu^\flat.
\end{align}
That is, $m_\mu \cdot \bs \om^\mu = i x_\mu m_\mu^\flat, m_\mu \cdot \bs \om^\nu = 0$ if $\nu \neq \mu$. 

As a consequence, $m_\mu$ is an eigenvector of $\bs h$, with $i x_\mu$ as its associated eigenvalue:
\begin{align}
    m_\mu \cdot \bs h &= i x_\mu m_\mu^\flat.
\end{align}
(In component notation, $m_\mu^a \bs h_{a b} = i x_\mu (m_\mu)_b$.)  

We can very similarly define $m_\mu^*$ by
\begin{align}
    m_\mu^* &\equiv \f{-i e_\mu + e_{\hat \mu}}{\sqrt{2}}.
\end{align}
The star convention of course evokes complex conjugation, and indeed if $e_\mu$ and $e_{\hat \mu}$ are real $m_\mu^*$ is the complex conjugate of $m_\mu$. In our situation, $e_{\hat n}$ is not real, and so we cannot interpret $m_n^*$ as the actual complex conjugate of $m_n$, since we are not additionally complex conjugating $e_n$. Like $m_\mu$, $m_\mu^*$ is null and an eigenvector of $\bs h$ with
\begin{align}
    m_\mu^* \cdot \bs h = - i x_\mu (m_\mu^*)^\flat.
\end{align}

$m_\mu$ and $m_\nu^*$ satisfy
\begin{align}
    m_\mu \cdot m_\nu^* &= \de_{\mu \nu}.
\end{align}

As stated in in \cite{KrtousFrolov},
\begin{align}
    m_\mu \cdot \beta = i \sqrt{\frac{Q_\mu}{2}}.
\end{align}

It is useful to define a rescaled version of $m_\mu$. I will call it $\tilde m_\mu$. 
\begin{align}
    \tilde m_\mu \equiv \sqrt{2/Q_\mu} m_\mu = (Q_\mu)^{-1/2} (i e_\mu + e_{\hat \mu}).
\end{align}
$\tilde m_\mu$ has a constant unit interior product with $\beta$:
\begin{align}
    \tilde m_\mu \cdot \beta = 1. \label{tildemmudotbeta}
\end{align}
$\tilde m_\mu$, as shown by Hamamoto \cite{Hamamoto}, is tangent to an affinely parametrized geodesic:
\begin{align}
    \nabla_{\tilde m_\mu} \tilde m_\mu = 0.
\end{align}

Explicitly, the vector and one-form components of the $\tilde m_\mu$ are
\begin{align}
    \tilde m_\mu &= i \partial_{x_\mu} + \frac{1}{X_\mu} \sum_{j = 0}^{n-1+\ve} (-x_\mu^2)^{n-1-j} \partial_{\psi_j} \nonumber \\
    \tilde m_\mu^\flat &= \frac{i dx_\mu}{Q_\mu} + \sum_{j = 0}^{n-1} A_\mu^{(j)} d\psi_j. \label{tildempsicoords}
\end{align}

Similarly, define $\tilde m_\mu^* = \sqrt{2/Q_\mu} m^*_\mu$. $\tilde m_\mu^*$ is also tangent to an affinely parametrized geodesic and has $\tilde m_\mu^* \cdot \beta = 1$. Of course that means that $m_\mu, m_\mu^*$ are tangent to non-affinely parametrized geodesics. 

Hamamoto et al.~\cite{Hamamoto} showed that the metrics with a PCKY tensor are also Petrov Type D, and that they have Weyl Aligned Null Directions (WANDs). These WANDs are none other than the $m_\mu$ and $m^*_\mu$ (or, alternatively, the $\tilde m_\mu, \tilde m^*_\mu$). Let $e_A,e_B,e_C$ be (for the following equation only) any basis vectors which are not $e_\mu$ or $e_{\hat \mu}$. Then 
\begin{align}
    W(m_\mu, e_A, e_B, e_C) &= W(m^*_\mu, e_A, e_B, e_C) = 0 \nn
    W(m_\mu, e_A, m_\mu, e_B) &= W(m^*_\mu, e_A, m^*_\mu, e_B) = 0 \nn 
    W(m_\mu, m^*_\mu, m_\mu, e_A) &= W(m_\mu, m^*_\mu, m^*_\mu, e_A) = 0, \label{TypeD}
\end{align}
where $W(u,v,w,p) = C_{a b c d} u^a v^b w^c p^d$ for vectors $u,v,w,p$.

$m_\mu$ is an eigenvector of the Ricci tensor, which follows from the Ricci tensor decomposition \eqref{Riccicanonical}:
\begin{align}
    R_{a b} m_\mu^b &= \sum_{\nu = 1}^n \mc R_{\nu \nu} (e^\nu_a e^\nu_b + e^{\hat \nu}_a e^{\hat \nu}_b) m_\mu^b + \varepsilon \mc R_{\hat 0 \hat 0} e^{\hat 0}_a e^{\hat 0}_b m_\mu^b \nn 
    &= \mc R_{\mu \mu} \left(\frac{i e^\mu_a + e^{\hat \mu}_a}{\sqrt 2}\right) \nn 
    &= \mc R_{\mu \mu} (m_\mu)_a.
\end{align}
No sum over $\mu$ is implied. Similarly, $R_{a b} \tilde m_\mu^b = \mc R_{\mu \mu} (\tilde m_\mu)_a$, and also similarly for $m^*_\mu, \tilde m^*_\mu$.

In addition to being eigenvectors of $\bs h$ and of the Ricci tensor, the $m_\mu, m^*_\mu$ (and thus $\tilde m_\mu, \tilde m^*_\mu)$ are also eigenvectors of $d \beta^\flat$, which follows from \eqref{dbetaflat}.  Explicitly, 
\begin{align}
    m_\mu \cdot d \beta^\flat &= i\f{\pa \beta^2}{\pa x_\mu} m_\mu^\flat \nn 
    m^*_\mu \cdot d \beta^\flat &= -i \f{\pa \beta^2}{\pa x_\mu} (m^*_\mu)^\flat.
\end{align}
Because $\beta$ is a Killing vector, this also implies
\begin{align}
    (\na_{m_\mu} \beta)_b &= m_\mu^a \na_a \beta_b \nn 
    &= m_\mu^a \na_{[a}\beta_{b]} \nn 
    &= \f12 m_\mu^a (d \beta^\flat)_{a b} \nn 
    &= \f12 (m_\mu \cdot d \beta^\flat)_b \nn 
    &= \f i2 \f{\pa \beta^2}{\pa x_\mu} (m_\mu)_b.
\end{align}
Because $\lie_\beta e_\mu = \lie_\beta e_{\hat \mu} = 0$, we also have $\lie_\beta m_\mu = 0$, from which we conclude that 
\begin{align}
    \na_\beta m_\mu &= \na_{m_\mu} \beta = \f i2 \f{\pa \beta^2}{\pa x_\mu} m_\mu.
\end{align}
Similarly,
\begin{align}
    \na_\beta m^*_\mu &= \na_{m^*_\mu} \beta = -\f i 2 \f{\pa \beta^2}{\pa x_\mu} m^*_\mu.
\end{align}
Because $\lie_\beta Q_\mu = \pa Q_\mu/\pa \psi_0 = 0$, we also have
\begin{align}
    \na_\beta \tilde m_\mu = \na_{\tilde m_\mu} \beta &= \f{i}{2} \f{\pa \beta^2}{\pa x_\mu} \tilde m_\mu \nn
    \na_\beta \tilde m^*_\mu = \na_{\tilde m^*_\mu} \beta &= -\f{i}{2} \f{\pa \beta^2}{\pa x_\mu} \tilde m^*_\mu. \label{nabetatildem}
\end{align}

As explained in Appendix Section \ref{dbetacurvature}, the fact that $m_\mu$ is an eigenvector of $d\beta^\flat$ can also be shown to be a consequence of the expansion of $d \beta^\flat$ in terms of $\bs h$ and the curvature tensors as well as $m_\mu$ being a WAND and an eigenvector of both $\bs h$ and the Ricci tensor. This gives some sense of how deeply the $m_\mu$ (and $m_\mu^*$, and the directions proportional to them) are entwined with the symmetries of the spacetime. 

To summarize: $\tilde m_\mu$ is a null, affinely parametrized geodesic, which is a WAND and an eigenvector of the PCKY tensor $\bs h$, the Ricci tensor, and $d \beta^\flat$, and its inner product with $\beta$ is a constant. The same statements are all true for $\tilde m^*_\mu$. 

It turns out that the Kerr--Schild null vector $k$ can be written as $\tilde m_n$, and so this allows us to transition into discussing the Kerr--anti-de Sitter spacetime and the Kerr--Schild form.

\section{Generalized Kerr--anti-de Sitter: Kerr--Schild Form and Black Hole Horizons} \label{KadSKSForm}

We return to the GKAdS spacetimes, which are specifically those for which $\bar X_\alpha = Y(y_\alpha)$. There are really two spacetimes involved, a background pure AdS spacetime for which $X_n = -\bar X = Y(x_n) = Y(i r)$, and a full spacetime where $X_n = -X$, an arbitrary function of $r$, satisfying
\begin{align}
    X = \bar X - 2 \mu(r) r^{1-\ve},
\end{align}
where $\mu(r) = m$ is the Kerr--AdS solution. Of course, both the full and background spacetimes are GKAdS and thus also part of the Generalized Kerr--NUT--AdS spacetime, and so both have the associated symmetries.

Consider now the Kerr--Schild null vector $k$. Chen and L\"u \cite{ChenLu08} demonstrated that the Kerr--NUT--AdS spacetimes can be written in multi-Kerr--Schild form, including the specific case of the Kerr--AdS spacetime. In \cite{FrolovReview}, it is made explicit that the Kerr--Schild null vectors are the WANDs of the spacetime (that is, parallel to the $m_\mu$ or $m^*_\mu$). I use slightly different notation but my argument is essentially the same, though I will go through the derivations because I use some of the results (in my notation) going forward. 

\subsection{Kerr--Schild Null Vector} \label{KSNullVector}

Focusing on the full spacetime for the moment, using the expansion of $k^a \pa_a$ in $(t,r,\phi_i)$ coordinates as shown in Appendix \ref{kCanonicalFrame}, the Kerr--Schild null vector $k$ can be written as $\tilde m_n$:
\begin{align}
    k &= \tilde m_n = \f{i e_n + e_{\hat n}}{\sqrt{Q_n}}. \label{kexp}
\end{align}
This shows a close relationship between the Kerr--Schild form of the spacetime and the canonical basis, which I will elaborate on momentarily. Because $k = \tilde m_n$, $k$ has all the properties that we discussed in the last section. Note especially that $k$ is an affinely parametrized null geodesic, that, from \eqref{tildemmudotbeta},
\begin{align}
    k \cdot \beta = 1, \label{kdotbeta}
\end{align}
that $k$ is an eigenvector of $\bs h$ and $d \beta^\flat$,
\begin{align}
    k \cdot \bs h &= -r k^\flat \nn 
    k \cdot d \beta^\flat &= \f{\pa \beta^2}{\pa r} k^\flat,
\end{align}
and that from \eqref{nabetatildem} $k$ satisfies the covariant derivative expressions
\begin{align}
    \na_\bt k &= \na_k \bt = \f{1}{2} \f{\pa \bt^2}{\pa r} k. \label{nabetak}
\end{align}

In $(r,y_\alpha,\psi_j)$ coordinates, $k^a$ expands to (from \eqref{tildempsicoords}) 
\begin{align}
    k&=\f{\pa}{\pa r} - \f{1}{X} \sum_{j=0}^{n-1+\ve} r^{2(n-1-j)} \f{\pa}{\pa \psi_j}, \label{kpsicoords}
\end{align}
so that the leading order term in the time--azimuthal sector is proportional to $\pa/\pa\psi_0 = \beta$. 

In coordinates $(t,r,y_\alpha,\phi_i)$, the covariant components $k_a dx^a$ can be found by using $k_a = \bar g_{a b} k^b$ (from \eqref{kmu} and \eqref{backgroundds2rytphi})
\begin{align}
    k_a dx^a &= \f{\prod_{\alpha=1}^{n-1}(1-y_\alpha^2/l^2)}{\prod_{j=1}^n \Xi_j} dt + \f{U_n}{\bar X} d r - \sum_{i=1}^{n-1+\ve} \f{\G_i}{\Xi_i(-1)^{n+\ve} a_i^{1-2\ve} \Ups_i (r^2+a_i^2)} d \phi_i. \label{kadxatryphi}
\end{align}
The contravariant components in coordinates $(r,y_\alpha,\psi_j)$ has the pleasing form \eqref{kpsij}.

Let $\ell$ be the vector in the direction $m^*_n$, normalized so that $k \cdot \ell = -1$. This will also be a null geodesic (though not affinely parametrized) and a WAND of the spacetime. This is
\begin{align}
    \ell &= -\f{\sqrt{ Q_n}}{2} m_n^* = \f{\sqrt{Q_n} ( i e_n - e_{\hat n})}{ 2}. \label{elldef}
\end{align}
$\ell$ is an eigenvector of $\bs h$ with eigenvalue $+r$:
\begin{align}
    \ell \cdot \bs h = +r \ell^\flat. \label{elldoth}
\end{align}

Using the expressions for the vectors in Section \ref{basisvectorsKS}, in $(t,r,y_\alpha,\phi_i)$ coordinates $\ell$ is written
\begin{align}
    \ell &= \f{1}{2 U_n} \left( X \f{\pa}{\pa r} - (X-2\bar X)\sum_{i=0}^{n-1+\ve} \f{a_i}{r^2+a_i^2} \f{\pa}{\pa \phi_i} \right) \nn 
    &= \f{1}{2 U_n} \left( X \f{\pa}{\pa r} - (X-2 \bar X) \left[\f{1}{1+r^2/l^2} \f{\pa}{\pa t}+ \sum_{i=1}^{n-1+\ve} \f{a_i}{r^2+a_i^2} \f{\pa}{\pa \phi_i}\right]\right).
\end{align}
If $X = 0$, $\ell$ becomes
\begin{align}
    \ell &= \f{\bar X}{U_n} \sum_{i=0}^{n-1+\ve} \f{a_i}{r^2+a_i^2} \f{\pa}{\pa \phi_i}, \qquad \textrm{if }X = 0.\label{ellifXis0}
\end{align}
If we consider some $r = r_1$ such that $X = 0$, $(U_n/\bar X) \ell|_{r=r_1}$ is then a Killing vector, equal to $\sum_{i=0}^{n-1+\ve} \f{a_i}{r_1^2+a_i^2} \pa_{\phi_i}$. 

Note,
\begin{align}
    k^\flat \wedge \ell^\flat &= - i e^n \wedge e^{\hat n} = - i \bs \om^n.
\end{align}
Consequently, 
$\bs h$ can then be rewritten as
\begin{align}
    \bs h &= \sum_{\alpha=1}^{n-1} y_\alpha \bs \om^\alpha - r k^\flat \wedge \ell^\flat. \label{hkl}
\end{align}

For the Kerr--Newman metric in four dimensions, it will be convenient to choose a gauge in which $\bs A$ is proportional to $k^\flat$. This can be achieved by taking the $\bs A$ from \eqref{AABL} and sending $\bs A \to \bs A' = \bs A + \f{Q r dr}{\Delta}$, an allowed gauge transformation (since $\Delta$ is a function of $r$ only, so that $Q r dr /\Delta = d \int \f{Q r dr}{\Delta}$). Comparing to \eqref{k4DABL}, we have
\begin{align}
    \bs A &= -\f{Q r}{\rho^2} k^\flat. \label{Aproptok}
\end{align}

It will now be useful to introduce an orthonormal basis for the background spacetime in order to further tease out some properties for the full and background spacetimes.

\subsubsection{Canonical Basis for Background Spacetime} \label{canonicalbasisbackground}

The transformations from $\mu_i$ coordinates to $y_\alpha$ coordinates do not depend on $\mu(r)$, and so can be said to be the same for both the full and background spacetime. On the other hand, the transformations between the $(t,\phi_i)$ and $\psi_j$ coordinates manifestly depend on $\mu(r)$. It is useful then to define coordinates $\bar \psi_j$ which are specifically associated with the \emph{background} AdS spacetime, and indeed to define a new orthonormal basis associated with the background, $\bar e^A$, written in terms of $(x_\mu, \bar \psi_j$). We have
\begin{align}
    \bar e^\mu &= \f{d x_\mu}{\sqrt{\bar Q_\mu}} \nn 
    \bar e^{\hat \mu} &= \sqrt{\bar Q_\mu} \sum_{j=0}^{n-1} A_\mu^{(j)} d \bar \psi_j \nn 
    \bar e^{\hat 0} &= \sqrt{S} \sum_{j=0}^n A^{(j)} d \bar \psi_j,
\end{align}
with $\bar e^{\hat 0}$ defined only in odd dimensions, where
\begin{align}
    \bar Q_\mu &= \f{\bar X_\mu}{U_\mu} = \f{Y(x_\mu)}{U_\mu},
\end{align}
and any unbarred quantities maintain their meanings. We then can write
\begin{align}
    d\bar s^2 &= \sum_{A=1}^D (\bar e^A)^2 \nn 
    &= \sum_{\mu=1}^n \left( \f{dx_\mu^2}{\bar Q_\mu} + \bar Q_\mu \left(\sum_{j=0}^{n-1} A_\mu^{(j)} d \bar \psi_j\right)^2\right) + \ve S \left(\sum_{j=0}^n A^{(j)} d \psi_j\right)^2.
\end{align}
Because $\psi_j$ are just scalar functions on the manifold, the one-forms $e^A$ and $\bar e^A$ both ``exist'' in both the background and full spacetime, in an unambiguous way. However, we must state that the one-forms are only orthonormal when we make use of the corresponding metric: $e^A$ are only orthonormal for the full metric, and $\bar e^A$ are only orthonormal for the background. I will presume further that the corresponding dual basis vectors $e_A$ are raised using the full metric and $\bar e_A$ are raised using the background metric. 

To transform to the form \eqref{adsrytphi} requires, as usual, $x_n = i r, x_\alpha = y_\alpha$, and for the $\bar \psi_j$ we can write them in terms of $(t = l \phi_0,\phi_i)$ by simply adopting the expressions for $\psi_j$ in terms of $\hat \vp_i$ and substituting $\hat \vp_i \to \phi_i$:
\begin{align}
    d\bar \psi_j &= \sum_{i=0}^{n-1+\ve} \hat D^i_{(j)} \f{l^2}{a_i} d \phi_i.
\end{align}
This is because if the GKNAdS metric we \emph{begin} with is already AdS, then we simply have $\phi_i = \hat \vp_i$ (since $\phi_i$ and $\hat \vp_i$ differ only by a function of $r$ proportional to $\mu(r)$). Using \eqref{dphidphihat} (with $m \to \mu(r)$) as well as \eqref{sumDhatbyr2pa2}, we find
\begin{align}
    d\psi_j &= d \bar \psi_j - \f{X_n-\bar X_n}{X_n \bar X_n} r^{2(n-1-j)}dr.
\end{align}
This is integrable since $X,\bar X$ are functions of $r$ only. It is now worth checking how the $e^{\hat \mu}$ and $\bar e^{\hat \mu}$ relate to one another.
\begin{align}
    \sum_{j=0}^{n-1} A_\mu^{(j)} r^{2(n-1-j)} &= \sum_{j=0}^{n-1} A_\mu^{(j)} B^n_{(j)} U_n \nn 
    &= U_n \de^n_\mu.
\end{align}
Relatedly, for $\ve = 1$,
\begin{align}
    \sum_{j=0}^n A^{(j)} r^{2(n-1-j)} &= \f{1}{r^2} \prod_{\mu=1}^n (x_\mu^2+r^2) \nn 
    &= 0
\end{align}
since $x_n^2 = -r^2$. 

Because we also have $X_\alpha = \bar X_\alpha$, we have that \emph{most} of the basis one-forms for the full and background metrics are the same:
\begin{align}
    e^{\alpha} = \bar e^\alpha &= \f{d y_\alpha}{\sqrt{Q_\alpha}} = \f{d y_\alpha}{\sqrt{\bar Q_\alpha}} \nn 
    e^{\hat \alpha} = \bar e^{\hat \alpha} &= \sqrt{Q_\alpha} \sum_{j=0}^{n-1} A_\alpha^{(j)} d \psi_j = \sqrt{\bar Q_\alpha} \sum_{j=0}^{n-1} A_\alpha^{(j)} d \bar \psi_j \nn 
    e^{\hat 0} = \bar e^{\hat 0} &= \sqrt{S} \sum_{j=0}^{n} A^{(j)} d \psi_j = \sqrt{S} \sum_{j=0}^n A^{(j)} d \bar \psi_j.
\end{align}
The cases where the one-forms differ are $e^n \neq \bar e^n$ and $e^{\hat n} \neq \bar e^{\hat n}$. Because $e^n \propto X_n^{-1/2}$ and $\bar e^{ n} \propto \bar X_n^{-1/2}$, we have
\begin{align}
    e^n &= \sqrt{ \f{\bar X_n}{X_n}} \bar e^n.
\end{align}
We can expand $e^{\hat n}$ as (using $dr = -i \sqrt{\bar Q_n} \bar e^n$)
\begin{align}
    e^{\hat n} &= \sqrt{Q_n} \sum_{j=0}^{n-1} A_n^{(j)} d \psi_j\nn 
    &= \sqrt{Q_n} \left( \sum_{j=0}^{n-1} A_n^{(j)} d \bar \psi_j - \f{(X_n - \bar X_n) U_n d r}{X_n \bar X_n}\right) \nn 
    &= \sqrt{Q_n} \left( \f{\bar e^{\hat n}}{\sqrt{\bar Q_n}} + \f{(X_n - \bar X_n) U_n}{X_n \bar X_n} i \bar e^n \sqrt{\bar Q_n}\right) \nn 
    &= \sqrt{ \f{X_n}{\bar X_n}} \bar e^{\hat n} + \f{(X_n - \bar X_n)}{\sqrt{X_n \bar X_n}} i \bar e^n.
\end{align}
Letting (for this section only) $f = \sqrt{X_n/\bar X_n}$ we can write these relations compactly as
\begin{align}
    e^n &= f^{-1} \bar e^n \nn 
    e^{\hat n} &= f \bar e^{\hat n} + i (f-f^{-1}) \bar e^n.
\end{align}
This inverts to
\begin{align}
    \bar e^{\bar n} &= f e^n \nn 
    \bar e^{\hat n} &= f^{-1} e^{\hat n} - i (f^{-1} - f)e^n
\end{align}
Similarly, using $\bar e_A \cdot \bar e^B = 0$ for the dual basis, we conclude that the barred dual basis must satisfy
\begin{align}
    \bar e_A &= e_A, A \neq n,\hat n \nn 
    \bar e_n &= f^{-1} e_n + (f-f^{-1}) e_{\hat n} \nn 
    \bar e_{\hat n} &= f^{-1} e_{\hat n} \nn 
    e_n &= f \bar e_n + (f^{-1}-f) \bar e_{\hat n} \nn 
    e_{\hat n} &= f^{-1} \bar e_{\hat n}.
\end{align}

We then can write the combination $(e^n)^2 + (e^{\hat n})^2$ as
\begin{align}
    (e^n)^2 + (e^{\hat n})^2 &= \f{\bar X_n}{X_n} (\bar e^n)^2 + \f{X_n}{\bar X_n} (\bar e^{\hat n})^2 - \f{(X_n-\bar X_n)^2}{X_n \bar X_n} (\bar e^n)^2 + 2 i \f{(X_n-\bar X_n)}{\bar X_n} \bar e^{\hat n} \bar e^{n} \nn 
    &= (\bar e^n)^2 + (\bar e^{\hat n})^2 + \f{X_n-\bar X_n}{\bar X_n} (\bar e^{\hat n} + i \bar e^n)^2.
\end{align}
Note that $\bar e^{\hat n} + i \bar e^n$ is a null one-form (in the background metric).

Since $e^A = \bar e^A$ for $A \neq n,\hat n$, we can then write the relationship between the full and background spacetime as
\begin{align}
    ds^2 &= \sum_{A=1}^D (e^A)^2 \nn 
    &= \sum_{A \neq n,\hat n} (e^A)^2 + (e^n)^2+(e^{\hat n})^2 \nn 
    &= \sum_{A\neq n, \hat n} (\bar e^A)^2 + (\bar e^n)^2 + (\bar e^{\hat n})^2 +  \f{X_n-\bar X_n}{\bar X_n} (\bar e^{\hat n} + i \bar e^n)^2 \nn 
    &= d\bar s^2 + \f{X_n-\bar X_n}{\bar X_n} (\bar e^{\hat n} + i \bar e^n)^2.
\end{align}
We have thus recovered the Kerr--Schild decomposition in terms of the new basis. We can run the above argument again, writing $\bar e^n, \bar e^{\hat n}$ in terms of $e^n, e^{\hat n}$ and similarly find
\begin{align}
    ds^2 &= d \bar s^2 + \f{X_n - \bar X_n}{ X_n} ( e^{\hat n} + i e^n)^2,
\end{align}
where $e^{\hat n} + i e^n$ is a null one-form (in the full metric). Using $k = m_n$ we then have
\begin{align}
    ds^2 &= d \bar s^2 + \f{X_n - \bar X_n}{U_n} (k^\flat)^2.
\end{align}
That we recover the Kerr--Schild decomposition is not a surprise. The scalar function $H$ is now (written in various forms)
\begin{align}
    H &= \f{X_n - \bar X_n}{U_n} \nn 
    &= Q_n - \bar Q_n \nn 
    &= \f{2 \mu(r) r^{1-\ve}}{U_n}. \label{HintermsofQn}
\end{align}
Comparing to \eqref{His2mbyU} (with $m \to \mu(r)$), we have
\begin{align}
    U &= \f{U_n}{r^{1-\ve}}. \label{UUn}
\end{align}
(This relation is also shown explicitly in Appendix \ref{UUnrelation}.) 

We can confirm, using the relations between the various one-forms and vectors, that $k, k^\flat$ can indeed be raised and lowered with either metric without problem, and that it has the form
\begin{align}
    k &= \f{\bar e_{\hat n} + i \bar e_n}{\sqrt{\bar Q_n}} \nn 
    k^\flat &= \f{\bar e^{\hat n} + i \bar e^n}{\sqrt{\bar Q_n}},
\end{align}
so that, extending the definition of $m_n$ to the background spacetime by defining $\bar m_n = (\bar e_{\hat n} + i \bar e_n)/\sqrt{\bar Q_n}, \bar m^\flat_n = (\bar e^{\hat n} + i \bar e^n)/\sqrt{\bar Q_n}$, $\bar k = \bar m_n, \bar k^\flat = \bar m^\flat_n$. In general we should be wary of using the musical isomorphisms when talking about the background spacetime, but the fact that $k$ can itself be raised and lowered using either metric means there is no ambiguity.

Note also, defining $\bar{\bs \om}^\mu = \bar e^\mu \wedge \bar e^{\hat \mu}$, of course $\bar{\bs \om}^\alpha = \bs \om^\alpha$ for $1 \leq \alpha \leq n-1$. For $\mu = n$ we have
\begin{align}
    {\bs \om}^n &= e^n \wedge e^{\hat n} \nn 
    &= f^{-1} \bar e^n \wedge ( (f-f^{-1}) \bar e^n + f \bar e^{\hat n}) \nn 
    &= \bar e^n \wedge \bar e^{\hat n} \nn 
    &= \bar{\bs \om}^n.
\end{align}
This means that $\bs \om^\mu = \bar{\bs \om}^\mu$ for all $\mu$. This means that, using $\bs h$ for the PCKY tensor for the full spacetime, it can be expressed as
\begin{align}
    \bs h &= \sum_{\mu=1}^n x_\mu \bs \om^\mu \nn 
    &= \sum_{\mu=1}^n x_\mu \bar{\bs \om}^\mu.
\end{align}
This means that $\bs h$ not only has the same form (in terms of the canonical bases) for the full and background spacetime, but that $\bs h$ is still a PCKY tensor for pure AdS. (Again, that $\bs h$ has the same form in the full and background spacetime appears in \cite{FrolovWeaklyCharged,FrolovReview}.) We also show in Appendix \ref{RaisingAndLoweringh} that, in any given coordinates, $\bs h_{ab}$ can be raised to $\bs h^{ab}$ using either metric, because of the antisymmetry of $\bs h$ and the fact that $k$ is an eigenvector of $\bs h$. This means that $\bs h^{ab}$ and $\bs h_{ab}$ unambiguously can refer to the value of $\bs h$ (in fully covariant and fully contravariant form) for both the full and background metric. This, along with the fact that the metric determinant is the same for both the full and background metric, implies that the associated vector $\beta^a$ must have the same (contravariant) form whether the full or background metric is used to calculate it:
\begin{align}
    \beta^a &= \f{1}{D-1} \na_b \bs h^{ba} \nn 
    &= \f{1}{D-1} \f{1}{\sqrt{-g}} \pa_b (\sqrt{-g} \bs h^{ba}) \nn 
    &= \f{1}{D-1} \f{1}{\sqrt{-\bar g}} \pa_b (\sqrt{-\bar g} \bs h^{ba}) \nn 
    &= \f{1}{D-1} \bar \na_b \bs h^{ba},
\end{align}
where $\bar \na_a$ is the covariant derivative operator associated with $\bar g_{ab}$. This means that in $(x_\mu, \bar \psi_j)$ coordinates,
\begin{align}
    \beta &= \f{\pa}{\pa \bar \psi_0},
\end{align}
which is also simple to verify directly using $\beta = \pa/\pa \psi_0$ and the coordinate transformations between $(x_\mu,\psi_j)$ and $(x_\mu,\bar \psi_j)$ coordinates.

We must be careful when discussing the \emph{covariant} form of the $\beta$ vector. Let $\beta^\flat$ and $\overline{\beta^\flat}$ be, respectively,
\begin{align}
    \beta^\flat &= (g_{ab} \beta^b) dx^a \nn 
    \overline{\beta^\flat} &= \overline{(g_{ab} \beta^b)} dx^a \nn 
    &= \bar g_{ab} \beta^b dx^a.
\end{align}
The two differ by
\begin{align}
    \beta^\flat - \overline{\beta^\flat} &= (g_{ab} - \bar g_{ab}) \beta^b dx^a \nn 
    &= H k_a k_b \beta^b dx^a \nn 
    &= H k^\flat
\end{align}
(using $k \cdot \beta = 1$). For pure AdS, we have \eqref{dbetaflatAdS}, so that we then have
\begin{align}
    \overline{d\beta^\flat} &= \f{2}{l^2} \bs h. \label{dbetaflatAdShv2}
\end{align}
We know that $k \cdot \overline{d\beta^\flat} \propto k^\flat$ because $k \cdot d\beta^\flat \propto k^\flat$ and all the properties we discover for the full spacetime (as a member of the class of Generalized Kerr--NUT--AdS spacetimes) must also apply to the background spacetime (as another such member of the class). We can evaluate $k \cdot \overline{d\beta^\flat}$ directly using \eqref{dbetaflatAdShv2} and rearranging:
\begin{align}
    k^a \overline{\na_a \beta_b} &= \f{1}{l^2} k^a \bs h_{a b} \nn 
    &= -\f{r}{l^2} k_b,
\end{align}
which also implies (also using the fact that $\lie_\beta k = 0$)
\begin{align}
    k^a \bar \na_a \beta^b &= \beta^a \bar \na_a k^b = - \f{r}{l^2} k^b. \label{kdbeta}
\end{align}
We also verify this simple relationship directly using the coordinate expressions for $k$ and $\beta$ in Appendix \ref{kdbetaSection}. 

While we can refer to $\bs h$ associated with the background AdS spacetime as a PCKY tensor, it is worth noting that it is not the unique PCKY tensor of the AdS background, as there are a large number of distinct Closed Conformal Killing--Yano tensors for pure AdS. For our purposes all that is required to see this is to note that a different set of choices for the $a_i$ terms would lead to a different initial set of spheroidal coordinates and consequently a different set of Jacobi transformed coordinates, and thus a different PCKY tensor, which would not be equal to $\bs h$ even up to coordinate transformations. Another way to say this is that the AdS spacetime has such a high degree of symmetry that it has an infinite array of PCKY tensors. I will say more about this in Section \ref{pseudoCartesian}. Let it be understood that when I discuss $\bs h$ or $\beta$ associated with the background AdS spacetime, it is the $\bs h, \beta$ defined above: the ones that are equal to the $\bs h, \beta$ in the full spacetime which is related to the background spacetime by a Kerr--Schild relationship. 

\subsubsection{Generalized Kerr--NUT--AdS as Multiple Kerr--Schild Perturbation} \label{multiplekerrschild}

Chen and L\"u \cite{ChenLu08} demonstrate that the Kerr--NUT--AdS spacetime can be written in multiple Kerr--Schild form, just as Kerr--AdS can be written in Kerr--Schild form (see also \cite{FrolovReview}). The disadvantage of the multiple Kerr--Schild form is that it requires a Wick rotation of some of the coordinates, so that some of the coordinates are now imaginary. 

The argument is as follows. The result from Section \ref{canonicalbasisbackground} is that it is possible to take an AdS spacetime written in terms of canonical basis vectors $\bar e^A$ and use one of the eigenvectors of the associated PCKY tensor $\bs h$ to generate a GKAdS spacetime, with (in certain senses) the same $\bs h$. The eigenvector we chose, $k$, involved only the $\bar e^n, \bar e^{\hat n}$ basis one-forms. We can easily generalize the process to involve other eigenvectors of $\bs h$---creating a multiple Kerr--Schild form, as follows. Requiring $\bar X_\mu = Y(x_\mu)$ and letting $X_\mu(x_\mu)$ be, in general, arbitrary functions of $x_\mu$ (but not of any $x_\nu$ where $\nu \neq \mu$), 
\begin{align}
    d \bar s^2 &= \sum_{A=1}^D (\bar e^A)^2 \nn 
    d s^2 &= \sum_{A=1}^D (e^A)^2 \nn 
    Q_\mu &= \f{X_\mu}{U_\mu} \nn 
    \bar Q_\mu &= \f{\bar X_\mu}{U_\mu} \nn 
    e^\mu &= \f{dx_\mu}{\sqrt{Q_\mu}} \nn 
    \bar e^\mu &= \f{dx_\mu}{\sqrt{\bar Q_\mu}} \nn 
    e^{\hat \mu} &= \sqrt{Q_\mu} \sum_{j=0}^{n-1} A_\mu^{(j)} d \psi_j \nn 
    \bar e^{\hat \mu} &= \sqrt{\bar Q_\mu} \sum_{j=0}^{n-1} A_\mu^{(j)} d \bar \psi_j \nn
    e^{\hat 0} &= \sqrt{S} \sum_{j=0}^n A^{(j)} d \psi_j \nn 
    \bar e^{\hat 0} &= \sqrt{S} \sum_{j=0}^n A^{(j)} d \bar \psi_j \nn 
    \bs h&=\sum_{\mu=1}^n x_\mu e^\mu \wedge e^{\hat \mu} = \sum_{\mu=1}^n x_\mu \bar e^\mu \wedge \bar e^{\hat \mu},
\end{align}
where $\bs h$ is the PCKY tensor for both spacetimes. Under these definitions, we can decompose $ds^2$, the metric for the full spacetime, as
\begin{align}
    d s^2 &= d \bar s^2 + \sum_{\mu=1}^n \f{X_\mu - \bar X_\mu}{\bar X_\mu} \left( \bar e^{\hat \mu} + i \bar e^\mu\right)^2. 
\end{align}
The $\psi_j, \bar \psi_j$ coordinates are related by
\begin{align}
    d \psi_j &= d \bar \psi_j + i \sum_{\mu=1}^n \f{X_\mu-\bar X_\mu}{X_\mu \bar X_\mu} (-x_\mu^2)^{n-1-j} d x_\mu.
\end{align}
This tells us that we can write any member of the GKNAdS (note the N) class as a multiple Kerr--Schild perturbation to a pure AdS spacetime, with the significant caveat that, if the $\psi_j$ are real, in general the $\bar \psi_j$ will not be real. It is unclear whether there is any benefit to this decomposition. The reason that the GKAdS (no N) are special is because the fact that $x_n$, and only $x_n$, is imaginary in those spacetimes means that the vector $k = \tilde m_n$ associated with the $x_n$ eigenvalue is real-valued whereas the other $\tilde m_\mu$ are not. While my work focuses on the non-NUT case, I mention this here because (as I mention in the Future Work section) it may be possible to apply some of the results associated with the GKAdS spacetimes to the GKNAdS ones, using the multiple-Kerr--Schild form. 

Note again nomenclature: what I call the Generalized Kerr--NUT--AdS is referred to in the literature as the ``off-shell Kerr--NUT--(A)dS'' in, e.g.,~\cite{FrolovReview}.  

\subsection{Black Hole Horizon} \label{BHHorizonGKAdS}

The Kerr--AdS spacetimes are stationary black holes, with an event horizon located where $V = 2m$, as discussed in \cite{GibbonsLu,GibbonsLu2}. Their results generalize readily to the GKAdS spacetimes where $m \to \mu(r)$. In either form of the Boyer--Lindquist coordinates, $g_{rr}$ diverges when $V = 2\mu(r)$, which also corresponds to $X = 0$. This is the horizon location. In general there are multiple solutions to $V = 2\mu(r)$. Let the largest solution of $V = 2\mu(r)$ be $r = r_+$. If $r_+$ is real and positive, then we will consider this the location of the event horizon, $\mc H$, of the Generalized Kerr--AdS black hole. In four dimensions, this is the solution to
\begin{align}
    (r_+^2+a^2)(1 + r_+^2l^{-2}) - 2 \mu(r) r_+ = 0.
\end{align}
We will be focused on the outer horizon $(r_+)$ rather than inner horizons (labelled something like $r_-$). Because we are only concerned with one horizon I will just write $\mc H$ instead of $\mc H^+$ (the $+$ label on $r_+$ is to distinguish it from the coordinate $r$ in general).

Because the horizon is located at constant $r$, $g^{rr} = 0$ (on $\mc H$, or any other horizon). Using the Kerr--Schild coordinates and decomposition, $g^{rr} = \bar g^{rr} - H k^r k^r = \bar g^{rr} - H$ (since $k^r = 1$), so the horizon can also be found where $H = \bar g^{rr} = \bar g_{rr}^{-1}$ (which is consistent with $V = 2 \mu(r)$).

Because the Generalized Kerr--AdS black holes are stationary, the event horizon is also a Killing horizon, where a Killing vector field becomes null. I will call this Killing vector field, normalized so that its $t$ or $\tau$ component is 1, $\zeta$ (called $l$ in \cite{GibbonsLu,GibbonsLu2}). It is given by \cite{GibbonsLu,GibbonsLu2}
\begin{align}
    \zeta &= \xi + \sum_{i = 1}^{n-1+\ve} \Omega_i \eta_i \nn 
    &= \beta + \sum_{i = 1}^{n-1+\ve} \omega_i \eta_i, \label{zetabreakdown}
\end{align}
where the angular velocities $\Omega_i$ and $\omega_i$ are given by
\begin{align}
    \Omega_i &= \frac{a_i (1+r_+^2l^{-2})}{r_+^2+a_i^2} \nn 
    \omega_i &= \frac{a_i \Xi_i}{r_+^2+a_i^2}. \label{Omegaomega}
\end{align}
The $\Om_i$ are angular velocities as measured in the asymptotically-static frame, and the $\om_i$ are the angular velocities as measured in the asymptotically-rotating frame. The Kerr--Schild vector field $-k^a \partial_a$ is future-directed and inward-pointing. (Of course, $k^a \partial_a$ is past-directed and outward-pointing.) It crosses the horizon, and so the KS coordinates extend through the future horizon, whereas the BL and ABL coordinates are only valid outside the horizon. In four dimensions,
\begin{align}
    \z &= \xi + \Om \eta = \bt + \om \eta \nn 
    \Om &= \f{a(1+r_+^2l^{-2})}{r_+^2+a^2}\nn 
    \om &= \f{a \Xi}{r_+^2+a^2}. \label{zeta4D}
\end{align}

We note that, on the horizon, which I will denote by $|_{\mc H}$, $\zeta$ is in the direction of $\ell$, given by \eqref{ellifXis0}, verifiable in the KS coordinates which extend through the horizon,
\begin{align}
    \zeta|_{\mc H} &= \left.\left(\f{(1+r^2/l^2)U_n}{\bar X} \ell\right)\right|_{\mc H}. \label{zetaproptol}
\end{align}
The Killing vector tangent to the generators of the horizon thus points along one of the WANDs of the spacetime. 

Gibbons et al.~\cite{GibbonsLu,GibbonsLu2} also give the horizon area $ A$ as 
\begin{align}
     A &= r_+^{-\ve} \mc A_{D-2} \prod_{i = 1}^{n-1+\ve} \frac{r_+^2+a_i^2}{\Xi_i}, \label{AGibbonsLu}
\end{align}
where $\mc A_{D-2}$ is the $(D-2)$-volume of the unit $(D-2)$-sphere (see Section \ref{AreaRevisited} for more details). 

The surface gravity $\kappa$ of a Killing horizon is given by (e.g.~\cite{Poisson})
\begin{align}
    \z^a \na_a \z^b &= \ka \z^b, \label{horizonzetakappa}
\end{align}
where the equality holds only on the horizon. Gibbons et al.~give $\kappa$ as, \emph{specifically for Kerr--AdS},
\begin{align}
    \kappa &= \f12 (1 + r_+^2 l^{-2}) \frac{V'(r_+)}{V(r_+)}, \label{kappaeqVform}
\end{align}
treating $V$ as a function of $r$. This corresponds to
\begin{align}
    \kappa &= r_+ (1 + r_+^2l^{-2}) \sum_{i = 1}^{n-1+\ve} \frac{1}{r_+^2+a_i^2} - \f1{r_+}, \qquad \qquad D\textrm{ odd}\nn
    \kappa &= r_+ (1 + r_+^2 l^{-2}) \sum_{i = 1}^{n-1+\ve} \frac{1}{r_+^2+a_i^2} - \f{1 - r_+^2l^{-2}}{2 r_+}, \qquad D\textrm{ even}.
\end{align}

I repeat the calculations to check the effect of sending $m \to \mu(r)$. The area expression is unchanged; the calculation appears in Section \ref{AreaRevisited}. To calculate $\kappa$, use the expression from \cite{Poisson}, valid on the horizon,
\begin{align}
    -\pa_b(\zeta^a \zeta_a) = 2 \kappa \zeta_b.
\end{align}
To simplify the calculation, let $Z$ be the vector given by
\begin{align}
    Z &= \sum_{i=0}^{n-1+\ve} \f{a_i}{r^2+a_i^2} \f{\pa}{\pa \phi_i} \nn  
    &= \f{i \sqrt{ X U_n}}{\bar X} e_{\hat n}. \label{Zdefinition}
\end{align}
Its covariant components are found by writing $Z^\flat = \f{i \sqrt{X U_n}}{\bar X} e^{\hat n}$. Comparing to the expression for $e^{\hat n}$ from \eqref{canonicalbasisKScoords} and noting that $Q_n = X_n/U_n = -X/U_n$, we have that the covariant components $Z_t$ and $Z_{\phi_i}$ of $Z$ vanish on $\mc H$ (in $(t,r,y_\alpha,\phi_i)$ coordinates), leaving only a nonzero $Z_r$ component, equal to $2\mu(r) r^{1-\ve} U_n/\bar X^2 = U_n/\bar X = -\bar Q_n^{-1}$ (since $2 \mu(r) r^{1-\ve} = \bar X$ on $\mc H$). This also tells us that, on $\mc H$,
\begin{align}
    \zeta^\flat|_{\mc H} &= \f{(1+r_+^2/l^2) U_n}{\bar X} dr \nn 
    &=(1+r_+^2/l^2) \bar g_{rr} dr. \label{zetaflatH}
\end{align}

We have that, on $\mc H$, $\zeta = (1+r_+^2/l^2) Z$. In general we can write
\begin{align}
    \zeta&= \left(1 + \f{r_+^2}{l^2}\right) \left(Z + \delta\right),
\end{align}
where 
\begin{align}
    \delta &= \sum_{i=0}^{n-1+\ve} \f{a_i (r^2-r_+^2)}{(r^2+a_i^2)(r_+^2+a_i^2)} \f{\pa}{\pa \phi_i},
\end{align}
which is obviously zero on $\mc H$. We can then write, evaluating on $\mc H$,
\begin{align}
    \zeta^a \zeta_a &= (1+r_+^2/l^2)^2 (Z^a Z_a + 2 Z_a \de^a + \de^a \de_a) \nn
    \pa_b (\zeta^a \zeta_a) &= (1+r_+^2/l^2)^2 \pa_b (Z^a Z_a + 2 Z_a \de^a + \de^a \de_a) \nn 
    &= (1+r_+^2/l^2)^2 \left( \pa_b (Z^a Z_a) + 2 \de^a \pa_b Z_a + 2 Z_a \pa_b \de^a + \de^a \pa_b \de_a + \de_a \pa_b \de^a\right).
\end{align}
Since $\de^a = \de_a = 0$ on $\mc H$, and additionally since $Z_t = Z_{\phi_i} = 0$ on $\mc H$, $(2 \de^a \pa_b Z_a + 2 Z_a \pa_b \de^a + \de^a \pa_b \de_a + \de_a \pa_b \de^a)|_{\mc H} = 0$. We are thus left with
\begin{align}
    \pa_b(\zeta^a \zeta_a)|_{\mc H} &= (1+r_+^2/l^2)^2 \pa_b (Z^a Z_a)|_{\mc H}.
\end{align}
Using \eqref{Zdefinition}, $Z^aZ_a= - X U_n/\bar X^2$. We then have, letting a prime denote differentiation with respect to $r$,
\begin{align}
    \pa_b (\zeta^a \zeta_a)|_{\mc H} &= (1+r_+^2/l^2)^2 \pa_b ( - X U_n/\bar X^2)_{\mc H} \nn 
    &= - (1 +r_+^2/l^2)^2 \left.\left[ (\pa_b X) \f{U_n}{\bar X^2} + X \pa_b (U_n/\bar X^2)\right]\right|_{\mc H} \nn 
    &= -(1+r_+^2/l^2)^2 \left.\left( (\pa_b X) \f{U_n}{\bar X^2}\right)\right|_{\mc H}\nn 
    &= -\f{(1+r_+^2/l^2) X'}{\bar X} \zeta_b,
\end{align}
so that
\begin{align}
    \kappa &= \f12 (1+r_+^2/l^2) \f{X'(r_+)}{\bar X(r_+)}.
\end{align}
Using $X = (V-2\mu(r))r^{1-\ve}$ and $\bar X = V r^{1-\ve}$, this is equivalent to
\begin{align}
    \kappa &= \f12 (1+r_+^2/l^2) \f{r^{1-\ve}(V'(r_+) - 2 \mu'(r_+)) + (1-\ve) (V(r_+)-2\mu(r_+))}{r^{1-\ve} V(r_+)}.
\end{align}
Since $V(r_+) = 2\mu(r_+)$, this reduces to
\begin{align}
    \kappa &= \f12 (1+r_+^2/l^2) \f{V'(r_+) - 2 \mu'(r_+)}{V(r_+)}. \label{kappawithmu}
\end{align}

\subsection{Ring Singularity} \label{ringsingularity}

As pointed out by CGKP~\cite{Cvetic} (and possibly earlier), the ring singularity in the Kerr--AdS spacetime is located at $r = 0$ in even dimension, but $r^2 + a_n^2 = 0$ (where $a_n$ is the smallest of the $a_i$ parameters) in odd dimension. As a result, when discussing the region below the event horizon, in the KS coordinates which extend below said horizon, we will always be discussing the region $r > 0$ for $D$ even and $r^2 > -a_n^2$ for $D$ odd. 

This means that $r$ actually takes on imaginary values if $D$ is odd; we really require $r^2 > - a_n^2$ rather than the more usual $r^2 > 0$. In that sense, $r^2$ might be a better coordinate than $r$ to use, at least for small values of $r^2$, but for our purposes it is sufficient to keep this in mind.

Note that this is consistent with the spheroidal transformation \eqref{spheroidaltransformation}. In even dimension, $a_n^2 = 0$, and so the equation $y^2 \hat \mu_n^2 = r^2 \mu_n^2$ implies that $r = 0$ corresponds to $y = 0$ (except possibly if $\hat \mu_n = 0$). In odd dimensions, $r= 0$ simply gives
\begin{align}
    y^2 \hat \mu_i^2 &= \f{a_i^2 \mu_i^2}{\Xi_i},
\end{align}
which does not in general imply $y = 0$. It requires $r^2+a_i^2 = 0$ to give $y = 0$, and so $r^2$ extends down to $-a_n^2$ in the spheroidal coordinate transformation. 

\section{Pseudo-Cartesian Coordinates for Generalized Kerr--anti-de Sitter} \label{pseudoCartesian}

In this section I introduce a set of pseudo-Cartesian coordinates for Generalized Kerr--anti-de Sitter in its Kerr--Schild form, which I believe to be novel (including for Kerr--anti-de Sitter). These are based on a well-known set of coordinates describing anti-de Sitter. 

It is known that it is possible to embed $D$-dimensional anti-de Sitter into a flat space of dimension one higher with signature $(D-3)$, or, specifically, with signature $(--+ \ldots +)$ (see, for instance, \cite{Stephani}; see also \cite{HenneauxTeitelboim85} which considers the $O(3,2)$ charges associated with asymptotically anti-de Sitter spacetimes). Let AdS space be expressed \emph{locally} in terms of coordinates $(u,v,x_i,y_i, z)$ for even dimensions and $(u,v,x_i,y_i)$ for odd dimensions. Here $1 \leq i \leq n-1+\ve$. I will call these coordinates pseudo-Cartesian. The metric is
\begin{align}
    d\bar s^2 &= -du^2 - dv^2 + \sum_{i = 1}^{n-1+\ve} (dx_i^2+dy_i^2) + (1-\ve) d z^2, \label{UVXYZ}
\end{align}
subject to the constraint 
\begin{align}
    - u^2 - v^2 + \sum_{i = 1}^{n-1+\ve} (x_i^2 + y_i^2) + (1-\ve) z^2 &= -l^2. \label{UVXYZConstraint}
\end{align}
This is related to the standard spherical polar coordinates \eqref{dbarsspherical} by
\begin{align}
    u &= \sqrt{y^2+l^2} \cos (t/l) \nn 
    v&= \sqrt{y^2+l^2} \sin (t/l) \nn 
    x_i &= y \hat \mu_i \cos \phi_i \nn 
    y_i &= y \hat \mu_i \sin \phi_i \nn 
    z&= y \hat \mu_{n}. \label{UVXYZtransformations}
\end{align}
It is understood (here and for the rest of Section \ref{pseudoCartesian}) that the $z$ coordinate transformation is only meaningful if $z$ exists (if $D$ is even). (Note that the $y_i$ coordinates represent coordinates associated with the 2-plane with associated coordinates $\mu_i, \hat \mu_i, \phi_i$ etc.~and that $y$ represents the spherical polar radius. $y_i$ is also not the same as the $y_\alpha$ associated with the Jacobi transformations. Similarly, the $x_i$ are not to be confused with the $x_\mu$.) 

To get to an AdS which does not have closed timelike curves, AdS space must be ``unwound'' from the embedding I have just described, in the sense that the $(D+1)$-dimensional flat space with coordinates $(u,v,x_i,y_i,z)$ has $t/l$ acting as a periodic coordinate, where the points $t = t_0$ and $t = t_0+2\pi l$ are identified in that they lead to the same values of $u$ and $v$. I consider AdS to be the universal covering space, where $t$ varies from $-\infty$ to $\infty$ (without the periodic identification above). The unwinding issue is a disadvantage of this embedding into $(D+1)$-dimensional flat coordinates, which are not valid for describing the global spacetime. 

The way the pseudo-Cartesian coordinates relate to the spheroidal coordinates with $(r,\mu_i,t,\phi_i)$ is (just applying the coordinate transformation from the $(y,\hat \mu_i)$ coordinates to $(r,\mu_i)$)
\begin{align}
    u &= \left( (r^2+l^2) \sum_{i = 1}^{n} \frac{\mu_i^2}{\Xi_i} \right)^{1/2} \cos(t/l) \nn
    v &= \left( (r^2+l^2) \sum_{ i = 1}^{n} \frac{\mu_i^2}{\Xi_i}\right)^{1/2} \sin(t/l) \nn 
    x_i &= \sqrt{ \frac{r^2+a_i^2}{\Xi_i}} \mu_i \cos \phi_i \nn 
    y_i &= \sqrt{ \frac{r^2+a_i^2}{\Xi_i}} \mu_i \sin \phi_i \nn
    z &= r \mu_{n}, \label{uvxyzspheroidal}
\end{align}
where it is understood that the $z$ definition is only meaningful if $D$ is even.

We can interpret $r$ as a function of the pseudo-Cartesian coordinates and express it implicitly as
\begin{align}
    -\frac{u^2+v^2}{r^2+l^2} + \sum_{i = 1}^{n-1+\ve} \frac{x_i^2+y_i^2}{r^2+a_i^2} + (1-\ve) \frac{z^2}{r^2} &=0, \label{requationuvxyz}
\end{align}
which is verified by noting that 
\begin{align}
    \frac{u^2+v^2}{r^2+l^2} &= \sum_{i = 1}^{n} \frac{\mu_i^2}{\Xi_i} \nn 
    \frac{x_i^2 + y_i^2}{r^2+a_i^2} &= \frac{\mu_i^2}{\Xi_i} \nn 
    \frac{z^2}{r^2} &= \mu_{n}^2.
\end{align}
Using the $(D+1)$ coordinates $(u,v,x_i,y_i,z)$, the surfaces of constant $r$ are hyperboloids. 

If we also apply the constraint \eqref{UVXYZConstraint}, we can eliminate the $u^2+v^2$ term and so find, after some rearranging,
\begin{align}
    1 &= \sum_{i = 1}^{n-1+\ve} \frac{x_i^2+y_i^2}{\Xi_i^{-1} (r^2+a_i^2)} + (1-\ve) \frac{z^2}{r^2}. \label{rspheruvxyzequation}
\end{align}
If we treat the coordinates $(x_i,y_i, z)$ as a set of $(D-1)$ Cartesian coordinates, then surfaces of constant $r$ are ellipsoids in these coordinates, with two semi-major axes each of size $\sqrt{(r^2+a_i^2)/\Xi_i}$ for $i$ from 1 to $n-1+\ve$, and, if $D$ is even, one additional semi-major axis of size $r$. Once again recall that in even dimensions, $r$ extends down to $r = 0$ (which represents a completely flat ellipsoid, with $z = 0$) but in odd dimensions $r^2$ extends down to $r^2 + a_n^2 = 0$; $r = 0$ is just any other higher-dimensional ellipsoid with semi-major axes $a_i/\Xi_i$, but $r^2 + a_n^2$ gives the (zero-volume) ellipsoid where two of the semi-major axes are 0. \eqref{rspheruvxyzequation} will aid in volume calculation; see Section \ref{VolumeSimplified}.

It is not only pure anti-de Sitter that can be expressed in these pseudo-Cartesian coordinates, but also Generalized Kerr--anti-de Sitter. Beginning with the KS form of the GKAdS spacetimes, starting with coordinates $(t,r,\mu_i, \phi_i)$, the coordinate transformations \eqref{uvxyzspheroidal} imply that the background AdS spacetime has metric \eqref{UVXYZ} subject to the constraint \eqref{UVXYZConstraint}, and that the Kerr--Schild null vector $k^a \partial_a$ and corresponding one-form $k_a dx^a$ take the form
\begin{align}
    k^a \partial_a &= +\frac{(ru+lv) \partial_u + (rv-lu) \partial_v}{r^2+l^2} + \sum_{i = 1}^{n-1+\ve}\frac{(rx_i + a_i y_i) \partial_{x_i}+ (ry_i-a_ix_i) \partial_{y_i}}{r^2+a_i^2} + (1-\ve) \frac{z \partial_z}{r} \nn
    k_a dx^a &= -\frac{(ru+lv) du + (rv-lu) dv}{r^2+l^2} + \sum_{i = 1}^{n-1+\ve}\frac{(rx_i + a_i y_i) d{x_i}+ (ry_i-a_ix_i) d{y_i}}{r^2+a_i^2} + (1-\ve) \frac{z dz}{r}. \label{kmudxmuuvxyz}
\end{align}
This is simple and very similar to the form for the Cartesian coordinates for the Kerr--Schild background and metric for the $(\Lambda = 0)$ Kerr solutions. I have not seen this expressed this way in the literature, though it is similar to a form given for the Myers--Perry black holes, as I will show below.

To show that this is the form that $k^a \pa_a$ takes, start with \eqref{kmu} and perform the coordinate transformations \eqref{uvxyzspheroidal}. The $t$ and angular terms are straightforward, since $\partial_{\phi_i} = x_i \partial_{y_i} - y_i \partial_{x_i}$ and $\partial_t = l^{-1} (u \partial_v - v \partial_u)$. For the $\partial_r$ term,
\begin{align}
    \frac{\partial}{\partial r} &= \frac{\partial u}{\partial r} \frac{\partial}{\partial u} + \frac{\partial v}{\partial r} \frac{\partial}{\partial v} + \sum_{i = 1}^{n-1+\ve} \left( \frac{\pa x_i}{\pa r} \f{\pa}{\pa x_i} + \f{\pa y_i}{\pa r} \f{\pa}{\pa y_i}\right) + (1-\ve) \f{\pa z}{\pa r} \f{\pa}{\pa z}.
\end{align}
All the $\pa x^a/\pa r$ derivatives are taken at constant $\mu_i, t, \phi_i$. 

Taking $u$ as a representative example, since $u \propto \sqrt{r^2+l^2}$ and $\pa \sqrt{r^2+l^2}/\pa r = r/\sqrt{r^2+l^2}$, we have
\begin{align}
    \f{\pa u}{\pa r} = \f{r u}{r^2+l^2}.
\end{align}
Similarly,
\begin{align}
    \f{\pa v}{\pa r} &= \f{r v}{r^2+l^2} \nn 
    \f{\pa x_i}{\pa r} &= \f{r x_i}{r^2+a_i^2} \nn 
    \f{\pa y_i}{\pa r} &= \f{r y_i}{r^2+a_i^2} \nn 
    \f{\pa z}{\pa r} &= \f{z}{r}.
\end{align}

The result for $\pa_r$ is 
\begin{align}
    \f{\pa}{\pa r} &= r \left(\f{u \pa_u + v \pa_v}{r^2+l^2} + \sum_{i = 1}^{n-1+\ve} \f{x_i \pa_{x_i} + y_i \pa_{y_i}}{r^2+a_i^2} + (1-\ve)\f{z \pa_z}{r^2}\right).
\end{align}
The expression for $k^a$ follows.

Interestingly, $k_a k^a$ (without applying the constraint) is then simply equal to the left-hand side of \eqref{requationuvxyz}. This tells us that the statement that $k_a k^a$ is null in the pseudo-Cartesian coordinates is essentially the same condition as the implicit definition of $r$. 

The immediate interpretation here is that $k^a \pa_a$ is the Kerr--Schild null vector in the $D$-dimensional anti-de Sitter space, with $(D+1)$ coordinates and also the constraint \eqref{UVXYZConstraint}. It is also worth checking what happens if we consider $k^a \pa_a$, with the same coordinates, in the $(D+1)$-dimensional flat space with no constraint. Interpreting $k^a\pa_a$ as being a vector in the flat $(D+1)$-dimensional spacetime, it is tangent to the surface $l^2 = u^2+v^2-\sum_{i=1}^{n-1+\ve} (x_i^2+y_i^2) - (1-\ve) z^2$---that is, it is tangent, in the full $(D+1)$-dimensional spacetime, to the hyperboloid which satisfies the constraint for the $D$-dimensional space. $k_a k^a$ is equal to the left-hand side of \eqref{requationuvxyz} so that $k^a$ is null in the $(D+1)$-dimensional spacetime. 

Returning to the $D$-dimensional spacetime, the existence of one extra coordinate along with a constraint does not cause any more problems than it did when there was one extra $\mu_i$ and associated constraint $\sum_{i = 1}^{n} \mu_i^2 = 1$. The coordinates are helpful in showing one way in which the Kerr--anti-de Sitter solutions connect to the Kerr/Myers--Perry solutions. 

Applying the constraint $u^2 + v^2 = l^2 + \sum_{i = 1}^{n-1+\ve} (x_i^2+y_i^2) + (1-\ve) z^2$ as well as $u = \sqrt{u^2+v^2} \cos (t/l), v = \sqrt{u^2+v^2} \sin (t/l)$ gives a way of expressing the AdS background in terms of coordinates $(t,x_i,y_i,z)$. Let (for this section only) $Z_\alpha = (x_i,y_i,z)$---that is to say, all coordinates except for $t$; it does not matter particularly how they are ordered, because they will only be used to state the metric compactly, with $\alpha = 1, \ldots, D-1$. The AdS metric in these coordinates can be written as  \cite{Stephani}
\begin{align}
    d\bar s^2 = \bar g_{a b} dx^a dx^b &= -\left(1 + \sum_{\alpha=1}^{D-1} \frac{Z_\alpha^2}{l^2}\right)dt^2 + \sum_{\alpha = 1}^{D-1} d Z_\alpha^2 - \sum_{\alpha=  1}^{D-1} \sum_{\beta = 1}^{D-1} \frac{Z_\alpha Z_\beta dZ_\alpha d Z_\beta}{l^2 + \sum_{\alpha=1}^{D-1} Z_\alpha^2}. \label{gabpseudoCartesianwitht}
\end{align}
Using the spherical radius $y$ satisfying $y^2 = \sum_{\alpha=1}^{D-1} Z_\alpha^2$, we then have $\bar g_{tt} = -(1+y^2/l^2), \bar g_{Z_\alpha Z_\beta} = \delta_{\alpha \beta} - Z_\alpha Z_\beta / (l^2 +y^2)$. This allows the metric determinant to be calculated. The metric determinant will be $\bar g_{tt}$ times the determinant of the submatrix $\bar g_{Z_\alpha Z_\beta}$. Letting $I_{D-1}$ represent the $(D-1)\times (D-1)$ identity matrix and let $u$ be the column vector with 
\begin{align}
    u &= \begin{bmatrix} \f{Z_1}{\sqrt{l^2 + y^2}} \\ \vdots \\ \f{Z_{D-1}}{\sqrt{l^2+y^2}}\end{bmatrix}.
\end{align}
Then we can represent $\mathrm{det}(g_{Z_\alpha Z_\beta})$ by the matrix $I_{D-1} - u u^T$ where $I_{D-1}$ is the $(D-1)\times (D-1)$ identity matrix. From the Matrix Determinant Lemma \cite{Harville}, the determinant of $I_{D-1} - u u^T$ is $1 - u^T u$. Here, $u^T u = \sum_{\alpha=1}^{D-1} Z_\alpha^2 /(l^2+y^2) = y^2/(l^2+y^2)$. We then have
\begin{align}
    \mathrm{det}(g_{Z_\alpha Z_\beta}) &= \f{l^2}{l^2+y^2} = \f{1}{1+y^2/l^2}.
\end{align}
The combination then implies that, in coordinates $(t,x_i,y_i,z)$,
\begin{align}
    \mathrm{det} \bar g_{ab} &= -1. \label{determinantpseudoCartesianwitht}
\end{align}
(The fact that the metric determinant is the same for Minkowski space is a consequence of the fact that AdS can be expressed as a Kerr--Schild correction to Minkowski space, as discussed in Section \ref{KNAdSPaper}.)

In $(t,x_i,y_i,z)$ coordinates, the one-form version of $k$ takes the form
\begin{align}
    k_a dx^a &= \f{l^2 + \sum_{i = 1}^{n-1+\ve}(x_i^2+y_i^2) + (1-\ve) z^2}{l^2+r^2} dt +  (1-\ve) \f{ z dz}{r(1+r^2/l^2)} \nn
    &\qquad + \sum_{i = 1}^{n-1+\ve} \f{1}{r^2+a_i^2} \left( \f{r \Xi_i}{1+r^2/l^2} (x_i dx_i + y_i dy_i) + a_i (y_i dx_i - x_i dy_i)\right). \label{ktxyz}
\end{align}

The function $U$ is, in terms of $(x_i,y_i,z)$,
\begin{align}
    U &=  \left[\sum_{i = 1}^{n-1+\ve} \frac{\Xi_i (x_i^2+y_i^2)}{(r^2+a_i^2)^2} + (1-\ve) \frac{z^2}{r^4}\right] r^{1-\ve} \prod_{j = 1}^{n-1+\ve} (r^2+a_j^2),
\end{align}
by plugging in $\mu_i^2 = \Xi_i(x_i^2+y_i^2)/(r^2+a_i^2)$. With a little rearranging, this is expressible in the more ``democratic'' form (allowing $u$ and $v$ to appear on a similar footing to the other coordinates),
\begin{align}
    U &= \frac{1}{l^2}\left[ - \frac{u^2+v^2}{(r^2+l^2)^2} + \sum_{i = 1}^{n-1+\ve} \frac{x_i^2+y_i^2}{(r^2+a_i^2)^2} + (1-\ve) \frac{z^2}{r^4}\right] (r^2+l^2) r^{1-\ve} \prod_{i = 1}^{n-1+\ve} (r^2+a_i^2). \label{Udemocratic}
\end{align}
The intermediate steps are as follows. As usual, let $a_{n} = 0, \Xi_{n} = 1$ if $\ve = 0$.
\begin{align}
    \frac{u^2+v^2}{(r^2+l^2)^2} &= \frac{1}{r^2+l^2} \sum_{i = 1}^{n} \frac{\mu_i^2}{\Xi_i} \nn 
    \frac{x_i^2+y_i^2}{(r^2+a_i^2)^2} &= \frac{ \mu_i^2}{\Xi_i (r^2+a_i^2)} \nn 
    -\frac{u^2+v^2}{(r^2+l^2)^2} + \sum_{i = 1}^{n-1+\ve} \frac{x_i^2+y_i^2}{(r^2+a_i^2)^2} + (1-\ve) \frac{z^2}{r^4} &= -\f{1}{r^2+l^2} \sum_{i = 1}^{n} \frac{\mu_i^2}{\Xi_i} + \sum_{i = 1}^{n} \frac{\mu_i^2}{\Xi_i(r^2+a_i^2)} \nn 
    &= \sum_{i = 1}^{n} \frac{\mu_i^2}{\Xi_i} \left( - \frac{1}{r^2+l^2} + \frac{1}{r^2+a_i^2}\right) \nn 
    &= \sum_{i = 1}^{n} \frac{ \mu_i^2}{\Xi_i} \frac{l^2-a_i^2}{(r^2+l^2)(r^2+a_i^2)} \nn 
    &= \frac{l^2}{r^2+l^2} \sum_{i = 1}^{n} \frac{\mu_i^2}{r^2+a_i^2}
\end{align}
and the equality of the two expressions for $U$ follows.

The metric for Kerr--AdS in these pseudo-Cartesian coordinates follows from $ds^2 = d\bar s^2 + (2m/U) (k_a dx^a)^2$, again subject to the constraint \eqref{UVXYZConstraint}. Generalized Kerr--AdS is the same except $m \to \mu(r)$.

In the pseudo-Cartesian coordinates, the PCKY tensor associated with the GKAdS spacetime takes the simple form
\begin{align}
    \bs h &= - l du \wedge dv + \sum_{i = 1}^{n-1+\ve} a_i dx_i \wedge d y_i. \label{hinpseudoartesiancoords}
\end{align}
An associated potential one-form $\bs b'$ for which $\bs h = d \bs b'$ is
\begin{align}
    2 \bs b' &= - l (u dv - v du) + \sum_{i = 1}^{n-1+\ve} a_i (x_i dy_i - y_i dx_i).
\end{align}
$\bs b' = \bs b + d f$ for a scalar $f$, and thus is the same as $\bs b$ up to a gauge transformation. 
The associated vector $\beta$ takes the form
\begin{align}
    \beta = \frac{1}{l}\left( u \f{\pa}{\pa v} - v \f{\pa}{\pa u}\right) + \sum_{i = 1}^{n-1+\ve} \frac{a_i}{l^2} \left( x_i \f{\pa}{\pa y_i} - y_i \f{\pa}{\pa x_i}\right).
\end{align}
These forms are obtained in Appendix \ref{alternatebh} (which use some results from Section \ref{JacobiTransformationsSection}). I have not seen these forms, which are quite simple, in the literature, though some of the forms in \cite{KubiznakThesis} for the higher-dimensional Myers--Perry black holes are similar, as discussed below. 

One reason for mentioning this is to share these novel, appealing forms of the metric, Kerr--Schild vector, and PCKY form, and to impart the high degree of symmetry implicit in the GKAdS spacetime. More importantly for this work, we will use the expressions for the $x_i,y_i, z$ to simplify the calculations of the volumes associated with the spacetime.

We note the similarity of these results to the $D$-dimensional ($D = 2n+\ve$) Myers--Perry black holes, representing rotating black holes in a Minkowski background. They are given in a convenient form by \cite{KubiznakThesis} in Cartesian coordinates ($t,x_i,y_i,z)$, where $z$ is defined only for even dimensions. The metric $g_{a b}$ for the black holes takes the Kerr--Schild form
\begin{align}
    g_{a b} &= \eta_{a b} + \f{2m}{U} k_a k_b \nn 
    \eta_{a b} dx^a dx^b &= - dt^2 + \sum_{i = 1}^{n-1+\ve} (dx_i^2 + dy_i^2) + (1-\ve) dz^2 \nn 
    k_a d x^a &= dt + \sum_{i = 1}^{n-\ve+1} \frac{(r x_i - a_i y_i) dx_i + (r y_i + a_i x_i) dy_i}{r^2+a_i^2} + (1-\ve) \f{z dz}{r^2} \nn 
    U &= \left[r^{-1-\ve} \prod_{i = 1}^{n-1+\ve} (r^2+a_i^2) \right] \left(1 - \sum_{i = 1}^{n-1+\ve} \frac{a_i^2 \mu_i^2}{r^2+a_i^2}\right) \nn 
    \mu_i^2 &= \f{x_i^2+y_i^2}{r^2+a_i^2}. \label{MyersPerry}
\end{align}
Note the close similarity of the form of $k_a dx^a$ for the Myers--Perry black holes to \eqref{kmudxmuuvxyz} (in terms of general structure, ignoring the $dt$ component) and \eqref{ktxyz} (which approaches the \eqref{MyersPerry} form for $l \to \infty$). We can also see that $U$ takes a form similar to the expressions we have for Kerr--AdS in either $(u,v,x_i,y_i,z)$ or $(t,x_i,y_i,z)$ coordinates if we use $1 = \sum_{i = 1}^{n} \mu_i^2$ and the expression for $\mu_i^2$ in terms of $x_i,y_i$ to rewrite the Myers--Perry $U$ as 
\begin{align}
    U &= r^{1-\ve} \prod_{i = 1}^{n-1+\ve} (r^2+a_i^2)\left( \sum_{i = 1}^{n-1+\ve} \frac{\mu_i^2}{r^2+a_i^2}+ (1-\ve) \f{\mu_{n}^2}{r^2}\right) \nn
    &= r^{1-\ve} \prod_{i = 1}^{n-1+\ve} (r^2+a_i^2)\left( \sum_{i = 1}^{n-1+\ve} \frac{x_i^2+y_i^2}{(r^2+a_i^2)^2} + (1-\ve)\f{z^2}{r^4}\right).
\end{align}

\cite{KubiznakThesis} also gives a form of the PCKY tensor for the Myers--Perry spacetimes in the form
\begin{align}
    \bs h = \sum_{i=1}^{n-1+\ve} \left[ (x_i dx_i + y_i dy_i) \wedge d t + a_i d x_i \wedge d y_i\right] + (1-\ve) z d z \wedge d t,
\end{align}
which is similar to \eqref{hinpseudoartesiancoords}.

\subsection{Jacobi Transformations} \label{JacobiTransformationsSection}

The Jacobi transformations used to obtain the higher-dimensional Kerr--AdS metric, and the relationship between the $ir$ and $y_\alpha$ coordinates, become clearer when using the pseudo-Cartesian coordinates. The development here is similar to the development in Section 4.3.1~in \cite{FrolovReview}. 

Explicitly, let $\nu_i$, where $i = 0, \ldots, n$, be a set of variables satisfying the constraint 
\begin{align}\sum_{i = 0}^n \nu_i^2 = 1. \label{nusum1}
\end{align}
Let $\nu_0$ be a real number and let the other $\nu_i$ be imaginary. Then define
\begin{align}
    u &= l \nu_0 \cos \phi_0 \nn 
    v &= l \nu_0 \sin \phi_0 \nn 
    x_i &= -i l \nu_i \cos \phi_i \nn 
    y_i &= -i l \nu_i \sin \phi_i \nn 
    z &= -i l \nu_n, \label{pseudoCartesiansnu}
\end{align}
letting $\phi_0 \equiv t/l$ and where it is understood that the $z$ definition only applies when $D$ is even. The statement $d\bar s^2 = -du^2 - dv^2 + \sum_{i=1}^{n-1+\ve} (dx_i^2+dy_i^2) + (1-\varepsilon) dz^2$ is now equivalent to the statement 
\begin{align}
    d\bar s^2 &= -l^2 \left[ \sum_{i =0}^{n} d \nu_i^2 + \sum_{i = 0}^{n-1+\ve} \nu_i^2 d \phi_i^2\right],
\end{align}
subject of course to the constraint $\sum_{i=0}^{n} \nu_i^2 = 1$. 

Then the transformations \eqref{uvxyzspheroidal} are equivalent to, for $\nu_i$, $i = 1, \ldots, n$,
\begin{align}
    -l^2 \nu_i^2 &= \frac{(r^2+a_i^2)\mu_i^2}{\Xi_i} \nn 
    \nu_i^2 &= \frac{(r^2+a_i^2) \mu_i^2}{a_i^2 - l^2}.
\end{align}
When the Jacobi transformations \eqref{Jacobi} are applied to $\mu_i$, this becomes
\begin{align}
    \nu_i^2 &= \frac{(a_i^2+r^2) \prod_{\alpha = 1}^{n-1} (a_i^2-y_\alpha^2)}{(a_i^2-l^2) {\prod}_{j=1, j \neq i}^{n} (a_i^2-a_j^2)}. \label{nui2}
\end{align}
Just as CLP extended the $y_\alpha$  to $x_\mu$ by setting $x_n = i r$, let $a_0 \equiv l$. Then $\nu_i^2$ can be written
\begin{align}
    \nu_i^2 &= \frac{\prod_{\mu = 1}^n (a_i^2 - x_\mu^2)}{{\prod}_{j=0, j \neq i}^n (a_i^2 - a_j^2)}, \qquad 1 \leq i \leq n. \label{nui21ton}
\end{align}
This is just \eqref{Jacobi} again, with $(\nu_i,x_\mu)$ instead of $(\mu_i,y_\alpha)$ and with the larger set of $a_i$. This can be broken down further into
\begin{align}
    \nu_i^2 &= \f{(-1)^{n-1+\ve}\G_i}{\hat \Ups_i a_i^{2(1-\ve)}}, \qquad 1 \leq i \leq n-1+\ve \nn 
    \nu_n^2 &= \f{\prod_{\mu=1}^n x_\mu^2}{\prod_{j=0}^{n-1} a_j^2}, \qquad \textrm{if }\ve = 0.
\end{align}

Along similar lines, for $\nu_0$ we have
\begin{align}
    l^2\nu_0^2 &= (r^2+l^2) \sum_{i = 1}^{n} \frac{\mu_i^2}{\Xi_i} \nn 
    \nu_0^2 &= \frac{l^2+r^2}{l^2} \sum_{i = 1}^{n} \frac{\mu_i^2}{\Xi_i} \nn
    &= -(r^2+l^2)\sum_{i=1}^n \frac{\nu_i^2}{r^2+a_i^2}. \label{nu02v1}
\end{align}
Note that this equality is a restatement of \eqref{requationuvxyz}. We can also write $\nu_0^2$ as follows. We can posit temporarily that $\nu_0^2$ will have the same form as \eqref{nui21ton} with $i = 0$ and then check if we recover $\sum_{i=0}^n \nu_i^2 = 1$. To do so, note
\begin{align}
    \sum_{i=0}^n \f{(-a_i^2)^{n-l}}{\prod_{j=0,j\neq i}^n (a_j^2-a_i^2)} &= \de^l_0 \qquad\textrm{if } 0 \leq l \leq n,
\end{align}
analogously to $\sum_{\mu=1}^n (-x_\mu^2)^{n-1-j}/U_\mu = \de^0_j$.  We then have, assuming that we can extend \eqref{nui21ton} down to $i = 0$,
\begin{align}
    \sum_{i=0}^n \nu_i^2 &= \sum_{i=0}^n \f{\prod_{\mu=1}^n (a_i^2 - x_\mu^2)}{\prod_{j=0,j\neq i}^n (a_i^2-a_j^2)} \nn 
    &= \sum_{i=0}^n \f{\prod_{\mu=1}^n (x_\mu^2 - a_i^2)}{\prod_{j=0,j\neq i}^n (a_j^2-a_i^2)} \nn 
    &= \sum_{i=0}^n \f{\sum_{l=0}^n A^{(l)} (-a_i^2)^{n-l}}{\prod_{j=0,j\neq i}^n (a_i^2-a_j^2)} \nn 
    &= \sum_{l=0}^n A^{(l)} \sum_{i=0}^n \f{(-a_i^2)^{n-l}}{\prod_{j=0,j\neq i}^n (a_i^2-a_j^2)} \nn 
    &= \sum_{l=0}^n A^{(l)} \de^l_0 \nn 
    &= A^{(0)} \nn 
    &= 1,
\end{align}
as required. Thus we can state
\begin{align}
    \nu_i^2 &= \frac{\prod_{\mu = 1}^n (a_i^2 - x_\mu^2)}{{\prod}_{j=0, j \neq i}^n (a_i^2 - a_j^2)}, \qquad 0 \leq i \leq n.
\end{align}

It appears that the reason that the Jacobi transformations, expected to transform a sphere into a diagonal form, work when applied to the pure AdS metric in \emph{spheroidal} form  \eqref{dbarsspheroidal}. This section gives the interpretation that the Jacobi transformations work by ``completing'' the Jacobi transformations which were begun by introducing $r$.

We can now explicitly write
\begin{align}
    u &= l \left( \f{(-1)^{n-1+\ve} \G_0}{\hat \Ups_0 l^{2(1-\ve)}}\right)^{1/2} \cos(t/l) \nn 
    &= l \left( \f{(1+r^2/l^2) \prod_{\alpha = 1}^{n-1} (1-y_\alpha^2/l^2)}{\prod_{i = 1}^n \Xi_i}\right)^{1/2} \cos(t/l) \nn
    v &= l \left( \f{(-1)^{n-1+\ve} \G_0}{\hat \Ups_0 l^{2(1-\ve)}}\right)^{1/2} \sin(t/l)\nn 
    &= l \left( \f{(1+r^2/l^2)\prod_{\alpha=1}^{n-1}(1-y_\alpha^2/l^2)}{\prod_{i=1}^n \Xi_i}\right)^{1/2} \sin(t/l) \nn 
    x_i &= l \left( \f{(-1)^{n+\ve} \G_i}{\hat \Ups_i a_i^{2(1-\ve)}}\right)^{1/2} \cos \phi_i \nn 
    &= \left( \f{ (r^2+a_i^2)\prod_{\alpha = 1}^{n-1} (a_i^2 - y_\alpha^2)}{\Xi_i {\prod}_{j=1, j \neq i}^{n} (a_i^2-a_j^2)}\right)^{1/2} \cos \phi_i \nn 
    y_i &= l \left( \f{(-1)^{n+\ve} \G_i}{\hat \Ups_i a_i^{2(1-\ve)}}\right)^{1/2} \sin \phi_i \nn 
    &= \left( \f{(r^2+a_i^2)\prod_{\alpha=1}^{n-1}(a_i^2-y_\alpha^2)}{\Xi_i \prod_{j=1,j\neq i}^n (a_i^2-a_j^2)}\right)^{1/2} \sin \phi_i \nn 
    z &= l \f{-i \prod_{\mu = 1}^n x_\mu}{\prod_{i = 0}^{n-1} a_i} \nn 
    &= \f{r \prod_{\alpha=1}^{n-1} y_\alpha}{\prod_{i = 1}^{n-1} a_i}, \label{uvxyzjacobi}
\end{align}
with $z$ of course defined only if it exists. $l$ effectively plays a dual role as both the overall scale factor for the coordinates (so that the constraint \eqref{UVXYZConstraint} is satisfied) and as the value of the constant parameter associated with $(u,v)$ analogous to the way $a_i$ is associated with $(x_i,y_i)$. The variables are $(r, y_\alpha, t, \phi_i)$. Under \eqref{uvxyzjacobi}, not only is the constraint \eqref{UVXYZConstraint} is automatically satisfied but also the flat metric \eqref{UVXYZ} for $d\bar s^2$ becomes that in \eqref{adsrytphi}.

Interestingly, because of the highly symmetric way the $x_\mu^2$ come into the expressions, we can conclude that the equations \eqref{requationuvxyz} and \eqref{rspheruvxyzequation} must also be satisfied if $r^2$ is replaced by $-x_\mu^2$, where $\mu$ can take on any value from 1 to $n$,
\begin{align}
    -\f{u^2+v^2}{l^2-x_\mu^2} + \sum_{i=1}^{n-1+\ve} \f{x_i^2+y_i^2}{a_i^2-x_\mu^2} - (1-\ve) \f{z^2}{x_\mu^2} &= 0 \nn 
    \sum_{i=1}^{n-1+\ve} \f{x_i^2+y_i^2}{\Xi_i^{-1}(a_i^2-x_\mu^2)} - (1-\ve) \f{z^2}{x_\mu^2} &= 1.
\end{align}
The latter form can be rewritten (after multiplying by $x_\mu^2\prod_{i=1}^{n-1+\ve}(a_i^2-x_\mu^2)$) as a polynomial of degree $n$ in $x_\mu^2$, which naturally has $n$ solutions in general, and thus can be used to define the values of the $x_\mu^2$ and thus $x_\mu$ (up to a sign choice when taking the square root).

\subsection{Embedding in Higher Dimensions} \label{embedding}

Interestingly, there is a natural embedding of these GKAdS solutions in higher dimensions, which is achieved by taking the for GKAdS as described earlier in this section, using the Kerr--Schild form $ds^2 = d\bar s^2 + \f{2\mu(r)}{U} (k_a dx^a)$ with background metric \eqref{UVXYZ},  $U$ as given by \eqref{Udemocratic} and $k_a$ as given by \eqref{kmudxmuuvxyz}. The only change required to make this an embedding in a higher dimension is to \emph{remove} the constraint \eqref{UVXYZConstraint}. The resultant metric is $(D+1)$-dimensional with (non-Lorentzian) signature $(D-3)$. It is not expected to have physical significance, but is mathematically interesting and may yield more insights about the structure of the Kerr--AdS metrics. Kerr--AdS is recovered by applying the constraint.

The form that the metric takes is again very similar to the (Kerr--Schild) form of the Myers--Perry metrics in Cartesian coordinates. It also remarkably has the property that, if $\mu(r) = m$, the background metric is completely flat and that the full metric is Ricci-flat: $R_{a b} = 0$. Because the embedding is not used in the rest of this work, I will place this in Appendix \ref{embeddingcalcs}.

\subsection{Note on Kerr--de Sitter} \label{noteonkerrdesitter}

This brief note does not appear in the thesis. 

Before continuing, I will note that while my focus has been on Kerr--AdS, the Kerr--de Sitter (Kerr--dS) case is mathematically very similar, and essentially results from sending $l \to i l$ in the metric, so that $l$ becomes the de Sitter radius of curvature instead of the AdS radius of curvature. The pseudo-Cartesian coordinates associated with Kerr--dS can be found by taking the Kerr--AdS pseudo-Cartesian coordinates and sending $v \to i w, l \to il$, giving: 
\begin{align}
    ds^2 &= d \bar s^2 + H (k_a dx^a)^2 \nn 
    d\bar s^2 &= -du^2 +dw^2 + \sum_{i = 1}^{n-1+\ve} (dx_i^2+dy_i^2) + (1-\ve) d z^2 \nn 
     - u^2 +w^2 + \sum_{i = 1}^{n-1+\ve} (x_i^2 + y_i^2) + (1-\ve) z^2 &= +l^2 \nn 
     -\frac{u^2-w^2}{r^2-l^2} + \sum_{i = 1}^{n-1+\ve} \frac{x_i^2+y_i^2}{r^2+a_i^2} + (1-\ve) \frac{z^2}{r^2} &=0
\end{align}
along with $H = 2m/U$ with $U$ now given by
\begin{align}
    U &= -\frac{1}{l^2}\left[ - \frac{u^2-w^2}{(l^2-r^2)^2} + \sum_{i = 1}^{n-1+\ve} \frac{x_i^2+y_i^2}{(r^2+a_i^2)^2} + (1-\ve) \frac{z^2}{r^4}\right] (r^2-l^2) r^{1-\ve} \prod_{i = 1}^{n-1+\ve} (r^2+a_i^2)
\end{align}
and null vector
\begin{align}
     k^a \partial_a &= +\frac{(ru-lw) \partial_u + (rw-lu) \partial_w}{r^2-l^2} + \sum_{i = 1}^{n-1+\ve}\frac{(rx_i + a_i y_i) \partial_{x_i}+ (ry_i-a_ix_i) \partial_{y_i}}{r^2+a_i^2} + (1-\ve) \frac{z \partial_z}{r} \nn
    k_a dx^a &= -\frac{(ru-lw) du - (rw-lu) dw}{r^2-l^2} + \sum_{i = 1}^{n-1+\ve}\frac{(rx_i + a_i y_i) d{x_i}+ (ry_i-a_ix_i) d{y_i}}{r^2+a_i^2} + (1-\ve) \frac{z dz}{r}.
\end{align}

\chapter{The Covariant Phase Space Formalism and the Introduction of a Conserved Charge for Kerr--Schild Spacetimes} \label{NoetherChapter}

In this chapter, I use the covariant phase space methods of Wald and others, often referred to as the Iyer--Wald formalism from \cite{IyerWald94}, as well as a particular method by Barnich and Comp\`ere \cite{BarnichCompere} involving integration through solution space (the part of field configuration space corresponding to solutions to a given theory), to define a conserved charge associated with a Kerr--Schild spacetime. This conserved charge can be calculated by an integration over an arbitrary $(D-2)$-surface, and is related both to conserved quantities and to the variations of quantities on the horizon. It is thus a natural way to approach defining thermodynamic quantities which should obey the first law of black hole mechanics. 

The breakdown of the chapter is as follows. I begin in Section \ref{BHThermoforKAdSBHs} by reviewing the results for thermodynamics of Kerr--AdS black holes from \cite{GibbonsPerry,Cvetic}. In Section \ref{NoetherChargeFormalism} I then describe the covariant phase space formalism, in the manner of various papers by Wald and others, for defining conserved quantities associated with spacetimes, and the way that these relate to black hole mechanics. The result is that a $(D-2)$-form (using notation matching that of \cite{BarnichCompere,Compere,HajianSheikh-Jabbari} and others) $\bs k_\chi[\de g;g]$ is found, associated to each Killing vector $\chi$, and depending on the metric and its variation between solutions ``nearby'' in solution space. When $\bs k_\chi[\de g;g]$ is integrated over a suitable $(D-2)$-surface, it gives the variation in a conserved quantity associated with $\chi$. 

Because the covariant phase space formalism generally involves quantities ``at infinity,'' I will make use of some ideas from Barnich and Comp\`ere (BC) \cite{BarnichCompere} who develop a method of ``integrating through solution space'' in order to define conserved quantities at finite radius, including at a black hole horizon. This allows for direct evaluation of terms involved in the black hole first law and Smarr relation. The introduction of the BC method is in Section \ref{BCsection} and the application to black hole mechanics is in Section \ref{BHMechanicsSection}, with some additional notes about the Smarr relation in Section \ref{smarrrevisited}.

In Section \ref{KSBC}, I adapt BC's method to Kerr--Schild spacetimes in what I believe to be a novel way. I use this adaptation to define a conserved charge associated with a Kerr--Schild spacetime at arbitrary radius, rather than merely at infinity. I make some comments on black hole mechanics and the Smarr relation. I relate the new conserved charge associated with the Kerr--Schild decomposition to the stress--energy tensor in Section \ref{ComparisontoStressEnergyTensor}. I conclude the chapter with some notes on the connection of the Kerr--Schild charge to the Killing potential (Section \ref{relationshiptokillingpotential}) and to the KBL superpotential (Section \ref{KBLSection}).

The application of this new Kerr--Schild charge to the Kerr--AdS spacetimes will be the subject of Chapter \ref{ExplicitGKAdSChapter}.

\section{Black Hole Thermodynamics for Kerr--AdS Black Holes} \label{BHThermoforKAdSBHs}

To motivate this chapter, I describe the results from two key papers, the first by Gibbons, Perry and Pope (2005) \cite{GibbonsPerry} (hereafter GPP) and the second by Cveti\v{c}, Gibbons, Kubiz\v{n}\'ak and Pope (2011)~\cite{Cvetic}, hereafter CGKP. I discussed the latter in Chapter \ref{PRDPaper} to an extent but review it in greater detail here.  

\subsection{Gibbons, Perry and Pope (2005)} \label{GPPSection}

The goal of GPP was to establish a first law of black hole thermodynamics for KAdS black holes. Specifically, they wanted to find an expression of the form (slightly modifying notation)
\begin{align}
    d \mc E = T d S + \sum_i \Om_i d \mc J_i, \label{firstlawETSOJ}
\end{align}
for KAdS black holes (in four or higher dimensions), where $\mc E$ is the black hole's mass (or energy), $T$ is the black hole temperature, $S$ is the entropy, and $\Om_i$ and $\mc J_i$ are the angular velocity and angular momenta respectively associated with azimuthal symmetry vectors $\eta_i$. They pointed out that whereas there was agreement in the literature on expressions for the angular momenta, entropy and temperature of KAdS black holes, there was some controversy over the value for the mass and angular velocities. Their approach was based on the idea that the first law of black hole mechanics must be satisfied by  physically reasonable definitions for mass and angular velocity, and so they used the uncontroversial expressions for other terms in \eqref{firstlawETSOJ} to integrate the first law and to find the required expressions for $E$ and $\Om_i$. They then compared these expressions with others in the literature. If the first law is satisfied, then $T d S + \sum_i \Om_i d \mc J_i$ must be an exact differential ($d \mc E$).

In natural units $(\hbar = 1)$, GPP use the standard relation $S = \f14 A$ to relate the entropy to the area of the black hole outer horizon. The temperature $T$ was identified as $\kappa/2\pi$, where $\kappa$ is the surface gravity (again of the outer horizon). GPP gave these values in arbitrary dimension; they are written in this thesis in Section \ref{BHHorizonGKAdS}. $A$ and thus $S$ are unambiguous, and $\kappa$ and thus $T$ are unambiguous given the requirement that $\zeta^a$, the Killing vector tangent to the null generators of the horizon, must have a ``time component'' of 1. 

For the angular momenta, they calculated the Komar integrals associated with the normalized azimuthal symmetry vectors $\eta_i$. For a generic Killing vector $\chi$, let the Komar differential form $\bs K_\chi^K$ be given by
\begin{align}
    \bs K_\chi^K &= \f{1}{32\pi} * d \chi^\flat, \label{dKxiKdefinition}
\end{align}
where $*$ represents the Hodge dual; this means
\begin{align}
    (\bs K^K_\chi)_{a_1 \ldots a_{D-2}} &= \f1{16\pi} \na^c \chi^d \bs \ep_{c d a_1 \ldots a_{D-2}}. \label{dKxidefinition2}
\end{align}
Later on I will be calculating Komar integrals for different spacetimes. If there is some ambiguity of which spacetime is being used to evaluate $\bs K^K_\chi$ I will sometimes specify it with square brackets, such as $\bs K^K_\chi[g]$. (This Komar integrand convention largely follows \cite{BarnichCompere}.)

The angular momenta $\mc J_i$ are then identified as
\begin{align}
    \mc J_i &= \oint_{S_\infty} \bs K_{\eta_i}^K, \label{JiKomar}
\end{align}
where $S_\infty$ is a $(D-2)$-surface at infinity. Of course the $\eta_i$ are normalized (as throughout Chapter~\ref{GKAdSChapter}) so that the associated orbits close after parameter distance $2 \pi$. In KAdS, the result is independent of the choice of such surface. The Komar integrals were calculated explicitly up to $D \leq 7$, and the values found were
\begin{align}
    \mc J_i &= \f{m a_i \mc A_{D-2}}{4 \pi \Xi_i \prod_{j} \Xi_j}, \label{JivaluesGPP}
\end{align}
where $\mc A_{D-2}$ is the area of the $(D-2)$-sphere (see Section \ref{volumeandareaformulas}), and $m, a_i$ and $\Xi_i$ take their usual meanings. In $\prod_j \Xi_j$ the product is over all values of $j$ for which $a_j$ is nonzero. (Because the calculations are only explicitly done up to $D \leq 7$ I revisit these integrals and perform them explicitly in arbitrary dimensions in Section \ref{AngularMomentumKomarTerm}.) In four and five dimensions these expressions agree with the angular momenta that had been calculated in the literature up to that point through various methods, and in higher dimensions also coincided with the angular momenta developed through conformal methods by Ashtekar, Magnon and Das. 

There is an ambiguity in the definition of the angular velocity, as I alluded in Section \ref{BHHorizonGKAdS}, in equations \eqref{zetabreakdown} and \eqref{Omegaomega}, and which GPP address. The angular velocities $\Om_i$ from \eqref{Omegaomega} are the values of $\zeta^{\hat \vp_i}$ in the ABL coordinates, and represent the angular velocities as measured with respect to an asymptotically nonrotating frame, where $\xi = \pa/\pa \tau$ (in those coordinates). On the other hand, the angular velocities $\om_i$ (called $\Om'_i$ in GPP) from \eqref{Omegaomega} are the values of $\zeta^{\vp_i}$ in the BL coordinates, and represent the angular velocities as measured with respect to the frame where the time coordinate is such that $\beta = \pa/\pa \tau$ (in those coordinates), which is rotating at infinity. 

GPP emphasize the importance of using the angular velocities associated with a frame which is asymptotically nonrotating. They demonstrate this by pointing out that $T d S + \sum_i \Om_i d \mc J_i$ is an exact differential, but that $T d S + \sum_i \om_i d \mc J_i$ is not. This means that it is impossible for a first law \eqref{firstlawETSOJ} to be satisfied with $\om_i$ as the angular velocities instead of $\Om_i$. Having found that $T d S + \sum_i \Om_i d \mc J_i$ is an exact differential, they integrate the expression to find $\mc E$, choosing the integration constant so that $\mc E = 0$ when $m = 0$. The result for $\mc E$ is (slightly modifying notation)
\begin{align}
    \mc E &= \frac{m \mc A_{D-2}}{4\pi \prod_{j=1}^{n} \Xi_j}\left( \sum_{i = 1}^{n-1+\ve} \Xi_i^{-1} - \frac{\ve}{2}\right). \label{mcE}
\end{align}
This value of $\mc E$ automatically satisfies \eqref{firstlawETSOJ}. In four dimensions, this gives
\begin{align}
    \mc E &= \f{m}{\Xi^2}. 
\end{align}

Unlike the angular momenta, there was some controversy in the literature in the definition of $\mc E$. One issue is that although the Komar integrals for the angular momenta are finite and independent of integration surface, the Komar integral associated with $\xi$ (the asymptotically nonrotating time Killing vector) depends on integration surface, and has $\left |\oint_{S_\infty} \bs K_\xi^K\right| \to \infty$ where $S_\infty$ is a $(D-2)$-surface at spatial infinity.

GPP point out that the Ashtekar--Magnon--Das (AMD) definition of energy, using conformal methods, as defined in \cite{AshtekarMagnon,AshtekarDas} corresponds to $\mc E$, provided $\mc E$ is calculated as the conserved charge associated with the asymptotically nonrotating vector $\xi$ rather than the asymptotically rotating vector $\beta$. The AMD prescription associates to each asymptotic Killing vector a conserved quantity, which I will write as $Q_C[\chi]$ for a generic Killing vector $\chi$. Using $\xi$ for the asymptotically nonrotating Killing vector $\xi$ as usual, it turns out
\begin{align}
    \mc E &= Q_C[\xi] \nn 
    \mc J_i &= -Q_C[\eta_i]. \label{EJiQC}
\end{align}
(GPP have a positive sign in their expressions for both $\mc E$ and $\mc J_i$ but \eqref{EJiQC} is correct up to choice of orientation; it is because of the Lorentzian metric signature that the energy and angular momentum expressions have opposite signs. An alternate orientation for integration surface for which $\mc E = - Q_C[\xi]$ and $\mc J_i = +Q_C[\eta_i]$ would also be possible.) 

GPP present the Smarr--Gibbs--Duhem relation, where the thermodynamic potential $\Phi$ can be written as
\begin{align}
    \Phi &= \mc E - TS - \sum_i \Om_i \mc J_i. \label{SmarrGibbsDuhem}
\end{align}
The quantum statistical relation states that 
\begin{align}\Phi = T I_D, \label{PhiTID}
\end{align}
where $I_D$ is the Euclidean action, written in $D$ dimensions. The calculation of the Euclidean action is somewhat complicated and I return to this point, and how the Euclidean action relates to black hole volume, in Section \ref{EuclideanAction}.

GPP emphasize the importance of using the conserved charge associated with $\xi$, the vector nonrotating at infinity, rather than $\bt$, the one appearing in the BL coordinates. They state, ``The moral to be extracted from the above seems to be that one should always use the angular velocities measured relative to a non-rotating frame at infinity when discussing the thermodynamics of rotating black holes.'' They also state, ``The importance of measuring the angular velocity relative to a non-rotating frame was emphasized by Caldarelli,
Cognola and Klemm \cite{CaldarelliCognola}.'' Caldarelli et al.~point out that the angular velocity $\Om$ (relative to the nonrotating frame) is the same as the angular velocity of the rotating Einstein universe at infinity, and argue ``The fact that the angular velocity relevant to Kerr–Newman–AdS black hole
thermodynamics turns out to be that of the rotating Einstein universe at the AdS boundary,
agrees nicely with the AdS/CFT correspondence: if the KNAdS black hole in the bulk is
described by a conformal field theory living on the boundary, then the relevant angular velocity entering the thermodynamics should be that of the rotating Einstein universe at infinity.'' 

Part of this thesis is my attempt to provide another explanation for why it is so important to use the asymptotically nonrotating vector in order to have a first law be satisfied. I address it directly in Section \ref{explicitIxi}.

GPP point out that the values of $\mc E$ and $\mc J$ for KAdS black holes are, in four dimensions, those found by Henneaux and Teitelboim \cite{HenneauxTeitelboim85} consistent with the $SO(3,2)$ symmetry of AdS, Abbott and Deser \cite{AbbottDeser}, and Caldarelli, Cognola and Klemm \cite{CaldarelliCognola}, whereas the energy reported by Hawking, Hunter and Taylor-Robinson \cite{HHTR} and Silva \cite{Silva}, labelled $E'$ by GPP, is given as
\begin{align}
    E' &= \f{m}{\Xi},
\end{align}
which is off by a factor of $\Xi$. Hawking et al.~also use $\om$ rather than $\Om$ for the angular velocity. GPP point out that whereas the quantum statistical relation is satisfied using these quantities,
\begin{align}
    T I_4 &= E' - TS - \om \mc J,
\end{align}
a consequence of the fact that $E' - \om \mc J = \mc E - \Om \mc J$, the first law is not satisfied with $E'$ and $\om$,
\begin{align}
    d E' \neq T d S + \om d \mc J.
\end{align}
Similar results are obtained in higher dimensions for other definitions of energy. Again, it is possible to find quantum statistical relationships with alternate definitions of energy and using $\om_i$, but these alternate energy and $\om_i$ values do not satisfy the first law. In particular, Das and Mann \cite{DasMann} calculated the conserved energy associated with the AMD method in certain specific cases for $D \leq 7$ with a single rotation parameter, but using the BL coordinates, which are asymptotically rotating at infinity and for which $\pa/\pa \tau = \beta$. GPP point out that the AMD or conformal energy associated with $\xi$ will be given by (using the coordinates of \eqref{BLmetric})
\begin{align}
    \mc E &= Q_C \left[ \xi\right] \nn 
    &= Q_C\left[ \f{\pa}{\pa \tau} - \sum_{i} \f{a_i}{l^2} \f{\pa}{\pa \vp_i}\right] \nn 
    &= Q_C\left[ \f{\pa}{\pa \tau}\right] + \sum_i \f{a_i}{l^2} \mc J_i.
\end{align}
(In fact GPP have different signs here on some terms than I have here, but this discrepancy compensates for the discrepancy alluded to earlier below \eqref{EJiQC}.) GPP then calculate $Q_C\left[\pa/\pa \tau\right] = Q_C [\beta]$. I will define $\mc F \equiv Q_C[\beta]$ to be this quantity. (My use of $\mc F$ is simply because $\mc F$ is the letter following $\mc E$, and is not related to the Helmholtz free energy or any other quantities.) GPP show
\begin{align}
    \mc F &\equiv Q_C [\beta] = \f{(D-2) m \mc A_{D-2}}{8 \pi \prod_j \Xi_j},\label{Fdefinition}
\end{align}
satisfying, as a result of the relationship between $\beta$ and $\xi, \eta_i$,
\begin{align}
    \mc F &= \mc E - \sum_i \f{a_i}{l^2} \mc J_i. \label{Fexpression}
\end{align}
It is interesting to note that $\mc F$ has a simpler form than $\mc E$. It is worth noting that the $E'$ of \cite{HHTR,Silva} and some of the five-dimensional results are equal to $\mc F$, and that, similar to the $E'$ case, we have
\begin{align}
    \mc F - \sum_i \om_i \mc J_i = \mc E - \sum_i \Om_i \mc J_i. \label{FminusEminus}
\end{align}

See Ashtekar et al.~(2007) \cite{AshtekarPawlowski} for a detailed treatment of the relationship between the AMD conserved quantities and the first law of black hole mechanics. See also Hollands et al.~\cite{Hollands}, which compares various methods for developing conserved quantities associated with AdS using Wald's covariant phase space formalism (which I discuss in Chapter \ref{NoetherChargeFormalism}) and shows how this relates back to other definitions of conserved charges, including the AMD method and the method of Henneaux and Teitelboim.

\subsection{Cveti\v{c}, Gibbons, Kubiz\v{n}\'ak and Pope (2011)} \label{CveticSection}

CGKP modify the first law as it appears in GPP by also allowing $\tilde \La$ to vary between different KAdS solutions. 

(I note before continuing that CGKP use the convention where ``$\La$'' is the constant of proportionality in vacuum between the Ricci tensor and the metric tensor. I call this constant $\tilde \La = 2 \La/(D-2)$. Except for a few factors of $2/(D-2)$, the results from varying $\La$ instead of $\tilde \La$ are the same.)

Additionally, the possible existence of charges (such as electric charge) $\mc Q_\alpha$ with conjugate potentials $\Phi_\alpha$ is considered. Following a line of argument by Kastor, Ray and Traschen \cite{KastorEtal:2009}, $\mc E$ is now interpreted not as an energy but an enthalpy as the spacetime, with $\ti \La$ being identified with a pressure according to
\begin{align}
    P &= - \f{D-2}{16\pi} \tilde \La. \label{PLambdaRelation}
\end{align}
The variation in the enthalpy of an ordinary thermodynamic system includes a term proportional to $V d P$, where $V$ is the volume. The first law of black hole mechanics is now interpreted as having the form
\begin{align}
    d \mc E = T d S + \sum_i \Om_i d \mc J_i + \sum_\alpha \Phi_\alpha d \mc Q_\alpha + \Th d \ti \La, \label{firstlawwithThetadLambda}
\end{align}
where $\Th$ is a term conjugate to the variation in $\ti \La$. The $\Th d \ti \La$ term can be interpreted as a $V_{th} d P$ term, where $P$ is given by \eqref{PLambdaRelation} and where 
\begin{align}
    V_{th} &\equiv -\f{16 \pi \Th}{D-2} \label{Vth}
\end{align}
is called the \emph{thermodynamic volume}, which is associated with the black hole's thermodynamic properties, and can be written as
\begin{align}
    V_{th} &= -\f{16 \pi}{D-2} \left.\f{\pa \mc E}{\pa \ti \La}\right|_{S,\mc J_i, \mc Q_\alpha}.
\end{align}

Using \eqref{firstlawwithThetadLambda} and dimensional scaling arguments, the generalized Smarr relation
\begin{align}
    \mc E &= \f{D-2}{D-3} (T S + \sum_i \Om_i \mc J_i) + \sum_\alpha \Phi_\alpha \mc Q_\alpha - \f{2}{D-3} \Th \ti \La \label{Smarr}
\end{align}
automatically follows. 

(To elaborate, relaxing the $\hbar = 1$ condition, if $L$ is the dimension of length, the Smarr relation is a consequence of $\mc E$ and $\mc Q_\alpha$ having dimensions of $L^{D-3}$, $A$ and $\mc J_i$ having dimensions $L^{D-2}$, and $\ti \La$ having dimensions $L^{-2}$, as well as the idea that rescaling each in proportion to the rescaling of a length parameter will take one solution to Einstein's equations, with one value of $\ti \La$, into another solution to Einstein's equations with a different $\ti \La$. We also have that $\ka$ and $\Om_i$ have dimensions $L^{-1}$, $\Phi_\alpha$ is dimensionless, and $\Th$ has dimensions $L^{D-1}$. We can then set $T = \hbar \ka/2\pi$, $S = A/4\hbar$. Let $c$ be a parameter with dimension length and let $\mc E = c^{D-3} \hat{\mc E}$, $S = c^{D-2} \hat S$ and so on, so that $\hat{\mc E}$, $\hat S$ and so on are dimensionless constants.  Then, keeping $\hat {\mc E}$ and so on fixed and varying only $c$, the first law relation \eqref{firstlawwithThetadLambda} gives, moving all the terms to the left-hand side,
\begin{align}
    \left[(D-3) \hat{\mc E} - (D-2) \hat T \hat S - (D-2) \sum_i \hat \Om_i \hat{\mc J}_i - (D-3) \sum_\alpha \hat \Phi_\alpha \hat {\mc Q}_\alpha  + 2 \hat \Th \hat{\ti \La} \right] c^{D-4} d c &= 0 \nn 
    \left[ (D-3) \mc E - (D-2) T S - (D-2) \sum_i \Om_i \mc J_i - (D-2) \sum_\alpha \Phi_\alpha \mc Q_\alpha + 2 \Th \ti \La \right] d \ln c &= 0.
\end{align}
\eqref{Smarr} follows.)

The thermodynamic volume $V_{th}$ consequently also follows from the Smarr relation \eqref{Smarr}, provided that the first law \eqref{firstlawwithThetadLambda} is satisfied \emph{with constant $\ti \La$}. 

CGKP define a ``naive'' geometric volume $V_{geo}$ according to \begin{align}
    V_{geo} &= \int_{r_0}^{r_+} \sqrt{-g} dr \int d \Om,
\end{align}
where $d \Om$ is the integral over the $(D-2)$-sphere and the integral over a radial coordinate is from the singularity at $r = r_0$ (not necessarily real) to the horizon at $r = r_+$. This can also be written
\begin{align}
    V_{geo} &= \int \sqrt{-g} d^{D-1} x,
\end{align}
where the integral is over the region below $r_+$ and above $r_0$, where $d^{D-1} x$ is the product of differentials excluding the time component differential $dt$ and where $g$ is the metric determinant. To make connection with the relationship between $V_{th}$ and $\Th$, $\Th'$ is written as
\begin{align}
    \Th' &= - \f{D-2}{16\pi} V_{geo}. \label{ThetaprimeVgeo}
\end{align}

This quantity $V_{geo}$ was pointed out in \cite{Ballik} (appearing here as Chapter \ref{PRDPaper}) to be equal to the vector volume associated with the Killing vector $\xi$ (or $\beta$) and the black hole region. One point to emphasize here is that the black hole region is taken to be between the horizon at $r = r_+$ and a lower value of $r$, which is zero if $D$ is even or $r^2+a_n^2 = 0$ if $D$ is odd. I will call the $(D-2)$-surface of constant $t$ (in KS coordinates) with $r = 0$ (even dimension) or $r^2+a_n^2 = 0$ (odd dimension) $S_0$. Letting $H$ be a surface of constant-$t$ (in KS coordinates) with $r=r_+$ and letting $\Si$ be the constant-$t$ hypersurface with boundary $S_0 \cup H$, we have
\begin{align}
    V_{geo} = \int_\Si \xi^a d \Si_a = \mc V_{\xi,\mc B}, 
\end{align}
where $\xi = \pa/\pa t$. This is the canonical black hole volume $\mc V_C$ from Section \ref{canon}. I use the symbol $\mc V_{\xi,\mc B}$ to remind the reader that it is the vector volume associated with the vector $\xi$ and the black hole region $\mc B$. 

In spherical symmetry or in the Kerr--Newman--AdS black holes, the geometric volume $V_{geo}$ satisfies the relation
\begin{align}
    V_{geo} &= \f{r_+ A}{D-1}, \label{VrArelationship}
\end{align}
where $A$ is the horizon radius and $r_+$ is the value of the radius parameter $r$ on the horizon. For definiteness, in the spherical symmetry cases under consideration, the metric can be taken to be
\begin{align}
    ds^2 &= -f(r) dt^2 + \f{dr^2}{f(r)} + r^2 d\Om_{D-2}^2,
\end{align}
where the horizon is located on the outermost root $f(r_+) = 0$. In Kerr--Newman--AdS, $r$ is the spheroidal radial coordinate that appears in the various forms of the metric given in Chapter \ref{GKAdSChapter}. 

What was then noted by CGKP is that in spherical symmetry (in the Schwarzschild--AdS and Reissner--Nordstr\"om--AdS cases) $V_{th}$ takes the simple form 
\begin{align}
    V_{th} &= V_{geo} \qquad\textrm{(spherical symmetry).}
\end{align}
Away from spherical symmetry, for the Kerr--Newman--AdS black holes, $V_{th}$ and $V_{geo}$, and thus $\Th$ and $\Th'$, differ according to
\begin{align}
    \Th' &= \Th + \f{1}{2(D-1)} \sum_i a_i \mc J_i \nn 
    V_{geo} &= V_{th} -\f{8\pi}{(D-1)(D-2)} \sum_i a_i \mc J_i.
\end{align}
Interestingly, CGKP point out (using different notation) if $\mc F$ and $\omega_i$, the quantities for the energy and angular momentum which are related to the asymptotically rotating frame (the one which is adapted to the vector $\beta$) are used instead of $\mc E$ and $\Om_i$, the Smarr relation (including charge) takes the form
\begin{align}
    \mc F &=  \f{D-2}{D-3} (T S + \sum_i \om_i \mc J_i) + \sum_\alpha \Phi_\alpha \mc Q_\alpha - \f{2}{D-3} \Th' \ti \La  \nn 
     &= \f{D-2}{D-3} (T S + \sum_i \om_i \mc J_i) + \sum_\alpha \Phi_\alpha \mc Q_\alpha - \f{2}{D-3} V_{geo} P.\label{SmarrwithF}
\end{align}
The geometric volume then appears naturally in the Smarr relation if $\beta$, rather than $\xi$, is used to define the energy.

\subsubsection{Komar Integrals and Killing Potentials} \label{KomarKillingPotential}

In in a Ricci-flat spacetime, the Komar integrand associated with a Killing vector $\chi$, $\bs K^K_\chi$, satisfies $d \bs K^K_\chi = 0$. Consequently it is possible to define conserved quantities (independent of integration surface) according to integrals of $\oint_S \bs K^K_K$ over $(D-2)$-surfaces. For the Myers--Perry spacetimes (including Kerr in four dimensions), the angular momenta $\mc J_i$ are given by \eqref{JiKomar} and the mass $\mc E$ by
\begin{align}
    \mc E &= -\f{D-2}{D-3}\oint_S \bs K^K_\xi,
\end{align}
where $\xi$ is the asymptotically static Killing vector normalized so that $\xi^a \xi_a \to -1$ at infinity. $S$ is an arbitrary surface. Using $\zeta = \xi + \sum_i \Om_i \eta_i$ for the Killing vector tangent to the null generators of the horizon, if $H$ is a $(D-2)$-surface on the horizon,
\begin{align}
    \oint_H \bs K^K_\zeta &= -\f{\kappa A}{8 \pi} \nn 
    &= -TS,
\end{align}
which holds in general, regardless of whether the spacetime is Ricci flat or not. (I revisit this in Section \ref{BCSmarrSection}.) That $d \bs K^K_\z = 0$ means that $\oint_S \bs K^K_\z = -TS$ on an arbitrary surface $S$ as well. The linearity of Komar integrals with respect to their corresponding Killing vector implies the Smarr relation 
\begin{align}
    \mc E &= - \f{D-2}{D-3} \oint_S \bs K^K_\xi \nn 
    &= - \f{D-2}{D-3} \oint_S \left( \bs K^K_{\zeta} - \sum_i \Om_i \bs K^K_{\eta_i}\right) \nn 
    &= \f{D-2}{D-3} \left( T S + \sum_i \Om_i \mc J_i\right). \label{SmarrZeroLambda}
\end{align}

The KAdS spacetimes are not Ricci-flat, and have $R^a_b = \ti \La \de^a_b = -(D-1)l^{-2} \de^a_b$. The Komar integrals for angular momentum are still independent of integration surface, but $\oint_S \bs K_\xi^K$ depends on the surface $S$. Let $\Si$ be a $(D-1)$-dimensional hypersurface bounded by two $(D-2)$-surfaces $S_1$ and $S_2$, with $S_2$ at larger radius. Then, using \eqref{dKKxi}, and choosing orientation so that the normal to $S_2$ and $S_1$ are both pointing toward larger radius,
\begin{align}
    \oint_{S_2} \bs K^K_\xi - \oint_{S_1} \bs K^K_\xi &= \int_\Si d \bs K^K_\xi \nn 
    &= -\int_\Si \f{D-1}{8\pi l^2} * \xi^\flat. \label{KomarDifferenceXi}
\end{align}
We recognize the integral over $* \xi^\flat$ as being the vector volume associated with $\xi$ as defined in Chapter \ref{PRDPaper}. 

Kastor, Ray and Traschen \cite{KastorEtal:2009} compensated for the non-constancy of the Komar integral associated with $\xi$ by using a Killing potential 2-form. For any Killing vector $\chi$, since $\na_a \chi^a = 0$ there must always exist at least locally a 2-form $\bs \om_\chi$, called the Killing potential, such that
\begin{align}
    \chi^b &= \na_a \bs \om_\chi^{a b}. \label{omegachidef}
\end{align}
In general $\bs \om_\chi$ is not unique and can be modified by the addition of a co-closed form $\bs \nu$ (i.e.~a form for which $\na_a \bs \nu^{ab} = 0$). Then, $\tilde{\bs \om}_\chi$ defined by
\begin{align}
    \tilde{\bs \om}_\chi &= \bs \om_\chi + \bs \nu
\end{align}
will also be a Killing potential for $\chi$. 

Since $\na_a \na^a \chi^b = -R^b_a \chi^a = -\tilde \La \chi^b$ (if $R_{ab} = \tilde \La g_{ab}$) for a Killing vector $\chi^a$, we then have
\begin{align}
    \na_a ( \na^a \chi^b + \tilde \La \bs \om_\chi^{ab}) &= 0.
\end{align}
This means by the Gauss--Stokes law, if there is a $(D-1)$-dimensional hypersurface $\Si$ with boundary $\pa \Si$,
\begin{align}
    \oint_{\pa \Si} d S_{ab} (\na^a \xi^b + \tilde \La \bs \om_\chi^{ab}) &= 0.
\end{align}
This replaces the Komar integrand with a term also involving the Killing potential. 

Now consider vacuum (with $\La$). Letting $\bs \om_\zeta$ be a Killing potential associated with $\zeta$ (which is, recall, the Killing vector, with ``time component'' 1, which is tangent to the null generators of the horizon, called $\xi$ by CGKP), CGKP state that they can write the energy $\mc E$ as
\begin{align}
    \mc E &= -\f{D-2}{D-3} \oint_{S_\infty} \left( \bs K_\xi^K + \f{\tilde \La}{16 \pi} * \bs \om_\zeta\right),
\end{align}
where $\xi$ is the asymptotically nonrotating Killing vector as usual. (CGKP are using a different convention for the Hodge dual; I am converting into my notation.) 

Because of the freedom to change $\bs \om_\zeta$ by a co-closed form $\bs \nu$, this expression is not sufficient to define $\mc E$, but if $\mc E$ is known by other means then this can be used to fix $\bs \om_\zeta$. The integrals $\oint_{S_\infty} \bs K^K_\xi$ and $\oint_{S_\infty} * \bs \om_\zeta$ \emph{both} diverge, but the combination does not. We have 
\begin{align}
    \mc E &= -\f{D-2}{D-3} \oint_{S_\infty} \left( \bs K_\xi^K + \f{\ti \La}{16\pi} * \bs \om_\zeta\right). \label{EKillingPotentialForm}
\end{align}
We can then repeat the Smarr argument in the $\La = 0$ case armed with these new definitions, and find
\begin{align}
    \mc E &= -\f{D-2}{D-3} \oint_{S_\infty}\left( \bs K^K_\xi + \f{\ti \La}{16\pi} * \bs \om_\zeta\right) \nn 
    &= -\f{D-2}{D-3} \oint_{S_\infty} \left( \bs K^K_\zeta - \sum_i \Om_i \bs K^K_{\eta_i} + \f{\ti \La}{16\pi} * \bs \om_\z\right) \nn 
    &= \f{D-2}{D-3}  \sum_i \Om_i \mc J_i- \f{D-2}{D-3} \oint_{S_\infty} \left( \bs K^K_\zeta + \f{\ti \La}{16\pi} * \bs \om_\z\right).
\end{align}
Because $d (\bs K^K_\zeta + (16\pi)^{-1} \ti \La * \bs \om_\z) = 0$, the integral can be moved from $S_\infty$ to a $(D-2)$-surface on the horizon, $H$. Then we have
\begin{align}
    \mc E &= \f{D-2}{D-3} \sum_i \Om_i \mc J_i - \f{D-2}{D-3} \oint_H \left( \bs K^K_\z + \f{\ti \La}{16\pi} * \bs \om_\z\right) \nn 
    &= \f{D-2}{D-3} \left( T S + \sum_i \Om_i \mc J_i\right) - \f{D-2}{D-3} \f{\ti \La}{16\pi} \oint_H * \bs \om_\z.
\end{align}
Comparing with \eqref{Smarr} with zero charge, they can then define $\Th$ to be
\begin{align}
    \Th &= \f{D-2}{32\pi} \oint_H * \bs \om_\z \nn 
    V_{th} &= -\f12 \oint_H * \bs \om_\z. \label{ThetaVth}
\end{align}
(Their expression for $\Th$ has $16\pi$ in the denominator, but I have $32\pi$ because of the different conventions we are using for the Hodge dual.)

CGKP also present an argument for the construction of a potential for $\zeta$ using the PCKY tensor $\bs h$, in the case of the Kerr--AdS spacetimes. Following \cite{KrtousKubiznak,Frolov} (and others), define $\bs h^{(j)} = \bs h^{\wedge j}$ to be 
\begin{align}
    \bs h^{(j)} = \bs h^{\wedge j} &\equiv \bigwedge_{i=1}^j \bs h, \label{hj}
\end{align}
i.e.~the object resulting from wedging together $j$ copies of $\bs h$. (By convention, $\bs h^{\wedge 0} = 1, \bs h^{\wedge 1} = \bs h$. We then have $\bs h^{\wedge 2} = \bs h \wedge \bs h, \bs h^{\wedge 3} = \bs h \wedge \bs h \wedge \bs h$ and so on.) $\bs h^{(j)}$ is also a closed conformal Killing-Yano tensor. Because $\bs h$ is nondegenerate, these will be nonzero for $0 \leq j \leq \lfloor{D/2}\rfloor$. The expression in terms of the variables $x_\mu$ is \cite{KrtousKubiznak}
\begin{align}
    \bs h^{(j)} &= j! \sum_{\nu_1 \leq \ldots \leq \nu_j} x_{\nu_1} \ldots x_{\nu_j} \bs \om^{\nu_1} \wedge \ldots \wedge \bs \om^{\nu_j}.
\end{align}
Let $\bs f^{(j)}$ be the Hodge dual of $\bs h^{(j)}$, possibly up to a numerical factor. The $\bs f^{(j)}$ are then Killing--Yano tensors. In odd dimensions, $\bs f^{(n)}$ is a Killing one-form equal, also possibly up to a numerical factor, to $(\pa/\pa \psi_n)^\flat$, in the $(x_\mu, \psi_j)$ coordinates. Then the symmetric, two-index Killing tensor $K^{(j)}_{ab}$ is given by, also up to a multiplicative factor, 
\begin{align}
    (K^{(j)})^a_b &\propto \bs f^{(j)}_{b c_1 \ldots c_{D-2j-1}} (\bs f^{(j)})^{a c_1 \ldots c_{D-2j-1}}.
\end{align}
The result for $K^{(j)}_{ab}$ is given in \cite{Frolov} to be
\begin{align}
    K^{(j)}_{a b}&=\sum_{\mu=1}^n A_\mu^{(j)} (e^\mu_a e^\mu_b + e^{\hat \mu}_a e^{\hat \mu}_b) + \ve A^{(j)} e^{\hat 0}_a e^{\hat 0}_b. \label{Kjab}
\end{align}
(In CGKP, they let $K^{(0)}_{ab} = -g_{ab}$ rather than what I have here, $K^{(0)}_{ab} = +g_{ab}$, but I think this sign difference is also resulting from a difference in convention---in CGKP their $\bs h$ is the negative of the $\bs h$ I am using. I am omitting the multiplicative constants on $\bs f^{(j)}$ and $K^{(j)}_{ab}$ as constructed from $\bs f^{(j)}$ because they depend on the Hodge dual in use by CGKP and I am interested only in the resulting expression for $K^{(j)}_{ab}$ which is \eqref{Kjab}.) They further introduce $\bs \om^{(j)}$ to be 2-forms defined by
\begin{align}
    (\bs \om^{(j)})_{ab} &= \f{1}{D-2j-1} K^{(j)}_{ac} {\bs h^c}_b,
\end{align}
or, using \eqref{Kjab},
\begin{align}
    \bs \om^{(j)} &= \f{1}{D-2j-1} \sum_{\mu=1}^{n} A_\mu^{(j)} x_\mu \bs \om^\mu, \label{omegaj}
\end{align}
for $0 \leq j \leq n-1$, along with, if $D$ is odd,
\begin{align}
    \bs \om^{(n)} &= \f{\sqrt{c}}{n! (2n-1)!} * \bs b^{(n)} \nn  
    &= \f{l^2}{2n} \sum_{\mu=1}^n \f{1}{x_\mu} \left(  c \bs \om^\mu + \f{\sqrt{-cA^{(n)}} \bar Q_\mu}{\sqrt{Q_\mu}} e^\mu \wedge e^{\hat 0}\right). \label{omegan}
\end{align}
where $\bs b^{(n)} = \bs b'' \wedge \bs h^{(n-1)}$. Here, $\bs b''$ is the potential used by CGKP for $\bs h$, which differs from my $\bs b$ by a closed form and is (up to a sign) given by \eqref{bprimeprime}. The multiplicative constant is different from the one stated by CGKP because of my different convention for the Hodge dual; see Appendix \ref{potentialomegan} for derivation. Because the potential $\bs b''$ can be shifted by a closed form, there is an ambiguity in the ``natural'' $\bs \om^{(n)}$ constructed from $\bs b''$. 

These $\bs \om^{(j)}$ then act as potentials for the vectors $\beta^{(j)}$, given by:
\begin{align}
    (\beta^{(j)})^a &= (K^{(j)})^a_b \beta^b \nn 
    (\beta^{(j)})^a &= \na_b (\bs \om^{(j)})^{b a}.
\end{align}
In fact, in coordinates $(x_\mu,\psi_j)$, 
\begin{align} 
    \beta^{(j)} = \f{\pa}{\pa \psi_j}.
\end{align}
This means that any vector which is a linear combination of the $\beta^{(j)}$ can have a Killing potential constructed out of the $\bs \om^{(j)}$: if $\chi = \sum_j c_j \bt^{(j)}$, then $\bs \om_\chi = \sum_j c_j \bs \om^{(j)}$.

The result for $\zeta$ is
\begin{align}
    \bs \om_\zeta &= \f{r_+^{2n}}{\prod_{i=1}^{n-1+\ve} (r_+^2+a_i^2)}\sum_{j=0}^{n-1+\ve} \f{1}{r_+^{2j}} \bs \om^{(j)}.
\end{align}
(This follows directly from the decomposition of $\zeta$ in terms of the $\pa/\pa \psi_j$.) 

This construction is unique, up to a freedom to add an exact form to $\bs \om_\z$: $\bs \om_\z \to \bs \om_\z + \bs \nu$, where $d * \bs \nu = 0$. If no such form $
\bs \nu$ is added, then in fact the integral $\f{D-2}{16\pi} \oint_H \bs \om_\zeta$ gives $\Th'$, so that
\begin{align}
    \oint_H * \bs \om_\z &= - 2 V_{geo}. \label{ointHomegazeta}
\end{align}
CGKP state, ``it is unclear if there is a simple geometric explanation for this'' and speculate that the fact that this gives the geometric volume and also the volume--area relationship might be a consequence of the existence of hidden symmetries of the Kerr--AdS spacetime (i.e.,~the existence of $\bs h$). 

Using $\mc E$ and $\Th$ instead of $\mc F, \Th'$, the form $\bs \nu$ must be added to $\bs \om_\z$, taking the form
\begin{align}
    \bs \nu &= \alpha \Om_{D-2},
\end{align}
where $\Om_{D-2}$ is the volume element of the unit $(D-2)$-sphere and $\alpha$ is the constant
\begin{align}
    \alpha &= -\f{2m}{(D-1)(D-2) \prod_j \Xi_j} \sum_i \f{a_i^2}{\Xi_i}.
\end{align}
(Again, I am following CGKP closely, and because their conventions for the Hodge dual are different from mine there may be some numerical factors which differ.) 

\subsubsection{Commentary on Komar Integrals and Killing Potentials}

Before continuing I want to make a few notes about Komar integrals and the Killing potential method, which will be useful throughout the remainder of this thesis. 

It is worth noting before continuing that the Komar integral associated with the \emph{azimuthal} Killing vectors is surface-independent, and does not require modification with a Killing potential in order to achieve a constant result. This is the result of the vanishing of the vector volume contribution associated with the azimuthal vectors $\eta_i$, as in \eqref{cyclicvectorvolumevanishing}. Explicitly, we have, using \eqref{KomarDifferenceXi},
\begin{align}
    \oint_{S_2} \bs K^K_{\eta_i} - \oint_{S_1} \bs K^K_{\eta_i} &= -\int_\Sigma \f{D-1}{8\pi l^2} * \eta_i^\flat \nn 
    &= 0.
\end{align}
(As an example, consider where $\Sigma$ is a constant-$t$ surface. Then $(* \eta_i^\flat)|_\Si = (\eta_i^a) d \Sigma_a = (\eta_i^a) \sqrt{-g} \de_a^t d^{D-2} x = \de^t_{\phi_i} \sqrt{-g} d^{D-2} x = 0.$) 

This surface-independence also extends to the Killing potential for $\eta_i$. Let $\bs \om_{\eta_i}$ be a Killing potential with $\mathrm{div} \bs \om_{\eta_i} = \eta_i^\flat$, not necessarily the one constructed using the $\bs \om^{(j)}$. Then we have, similarly,
\begin{align}
    \oint_{S_2} * \bs \om_{\eta_i} - \oint_{S_1} * \bs \om_{\eta_i} &= - 2 \int_\Sigma * \eta_i^\flat  \nn 
    &= 0.
\end{align}
(Here I used \eqref{StokesLawCodim1}.) 

Using the linearity of the Killing potential, we can write
\begin{align}
    \bs \om_\z &= \bs \om_\xi + \sum_{i=1}^{n-1+\ve} \Om_i \bs \om_{\eta_i},
\end{align}
and then rewrite $\mc E$ from \eqref{EKillingPotentialForm} as
\begin{align}
    \mc E &= -\f{D-2}{D-3} \oint_{S_\infty} \left( \bs K_\xi^K + \f{\ti \La}{16\pi} \left[* \bs \om_\xi + \sum_{i=1}^{n-1+\ve} * \bs \om_{\eta_i}\right]\right).
\end{align}
The integrals over $* \bs \om_{\eta_i}$ are surface-independent, as is the combination $\oint \left(\bs K^K_\xi + \f{\ti \La}{16\pi} * \bs \om_\xi\right)$. Thus the whole integral is surface-independent and need not necessarily be performed at $S_\infty$. 

Along similar lines, it would also be possible to construct $\mc E$ using a Killing potential for $\xi$ instead of one for $\z$, because then the combination $\oint \left(\bs K^K_\z + \f{\ti \La}{16\pi} * \bs \om_\xi\right)$ would be surface-independent. In this case we would have
\begin{align}
    \mc E &= - \f{D-2}{D-3} \oint_S \left( \bs K^K_\xi + \f{\ti \La}{16\pi} * \bs \om_\xi\right). \label{mcEKillingPotentialWithXi}
\end{align}
As with $\bs \om_\z$, there would be freedom to add a closed form to $\bs \om_\xi$, \emph{a priori}. I will return to this point.

\subsubsection{Kastor's Derivation}  \label{kastorderivation} 

There is a proof of the first law in the context of a variable $\ti \La$ given in \cite{KastorEtal:2009} and then subsequently in \cite{DolanKastor,KubiznakMannTeo,Mann} in which the thermodynamic volume $V_{th}$ is written explicitly in terms of the Killing potential at the horizon and at infinity, along with a \emph{background} Killing potential $\bs \om_{\textrm{AdS}}$. The details of the derivation are in, for instance, \cite{KubiznakMannTeo}. $V_{th}$ is then given by
\begin{align}
    V_{th} &= \int_\infty d A r_c n_b (\bs \om^{cb} - \bs \om^{cb}_{\textrm{AdS}}) - \int_H d A r_c n_b \bs \om^{cb}, \label{KastorVth}
\end{align}
where $\bs \om^{cb}$ is the Killing potential associated with $\z$ and $\bs \om^{cd}_{\textrm{AdS}}$ is a Killing potential associated with the \emph{background AdS} spacetime, the integrals are carried out on $(D-2)$-surfaces at $r \to \infty$ and at the horizon, $n_b$ is a unit timelike normal, $r_c$ is a unit spacelike normal and $dA$ is the area element. Note however that not only is a gauge choice necessary to fix $\bs \om^{cb}$, but also to fix $\bs \om^{cd}_{\textrm{AdS}}$; the choice required, as pointed out in \cite{KubiznakMannTeo}, is such that
\begin{align}
    \mc E &= - \f{D-2}{16\pi(D-3)} \oint_\infty d A r_c n_b (\na^{[c} \xi^{b]} + 2 \ti \La \bs \om^{cb}_{\textrm{AdS}}),
\end{align}
which again fixes 
\begin{align}
    V_{th} &= -\oint_H d A r_c n_b \bs \om^{cb},
\end{align}
which matches with \eqref{ThetaVth}. This means that this first-law argument does not by itself tell us the values of the Killing potentials or the thermodynamic volume, but requires some other method to produce $\mc E$ and thus $V_{th}$.

There is a related derivation by Jacobson and Visser \cite{JacobsonVisser}, who follow Kastor, for the most part, in deriving the AdS Smarr relation, recovering a form of the expression \eqref{KastorVth} (up to a difference in sign), as well as a related expression for a thermodynamic volume
\begin{align}
    \bar V_\z &= \int_\Si \z \cdot \bs \ep - \int_{\Si'} \z^{\textrm{AdS}} \cdot \bs \ep^{\textrm{AdS}},
\end{align}
where $\z$ is the horizon-generating Killing field. $\Si$ extends from the black hole horizon to infinity, and $\Si'$ extends from $r = 0$ to $\infty$ in the background spacetime. Note that $\int_\Si \z \cdot \bs \ep$ and $\int_{\Si'} \z^{\textrm{AdS}} \cdot \bs \ep^{\textrm{AdS}}$ are instances of the Parikh/vector volume. This definition ensures that the Smarr relation
\begin{align}
    \f{D-3}{D-2} \mc E - \Om_H \mc J &= \f{\ka A}{8\pi} - \f{2 \bar V_\z \La}{(D-2) 8\pi}
\end{align}
is obeyed. (I have slightly modified notation here. Jacobson and Visser consider only one angular momentum.) To deal with the infinite volume integrals, the Killing potentials can be introduced, but the same general issue having to do with fixing the gauge in order to ensure that the correct value of $\mc E$ is determined in terms of the Killing potentials or volume integral comes up here as well.

\subsection{Commentary: ``Open Questions'' Inspired By GPP and CGKP} 

Here I make some comments on these papers to motivate my approach for the rest of this thesis.

CGKP show that the conserved charge $\mc E$ associated with $\xi$ is the one which satisfies the first law of black hole mechanics, whereas the conserved charge $\mc F$ associated with $\beta$ does not. However, not only does $\mc F$ have a simpler form than $\mc E$, but the geometric volume $V_{geo}$ appears in the Smarr relation associated with $\mc F$, rather than the one associated with $\mc E$. The geometric volume also satisfies the simple volume--area relationship $V_{geo} = r_+ A/(D-1)$, and appears naturally in the Killing potential construction as (up to a numerical factor) the integral of $\oint_H * \bs \om_\z$, where $\om_\z$ is the Killing potential constructed in the unique manner from $\bs h$ (and $\bs b''$ in odd dimensions). This all suggests that both $\xi$ and $\beta$ ($\mc E$ and $\mc F$) play some sort of important role.

$V_{geo}$ is, again as pointed out in \cite{Ballik}, also equal to the vector volume associated with either $\xi$ or $\beta$,
\begin{align}
    V_{geo} &= \mc V_{\xi, \mc B} = \mc V_{\beta, \mc B} \nn 
    &= \int_\Sigma \xi^a d \Si_a = \int_\Sigma  \beta^a d \Si_a,
\end{align}
where $\Si$ is a hypersurface extending from $r = 0$ (even dimension) or $r^2+a_n^2 = 0$ (odd dimension) up to the horizon $r = r_+$. Thus $V_{geo} = \mc V_{\xi,\mc B}$ is closely connected with our work. 

This raises the following questions, which I address in the remainder of this thesis:
\begin{enumerate}
    \item Why does the first law work with $\xi$ but not $\beta$? (Or: is there more we can say about why it is necessary to use an asymptotically nonrotating frame?) 
    \item Why is it that the geometric volume appears in the Smarr relation associated with $\beta$, but not the one associated with $\xi$?
    \item Why does the Killing potential method, constructing $\bs \om^{(j)}$ purely out of $\bs h$ (and $\bs b''$ for $\bs \om^{(n)}$ in odd dimensions), produce $\oint_H * \bs \om_\z = -2 V_{geo}?$
    \item What is the reason for the simple relationship $V_{geo} = r_+A /(D-1)$ between the geometric volume and the horizon area?
\end{enumerate}

In this chapter I will use the covariant phase space methods of Wald, along with some specific methods of Barnich and Comp\`ere, to develop a definition of conserved charge which applies to Kerr--Schild spacetimes such as Kerr--AdS. This gives a conserved charge associated with each Killing vector which is expressible in terms of Komar integrals and the Kerr--Schild correction to the spacetime, and which can be evaluated on any surface (not just at infinity). 

This charge is constructed to satisfy the first law, but there are some subtleties about why it is that this charge will satisfy the first law associated with $\xi$ and not with $\beta$. A discussion of the first question above, why the charge associated with $\xi$ satisfies the first law and that associated with $\beta$ does not, appears later on, in Section \ref{explicitIxi}. Only the $D = 4$ case is treated directly but I believe the results generalize naturally to higher dimensions. 

The second question, having to do with why the vector volume appears more naturally in the Smarr relation associated with $\beta$ rather than the one associated with $\xi$, is addressed in Section \ref{chargebreakdown}.

For the first two questions, various approaches have been taken in the literature. For the first question, authors such as \cite{HajianSheikh-Jabbari,Chrusciel,Blagojevic:2020edq,Jing:2017jxw} identify the breakdown of the first-law when using a frame adapted to $\bt$ rather than $\xi$ as coming down to the non-integrability of the Hamiltonian associated with $\bt$. I return to this point. My work is related to this, though the approach is distinctive. 

For the second question, 
there has been a very recent interest in the difference between the thermodynamic and geometric volumes, which has appeared in papers such as \cite{Xiao,Xiao25,Campos24,Campos25,Campos26}. Xiao et al.~\cite{Xiao,Xiao25} consider the extended Iyer--Wald formalism, related to the covariant phase space formalism I describe below, but extended to allow variations in $\de \La$, and identify the difference between the geometric and thermodynamic volumes through this method in \cite{Xiao} and, in \cite{Xiao25}, providing a direct definition for the thermodynamic volume through this method.  In \cite{Xiao}, they additionally identify the geometric volume as naturally arising in the Smarr relation due to the difference between integrals of the form $\int \xi\cdot \bs \ep$ (in other words, vector volumes) in the full and background spacetime, and additionally relate it to the difference between the Komar integral associated with $\xi$ for the full and background spacetime, in a way that is somewhat similar to my approach below. Campos et al.~\cite{Campos24,Campos25,Campos26} note that the difference between the geometric/vector volume and the ``potential volume'' or thermodynamic volume (associated with the co-potential given above) is given by the integral of a gauge term. My investigation proceeds along different lines. Additionally, to my knowledge, a close investigation of why it is that the geometric volume appears in the Smarr relation associated with $\mc F$ (the conserved quantity associated with $\bt$), specifically, has not since been done.

Related to this, there is intriguing work by  Gao et al.~\cite{Gao} and Campos et al.~\cite{Campos24,Campos25,Campos26} (see also \cite{Chrusciel} for the Kerr--de Sitter case), in which a variation law (in four dimensions) based not on $\xi$ or $\bt$ but $\f{1}{\sqrt{\Xi}} \bt$ is used (where $\Xi = 1 - a^2/l^2$). In this case, in my notation, the first law takes the form
\begin{align}
    \de \left( \f{1}{\sqrt{\Xi}} \mc F\right) &= \f{1}{\sqrt{\Xi}} \om \de \mc J + \f{1}{8\pi} \f{1}{\sqrt{\Xi}} \ka \de A + \f{1}{\sqrt{\Xi}} \mc V_{\bt/\sqrt{\Xi},\mc B} \de P 
\end{align}
and the Smarr relation takes the form
\begin{align}
    \f{1}{\sqrt{\Xi}} \mc F &= \f{1}{4\pi} \f{\ka}{\sqrt{\Xi}} A + 2 \f{\om}{\sqrt{\Xi}} \mc J - 2 \mc V_{\bt/\sqrt{\Xi},\mc B} P,
\end{align}
where $V_{\bt/\sqrt{\Xi}, \mc B}$ is the vector volume (and thus the geometric volume) of the black hole associated with vector $\bt/\sqrt{\Xi}$. Campos et al.~then state that this formulation, which they dub the alternative thermodynamic theory (ATT), ``is the unique description whose thermodynamic volume precisely coincides with the geometric volume, defined as the vector volume associated to the Killing generator of the ATT temperature.'' I do not examine this in detail, though I note that the fact that the geometric and thermodynamic volumes coincide is consistent with the appearance of the geometric volume in the Smarr relation associated with $\bt$; the Smarr relation here is the one associated with the linearly rescaled vector $\bt/\sqrt{\Xi}$. (While the particular form that this vector takes is surprising, it is no contradiction that the first law can be satisfied with $\bt/\sqrt{\Xi}$ but not with $\bt$, since $\Xi$ depends on $a$ and $l$, and thus will be modified under transformation between spaces with different $a,l$ values.) (The higher-dimensional generalization appears in \cite{Campos24}, and is the charge associated with the vector $\bt /  \prod \Xi_i^{1/(D-2)}$.)

The third and fourth questions do not require the new charge construction. These can be traced back to the properties of $\bs h$ and of black hole horizons. These appear as, respectively, Sections \ref{killingpotentialvectorvolumerelationship} and \ref{AreaVolumeRelationship}.

I will make an important point about the AMD energy and angular momentum. In \cite{Magnon}, Magnon argues that the AMD definition is equivalent, up to a possible numerical factor, to the Komar integral for angular momentum, and, for the mass, to a Komar integral with a background subtraction. In repeating the calculation, whereas I reproduce her results for the angular momentum, I find that the AMD energy does not correspond to the value found by Komar subtraction method for $\xi$, although the AMD and Komar-with-background-subtraction methods agree if $\beta$ is used. 

\section{The Covariant Phase Space Formalism} \label{NoetherChargeFormalism}

In this section I review some important points of the Covariant Phase Space formalism for the definition of conserved quantities. The explanation and notation here mostly follows the readable review article \cite{Rossi} (and references therein), which itself is largely based on several papers by Wald and others, such as \cite{WaldZoupas,GaoWald,IyerWald94,Wald90,IyerWald95,Wald93}. Whereas these papers describe asymptotically flat spacetimes, the covariant phase space approach generalizes to asymptotically anti-de Sitter spacetimes, as described in, for instance, \cite{Hollands,AshtekarPawlowski,HajianSheikh-Jabbari}. My notation more closely matches that from the Wald papers (Ashtekar et al.~\cite{AshtekarPawlowski} use the Palatini framework, for instance).

\subsection{Action Principle} \label{ActionPrinciple}

Consider a theory of gravity which is diffeomorphism-invariant, in the sense that the laws of physics do not change under applying a diffeomorphism to the manifold, metric and matter fields. Assume that the action $I$ can be expressed as an integral over the manifold $M$ of a Lagrangian density $\bs L = L \bs \epsilon$, where $L$ is the Lagrangian scalar:
\begin{align}
    I &= \int_M \bs L. \label{IisintL}
\end{align}
Consider a $D$-dimensional pseudo-Riemannian manifold $M$ with metric $g_{a b}$, equipped with a Levi-Civita connection. Assume that the action depends on the metric, the Riemann tensor and a finite number of (symmetrized) derivatives thereof, and on matter fields and a finite number of their (symmetrized) covariant derivatives. The matter fields are denoted collectively by $\psi$. (The point about symmetrization is because antisymmetrized covariant derivatives can always be replaced with the Riemann tensor.) Denote the collection of all fields, including the metric, Riemann tensor and derivatives, and matter fields and derivatives thereof, by $\phi$. 

Diffeomorphism-invariant theories are ones for which the Lagrangian transforms covariantly under diffeomorphism $f: M \to M$ mapping the manifold into another manifold, so that the integral $I$ is unchanged. In technical terms, if $f^*$ is the pull-back associated with $f$, a diffeomorphism-covariant Lagrangian $\bs L$ satisfies
\begin{align}
    \bs L(f^* \phi) = f^* \bs L(\phi).
\end{align}

It is shown by Iyer and Wald \cite{IyerWald94} that $\bs L$ is diffeomorphism-covariant if the Lagrangian takes the form
\begin{align}
    \bs L = \bs L(g_{a b}, R_{a b c d}, \na_{a_1} R_{a b c d}, \ldots, \na_{(a_1} \ldots \na_{a_m)} R_{a b c d}, \psi, \na_{a_1}\psi, \ldots,  \na_{(a_1} \ldots \na_{a_l)} \psi),
\end{align}
where $\psi$ could in principle include other (covariant) indices, contractions and so on in some way, and $l$ and $m$ are finite. $\bs L$, being a differential form, behaves as a covariant tensor under diffeomorphisms, as do the $\phi$ fields it depends on. For an example, in pure Einstein--Hilbert gravity, $\bs L = (R-2\La) \bs \ep$ where $R$ is the Ricci scalar. $R$ depends on $R_{abcd}$ and the metric, and $\bs \ep$ depends on the metric only; there are no matter fields. 

A variation $\de \phi$ of linear order of the fields results in a deviation of $\bs L$ which can be written as
\begin{align}
    \de \bs L = \bs E \de \phi + d \bs \Th(\phi, \de \phi). \label{deltaL}
\end{align}
$\bs E$, which is treated as a $D$-form but may involve other indices to contract with possible indices in $\phi$, depends on $\phi$ but not on $\de \phi$. In particular, we can break $\bs E \de \phi$ into terms related to $g$ and terms related to the fields $\psi$,
\begin{align}
    \bs E \de \phi = \bs E_g^{a b} \de g_{a b} + \bs E_\psi \de \psi,
\end{align}
where $\bs E^{a b}_g = E^{a b}_g \bs \epsilon$ (an object with two symmetric contravariant indices and $D$ antisymmetric covariant indices). The assumption of linear order in $\de \phi$ usually means that $\de \phi$ is taken to be very small. 

$\bs \Th$, called the symplectic potential form, depends on the variations to linear order. The $d \bs \Th$ term in \eqref{deltaL} appears due to terms of the form $\na_a \de \phi$. The $(D-1)$-form $\bs \Th$ is defined up to the addition of a closed $(D-2)$-form, since the exterior derivative of a closed form is zero.

Under a variation of compact support but otherwise arbitrary, for which $\de \phi \to 0$ on the boundary (if it exists) or vanishes asymptotically sufficiently quickly so that $\bs \Th \to 0$ at the boundary or asymptotically, $\de I$ is
\begin{align}
    \de I &= \int_M ( \bs E \de \phi + d \bs \Th) \nn
    &= \int_M \bs E \de \phi + \int_{\partial M} \bs \Th \nn
    &= \int_M \bs E \de \phi.
\end{align}
The condition that the action remain stationary under perturbations $\de \phi$ which are arbitrary and with compact support within the bulk of the spacetime implies the equations of motion, $\bs E = 0$.

The Einstein--Hilbert Lagrangian density is
\begin{align}
    \bs L_{EH} &= \f1{16\pi} (R-2\Lambda) \bs \ep, \label{LEH}
\end{align}
which leads to variation (see Appendix \ref{EHterms})
\begin{align}
    \de \bs L_{EH} &= -\f1{16\pi} (G^{ab} + \La g^{ab}) \de g_{ab} \bs \ep + d \bs \Th^{EH} (g_{ab}, \de g_{ab}),
\end{align}
where 
\begin{align}
    \bs \Th^{EH} &= \f1{16\pi} v^{EH} \cdot \bs \ep \nn
    v^{EH}_a &= \na^b \de g_{ab} - g^{bc} \na_a \de g_{bc}. \label{ThetaEH}
\end{align}

The Lagrangian density associated with the Maxwell electromagnetic (EM) field (see Appendix \ref{MaxwellTerms}) is 
\begin{align}
    \bs L_{EM} &= -\f{1}{16\pi} (g^{ac} g^{bd} \bs F_{ab} \bs F_{cd}) \bs \ep, \label{LEM}
\end{align}
where $\bs F = d \bs A$ is the EM field with vector potential $\bs A$. Its variation is 
\begin{align}
    \de \bs L_{EM} &= \f12 T^{ab}_{EM} \de g_{ab} \bs \ep + \f1{4\pi} \na_c \bs F^{cd} \de \bs A_d \bs \ep + d \bs \Th^{EM},
\end{align}
where $T^{EM}_{ab}$ is the electromagnetic stress--energy tensor,
\begin{align}
    T_{ab}^{EM} &= \f1{4\pi} \left( \bs F_{ac} {\bs F_b}^c - \f14 g_{ab} \bs F_{de} \bs F^{de} \right)
\end{align}
and, focusing on four dimensions,
\begin{align}
    \bs \Th^{EM}_{abc} &= - \f1{4\pi} \bs F^{de} \de \bs A_e \bs \ep_{dabc}.  \label{ThetaEM}
\end{align}

\subsection{Alternate Forms of Lagrangian and Symplectic Potential} \label{alternateLagrangian}

The conventions and explanations of this subsection again follow \cite{Rossi}. The equations of motion associated with a Lagrangian form $\bs L$ are preserved under a transformation $\bs L \to \bs L + d \bs \mu$ for some form $\bs \mu$. This transformation may be warranted or necessary to have a well-posed variational problem, to ensure that the $\partial M$ contribution to the variation in action has compact support. 

In a diffeomorphism-invariant theory of gravity, Iyer and Wald showed that $\bs \Th$ can always be chosen to have a diffeomorphism-invariant form, which can be written in the form
\begin{align}
    \bs \Th &= 2 \bs E^{b c d}_R \na_d \de g_{bc} + \bs \Th', \label{Thetaform}
\end{align}
where $\bs \Th'$ is of the general form (somewhat abbreviating the notation in \cite{Rossi})
\begin{align}
    \bs \Th' = \bs S(\phi) \de \phi,
\end{align}
where $\de \phi$ could in principle include the variation of derivatives of the metric or fields, but where there are no terms of the form $\na \de \phi$ (with the covariant derivative outside the variation) in $\bs \Th'$. 
\begin{align}
    (\bs E^{b c d}_{R})_{b_2 \ldots b_D} &= E^{a b c d}_R \bs \ep_{a b_2 \ldots b_D}
\end{align}
where $E_R^{abcd}$ is the equation of motion term that results from treating $R_{abcd}$ as an independent term to be varied in the Lagrangian,
\begin{align}
    E^{a b c d}_R &= \frac{\pa L}{\pa R_{ab c d}} - \na_{a_1}\f{\pa L}{\pa \na_{a_1} R_{abcd}} + \ldots + (-1)^m \na_{(a_1} \ldots \na_{a_m)} \f{\pa L}{\pa \na_{(a_1} \ldots \na_{a_m)} R_{abcd}}.
\end{align}

Under $\bs L \to \bs L + d \bs \mu$, $\bs \Th \to \bs \Th + \de \bs \mu$. $\bs \Th$ can also always be shifted by a closed form, $\bs \Th \to \bs \Th + d \bs Y$. It is important to emphasize this freedom in defining $\bs \Th$.

\subsection{Noether Current and Charge Forms} \label{NoetherCurrentCharge}

Let $\chi$ be a vector field (in general not Killing). The variation in the fields (including the metric tensor) associated with a diffeomorphism generated by a vector $\chi$ is the Lie derivative of $\phi$ with respect to $\chi$, $\lie_\chi \phi$. Define the Noether current $(D-1)$-form $\bs J_\chi$ associated with $\chi$ to be
\begin{align}
    \bs J_\chi &= \bs \Th(\phi,\lie_\chi \phi) - \chi \cdot \bs L, \label{Jchi}
\end{align}
where $\bs \Th(\phi, \lie_\chi \phi)$ is the value of the $\bs \Th$ form associated with this specific variation in the fields. 

The attentive reader will note that the argument $\de \phi$ in $\bs \Th(\phi,\de \phi)$ was assumed to be infinitesimal, whereas, for a finite vector field $\chi$, $\lie_\chi\phi$ is finite. The idea is that an infinitesimal diffeomorphism is generated by $\chi$, so under $\de \phi = \lie_\chi \phi \de \la$ (for infinitesimal $\de \la$), $\de \bs L = \lie_\chi \bs L \de \la$, and so
\begin{align}
    \lie_\chi \bs L \de \la = \bs E \lie_\chi \phi \de \la + \bs \Th(\phi,\lie_\chi \phi \de \la).
\end{align}
Then define the ``finite argument'' version of $\bs \Th$ to be
\begin{align}
    \bs \Th(\phi,\lie_\chi \phi) &= \f{\bs \Th(\phi,\lie_\chi \phi \de \la)}{\de \la}
\end{align}
for sufficiently small $\de \la$, which is allowed because $\bs \Th(\phi,\de \phi)$ is linear for sufficiently small $\de \phi$. 

Similar arguments will apply, going forward, for any other terms where a finite argument is placed inside some variation---that is, if the fields $\phi$ depend on some parameter $\la$ in some way, as $\phi = \phi^{(\la)}$, where $\phi^{(0)}$ are the unvaried fields, then we can write
\begin{align}
    \de \phi &= \f{d \phi^{(\la)}}{d \la} \de \la,
\end{align}
and so have
\begin{align}
    \bs \Th(\phi, \de \phi) &= \bs \Th\left(\phi, \f{d \phi^{(\la)}}{d \la} \de \la\right)  \nn 
    &\equiv \bs \Th\left( \phi, \f{d \phi^{(\la)}}{d \la} \right) \de \la,
\end{align}
where we use linearity on the last line to define $\bs \Th(\phi, d\phi^{(\la)}/d\la$), and evaluations are taken at $\la = 0$. 

Taking the exterior derivative,
\begin{align}
    d \bs J_\chi &= d \bs \Th(\phi,\lie_\chi\phi) - d (\chi \cdot \bs L) \nn 
    &= d \bs \Th(\phi,\lie_\chi\phi) - \lie_\chi \bs L \nn
    &= -\bs E \lie_\chi \phi.
\end{align}
Cartan's identity \eqref{Cartan} and the fact that $d \bs L = 0$ because $\bs L$ is fully $D$-dimensional and there are no nonvanishing $(D+1)$-forms on a $D$-dimensional manifold were used on the second line. For the last line, we used \eqref{deltaL} where the variation is $\lie_\chi \phi$. 

Thus if the equations of motion are satisfied, $d \bs J_\chi = 0$, and so (at least locally) $\bs J_\chi$ can be taken as the exterior derivative of some $(D-2)$-form, say $\bs Q_\chi$. $\bs Q_\chi$ is called the Noether charge $(D-2)$-form associated with the Noether current form $\bs J_\chi$. 

Define the symplectic current $(D-1)$-form $\bs \om(\phi,\de_1 \phi, \de_2\phi)$ by
\begin{align}
    \bs \om(\phi,\de_1 \phi, \de_2\phi) &= \de_1 \bs \Th(\phi,\de_2 \phi) - \de_2 \bs \Th(\phi,\de_1\phi),
\end{align}
where $\de_1\phi$ and $\de_2\phi$ are two independent variations of the fields. $\bs \om$ is linear in the variations of the fields, so that if one of the variations is zero, $\bs \om = 0$. 

Now assume that the equations of motion are satisfied, $\bs E = 0$. Consider a variation $\de \phi$ for which $\de \chi^a = 0$ (the vector $\chi$ in its contravariant form does not change). Then we have
\begin{align}
    \de \bs J_\chi &= \de \bs \Th (\phi, \lie_\chi \phi) - \de (\chi \cdot \bs L) \nn
    &= \de \bs \Th (\phi, \lie_\chi \phi) - \chi \cdot \de \bs L \nonumber \\
    &= \de \bs \Th (\phi, \lie_\chi \phi) - \chi \cdot d \bs \Th (\phi, \de \phi) \nn
    &= \de \bs \Th(\phi, \lie_\chi\phi) - \lie_\chi \bs \Th(\phi,\de \phi) + d (\chi \cdot \bs \Th(\phi,\de \phi)) \nn
    &= \bs \om (\phi, \de \phi, \lie_\chi \phi) + d (\chi \cdot \bs \Th(\phi,\de \phi)). \label{deltaJ}
\end{align}
We applied $\bs E = 0$ between the second and third lines. Keeping all terms if $\bs E \neq 0$ and $\de \chi^a \neq 0$, then 
\begin{align}
    \de \bs J_\chi &= \bs \om(\phi,\de \phi, \lie_\chi \phi) - d (\chi \cdot \bs \Th(\phi,\de \phi)) - \chi \cdot \bs E \de \phi - \de \chi \cdot \bs L,
\end{align}
where $(\de \chi \cdot \bs L)_{a_1 \ldots a_{D-1}} = \de \chi^a \bs L_{a a_1 \ldots a_{D-1}}$. We will now set $\bs E = 0, \de \chi^a = 0$ again. Using $\bs J_\chi = d \bs Q_\chi$ (valid if $\bs E = 0$) and since $d$ and the variation $\delta$ commute, we have
\begin{align}
    \bs \om(\phi, \de \phi, \lie_\chi \phi) &= d ( \delta \bs Q_\chi - \chi \cdot \bs \Th(\phi,\de\phi)). \label{omegaintermsofQ}
\end{align}
If $\chi$ is a symmetry of the fields so that $\lie_\chi \phi = 0$, then this implies
\begin{align}
    d (\de \bs Q_\chi - \chi \cdot \bs \Th (\phi, \de \phi)) &= 0. \label{ddeltaQmchidotTheta}
\end{align}
Integrating this over a closed hypersurface $\Sigma$ with boundary $\partial \Sigma$ and using the Gauss--Stokes theorem gives
\begin{align}
    \oint_{\pa \Si} \left(\de \bs Q_\chi - \chi \cdot \bs \Th(\phi, \de \phi)\right) &= 0. \label{Wald77}
\end{align}
Equation \eqref{Wald77} is valid provided that $\phi$ is a solution to the equations of motion ($\bs E(\phi) = 0$), that $\chi$ is a symmetry of the fields $(\lie_\chi \phi = 0$), that $\de \chi^a = 0$, and that $\de \phi$ is a variation which satisfies the linearized equations of motion $\de \bs E = 0$. It is not necessarily required that $\lie_\chi\de\phi = 0$. 

On a ``single'' spatial section, we have 
\begin{align}
    \oint_S (\de \bs Q_\chi - \chi \cdot \bs \Th(\phi, \de \phi)),
\end{align}
which is independent of the choice of surface $S$ (up to orientation). This follows from \eqref{Wald77}. For example, $\partial \Sigma$ could be composed of two surfaces $S_1$ and $S_2$. Let $S_2$ be ``further outside'' (toward spatial infinity) than $S_1$, and let $S_2$ and $S_1$ both have orientation so that they are pointing away from $\Si$. Then $\pa \Si$ is the union of $S_2$ (with positive orientation) and $S_1$ (with negative orientation) so that 
\begin{align}
    \oint_{\partial \Sigma}\left( \de \bs Q_\chi - \chi \cdot \bs \Th(\phi,\de\phi)\right) = \oint_{S_2} \left( \de \bs Q_\chi - \chi \cdot \bs \Th(\phi,\de\phi)\right) - \oint_{S_1} \left( \de \bs Q_\chi - \chi \cdot \bs \Th(\phi,\de\phi)\right) = 0.
\end{align}
Using notation from Barnich and Comp\`ere \cite{BarnichCompere,Compere} (as well as previous papers, such as \cite{BarnichBrandt}, and subsequent ones such as \cite{HajianSheikh-Jabbari}), 
let $\bs k_\chi[\de \phi;\phi]$ be
\begin{align}
    \bs k_\chi [\de \phi;\phi] &\equiv \de \bs Q_\chi - \chi \cdot \bs \Th(\phi,\de \phi). \label{kchi}
\end{align}
(In fact, BC include a generalization to allow $\chi^a$ to have varying components, which I will revisit in Section \ref{BCsection}.) When $\bs Q_\chi$ and $\bs \Th$ take on their values for Einstein--Hilbert gravity specifically, I refer to $\bs k_\chi$ as $\bs k_\chi^{EH}[\de g;g]$ (the only field here is $\phi = g_{ab}$, and I omit the covariant indices for compactness). If $\chi$ is a symmetry of the fields $(\lie_\chi \phi = 0$), \eqref{ddeltaQmchidotTheta} implies
\begin{align}
    d \bs k_\chi[\de \phi;\phi] &= 0. \label{dkiszero}
\end{align}

Because of the ambiguity in the definition of $\bs \Th$, there is a related ambiguity under the definition of $\bs J_\chi$ and $\bs Q_\chi$. Under $\bs L \to \bs L + d \bs \mu$, $\bs \Th$ has the ambiguity $\bs \Th \to \bs \Th + \de \bs \mu + d \bs Y(\phi, \de \phi)$, where the $d \bs Y$ is an additional ambiguity since $\bs \Th$ is defined only up to a closed form. (I will, for our purposes, assume that any closed forms are also exact.) Under those transformations of $\bs L$ and $\bs \Th$, $\bs J_\chi$ transforms as
\begin{align}
    \bs J_\chi \to \bs J_\chi + d (\chi \cd \bs \mu) + d \bs Y(\phi, \lie_\chi \phi).
\end{align}
This change plus the ambiguity that an exact form can be added to $\bs Q_\chi$ means that $\bs Q_\chi$ has the ambiguity
\begin{align}
    \bs Q_\chi \to \bs Q_\chi + \chi \cd \bs \mu + \bs Y (\phi, \lie_\chi \phi) + d \bs Z.
\end{align}
This ambiguity means that one cannot make a single definitive choice of the Noether charge or current or the symplectic potential form. However, Iyer and Wald \cite{IyerWald94} proved that they can always be written in a particular form. The argument proceeds as follows. Consider a general vector field $\chi$ (not necessarily a symmetry of the fields). Then $\lie_\chi g_{a b} = \na_a \chi_b + \na_b \chi_a$. Given the choice \eqref{Thetaform}, $\bs J_\chi$ can be written
\begin{align}
    \bs J_\chi &= 2 \bs E_R^{b c d}\na_d (\na_b \chi_c + \na_c \chi_b) + \bs \Th'(\phi,\lie_\chi \phi) - \chi \cdot \bs L.
\end{align}
Then an algorithm from \cite{IyerWald94} can be applied to find $\bs Q_\chi$ from $\bs J_\chi$, which yields the result
\begin{align}
    \bs Q_\chi = \bs W_c(\phi) \chi^c + \bs X^{cd} \na_{[c} \chi_{d]},
\end{align}
where each component $\bs W_c$ is a covariant $(D-2)$-form constructed out of the fields $\phi$ and their derivatives but does not otherwise depend on $\chi$, and 
\begin{align}
    (\bs X^{c d})_{c_3 \ldots c_n} &= - E^{abcd}_R \bs \ep_{ab c_{3} \cds c_n}. \label{Xcd}
\end{align}
Up to the additional ambiguity in choosing $\bs Q_\chi$, this means $\bs Q_\chi$ can be written in the general form
\begin{align}
    \bs Q_\chi &= \bs W_c(\phi) \chi^c + \bs X^{cd}(\phi) \na_{[c}\chi_{d]} + \bs Y(\phi,\lie_\chi \phi) + d \bs Z(\phi, \chi),
\end{align}
with $\bs Z$ linear in $\chi$ and $\bs Y$ linear in $\lie_\chi$. $\bs Y$ and $\bs Z$ can be chosen to be zero. Note that $\bs X^{c d}$ need not have the form \eqref{Xcd}, though it can be chosen to have that form.

\subsection{Hamiltonian} \label{HamiltonianNoetherSection}
There is a prescription for defining the Hamiltonian for the system $H_\chi$ associated with vector $\chi$, if the Hamiltonian exists, according to \cite{IyerWald94,IyerWald95,WaldZoupas} (and related papers). The explanation here most closely follows \cite{IyerWald94}. 

Letting $\Si$ be a Cauchy surface, define the pre-symplectic form $\Om(\phi,\de_1 \phi, \de_2\phi)$ by
\begin{align}
    \Om(\phi,\de_1 \phi,\de_2 \phi) &= \int_\Si \bs \om(\phi,\de_1 \phi, \de_2\phi).
\end{align}
Let the field configuration space of $\phi$ be $\mathscr{F}$ (that is, the space of possible values of $\phi$). Like $\bs \om$, $\Om$ is antisymmetric and linear in its arguments $\de_1 \phi, \de_2 \phi$, and thus at each point $\phi$ acts as a 2-form, treating $\de_1 \phi, \de_2\phi$ as infinitesimal vectors. The details are technical (see \cite{IyerWald94,IyerWald95,WaldZoupas,Rossi}), but in short $\mathscr{F}$ can be used to define the phase space of the theory, $\G$, and thus the symplectic manifold, on which $\Om$ acts as a symplectic 2-form, which can then be used to define the Hamiltonian. 

On a symplectic manifold, with generalized coordinates $x^A$ (with indices denoted by capital letters, where $x^A$ includes both generalized space coordinates and generalized momenta), given a \emph{phase space} time evolution vector field $T^A$ and symplectic form $\Om_{AB}$, the Hamiltonian $H$ satisfies
\begin{align}
    \f{\pa H}{\pa x^A} &= \Om_{AB} T^B. \label{phasespaceHamiltonian}
\end{align}

The equivalent of  \eqref{phasespaceHamiltonian} for the phase space $\G$ for a field theory of gravity is the following. Let $\chi^a$ (not necessarily Killing) be a ``time-evolution vector field'' for the manifold, which, despite the name, need not necessarily be timelike (for example, it could be the azimuthal symmetry Killing vector $\eta^a$). Then associated with this vector is a Hamiltonian $H_\chi$, for which, \emph{if it exists}, given a vacuum solution $\phi$ and a variation of the fields $\de \phi$ (which does not necessarily satisfy the equations of motion),
\begin{align}
    \de H_\chi &= \Om(\phi,\de \phi, \lie_\chi \phi).
\end{align}
$\lie_\chi \phi$ plays the role of the phase space time-evolution vector $T^A$. 

In its definition, $\Om(\phi,\de_1 \phi,\de_2 \phi)$ depends explicitly on the surface $\Si$, but if $\phi$ satisfies the equations of motion and $\de_1 \phi, \de_2 \phi$ satisfy the linearized equations of motion, the value of $\Om(\phi,\de_1 \phi, \de_2\phi)$ is independent of Cauchy surface $\Si$ for ``typical'' \cite{IyerWald94} fall-off conditions on the dynamical fields. 

Using \eqref{omegaintermsofQ} we can then write
\begin{align}
    \de H_\chi &= \de \int_\Si d (\de \bs Q_\chi - \chi \cdot \bs \Th) \nn 
    &= \oint_{\pa \Si} \left( \de\bs Q_\chi - \chi \cdot \bs \Th(\phi,\de \phi)\right).
\end{align}

Now if $\Si$ has no interior boundaries (so, excluding the case of black holes), the integral over $\pa \Si$ consists only of a term ``at infinity,'' which we denote by $S_\infty$. In this case we can write
\begin{align}
    \de H_\chi &= \oint_{S_\infty} \left( \de\bs Q_\chi - \chi \cdot \bs \Th(\phi,\de \phi)\right). \label{deHchi0}
\end{align}

At this point \cite{IyerWald94} state that if it is possible to find some $(D-1)$-form $\bs B$ (not necessarily diffeomorphism-covariant) such that
\begin{align}
    \de \oint_{S_\infty} \chi \cdot \bs B &= \oint_{S_\infty} \chi \cdot \bs \Th(\phi,\de \phi),
\end{align}
then $\de H_\chi$ can be rewritten
\begin{align}
    \de H_\chi &= \de \oint_{S_\infty} (\bs Q_\chi - \chi \cdot \bs B),  \label{deHchi1}
\end{align}
or
\begin{align}
    H_\chi &= \oint_{S_\infty} (\bs Q_\chi - \chi \cdot \bs B). \label{Hchi}
\end{align}
Iyer and Wald then define the Hamiltonian associated with a spacetime to be given by \eqref{Hchi}, \emph{even if there is an interior boundary} (so, including a black hole spacetime). 

I want to emphasize here that it is specifically at infinity that $\bs B$ is defined. Conceptually, it is possible that such a $\bs B$ will exist at infinity but not ``in the bulk'' due to the way in which fields behave/fall off asymptotically. To go ``into the bulk'' is something that appears in Barnich and Comp\`ere \cite{BarnichCompere} and will appear in Section \ref{BCsection}.

For my purposes, I will expand the argument slightly and allow \eqref{deHchi0} to define the variation in the Hamiltonian, at infinity, again \emph{even if there is an interior boundary}. Using the definition of $\bs k_\chi[\de \phi;\phi]$ from \eqref{kchi}, we then have
\begin{align}
    \de H_\chi &= \oint_{S_\infty} \bs k_\chi[\de \phi;\phi]. \label{deHchiInfinity}
\end{align}

So far we have not required that $\chi$ is a Killing vector and symmetry of the $\lie_\chi \phi = 0$. If we do require $\lie_\chi \phi = 0$, then $\bs \om(\phi,\de_1 \phi, \lie_\chi \phi) = 0$, so that the integral of $\de \bs Q_\chi - \chi \cdot \bs \Th(\phi,\de \phi)$ is surface-independent, provided that the equations of motion of the theory are satisfied $\bs E = 0$. In this case we can write
\begin{align}
    \de H_\chi &= \oint_S (\de \bs Q_\chi - \chi\cdot \bs \Th(\phi,\de \phi)) \nn 
    &= \oint_S \bs k_\chi[\de \phi;\phi] \label{deHchi}
\end{align}
for any $(D-2)$-surface $S$, which can be connected to $S_\infty$ by a hypersurface $\Si$ with no holes.

Iyer and Wald apply \eqref{deHchi1} to asymptotically flat solutions but the development can still be used for spacetimes with other asymptotic properties. The expression is used for the asymptotically anti-de Sitter case in \cite{Hollands} and later \cite{HajianSheikh-Jabbari} and many other subsequent papers. Here and elsewhere when I refer to $S_\infty$ as being the surface at infinity I mean that a surface is taken at large radius and then moved toward spatial infinity in some sort of limiting process. Specifically when considering the GKAdS spacetimes introduced in Chapter 3, I will generally let $S_\infty$ be a surface of either constant spheroidal radius $r$ or constant spherical radius $y$, then allow $r$ or $y$ to approach $\infty$. I will also take constant time parameter (usually $t$ or $\tau$).

In the case of a stationary, azimuthally symmetric spacetime, as in Chapter \ref{GKAdSChapter}, let $\xi^a$ be the asymptotically timelike Killing vector (the asymptotically static Killing vector, if it exists) and let $\eta_i$ be Killing vectors corresponding to azimuthal symmetry, normalized so that the integral curves of $\eta_i$ close after $2\pi$ parameter distance. Let there be a suitable normalization of $\xi^a$, which may need to be applied on a case-by-case basis. I will call these the canonically normalized stationarity and azimuthal symmetry Killing vectors. Then define the canonical energy $\mc E$ (also called the \emph{mass}) to be the Hamiltonian associated with $\xi$ and $\mc J_i$ to be the negative of the Hamiltonian associated with $\eta_i$, if they exist, so that
\begin{align}
    \de \mc E &= \de H_\xi = \oint_{S_\infty} \bs k_\xi [\de \phi;\phi] = \oint_{S_\infty} (\de \bs Q_\xi - \xi \cdot \bs \Th(\phi,\de\phi)) \label{deE} \\ 
    \de \mc J_i &= - \de H_{\eta_i} = -\oint_{S_\infty} \bs k_{\eta_i} [\de \phi;\phi] = -\oint_{S_\infty} (\de \bs Q_{\eta_i} - \eta_i \cdot \bs \Th(\phi,\de \phi)) . \label{deJi}
\end{align}
In four dimensions there is only one azimuthal symmetry Killing vector, identified as $\eta$, and in this case the angular momentum is identified as $\mc J$,
\begin{align}
    \de \mc J &= - \de H_\eta = -\oint_{S_\infty} \bs k_\eta[\de \phi;\phi] = - \oint_{S_\infty} (\de \bs Q_\eta - \eta \cdot \bs \Th (\phi,\de\phi)).
\end{align}

If there exists some $\bs B$ such that $\de \oint_{S_\infty} \xi \cdot \bs B = \oint_{S_\infty} \xi \cdot \bs \Th(\phi,\de \phi)$, then
\begin{align}
    \mc E &= H_\xi = \oint_{S_\infty} (\bs Q_\xi - \xi \cdot \bs B). \label{mcEHxi}
\end{align}
In the case of an asymptotically AdS spacetime, it may be that $\bs Q_\xi$ and $\xi \cdot \bs B$ individually diverge as $r \to \infty$ (or $y \to \infty)$, but that their difference remains finite. Similarly for the angular momenta, for some (possibly different) $\bs B$ for which $\de \oint_{S_\infty} \eta_i \cdot \bs B = \oint_{S_\infty} \eta_i \cdot \bs \Th$,
\begin{align}
    \mc J_i&= - H_{\eta_i} \nn
    &= - \oint_{S_\infty} (\bs Q_{\eta_i} - \eta_i \cdot \bs B).
\end{align}
Because the $(D-2)$-surface can always be chosen to have $\eta_i$ tangent to the surface (as is the case for the GKAdS solutions where $S_\infty$ is constant $r$ or $y$ and constant $t$ or $\tau$), $(\eta_i \cdot \bs B)|_{S_\infty} = 0$ for any $(D-2)$-form $\bs B$, where by $|_{S_\infty}$ I mean the restriction to $S_\infty$. As a result, we can write
\begin{align}
    \mc J_i = - \oint_{S_\infty} \bs Q_{\eta_i},
\end{align}
with, in the case of only a single azimuthal symmetry vector, 
\begin{align}
    \mc J = - H_\eta = - \oint_{S_\infty} \bs Q_\eta.
\end{align}
The reason for the difference in sign ($\mc E = + H_\xi, \mc J_i = - H_{\eta_i}$) is a consequence of the Lorentzian signature of the metric.

(I will explain further the argument for the vanishing of the $\oint_{S_\infty} \eta_i \cdot \bs B$ term as follows. Consider four dimensions and the vanishing of the $\oint_{S_\infty} \eta \cdot \bs B$ term, and consider a situation where a chart can be found with coordinates $(t,r,\tht,\varphi)$, where $t$ and $\varphi$ are coordinates such that $\xi^a = (\partial_t)^a$ and $\eta^a = (\partial_\varphi)^a$. $r$ will be a coordinate for which the $S_\infty$ location corresponds to $r \to \infty$. $\tht$ is an angular coordinate such that the poles $\tht = 0$ and $\tht = \pi$ correspond to the vanishing of the norm of $\eta^a$. Then $S_\infty$ can be chosen to be the two-surface with constant $t$ and constant $r \to \infty$. Since $\eta^a = \delta^a_\varphi$, $\eta \cdot \bs B$ will have no terms involving wedge products of $d \vp$. Thus the most general form of $\eta \cdot \bs B$ in these coordinates is
\begin{align}
    \eta \cdot \bs B = A dt \wedge dr  + B dt \wedge d\tht + C dr \wedge d \tht.
\end{align}
When restricted to $S_\infty$, this will be zero, since the restriction of $dt$ and $dr$ onto the $t, r$ constant surface will be zero. Thus $\oint_{S_\infty} \eta \cdot \bs B = 0$ for any $\bs B$.)

I want to emphasize again that this derivation assumes that $\de \chi^a = 0$---that the components of the Killing vector are unchanged in the variation of the spacetime. 

The Hamiltonian can also be found by beginning with the action of general relativity in Hamiltonian form, which involves decomposing the spacetime into $(D-1)$ spatial and 1 time dimensions. As pointed out by \cite{BarnichCompere}, general results from, e.g.,~\cite{HenneauxTeitelboim92} on how Hamiltonian and Lagrangian charges relate suggest that the Hamiltonian charges computed in both methods will coincide.

If the spacetime is a solution to the vacuum field equations, then $d \bs k_\chi[\de g;g] = 0$ by \eqref{ddeltaQmchidotTheta} for a Killing $\chi$, so that we also have that $\de \mc E$ and $\de \mc J_i$ are defined the same way as \eqref{deE} and \eqref{deJi} over \emph{any} closed surface $S$ which can be connected to $S_\infty$ via a hypersurface $\Si$. This gives us expressions for $\de H_\chi$ on any $(D-2)$-surface $S$, but does not necessarily give us $H_\chi$ itself, because we might not necessarily be able to find a $\bs B$ such that $\de \oint_S \chi \cdot \bs B = \oint_S \chi \cdot \bs \Th(\phi,\de \phi)$ for an arbitrary $(D-2)$-surface $S$ even if we can find one on $S_\infty$. 

\subsection{Noether Current and Charge for Einstein Gravity} \label{EinsteinGravityNoetherCharges}

Using \eqref{ThetaEH}, we can calculate the Noether current and charge associated with the Einstein--Hilbert Lagrangian.  It turns out that the Noether charge $\bs Q_\chi$ can be written in a way that is not (directly) dependent on $\La$ and will be related to the Komar integrals. 

I will follow the calculation as shown in Rossi \cite{Rossi}, slightly adapting it to include the nonzero $\La$ term. I will work in four dimensions but the generalization to higher dimensions is obvious.

First, we calculate $\bs \Th^{EH}(\phi,\lie_\chi \phi)$ (substituting in that the variation in $\phi$ is $\lie_\chi \phi$ specifically), on the way to finding an expression for $\bs J_\chi$ and thus $\bs Q_\chi$. Using \eqref{ThetaEH},
\begin{align}
    \bs \Th^{EH} (\phi, \lie_\chi \phi) &= \f1{16\pi} v \cdot \bs \ep \nn
    v_a &= \na^b (\na_a \chi_b + \na_b \chi_a) - g^{b c} \na_a (\na_b \chi_c + \na_c \chi_b)
\end{align}
$v^d$ can be rearranged to
\begin{align}
    v^d &= g^{d e} g^{fh} \left(\na_f (\na_e \chi_h + \na_h \chi_e) - \na_e (\na_f \chi_h + \na_h \chi_f) \right) \nn
    &= g^{d e} g^{f h} (\na_f \na_e \chi_h - \na_e \na_f \chi_h) + g^{de} g^{f h} (\na_f \na_h \chi_e - \na_e \na_h \chi_f).
\end{align}
Using the definition of the Riemann curvature,
\begin{align}
    \na_f \na_e \chi_h - \na_e \na_f \chi_h &= {R^k}_{hef} \chi_k = R_{khef} \chi^k \nn
    \na_f \na_h \chi_e &= R_{kehf} \chi^k + \na_h \na_f \chi_e \nn
    \na_e \na_h \chi_f &= R_{kfhe} \chi^k + \na_h \na_e \chi_f.
\end{align}
Consequently,
\begin{align}
    v^d &= g^{de} g^{fh} \left( R_{khef} \chi^k + \na_h \na_f \chi_e + R_{kehf} \chi^k - \na_h \na_e \chi_f - R_{kfhe} \chi^k  \right) \nn
    &=2 R_k^d \chi^k + \na_h (\na^h \chi^d - \na^d \chi^h).
\end{align}
Thus we have
\begin{align}
    \bs \Th_{a b c}^{EH}(g_{ab},\lie_\chi g_{ab}) &= \f1{8\pi} \bs \ep_{d a b c} \left(  R^d_e \chi^e + \na_e  \na^{[e} \chi^{d]}\right).
\end{align}
We have not yet applied the equations of motion. The current $\bs J^{EH}_\chi$ is
\begin{align}
    \bs J^{EH}_\chi &= \bs \Th^{EH}(g_{ab}, \lie_\chi g_{ab}) - \chi \cdot \bs L_{EH}.
\end{align}
Using $L_{EH} = (16\pi)^{-1} (R - 2 \La)$, $\bs J_\chi$ is then
\begin{align}
    (\bs J^{EH}_{\chi})_{a b c} &= \bs \Th_{a b c} - \frac{1}{16\pi}\chi^d \bs \epsilon_{d a b c} (R - 2 \La) \nn 
    &= \frac{1}{8 \pi} \bs \epsilon_{d a b c} \left( \na_e \na^{[e}\chi^{d]} + \left(R^d_e - \frac{1}{2} R \de^d_e + \La \de^d_e\right) \chi^e\right).
\end{align}
The term $R^d_e - \f12 R \de^d_e + \La \de^d_e = 0$ when the equations of motion for pure Einstein gravity (with nonzero $\Lambda$) are applied, so we recover 
\begin{align}
    (\bs J^{EH}_{\chi})_{a b c} &= \f1{8\pi} \bs \ep_{d a b c} \na_e \na^{[e} \chi^{d]}
\end{align}
as in the $\La = 0$ case, for a general diffeomorphism along a vector $\chi$ (not necessarily Killing), if Einstein's equations are satisfied.

It then follows that
\begin{align}
    \bs J^{EH}_\chi = d \bs Q^{EH}_\chi
\end{align}
where
\begin{align}
    (\bs Q^{EH}_\chi)_{a b} &= - \f{1}{16\pi} \bs \ep_{a b c d} \na^c \chi^d. \label{QEH}
\end{align}
The generalization to higher dimensions is obvious; the result is
\begin{align}
    (\bs Q^{EH}_\chi)_{a_1 \ldots a_{D-2}} &= -\f{1}{16\pi} \bs \ep_{a_1 \ldots a_{D-2} c d} \na^c \chi^d.
\end{align}

Note again that there is ambiguity in $\bs Q^{EH}_\chi$ resulting from the ambiguity in $\bs \Th^{EH}$. However, this is one possible choice that can be made, in pure Einstein gravity. This form appears in \cite{Hollands} for Einstein gravity with negative $\La$. 

For a Killing vector $\chi$, we recognize $\bs Q^{EH}_\chi$ as the negative Komar form \eqref{dKxidefinition2},
\begin{align}
    \bs Q^{EH}_\chi = - \bs K^K_\chi,
\end{align}
using the form of $\bs Q^{EH}_\chi$ given by \eqref{QEH}. The reason for having a separate definition of $\bs K^K_\chi$ is, in addition to connecting to the notation of Comp\`ere \cite{Compere}, to make clear that $\bs K^K_\chi$ is unambiguously defined and is not necessarily a Noether charge, whereas $\bs Q_\chi^{EH}$, the Noether charge, could in principle have a different form from \eqref{QEH}, even if \eqref{QEH} is the choice we will make.

In the case of Einstein--Maxwell gravity, the Noether charge has an additional contribution $\bs Q^{EM}_\chi$. In four dimensions \cite{GaoWald},
\begin{align}
    \bs Q_\chi &= \bs Q^{EH}_\chi + \bs Q^{EM}_\chi \nn 
    (\bs Q^{EM}_\chi)_{ab} &= -\f{1}{8\pi} \bs \ep_{abcd} \bs F^{cd} \bs A_e \chi^e. \label{NoetherChargeEM}
\end{align}
The expression for higher dimensions is analogous, but I will only discuss Einstein--Maxwell gravity in four dimensions. The electric charge $\mc Q$ is given by \cite{Poisson}
\begin{align}
    \mc Q &= \f1{8\pi} \oint_{S_\infty} \bs F^{ab} d S_{ab} \nn 
    &= \f1{8\pi} \oint_{S_\infty} * \bs F. \label{mcQdef}
\end{align}
In electrovacuum, $d * \bs F = 0$ and so $\mc Q$ can be calculated on any $(D-2)$-surface $S$.

The Hamiltonian associated with $\chi$ is defined if $\chi$ is a symmetry of all fields, which means that for Einstein--Maxwell theory we require not just that $\chi$ be a Killing vector in this case but also
\begin{align}
    \lie_\chi \bs A = 0.
\end{align}

\subsection{Vacuum} \label{vacuum}

If $\chi$ is a Killing vector, then $\na^{[e} \chi^{d]} = \na^e \chi^d$. Using the identity $\na_e \na^e \chi^d = -R^d_e \chi^e$ \cite{Poisson}, this means that for a Killing vector,
\begin{align}
    (\bs J^{EH}_\chi)_{abc} &= -\f1{8\pi} \bs \ep_{d a b c} R^d_e \chi^e
\end{align}
In higher dimensions, this can be written as
\begin{align}
    \bs J^{EH}_\chi &= w\cdot\bs \ep,
\end{align}
where
\begin{align}
    w^a &= -\f1{8\pi} R^a_b \chi^b.
\end{align}

In pure Einstein gravity, with no matter and no additional fields, the equations of motion are $G_{ab} + \La g_{ab} = 0$. In that case, $\bs Q^{EH}_\chi$ is the ``complete'' Noether charge associated with $\chi$ (i.e.,~there are no additional corrections). Furthermore, if the equations of motion are satisfied, $R_{ab} = \left(2\La/(D-2)\right) g_{ab}$. In this case, $\bs J^{EH}_\chi$ reduces to
\begin{align}
    \bs  J^{EH}_\chi &= -\f{\La}{4\pi (D-2)} \chi \cdot \bs \ep.
\end{align}
(We could also find this by noting that $\bs \Th^{EH}(\phi,\lie_\chi \phi) = 0$ if $\chi$ is a Killing vector, so that $\bs J_\chi^{EH}=-\chi \cdot \bs L_{EH}$ in that case.) If $\La = 0$, then the Noether current associated with a Killing vector when the equations of motion are satisfied is not only closed, but actually zero. 

This means that the Komar integral $\oint_S \bs Q_\chi^{EH}$, if $\La = 0$ and the equations of motion hold, is independent of integration surface (since $d\bs Q_\chi^{EH} = \bs J_\chi^{EH} = 0$). In $\La = 0$ vacuum (four-dimensional) spacetimes, $\oint_S \bs Q_\xi^{EH}$ (as given by \eqref{QEH}) typically gives ``half the mass'' if $\xi = \partial_t$ is the Killing vector corresponding to the time symmetry, as pointed out in, e.g.,~\cite{Wald84}. The reason for the factor of $1/2$ is that the Noether charge, while surface-independent in $\La=0$ vacuum, is still not itself the Hamiltonian, which satisfies \eqref{mcEHxi}; $-\oint_{S_\infty} \xi \cdot \bs B$ supplies the ``other half'' of the mass. 

If $\Lambda \neq 0$, the integral $\oint_S \bs Q^{EH}_\xi$ in general depends on the integration surface. Not only that, but in anti-de Sitter space (or asymptotically anti-de Sitter space---vacuum which approaches AdS under some fall-off conditions), the Komar integral for $\xi = \pa/\pa t$ diverges as the integration surface $S$ is taken to infinity: 
\begin{align}
    \left|\lim_{S\to S_\infty} \oint_S \bs K^K_\xi\right| = \infty.
\end{align}
However, using the definitions above, the combination $\de \bs Q_\xi - \xi \cdot \bs \Th^{EH}(\phi,\de \phi)$ is finite under a transformation, say, from one Kerr--anti-de Sitter state to another. 

In asymptotically flat vacuum as defined by Iyer and Wald \cite{IyerWald94}, the Arnowitt--Deser--Misner (ADM) energy (or mass) and angular momentum can be recovered through the covariant phase space method. Finding some differential form $\bs B$ such that, asymptotically, $\de \oint_{S_\infty} (\xi \cdot \bs B) = \oint_{S_\infty} \xi \cdot \bs \Th^{EH}(\phi,\de \phi)$, the canonical energy $\mc E$ is defined by
\begin{align}
    \mc E &= H_\xi \equiv \oint_{S_\infty} (\bs Q^{EH}_\xi - \xi \cdot \bs B).
\end{align}
$\mc E$ corresponds then to the ADM energy. 

The situation is more complicated in asymptotically anti-de Sitter spacetimes, and there are many different definitions of conserved energy and angular momentum, derived from a Lagrangian or Hamiltonian method. Much of the crux of the problem comes down to the divergence of the Komar integral. A full survey is beyond the scope of this thesis. Now I will discuss one particular paper of interest, by Barnich and Comp\`ere \cite{BarnichCompere}, which is useful for my purposes. 

\section{Barnich and Comp\`ere} \label{BCsection}

In Barnich and Comp\`ere's paper \cite{BarnichCompere} (hereafter referred to as BC), they find a way to ``go into the bulk'' (finite radius) for conserved quantities rather than to rely only on definitions at infinity. They use this technique to comment on the Smarr relation for black holes in asymptotically AdS spacetime. This paper appeared shortly after the publication of GPP and several years before the publication of CGKP. I will present how BC approach the conserved charges in the next two sections and then present how I adapt it beginning in Section \ref{smarrrevisited}.

I will introduce some notation. BC allow the Killing vector to have varying contravariant components. Slightly adapting their notation, let the initial field configuration (including the metric) be represented by $\phi$ and the varied fields be represented by $\phi + \de \phi$, with the metric in particular being represented by $g_{ab}, g_{ab}+\de g_{ab}$. The Killing vector field $\chi^a$ (also a symmetry of all other fields: $\lie_\chi \phi=0$) is now allowed to have varying metric components $\de \chi^a$. Let $\bar \chi^a$ be the contravariant components of $\chi^a$ in the unvaried spacetime, so that we write $\chi^a|_g = \bar \chi^a$, and now let the components of $\chi^a$ in the varied spacetime with metric $g_{ab} + \de g_{ab}$ be $\chi^a|_{g+\de g} = \bar \chi^a + \de \chi^a$. (I am modifying BC's notation somewhat here.) The expressions $|_g, |_{g+\de g}$ are to denote whether the evaluation is in the unvaried or varied metric. BC require that the Killing vector $\chi$ satisfies the linearized Killing equation of the metric, as well as its generalization to general fields $\phi$,
\begin{align}
    \lie_\chi \de g_{ab} + \lie_{\de \chi} g_{ab} &= \lie_\chi \de \phi + \lie_{\de \chi} \phi = 0. \label{variationKillingequation}
 \end{align}
Terms involving a variation $\de$ are assumed to be small, so that terms quadratic in the variation vanish. 

With this stipulation, define $\bs k_\chi[\de g;g]$ to be equal to
\begin{align}
    \bs k_\chi[\de g;g] &= \de \bs Q_\chi - \bs Q_{\de \chi} - \chi \cdot \bs \Th[\de \phi;\phi]. \label{kchiwithdeltachi}
\end{align}
Here I am switching to BC's notation for $\bs \Th$: $\bs \Th[\de \phi;\phi] = \bs \Th(\phi,\de \phi)$, but the object is otherwise the same. Here $\de \bs Q_\chi$ is the difference between the Noether charge form associated with the fields $\phi+\de \phi$ and vector $\chi^a|_{g+\de g} = \bar \chi^a + \de \chi^a$ minus that associated with $\phi$ and $\chi^a|_g = \bar \chi^a$, or, including indices for clarity,
\begin{align}
    (\de \bs Q_\chi)_{a_1 \ldots a_{D-2}} &= \left(\bs Q_{\chi}[\phi+\de \phi]\right)_{a_1 \ldots a_{D-2}} - \left(\bs Q_\chi[\phi]\right)_{a_1 \ldots a_{D-2}}.
\end{align}
Meanwhile, $\bs Q_{\de \chi}$ is the value of the Noether charge evaluated on the value of the fields $\phi$ where $\de \chi^a$ is the vector argument. 

In particular, with $\bs k^{EH}_\chi[\de g;g]$ we have
\begin{align}
    \bs k^{EH}_\chi[\de g;g] &= -\de \bs K^K_\chi + \bs K^K_{\de \chi} - \chi \cdot \bs \Th[\de g;g], \label{kEHincludingdeltachi}
\end{align}
where
\begin{align}
    \de \bs K^K_\chi &= \bs K^K_{\chi}[g+\de g] - \bs K^K_\chi[g] \nn 
    (\de \bs K^K_\chi)_{a_1 \ldots a_{D-2}} &= \f1{16\pi} \left(  \left.\left( \na^a \chi^b \bs \ep_{a b a_1 \ldots a_{D-2}}\right)\right|_{g+\de g} - \left.\left( \na^a \chi^b \bs \ep_{a b a_1 \ldots a_{D-2}}\right)\right|_g \right),
\end{align}
where the $|_{g+\de g}$ and $|_g$ terms indicate with which metric the covariant derivative and $\bs \ep$ are calculated as well as in what spacetime $\chi^a$ is evaluated. We also have
\begin{align}
    \bs K^K_{\de \chi} &= \bs K^K_{\de \chi} [g] \nn 
    (\bs K^K_{\de \chi})_{a_1 \ldots a_{D-2}} &= \f{1}{16\pi} \left. \left( \na^a \de \chi^b \bs \ep_{a b a_1 \ldots a_{D-2}}\right)\right|_{g}.
\end{align}
By assumption of linearity in the differentials, we could also evaluate $ \bs K^K_{\de \chi}$ in the metric $g + \de g$ to get the same result (the quadratic terms dropping out).

BC argue that this form of $\bs k_\chi[\de \phi;\phi]$ accounts for the possibility of $\chi$ changing its contravariant components. Obviously if $\de \chi^a = 0$ then the form \eqref{kchi} is recovered. One way to justify this form is the following. Since $\bar \chi^a$ is unvarying, $\de \bar \chi^a = 0$. Then when comparing components of $\bar \chi^a$ and $\chi^a$ in the spacetime associated with the metric $g_{ab} + \de g_{ab}$, $\bar \chi^a = \chi^a - \de \chi^a$. Then we have,
\begin{align}
    \de \bs K^K_{\bar \chi} &= \bs K^K_{\bar \chi}[g+ \de g] - \bs K^K_{\bar \chi}[g] \nn 
    &= \bs K^K_{\chi - \de \chi}[g+\de g] - \bs K^K_\chi,
\end{align}
applying $\bar \chi^a = \chi^a - \de \chi^a$ for $g + \de g$ and $\bar \chi^a = \chi^a$ for $g$. Using the linearity of $\bs K^K_\chi$ with respect to $\chi$, this expands to
\begin{align}
    \de \bs K^K_{\bar \chi} &= \bs K^K_{\chi}[g + \de g] - \bs K^K_{\de \chi}[g+\de g] - \bs K^K_{\chi}[g] \nn 
    &= \de \bs K^K_\chi - \bs K^K_{\de \chi}.
\end{align}
Thus the combination $\de \bs K^K_\chi - \bs K^K_{\de \chi} = \de \bs K^K_{\bar \chi}$ where $\bar \chi^a$ is the vector with unvarying components. The analogous argument applies for $\bs Q_\chi$. Thus we can really write
\begin{align}
    \bs k_\chi[\de \phi;\phi] = \bs k_{\bar \chi}[\de \phi;\phi], \label{kchiequalskbarchi}
\end{align}
so that $\bs k_\chi[\de \phi;\phi]$ is defined, if $\de \chi \neq 0$, to be equal to the $\bs k_{\bar \chi}[\de \phi;\phi]$ associated with the vector $\bar \chi^a$ which is equal to $\chi^a$ in the unvaried spacetime and which has unvarying contravariant components. Because of \eqref{kchiequalskbarchi}, there is no distinction between $\bs k_\chi[\de \phi;\phi]$ and $\bs k_{\bar \chi}[\de \phi;\phi]$, but I will sometimes include the bar in order to emphasize that the quantity $\bs k_{\bar \chi}[\de \phi;\phi]$ implicitly assumes that the components $\bar \chi^a$ do not change (and $\bs k_\chi[\de \phi;\phi]$ corrects for changing components in $\chi^a$). 

BC use $Q$ (with various superscripts) to refer to the $H_\chi$ conserved charges associated with the Hamiltonian. I will use $H_\chi$ to connect to earlier sections, where possible. BC use Greek indices, and I will convert to Roman indices (except in direct quotes from BC). I will also focus here on Einstein--Hilbert gravity and make additional notes for other cases (such as in the presence of a Maxwell field) later on.

Consider now the somewhat different, but related scenario, where there is a background metric $\bar g_{ab}$ and a full metric $g_{ab}$, both of which are vacuum solutions to Einstein's equations. Let $h^{BC}_{ab} = g_{ab} - \bar g_{ab} = \de g_{ab}$. In general we do not assume a Kerr--Schild form for $h^{BC}_{ab}$. (BC use $h_{\mu \nu}$ here, but I will append the BC superscript to distinguish it from the Kerr--Schild correction $h_{ab}$ I am using.) $h^{BC}_{ab}$ is assumed to be small enough \emph{at infinity} ($S_\infty$) that terms of order quadratic or higher in $h^{BC}_{ab}$ drop out of calculations of conserved charges \emph{at infinity}. The linearized approximation is the one for which we can raise and lower $h^{BC}_{ab}$ using either metric. For our purposes we will mostly take $\bar g_{ab}$ to represent pure anti-de Sitter, though the procedure here is more general and could apply to a situation where the background is something else (such as Minkowski space). Away from infinity, $h^{BC}_{ab}$ need not be small and terms quadratic and higher may be necessary to account for in order to calculate conserved charges. (This last point is why the scenario I am now introducing is not precisely the one introduced at the start of this section, which will recur later.)

Let $\chi^a$ be a Killing vector, which is allowed to have varying components. For clarity let its components in the background spacetime with metric $\bar g_{ab}$ be given by $\bar \chi^a$, with components in the full spacetime with metric $g_{ab}$ given by $\chi^a|_{\bar g+ \de g} = \bar \chi^a + \de \chi^a$. $\chi^a$ is a Killing vector in both the full and background spacetime, and so satisfies the linearized Killing equation 
\begin{align}
    \lie_{\de \chi}\bar g_{ab} + \lie_{\bar \chi} \de g_{ab} = 0,
\end{align}
evaluated in the background spacetime. BC use slightly different notation in their Sections 2 and 3; I attempt to have a unified notation without sacrificing clarity.

For Einstein--Hilbert gravity, the conserved (Hamiltonian) quantity calculated at a $(D-2)$-surface at (spatial) infinity, found through the so-called descent equations, is given by
\begin{align}
    H_{\chi}^\infty &= \oint_{S_\infty} \bs k^{EH}_{\chi}[h^{BC};\bar g], \label{Hchiinfinity}
\end{align}
where $\bs k^{EH}_{\chi}[h^{BC};\bar g]$ is equal to
\begin{align}
    \bs k^{EH}_{\chi}[h^{BC};\bar g] = \f{1}{8 \pi} (d^{D-2}x)_{ab} \sqrt{-\bar g} &\left( \bar \chi^{[b} \bar \na^{a]} h^{BC} + \bar \chi^{[a} \bar \na_c (h^{BC})^{b]c} + \bar \chi_c \bar \na^{[b} (h^{BC})^{a]c} + \f 1 2 h^{BC} \bar \na^{[b} \bar \chi^{a]} +\right. \nn
    &\qquad \left.\f12 (h^{BC})^{c[a} \bar \na_c \bar \chi^{b]} + \f12 (h^{BC})^{c[b} \bar \na^{a]} \bar \chi_c\right), \label{kbarxiterm}
\end{align}
where as usual $\bar g$ is the determinant of $\bar g_{ab}$, $\bar \na_a$ is the covariant derivative associated with the background metric, and the raising and lowering of indices is with respect to the background metric. The appearance of bars on all the $\chi^a$ terms are to emphasize that the Killing vector is evaluated with respect to its components in the background metric. Note also $h^{BC} =  \bar g^{ab} h^{BC}_{ab}$, although the raising could also be done with $ g^{ab}$ if $h^{BC}_{ab}$ is assumed to be sufficiently small (in fact, the raising appears to be defined with respect to $g^{ab}$ in BC but $\bar g^{ab}$ in \cite{BarnichBrandt} where the formula is derived). Finally, the term
\begin{align}
    (d^{D-p} x)_{a_1 \ldots a_p} &= \f1{p!(D-p)!} e_{a_1 \ldots a_D} d x^{a_{p+1}} \ldots d x^{a_D}, 
\end{align}
where $e_{a_1 \ldots a_n}$ is the completely antisymmetric \emph{symbol} (not tensor) with $e_{0 \ldots (D-1)} = 1$. 

BC state that if $h^{BC}_{ab} = \de g_{ab}$, this form of $\bs k^{EH}_\chi$ \eqref{kbarxiterm} is equivalent to \eqref{kEHincludingdeltachi} with the background metric as its argument. I perform my own calculation to show that these two expressions for $\bs k_\chi^{EH}$ are equal in Appendix \ref{kxicomparison}. 

$H_{\chi}^\infty$ plays the same role as $H_\chi$ in the previous section, in the specific case of the Einstein--Hilbert Lagrangian. (Specifically, the result \eqref{Hchiinfinity} is equivalent to \eqref{deHchiInfinity}.) The $\infty$ superscript is to specify that it is calculated at infinity. Using \eqref{kEHincludingdeltachi} and $h^{BC}_{ab} = \de g_{ab}$,
\begin{align}
    \bs k^{EH}_{\chi}[h^{BC};\bar g] &= - \de \bs K^K_{\chi} + \bs K^K_{\de \chi} - \bar \chi \cdot \bs \Th^{EH}[h^{BC};\bar g]. \label{kKomarForm}
\end{align}

Comp\`ere \cite{Compere} also writes for the scenario where $\de \chi^a = 0$,
\begin{align}
    \bs k^{EH}_{\chi}[\de_{h^C},\bar g] &= -\de_{h^C} \bs K^K_\chi - \chi \cdot \bs \Th^{EH}[h^C,\bar g], \label{kxihg1}
\end{align}
where $\de_{h^C}$ is the variation induced by $\de g_{ab} = g_{ab} - \bar g_{ab} = h^{C}_{ab}$. Note that to first order in $h^C$, $\bs k_\chi^{EH}[\de_{h^C}, \bar g] = \bs k_\chi^{EH}[\de_{h^C},g]$.

The development of \eqref{kbarxiterm} expression appears in Barnich and Brandt \cite{BarnichBrandt}. Significantly, what is written here is not $\de H_{\chi}^\infty$ but $H^\infty_{\chi}$ itself. The conserved mass and angular momentum can be calculated at infinity by simply taking $\bar g_{ab}$ as the metric for pure anti-de Sitter and taking $\de g_{ab} = h^{BC}_{ab} = g_{ab} - \bar g_{ab}$, and then identifying $H_{\chi}$ with $\de H_{\chi}$ under $\de g_{ab} = h_{ab}^{BC}$, taking BC's $H^\infty_{\chi}$ to be equal to $H_{\chi}$. This only can work at infinity and under sufficiently steep fall-off conditions that it is valid to take the linear approximation, i.e.~to write
\begin{align}
    \Delta H_{\chi} &= \de H_{\chi},
\end{align}
where $\Delta H_{\chi} = H_{\chi}[g_{ab}] - H_{\chi}[\bar g_{ab}]$, with normalization so that $H_{\chi} [\bar g_{ab}] = 0$. This would, in general, not work ``in the bulk'' where the changes in the metric components are more significant and so nonlinear effects from the change in metric components will be significant. 

BC also point out that their expression agrees with the one derived by Abbott and Deser \cite{AbbottDeser}, which in (slightly modified) BC notation is
\begin{align}
    \bs k^{AD}_{\chi} [h;\bar g] &= - \f1{16\pi} (d^{D-2}x)_{ab} \sqrt{-\bar g} \left( \bar \chi_c \bar \na_d H^{cdab} + \f12 H^{cdab} \bar \na_c \bar \chi_d\right),
\end{align}
where
\begin{align}
    H^{manb} [h;g] &= - \hat h^{ab} \bar g^{mn} -  \hat h^{mn} \bar g^{ab} + \hat h^{an} \bar g^{mb} + \hat h^{mb} \bar g^{an} \nn
    \hat h_{mn} &= h^{BC}_{mn} - \f12 \bar g_{mn} h^{BC}.
\end{align}
They also show that their expression is equivalent to the KBL superpotential (under certain circumstances). I will discuss the KBL superpotential, and how it relates to my work, in Section \ref{KBLSection}.

\subsection{Entering the Bulk} \label{enteringthebulk}

Barnich and Comp\`ere argue that it is possible to develop expressions for the conserved charges in the bulk (i.e.,~not at infinity) in addition to the ones at infinity, by ``integrating through solution space.'' Solution space refers to the space of solutions to a given theory, referred to as $\bar{\mathscr F}$ in \cite{WaldZoupas}, a submanifold of the field configuration space $\mathscr F$, and is also known as the covariant phase space of the theory.

Consider a path $\gamma$ in the space of solutions to Einstein's equations, interpolating between the background $\bar g_{ab}$ and a given solution $g_{ab}$. Specifically, let $g^{(s)}_{ab}$ be a series of metric functions with $s \in [0,1]$, each satisfying Einstein's equations (in vacuum), with $g^{(0)}_{ab} = \bar g_{ab}$ and $g^{(1)}_{ab} = g_{ab}$. Each $g_{ab}^{(s)}$ is defined at each point in coordinate space (by which I mean, in the physical manifold corresponding to constant $s$). Let $\gamma$ be the path in solution space, between $\bar g_{ab}$ and $g_{ab}$, here equal to the real line segment $[0,1]$. Let $\chi^{(s)}$ be a family of Killing vector fields of the equivalent metrics, $\lie_{\chi^{(s)}} g_{ab}^{(s)} = 0$. Define $h_{ab}^{(s)} = \f{d}{ds} g_{ab}^{(s)}$ to be the tangent vector to $g_{ab}^{(s)}$ in solution space. Assume $g_{ab}^{(s)}$ varies smoothly with $s$.

Let $d_V g^{(s)}_{ab}$ denote ``a 1-form on the space of metrics,'' taking $d_V$ essentially to be the exterior derivative in solution space. This is equal, in the above prescription, to
\begin{align}
    d_V g^{(s)}_{ab} &= \f{d}{ds} g^{(s)}_{ab} ds \nn 
    &= h^{(s)}_{ab} ds.
\end{align}
$d_V$ more or less has the role that $\delta$ has played previously, as a descriptor for a variation between two solutions, but here it is more specified to be related to the variation between two solutions within the path $\gamma$. 

Then define the $(D-2)$-form in coordinate space $\bs K^{EH}_{\chi;\gamma}$ (not to be confused with the Komar integrand $\bs K^K_\chi$) by 
\begin{align}
    \bs K^{EH}_{\chi;\gamma} &= \int_\gamma \bs k^{EH}_{\chi_\gamma} [d_V g; g],
\end{align}
where $\gamma$ is a path in solution space. Here $\chi_\gamma$ refers to the value of $\chi$ at each point along the curve $\gamma$. $\bs k^{EH}_{\chi_\gamma} [d_V g;g]$ is a $(D-2)$-form in coordinate space but, due to the linearity of $\bs k^{EH}_\chi[h^{BC};g]$ with respect to $h^{BC}$, is a one-form in the solution space, which can then be integrated through solution space. The resulting expression $\bs K^{EH}_{\chi;\gamma}$ is a $(D-2)$-form in coordinate space which depends on the path $\gamma$ in solution space. 

In the one-parameter group of solutions we have described, $\gamma = [0,1], d_V g_{ab} = h^{(s)}_{ab} ds$. So
\begin{align}
    \bs K^{EH}_{\chi;\gamma} &= \int_0^1 ds \bs k_{\chi^{(s)}} [ h^{(s)}_{\mu \nu},g^{(s)}_{\mu \nu}].
\end{align}
$\bs K^{EH}_{\chi;\gamma}$, according to BC, ``can be shown to be closed wherever the interpolation is meaningful''
\begin{align}
    d \bs K^{EH}_{\chi;\gamma} &= 0,
\end{align}
where $d$ is the usual exterior derivative in coordinate space. From the Gauss--Stokes theorem, the charges
\begin{align}
    Q_{\chi;\gamma} &= \oint_S \bs K^{EH}_{\chi;\gamma}
\end{align}
do not depend on the closed ($D-2$)-dimensional hypersurface $S$ used for their evaluation:
\begin{align}
    Q_{\chi;\gamma} &= \oint_S \bs K^{EH}_{\chi;\gamma} = \oint_{S'} \bs K^{EH}_{\chi;\gamma}.
\end{align}

Another way of describing $\bs K^{EH}_{\chi;\gamma}$ (which does not appear in BC) is the following. Let $\gamma^{(s')}$ ($s' \in [0,1]$) be the segment $s \in [0,s']$ of the path $\gamma$, and let (dropping the EH superscript for compactness) $\bs K^{(s')}_{\chi;\gamma}$ be given by
\begin{align}
    \bs K^{(s')}_{\chi;\gamma} = \int_0^{s'} ds \bs k_{\chi^{(s)}}^{EH}[h^{(s)};g^{(s)}].
\end{align}
We have then $\bs K^{(1)}_{\chi;\gamma} = \bs K_{\chi;\gamma}$ and $\bs K^{(0)}_{\chi;\gamma} = 0$. Then we have
\begin{align}
    d_V \bs K_{\chi;\gamma}^{(s')} &= \frac{d}{ds'} \bs K_{\chi;\gamma}^{(s')} ds' \nn 
    &= \bs k^{EH}_{\chi^{(s')}}[d_V g_{ab}^{(s')};g_{ab}^{(s')}].
\end{align}
The $(D-2)$-form $\bs K_{\chi;\gamma}^{(s')}$ is the $(D-2)$-form which has a variation along $\gamma$ which corresponds to the term $\bs k_{\chi^{(s')}}[d_V g_{\mu \nu}^{(s')};g_{\mu \nu}^{(s')}]$. I am allowing $d_V$ here to act as $ds' (d/ds')$; hopefully the notation is clear. If we let
\begin{align}
    Q_{\chi;\gamma^{(s')}} &= \oint_S \bs K^{(s')}_{\chi;\gamma},
\end{align}
then at each point along $\gamma$,
\begin{align}
    d_V Q_{\chi;\gamma^{(s')}} &= \oint_S \bs k^{EH}_{\chi^{(s')}} [ d_V g^{(s')}_{ab};g^{(s')}_{ab}]. \label{deQtech}
\end{align}
If we let $\de = d_V$, this statement is essentially that
\begin{align}
    \de Q_{\chi;\gamma^{(s')}} = \oint_S \bs k_{\chi}^{EH} [\de g; g^{(s')}] \label{deQ}
\end{align}
at each point along the curve $\gamma$, where $\de = \de s' \f{d}{ds'}$. This is the defining equation for the variation in the conserved charge corresponding to the Hamiltonian (see \eqref{deHchi}). If the Hamiltonian associated with a Killing vector $\chi^{(s')}$ is well-defined, then it continues to be well-defined all along the curve in solution space. This is a method to construct a well-defined Hamiltonian charge based on a background, which always locally satisfies \eqref{deQ} (or, more technically, \eqref{deQtech}). Thus this method constructs conserved charges along a path in solution space such that the variation between two nearby points on the solution space gives the difference in the Hamiltonian, between those two points. This can be thought of as, essentially, integrating the $\de H_\chi$ from the background to the full metric, in such a way as can be done at finite radius even where the difference between the background and full metric are very large.

It is further worth noting here that instead of the term $\bs k_\chi^{EH}[h^{BC};g]$ given above, a different term could be introduced associated with a different Noether charge $\bs Q_\chi$ and symplectic potential form $\bs \Th$, corresponding to a different theory (say Einstein--Maxwell theory). 

\subsection{Path Independence and Equivalence to Charge at Infinity} \label{BCPathIndependence}

The next section in BC shows that the charges $Q_{\chi;\gamma}$ do not depend on the path $\gamma$ in solution space and that the charges correspond to the charges $H_{\bar \chi}^\infty$ defined at infinity. Here is their argument:

Let $g_{ab}^{(s)}(x)$ and $(\chi^{(s)})^a$ depend on $s$ in an analytical way and further demand that, in an expansion in $s$, all terms of order $s$ or higher in $h^{(s)}_{ab}$ vanish at infinity. Specifically this implies that, at the $(D-2)$-surface at infinity $S_\infty$,
\begin{align}
    g_{ab}^{(s)} = \bar g_{ab} + s h^{BC}_{ab} + \mc O(s^2), \label{gabsatinfinity}
\end{align} 
with $g_{ab} = \bar g_{ab} + h^{BC}_{ab}$ as before, so that $h^{(s)}_{ab} = h^{BC}_{ab} + \mc O(s)$. The consequence is that, on $S_\infty$,
\begin{align}
    \bs K^{EH}_{\chi;\gamma} = \bs k_{\chi}^{EH} [h^{BC};\bar g].
\end{align}
This implies that, on $S_\infty$,
\begin{align}
    Q_{\chi;\gamma} &= \oint_{S_\infty} \bs k_{\chi}^{EH} [h^{BC};\bar g] \nn 
    &= H_{\chi}^\infty.
\end{align}
This implies that, provided $h^{(s)}_{ab}$ at infinity can be approximated by $h^{BC}_{ab} + \mc O(s)$ and that those $\mc O(s)$ terms vanish at infinity, $Q_{\chi;\gamma}$ do not depend on the path $\gamma$. Since they also do not depend on integration surface, we conclude that $Q_{\chi;\gamma} = H_{\chi}^\infty$, even if evaluated at surfaces in the bulk where the $s$-dependence of $\bs K^{EH}_{\chi;\gamma}$ does not vanish.

The alternate method presented by BC is to show the path-independence of the charges $Q_{\chi;\gamma}$ by studying the strong integrability conditions
\begin{align}
    \oint_S d_V \bs k^{EH}_\chi[d_V g;g] &= 0, \label{strongintegrability}
\end{align}
corresponding to \cite{Compere}
\begin{align}
    \oint_S \left( \de_1 \bs k^{EH}_\chi[\de_2 g;g] - \de_2 \bs k^{EH}_\chi[\de_1 g;g]\right) = 0, \label{strongintegrability2}
\end{align}
which amount to the idea that under two variations (which both satisfy the linearized equations of motion), the resulting variations in the field $\bs k_\chi$ are closed (in solution space). If this is the case, then the Gauss--Stokes theorem \emph{in solution space} implies that the integral for $Q_{\chi;\gamma}$ is path-independent. 

To clarify further: consider a two-parameter set of metrics $g^{(q,s)}_{ab}$, each representing a vacuum solution to Einstein's equations, with $q \in [0,1]$ and $s \in [0,1]$. Let $g^{(q,0)}_{ab} = \bar g_{ab}$ and $g^{(q,1)}_{ab} = g_{ab}$ for all $q$. For $s$ between $0$ and $1$, $g^{(q,s)}_{ab}$ depends on $q$. Then the curves consisting of $s \in [0,1]$, constant $q$ correspond to different paths in solution space interpolating between $\bar g_{ab}$ and $g_{ab}$. Assume that the functions $g^{(q,s)}_{ab}$ are analytic in $q,s$. Then let $\de_1 s = \de_1 s, \de_1 q = 0; \de_2 s = 0, \de_2 q = \de_2 q$. In this case, $\de_1 = \de_1 s \pa/\pa s$ (at constant $q$), $\de_2 = \de_2 q \pa/\pa q$ (at constant $s$). In this case \eqref{strongintegrability2} becomes
the condition
\begin{align}
    \oint_S \left( \f{\pa}{\pa s} \bs k_\chi^{EH}\left[ \f{\pa g}{\pa q};g\right] - \f{\pa}{\pa q} \bs k_\chi^{EH}\left[ \f{\pa g}{\pa s};g\right]\right) \de_1 s \de_2 q = 0,
\end{align}
or simply
\begin{align}
    \oint_S \left( \f{\pa}{\pa s} \bs k_\chi^{EH}\left[ \f{\pa g}{\pa q};g\right] - \f{\pa}{\pa q} \bs k_\chi^{EH}\left[ \f{\pa g}{\pa s};g\right]\right) = 0.
\end{align}
The strong integrability conditions require that this statement hold at any particular point $(q,s)$ in $q,s$ space between $s=0$ and $s=1$. Whether the strong integrability conditions hold or not depends on the category of solutions under consideration. 

The strong integrability conditions are related to the weak integrability conditions,
\begin{align}
    d (d_V \bs k^{EH}_\chi [d_V g;g])|_{\de g, \de \chi;g,\chi} &= 0. \label{weakintegrability}
\end{align}
The weak integrability conditions are shown by BC to hold for any set $g_{ab}, \de_1 g_{ab}, \de_2 g_{ab}, \chi^a, \de_1\chi^a, \de_2\chi^a$, provided that $g_{ab}(x)$ is a solution to Einstein's equations, the $\de g_{ab}$ are solutions to the linearized Einstein's equations, $\chi^a(x)$ is a Killing vector field for $g_{ab}(x)$, and $\de \chi^a(x)$ satisfies the linearized Killing equation $\lie_\chi \de g_{ab} + \lie_{\de \chi} g_{ab} = 0$ (for both $\de = \de_1, \de = \de_2$). (Intuitively, assuming that the physical spacetime exterior derivative $d$ commutes with the solution space exterior derivative $d_V$, then \eqref{weakintegrability} is satisfied because $d \bs k^{EH}_\chi[d_V g;g] = 0$, according to \eqref{dkiszero}.)

If the weak integrability conditions hold, from the Gauss--Stokes theorem, the left-hand side of \eqref{strongintegrability} is independent of surface $S$. This still does not guarantee that the integral is zero. (In technical terms this is a consequence of the existence of the nonvanishing De Rham cohomology class $\bs c^{D-2}$ of $(D-2)$-forms, where $\bs c^{D-2}$ is a class of closed but not exact $(D-2)$-forms in the physical space. This implies that $(d_V \bs k^{EH}_\chi[d_Vg;g])|_{\de g,\de \chi;g,\chi} = k \bs c^{D-2}$ plus an exact form, with $k$ a 2-form in solution space.) BC assume that nevertheless the strong integrability conditions hold in what follows, and verify that it holds in the Kerr/Myers--Perry and Kerr--AdS cases they apply their charge to, provided that the vector used is $\xi$ or $\eta$. 

Assuming the strong integrability conditions hold, the charge $Q_{\chi;\gamma}$ is path-independent and is identified by BC as simply $Q_\chi$ (associated with $g_{ab},\chi^a$ measured with respect to the background $\bar g_{ab},\bar \chi^a$). If the strong integrability conditions do not hold, the charge is not path-independent and so is not meaningful. Assuming the strong integrability conditions hold, then, BC showed that this corresponds to the Hamiltonian charge $H_{\chi}^\infty$. I will use $H_\chi$ to refer to this charge from now on, and so write
\begin{align}
    H_\chi &= \int_0^1 d s \oint_{S} \bs k^{EH}_{\chi^{(s)}} [h^{(s)};g^{(s)}]. \label{Hxiintegral}
\end{align}

\subsection{Form for BC Conserved Charge}

The charges that are generated by integrating through solution space can then be found by integrating any individual expression for the local charge variations $\bs k^{EH}_\chi [\de \phi,\phi]$. The one they use is from \eqref{kKomarForm}, which is tied only to the Einstein--Hilbert Lagrangian solution.  For an arbitrary theory of gravity, \eqref{kchiwithdeltachi} could be used.

The integral \eqref{Hxiintegral} calculated using \eqref{kEHincludingdeltachi} is
\begin{align}
    H_\chi &= -\oint_S \bs K_\chi^K + \oint_S \overline{\bs K^K_{\chi}} + \oint_S \bs C_{\chi;\gamma} \label{BC3.9H} \\
    \bs C_{\chi;\gamma} 
    &=\int_0^1 ds \oint_S \left( \bs K^K_{\f{d}{ds} \chi^{(s)}}\left[g^{(s)}\right] - \chi^{(s)} \cdot \bs \Th^{EH}\left[ h^{(s)};g^{(s)}\right]\right). \label{BC3.9}
\end{align}
The term $\oint_S \bs K^K_\chi$ is the Komar integral associated with the full spacetime $g_{ab}$. By $\overline{\bs K_\chi^K}$ I mean that all quantities are evaluated in the background metric; it can also be written $\bar{\bs K}_{\bar \chi}^K$, where $\bar \chi^a$ are the components of $\chi^a$ in the background spacetime. By $\bs K^K_{d \chi^{(s)}/ds}[g^{(s)}]$ I mean that the Komar $(D-2)$-form associated with the the vector $d \chi^{(s)}/ds$ is evaluated with respect to the metric $g^{(s)}_{ab}$. Similarly, $\bs \Th^{EH}[h^{(s)};g^{(s)}]$ is the value for $\bs \Th^{EH}$ where the background metric is taken to be $g^{(s)}_{ab}$ and $h^{(s)}_{ab}$ is plugged in for the metric variation.

BC further point out that $\oint_S \bs C_{\chi;\gamma}$ is independent of the path $\gamma$. This is because $H_\chi$ (BC's $Q_{\chi;\gamma}$) is path-independent and so are $\bs K^K_\chi$ and $\overline{\bs K^K_{\chi}}$, which depend only on the endpoints for the metric and $\chi$.

In the case where $\chi$ has the same contravariant components throughout ($\chi^a = \bar \chi^a, \de \chi^a = 0$),  \eqref{BC3.9} reduces to
\begin{align}
   \bs C_{\chi;\gamma} &= -\int_0^1 ds \: \chi \cdot \bs \Th^{EH} \left[ h^{(s)};g^{(s)}\right]. 
\end{align}

BC consider only Einstein--Hilbert gravity with the specific choice $\bs Q_\chi = -\bs K^K_\chi$ and the choice \eqref{ThetaEH} for $\bs \Th^{EH}$. The obvious generalization to a case where $\bs Q, \bs \Th$ can be distinct from the above (either an alternate theory of gravity, or in the presence of fields, or even just using a different charge compatible with a transformation such as $\bs \Th \to \bs \Th + d \bs Y$ or some such) is
\begin{align}
    H_\chi &= \oint_S \bs Q_\chi - \oint_S \overline{\bs Q_{\chi}} + \oint_S \bs C_{\chi;\gamma} \label{HchiQchiversion} \\
    \bs C_{\chi;\gamma} &= \int_0^1 ds \left( - \bs Q_{\f{d}{ds} \chi^s}\left[g^{(s)}\right] - \chi^{(s)} \cdot \bs \Th \left[ h^{(s)};g^{(s)}\right]\right),
\end{align}
in general, and, if $\chi^a = \bar \chi^a$, 
\begin{align}
    H_\chi &= \oint_S \bs Q_\chi - \oint_S \overline{\bs Q_\chi} + \oint_S \bs C_{\chi;\gamma} \nn 
    \bs C_{\chi;\gamma} &= - \int_0^1 ds \: \chi \cdot \bs \Th\left[ h^{(s)};g^{(s)}\right]. \label{generalizationC}
\end{align}

BC then apply their formula to the case of black hole mechanics, and derive a Smarr relation. I will review some salient features of black hole mechanics and get to the formulas that appear in BC.

Before continuing, I will note that a similar development occurs in papers by Hajian et al.~\cite{HajianSheikh-Jabbari,HajianGRG,HajianOzsahin} who refer to the ``solution phase space method,'' and integrate charges via \eqref{Hxiintegral}.  The argument in \cite{HajianSheikh-Jabbari} relates the conserved quantities to both the Iyer--Wald formalism as well as to Barnich et al.~\cite{BarnichBrandt,BarnichCompere,Compere}. The idea is that the space of solutions is defined by different values of the parameters $m$ and $a_i$, and a variation consists of varying $m$ and $a_i$. Then, $\de H_\chi = \oint_S \bs k^{EH}_\chi[\de g;g]$ is calculated, with the expression on the right-hand side involving terms $\de m$ and $\de a$, and then, if the right-hand side is an exact differential in terms of $\de$, $H_\chi$ is integrable. This is equivalent to 
\begin{align}
    (\de_1 \de_2 - \de_2 \de_1) H_\chi = 0, \label{integrability3}
\end{align}
or the strong integrability conditions of BC. \cite{HajianSheikh-Jabbari} note that (using my notation), $H_\xi$ and $H_{\eta_i}$ are integrable, but that $H_\bt$ is not (even calculated in BL coordinates). Additionally, they note that $H_{2\pi \z/\kappa}$ is integrable (where both $\z$ and $\ka$ are allowed to vary). They comment, ``Similarly, it is not trivial that the charge associated with exact symmetries $\pa_t$ or $\pa_{\varphi_i}$ is integrable. This point needs explicit examination,'' leaving it, for the moment, open why it is that the charges associated with $\xi$ and $\eta_i$ are integrable, but that those associated with $\bt$ are not. See also other papers by Hajian et al.~in \cite{Hajian:2025hxf} and references therein, which continue the work in interesting directions, describing the Smarr relation involving the the variation in $\La$ and other coupling constants. In these papers, $\La$ for Kerr--AdS (and similar spacetimes) is identified as a conserved charge, and the thermodynamic volume term is identified as conjugate to that charge. I take a different approach, though there are similarities. (See also \cite{Chrusciel}, which considers the integrability of Hamiltonian charges without using the Iyer--Wald formalism, for Kerr--de Sitter and Kerr--AdS.)

\section{Black Hole Mechanics} \label{BHMechanicsSection}

The conventions here are mostly designed to match Comp\`ere \cite{Compere}, with a few modifications. As in Section \ref{BHHorizonGKAdS}, let $\zeta^a$ be the Killing vector which is tangent to the null generators of a stationary black hole horizon. (There is ambiguity in $\zeta^a$, since it can always be rescaled by a constant $\zeta^a \to C \zeta^a$. As in Section \ref{BHHorizonGKAdS}, the choice will be made to have $\zeta^t = 1$ in KS coordinates for the GKAdS solutions. $\zeta^a$ is thus both parallel to and orthogonal to the horizon $H$, which is a null hypersurface.) Let $n^a$ be a transverse null vector, which is not tangent to (or orthogonal to) $H$, and which satisfies $n_a \zeta^a = -1$. 

$\zeta$ satisfies \eqref{horizonzetakappa}. This expression after contraction with $n^a$ gives
\begin{align}
    \ka &= \left(-n_a \z^b \na_b \z^a\right)|_H \nn
    &= \left( n_a \z^b \na_a \z_b\right)|_H \nn 
    &= \f12 \left( n^a \na_a \z^2 \right)|_H, \label{kappaBC}
\end{align}
where $|_H$ means evaluation at the horizon and $\z^2 \equiv \z_a \z^a$. 

The horizon has an area formula which is found in, e.g.,~Poisson \cite{Poisson}.  Let $dS_{a b}$ be the area element on the horizon, in the notation of Poisson,
\begin{align}
    dS_{a b} = 2 \z_{[a} n_{b]} dS, \label{horizonbinormal}
\end{align}
where $dS$ is the area element on the horizon and $\z$ and $n$ are as above. Let $H$ be $(D-2)$-dimensional spacelike surface on the horizon. (We use $H$ for both the full $(D-1)$-dimensional null surface and for the spacelike $(D-2)$-dimensional surface; this should hopefully not cause confusion.) Then
\begin{align}
    \oint_H \na^a \zeta^b d S_{ab} &= 2\oint_H \na^a \zeta^b \z_{[a} n_{b]} d S \nn 
    &= \oint_H (\na^a \z^b \z_a n_b - \na^a \z^b \z_b n_a ) dS \nn 
    &= -2 \oint_H \kappa dS \nn 
    &= -2 \kappa A,
\end{align}
where $A$ is the area of the horizon, using the fact that $\ka$ is constant on the horizon. 

This implies we have
\begin{align}
    \oint_H \bs K^K_\z &= - \f{\ka A}{8 \pi}. \label{KomarArea}
\end{align}

Now consider a variation between two black hole solutions. The initial calculation was performed by Bardeen, Carter and Hawking \cite{BCH}. Here I continue to (mostly) follow the notation of Comp\`ere \cite{Compere}. An overview is provided here and more technical details can be found in \cite{BCH,Compere}.

It is conventional to locate the horizon at the same coordinates in both systems. This implies that the $(D-2)$-surface $H$ will be represented by the same set of coordinates (typically constant-$t$ and constant-$r$). Let $g_{ab}$ be a black hole metric with horizon located at $H$, and perform a small perturbation to another black hole spacetime with a horizon also at $H$, with metric $g_{ab} + \de g_{ab}$. Let $\z^a$ represent the Killing vector tangent to the null generators of the horizon, for both the original and unvaried metric, where $\bar \z^a$ represent the unvaried components of $\bar \z^a$ (with $\de \bar \z^a = 0$), and the varied components are $\z^a|_{g+\de g} = \bar \z^a + \de \z^a$. We do not require that the metrics $g_{ab}$ and $g_{ab} + \de g_{ab}$ be vacuum solutions to Einstein's equations. Using \eqref{KomarArea},
\begin{align}
    \de \oint_H \bs K^K_{\z} &\equiv \oint_H \bs K^K_{\bar \z+\de \z}[g + \de g] - \oint_H \bs K^K_{ \bar \z} [ g] \nn 
    &= - \de \left(\frac{\ka A}{8\pi}\right). \label{dKz}
\end{align}

Meanwhile, it turns out
\begin{align}
    \oint_H \bs K^K_{\de \z} - \oint_H \z \cdot \bs \Th^{EH}[\de g;g] &= - \frac{A}{8\pi} \de \kappa, \label{dkappaA}
\end{align}
where $\de \kappa$ is the variation in the surface gravity $\ka$, through a more involved calculation, as given in \cite{Compere} (based in part on previous derivations such as \cite{CarterLesHouches,BCH}). On $H$ we can write $\bs K^K_{\de \z}-\z \cdot \bs \Th^{EH}$ as
\begin{align}
    \left.\left(\bs K^K_{\de \z} - \z \cdot \bs \Th^{EH}\right)\right|_H &= \f{1}{16\pi} \left( \na^a \de \z^b + \z^a ( g^{bc} \na^d \de g_{c d} - g^{cd} \na^b \de g_{cd})\right) dS_{ab}. \label{dekappaALHS}
\end{align}
Using \eqref{horizonbinormal} along with $\z^a \z_a = 0, \z^a n_a = -1$, this can be rewritten as
\begin{align}
    \left. \left( \bs K^K_{\de \z} - \z \cdot \bs \Th^{EH}\right)\right|_H &= -\f{1}{8\pi} \left(-\f12 \na^a \de g_{ab} \z^b + \f12 \z^a g^{bc} \na_a \de g_{bc} + \f12 \na^b \de \z^a ( \z_a n_b - \z_b n_a)\right) dS .
\end{align}

The calculation in \cite{Compere}, which I won't reproduce here, begins with \eqref{kappaBC} and then varies it. Using \eqref{variationKillingequation}, the restriction that the horizon coordinate location is the same initial and perturbed black hole spacetimes, and some other properties of the black hole horizons, Comp\`ere shows
\begin{align}
    \de \ka &= -\f12 \na^a \de g_{ab} \z^b + \f12 \z^a g^{bc} \na_a \de g_{bc} + \f12 \na^b \de \z^a ( \z_a n_b - \z_b n_a) + \f12 (\gamma^a_b \de \z^b)_{|a}, \label{deltakappa}
\end{align}
where $\gamma_{ab} = g_{ab} + n_a \z_b + \z_a n_b$ is the metric intrinsic to the $(D-2)$-surface $H$ (expressed as a tensor in the full $D$-dimensional spacetime) and $_{|a}$ is the associated intrinsic covariant derivative. Interestingly, $\de \ka$ is constant on the horizon. It is nice to note that this partitions $\de \kappa$ into terms related to the derivatives of $\de \z^a$ (which are related to $\bs K^K_{\de \z}$) and terms related to the derivatives of $\de g_{ab}$ (which are related to $\z \cdot \bs \Th^{EH}$). Because of the Gauss--Stokes theorem, $\oint_H (\gamma^a_b \de \z^b)_{|a} = 0$ since $H$ is closed. Comparing with \eqref{dekappaALHS}, \eqref{dkappaA} follows. (In fact, if $(\gamma^a_b \de \z^b)_{|a} = 0$, we have $(\bs K^K_{\de \z} - \z \cdot \bs \Th^{EH})|_H = - \f1{8\pi} \de \ka dS$.) 

Combining  \eqref{dKz} and \eqref{dkappaA} gives
\begin{align}
    -\de \oint_H \bs K^K_{\z} + \oint_H \bs K^K_{\de \z} - \oint_H \z \cdot \bs \Th^{EH}[\de g;g] &= \f{\ka}{8\pi} \de A. 
\end{align}

Since $\de \bar \z^a = 0$, we can write
\begin{align}
    \de \bs K^K_{\bar \z} &= \bs K^K_{\bar \z}[g+\de g] - \bs K^K_{\bar \z}[g] \nn 
    &= \de \bs K^K_{\z} - \bs K^K_{\de \z}.
\end{align}
We also have $\z^a = \bar \z^a$ in the unvaried metric. We can then rewrite \eqref{dkappaA} in terms of $\bar \z$ as
\begin{align}
    \f{\ka}{8\pi} \de A &= -\de \oint_H \bs K_{\bar \z}^K - \oint_H \bar \z \cdot \bs \Th^{EH}[\de g; g]\label{kdarule}
\end{align}
This formula \emph{does not require} vacuum or for the field equations to be satisfied or for the Lagrangian to be the Einstein--Hilbert Lagrangian; we use the term $\bs \Th^{EH}[\de g; g]$ for its utility. This is important in that it gives us a formula for describing the change in black hole surface area in terms of Komar integrals and the term $\bs \Th^{EH}$ under arbitrary perturbations. Any variation from one quasi-stationary state to another will follow \eqref{kdarule}, in any theory of gravity. There is also a ``physical process'' version where the Einstein tensor (or, usually equivalently, the stress-energy tensor) is used to define small currents of mass and angular momentum (see, e.g.~\cite{Poisson}), which I will return to in Section \ref{physicalprocessversion}.  

Interestingly, as I will show in Section \ref{explicitIxi}, the restriction that the horizon be located at the same coordinates in both the unperturbed and perturbed spacetime can be relaxed somewhat, in that if the horizon is located at some value $r = r_+^{(0)}$ in the unperturbed spacetime, even if the horizon in the perturbed spacetime is located at some $r = r_+^{(0)}+\de r_+$ ($\de r_+ \neq 0$) the formula \eqref{kdarule} can still sometimes hold.

\emph{In the particular case} where we have Einstein gravity and vacuum, we can choose $\bs Q_{\z} [g]= -\bs K^K_{\z}[g]$ and $\bs \Th[\de g;  g] = \bs \Th^{EH}[\de g; g]$. In this case, we have
\begin{align}
    \f{\ka}{8\pi} \de A &= \oint_H \bs k^{EH}_{\bar \z} [\de g;g] =\oint_H \bs k^{EH}_{\z} [\de g;g] .\label{kappadaBC2}
\end{align}
(The second equality is by \eqref{kchiequalskbarchi}.)

I want to make the following point clear. I will always use the symbol $\bs k^{EH}_\chi[\de g;g]$ to refer to the combination
\begin{align}
    \bs k^{EH}_\chi [\de g;g] &= -\de \bs K^K_\chi - \bs K^K_{\de \chi} - \chi \cdot \bs \Th^{EH}[\de g;g],
\end{align}
whether or not the spacetimes under consideration are actually vacuum solutions to Einstein--Hilbert gravity. If, additionally, the spacetimes under consideration are vacuum Einstein--Hilbert gravity solutions with the same $\La$, then we also have $\de H_\chi = \oint_S \bs k^{EH}_\chi[\de g;g]$. This means that \eqref{kappadaBC2} always holds, for variation between two stationary black hole spacetimes where the horizon is held at the same coordinates, even if the black hole solutions are not vacuum.

Now consider a vacuum solution to Einstein--Hilbert gravity. In the case where $\bar \z^a$, the value of the contravariant components of the vector $\z$ in the unvaried spacetime, can be decomposed into $\bar \z^a = \xi^a + \sum_i \Omega_i \eta_i^a$, with $\xi$ the asymptotically-static Killing vector, $\eta_i$ the azimuthal symmetry Killing vectors, and $\Om_i$ the angular momenta, consider the situation where $\de \xi^a = \de \eta_i^a = 0$. We can then write, using \eqref{kdarule},
\begin{align}
    \f{\ka}{8\pi} \de A &=  \oint_H \left( - \de \bs K_{\bar \z}^K - \bar \z \cdot \bs \Th^{EH}[\de g;g]\right) \nn 
    &= \oint_H \left( - \de \bs K^K_{\xi} - \sum_i \Om_i \bs K_{\eta_i}^K - (\xi + \sum_i \Om_i \eta_i)\cdot \bs \Th^{EH}[\de g;g]\right) \nn 
    &= \oint_H \left(\bs k_{\xi}^{EH} [\de g;g] + \sum_i \Omega_i \bs k_{\eta_i}^{EH} [\de g;g]\right) \nn 
    &= \de \mc E - \sum_i \Omega_i \de \mc J_i, \label{firstlaw}
\end{align}
which gives the first law of black hole mechanics in vacuum Einstein gravity. (When $\de \xi^a = 0$, $\bs k_\xi^{EH}[\de g;g] = - \de \bs K^K_\xi - \xi \cdot \bs \Th$, and similarly for $\de \eta_i^a = 0$.) The $\mc E$ and $\mc J_i$ are as defined in \eqref{deE} and \eqref{deJi}. They are equal to the quantities which are defined at infinity according to the definitions in Section \ref{vacuum}. We now have that the conserved charges automatically satisfy the first law, which was the requirement that GPP applied.

So far the argument I am presenting is one that reproduces (with some notational changes) that of BC. One question, not posed by BC, is: could we instead take coordinates where it is $\bt^a$, rather than $\xi^a$, which has unchanging components, and so decompose $\bar \z^a =\bt^a + \om_i \eta_i^a$? In this case, by the above argument, we would have
\begin{align}
    \f{\ka}{8\pi} \de A &= \oint_H \left(\bs k_{\bt}^{EH} [\de g;g] + \sum_i \om_i \bs k_{\eta_i}^{EH} [\de g;g] \right) \nn 
    &= \oint_H \bs k_\bt^{EH}[\de g;g] - \sum_i \om_i \de \mc J_i.
\end{align}
The answer is yes, we could instead take coordinates where $\bt^a$ has unchanging components---but, at least for Kerr--AdS, we cannot interpret $\oint_H \bs k_\bt^{EH}[\de g;g]$ as being the variation in some ``energy'' quantity. I consider variations in four dimensions, including in BL coordinates, in Section \ref{variationsinfourdimensions}. (Again, this is related to the observation that the Hamiltonian associated with $\bt$ is not integrable, as observed by papers such as \cite{HajianSheikh-Jabbari,Chrusciel,Blagojevic:2020edq,Jing:2017jxw}. This also matches up with BC's observation that the strong integrability conditions must hold for us to interpret $Q_{\chi;\g}$ as the Hamiltonian.) 

\subsection{BC Smarr Relation} \label{BCSmarrSection}

We can also use the Komar integral over the horizon to set up a Smarr relation for the area. I will continue to work, for now, in Einstein gravity, so that we can use \eqref{BC3.9}. I want to emphasize that while the expressions leading to \eqref{firstlaw} were comparing two black hole solution states, in this section we will be looking at a single black hole state of interest with metric $g_{ab}$ as well as a path through solution space from a reference background $\bar g_{ab}$. In this section then we take the final metric $g_{ab}$ to be a black hole spacetime, with horizon located at $H$ and with $\z$ a Killing vector tangent to the null generators of the horizon. We assume that $\z$ has constant components and is a Killing vector in $\bar g_{ab}$ and all other spacetimes $g^{(s)}_{ab}$ interpolating between $\bar g_{ab}$ and $g_{ab}$ (though it is only restricted to be tangent to the null generators of a Killing horizon in $g_{ab}$ specifically). $H$ is not required to be the horizon in any of the interpolating spacetimes (which are not required to be black hole spacetimes, though they may be). 

BC state how to evaluate $\mc E$ and $\mc J_i$, using $\mc E = H_\xi, \mc J_i = -H_{\eta_i}$. For the angular momenta $\mc J_i$, they can be integrated on any surface $S$ so that the Killing vector $\eta_i$ is tangent to $S$. Assume further that $\de \eta_i^a = 0$. Then the $\bs C_{\eta_i,\gamma}$ term is zero. This leaves
\begin{align}
    \mc J_i &= - H_{\eta_i} = \oint_S \bs K^K_{\eta_i} - \oint_S \overline{\bs K^K_{\eta_i}}.
\end{align}
BC point out that in the particular case of Kerr--AdS, the background Komar integral does not contribute. (I will elaborate on this point; see \eqref{angularmomentumPureAdS}.) This means that we recover
\begin{align}
    \mc J_i &= \oint_S \bs K_{\eta_i}^K,
\end{align}
on any integration surface (which encloses the singularity). Of course this includes $S\to S_\infty$, which is how GPP calculated theirs. 

For $\mc E$ we have, similarly making the assumption that $\de \xi^a = 0$ (that there is no change in the coordinate expression of the vector $\xi^a$), 
\begin{align}
    \mc E &= H_\xi = -\oint_S \bs K_\xi^K + \oint_S \overline{\bs K_\xi^K} + \oint_S \bs C_{\xi,\gamma}.
\end{align}
We can then combine $\mc E - \sum_i \Om_i \mc J_i$ to give
\begin{align}
    \mc E - \sum_i \Om_i \mc J_i &= H_\xi + \sum_i \Om_i H_{\eta_i} \nn 
    &= -\oint_H \bs K^K_\xi + \oint_H \overline{\bs K^K_\xi} + \oint_H \bs C_{\xi;\g} + \sum_i \Om_i \left( -\oint_H \bs K^K_{\eta_i} + \oint_H \overline{\bs K^K_{\eta_i}} + \oint_H \bs C_{\eta_i;\g}\right) \nn 
    &= - \oint_H \bs K^K_\z + \oint_H \overline{\bs K^K_\z} + \oint_H \bs C_{\z;\g}.
\end{align}
Using \eqref{KomarArea}, this becomes
\begin{align}
    \mc E - \sum_i \Om_i \mc J_i &= \f{\kappa A}{8\pi} + \oint_H \overline{\bs K^K_\z} + \oint_H \bs C_{\z;\g}. \label{BCSmarr}
\end{align}
They refer to this as the generalized Smarr relation. 

BC then go on, after doing an example calculation for the Schwarzschild black hole, to calculate the Smarr relation associated with the first law for the Kerr--AdS black hole. Since the combination $\oint_S \bs K_{\z;\g}$ is independent of integration surface $S$, if the combination $-\oint_S \bs K^K_\zeta + \oint_S \overline{\bs K^K_\zeta}$ is independent of the integration surface, then $\oint_S \bs C_{\zeta,\gamma}$ will also independent of $S$. This means that it can be calculated anywhere, the most obvious locations being either on the horizon itself (possibly even at the bifurcation two-sphere, as BC do for the Schwarzschild BH) or at infinity. In the case of Kerr--AdS, they find that this is the case, and perform the calculation of $\bs C_{\xi;\gamma}$ at infinity in calculating the energy $\mc E$, and also show (see \eqref{CgammaCxi} below) that $\oint_H \bs C_{\z;\g} = \oint_{S_\infty} \bs C_{\xi;\g}$.

Here I will briefly show the calculation they perform for the Kerr--AdS charges/Smarr relation. They use the form of the Kerr--anti-de Sitter metrics in higher dimensions as obtained in \cite{GibbonsLu2}, which I refer to as the ABL form \eqref{ABL}. I modify BC's notation somewhat to conform with mine. I will refer to their metric correction as $h_{ab}$, dropping the BC superscript for compactness, but it is important to note that they are not using the Kerr--Schild form of the metric and that $h_{ab}$ is consequently not of Kerr--Schild form. Their background-and-correction decomposition is as follows:
\begin{align}
    g_{ab} &= \bar g_{ab} + h_{ab} \nn 
    d \bar s^2 &= \bar g_{ab} dx^a dx^b \nn 
    d s^2 &= g_{ab} dx^a dx^b.
\end{align}
$d \bar s^2$ is the metric for pure AdS in spheroidal coordinates as given by \eqref{dbarsspheroidal}. They keep the value of $a_i$ constant, and so are effectively comparing the solution with $m, a_i$ to that with $m=0,a_i$. The correction term $h_{ab}$ is given by
\begin{align}
    h_{ab} dx^a dx^b &= \f{2m}{U} \left( W d\tau - \sum_{i=0}^{n-1+\ve} \f{a_i \mu_i^2}{\Xi_i} d \hat \vp_i\right)^2 + \f{2mU}{V(V-2m)} dr^2 
\end{align}
The values of $U,V,W$ are as in Section \ref{KAdSForms}.

BC choose, as an integration path, $g^{(s)}_{ab} = \bar g_{ab} + s h_{ab}$, where $s \in [0,1]$. They then calculate $\mc E$ and $\mc J_i$ (given simply by the Komar integration, without background subtraction) at $S_\infty$. Defining $h^a_b \equiv \bar g^{ac} h_{cb}$, the inverse metric has expansion
\begin{align}
    g^{mn} &= \bar g^{ma}\left( \de^n_a - h^n_a + h^b_a h^n_b - h^c_a h^c_b h^n_b + \ldots \right).
\end{align}
The asymptotic behaviour of $h_{ab}$ guarantees $h^a_b = \mc O(r^{-D+1})$ at large $r$. This means that, at infinity, the only terms which will contribute to $g^{mn}$ will be $\bar g^{ma}$ and the term linear in $h^n_a$. They also show that in the calculation of $\oint_{S_\infty} \bs C_{\xi;\gamma}$ only the terms linear in $h_{ab}$ contribute. Away from infinity, more care would be required in integrating through solution space due to the fact that $g^{ab} - \bar g^{ab}$ is not linear in $h_{ab}$. 

The results are that $\mc J_i$ is as in \eqref{JiKomar}, as well as
\begin{align}
    \oint_{S_\infty} (-\bs K_\xi^K + \overline{\bs K^K_\xi}) &= \f{m \mc A_{D-2}}{4\pi \prod_i \Xi_i} \left( \sum_{i=1}^{n-1+\ve} \f{1}{\Xi_i} - \f{1+\ve}{2}\right) \nn
    \oint_{S_\infty} \bs C_{\xi;\gamma} &= \f{m \mc A_{D-2}}{8\pi \prod_i \Xi_i}. \label{CzetagammaBC}
\end{align}

For the calculation of $\bs C_{\z;\gamma}$, BC point out that since $\sqrt{-g} = \sqrt{-\bar g}$, $d (\bs K^K_\z -\overline{\bs K^K_\z}) = 0$, so that $\oint_S (\bs K^K_\z - \overline{\bs K^K_\z}) = 0$. This implies that $\oint_S \bs C_{\z;\g}$ is independent of integration surface $S$. This means it can be evaluated at $S_\infty$, so that
\begin{align}
    \oint_H \bs C_{\z;\g} &= \oint_{S_\infty} \bs C_{\z;\g} \nn 
    &= \oint_{S_\infty} \left( \bs C_{\xi;\g} + \sum_i \Om_i \bs C_{\eta_i;\g}\right)\nn 
    &= \oint_{S_\infty} \bs C_{\xi;\g}. \label{CgammaCxi}
\end{align}

Finally calculating 
\begin{align}
    \oint_H \overline{\bs K^K_\xi} &= -\f{\mc A_{D-2}}{8\pi l^2 \prod_i \Xi_i} r_+^{1-\ve} \prod_{i=1}^{n-1+\ve} (r_+^2+a_i^2),
\end{align}
BC find
\begin{align}
    \mc E - \sum_i \Om_i \mc J_i &= \f{\kappa A}{8\pi} + \f{\mc A_{D-2}}{\prod_i \Xi_i} \left( m - \f{r_+^{1-\ve}}{l^2} \prod_{i=1}^{n-1+\ve} (r_+^2+a_i^2)\right). \label{SmarrGibbsDuhem2}
\end{align}
This recovers \eqref{SmarrGibbsDuhem}, matching the results calculated by GPP using Euclidean methods, with the final term being equal to the thermodynamic potential $\Phi$; see Section \ref{EuclideanAction}. 

The Kerr/Myers--Perry black holes are addressed as the $l \to \infty$ limit of the KAdS BHs. In this case, $\oint_S \overline{\bs K_\xi^K} = 0$ and $\oint_S \bs C_{\xi,\gamma} = (D-2)^{-1} \mc E$, and this leads to the usual Smarr relation for the Kerr/Myers--Perry black holes, \eqref{SmarrZeroLambda}.

\subsection{Generalization to Einstein--Maxwell}

In Comp\`ere \cite{Compere} the generalization of \eqref{firstlaw} to the Einstein--Maxwell case in four dimensions is included. In this case, contributions from the Einstein--Hilbert (EH) Lagrangian and from the electromagnetic (EM) terms can be separated out. We will consider solutions to the Einstein--Maxwell equations, which correspond to electrovacuum (possibly with $\La$).

In Einstein--Maxwell theory, we have, where $\chi$ represents a generic Killing vector, letting $\phi$ represent the fields (both $g_{ab}$ and $\bs A_a$),
\begin{align}
    \de H_\chi &= \oint_S \bs k_\chi[\de \phi;\phi] \nn 
    &= \oint_S \left( \de \bs Q_\chi - \bs Q_{\de \chi} - \chi \cdot \bs \Th[\de \phi;\phi]\right)
\end{align}
Define 
\begin{align}
    \bs k_\chi^{EM}[\de \phi;\phi] &= \de \bs Q^{EM}_\chi - \bs Q^{EM}_{\de \chi} - \chi \cdot \bs \Th^{EM}[\de \phi;\phi]
\end{align}
using \eqref{NoetherChargeEM} and \eqref{ThetaEM}. Then
\begin{align}
    \bs k_\chi[\de \phi;\phi] &= \bs k_\chi^{EH}[\de g;g] + \bs k_\chi^{EM}[\de \phi;\phi].
\end{align}

Using \eqref{kappadaBC2} we can then write
\begin{align}
    \f{\kappa}{8\pi} \de A &= \oint_H \bs k^{EH}_{\z}[\de g;g] \nn 
    &= \oint_H \left(\bs k_{ \z} [\de \phi;\phi]- \bs k_{ \z}^{EM} [\de \phi;\phi]\right) \nn 
    &= \oint_H \left(\bs k_{\bar \z}[\de \phi;\phi] - \bs k_\z^{EM}[\de \phi;\phi]\right).
\end{align}
Using $\bar \z = \xi + \sum_i \Om_i \eta_i$, as in \eqref{firstlaw},
\begin{align}
    \f{\kappa}{8\pi} \de A &= \oint_H \left(\bs k_\xi [\de \phi;\phi] + \sum_i \Om_i \bs k_{\eta_i} [\de \phi;\phi] - \bs k_{ \z}^{EM}[\de \phi;\phi]\right) \nn 
    &= \de \mc E - \sum_i \Om_i \de \mc J_i - \oint_S \bs k_{ \z}^{EM} [\de \phi;\phi].
\end{align}

On the horizon $H$, the electric potential $\Phi^{EM}$ is given by
\begin{align}
    \Phi^{EM} &\equiv -\bs A_a \z^a, \label{PhiEM}
\end{align}
constant on $H$. Thus $\bs Q^{EM}_\z$ reduces to
\begin{align}
    \bs Q^{EM}_\z|_H &= \f{1}{8\pi} * \bs F \: \Phi^{EM},
\end{align}
so that, using \eqref{mcQdef} and the fact that $d * \bs F = 0$ (electrovacuum),
\begin{align}
    \oint_H \bs Q^{EM}_\z &= \mc Q \Phi^{EM}, \label{intHQEM}
\end{align}
where $\mc Q$ is the electric charge. 

We then can write
\begin{align}
    \oint_H \bs k^{EM}_{\z}[\de \phi;\phi] &= \de \oint_H \bs Q^{EM}_\z - \oint_H \bs Q^{EM}_{\de \z} - \z \cdot \bs \Th^{EM} \nn 
    &=  \de (\mc Q \Phi^{EM}) - \oint_H \bs Q^{EM}_{\de \z} - \z \cdot \bs \Th^{EM}. \label{intHkEMzeta}
\end{align}

After some algebra,
\begin{align}
    (-\bs Q^{EM}_{\de \z} - \z \cdot \bs \Th^{EH})_{ab} &= \f{1}{8\pi} \bs \ep_{abcd} \left( \bs F^{cd} \bs A_e \de \z^e + 2 \z^c \bs F^{de} \de \bs A_e\right) \nn 
    \oint_H (-\bs Q^{EM}_{\de \z} - \z \cdot \bs \Th^{EH}) &= \f{1}{8\pi} \oint_H \left( \bs F^{cd} \bs A_e \de \z^e + 2 \z^c \bs F^{de} \de \bs A_e\right) dS_{cd}.
\end{align}
Using the decomposition \eqref{horizonbinormal}, this becomes
\begin{align}
    \oint_H (-\bs Q^{EM}_{\de \z} - \z \cdot \bs \Th^{EH}) &= \f{1}{4\pi} \oint_H \left( \bs F^{cd} \bs A_e \de \z^e + 2 \z^c \bs F^{de} \de \bs A_e\right) \z_{[c}n_{d]} dS \nn 
    &= \f{1}{4\pi} \oint_H \z_c \bs F^{cd} \left( n_d \bs A_e \de \z^e - \de \bs A_d\right).
\end{align}
As stated by Comp\`ere, the zeroth law of black hole mechanics extended to electromagnetism includes $\Phi^{EM} = -\bs A_a \z^a$ is constant. This implies that $\lie_\z \bs A = 0$, from which (on the horizon) $\z^b \bs F_{b a} = - \na_b (\z^a \bs A_a)$. The fact that $\z^a \bs A_a$ is constant on the horizon implies that $\z^b \bs F_{b a}$ contracted with any vector tangent to the horizon will be zero, which implies that $\z^b \bs F_{ba}$ is parallel to $\z_a$. Specifically Comp\`ere gives $\bs F^{ab} \z_b = \bs F^{b c} \z_b n_c \z^a$ on the horizon. Plugging this in gives
\begin{align}
    \oint_H (-\bs Q^{EM}_{\de \z} - \z \cdot \bs \Th^{EH}) &= -\f{1}{8\pi}\oint_H \bs F^{bc} \z_b n_c \z^d (n_d \bs A_e \de \z^e + \de \bs A_d) dS \nn 
    &= +\f{1}{8\pi} \oint_H \bs F^{bc} \z_b n_c dS \left( \bs A_e \de \z^e + \z^d \de \bs A_d\right) \nn 
    &= -\f{1}{8\pi} \oint_H \bs F^{ab} d S_{ab} \de \Phi^{EM} \nn 
    &= - \mc Q \de \Phi^{EM}. \label{QdPhi}
\end{align}
As with the way $\de \kappa$ reduces nicely to terms related to $\de \z^a$ and terms related to $\de g_{ab}$, $\de \Phi^{EM}$ reduces nicely to terms related to $\de \z^a$ and $\de \bs A_a$, and those terms match up to the $\bs Q^{EM}_{\de \z}$ and $\z \cdot \bs \Th^{EH}$ terms (respectively). Plugging \eqref{QdPhi} into \eqref{intHkEMzeta}, we have
\begin{align}
    \oint_H \bs k^{EM}_{\z}[\de \phi;\phi] &= \Phi^{EM} \de \mc Q.
\end{align}
We thus recover
\begin{align}
    \f{\kappa}{8\pi} \de A &= \de \mc E - \sum_i \Om_i \de \mc J_i - \Phi^{EM} \de \mc Q,
\end{align}
matching \eqref{firstlawwithThetadLambda} with a single charge (with no variation in $\La$).

\subsection{Bifurcation Surface}

The bifurcation surface of a black hole with non-degenerate horizons ($\kappa \neq 0$) is the $(D-2)$-surface on the horizon where the Killing vector tangent to the null generators vanishes. In various papers (for example, \cite{Wald93,IyerWald95}), Wald and others  use the bifurcation surface to make statements and proofs about black hole thermodynamics, since many otherwise cumbersome terms are exactly zero there. Specifically, in these papers Wald et al.~show that the first law of black hole mechanics is satisfied given the conserved charges $\mc E, \mc J_i$ defined using the covariant phase space methods, using the bifurcation surface to make statements about horizon-related quantities such as $\kappa$ and $A$. The disadvantage of using the bifurcation surface for calculations for my purposes is that I will generally want to use properties of the metric in a more manifestly stationary form, in which the Killing vectors are coordinate-independent. In this form of the metric the Killing vectors do not vanish anywhere, which is to say that the bifurcation surface is excluded by the coordinate charts I am considering (though not by the manifold). Additionally, the statements about the Smarr relation made by BC which will make up the following section do not rely on using the bifurcation surface.

(Another approach, taken by Hajian and Sheikh-Jabbari \cite{HajianSheikh-Jabbari}, is to show that the charge associated with the Killing vector $2 \pi \z/\kappa$ (where $\z$ and $\ka$ are allowed to vary) is integrable, even away from the horizon, and in so doing they, to a degree, ``do away with the horizon.'' My focus is still on the horizon proper.)

\section{Smarr Relation Revisited} \label{smarrrevisited}

In this section I provide my own commentary on the Smarr relation, first using the expressions associated with the conserved charges for Killing vectors $(\xi,\eta_i$) corresponding to the asymptotically-static coordinates, and then to the Killing vectors $(\beta, \eta_i)$ corresponding to the asymptotically-rotating coordinates.  

\subsection{Asymptotically Static Killing Vector} \label{SmarrAsymptoticallyStatic}

In order for a Smarr relation to be able to coincide with the scaling relation, the coefficients on the energy, angular momenta etc.~must correspond to their coefficients in the first law. Thus one way to rewrite \eqref{BCSmarr} is
\begin{align}
    (D-2) \left(\mc E - \sum_{i} \Omega_i \mc J_i\right) &= (D-2) \frac{\kappa A}{8\pi} + (D-2) \oint_H \overline{\bs K^K_\zeta} + (D-2) \oint_H \bs C_{\zeta;\gamma} \nn 
    (D-3) \mc E - (D-2) \sum_i \Omega_i \mc J_i &= (D-2) \frac{\kappa A}{8\pi} + (D-2) \oint_H \overline{\bs K^K_\zeta} + \left( (D-2) \oint_H \bs C_{\zeta;\gamma} - \mc E\right) \label{BCSmarrDm3version}
\end{align}

Now assume that we have made a coordinate choice so that $\sqrt{-\bar g} = \sqrt{-g}$, that is, the way the background and full spacetimes are related to each other preserves the metric determinant. For Kerr--AdS with background AdS, this is the case for the use of the ABL coordinates that BC use, and is also the case if we use KS coordinates or (as I stated in Chapter \ref{PRDPaper}) a Kerr--Schild decomposition of the form $g_{ab} = \bar g_{ab} + H k_a k_b$ in general. In the case of both ABL and KS coordinates the statement that the metric determinant is unchanged relies on using the spheroidal coordinates for the background AdS spacetime that uses the same values of $a_i$ as in the full Kerr--AdS spacetime. 

At this point, we note that, if the background spacetime is pure AdS, that $\oint_H \overline{\bs K_\zeta^K}$ has a clear interpretation in terms of the volume. Consider a region $\Si$ of the background AdS spacetime with no curvature singularities. Using \eqref{KomarDifferenceXi} and $\tilde \La = -(D-1)/l^2$, we have
\begin{align}
    \oint_{\pa \Si} \overline{\bs K^K_\xi} &= \f{1}{8\pi} \tilde \La \mc V_{\xi,\Sigma},
\end{align}
for a generic Killing vector $\xi$, where $\mc V_{\xi,\Si}$ is the vector volume of the region $\Si$. Here I also used the fact that the vector volume as calculated in the full and background spacetimes are equal (since $\sqrt{-g} = \sqrt{-\bar g}$).

Note further that the vector volume associated with the azimuthal symmetry Killing vectors $\eta_i$ vanishes under most circumstances. Consider, in particular, the case where we represent the AdS background in KS coordinates, and the coordinate $t$ is constant on $\Si$. Let the boundary of $\Si$ be $C \equiv \pa \Si$. In this case 
\begin{align}
    \oint_{\pa \Si} \overline{\bs K_{\eta_i}^K} \propto \int_\Si \eta^a d \Si_a = 0. \label{angularmomentumPureAdS}
\end{align}
This follows from the fact that $\int_\Sigma \eta_i^a d \Si_a = 0$ in this case, which was discussed in Section \ref{twovectorsSection} in terms of the non-contribution of azimuthal terms. This can be seen by the fact that the normal $\bar n_a$ to $\Si$ in the AdS background only has a nonzero $\bar n_t$ component, and so $\overline{\eta_i^a n_a} = 0$ and so $\overline{\eta_i^a d \Sigma_a} = 0$. This tells us that angular momentum Komar integrals for the pure AdS background are zero, which makes sense physically: by symmetry, one would not expect the AdS background to be spinning. In fact the constant-$t$ requirement is more restrictive than is strictly necessary, but is sufficient for our purpose. 

Let $S_0$ be $(D-2)$-surface in KS coordinates consisting of constant $t$ and with $r = 0$ (if $D$ is even) or $r^2+a_n^2 = 0$ (if $D$ is odd). The argument will be made in more detail in Chapter \ref{VolumeAreaChapter}, but it turns out that the region ``interior to $S_0$'' has zero spatial volume and thus zero volume in the AdS background. This means that we expect $\oint_{S_0} \overline{\bs K^K_\xi} = 0$. I will show that $\oint_{S_0} \overline{\bs K^K_\beta} = 0$ as follows. Equation \eqref{dbetaflatAdShv2} states that $\overline{d \beta^\flat} = 2 l^{-2} \bs h$ and Appendix \ref{RaisingAndLoweringh} states that $\bs h^{a b}$ can be found from $\bs h_{ab}$ while raising and lowering using either the full or background metric. Consequently we have
\begin{align}
    \overline{ \na^a \beta^b} &= \f{1}{l^2} \bs h^{a b} = \f{1}{l^2} \overline{\bs h^{a b}}.
\end{align}
We then have that $\overline{\bs K_\beta^K}$ is given by
\begin{align}
    (\overline{\bs K_\beta^K})_{c_1 \ldots c_{D-2}} &= \f{1}{16\pi} \overline{\na^a \beta^b} \bs \ep_{a b c_1 \ldots c_{D-2}} \nn 
    &= \f{1}{16\pi l^2} \bs h^{a b} \bs \ep_{a b c_1 \ldots c_{D-2}}.
\end{align}
I verify that the integral of $*\bs h$ on the surface $S_0$ is zero in Section \ref{killingpotentialvectorvolumerelationship}, so that $\oint_{S_0} \overline{\bs K^K_\beta} = 0$. Due to the linearity of the Komar integral with respect to the Killing vector, $\overline{\bs K^K_\xi} = \overline{\bs K^K_\beta} - \sum_{i=1}^{n-1+\ve} \f{a_i}{l^2} \overline{\bs K^K_{\eta_i}}$, and so $\oint_{S_0} \overline{\bs K^K_\xi} = 0$ as well.

Let $\mc B$ refer to a constant-$t$ hypersurface in KS coordinates extending from $S_0$ to the BH horizon at $H$. Then we have
\begin{align}
    \oint_H \overline{\bs K_\xi^K} &= \oint_H \overline{\bs K_\z^K} = \f{1}{8\pi} \tilde \La \mc V_{\xi,\mc B}, \label{backgroundKomarVolumeRelation}
\end{align}
using that $\oint_{S_0} \overline{\bs K_\xi^K} = 0$ as well as the vanishing of the vector volume for the azimuthal Killing vectors. 

$\mc V_{\xi,\mc B}$ is also equal to the geometric volume $V_{geo}$ as defined in \cite{Cvetic}. We can use \eqref{ThetaprimeVgeo} to replace the $V_{geo}$ term with $\Th'$ and so \eqref{BCSmarrDm3version} becomes
\begin{align}
    (D-3) \mc E = (D-2) \left(\sum_{i} \Omega_i \mc J_i + \f{\ka A}{8\pi}\right) - 2 \Th' \tilde \La + \left( (D-2) \oint_H \bs C_{\zeta;\gamma} - \mc E\right), \label{Smarrwithextraterm}
\end{align}
in the absence of Maxwell charges. We have now isolated the term which distinguishes between the geometric volume and the thermodynamic volume in the Smarr relation as
\begin{align}
    (D-2) \oint_H \bs C_{\zeta;\gamma} - \mc E,
\end{align}
from which we conclude that the Smarr relation \eqref{Smarr} is satisfied with
\begin{align}
    \Th = \Th' - \frac{(D-2) \oint_H \bs C_{\zeta;\gamma} - \mc E}{2 \tilde \La}. \label{ThetaminusThetaPrime}
\end{align}
Generalization to include (Maxwell) charge can also be done and I will revisit this point in the particular case of four-dimensional Kerr--Newman--AdS.

Thus the appearance of the geometric/vector volume in the Smarr relation comes down to the existence of the background Komar integral. The connection between the geometric volume and the difference between the Komar integral for the full and background spacetime is also  discussed in \cite{Xiao}. 

\subsection{Using the Principal Vector} \label{SmarrPrincipal}
Of course $\mc E$ is the conserved charge associated with the vector $\xi = \pa_t$ and $\mc J_i$ are the charges associated with the vectors $-\eta_i = -\pa_{\phi_i}$. If instead of $\mc E$ we base our expressions on $\beta = \partial_t + \sum_i a_i l^{-2} \partial_{\phi_i}$, we get somewhat different results. 

Recall $\mc F$ defined by \eqref{Fdefinition}. $\mc E$ and $\mc J_i$ are, in BC's method, defined differently from the AMD prescription in \eqref{Fdefinition}, but because the BC charge definitions are also linear in the Killing vector, we can similarly define $\mc F$ in the BC method so that it also satisfies \eqref{Fexpression},
\begin{align}
    \mc F &\equiv -\oint_S \bs K^K_\beta + \oint_S \overline{\bs K^K_\beta} + \oint_S \bs C_{\beta;\gamma},
\end{align}
so that $\mc F$ also automatically satisfies \eqref{FminusEminus}. We can also expand $\z$ as \eqref{zetabreakdown} in terms of the $\om_i$, given in \eqref{Omegaomega}. I will demand for the moment that $\bt^a$ has unvarying components when comparing different spacetimes, and further demand as before that we make comparisons between the full Kerr--AdS and background AdS metrics with the same $a_i$ parameters. 

Using the same argument as for $\mc E$, but with $\mc E \to \mc F, \Om_i \to \om_i$, we then arrive at 
\begin{align}
    (D-3) \mc F - (D-2) \sum_i \om_i \mc J_i &= (D-2) \f{\ka A}{8\pi} + (D-2) \oint_H \overline{\bs K^K_\z} + \left( (D-2) \oint_H \bs C_{\z;\gamma} - \mc F\right),
\end{align}
or, following \eqref{Smarrwithextraterm},
\begin{align}
    (D-3) \mc F &= (D-2) \left( \sum_i \om_i \mc J_i + \f{\ka A}{8\pi}\right) - 2 \Th' \tilde \La + \left( (D-2) \oint_H \bs C_{\z;\gamma} - \mc F\right). \label{SmarrF}
\end{align}
Comparing to \eqref{SmarrwithF} (with no Maxwell charges), we then conclude that, for Kerr--AdS specifically,
\begin{align}
    (D-2) \oint_H \bs C_{\z;\gamma} &= \mc F \qquad \textrm{(Kerr--AdS)}. \label{D-2C=F}
\end{align}
It is remarkable that $\oint_H \bs C_{\z;\gamma}$ takes such a simple form. We can now rewrite \eqref{ThetaminusThetaPrime} as
\begin{align}
    \Th &= \Th' - \f{\mc F - \mc E}{2 \tilde \La} \nn 
    &= \Th' + \f{1}{2 \tilde \La} \sum_i a_i \mc J_i.
\end{align}
Thus the deviation of $\Th$ from the value $\Th'$ which is proportional to the black hole volume is given by the above. We can say that it is a consequence of the fact that the relationship \eqref{D-2C=F} holds, which leads to the black hole volume $\mc V_{\xi,\mc B} = V_{geo}$ appearing in the Smarr relationship for $\mc F$ but not the one for $\mc E$.

If $(D-2) \oint_H \bs C_{\z;\gamma}$ should be equal to an ``energy'' term, why should it be equal to $\mc F$ rather than $\mc E$?  I will revisit this question in Section \ref{ComparisonBetweenKomarAndNonKomarTerms}. 

\section{Use of the Kerr--Schild Form for the BC Method} \label{KSBC}

Assume now, in the BC method, that a Killing vector is chosen to have constant contravariant components, $\chi^a = \bar \chi^a = (\chi^{(s)})^a$. The calculation of $\bs Q_\chi$ and $\overline{\bs Q_\chi}$ is relatively straightforward. In general, $\bs C_{\chi;\gamma} = -\int_0^1 ds\: \chi \cdot \bs \Th\left[h^{(s)};g^{(s)}\right]$ might be a complicated expression. This is why BC do their calculation at infinity for the Kerr--AdS metric in the form they worked with. Under their program of sending $m \to s m$ and allowing $s$ to vary from 0 to 1, the values $g^{(s)}_{ab} = \bar g_{ab}$ and $(g^{(s)})^{ab} - \bar g^{ab}$ are generally nonlinear in $m$ (and thus in $s$), except at infinity. This complicates the calculation and interpretation of the integral at finite radius. 

One workaround I devised is to work with a metric where the correction to the background is exactly linear in $s$. This occurs for the Kerr--Schild class of metric. As before, say we have a Kerr--Schild metric with 
\begin{align}
    g_{ab} &= \bar g_{ab} + h_{ab} \nn 
    g^{ab} &= \bar g^{ab} - h^{ab},
\end{align}
where $h_{ab} = H k_a k_b$ for some scalar function $H$ and some geodesic null vector $k_a$. (We will, for later convenience, impose that $k_a$ is affinely parametrized.) Then $h_{ab}$ can be raised and lowered with either metric $g_{ab}$ or $g^{ab}$, with $h^a_b = g^{ac} h_{cb} = \bar g^{ac} h_{cb}$ and $h^{ab} = g^{cb} h^a_c = \bar g^{cb} h^a_c$. Furthermore, if $k^a$ is an affinely parametrized geodesic vector with respect to $\bar g_{ab}$, it will be an affinely parametrized geodesic vector with respect to $g_{ab}$.

If $\chi$ is a Killing vector with respect to both the background and full metric, then to calculate conserved EH charges we can choose the path through solution space to be 
\begin{align}
    g^{(s)}_{ab} &= \bar g_{ab} + s h_{ab} \nn 
    &= \bar g_{ab} + s H k_a k_b.
\end{align}
For simplicity we demand that the same coordinates are used for the background, intermediate, and full spacetime in the sense that $\bar g_{ab}$ and $h_{ab}$ are independent of $s$. Thus $g^{(s)}_{ab}$ is a Kerr--Schild perturbation to $\bar g_{ab}$. Under the assumption that $k^a$ is tangent to affinely parametrized null geodesics in the background spacetime, $k^a$ will also be tangent to affinely parametrized geodesics not just in the full spacetime ($s = 1$) but also in all the intermediate spacetimes ($s \in [0,1]$), which after all just amounts to replacing the scalar function $H$ in $g_{ab}$ with $s H$. 

We thus must also have
\begin{align}
    (g^{(s)})^{a b} &= \bar g^{a b} - s h^{a b}.
\end{align}
We then have
\begin{align}
    h^{(s)}_{ab} = \frac{d}{ds} g^{(s)}_{ab} = h_{ab}.
\end{align}
Thus we have (again specified for $\de \chi^a = 0$)
\begin{align}
    \bs C_{\chi;\gamma} &= - \chi\cdot \int_0^1 ds \bs \Th \left[h;g^{(s)}\right]. \label{Cintegraldeltaxi0}
\end{align}

In pure Einstein gravity, we can choose $\bs \Th = \bs \Th^{EH}$. Then we have 
\begin{align}
    \bs \Th[h; g^{(s)}] &= (d^{D-1} x)_a \frac{\sqrt{-g^{(s)}}}{16 \pi} \left( \na^{(s)}_c h^{ac} - (\na^{(s)})^a h\right),
\end{align}
where I am letting $\na^{(s)}_a$ be the covariant derivative adapted to the metric $g^{(s)}_{ab}$. $h = (g^{(s)})^{ab} h_{ab} = 0$, so that the second term in the parentheses vanishes. The first term at first appears to be tricky, but we will show that we can replace $\na^{(s)}_a$ with either $\bar \na_a$ or $\na_a$. It is also true that in a Kerr--Schild situation, the background and full metric determinants are the same, so $\sqrt{-\bar g} = \sqrt{-g} = \sqrt{-g^{(s)}}$. We then have,
\begin{align}
    \na^{(s)}_a h^{ab} &= \na^{(s)}_a (H k^a k^b) \nn 
    &= k^b \na^{(s)}_a(Hk^a) + H k^a \na^{(s)}_a k^b.
\end{align}
The last term vanishes because $k^a$ is an affinely parametrized geodesic vector with respect to $g^{(s)}_{ab}$.  What remains is
\begin{align}
    \na^{(s)}_a h^{ab} &= k^b \f{1}{\sqrt{-g^{(s)}}} \pa_a (\sqrt{-g^{(s)}} k^a) \nn 
    &= k^b \f{1}{\sqrt{-\bar g}} \pa_a \sqrt{-\bar g} k^a) \nn 
    &= k^b \bar \na_a (H k^a) \nn 
    &= k^b \bar \na_a (Hk^a) + H k^a \bar \na_a k^b \nn 
    &= \bar \na_a h^{ab}.
\end{align}
Thus $\na^{(s)}_a h^{ab} = \bar \na_a h^{ab}$, for any $s$, including $s = 1$. Let us call this quantity $(V^{EH})^b$:
\begin{align}
    (V^{EH})^b &\equiv \bar \na_a h^{ab}, \label{VEHdef} 
\end{align}
satisfying
\begin{align}
    (V^{EH})^b &= k^b \bar \na_a(H k^a). \label{VEHproptok}
\end{align}
We could replace $\bar \na_a$ with $\na^{(s)}_a$ or $\na_a$ in \eqref{VEHdef} or \eqref{VEHproptok}.

Consequently, this expression reduces to
\begin{align}
    \bs \Th^{EH} [h ;g^{(s)}] = \bs \Th^{EH} [h;g] = \bs \Th^{EH} [h;\bar g] = (d^{D-1}x)_a \frac{\sqrt{-g}}{16\pi} \na_b h^{ab}.
\end{align}
Let us call this quantity $\bs \tht^{EH}$ and emphasize that it is valid not just for infinitesimal $h_{ab}$ values but for larger ones as well. Explicitly,
\begin{align}
    \bs \tht^{EH}[h,g] &= (d^{D-1} x)_a \frac{\sqrt{-g}}{16\pi} \na_b h^{ab},
\end{align}
where we could replace $g$ with $\bar g$ in the above as well (because of the Kerr--Schild form). We can also write
\begin{align}
    \bs \tht^{EH}[h;g] &= \f{1}{16\pi} V^{EH} \cdot \bs \ep \nn 
    &= \f{1}{16\pi} * (V^{EH})^\flat. \label{thetaEH}
\end{align}
Recall that $\bs \ep = \bar{\bs \ep} = \bs \ep^{(s)}$ since the metric determinant is the same for all three because of the Kerr--Schild form. 

This is constant in $s$ so the integral just gets a factor of 1, and so \eqref{Cintegraldeltaxi0} becomes
\begin{align}
    \bs C_{\chi;\gamma} &= - \chi \cdot \bs \tht^{EH}[h;g], \label{CchigammathetaEH}
\end{align}
or
\begin{align}
    (\bs C_{\chi;\gamma})_{a_1 \ldots a_{D-2}} &=  \f{1}{16\pi} \chi^c (V^{EH})^d \bs \ep_{c d a_1 \ldots a_{D-2}} \nn 
    \bs C_{\chi;\gamma} &= \f{1}{32 \pi} * (\chi ^\flat \wedge (V^{EH})^\flat).
\end{align}
(The sign results from the order in which the contractions are performed.)

For the actual Kerr--Schild forms of the Kerr--AdS metrics (in arbitrary dimensions), the idea would then be to start with the metric of the background in spheroidal coordinates, with the same values of $a_i$ as in the full Kerr--AdS metric. Thus the background metric will in general depend on the rotational parameters $a_i$ but not the mass $m$. $h_{ab}$ is linear in $m$, and so if $g_{ab} = \bar g_{ab} + h_{ab}$ is a vacuum solution to Einstein's equations with $\La$ with Kerr--AdS parameters $(m,a_i)$, the solution corresponding to $g^{(s)}_{ab} = \bar g_{ab} + s h_{ab}$ will be a vacuum solution to Einstein's equations with $\La$ with Kerr--AdS parameters $(s m, a_i)$. While this will be a vacuum solution to Einstein's equations, it is not guaranteed to have all the properties we might desire a physically reasonable solution to have. (For instance, for sufficiently small nonzero $m$ but fixed $a$, $(r^2+a^2)(1+r^2/l^2)-2mr = 0$ could have no positive real roots, corresponding to a solution with no event horizon and thus a naked singularity, and so this will be achieved for small values of $s$.) It might be that some components of the metric are not real.  I will assume that we can integrate through \emph{possibly physically unreasonable} spacetimes which nevertheless are vacuum solutions to Einstein's equations, in integrating to define the Hamiltonian.

The advantage of this choice of path through solution space is that it allows for an expression for $\bs C_{\chi;\gamma}$ which is relatively simple and easily calculated, and which is defined everywhere in the manifold (not just at infinity). This allows for conserved quantities associated with a given Killing vector to be calculated on an arbitrary integration surface. (This was true in principle for $\bs C_{\chi;\g}$ as well, but it is now easier to calculate the result.) Because it depends on the existence of a Kerr--Schild solution, this choice for path through solution space  has limited applicability, but fortunately the Kerr--AdS (and Generalized Kerr--AdS) spacetimes fall within it. Choosing a different path through solution space will in general lead to a different $\bs C_{\chi;\gamma}$, but if BC's strong integrability conditions hold, $\oint_S \bs C_{\chi;\gamma}$ will be independent of path. This method gives a particularly simple form for $\bs C_{\chi;\gamma}$.

(Because $h_{ab} = H k_a k_b$, we could, conceptually, consider $g_{ab}^{(s)}$ either as
\begin{align}
    g_{ab}^{(s)} = \bar g_{ab} + (s H) k_a k_b,
\end{align}
where the null vector is kept constant and the scalar function $H$ is rescaled, or as
\begin{align}
    g_{ab}^{(s)} &= \bar g_{ab} + H (\sqrt s k_a) (\sqrt s k_b),
\end{align}
where the scalar function is kept constant and the null vector is rescaled by $\sqrt s$. The latter perspective results in the exact same result for $\bs C_{\chi;\g}$.)

It is worth writing the expression we have out explicitly. Define $\bs I_\chi$ to be
\begin{align}
    \bs I_\chi &= - \bs K^K_\chi + \overline{\bs K^K_\chi} - \chi \cdot \bs \tht^{EH} \label{Ixidef1}
\end{align}
This can be expanded into components as
\begin{align}
    (\bs I_\chi)_{c_1 \ldots c_{D-2}} &= \frac{1}{16 \pi} \left(- \na^a \chi^b + \overline{\na^a \chi^b} + \chi^a \na_e h^{b e}\right) \bs \ep_{abc_1 \ldots c_{D-2}}, \label{Ixidifferenceform}
\end{align}
where calculations are made in the full metric except for $\overline{\na^a \chi^b}$ which is done in the background metric. Then from \eqref{BC3.9H} and \eqref{CchigammathetaEH},
\begin{align}
    H_\chi &= \oint_S \bs I_\chi.
\end{align}
This can be rewritten by noting that the difference between the two Komar terms can be expressed in terms of only one of the metrics (say the background) and the difference $h_{a b}$, and I will do so in Section \ref{KSChargeAlternateForms}. To specify that $H_\chi$ is calculated by integrating $\bs I_\chi$ I will denote it by
\begin{align}
    H^{\bs I}_\chi &= \oint_S \bs I_\chi. \label{HIchi}
\end{align} 

I note now (in a point I did not make sufficiently clearly in the thesis) that calling $\oint_S \bs I_\chi$ ``$H_\chi^{\bs I}$'' may be a misnomer, because $H_\chi$ is really only defined if the integrability conditions \eqref{strongintegrability2}, \eqref{integrability3} hold, and all we have done so far is to establish that under a specific integral through solution space, $\oint_S \bs I_\chi = \oint_S \bs k_\chi^{EH}\left[ \f{d g^{(s)}}{ds},g^{(s)}\right]$. As I will show in Section \ref{varyingk}, we also have that $\de \oint_S \bs I_\chi = \oint_S k_\chi^{EH}\left[ \f{d g^{(s)}}{ds},g^{(s)}\right]$ if $\de \bar g_{ab} = 0$, but not necessarily in a more general setting. Nevertheless, I will continue to use this notation going forward. I will also examine more general situations in which we expect the charges to be integrable even if $\de \bar g_{ab} \neq 0$, in Chapter \ref{ExplicitGKAdSChapter}.

This means we can write, in Einstein gravity,
\begin{align}
    \mc E &= H^{\bs I}_\xi = \oint_S \bs I_\xi \nn 
    \mc J_i &= -H^{\bs I}_{\eta_i} = -\oint_S \bs I_{\eta_i}. \label{EJiIintegrals}
\end{align}
Note that while the main application we have in mind is to Kerr--Schild spacetimes with an AdS background, there is no reason not to use it with a Minkowski background. This means that the Kerr/Myers--Perry black holes can also be expressed by \eqref{EJiIintegrals}. In this case, the Komar integrals associated with the Minkowski background vanish.

Armed with this new method for calculating the charge, I will, in Chapter \ref{ExplicitGKAdSChapter}, revisit some of the ``open questions'' described at the beginning of the chapter. Because the method is a specific form of the method developed by BC, which reproduces the same values for $\mc E, \mc J_i$ as are found by GPP, we expect that the conserved quantities in Kerr--AdS will match those of GPP. As I will discuss in Chapter \ref{ExplicitGKAdSChapter}, this is the case. The advantage of the new definition in terms of the Kerr--Schild decomposition is that it is easier to work with and its generalization to the Generalized Kerr--AdS case is straightforward.

I will now turn to elucidating some of the properties of this new definition of charge.

\subsection{Einstein--Maxwell} \label{KSFormEinsteinMaxwell}

It is worth considering the case of Einstein--Maxwell theory, as usual only in four dimensions. In addition to the metric there is a Maxwell potential one-form $\bs A$ which must also be varied. We must be careful to choose a parametrization of the path through solution space for both the metric and for $\bs A^{(s)}$. There is some freedom to make a gauge choice for $\bs A$, and one possible choice is to identify $\bs A = 0$ in pure AdS (or Minkowski, or whatever the background of the theory is). (Since pure AdS has $\bs F = d \bs A = 0$, we really require only that the background $\bs A$ be a closed form, but I will choose $\bs A = 0$.) A solution consists of a metric and potential one-form pair $(g_{a b},\bs A_a)$ and we will want to interpolate between solutions $(\bar g_{a b},0)$ and $(g_{a b},\bs A_a)$, with intermediate solutions given by $(g^{(s)}_{a b}, \bs A^{(s)}_a)$. 

The Kerr--Newman--AdS electric solution for four dimensions is described in Section \ref{GKAdSSection}. The electromagnetic potential one-form $\bs A$ from \eqref{Aproptok} can be written in the form $\bs A = \alpha k^\flat$, with $\alpha = -Qr/\rho^2$. Let us choose this form for $\bs A$. We then have the background set $(\bar g_{a b}, 0)$ and the full metric and Maxwell one-form is $(\bar g_{a b} + H k_a k_b, \alpha k_a)$. 

It is tempting to choose $(\bar g_{a b} + s H k_a k_b, s \alpha k_a)$ for the intermediate solution set $(g^{(s)}_{a b} , \bs A^{(s)}_{a})$, but this will not actually be a solution to the Einstein--Maxwell equations for intermediate values of $s$. A form of $H$ and $\alpha$ that we can use for intermediate solutions is to take $H \propto 2 m r - Q^2$, where $m$ is the mass parameter and $Q$ is the charge parameter, and $\alpha \propto Q$. If we choose $g^{(s)}_{a b} = \bar g_{a b} + s h_{a b}$ as above, then the $g^{(s)}_{a b}$ solution corresponds to $m \to s m, Q \to \sqrt{s} Q$. Thus we require that $\bs A_a \to \sqrt s \bs A_a$. So we shall choose 
\begin{align}
    (g^{(s)}_{a b}, \bs A^{(s)}_a) &= (\bar g_{a b} + s h_{a b}, \sqrt{s} \bs A_a) \nn 
    &= (\bar g_{a b} + s H k_a k_b, \sqrt{s} \alpha k_a).
\end{align}
(We can either think about this as sending $H \to s H$ and $\alpha \to \sqrt s \alpha$ when converting from $g_{a b}$ to $g^{(s)}_{a b}$, or as sending $k_a \to \sqrt{s} k_a$. The latter has the advantage that it makes clearer why $\bs A_a$ must be modified by the square root of the modification to $h_{a b}$, in that the latter is quadratic in $k_a$ and $\bs A_a$ is only linear in it. It might also be conceptually clearer to think of the $k_a$ one-form as fixed and the scalars as changing.) 

Assuming again for simplicity that we choose a Killing vector $\chi$ which does not change direction $(\de \chi = 0)$, its associated conserved quantity is given by \eqref{generalizationC}. Here we have $h_{ab}^{(s)} = h_{ab}$ as with Kerr--AdS. Here, $d \bs A_a^{(s)}/ds = (2\sq{s})^{-1} \bs A_a$. Consequently,
\begin{align}
    \bs C_{\chi;\gamma} &= - \int_0^1 ds \: \chi \cdot \bs \Th\left[ h^{(s)};g^{(s)}\right] \nn 
    &= -\oint_0^1 ds \: \chi \cdot \bs \Th \left[ h, (2 \sqrt s)^{-1} \bs A; \bar g + s h, \sqrt s \bs A\right] \nn 
    &= - \oint_0^1 ds \: \chi \cdot \left( \bs \Th^{EH}\left[ h;\bar g+s h\right] + \bs \Th^{EM}\left[h, (2 \sqrt s)^{-1} \bs A; \bar g + s h, \sqrt s \bs A\right] \right).
\end{align}
The calculation for the $\bs \Th^{EH}$ term is the same as before, so we end up with $\bs \Th^{EH} = \bs \tht^{EH}[h,g]$ (with an $h$ which now depends on both $m$ and $Q$). The calculation for $\bs \Th^{EM}$ proceeds as follows. We have
\begin{align}
    (v^{(s)})^a_{EM} &= - 4 (\bs F^{(s)})^{a b} \frac{d \bs A^{(s)}_b}{d s} \nn 
    &= - 4 (g^{(s)})^{a c} (g^{(s)})^{b d} \bs F^{(s)}_{c d} \frac{d \bs A^{(s)}_b}{d s} \nn
    &= - 4 (g^{(s)})^{a c} (g^{(s)})^{b d} \bs F^{(s)}_{c d} \f{1}{2 \sqrt s} \bs A_b.
\end{align}
For $\bs F^{(s)}$ we have
\begin{align}
    \bs F^{(s)} &= d \bs A^{(s)} \nn 
    &= d(\sqrt{s} \bs A) \nn 
    &= \sqrt{s} d \bs A \nn 
    &= \sqrt{s} \bs F.
\end{align}
Thus we have
\begin{align}
    (v^{(s)})^a_{EM} &= - 2 (g^{(s)})^{a c} (g^{(s)})^{b d} \bs F_{c d} \bs A_b.
\end{align}

Because $\bs A_b \propto k_b$, $h^{ab} \bs A_b = 0$. Thus $(g^{(s)})^{bd} \bs A_b = (g^{bd} + (1-s) h^{bd}) \bs A_b = \bs A^d$ (the contravariant components of $\bs A$ in the full metric). Thus we have
\begin{align}
    (v^{(s)})^a_{EM} = -2 (g^{(s)})^{a c} \bs F_{c d} \bs A^d.
\end{align}
Since $\bs F_{a b}$ is antisymmetric, it follows that $\bs F_{a b} \bs A^b k^a = 0$ (since $A^b \propto k^b$). Thus $\bs F_{a b} \bs A^b h^{a c} = 0$. Thus we similarly have $(g^{(s)})^{ac} \bs F_{cd} \bs A^d = ( g^{ac} + (1-s) h^{ac}) \bs F_{cd} \bs A^d = g^{ac} \bs F_{cd} \bs A^d = \bs F^{ab} \bs A_b$, where $\bs F^{ab} = g^{ac} g^{bd} \bs F_{cd}$. Consequently, 
\begin{align}
    (v^{(s)})^a_{EM} &= - 2 (g^{(s)})^{ac}\bs F_{cd} \bs A^d \nn 
    &= - 2 \bs F^{ab} \bs A_b.
\end{align}
Let us call this quantity $V_{EM}^a$:
\begin{align}
    V^a_{EM} &= -2 \bs F^{ab} \bs A_b.
\end{align}

(In fact we can go further and show that $\bs F_{a b} \bs A^b$ is proportional to $k_a$ by making use of $\bs A_a = \alpha k_a$. We have,
\begin{align}
    \bs F_{a b} \bs A^b &= \bs A^b \left( \na_a \bs A_b - \na_b \bs A_a\right).
\end{align}
The first term disappears because 
\begin{align}
    \bs A^b \na_a \bs A_b = \f 1 2 \na_a (\bs A^b \bs A_b) = \f 1 2 \na_a (\alpha^2 k^b k_b) = 0
\end{align}
since $k^b k_b = 0$. We are left with
\begin{align}
    \bs F_{a b} \bs A^b &= - \bs A^b \na_b \bs A_a \nn 
    &= - \alpha k^b \na_b (\alpha k_a) \nn 
    &= - \alpha k^b k_a \na_b \alpha \nn 
    &= - \f 1 2  k_a k^b \na_b (\alpha^2),
\end{align}
using $k^b \na_b k_a = 0$. So we have $V_{EM}^a = k^b \na_b(\alpha^2) k^a$.)

Let us call
\begin{align}
    \bs \tht^{EM} &= \f{1}{16\pi} V_{EM} \cdot \bs \ep \nn
    \bs \tht^{EM}_{a b c} &= - \f{1}{8\pi} \bs F^{d e} \bs A_e \bs \ep_{d a b c}. \label{thetaEM}
\end{align}
The conserved charge associated with a Killing vector $\chi$ in four-dimensional Einstein--Maxwell electrovacuum is from \eqref{HchiQchiversion}
\begin{align}
    H_\chi^{\bs I} &= \oint_S \bs Q_\chi - \oint_S \overline{\bs Q_\chi}  - \oint_S \chi \cdot (\bs \tht^{EH} + \bs \tht^{EM}) \nn 
    &= -\oint_S \bs K^K_\chi + \oint_S \overline{\bs K^K_\chi} + \oint_S \bs Q_\chi^{EM} - \oint_S \chi \cdot (\bs \tht^{EH} + \bs \tht^{EM}). \label{HchiEM}
\end{align}
Note that $\overline{\bs Q_\chi^{EM}} = 0$ since $\overline{\bs A} = 0$. $H^{\bs I}_\chi$ can be expressed as 
\begin{align}
    H^{\bs I}_\chi &= \oint_S  (\bs I^{EH}_\chi + \bs I^{EM}_\chi)
\end{align}
where $\bs I^{EH}_\chi$ is just the $\bs I_\chi$ from \eqref{Ixidifferenceform} and
\begin{align}
    (\bs I^{EM}_\chi)_{c_1 \ldots c_{D-2}} &= \f{1}{16\pi} \left(- 2 \bs F^{ab} \bs A_e \chi^e - 2 \chi^a \bs F^{b e} \bs A_e\right) \bs \ep_{ab c_1 \ldots c_{D-2}}. \label{EMI}
\end{align}

I will generally use $\bs I_\chi$ to mean $\bs I_\chi^{EH}$, i.e.~the differential form which gives the conserved charge associated with pure Einstein gravity, and, when considering Einstein--Maxwell specifically, write out the combination $\bs I_\chi^{EH} + \bs I_\chi^{EM}$ explicitly to show the differential form which gives the conserved charge in Einstein--Maxwell gravity. 

The expression \eqref{HchiEM} above allows us to write out the Einstein--Maxwell Smarr relation. Let $S$ be the horizon $H$ and let $\chi = \z$. Then we have
\begin{align}
    H^{\bs I}_\z &= - \oint_H \bs K^K_\z + \oint_H \overline{\bs K^K_\z} + \oint_H \bs Q_\z^{EM} - \oint_H \z \cdot (\bs \tht^{EH}+\bs \tht^{EM}) \nn 
    \mc E - \Om \mc J &= \f{\kappa A}{8\pi} + \f{\tilde \La \mc V_{\xi,\mc B}}{8\pi} + \mc Q \Phi^{EM} - \oint_H \z \cdot (\bs \tht^{EH} + \bs \tht^{EM}), \label{EMSmarr}
\end{align}
using \eqref{KomarArea}, \eqref{backgroundKomarVolumeRelation} and \eqref{intHQEM}. Here I am working with $D = 4$ only, hence the single $\Om$ and $\mc J$. (Also, in four dimensions, $\tilde \La = \La$.)

As with the pure Einstein--Hilbert gravity case, it is worth breaking this down in a way that matches with the expected Smarr relation \eqref{Smarr}, where the $\mc E$ and $\mc Q \Phi^{EM}$ have coefficients of 1 and $TS = \kappa A/8\pi$ and $\Om \mc J$ have coefficients of 2. Multiply by $2$ and rearrange. 
\begin{align}
    \mc E = 2 \left(\f{\kappa A}{8\pi} + \Om \mc J\right) + \Phi^{EM} \mc Q + \left( \f{\tilde \La \mc V_{\xi,\mc B}}{4\pi} - \mc E + \Phi^{EM} \mc Q - 2\oint_H \z \cdot (\bs \tht^{EH} + \bs \tht^{EM})\right).
\end{align}
Comparing to $\mc E = 2 (TS + \Om \mc J) + \Phi^{EM} \mc Q + \tilde \La \Th$ from \eqref{Smarr}, we expect
\begin{align}
    \Th &= \Th' + \f{\mc E - \Phi^{EM}\mc Q - 2 \oint_H \z \cdot (\bs \tht^{EH}+\bs \tht^{EM})}{2 \tilde \La}, \label{ThetaThetaprimeEM}
\end{align}
similar to \eqref{ThetaminusThetaPrime}. 

Finally we can rewrite this in terms of $\mc F$ instead of $\mc E$, with $\om$ instead of $\Om$, using $\mc F - \om \mc J = \mc E - \Om \mc J$, and the result is then
\begin{align}
    \mc F &= 2 \left( \f{\kappa A}{8\pi} + \om \mc J\right) + \Phi^{EM} \mc Q - 2 \Th' \tilde \La + \left( - \mc F + \Phi^{EM} \mc Q - 2\oint_H \z\cdot (\bs \tht^{EH} + \bs \tht^{EM})\right). \label{SmarrFEM}
\end{align}

\subsection{Alternate Forms} \label{KSChargeAlternateForms}

For a Killing vector $\chi$ in a Kerr--Schild metric form,
\begin{align}
    \na^a \chi^b - \overline{\na^a \chi^b} &= - h^{c[a} \bar \na_c \chi^{b]} + \chi^c \bar \na^{[a} h^{b]}_c \nn 
    &= -h^{c[a} \na_c \chi^{b]} + \chi^c  \na^{[a} h^{b]}_c. \label{komardifferenceformula}
\end{align}
The formula takes the same form (on the right hand side) using the covariant derivative associated with either of the metrics $\bar g_{ab}$ or $g_{ab}$. $\bar \na^a = \bar g^{a b} \bar \na_b$. 

This formula is an adaptation of terms which appear in a paper by Petrov and Katz \cite{PetrovKatz}, who wrote (changing notation to match with ours)
\begin{align}
    \sqrt{-g} \na^a \chi^b - \overline{\sqrt{- g} \na^a \chi^b} &= -\sqrt{-g} h^{c[a} \bar \na_c \chi^{b]} - \sqrt{-g} g^{c[a} (\bar \na_c g^{[b] d} g_{d e} \chi^e)
\end{align}
and eventually associate their terms, under certain circumstances, with
\begin{align}
    \sqrt{-g}\na^a \chi^b - \overline{\sqrt{- g} \na^a \chi^b} &= \sqrt{-g}(-h^{c[a} \bar \na_c \chi^{b]} - \chi^{[a} \bar \na_c h^{b] c}).
\end{align}
Note however that their calculations are under the general assumption of smallness of variation between the metrics. In Section \ref{KomarDifference} of the Appendix, I derive the formula \eqref{komardifferenceformula} for arbitrarily large, but still Kerr--Schild, differences between metrics, provided that $\chi$ is a Killing vector for both. 

With the formula \eqref{komardifferenceformula}, for EH gravity in Kerr--Schild form, we can write
\begin{align}
    (\bs I_\chi)_{c_1 \ldots c_{D-2}} &= \f{1}{16\pi} \left( h^{d[a}\bar \na_d \chi^{b]} - \chi^d \bar \na^{[a}h^{b]}_d + \chi^{[a} \bar \na_d h^{b] d}\right) \bs \ep_{c_1 \ldots c_{D-2} ab} \nn 
    &=  \f{1}{16\pi} \left( h^{d[a} \na_d \chi^{b]} - \chi^d  \na^{[a}h^{b]}_d + \chi^{[a}  \na_d h^{b] d}\right) \bs \ep_{c_1 \ldots c_{D-2} a b}. \label{Ixi}
\end{align}

We can also write $\bs I_\chi$ by defining, say, ${\bs i}_\chi^{a b}$ to be the inverse Hodge dual of $\bs I_\chi$, so that
\begin{align}
    \bs I_\chi = * \bs i_\chi. \label{Istari}
\end{align}
Then,
\begin{align}
    {\bs i}_\chi^{ab} &= \f{1}{16\pi} \left( h^{d[a} \na_d \chi^{b]} - \chi^d  \na^{[a}h^{b]}_d + \chi^{[a}  \na_d h^{b] d}\right) = \f{1}{16\pi} \left( h^{d[a}\bar \na_d \chi^{b]} - \chi^d \bar \na^{[a}h^{b]}_d + \chi^{[a} \bar \na_d h^{b] d}\right), \label{mcKab}
\end{align}
Of course we can also write (using the form \eqref{Ixidifferenceform})
\begin{align}
    {\bs i}_\chi^{a b} &= \f{1}{16\pi} \left( -\na^a \chi^b + \overline{\na^a \chi^b} + \chi^{[a} \na_e h^{b]e}\right). \label{Kxialt}
\end{align}
It will sometimes be useful to use ${\bs i}_\chi^{a b}$ instead of $\bs I_\chi$ for calculations.

\subsection{Varying the Kerr--Schild Correction} \label{varyingk}

(We will ignore the Maxwell term for this section.)

$\bs I_\chi$ was derived to give a conserved quantity which gives $\bs k_\chi^{EH}[\de g;g]$ when varied, under the specific situation where the initial and final spacetime are linearly related ($g^{(s)}_{ab} = \bar g_{ab} + s h_{ab}$ for $s$ varying from 0 to 1). It is now worth checking whether the variation holds in a more general setting: that is, under what circumstances does
\begin{align}
    \de \bs I_\chi = \bs k_\chi^{EH} [\de g;g] \label{varyingkcheck}
\end{align}
hold? 

The argument in BC is that, if the strong integrability conditions hold, the conserved charges are independent of the path found to define them, in the case where the conserved charges are vacuum solutions to Einstein's equations (more specifically, where the metric $g_{ab}$ satisfies Einstein's vacuum equations and the variation $\de g_{ab}$ satisfies the linearized Einstein's vacuum equations). I will now check the more general setting where we do not demand that Einstein's vacuum equations be satisfied. If Einstein's equations are not satisfied, we can no longer reliably interpret $\oint_S \bs I_\chi$ as referring to a conserved Hamiltonian, but the equation \eqref{kappadaBC2} relating $\oint_H \bs k^{EH}_{\bar \z}[\de g;g]$ to the variation in the area does not require Einstein's vacuum equations to be satisfied, and so it will be useful to know whether our $\bs I_\chi$ (specifically $\bs I_{\bar \z}$) can be related to the horizon variation law. 

Here I will show the validity of this directly by varying $\bs I_\chi$ between two spacetimes, which are both of Kerr--Schild form and with the same background spacetime, which has unchanging components. Specifically, let 
\begin{align}
    g_{ab} &= \bar g_{ab} + h_{ab},
\end{align}
and consider a variation that does not affect the background metric or the (contravariant) components $\chi^a$ and which leaves the final metric also in Kerr--Schild form:
\begin{align}
    \de \bar g_{ab} &= 0 \label{debargab} \nn
    \de \chi^a &= 0 .
\end{align}
In general $\de \chi_a = \de(g_{ab} \chi^b) = \chi^b \de g_{ab} \neq 0$. Finally, let $h_{ab}$ be of Kerr--Schild form $h_{ab} = H k_a k_b$, throughout the variation, allowing $H$ and $k_a$ to remain, respectively, a scalar and an affinely parametrized null geodesic vector throughout. \emph{However}, we do not require that either the covariant or contravariant components of $k$ be fixed under the variation, in that $k^a|_g$ and $k^a|_{g + \de g}$ need not be the same, so that $\de k^a = k^a|_{g+\de g} - k^a|_g$ is allowed to be nonzero (and similarly for the covariant components), though we do demand that $\de k^a$ be small enough to be treated as a linear variation. Similarly we do not require $\de H = 0$. Finally, we do not require the spacetimes to be ``on-shell'': they do not have to be solutions to Einstein's equations. 

That we consider a variation which preserves the Kerr--Schild character of the spacetime is important, because $\bs I_\chi$ is defined only for Kerr--Schild spacetimes. 

We will derive some properties we will need. The statement \eqref{debargab} also implies that any quantities derived exclusively from $\bar g_{ab}$ will have zero variation,
\begin{align}
    \de \bar g^{ab} &= 0 \nn 
    \de \bs \ep_{a_1 \ldots a_D} &= 0 \nn 
    \de \bar \G^a_{bc} &= 0.
\end{align}
The zero variation of $\bs \ep_{a_1 \ldots a_D}$ is because $\sqrt{-g} = \sqrt{-\bar g}$ due to the Kerr--Schild form. The zero variation in the Christoffel symbols implies that the variation commutes with the covariant derivative with respect to the background metric, since, using a vector $u^a$ as an example,
\begin{align}
    \de \bar \na_a u^b &= \de \left( \pa_a u^b + \bar \G^b_{a c} u^c\right) \nn 
    &= \pa_a \de u^b + \bar \G^b_{ac} \de u^c \nn 
    &= \bar \na_a \de u^b.
\end{align}
The generalization to arbitrary tensors is straightforward. By contrast, in general the variation does not commute with the covariant derivative with respect to $g_{ab}$ since $\de g_{ab} \neq 0$ so that $\de \G^a_{bc} \neq 0$.

The Kerr--Schild condition requires that $\bar g^{ab} k_a k_b = 0$ remain true under perturbation, so (discarding terms quadratic in the variation)
\begin{align}
    \de (\bar g^{ab} k_a k_b) &= 0 \nn 
    \bar g^{ab} \de (k_a k_b) &= 0 \nn 
    \bar g^{ab} (k_a \de k_b + k_b \de k_a) &= 0 \nn 
    k^a \de k_a &= 0.
\end{align}
Thus $k^a \de k_a$ is necessarily zero. (To be clear, $h_{ab}$ can be large, but $\de h_{ab}$ must be small.) We further have, since $k^a = \bar g^{ab} k_b$,
\begin{align}
    \de k^a &= \de(\bar g^{ab} k_b) \nn 
    &= \bar g^{ab} \de k_b,
\end{align}
so that we can use $\bar g^{ab}$ to raise and lower $\de k_b$. Note also,
\begin{align}
    g^{ab} \de k_a &= (\bar g^{ab} - H k^a k^b) \de k_a \nn 
    &= \bar g^{ab} \de k_a \nn 
    &= \de k^b,
\end{align}
so that we can raise and lower $\de k_a$ using either metric.

From $g_{ab} = \bar g_{ab} + h_{ab}$ and $\de \bar g_{ab} = 0$, we have 
\begin{align}
    \de g_{ab} &= \de h_{ab}.
\end{align}
Writing $h_{ab} = H k_a k_b$, we have
\begin{align}
    \de h_{ab} &= \de (H k_a k_b) \nn 
    &= k_a k_b \de H + H (k_b \de k_a +  k_a \de k_b).
\end{align}
Since both $k_a$ and $\de k_a$ can be raised and lowered using either $\bar g^{ab}$ or $g^{ab}$, so can $\de h_{ab}$ (just as $h_{ab}$ can). We also have 
\begin{align}
    \bar g^{ab} \de h_{ab} = 0,
\end{align}
from $\bar g^{ab} k_a k_b = \bar g^{ab} k_a \de k_b = 0$, and
\begin{align}
    g^{ab} \de h_{ab} &= (\bar g^{ab} - H k^a k^b) \de h_{ab} \nn 
    &= 0 \label{tracedeh}
\end{align}
since $k^a \de h_{ab} = 0$ (from $k^a k_a = k^a \de k_a = 0$).

Now consider the definition for $\bs I_\chi$ from \eqref{Ixidef1} and the definition of $\bs k^{EH}_\chi[\de g;g]$ from \eqref{kxihg1} (see also \eqref{kappadaBC2}). I will now show $\de \bs I_\chi = \bs k^{EH}_\chi[\de g;g]$. Expanding both sides, we wish to show
\begin{align}
    \de \left( - \bs K^K_\chi + \overline{\bs K^K_\chi} - \chi \cdot \bs \tht^{EH}\right) &= -\de \bs K^K_\chi - \chi \cdot \bs \Th^{EH}[\de g;g].
\end{align}
Because the background metric is unchanging under the variation, $\de \overline{\bs K^K_\chi} = 0$. We also have $\de \chi^a = 0$. We thus have
\begin{align}
    \de \bs I_\chi &= -\de \bs K^K_\chi - \chi \cdot \de \bs \tht^{EH}.
\end{align}
We wish to prove $\chi \cdot \de \bs \tht^{EH} = \chi \cdot \bs \Th^{EH}[\de g;g]$. We will do so by proving 
\begin{align}
    \de \bs \tht^{EH} = \bs \Th^{EH}[\de g;g]. \label{deltathetaEHisThetaEH}
\end{align}

We have,
\begin{align}
    \de \bs \tht^{EH}_{a_1 \ldots a_{D-1}} &= \de \left(\f{1}{16\pi} \bar \na_b h^{ab} \bs \ep_{a a_1 \ldots a_{D-1}}\right) \nn 
    &= \f{1}{16\pi} \bar \na_b \de h^{ab} \bs \ep_{a a_1 \ldots a_{D-1}}.
\end{align}

We take $\bs \Th^{EH}$ from \eqref{ThetaEH}. Since $\de g_{ab} = \de h_{ab}$ we have, also noting \eqref{tracedeh},
\begin{align}
    v_a^{EH} &= \na^b \de h_{ab} \nn 
    (v^{EH})^a &= \na_b \de h^{ab},
\end{align}
using on the last line that $\de h_{ab}$ can be raised and lowered using $g_{ab}$. 

Let $\Delta^a_{bc}$ be the difference between the Christoffel symbols associated with $g_{ab}$ and $\bar g_{ab}$,
\begin{align}
    \Delta^a_{bc} &\equiv \G^a_{bc} - \bar \G^a_{bc}.
\end{align}
We can then expand $(v^{EH})^a$ as
\begin{align}
    (v^{EH})^a &= \bar \na_b \de h^{ab} + \Delta^a_{bc} \de h^{cb} + \Delta^b_{bc} \de h^{ac}.
\end{align}
Because the determinants for the full and background spacetime are equal, $ \G^a_{ab} = \bar \G^a_{ab} = (-g)^{-1/2} \pa_a \sqrt{-g} = (-\bar g)^{-1/2} \pa_a \sqrt{-\bar g}$, so $\Delta^a_{ab} = 0$. Consequently, $\Delta^b_{bc} \de h^{ac} = 0$. Provided $k^a$ is tangent to an affinely parametrized geodesic, as is the case here, it was shown in \cite{Taub81} that
\begin{align}
    \Delta^a_{bc} k^b &= \f12 (k^b \pa_b H) k^a k_c.
\end{align}
We can then expand $\Delta^a_{bc} \de h^{cb}$ as
\begin{align}
    \Delta^a_{bc} \de h^{cb} &= \Delta^a_{bc} ( \de H k^c k^b + 2 H k^{(c} \de k^{b)}) \nn 
    &= \f{1}{2} \pa_b H k^a k_b k_c \left( \de H k^c k^b + 2 H k^{(c} \de k^{b)}\right) \nn 
    &= 0.
\end{align}
Thus we have
\begin{align}
    (v^{EH})^a &= \bar \na_b \de h^{ab} \nn 
    (\bs \Th^{EH}[\de g;g])_{a_1 \ldots a_{D-1}} &= \f{1}{16\pi} \bar \na_b \de h^{ab} \bs \ep_{a a_1 \ldots a_{D-1}}.
\end{align}
We conclude \eqref{deltathetaEHisThetaEH} and so we verify \eqref{varyingkcheck}.

The conclusion we can draw is that $\de \bs I_\chi$ actually does give $\bs k^{EH}_\chi[\de g;g]$ in the case where the variation is as considered in this section ($\de \chi^a = 0, \de \bar g_{ab} = 0$, Kerr--Schild form throughout the variation). Since $\bs I_\chi$ can be calculated anywhere, this means that in Einstein--Hilbert gravity (in vacuum, with constant $\La$), its variation should give the variation in $\mc E$ between two spacetimes. It also means that, even outside Einstein--Hilbert gravity, $\bs I_{\bar \z}$ can be calculated on the horizon of a black hole, and the variation in $\bs I_{\bar \z}$ \emph{on the black hole horizon, keeping the vector components $\bar \z^a$ constant, and evaluated at the location of the horizon in the initial coordinates}, should give $\kappa \de A/8\pi$ according to \eqref{kappadaBC2}, provided that $\de \bar g_{ab} = 0$), as long as the fact that the horizon's coordinates change does not affect the results of \eqref{kappadaBC2}.

It is important to emphasize that we have only shown \eqref{varyingkcheck} for the specific variation under consideration. A variation in which the components $\de \bar g_{ab} \neq 0$ will not necessarily satisfy \eqref{varyingkcheck} (and will in general not be expected to), even if $\bar g_{ab}$ and $\bar g_{ab} + \de \bar g_{ab}$ both are metrics for the same spacetime, as could occur, for example, if there is a coordinate transformation.

These important points will be applied to the case of a Generalized Kerr--AdS black hole in four dimensions in Section \ref{explicitIxi}.

Interestingly, the above argument does not rely on $\chi$ being a Killing vector, and only on its components not varying. (Recall that $\na^a \chi^b$ automatically appears as $\na^{[a} \chi^{b]}$ in $\bs K^K_\chi$ since it is contracted with the completely antisymmetric $\bs \ep$, so we did not need to use the Killing equation $\na^a \chi^b = \na^{[a} \chi^{b]}$.) However, $\bs k^{EH}_\chi[\de g;g]$ only has an interpretation either in terms of the variation of conserved charges or in terms of its relationship to the horizon area if $\chi$ is Killing.

It is worth reporting the full expression for $\de \bs I_\chi$ if the background metric and Killing vector are allowed to change components. The initial and perturbed spacetimes will both still be of Kerr--Schild form. Let $\bar \chi^a$ once again be the constant-component version of $\chi^a$, with $\de \bar \chi^a = 0$ and with $\bar \chi^a = \chi^a$ in the unvaried spacetime (that, is associated with $g_{ab}$ rather than $g_{ab} + \de g_{ab}$). Then we firstly have
\begin{align}
    \de \bs I_\chi &= \bs I_{\de \chi} + \de \bs I_{\bar \chi}. \label{deltaIchichivaries}
\end{align}
Here,
\begin{align}
    \bs I_{\de \chi} = - \bs K^K_{\de \chi} + \overline{\bs K^K_{\de \chi}} - \de \chi \cdot \bs \tht^{EH}.
\end{align}
The remaining terms are
\begin{align}
    \de \bs I_{\bar \chi} &=  - \de \bs K^K_{\bar \chi} + \de \left(\overline{\bs K^K_{\bar \chi}}\right) - \bar \chi \cdot \de \bs \tht^{EH}.
\end{align}

When the background metric is allowed to vary, $v^{EH}_a$ and thus $\bs \Th^{EH}[\de g;g]$ is more complicated, as is $\de \bs \tht^{EH}$, so that in general $\de \bs \tht^{EH} \neq \bs \Th^{EH}$. We have
\begin{align}
    \de \bs I_{\bar \chi} &= \bs k_{\bar \chi}^{EH}[\de g;g] + \de \left(\overline{ \bs K^K_{\bar \chi}} \right) - \bar \chi \cdot (\de \bs \tht^{EH} - \bs \Th^{EH}[\de g;g]).
\end{align}
It would be worth expanding out $\de \bs \tht^{EH}$ and $\bs \Th^{EH}$ when the background metric is allowed to vary, but this is left for future work.

\section{Comparison to Stress--Energy Tensor} \label{ComparisontoStressEnergyTensor}

$\bs I_\chi$ has an interesting relationship to the stress--energy tensor $T_{a b}$. 

Using the form ${\bs i}^{a b}_\chi$, we can write on some $(D-2)$-dimensional surface $S$,
\begin{align}
    \oint_S \bs I_\chi = \oint_S {\bs i}^{a b}_\chi d S_{a b}.
\end{align}

It turns out that, for a vector $\chi$ which is Killing with respect to both $g_{ab}$ and $\bar g_{ab}$, 
\begin{align}
    \na_a {\bs i}^{a b}_\chi &= \f{1}{2} T^a_b \chi^b, \label{Kxidivergence}
\end{align}
where $T_{ab} = (8\pi)^{-1} (G_{ab} + \La g_{ab})$ is the stress--energy tensor. This is specifically if the background is a constant curvature space (Minkowski, de Sitter or anti-de Sitter), but we can generalize even further: if $g_{a b} = \bar g_{a b} + h_{ab}$ where $h_{ab} = H k_a k_b$ is of Kerr--Schild form and $\bar g_{ab}$ is a space for which the background Ricci tensor and $k_a$ satisfy 
\begin{align}
    \bar R_{ab} k^a k^b = 0, \label{barRkkassumption}
\end{align}
and $\chi^a$ is a Killing field of both $\bar g_{ab}$ and $g_{a b}$, then 
\begin{align}
    \na_a {\bs{i}}^{a b}_\chi &=  \f1{16\pi} \left( G^a_b - \bar G^a_b\right)\chi^b, \label{KxiGxi}
\end{align}
where $G^a_b$ and $\bar G^a_b$ are the Einstein tensors (in mixed components) associated with $g_{a b}$ and $\bar g_{a b}$, respectively. If $\bar g_{ab}$ is vacuum, then $G^a_b - \bar G^a_b = 8 \pi T^a_b$. (Specifically, we assume $G^a_b = -\La \de^a_b + 8\pi T^a_b$ and $\bar G^a_b = -\La \de^a_b$, with $\La$ the same in the two metrics.) The assumption \eqref{barRkkassumption} is clearly satisfied if $\bar g_{ab}$ is an Einstein space, where $\bar R_{ab}k^a k^b \propto \bar g_{ab} k^a k^b = 0$.

From \eqref{dstaromega} and \eqref{divomega}, 
\begin{align}
    d \bs I_\chi &= - 2 * \textrm{div} \bs i_\chi \nn 
    (d \bs I_\chi)_{a_1 \ldots a_{D-1}} &= -\f{1}{8\pi} ( G^a_b - \bar G^a_b) \chi^b \bs \ep_{a a_1 \ldots a_{D-1}}. \label{dIchi}
\end{align}
The last line can be written, in the case where the background spacetime is vacuum,
\begin{align}
    (d \bs I_\chi)_{a_1 \ldots a_{D-1}} &= - T^a_b \chi^b \bs \ep_{a a_1 \ldots a_{D-1}}.
\end{align}

That $\bs i_\chi^{ab}$ has divergence proportional to the stress--energy tensor is similar to expressions involving the superpotentials of KBL \cite{KBL} and Abbott and Deser \cite{AbbottDeser}, both of which I will discuss (the Abbott and Deser superpotential in Section \ref{TxiGauss} and the KBL superpotential in Section \ref{KBLSection}). The key ingredient that appears to be new is the use of the Kerr--Schild solution to give an \emph{exact} result at finite radius, even for large fields. This is due to the fact that in the Kerr--Schild spacetime, the stress--energy tensor is exactly linear in the perturbation $h_{a b}$.

To show \eqref{KxiGxi} we use the form \eqref{Kxialt} to write
\begin{align}
    \na_a \bs i_\chi^{ab} &= \f{1}{16\pi} \na_a \left( -\na^a \chi^b + \overline{\na^a \chi^b} + \chi^{[a} \na_e h^{b]e}\right). \label{expansionofdivergenceichi}
\end{align}
From \eqref{Boxchi} we have $\na_a \na^a \chi^b = -R^a_b \chi^b$. Using \eqref{twoformdivergence}, we can write, interpreting $\overline{\na^a \chi^b}$ as an antisymmetric covariant tensor,
\begin{align}
    \na_a \overline{\na^a \chi^b} &= \frac{1}{\sqrt{-g}} \pa_a \left( \sqrt{-g} \overline{\na^a \chi^b} \right) \nn 
    &= \frac{1}{\sqrt{-\bar g}} \pa_a \left( \sqrt{-\bar g} \overline{\na^a \chi^b}\right) \nn
    &= \bar \na_a \overline{\na^a \chi^b} \nn 
    &= - \bar R^a_b \chi^b.
\end{align}
We also have, dropping the $EH$ superscript on $(V^{EH})^a = \na_b h^{ab}$ for the next few lines for compactness,
\begin{align}
    \na_a (\chi^{[a} \na_e h^{b]e}) &= \na_a (\chi^{[a}V^{b]}) \nn 
    &= \f12 \na_a (\chi^a V^b - \chi^b V^a) \nn 
    &= \f12 \left( (\na_a \chi^a) V^b + \chi^a \na_a V^b - (\na_a \chi^b) V^a - \chi^b \na_a V^a\right).
\end{align}
All Killing vectors are divergence free, so $\na_a \chi^a = 0$. The second and third terms together combine to form the Lie derivative of $V$ with respect to $\chi$,
\begin{align}
    \chi^a \na_a V^b - V^a \na_a \chi^b &= (\lie_\chi V)^b,
\end{align}
which is zero since we demand that $\lie_\chi h_{a b} = \lie_\chi (g_{ab} - \bar g_{ab}) = 0$ so that $\chi$ is a Killing vector of both metrics. We are left with
\begin{align}
    \na_a ( \chi^{[a} \na_e h^{b]e}) &= -\f12 \chi^b \na_a V^a \nn 
    &= -\f12 \chi^b \na_a \na_c h^{ac}.
\end{align}
We also have
\begin{align}
    \na_a \na_b h^{ab} &= \na_a V^a \nn 
    &= \f1{\sqrt{-g}} \pa_a (\sqrt{-g} V^a) \nn 
    &= \f{1}{\sqrt{-\bar g}} \pa_a (\sqrt{-\bar g} V^a) \nn 
    &= \bar \na_a V^a \nn 
    &= \bar \na_a \bar \na_b h^{ab},
\end{align}
recalling $\sqrt{-\bar g} = \sqrt{-g}$.

To interpret $\bar \na_a \bar \na_b h^{ab}$ we make use  of the relationship between the Ricci tensor for the full and background spacetimes, \eqref{KSRicciRelationship}, which, setting $a = b$ and summing, implies
\begin{align}
    R &= \bar R - h^a_c \bar R^c_a + \f12 \bar \na_c \bar \na_a h^{ac},
\end{align}
using $R = R^a_a, \bar R = \bar R^a_a$ and that $h^{ab}$ can be raised and lowered using either metric. $h^a_c \bar R^c_a = 0$ by assumption \eqref{barRkkassumption}. Then we have 
\begin{align}
    \bar \na_a \bar \na_b h^{ab} &= R - \bar R. \label{RequalsRbarplusdoublederivatives}
\end{align}
Consequently, we have
\begin{align}
    \na_a {\bs{i}}_\chi^{ab} &= \f{1}{16\pi} \left( R^b_a \chi^a - \bar R^b_a \chi^a - \f12 (R - \bar R) \chi^b\right),
\end{align}
or \eqref{KxiGxi}.

Another thing to note is that from \eqref{RequalsRbarplusdoublederivatives}, if $R = \bar R$, implying $T = 0$, which is a necessary but not sufficient condition for vacuum, then $\bar \na_c \bar \na_b h^{bc} = 0$. 

A way to think about the expression ${\bs{i}}_\chi$ (and consequently about $\bs I_\chi$) is that the divergences associated with the Komar terms give a term (up to a multiplicative constant) $(R^b_a - \bar R^b_a) \chi^a$. It requires the extra $\chi^{[a}\na_c h^{b] c}$ term to bring the extra $(R-\bar R)\chi^b$-proportional term required to give $(G^b_a - \bar G^b_a)\chi^a$.

An important point is that in vacuum, $T_{ab} = 0$, so that $d \bs I_\chi = 0$, as expected, so that we can indeed evaluate $\oint_S \bs I_\chi$ on a surface $S$ which can be connected to $S_\infty$, not just at infinity. Away from vacuum, if there are two $(D-2)$-surfaces $S_1,S_2$ connected by a hypersurface $\Si$ such that $\pa \Si = S_2 \cup S_1$, up to orientation on $S_2,S_1$ and $\Si$,
\begin{align}
    \oint_{S_2} \bs I_\chi - \oint_{S_1} \bs I_\chi = \int_\Si d \bs I_\chi, \label{S2S1dIchi}
\end{align}
with $\bs I_\chi$ given by \eqref{dIchi}. 

We can now make two more notes. The first is to show the generalization to Einstein--Maxwell theory in four dimensions. The second will be to use Gauss' law to interpret the conserved charge, expanding on \eqref{S2S1dIchi}. While showing the Gauss' law argument I will also make connection to the Abbott and Deser \cite{AbbottDeser} and KBL \cite{KBL} superpotentials.

\subsection{Einstein--Maxwell Stress--Energy Tensor}

In Einstein--Maxwell theory, we have \eqref{Istari} with
\begin{align}
    \bs i^{a b}_\chi &= {\bs{i}}^{ab}_{\chi,EH} + {\bs{i}}^{ab}_{\chi, EM},
\end{align}
where ${\bs{i}}^{ab}_{\chi,EH}$ is the Einstein--Hilbert part of the antisymmetric tensor \eqref{mcKab}. The Maxwell terms are
\begin{align}
    {\bs{i}}^{ab}_{\chi,EM} &= \f{1}{16\pi} \left( - 2 \bs F^{ab} \bs A_e \chi^e - 2 \chi^{[a}\bs  F^{b]e} \bs A_e\right).
\end{align}
Its divergence is (making use of $\na_a \chi^a = 0$)
\begin{align}
    \na_a {\bs{i}}^{ab}_{\chi,EM} &= \f{1}{16\pi} \left( - 2 \na_a \bs F^{ab}\bs  A_e \chi^e - 2 \bs F^{a b} \na_a (\bs A_e \chi^e) -  \chi^a \na_a (\bs F^{b e} \bs A_e) +  \bs F^{a e}\bs  A_e \na_a \chi^b + \chi^b \na_a (\bs F^{a e}\bs  A_e)\right)
\end{align}
We have
\begin{align}
    -\chi^a \na_a (\bs F^{be}\bs A_e) + \bs F^{a e}\bs  A_e \na_a \chi^b &= -\lie_\chi (\bs F^{be}\bs  A_e) = 0
\end{align}
provided $\lie_\chi \bs A_e = 0$. (If $\lie_\chi\bs  A_e = 0$, then $\lie_\chi\bs  F_{a b} = 0$ also, which follows from, making use of Cartan's identity \eqref{Cartan},
\begin{align}
    \lie_\chi \bs F &= \chi \cdot d \bs F + d( \chi \cdot \bs F) \nn 
    &= \chi \cdot d^2 \bs A + d ( \chi \cdot d \bs A) \nn 
    &= 0 + d (\lie_\chi \bs A - d (\chi \cdot \bs A)) \nn
    &= 0.
\end{align}
Then it follows that $\lie_\chi (\bs A_a \bs F^{ab}) = 0$.)

For simplicity, assume the source-free Maxwell equations hold. Then $\na_a \bs F^{a b} = 0$. The terms which remain are
\begin{align}
    \na_a {\bs{i}}^{ab}_{\chi,EM} &= \f{1}{16\pi} \left( - 2 \bs F^{a b} \na_a (\bs A_e \chi^e)  + \chi^b \bs F^{ae} \na_a \bs A_e\right).
\end{align}
Since $\bs F^{ae}$ is antisymmetric, $\bs F^{ae} \na_a \bs A_e =\bs  F^{ae} \na_{[a} \bs A_{e]} = \f12 \bs F^{ae} \bs F_{ae}$. Additionally, again using Cartan's identity,
\begin{align}
    \na_a (\chi^e \bs A_e) &= \left( d (\chi \cdot \bs A)\right)_a \nn 
    &= \left( \lie_\chi \bs A - \chi \cdot d \bs A \right)_a \nn 
    &= (-\chi \cdot \bs F)_a \nn 
    &= -\chi^e \bs F_{e a}.
\end{align}

Then we have
\begin{align}
    \na_a {\bs{i}}^{ab}_{\chi,EM} &= \f{1}{16\pi} \left( - 2 \bs F^{ab} \bs F_{a e} \chi^e - \f1 2 \chi^b \bs F^{ae} \bs F_{ae}\right) \nn 
    &= -\f{1}{8\pi} \left( \bs F^{ba } \bs F_{ca} - \f14 \bs F^{ad} \bs F_{ad} \de^b_c\right) \chi^c \nn
    &= - \f{1}{2} (T^{EM})^b_c \chi^c,
\end{align}
using \eqref{TEM}. 

In electrovacuum, $T^{a b} = (T^{EM})^{a b}$, so
\begin{align}
    \na_a {\bs{i}}^{a b}_\chi &= \na_a {\bs{i}}^{a b}_{\chi,EH} + \na_a {\bs{i}}^{a b}_{\chi, EM} \nn 
    &= \f{1}{2} T^b_a \chi^a - \f12 (T^{EM})^b_a \chi^a \nn 
    &= 0,
\end{align}
so that (as expected) the conserved charge will be surface-independent. 

In Einstein--Maxwell gravity in Kerr--Schild form (with positive, zero or negative $\La$) we can also write
\begin{align}
    \na_a {\bs{i}}^{a b}_\chi &= \f12 (T^{non-EM})^b_a \chi^a,
\end{align}
where $T^{non-EM}_{ab} = T_{ab} - T^{EM}_{ab}$ is the stress--energy tensor excluding the stress--energy tensor of the electromagnetic field. 

\subsection{The Gauss--Stokes Theorem} \label{TxiGauss}

We can rewrite \eqref{S2S1dIchi} in terms of the $\bs i^{ab}_\chi$ to consider the Gauss--Stokes theorem:
\begin{align}
    \oint_{\partial \Sigma} {\bs{i}}^{ab}_\chi d S_{a b} &= 2\int_\Sigma \na_b {\bs{i}}^{ab}_\chi d \Si_a.
\end{align}
For simplicity let ${\bs{i}}_\chi^{ab}$ be just the EH part of the tensor, and let the background be vacuum. In this case we have
\begin{align}
    \oint_{\partial \Sigma} {\bs{i}}^{a b}_\chi d S_{ab} &= -2 \int_\Sigma \na_b {\bs{i}}^{b a}_\chi d \Si_a \nn 
    &= - \int_\Si T^a_b \chi^b d\Sigma_a.
\end{align}
Thus the quantity $\oint_{\partial \Si} {\bs{i}}^{a b}_\chi d S_{ab}$ is the flux integral of $T^a_b \chi^b$ over the surface $\Sigma$. 

I will note now that without assuming any features of the spacetime except that $\chi$ is a Killing vector, that $T^a_b \chi^b$ is automatically divergence-free, as is well-known (see \cite{Poisson}):
\begin{align}
    \na_a (T^a_b \chi^b) &= \na_a (T^{a b} \chi_b) \nn 
    &= (\na_a T^{ab}) \chi_b + T^{a b} \na_a \chi_b.
\end{align}
The first term is zero since the stress--energy tensor is automatically divergence-free. The second vanishes if $\chi$ is Killing since $T^{a b}$ is symmetric and $\na_a \chi_b$ is antisymmetric.

Since $T^a_b \chi^b$ is divergence-free, its flux is conserved through a region---that is to say, if $\mc R$ is some closed $D$ dimensional region and $\pa \mc R$ is its boundary, $\oint_{\pa \mc R} T^a_b \chi^b d\Sigma_a = 0$. In particular, if $T^a_b$ vanishes at large radius, and there are no internal singularities, $\int_{\Sigma_1} T^a_b \chi^b d\Sigma_a = \int_{\Sigma_2} T^a_b \chi^b d\Sigma_a$ for two Cauchy surfaces $\Sigma_1, \Sigma_2$ (if their orientations match). $T^a_b \chi^b$ is thus a conserved current vector associated with a symmetry; because it involves the stress--energy tensor, it can be thought of as being associated with an energy, momentum or angular momentum, if $\chi$ is the generator of a time, space translation, or rotation symmetry (respectively). This argument appears in, for instance, \cite{Poisson}, specifically in the context of the transfer of mass and angular momentum. 

We could further extend this idea to associate the energy (or momentum or angular momentum) content in some closed region $\Sigma$ with the integral $\int_\Sigma T^a_b \chi^b d \Sigma_a$, which would then be equal to the integral of $\int_{\pa \Si} {\bs{i}}_\chi^{a b} d S_{a b}$. Let the ``Killing energy'' $E[\chi]$ (using Abbott and Deser's \cite{AbbottDeser} naming; more on Abbott and Deser later in this subsection) for such a region $\Sigma$ be
\begin{align}
    E[\chi] &= -\int_\Sigma T^a_b \chi^b d \Sigma_a. \label{KillingEnergy}
\end{align}
The sign is chosen so that the energy will be positive when $\chi$ is timelike (if the dominant energy condition holds $-T^a_b \chi^b$ must be future-directed, so that $T^a_b \chi^b d \Sigma_a < 0$ for usual orientation \cite{Poisson}). 
Then we have
\begin{align}
    E[\chi] &= \oint_{\pa \Si} {\bs{i}}_\chi^{a b} d S_{a b}.
\end{align}
If in fact $\Sigma$ is a Cauchy surface with no interior boundary, extending to infinity, then this reduces to
\begin{align}
    E[\chi] &= \oint_{S_\infty} {\bs{i}}_\chi^{ab} d S_{a b} \nn 
    &= \oint_{S_\infty} \bs I_\chi
\end{align}
where $S_\infty$ is some $(D-2)$-surface at infinity. This gives another reason for the integral of $\oint \bs I_\xi$ to be interpreted as energy. Similarly, $E[-\eta_i] =- \oint_{S_\infty} \bs I_{\eta_i}$ would correspond to an angular momentum.

Of course, the main situations of interest for our purposes are black holes, where there \emph{is} an inner boundary (the horizon, and inside the horizon is a singularity), and so we can only take this argument so far. In the case of Kerr and Kerr--anti-de Sitter, $T^a_b = 0$ everywhere the spacetime is nonsingular. This is, again, helpful in interpreting why $\oint \bs I_\chi$ is independent of integration surface, since $\oint_{\pa \Sigma} \bs I_\chi =- \int_\Sigma T^a_b \chi^b d\Sigma_a = 0$ for any region $\Sigma$ where the spacetime is regular. Nevertheless, the expression \eqref{KillingEnergy} is not as useful for interpreting the total conserved energy as being the result of the stress--energy tensor. (Possibly, the value of the stress--energy tensor could be expressed as some sort of delta function at the ring singularity, similar to \cite{Israel70}, though this could be difficult to perform and interpret.)

Additionally, in spacetimes which are not linear in the deviation from a vacuum background, it is not in general the case that the usual conserved charge (such as the ADM conserved charge) as calculated at infinity is equal to the integral $E[\chi] = -\int_\Sigma T^a_b\chi^b d\Sigma_a$, even if there are no inner singularities. As an example, consider the Schwarzschild interior spacetime matched to the Schwarzschild exterior spacetime in four dimensions \cite{LakeLectureNotes}.  The exterior $r\geq a$ solution
\begin{align}
    ds^2 &= -\left(1-\f{2m}{r}\right) dt^2 + \f{dr^2}{1-2m/r} + r^2 d \Omega_2^2
\end{align}
and the interior $r \leq a$ solution is
\begin{align}
    ds^2 &= -\left( \frac{3}{2} \sqrt{1-\f{2m}{a}} - \f12 \sqrt{1 - \f{2 m r^2}{a^3}}\right)^2 dt^2 + \f{dr^2}{1 - 2mr^2/a^3} + r^2 d \Omega_2^2,
\end{align}
where $a$ is the radius of the ``star.'' $T^t_t = - \rho \equiv - m/(4\pi a^3/3)$ is constant in the interior and zero outside. The conserved charge as calculated at infinity (through the ADM mass, say, which is equal to the Komar mass, and can be reproduced by the Noether charge method) is given by the Schwarzschild exterior mass parameter $m$, but the integral $-\int T^a_b \xi^b d \Sigma_a$ is, taking a constant-$t$ surface and $\xi = \partial_t$ (see Appendix \ref{schwint} for intermediate steps)
\begin{align}
    -\int T^a_b \xi^b d \Sigma_a &= \f74 m - \f 9 8 a \left(1 - \sqrt{\f{a}{2m}-1} \: \arcsin\sqrt{\f{2m}{a}} \right).
\end{align}

This is of course in general not equal to $m$. It turns out that if $a$ is allowed to become arbitrarily large, with finite $m$, then this quantity asymptotically approaches $m$. This gives us the intuition that $-\int T^a_b \xi^b d\Si_a$ is a measure of energy which will match the ADM energy in the limit where the actual stress--energy tensor term is very small (in this case, due to a highly diffuse matter distribution). 

It is interesting and perhaps surprising, then, that in a Kerr--Schild spacetime with a Killing vector and no interior singularity, the Killing energy will continue to correspond to the energy as calculated by covariant phase space methods, since we can use $\bs I_\xi$. This again has to do with the fact that the Kerr--Schild decomposition leads to the stress--energy tensor being exactly linear in the perturbation to the metric.

As alluded to above, the general argument presented in this subsection for the Killing energy is very similar to the one appearing in Abbott and Deser \cite{AbbottDeser}, who construct a superpotential associated with linear-order perturbations to a vacuum background in order to evaluate the stability of de Sitter and anti-de Sitter spacetimes under fluctuations. There are some subtleties that distinguish their argument from the one given above. They have a background $\bar g_{a b}$ which is vacuum and full metric $g_{a b} = \bar g_{a b} + h_{a b}$ and consider contributions to the stress--energy tensor and related quantities linear in $h_{a b}$, which is assumed to be small (not necessarily Kerr--Schild). They define a stress--energy tensor \emph{density} $\mc T^{a b}$ satisfying
\begin{align}
    (-\bar g)^{-1/2}\mc T^{a b} = R_L^{ab} - \f12 \bar g^{a b} R_L - \La h^{a b},
\end{align}
where $R_L^{a b}$ is the correction to the background Ricci tensor \emph{linear} in the perturbation $h^{a b}$. In our notation this means $T^{a b} = (-\bar g)^{-1/2} \mc T^{a b}/8\pi$ to linear order in $h^{ab}$. Their Killing energy is then defined as
\begin{align}
    E_{AD}[\bar \chi] &= \int_\Sigma T^{a b} \bar \chi_b \sqrt{-\bar g} d^{D-1}x,
\end{align}
for a Killing vector $\chi^a$ \emph{of the background}, where $\bar \chi_b = \bar g_{a b} \chi^b$. This is also equal (to first order in $h^{ab}$) to $\int T^{a b} \bar \chi_b d \bar \Sigma_a$ in the notation I am using. (Further, $T^{a b}$ is raised and lowered only using the background metric, by Abbott and Deser convention.) $\Sigma$ is a surface which extends across the whole space (for example, a constant-$t$ surface). Recall that there are no curvature singularities in the case Abbott and Deser were considering, because they were looking at small perturbations to a pure vacuum background. Note the sign difference from \eqref{KillingEnergy}. $E_{AD}[\bar \chi]$ is re-expressed as an integral over a boundary $(D-2)$ surface, using Gauss' law. The superpotential $\bs{i}_{\chi,AD}^{a b}$ (modifying notation to match with mine) is introduced, equal to
\begin{align}
    {\bs{i}}_{\chi,AD}^{am} &= \f{1}{16\pi} \left(\bar \na_b K^{m a n b} \bar \chi_n - K^{mbna} \bar \na_b \bar \chi_n\right),
\end{align}
where $\bar \chi_n = \bar g_{mn} \chi^m$ and 
\begin{align}
    K^{m a n b} &= \f{1}{2} \left( \bar g^{m b} H^{n a} + \bar g^{n a} H^{m b} - \bar g^{m n} H^{a b} - \bar g^{a b} H^{m n}\right) = \bar g^{b[m}H^{a]n} + \bar g^{n[a}H^{m]b} \nn 
    H^{m n} &= h^{m n} - \f{1}{2} \bar g^{m n} (h^{a b} \bar g_{a b}).
\end{align}
This satisfies
\begin{align}
    \bar \na_a \bar {{\bs{i}}}^{am}_{\chi,AD} &= \f{1}{2} T^{m n} \bar \chi_n.\label{ADderivative}
\end{align}
\emph{to linear order}. Consequently, Gauss' law implies
\begin{align}
    E_{AD}[\bar \chi] &= -\oint_S {\bs{i}}^{am}_{\chi,AD} dS_{am}, \label{EADxi}
\end{align}
where $S$ is a boundary surface. In asymptotically Minkowski or AdS, the boundary surface is taken to be at spatial infinity. 

Abbott and Deser point out that their formula \eqref{EADxi} reproduces the ADM mass in Minkowski space, and correctly gives the energy of the Schwarzschild mass $m$ in Schwarzschild--de Sitter (calculated at a boundary surface far from the Schwarzschild event horizon but still below the cosmological horizon) and Schwarzschild--AdS (calculated at infinity) cases. 

(The sign difference between Abbott and Deser's expressions and mine is because Abbott and Deser use a Killing vector $\bar \xi$ with a \emph{negative} contravariant $t$ component, and so their $\bar \chi$ corresponds to the negative of my $\chi$.)

As I mentioned in Section \ref{BCsection}, it was pointed out by BC that the Abbott and Deser superpotential matches with their $\bs k_\chi[\bar g;h]$ term. Of course my ${\bs{i}}_\chi^{a b}$ is based on integrating $\bs k_\chi[\bar g;h]$ in the case of a Kerr--Schild perturbation. It makes sense, then, that, if Abbott and Deser's $h_{ab}$ is taken to be of Kerr--Schild form,
\begin{align}
    {\bs{i}}_\chi^{a b} = {\bs{i}}_{\chi,AD}^{a b},
\end{align}
exactly. Further, the statement \eqref{ADderivative} holds exactly, in this Kerr--Schild case. Note that in this case, $h^{a b} \bar g_{a b} = 0$ so $H^{m n} = h^{m n}$. It takes a few lines to verify the equality:
\begin{align}
    \bar \na_b K^{m a n b} \bar \chi_n &= \bar \chi_n \bar \na^{[m} h^{a]n} + \chi^{[a} \bar \na_b h^{m]b} \nn 
    &= \chi^d \bar \na^{[m} h^{a]}_d + \chi^{[a} \bar \na_b h^{m] b} \nn 
    -K^{m b n a} \bar \na_b \bar \chi_n &= -\f12 \left( \bar g^{m a} h^{n b} + \bar g^{n b} h^{m a} - \bar g^{m n} h^{a b} - \bar g^{a b} h^{m n}\right) \bar \na_b \bar \chi_n \nn 
    &= \f12 \left( h^{a b} \bar \na_b \chi^m + h^{m n} \bar \na^a \bar \chi_n\right) \nn 
    &= \f12 \left( h^{a b} \bar \na_b \chi^m - h^{m n} \bar \na_n \chi^a\right) \nn 
    &= h^{b [a} \bar \na_b \chi^{m]}.
\end{align}
(We used $\bar g^{n b} \bar \na_b \bar \chi_n = h^{n b} \bar \na_b \bar \chi_n = 0$ since $\bar \na_b \bar \chi_n = \bar \na_{[b}\bar \chi_{n]}$.) 

My reason for emphasizing the equality between ${\bs{i}}_\chi^{a b}$ and ${\bs{i}}_{\chi,AD}^{ab}$ is that the latter was derived specifically for its property \eqref{ADderivative}, and we discovered the (exact) property \eqref{Kxidivergence} along the way.

The second note is the following. Consider the Generalized Kerr--AdS metric, in KS coordinates, with $\mu(r)$ given by some monotonic function which has
\begin{align}
    \mu(r) = \begin{cases} 0 & \text{if } r \leq r_1 \\ 
    0 < \mu(r) < M &\text{if } r_1 < r < r_2 \\
    M &\text{if } r \geq r_2.
    \end{cases}
\end{align}
For simplicity also demand that $\mu'(r_1) = \mu'(r_2) = 0$. 

We demand that the spacetime has no event horizon ($r_2$ is larger than the value of $r$ where a horizon would appear given $M$). Letting $S_{1,2}$ be (respectively) ($D-2)$-surfaces of constant $t$ and constant $r = r_{1,2}$, we can write
\begin{align}
    \oint_{S_1} \bs I_\xi &= 0 \nn 
    \oint_{S_2} \bs I_\xi &= \f{M}{\Xi^2},
\end{align}
respectively the values for $\mc E$ given the $m = 0$ and $m= M$ Kerr--AdS solutions. ($\xi = \pa_t$.) This is because, on $S_{1,2}$, the metric and its first derivatives are the same as those for Kerr--AdS with $m = 0$ and $m = M$, respectively. Using the Gauss--Stokes law, we can then write, letting $\Si$ be the hypersurface of constant $t$ over the two-sphere $(\tht,\phi)$ 
\begin{align}
    \f{M}{\Xi^2} &=  \oint_{S_2} \bs I_\xi \nn 
    &= \oint_{S_1} \bs I_\xi + \int_\Si d \bs I_\xi \nn 
    &= -\int_\Si T^a_b \xi^b d \Si_a \nn 
    &= - \int_\Si \sqrt{-g} dr d \tht d \phi T^t_t.
\end{align}
(The sign is up to choice of orientation.) (I also verified that this is true, using \emph{GRTensorIII} \cite{GrTensor}.) This same argument can also be used for Kerr, replacing the $m$ in the Kerr spacetime with a similar $\mu(r)$, to find that $M$ is given by an integral over $T^a_b \xi^b d \Si_a$. This shows a link between the energy associated with the black hole spacetime as calculated via $\bs I_\xi$ and the energy associated for a spacetime which is regular but at large radius matches a Kerr--AdS or Kerr spacetime, as calculated directly from $T_{ab}$. This relationship suggests that we are on the right track with the charge $\oint \bs I_\xi$. (Analogous statements exist regarding angular momentum.) 

\section{Relationship to Killing Potential} \label{relationshiptokillingpotential}

Recall \eqref{mcEKillingPotentialWithXi}, which is an adaptation of the argument in \cite{KastorEtal:2009,Cvetic} to define $\mc E$ using a Komar integral and a Killing potential, using a Killing potential associated with $\xi$ instead of $\zeta$. I will now find an expression for the Killing potential $\bs \om_\xi$ appearing in \eqref{mcEKillingPotentialWithXi} given the use of $\bs I_\xi$ for the calculation of $\mc E$. Note that this applies to the vacuum-with-$\La$ case specifically. Assume here that we are using the $\bs I_\xi$ for Einstein--Hilbert gravity only (no Maxwell field).
\begin{align}
    \mc E &= \mc E \nn 
    \oint_S \bs I_\xi &= -\f{D-2}{D-3} \oint_S \left( \bs K^K_\xi + \f{\tilde \La}{16\pi} * \bs \om_\xi\right) \nn 
    \oint_S *\bs \om_\xi &= -\f{16\pi}{\tilde \La} \oint_S \left( \f{D-3}{D-2} \bs I_\xi + \bs K_\xi^K\right). 
\end{align}
Of course this is satisfied if we write
\begin{align}
    * \bs \om_\xi &= -\f{16\pi}{\tilde \La} \left( \f{D-3}{D-2} \bs I_\xi + \bs K_\xi^K\right) \nn 
    \bs \om_\xi^{ab} &= -\f{16\pi}{\tilde \La} \f{D-3}{D-2} \bs{i}_\xi^{ab} - \f{1}{\tilde \La} \na^a \xi^b. \label{omegaxiab}
\end{align}
We shall now take the divergence of both sides to confirm that this has the required form. We have,
\begin{align}
    \na_a \bs \om_\xi^{ab} &= -\f{16\pi}{\tilde \La} \f{D-3}{D-2} \na_a \bs{i}_\xi^{ab} - \f{1}{\tilde \La} \na_a \na^a \xi^b \nn 
    &= -\f{8\pi(D-3)}{\tilde \La (D-2)} T^b_a \xi_b + \f{1}{\tilde \La} R^b_a \xi^a.
\end{align}
If the full spacetime is indeed vacuum, with $R_{ab} = \tilde \La g_{ab}$, then $T_{ab} = 0$ and so we recover the required $\na_a \bs \om_\xi^{ab} = \xi^b$. 

We can decompose $\bs \om_\xi^{ab}$ from \eqref{omegaxiab} using the decomposition for $\bs{i}_\xi^{ab}$ using \eqref{Kxialt} as
\begin{align}
    \bs \om_\xi^{ab} &= \f{(D-3)}{\tilde \La(D-2)} ( \na^a \xi^b - \overline{\na^a \xi^b} - \xi^{[a} \na_c h^{b] c}) - \f{1}{\tilde \La} \na^a \xi^b \nn 
    &= \f{1}{\tilde \La(D-2)}\left( -\na^a \xi^b - (D-3)\overline{\na^a \xi^b} - (D-3)\xi^{[a}\na_c h^{b] c}\right) \nn 
    &= -\f{1}{\tilde \La} \overline{\na^a \xi^b} -\f{1}{\tilde \La(D-2)} \left( \na^a \xi^b - \overline{\na^a \xi^b} + (D-3) \xi^{[a} \na_c h^{b] c}\right). \label{omegaxiabdecomposition}
\end{align}
The rationale for the final decomposition is as follows. We have, adapting \eqref{backgroundKomarVolumeRelation} and integrating on the horizon $H$, 
\begin{align}
    -\f{1}{\tilde \La} \oint_H \overline{\na^a \xi^b} d S_{ab} &= - 2 \mc V_{\xi,\mc B}.
\end{align}
The first term then reproduces the integral of $\bs \om_\z$ in \eqref{ointHomegazeta}, where the $\bs \om_\z$ integrated in that case is the one constructed from $\bs \om^{(j)}$ in the method described by CGKP throughout Section \ref{KomarKillingPotential}. I will give an argument in Section \ref{killingpotentialvectorvolumerelationship} why we would expect to recover \eqref{ointHomegazeta} given the method for constructing the Killing potential from the $\bs \om^{(j)}$. Thus the ``extra term'' required to give the Killing potential which gives the correct value for $\mc E$ is the remaining terms in \eqref{omegaxiabdecomposition}, which we note come down to the difference between the Komar terms for the full and background terms and the extra term $\xi \cdot \bs \tht^{EH}$.

\section{Comparison to KBL Superpotential} \label{KBLSection}

The Katz--Bi\v{c}\'ak--Lynden-Bell (KBL) superpotential \cite{KBL,Katz85,PetrovKatz} is a $2$-form associated with a vector $\chi$, where $\chi$ is at least asymptotically a Killing vector, as well as a full and background spacetime $g_{ab},\bar g_{ab}$ respectively. 

KBL use a convention where a caret represents multiplication by $\sqrt{-g}$. I will avoid this notation in this section for consistency with the rest of this thesis. The KBL superpotential $\bs J^{ab}_{KBL}[\chi]$ is given by
\begin{align}
    \sqrt{-g} \bs J_{KBL}^{ab}[\chi] &= \f{1}{8\pi} \left( \sqrt{-g} \na^{[a} \chi^{b]} - \overline{ \sqrt{-g} \na^{[a} \chi^{b]}} + \sqrt{-g} \chi^{[a} K_{KBL}^{b]}\right).
\end{align}
Here $K_{KBL}^a$ is the vector given by
\begin{align}
     K_{KBL}^a &= \left( g^{a b} \Delta^c_{bc} - g^{bc}\Delta^a_{bc}\right),
\end{align}
with $\Delta^a_{bc} = \G^a_{bc} - \overline{\G^a_{bc}}$. The integral over a $(D-2)$-surface $S$, at infinity,
\begin{align}
    \oint_S \bs J_{KBL}^{ab}[\chi] d S_{ab}
\end{align}
is then a conserved quantity under certain situations, in that it is independent of $S$, provided that $S$ is at spatial or null. Taking linear terms in the metric variation, KBL find (their equation (3.11))
\begin{align}
    \bs J_{KBL}^{ab}[\chi] &=\f{1}{8\pi} (l^{c[a} \na_c \bar \chi^{b]} + \bar \chi^{[a} \na_c l^{b]c} - \na^{[a}l^{b]}_c \bar \chi^c), \label{KBL3.11}
\end{align}
where $l^{ab} = -h^{ab} + \f12 g^{ab} (g^{cd} h_{cd})$ where $h_{ab} = g_{ab} - \bar g_{ab}$ and $g^{ab}$ is used to raise the metric. 

BC showed \cite{BarnichCompere} that the KBL superpotential is equivalent (to first order) to $\bs k^{EH}_\chi[h;g]$ (with $h_{ab} = g_{ab} - \bar g_{ab}$), provided that $\de \chi^a = 0$.

It was shown by Deruelle and Katz \cite{DeruelleKatz05} that the KBL superpotential reproduces the same values for the energy and angular momentum as were found by GPP for the higher-dimensional Kerr--AdS spacetimes, using 
\begin{align}
    \mc E &= -\oint_S \sqrt{-g} \bs J_{KBL}^{ab}[\xi] d^{D-2} x \nn 
    \mc J_i &= +\oint_S \sqrt{-g} \bs J_{KBL}^{ab}[\eta_i] d^{D-2}x. \label{KBLcharges}
\end{align}
Their background and full metric decomposition actually uses the Kerr--Schild form of the metric for the calculation. The calculation is performed at large radius, where the KBL superpotential method is valid. 

The KBL superpotential is also related to the stress--energy tensor. In fact,
\begin{align}
    \pa_b (\sqrt{-g} \bs J_{KBL}^{ab}[\chi]) &= \sqrt{-g} \left( \tht^a_b \chi^b + \sigma^{a[bc]}\pa_{[b}\chi_{c]} + \z_{KBL}^a\right).
\end{align}
$\z_{KBL}^a$ is a quantity which vanishes when $\chi$ is a Killing vector of the background spacetime. Additionally,
\begin{align}
    \sqrt{-g} \tht^a_b &= \f{1}{8\pi} \sqrt{-g} G^a_b - \f{1}{8\pi}\overline{\sqrt{-g} G^a_b} + \f{1}{16\pi} \sqrt{-g} l^{cd} \bar R_{cd} \de^a_b + \sqrt{-g}t^a_b \nn 
    16\pi t^a_b &= g^{cd}\left[(\Delta^e_{ce} \Delta^a_{db}+\Delta^a_{cd}\Delta^e_{eb} - 2 \Delta^a_{ce}\Delta^e_{db})-\de^a_b(\Delta^f_{cd}\Delta^e_{fe} - \Delta^f_{ce} \Delta^e_{fd}) \right] + g^{ae}(\Delta^d_{cd}\Delta^c_{eb} - \Delta^d_{ed}\Delta^c_{cb}) \nn 
    16\pi \sigma^{acd} &= \left( g^{ac}\bar g^{d b} + \bar g^{ad}g^{cb}-g^{ab}\bar g^{cd}\right) \Delta^e_{be} - \left( g^{bc}\bar g^{de}+\bar g^{bd}g^{ce}-g^{be}\bar g^{cd}\right)\Delta^a_{be}.
\end{align}
We can write $\sigma^{a[cd]}$ as
\begin{align}
    16\pi \sigma^{a[cd]} &= \left( g^{a[c}\bar g^{d]b} + \bar g^{a[d}g^{c] b}\right) \Delta^e_{be} - \left(g^{b[c}\bar g^{d]e} + \bar g^{b[d} g^{c]e}\right) \Delta^a_{be}.
\end{align}

In the case of a Kerr--Schild decomposition, the KBL superpotential is exactly equal to $\bs {i}_\chi^{ab}$. We can see it from \eqref{KBL3.11} by noting $l^{ab} = -h^{ab}$ since $g^{ab} h_{ab} = 0$, and then comparing to \eqref{mcKab}. Whereas the KBL superpotential is only known to be exact only for small perturbations (typically, at infinity), \eqref{mcKab} is valid for arbitrarily large Kerr--Schild perturbations. 

We can also show that in the Kerr--Schild case, the vector $K^a_{KBL}$ is equal to $-(V^{EH})^a$. If $g_{a b} = \bar g_{a b} + h_{ab}$, with $g^{ab} = \bar g^{ab} - h^{ab}$, then $\Delta^a_{bc} = \f12 (\bar \na_b h^a_c + \bar \na_c h^a_b - \bar \na^a h_{bc}) + \f12 k^d \pa_d H k^a h_{bc}$ \cite{Taub81}. I already argued that $\Delta^a_{ba} = 0$. In Appendix \ref{DeltaAppendix}, $\bar g^{cd} \Delta^b_{cd} = (V^{EH})^b$ \eqref{gbcDeltaabc} and $h^{cd} \Delta^b_{cd} = 0$ \eqref{hbcDeltaabc}. As a consequence, we have
\begin{align}
    K^a_{KBL} &= -(\bar g^{bc} - h^{bc})\Delta^a_{bc} \nn 
    &= -(V^{EH})^a.
\end{align}
From this as well as $\sqrt{-g} = \sqrt{-\bar g}$ we conclude that the KBL superpotential breaks down in the Kerr--Schild case to to
\begin{align}
    \bs J_{KBL}^{ab}[\chi] &= \f{1}{8\pi} \left(\na^a \chi^b - \overline{\na^a \chi^b} - \chi^{[a}V_{EH}^{b]}\right),
\end{align}
reproducing $-\bs{i}^{ab}_\chi$ (see \eqref{Ixidifferenceform}) and also showing that the vector $K^a_{KBL}$ term maps onto the $\bs \tht^{EH}$ term.

Further in the Kerr--Schild situation, since $\Delta^a_{ba} = 0$ and using $g^{bc} = \bar g^{bc}-h^{cd}$,
\begin{align}
    16\pi t^a_b &= g^{cd} \left[-2 \Delta^a_{ce} \Delta^e_{db} + \de^a_b \Delta^f_{ce} \Delta^e_{fd} \right] \nn 
    16\pi \sigma^{a[cd]} &=   \left(h^{b[c}\bar g^{d]e} + \bar g^{b[d} h^{c]e}\right) \Delta^a_{be}. \label{tab}
\end{align}
Using \eqref{kDelta}, 
\begin{align}
    16 \pi \sigma^{a[cd]} &= H \left( k^b k^{[c}\bar g^{d]e} + \bar g^{b[d}k^{c]}k^e\right) \Delta^a_{be} \nn 
    &= H \left( k^b \bar \na_b H k^a k_e k^{[c}\bar g^{d]e} + \bar g^{b[d}k^{c]} k^e \bar \na_e H k^a k_b\right) \nn 
    &= H \left( k^b \bar \na_b H k^a k^{[c}k^{d]} + k^{[d}k^{c]} k^e \bar \na_e H k^a\right) \nn 
    &= 0,
\end{align}
since $k^{[c}k^{d]} = 0$. I also show in Appendix \ref{DeltaAppendix} that $t^a_b = 0$. Thus we have
\begin{align}
    \pa_b (\sqrt{-g} \bs J^{ab}_{KBL}) &= \f{\sqrt{-g} }{8\pi} (G^a_b -\bar G^a_b),
\end{align}
recovering \eqref{KxiGxi}. The point of performing this calculation is to show how the relationship between the Kerr--Schild charge $\bs I_\chi$ and the Einstein tensor has precedent, with both the Abbott and Deser tensor and the KBL superpotential. Again I note that the result \eqref{KxiGxi} is exact (not approximate) in our case. 

\chapter{Notes on Black Hole Volume and Area} \label{VolumeAreaChapter}

In this chapter, I discuss some important results related to (horizon) area and volume for the Generalized Kerr--AdS black holes. 

I begin in Section \ref{volumeandareaformulas} by reviewing formulas for the area and volume, respectively, of higher dimensional unit spheres and balls. I perform the (well-known) calculation for the ``area'' ($N$-volume) of an $N$-sphere explicitly because it involves some integrals which will recur throughout Chapter \ref{ExplicitGKAdSChapter}. 

I then turn to the Generalized Kerr--AdS spacetimes. In Section \ref{VolumeSimplified}, I use the pseudo-Cartesian coordinates from Section \ref{pseudoCartesian} to calculate in an intuitive way the vector volume associated with $\xi = \pa_t$ inside a region $r \leq R$ for these spacetimes. I then calculate the area of the black horizon explicitly in Section \ref{AreaRevisited}, the content of which is moved to Appendix \ref{AreaCalculationAppendix}. In section \ref{EuclideanAction}, I review the Euclidean action method calculation as used by GPP and show how the vector volume arises naturally.

I then resolve two of the outstanding questions mentioned in the previous chapter. In Section \ref{killingpotentialvectorvolumerelationship}, I give an explanation for why the Killing potential for $\z$ constructed from the PCKY tensor as introduced by CGKP satisfies \eqref{ointHomegazeta}. Finally, in Section \ref{AreaVolumeRelationship}, I use the PCKY tensor to explain why the vector volume satisfies $\mc V_\xi = r_+ A/(D-1)$ for the GKAdS spacetimes.

\section{Volume and Area of Unit Balls and Spheres in Higher Dimensions} \label{volumeandareaformulas}

We can express various area and volume integrals in terms of the volume of unit $N$-balls and area of unit $N$-spheres, for dimension $N$. These formulas can be found in, e.g.,~\cite{SmithVamanamurthy,Apostol,Parks}.

Given an integer $N$, let $N$-dimensional Euclidean space be space with coordinates $x_i$, $i = 1, \ldots, N$, with metric
\begin{align}
    ds^2 &= \sum_{i=1}^N dx_i^2,
\end{align}
and where each coordinate is real and satisfies $-\infty < x_i < \infty$.

Let the unit $N$-sphere be the collection of points which can be embedded into $(N+1)$-dimensional Euclidean space on the surface satisfying 
\begin{align}
    \sum_{i=1}^{N+1} x_i^2 = 1.
\end{align}
Let the unit $N$-ball be the collection of points in $N$-dimensional Euclidean space satisfying
\begin{align}
    \sum_{i = 1}^N x_i^2 \leq 1.
\end{align}
The $N$-ball is an $N$-dimensional region. The $N$-sphere is the set of points on the boundary (``surface'') of the $(N+1)$-ball.

The metric on the $N$-sphere can be written as
\begin{align}
    ds^2 &= \sum_{i = 1}^{N+1} dx_i^2,
\end{align}
subject to the constraint $\sum_{i = 1}^{N+1} x_i^2 = 1$, which can also be written as, eliminating the last coordinate to use the coordinates $(x_1, \ldots, x_N)$,
\begin{align}
    ds^2 &= \sum_{i = 1}^N dx_i^2 + d x_{N+1}^2 \nn 
    &= \sum_{i = 1}^N dx_i^2 + \left(\frac{\sum_{i = 1}^N x_i dx_i}{x_{N+1}}\right)^2 \nn 
    &= \sum_{i = 1}^N dx_i^2 + \frac{\left(\sum_{i = 1}^N x_i dx_i\right)^2}{1 - \sum_{i = 1}^N x_i^2}.
\end{align}

Let the $N$-volume of the unit $N$-ball be given by $\mc V_N$, and let the area, or, really, $N$-volume of the unit $N$-sphere be given by $\mc A_N$. I will sometimes use the term area for the $N$-volume of the unit $N$-sphere, by analogy with the area of the 2-sphere, and to match with the notation $\mc A_N$.

The volume of a unit $N$-ball is
\begin{align}
    \mc V_N &= \frac{\pi^{N/2}}{\Gamma(N/2+1)}, \label{VN}
\end{align}
where $\Gamma(x)$ is the Gamma function which generalizes the factorial function:
\begin{align}
    \Gamma(n) &= (n-1)! \nn
    \Gamma\left(n + \f12\right) &= \left(n-\f12\right) \left( n - \f3 2\right) \ldots \left(\f1 2\right) \pi^{1/2} = \f{(2n-1)!!}{2^n} \sqrt{\pi}
\end{align}
if $n$ is a positive integer.  Note that $(n-1)!$ can be written as
\begin{align}
    (n-1)! &= \f{(2n-2)!!}{2^{n-1}}.
\end{align}

The area of the unit $N$-sphere is
\begin{align}
    \mc A_{N} &= (N+1) \mc V_{N+1} \nn 
    &= \frac{2 \pi^{(N+1)/2}}{\Gamma( (N+1)/2)}. \label{AsubD}
\end{align}

Later on it will turn out that we will want to know $\mc A_{D-2}$ where $D = 2n+\ve$, with $n$ an integer and $\ve$ either 0 or 1. With a little rearranging,
\begin{align}
    \mc A_{D-2} &= \f{2^{1-\ve} (2\pi)^{n-1+\ve}}{(D-3)!!}. \label{ADm2}
\end{align}

It will be convenient to have explicit expressions for the integrals over the latitude and azimuthal coordinates. I will thus go through explicitly to show one way to perform the calculation for the area (really $N$-volume) of the unit $N$-sphere. (I will not show the calculation for the unit $N$-ball volume, which is similar.) The calculation here is similar to some of the derivations given in \cite{SmithVamanamurthy,Apostol,Parks}, but my breakdown of the coordinates into latitude and azimuthal coordinates is different, in order to better match with \cite{GibbonsLu,GibbonsLu2} and in anticipation of some of the specific integrals I will encounter in Chapter \ref{ExplicitGKAdSChapter}.

Let us use $N+1$ (constrained) coordinates to describe the unit $N$-sphere: $n$ latitude angles $\mu_i$ and $k$ azimuthal angles $\phi_i$. If $N$ is even, $n = N/2+1, k = N/2$ and if $N$ is odd, $n = k = (N+1)/2$. ($n$, starting in this paragraph and for the rest of this section only, is not the same as the $n$ associated with $D = 2n+\ve$ in the generalized Kerr--NUT--AdS metrics.) The metric for the $N$-sphere is
\begin{align}
    ds_{S^N}^2 &= \sum_{i = 1}^{n} d \mu_i^2 + \sum_{i=1}^{k}\mu_i^2 d \phi_i^2,
\end{align}
subject to the constraint $\sum_{i=1}^{n} \mu_i^2 = 1$. 

We can replace $\mu_{n}$ with $\sqrt{1-\sum_{i=1}^{n-1}\mu_i^2}$ (as is done by \cite{GibbonsPerry}); then the metric becomes
\begin{align}
    ds^2_{S^N} &= \sum_{i=1}^{n-1} d \mu_i^2 + \frac{\left(\sum_{i=1}^{n-1}\mu_i d \mu_i\right)^2}{\mu_{n}^2} + \sum_{i=1}^{k} \mu_i^2 d \phi_i^2.
\end{align}
The latitude sector metric components are then (for $1 \leq i,j \leq n-1$)
\begin{align}
    g_{\mu_i \mu_j} &= \delta_{i j} + \frac{\mu_i \mu_j}{\mu_{n}^2}.
\end{align}
In matrix form, the metric for the purely latitude sector ($1\leq i,j\leq n-1$) can be rewritten as
\begin{align}
    g_{\mu\mu} &= I + v v^T
\end{align}
where $I$ is the identity matrix and $v$ is the column vector
\begin{align}
    v &= \begin{bmatrix} \mu_1/\mu_{n} \\ 
    \mu_2/\mu_{n} \nn 
    \vdots\nn
    \mu_{n-1}/\mu_{n}
    \end{bmatrix}.
\end{align}
Then by the Matrix Determinant Lemma \cite{Harville},
\begin{align}
    \mathrm{det}(g_{\mu \mu}) &= \mathrm{det}(I + v^T v) \nn
    &= 1 + (v^T v) \nn 
    &= 1 + \f{\sum_{i=1}^{n-1} \mu_i^2}{\mu_{n}^2}\nn 
    &= \f{1}{\mu_{n}^2}.
\end{align}
The square root of the determinant $|g_{\mu_i \mu_j}|$ of the purely latitude sector is then given by
\begin{align}
    \sqrt{|g_{\mu_i \mu_j}|} = \frac{1}{\mu_{n}} = \frac{1}{\sqrt{1 - \sum_{i = 1}^{n-1} \mu_i^2}}.
\end{align}
Thus the metric determinant (including the azimuthal sector) is
\begin{align}
    \sqrt{|g|} &= \frac{\prod_{i = 1}^{k} \mu_i}{\mu_{n}}. \label{sqrtgmuonly}
\end{align}
The azimuthal angles all vary over $2\pi$. If $N$ is odd, all latitude variables vary over $0 \leq \mu_i \leq 1$. If $N$ is even, $0 \leq \mu_i \leq 1$ for $i < n$, but $\mu_n$ varies as $-1 \leq \mu_n \leq 1$. 

(The combination $(\prod_{i=1}^{k} \mu_i) / \mu_{n}$ appears, in different notation, as a factor in the square root of the negative of the metric determinant in GPP \cite{GibbonsPerry}.) 

Consider first the case where $N$ is even. In this case $N = 2n$. Because we have as integration variables $\mu_1, \ldots , \mu_n$, we must multiply by 2 to account for the fact that $\mu_{n+1}$ has two values for each set of $\mu_1, \ldots, \mu_n$ (except for the trivial case where $\mu_{n+1} = 0$). The $N$-volume $\mc A_{N}$ of the $N$-sphere is calculated as follows. Let $\mu_n$ vary from 0 to 1. Then $\mu_{n-1}$ varies from 0 to $\sqrt{1-\mu_n^2}$, $\mu_{n-2}$ from 0 to $\sqrt{1-\sum_{i=n-1}^n \mu_i^2}$, and so on, all the way down to $\mu_1$ varying from 0 to $\sqrt{1-\sum_{i=2}^n \mu_i^2}$. 

\begin{align}
    \mc A_{2n} &= 2 \int_0^{2\pi} d \phi_{n} \ldots \int_0^{2\pi} d \phi_1 \int_0^1  d \mu_{n} \ldots \int_0^{\sqrt{1-\sum_{i= 2}^{n}\mu_i^2}} d\mu_1\sqrt{|g|} \nn 
    &= 2 (2\pi)^n \int_0^1\mu_{n} d \mu_{n} \ldots \int_0^{\sqrt{1-\sum_{i= 2}^{n}\mu_i^2}} \mu_1 d\mu_1 \frac{1}{\mu_{n}} \nn 
    &= 2 (2\pi)^n \int_0^1\mu_{n} d \mu_{n} \ldots \int_0^{\sqrt{1-\sum_{i= 2}^{n}\mu_i^2}} \mu_1 d\mu_1  \frac{1}{\sqrt{1-\sum_{i=1}^n \mu_i^2}}. \label{A2N}
\end{align}
Let (for this section only) $M_i = \mu_i^2$. Then $\mu_i d \mu_i = dM_i/2$, and so we have
\begin{align}
    \mc A_{2n} &= 2 \pi^n \int_0^1 d M_{n} \ldots \int_0^{1-\sum_{i = 2}^{n}M_i} d M_1 \frac{1}{\sqrt{1 - \sum_{i = 1}^n M_i}} \label{A2Nintegral}
\end{align}
The integral can be evaluated as follows.  We note, for $a \neq -1$, 
\begin{align}
    \int_0^C (C-X)^a d X &=  \f{C^{a+1}}{a+1}. \label{CmXintegral}
\end{align}
Then we have,
\begin{align}
    \int_0^{1 - \sum_{i = 2}^{n}} \frac{d M_1}{\sqrt{1 - \sum_{i=1}^n M_i}} &= \int_0^{1 - \sum_{i=2}^{n} M_i} \frac{d M_1}{\sqrt{\left(1 - \sum_{i = 2}^{n}M_i\right) - M_1}} \nn 
    &= 2 \sqrt{1 - \sum_{i = M_2}^{n} M_i}.
\end{align}
The next step is,
\begin{align}
    \int_0^{1-\sum_{i=3}^{n}} d M_2 2 \sqrt{1 - \sum_{i = 2}^{n} M_i} &= 2 \int_0^{1 - \sum_{i=3}^{n}} d M_2 \sqrt{(1-\sum_{i=3}^{n} M_i) - M_2} \nn 
    &= 2 \cdot \frac{2}{3} \left(1 - \sum_{i=3}^{n} M_i\right)^{3/2}. 
\end{align} 
Thus each successive step of the integral brings another power to the term involving $1$ and the remaining $M_i$ sum, while also multiplying in another power. We can conclude
\begin{align}
    \int_0^1 d M_{n} \ldots \int_0^{1-\sum_{i = 2}^{n}M_i} d M_1 \frac{1}{\sqrt{1 - \sum_{i = 1}^n M_i}} &= \frac{2}{1} \cdot \frac{2}{3} \ldots \frac{2}{2n-1} \nn 
    &= \frac{2^{n}}{(2n-1)!!}. \label{evendimmucalc}
\end{align}
We have then
\begin{align}
    \mc A_{2n} &= \f{2(2\pi)^n}{(2n-1)!!}
\end{align}
for $n \geq 1$. 

If $N$ is odd, $N = 2n-1$. In this case, 
\begin{align}
    \mc A_{2n-1} &= \int_0^{2\pi} d \phi_{n} \ldots \int_0^{2\pi} d \phi_1 \int_0^1 d \mu_{n-1} \ldots \int_0^{\sqrt{1-\sum_{i= 2}^{n-1}\mu_i^2}} d\mu_1\sqrt{|g|} \nn 
    &= (2\pi)^n \int_0^1 \mu_{n-1} d \mu_{n-1} \ldots \int_0^{\sqrt{1-\sum_{i= 2}^{n-1}\mu_i^2}}\mu_1 d\mu_1. \label{A2Nm1}
\end{align}
Again letting $M_i = \mu_i^2$ we have
\begin{align}
    \mc A_{2n-1} &= 2 \pi^n \int_0^1 d M_{n-1} \ldots \int_0^{{1 - \sum_{i = 2}^{n-1} M_i}} d M_1.
\end{align}
We can again use \eqref{CmXintegral}. We have,
\begin{align}
    \int_0^{1-\sum_{i=2}^{n-1} M_i} d M_1 &= 1- \sum_{i=2}^{n-1} M_i \nn \nn 
    \int_0^{1-\sum_{i=3}^{n-1} M_i} \left(1 - \sum_{i=2}^{n-1} M_i\right) &= \int_0^{1-\sum_{i=3}^{n-1} M_i} \left[\left(1 - \sum_{i=3}^{n-1} M_i\right) - M_2\right] d M_2 \nn 
    &= \f12 \left(1 - \sum_{i=3}^{n-1} M_i\right)^2,
\end{align}
and so on, so that each successive term brings in another power. The result is 
\begin{align}
    \int_0^1 d M_{n-1} \ldots \int_0^{1-\sum_{i=2}^{n-1} M_i} d M_1 &= \f{1}{(n-1)!}, \label{odddimensionMiintegrals}
\end{align}
and so
\begin{align}
    \mc A_{2n-1} = \f{2 \pi^n}{(n-1)!}.
\end{align}
These match with \eqref{AsubD}. We will encounter integrals of this sort at several points.

\section{Volume Calculations Simplified} \label{VolumeSimplified}

The volume calculations for the Generalized Kerr--AdS spacetimes can be simplified by showing that, when we use the pseudo-Cartesian coordinates, the vector volume with respect to $\xi$ is equal to the Euclidean volume of the region. This argument was given in the four-dimensional case in Section \ref{KNAdSPaper}, but the following generalizes to higher dimensions using the same principle.

Consider the GKAdS spacetime in KS coordinates (see Chapter \ref{GKAdSChapter}). Let $\xi = \pa/\pa t$ and let $\mc R$ be the region consisting of $-\infty < t < \infty, x^i \in \Sigma$ where $x^i$ represent all the $x^a$ coordinates except for $t$. The vector volume (see Chapter \ref{PRDPaper}) is then
\begin{align}
    \mc V_{\xi,\mc R} &= \int_\Sigma \sqrt{-g} d^{D-1} x.
\end{align}
$d^{D-1} x$ is the product of the coordinate differentials except for $dt$. Because $\sqrt{-g} = \sqrt{-\bar g}$, we can instead perform the volume calculation in the background metric, and so write
\begin{align}
    \mc V_{\xi,\mc R} &= \int_\Sigma \sqrt{-\bar g} d^{D-1} x.
\end{align}
It is now useful to find the most convenient coordinate system in which to evaluate the vector volume for the background spacetime. Rather than the spheroidal coordinates, it is useful to use the pseudo-Cartesian coordinates of Section \ref{pseudoCartesian}, in the coordinate system $(t,x_i,y_i,z) = (t,X_\alpha)$ with metric \eqref{gabpseudoCartesianwitht}. This gives metric determinant $-1$ from \eqref{determinantpseudoCartesianwitht}. In this case we simply have 
\begin{align}
    \mc V_{\xi,\mc R} &= \int_\Sigma d^{D-1} x,
\end{align}
that is to say, the $(D-1)$-dimensional \emph{Euclidean volume} of $\Sigma$, as expressed in coordinates $(x_i,y_i,z)$. 

Thus all we need to do to calculate the volume of some region is to calculate the Euclidean volume once we express the AdS background in $(t,x_i,y_i,z)$ coordinates. Note that the same argument applies but in an even more obvious way if the background is Minkowski rather than AdS (so that it will also apply to volume calculations for Kerr or Myers--Perry black holes).

The black hole horizon is at a surface of constant $r$. It is then useful to consider surfaces of constant $r$. 
The surface of constant $r$ is a $(D-1)$-dimensional ellipsoid in the coordinates $(x_i,y_i,z)$, satisfying \eqref{rspheruvxyzequation}. From \eqref{rspheruvxyzequation}, the semi-major axes are two copies each of $\sqrt{(r^2+a_i^2)/\Xi_i}$ (one corresponding to the $x_i$ direction and one to the $y_i$ direction), as well as a final semi-major axis $r$ if $D$ is even (corresponding to $z$). The Euclidean volume of a $(D-1)$-dimensional ellipsoid is the volume of the unit $(D-1)$-ball times the product of the $(D-1)$ semi-major axes (see e.g.~\cite{Jorgensen}). The reason is essentially that an ellipsoid is a linear rescaling of a sphere, where each dimension is rescaled by one of the semi-major axes. Thus the interior Euclidean volume of the ellipsoid to the interior of the $r$-constant surface is
\begin{align}
    &\mc V_{D-1} r^{1-\ve} \prod_{i=1}^{n-1+\ve} \f{r^2+a_i^2}{\Xi_i} \nn 
    &= (D-1) \mc A_{D-2} r^{1-\ve} \prod_{i=1}^{n-1+\ve} \f{r^2+a_i^2}{\Xi_i}. \label{ellipsoidvolume}
\end{align}
Note immediately that if $r = 0$ in even dimensions, or if $r^2+a_n^2 = 0$ in odd dimensions, this volume is zero. This makes sense: this corresponds to an ellipsoid with at least one zero semi-major axis, or a measure zero ``flattened ellipsoid''. 

Let $\mc R(R)$ be the region $-\infty < t < +\infty$ and with $x^i \in \Si(R)$ (where $x^i$ is the set of coordinates except for $t$), where $\Si(R)$ is defined as the region with $r \leq R$ as well as $r > 0$ ($D$ even) and $r^2+a_n^2 > 0$ ($D$ odd). The spacetime is regular (in the sense of a lack of curvature singularities) in this region for both the background AdS metric (which has no curvature singularities) and the full GKAdS metric (where the ring singularity at $r = 0$ or $r^2+a_n^2 = 0$ is excised). Define $\mc V(R)$ to be the volume associated with $\xi$ and this region,
\begin{align}
    \mc V(R) &\equiv \mc V_{\xi,\mc R}. \label{VofR}
\end{align}
In coordinates $(t,x_i,y_i,z)$, we have 
\begin{align}
    \mc V(R) &= \int_{\Si(R)} d^{D-1} x, 
\end{align}
where $d^{D-1} x$ is over the coordinates $x_i,y_i,z$, so that it is just the Euclidean volume of the region $\Si(R)$. This can be calculated in the background AdS spacetime, and in that case is equal to the difference in the volume interior to the ellipsoids with $r = R$ and the volume interior to the ellipsoid with $r = 0$ or $r^2+a_n^2 = 0$. The latter ellipsoid has zero volume, so the difference reduces to the volume of the ellipsoid with $r = R$, which results from substituting $r = R$ into \eqref{ellipsoidvolume}. Consequently we have
\begin{align}
    \mc V(R) &= (D-1) \mc A_{D-2} R^{1-\ve} \prod_{i=1}^{n-1+\ve} \f{R^2+a_i^2}{\Xi_i}. \label{VofRexpression}
\end{align}

This means that the black hole vector volume $\mc V_{\xi,\mc B}$, or CGKP's geometric volume, is equal to
\begin{align}
    \mc V_{\xi,\mc B} &= \mc V(r_+) \nn 
    &= (D-1) \mc A_{D-2} r_+^{1-\ve} \prod_{i=1}^{n-1+\ve} \f{r_+^2+a_i^2}{\Xi_i}.
\end{align}

\section{Area Calculation for Generalized Kerr--AdS} \label{AreaRevisited}

Here I find the black hole horizon area for Generalized Kerr--AdS metrics. The values for the horizon area for Kerr--AdS were found by Gibbons et al.~\cite{GibbonsLu,GibbonsLu2}. I thought it worthwhile to perform my own calculation, which additionally covers the Generalized Kerr--AdS case. Along the way I encounter an integral of a form which will recur in Chapter \ref{ExplicitGKAdSChapter}. This calculation appeared in the main body of my thesis, but it is sufficiently involved that I have relegated it to Appendix \ref{AreaCalculationAppendix} here. The result is that for the generalized Kerr--AdS spacetimes, \eqref{AGibbonsLu} is recovered.

\section{Euclidean Action} \label{EuclideanAction}

GPP \cite{GibbonsPerry} used the Euclidean Action method for calculating the black hole mass, in higher dimensions. The Euclidean Action method was initially applied to (four-dimensional) Schwarzschild--AdS thermodynamics by Hawking and Page in \cite{HawkingPage}. As I will show, the vector volume arises in a natural way from GPP's calculation. I will review GPP's calculation and then provide my own commentary in Section \ref{EuclideanCommentary}. They use Kerr--AdS in ABL coordinates, with metric given by \eqref{ABL}. Then the metric is Euclideanized, in the sense of changing the signature of the metric to a Euclidean (as opposed to Lorentzian) signature. Euclidean time $\tau_E$ and rotational parameters $\alpha_i$ are defined by $\tau = -i\tau_E, a_i = i \alpha_i$, so that the metric is recast in Euclidean form. From now on I'll drop the $E$ subscript. The horizon is at $V(r)-2m=0$, or $r = r_+$. The metric is smooth at the horizon, in the sense that conical singularities are avoided, if $\tau$ is identified with periodicity $\beta = 2\pi / \kappa$, where $\kappa$ is the surface gravity. $\beta$ is the inverse temperature of the black hole and is not to be confused with the $\beta$ Killing vector.

The Euclidean action must be supplemented by a surface contribution term, and so is given by
\begin{align}
    I_D &= - \f1{16\pi} \int_{\mc M} \sqrt{g} \left[ R + (D-1)(D-2)l^{-2}\right] d^D x - \f1{8\pi} \int_{\partial \mc M} \sqrt{h} K d^{D-1} x \nn 
    &= \frac{D-1}{8\pi l^2} \int_{\mc M} \sqrt{g} d^D x - \f1{8\pi} \int_{\partial \mc M} \sqrt h K d^{D-1} x,
\end{align}
where $\mc M$ is the space and $\partial \mc M$ is the boundary. (The negative sign to the left of the integral is GPP's convention.) $h_{i j}$ is the induced ($D-1)$-dimensional metric on the boundary and $K$ the trace of the second fundamental form on the boundary. $\tau$ is integrated only from 0 to $\beta$. 

$r$ still extends in principle from $r_+$ to infinity, and so the integral $\int_{\mc M} \sqrt{g} d^D x$ diverges. The method for regularization is to terminate the $r$ integration at some large value $r = R$ (not to be confused with the Ricci scalar $R$) and then to subtract the action for pure AdS adjusted so that the two metrics match at $r = R$. Then $R$ is sent to infinity. The surface terms will cancel under this prescription so the only concern are the bulk volume terms. The Euclidean action is thus relative to a pure AdS background.

The requirement that the metrics match at $r = R$ means that the background metric must have a rescaled coordinate $\tau$, called $\tau_0$, which satisfies
\begin{align}
    (V(R)-2m) \tau^2 = V(R) \tau_0^2.
\end{align}
$\tau_0$ consequently will be integrated from $0$ to $\beta_0$ where 
\begin{align}
    \beta_0 &= \left(1 - \frac{2m}{V(R)}\right)^{1/2} \beta \nn
    &= \left(1 - \f{m}{V(R)}\right) \beta. \label{beta0}
\end{align}
The latter applies for large $R$. Recall that this is the function $V$ from \eqref{GibbonsV}, not to be confused with the volume term $\mc V(R)$ (which will come up again in Section \ref{EuclideanCommentary} below).

There is no interior boundary for the background AdS metric. Letting $y$ be the spherical radius for AdS, $y$ extends down to 0. From \eqref{spheroidaltransformation}, this corresponds to $r = r_{y=0}$, where
\begin{align}
    r_{y=0}^2 &= - \f{\sum_{i=1}^{n-1+\ve} a_i^2\mu_i^2 \Xi_i^{-1}}{\sum_{j=1}^{n-1+\ve} \mu_j^2 \Xi_j^{-1}}. \label{ry=0}
\end{align}
We can simplify this by letting $r_{y=0} = 0$ in even dimensions and $r_{y=0} = r_0$ where $r_0^2+a_n^2 = 0$ in odd dimensions. (The argument is slightly more complicated, but the idea is that there is no difference between taking $r_{y=0}$ from \eqref{ry=0} and taking $r_{y=0} = 0$ or $r_{y=0}=r_0$ in the volume integrals, because the interior to $r = 0$ in even dimensions and $r = r_0$ in odd dimensions represents a zero-volume region in the AdS background.)

Let 
\begin{align}
    Y &= \frac{(D-1) \beta}{8\pi l^2} \int_{r_+}^R dr \int d^{D-2} x \sqrt{-g}
\end{align}
be the action for the Euclidean Kerr--AdS metric from $r = r_+$ up to radius $R$, and
\begin{align}
    Y_0 &= \frac{(D-1) \beta_0}{8 \pi l^2} \int_{r_{y=0}}^R dr \int d^{D-2} x \sqrt{-\bar g}
\end{align}
be the action of the Euclidean AdS metric integrated up to radius $R$. $d^{D-2} x$ is the integration over direction cosines $\mu_i$ and azimuthal angles $\phi_i$, with the final direction cosine eliminated by applying the constraint $\sum \mu_i^2 = 1$. (The use of $\sqrt{-g}$ above rather than $\sqrt{g}$, present in GPP, suggests that we have ``gone back into the Lorentzian-signature metric'' to perform the actual integral.) The Euclidean action is then expressed as
\begin{align}
    I_D &= \lim_{R \to \infty} (Y - Y_0).
\end{align}

The result that GPP give is
\begin{align}
    I_D &= \frac{\beta \mc A_{D-2}}{8\pi \prod_j \Xi_j} \left(m - r_+^{1-\ve}l^{-2} \prod_{i = 1}^{n-1+\ve} (r_+^2 +a_i^2)\right).
\end{align}
Gibbons et al.~state that they explicitly evaluated the integrals for even dimension up to 12 and odd dimension up to 9. In all dimensions they verify that these expressions satisfy the u,
\begin{align}
    \mc E - \sum_i \Om_i \mc J_i - T S = T I_D. \label{QuantumStatisticalRelation}
\end{align}

\subsection{Commentary} \label{EuclideanCommentary}

It is worth noting that it is possible to simplify their expressions somewhat, in the following way. We recognize immediately that $Y$ and $Y_0$ are proportional to volume integrals. Because the full and background metrics have the same determinant, using the definition $\mc V(R)$ from \eqref{VofR}, $\int_{\Si(R)} d^{D-1} x \sqrt{-g} = \int_{\Si(R)} d^{D-1} x \sqrt{-\bar g} = \mc V(R)$ where $\Si(R)$ represents the region from $r = r_{y=0}$ to $r = R$.

For $Y$, the integral is over the entire region below $r = R$, minus the region below $r = r_+$, and so can be written as
\begin{align}
    Y &= \f{(D-1)\beta}{8\pi l^2} \left( \int_{\Sigma(R)} d^{D-1} x \sqrt{-g} - \int_{\Sigma(r_+)} d^{D-1} x\sqrt{-g} \right) \nn 
    &= \f{(D-1) \beta}{8\pi l^2} (\mc V(R) - \mc V(r_+)).
\end{align}
For $Y_0$ we have, similarly,
\begin{align}
    Y_0 &= \f{(D-1)\beta_0}{8\pi l^2} \mc V(R).
\end{align}

Performing the subtraction, we have
\begin{align}
    I_D &= \lim_{R\to \infty} (Y - Y_0) \nn 
    &= \frac{D-1}{8\pi l^2}  \left( \lim_{R \to \infty}(\beta - \beta_0) \mc V(R) - \beta \mc V(r_+)\right) \nn 
    &= \frac{(D-1) \beta}{8\pi l^2} \left( \lim_{R \to \infty} \left(1 - \frac{\beta_0}{\beta}\right) \mc V(R) - \mc V(r_+)\right). 
\end{align}
In the large $R$ limit, the various terms $R^2 + a_i^2 \to R^2$. We then have, from \eqref{VofRexpression} and \eqref{GibbonsV},
\begin{align}
    \lim_{r\to\infty} \f{\mc V(R)}{V(R)} &= \f{l^2 \mc V_{D-1}}{\prod_{i=1}^n\Xi_i},
\end{align}
so that 
\begin{align}
    \lim_{R\to\infty} \left(1 - \frac{\beta_0}{\beta}\right) \mc V(R) &= \lim_{R\to\infty} \frac{m}{V(R)} \mc V(R) \nn 
    &= \frac{m l^2 \mc V_{D-1}}{\prod_{i = 1}^n \Xi_i}.
\end{align}
Thus we have
\begin{align}
    I_D &= \frac{(D-1) \beta}{8\pi l^2} \left( \frac{ m l^2 \mc V_{D-1}}{\prod_{i = 1}^n \Xi_i} - \mc V(r_+)\right).
\end{align}
$\mc V_{D-1}$ can be replaced with $\mc A_{D-2}/(D-1)$. Multiplying by $T = \beta^{-1}$,
\begin{align}
    T I_D &= \f{ m \mc A_{D-2}}{8 \pi \prod_{i=1}^n \Xi_i} - \f{(D-1) \mc V(r_+)}{8 \pi l^2} \nn 
    &= \f{m \mc A_{D-2}}{8\pi \prod_{i=1}^n \Xi_i} + \f{1}{8\pi} \tilde \La \mc V_{\xi,\mc B}.
\end{align}
Thus the vector volume of the black hole enters into the Euclidean action calculation in a natural way.

Now it is worth comparing to \eqref{QuantumStatisticalRelation}. Recall that we showed \eqref{SmarrGibbsDuhem2}, or that, in BC notation,
\begin{align}
    \mc E - \sum_i \Om_i \mc J_i - TS &= \f{1}{8\pi}\tilde \La \mc V_{\xi,\mc B}  + \oint_H \bs C_{\z;\g}.
\end{align}
Using the Kerr--Schild conserved charge from Section \ref{KSBC}, $\bs C_{\z;\g} = -\z \cdot \bs \tht^{EH}$, the integral of which indeed gives the expected result $m \mc A_{D-2}/(8\pi \prod_{i=1}^n \Xi_i)$. (See either \eqref{CzetagammaBC} for BC's calculation of $\bs C_{\z;\g}$ or Section \ref{nonKomarterm} for my explicit calculation.)

\section{The Relationship Between Killing Potentials and Vector Volume} \label{killingpotentialvectorvolumerelationship}

In this section I give an explanation for \eqref{ointHomegazeta}, $\oint_H * \bs \om_\z = - 2 V_{geo}$. As a reminder, the $\bs \om_\z$ in this equation is the one constructed out of $\bs \om^{(j)}$ given by \eqref{omegaj} and, in odd dimensions, $\bs \om^{(n)}$ given by \eqref{omegan}.

Let $S_0$ be the $(D-2)$-surface of constant $t$ with $r = 0$ in even dimension and $r^2+a_n^2 = 0$ in odd dimension. This is the $r$ location of the singularity. Technically $S_0$ is not part of the manifold, so in what follows assume that we perform some sort of limiting procedure (taking $S_0$ as the limiting case of some $S(r)$ of constant $t,r$ where $r$ approaches $0$ or $r^2+a_n^2 = 0$). I will now show that $\oint_{S_0} * \bs \om^{(j)} = 0$ for all $j$ (including $j = n$).

On a surface of constant $(t,r)$, we have,
\begin{align}
    (* \bs \om^{(j)})_{y_1 \ldots y_{n-1} \phi_1 \ldots \phi_{n-1+\ve}} \propto \sqrt{-g} (\bs \om^{(j)})^{t r}.
\end{align}
We thus wish to find $(\bs \om^{(j)})^{t r}$, in KS coordinates, in arbitrary dimension. With the exception of $\bs \om^{(n)}$ in odd dimension, we have \eqref{omegaj} so that $\bs \om^{(j)}$ is constructed out of $x_\mu \bs \om^\mu = x_\mu e^\mu \wedge e^{\hat \mu}$. The only canonical forms, in KS coordinates, with nonzero $e_A^t$ are ones where $A = \hat \mu$ (or $\hat 0$ in odd dimensions). We also have $e_\alpha^r = 0$ where $1 \leq \alpha \leq n-1$. So the only possible $(\bs \om^\mu)^{tr}$ will be from $\mu = n$. In this case we have, using \eqref{canonicalformsinryphicoords}
\begin{align}
    x_n (\bs \om^n)^{tr} &= -i r e_n^r e_{\hat n}^t \nn 
    &= +\f{r \bar X}{U_n (1+r^2/l^2)}.
\end{align}
Using the expression for $\sqrt{-g} = \sqrt{-\bar g}$ in KS coordinates $(t,r,y_\alpha,\phi_i)$ from \eqref{sqrtgrytphi} as well as $\bar X$ from \eqref{Xbar}, we find
\begin{align}
    \sqrt{-g} x_n (\bs \om^n)^{tr} &= \f{P r^{1-\ve} (\prod_{\alpha=1}^{n-1} y_\alpha)^\ve \prod_{i=1}^{n-1+\ve} (r^2+a_i^2)}{U_n C (\prod_{i=1}^n \Xi_i) (\prod_{i=1}^{n-1+\ve} a_i)^{1-\ve}}.
\end{align}
In even dimensions then, this expression is exactly zero when $r = 0$. In odd dimensions, it vanishes only when $r^2+a_i^2 = 0$. As stated in Section \ref{ringsingularity}, in odd dimensions, the ring singularity occurs when $r^2+a_n^2 = 0$ so we find that $\sqrt{-g} x_n (\bs \om^n)^{tr} = 0$ on this surface. (Precisely, the ring singularity is not part of the manifold, so we can instead take a limiting surface as $r \to 0$ in even dimensions and $r^2 \to -a_n^2$ in odd dimensions, as mentioned before.) The $A_n^{(j)}$ terms do not depend on $r$, and are finite (depending only on the $y_\alpha$). Thus we have that $\sqrt{-g} (\bs \om^{(j)})^{t r} = 0$ on $r = 0$ (even $D$) and $r^2+a_n^2 = 0$ (odd $D$). 

Similarly, for odd $D$, $\sqrt{-g} (\bs \om^{(n)})^{tr}$ vanishes on $r^2+a_n^2 = 0$, as shown in Appendix \ref{potentialomegan}. (Interestingly, this works if we use $\bs \om^{(n)} \propto * \bs b^{(n)} = * \left( \bs b'' \wedge \bs h^{(n-1)}\right)$, but not if we use $\bs \om^{(n)} \propto *\left( \bs b \wedge \bs h^{(n-1)}\right)$.) 

Consequently,
\begin{align}
    \oint_{S_0} * \bs \om^{(j)} &\propto \oint \sqrt{-g} (\bs \om^{(j)})^{tr} d y_1 \ldots dy_{n-1} d\phi_1 \ldots d\phi_{n-1+\ve} \nn 
    &= 0,
\end{align}
for $0 \leq j \leq n-1+\ve$.

Now let $S$ be another (closed) $(D-2)$-surface. Let $\Sigma$ be the $(D-1)$-surface with boundary $\pa \Sigma = S \cup S_0$. We then have, choosing orientation ``pointing outward'' (away from small $r$) on both $S$ and $S_0$ and with orientation pointing towards increasing $t$,
\begin{align}
    \oint_S * \bs \om^{(j)} &= -2 \int_\Sigma * \mathrm{div} \bs \om^{(j)}  + \oint_{S_0} * \bs \om^{(j)}.
\end{align}
Here we used \eqref{StokesLawCodim1}.

$\oint_{S_0} * \bs \om^{(j)} = 0$, as we pointed out. Of course $\mathrm{div} \bs \om^{(j)} = (\beta^{(j)})^\flat$. We recognize the integral over $\Sigma$ to be the vector volume associated with $\beta^{(j)}$ and the region spanned by $\Sigma$, which we can just write as $\mc V_{\beta^{(j)},\Sigma}$. 
\begin{align}
    \oint_S * \bs \om^{(j)} &= -2 \mc V_{\beta^{(j)},\Sigma}. \label{Vbetaj}
\end{align}

Because of the linearity of the vector volume with respect to vector and the vanishing of the vector volume contribution associated with azimuthal Killing vectors as in Section \ref{twovectorsSection}, the only term which will contribute comes from the $\xi$ component of $\beta^{(j)}$. Since $\beta^{(j)} = \f{\pa t}{\pa \psi_j}\f{\pa}{\pa t} + \sum_{i=1}^{n-1+\ve} \f{\pa \phi_i}{\pa \psi_j} \f{\pa}{\pa \phi_i} = \f{\pa t}{\pa \psi_j} \xi + \sum_{i=1}^{n-1+\ve} \f{\pa \phi_i}{\pa \psi_j} \eta_i$, we have
\begin{align}
    \mc V_{\beta^{(j)},\Sigma} &= \f{\pa t}{\pa \psi_j} \mc V_{\xi,\Sigma} \nn 
    &= \hat C_0^{(j)} \mc V_{\xi,\Si} \nn 
    \oint_S * \bs \om^{(j)} &= - 2 \hat C^{(j)}_0.
\end{align}

In particular, the Killing potential for $\bs \om_\z$ gives
\begin{align}
    \oint_S * \bs \om_\z &= - 2 \mc V_{\z,\Sigma} \nn 
    &= - 2 \mc V_{\xi,\Sigma}.
\end{align}
The first line follows from \eqref{Vbetaj} and $\z$'s decomposition into $\bt^{(j)}$. If evaluated on $H$, the result is, letting $\mc V_{\xi,\mc B}$ be the vector volume associated with $\xi$ for a region which extends from $r = 0$ (even) or $r^2+a_n^2$ (odd) out to the horizon at $r = r_+$,
\begin{align}
    \oint_H * \bs \om_\z &= - 2 \mc V_{\xi,\mc B},
\end{align}
recovering \eqref{ointHomegazeta}.

We can go further and state that if $\chi$ is a Killing vector given by
\begin{align}
    \chi &= \chi^t \xi + \sum_{i=1}^{n-1+\ve} \chi^{\phi_i} \eta_i,
\end{align}
and $\bs \om_\chi$ is its Killing potential constructed from the $\bs \om^{(j)}$ in the method proscribed by CGKP, then we must have
\begin{align}
    \oint_H * \bs \om_\chi &= - 2 \chi^t \mc V_{\xi,\mc B}.
\end{align}
Since $\mc V_{\xi,\mc B}$ is the vector volume, or geometric volume, associated with the black hole, this shows how the Killing potential method will \emph{automatically} recover the vector/geometric volume, up to a simple numerical constant, and that the result follows from the Gauss--Stokes law and from the vanishing of the Killing potentials $\bs \om^{(j)}$ at the $r = 0$ ($D$ even) or $r^2+a_n^2=0$ ($D$ odd) surface. This is also not unique to the horizon/black hole region. The integration on any $(D-2)$-surface $S$ will give the vector/geometric volume for the region $\Sigma$ bounded by $S$ and $S_0$.

Note further that since $\bs h = (D-1) \bs \om^{(0)}$,
\begin{align}
    \oint_H * \bs h &= -2 (D-1) \mc V_{\beta, \mc B} \nn 
    &= - 2 (D-1) \mc V_{\xi,\mc B}. \label{ointHstarh}
\end{align}

\section{The Area--Volume Relationship for Black Holes}
\label{AreaVolumeRelationship}

Here I show that the area--volume relationship \eqref{VrArelationship} is a simple consequence of the form of the PCKY tensor $\bs h$ as well as the result \eqref{ointHstarh}. We have,
\begin{align}
    V_{geo} &= \mc V_{\xi,\mc B} \nn 
    &= -\f{1}{2 (D-1)} \oint_H * \bs h \nn 
    &= - \f{1}{2(D-1)} \oint_H \bs h^{a b} d S_{a b}.
\end{align}
The area element $dS_{ab} = 2 \z_{[a} n_{b]} dS$ from \eqref{horizonbinormal}, where recall $n_a \z^a = -1$. We can thus write
\begin{align}
    V_{geo} &= -\f{1}{D-1} \oint_H \bs h^{a b} \z_a n_b dS
\end{align}
As in \eqref{zetaproptol} we find that $\zeta \propto \ell$ on the horizon. The null vector $\ell$ is one of the eigenvectors of $\bs h$, with eigenvalue $r$ (see \eqref{elldoth}). Consequently $\z$ is an eigenvector of $\bs h$ with eigenvalue $r$. We then have
\begin{align}
    V_{geo} &= -\f1{D-1} \oint_H r \z^b n_b dS \nn 
    &= +\f{1}{D-1} \oint_H r dS.
\end{align}
Since the horizon is at constant $r = r_+$,
\begin{align}
    V_{geo} &= \f{1}{D-1} r_+ \oint_H dS\nn 
    &= \f{r_+ A}{D-1},
\end{align}
thus demonstrating \eqref{VrArelationship}.

To emphasize, the chain of reasoning here uses the fact that $\bs h$ has $\oint_{S_0} * \bs h =0$, that $\mathrm{div} \bs h = (D-1)^{-1}\beta^\flat$ and that $\z$ is an eigenvector of $\bs h$ with, on the horizon, a constant eigenvalue $r_+$. Thus the simple area--volume relationship is very heavily entwined with the properties of $\bs h$. The appearance of $r_+$ in the relationship is a direct consequence of $r_+$ being the eigenvalue associated with $\z$'s role as eigenvector in $\bs h$. This is sufficiently involved that I will summarize it in one series of equations:
\begin{align}
     V_{geo} &= \int_\Si \xi^a d \Si_a &\textrm{(definition in terms of vector volume)} \nn
    &= \int_\Si \bt^a d \Si_a &\textrm{(noncontribution of angular Killing vector volumes)} \nn 
    &= \f{1}{D-1} \int_\Si \na_b \bs h^{ba} d \Si_a &\textrm{(relationship between }\bt^a\textrm{ and }\bs h) \nn 
    &= -\f1{2(D-1)} \left( \oint_{H} \bs h^{ab} d S_{ab} - \oint_{S_0} \bs h^{ab} d S_{ab}\right) & \textrm{(Gauss--Stokes)} \nn 
    &= -\f1{2(D-1)} \oint_H \bs h^{ab} d S_{ab} & \textrm{(vanishing of contribution from just outside singularity)} \nn 
    &= -\f1{2(D-1)} \oint_H \bs h^{ab} (2 \z_{[a} n_{b]} dS) & \textrm{(use of binormal)} \nn 
    &= -\f1{D-1} \oint_H r_+ \z^b n_b dS & (\bs h^{ab} \z_a = r \z^b = r_+ \z^b \textrm{on horizon}) \nn 
    &= + \f{1}{D-1} \oint_H r_+ dS & (\z^b n_b = -1) \nn 
    &= \f{r_+ A}{D-1} & (\oint_H dS = A). \label{chainofreasoningforrplus} 
\end{align} 

The focus on this thesis is on the Generalized Kerr--AdS solutions, but the same basic argument applies for Kerr (and Kerr--Newman), Myers--Perry, and Kerr--de Sitter black holes, which also have a PCKY tensor and also have $\mc V_{\xi,\mc B} = r_+A/(D-1)$ relationship. Moreover, the relationship is not unique to the outer horizon, but applies to the inner horizon(s) or, in the Kerr--dS case, the cosmological horizon. This relationship that $V_{geo} = r A/(D-1)$ was noted for the cosmological and inner horizons in \cite{DolanKastor}; I note here that it can be attributed to the existence of the PCKY tensor in Kerr--dS spacetimes. This follows from the fact that, if $X = 0$ for some value of $r = r_1$, $(U_n/\bar X) \ell|_{r=r_1}$ is a Killing vector (as pointed out after \eqref{ellifXis0}), and so there is a Killing horizon, where the vector tangent to the null generators of the horizon will be an eigenvector of $\bs h$ with eigenvalue $r_1$. In this case we will, by the same argument, have $\mc V_{\xi,\mc B_1} = r_1 A_1/(D-1)$, where $A_1$ is the area of the Killing horizon at $r = r_1$ and $\mc B_1$ is the region between $S_0$ and $r = r_1$. 

\chapter{Explicit Calculation of Charge Integrals for Generalized Kerr--AdS} \label{ExplicitGKAdSChapter}

In this chapter, I apply the quantity $\bs I_\chi$ defined in Chapter \ref{NoetherChapter} to the Generalized Kerr--AdS metrics to describe the energy (or enthalpy) and angular momentum. In doing so, I address the questions: ``why it is necessary to use a conserved charge based on $\xi$ rather than $\bt$ in order to recover a horizon variation law?'' and ``why does the volume occur naturally in the Smarr relation for the conserved quantity associated with $\bt$, rather than with $\xi$''? 

I give an overview of the values for the energy and angular momentum in Section \ref{overviewofconservedcharges}. Then in Section \ref{explicitIxi} I explicitly calculate the value of $\bs I_\chi$ for various values of $\chi$ in the Generalized Kerr--AdS case, including Kerr--Newman--AdS. I show that for solutions of Einstein's equations (Kerr--AdS and Kerr--Newman--AdS) the $\oint_H \bs I_\xi$ and $-\oint_H \bs I_\eta$ correspond to the $\mc E$ and $\mc J$, respectively, given in Chapter \ref{NoetherChapter} by either the GPP method or the AMD conformal methods. 

In Section \ref{variationsinfourdimensions}, I calculate the variations in the integrals of $\oint_H \bs I_\xi, \oint_H \bs I_\eta$ and $\oint_H \bs I_\bt$, and examine under what circumstances their variations reproduce the expected horizon variation law $\de \mc E - \Om \de \mc J = \kappa \de A/8\pi$ (in the Kerr--Newman--AdS case also including charge). I also examine under what circumstances $\oint_H \bs k^{EH}_{\bar \z}[\de g;g] = \kappa \de A/8\pi$. My finding is that whereas $\oint_H \bs k^{EH}_{\bar \z}[\de g;g] = \kappa \de A/8\pi$ holds in an arbitrary form of the GKAdS metric, even for metrics which are not solutions to Einstein's equations, $\de \mc E - \Om \de \mc J = \kappa \de A/8\pi$ only holds if we use the specific form $\mc E = \oint_H \bs I_\xi, \mc J = - \oint_H \bs I_\eta$ for the Kerr--AdS solution, a solution to Einstein's equations. I further show that in general $\de \mc F - \om \de \mc J \neq \kappa \de A/8\pi$ even in vacuum (where $\mc F = \oint_H \bs I_\bt$).

I then wish to address the fact that $\oint_H \bs k^{EH}_{\bar \z}[\de g;g] = \kappa \de A/8\pi$ has only been derived to apply in a situation where the horizon remains at the same coordinates in the initial and final spacetime, as well as the fact that we expect $\de \bs I_\chi = \bs k^{EH}_\chi [\de g;g]$ only if the background metric $\de \bar g_{ab} = 0$. To account for the fact that, in general, both the coordinates on the horizon and the components of the AdS background metric change when comparing nearby spacetimes, I introduce in Section \ref{infinitesimalcoordinatetransformations} the idea of using an infinitesimal coordinate transformation, using the Lie derivative, to a new set of closely related coordinates. This allows, when applied in different ways, for the horizon to be kept at constant coordinates and for the background metric to have constant components. Applying this allows me to show why it is that $\oint_H \bs k_{\bar \z}^{EH}[\de g;g] = \kappa \de A/8\pi$ can be used even when the horizon radius changes, as well as why, when $T_{ab} = 0$, $\de \oint_H \bs I_{\bar \z} = \oint_H \bs k^{EH}_{\bar \z}$ can be used even when the background metric varies. Finally I compare the cases where the metric is asymptotically static and asymptotically rotating, and show how it is the result of the fact that the difference between the vectors $\xi$ and $\bt$ is a function of rotation parameter $a$ that the variation law fails when using $\bt$ rather than $\xi$.

The overall thrust of these sections, specialized to four dimensions, is to show why $\oint_H \bs I_\chi$ makes sense as a conserved quantity constructed to produce a metric variation law, and why it applies particularly for vacuum and with $\xi,\eta$ rather than $\bt, \eta$---or, in other words, why it is that it is necessary to use charges constructed from $\xi,\eta$ rather than $\bt,\eta$. (In the context of the papers listed in Section \ref{clarifyingcomments}, this amounts to another demonstration that charges associated with $\xi$ and $\eta$ are integrable, but that the one associated with $\bt$ is not; in the language of BC, this is due to the failure of the strong integrability conditions associated with $Q_{\bt;\g}$.) 

In Section \ref{chargebreakdown}, I calculate the explicit values for $\oint \bs I_\chi$ for $\chi = \bt, \xi, \eta_i$ and $\z$ on surfaces of constant $t,r$ in arbitrary dimension $D \geq 4$. I show how the calculation of $\bs I_\bt$ simplifies considerably due to the properties of $\bt$ introduced in Chapter \ref{GKAdSChapter}. Breaking $\bs I_\bt$ down into terms related to the Komar terms (the Komar integral minus the Komar integral for the background) and the non-Komar integral, I simplify expressions down to show as cleanly as possibly why, for $\bt$ in particular as opposed to other Killing vectors, the Komar term is $D-3$ times the non-Komar term, in vacuum. This addresses, according to the arguments in Chapter \ref{NoetherChapter}, why the vector volume appears naturally in the Smarr relation constructed from $\bt$, but not the one constructed with $\xi$. 

In Section \ref{physicalprocessversion}, I discuss, in four dimensions, a physical process version of the first law and how this can also be used to describe the conserved quantities in Kerr--AdS, in a way that dovetails with other results in this chapter.

A note I want to make at the start of this chapter is that in Kerr (in four dimensions) or Myers--Perry spacetimes (in higher dimensions), which are the Kerr--AdS limit with $\La = 0$ or $l \to \infty$, the Komar integral associated with the background spacetime is zero, because Komar integrals are zero in Minkowski space. Further, since $\bt = \xi + l^{-2} a \eta$, in Kerr, $\bt = \xi$, and so the $\xi$ and $\bt$ conserved quantities coincide. The combination of the two simplifies matters greatly. The vanishing of the background Komar integral is the reason that there is no volume term appearing in the Smarr relation for Kerr/Myers--Perry, and the fact that $\xi = \bt$ explains why the distinction between $\xi$ and $\bt$, so important in the Kerr--AdS case, does not materialize in Kerr/Myers--Perry.

\section{Overview of Conserved Charges for Kerr--AdS} \label{overviewofconservedcharges}

The conserved charges for Kerr--AdS calculated using $\bs I_\xi$ and $\bs I_{\eta_i}$ are the same as the ones calculated using the Ashtekar--Magnon--Das method and which appear in Section \ref{BHThermoforKAdSBHs}. The aim of this chapter is to use my definition of charge to develop some insight into why the expressions take the form they do. Before doing so I want to make some general comments, which follow from the expressions in Section \ref{BHThermoforKAdSBHs}.

I wish to make a few comments on the results for $\mc E, \mc F$ and $\mc J_i$. I note that $\mc F$ has a somewhat simpler expression than $\mc E$ and $\mc J_i$, and so it is convenient to rewrite $\mc E$ and $\mc J_i$ in terms of $\mc F$:
\begin{align}
    \mc F &= \f{(D-2) m \mc A_{D-2}}{8 \pi \prod_j \Xi_j} \nn 
    \mc E &= \f{2}{D-2} \left( \sum_{i=1}^{n-1+\ve} \Xi_i^{-1} - \f{\ve}{2}\right) \mc F \nn 
    \mc J_i &= \f{2 a_i}{(D-2) \Xi_i} \mc F.
\end{align}

As stated previously, the $Q_C$ charges calculated using the AMD conformal methods are linear with respect to the Killing vector in use. These are also the quantities which were found by GPP to satisfy the first law of black hole mechanics. They are also, as I will show, the values that result from calculating $\oint_S \bs I_\xi$ and $-\oint_S \bs I_{\eta_i}$, respectively, for arbitrary $(D-2)$-surface $S$. Because they are associated with the Hamiltonian, I will use the notation $\mc E = H^{\bs I}_\xi, \mc J_i = -H^{\bs I}_{\eta_i}$, and further note that $H^{\bs I}_\chi$ is linear with respect to $\chi$ for Killing vectors $\chi$. We expect that $H^{\bs I}_\chi = Q_C[\chi]$ where $Q_C$ is the AMD charge, as well as $H^{\bs I}_\chi = \oint_S \bs I_\chi$. 

Because $(\xi,\eta_i)$ can be expressed in terms of basis vectors $\pa/\pa \psi_j$, and the conserved charge associated with $\beta = \pa/\pa \psi_0$ has a simple form, I wanted to check the value for $H^{\bs I}_{\pa/\pa \psi_j}$ for arbitrary $j$. We have, using \eqref{psivarphiequations},
\begin{align}
    \f{\pa}{\pa \psi_j} &= \sum_{i=0}^{n-1+\ve} \hat C_i^{(j)} \f{a_i}{l^2} \f{\pa}{\pa \hat \vp_i} \nn 
    &= \hat C_0^{(j)} \xi + \sum_{i=1}^{n-1+\ve} \hat C_i^{(j)} \f{a_i}{l^2} \eta_i \nn 
    H^{\bs I}_{\pa_{\psi_j}} &= \hat C_0^{(j)} Q_C[\xi] + \sum_{i=1}^{n-1+\ve} \hat C_i^{(j)} \f{a_i}{l^2} Q_C[\eta_i] \nn 
    &= \hat C_0^{(j)} \mc E - \sum_{i=1}^{n-1+\ve} \hat C_i^{(j)} \f{a_i}{l^2} \mc J_i \nn 
    &= \f{2 \mc F}{D-2} \left\{ \hat C_0^{(j)} \left( \sum_{i=1}^{n-1+\ve} \Xi_i^{-1} - \f{\ve}{2}\right) - \sum_{i=1}^{n-1+\ve} \hat C_i^{(j)} \f{a_i^2}{l^2 \Xi_i}\right\} \nn 
    &= \f{\mc F}{D-2} \left\{ - \ve \hat C_0^{(j)} + 2\sum_{i=1}^{n-1+\ve} \f{l^2 \hat C_0^{(j)} - a_i^2 \hat C_i^{(j)}}{l^2-a_i^2}\right\}.
\end{align}
Recalling the definition of $\hat C_i^{(j)}$ \eqref{Chatil} and of $C_i^{(j)}$ (no hat) from \eqref{Cil}, we note that, for $j > 1$, $\hat C_0^{(j)}$ can be expanded into terms which include $a_i^2$ and terms which do not, as
\begin{align}
    \hat C_0^{(j)} &= C_i^{(j)} + a_i^2 C_i^{(j-1)},
\end{align}
and similarly
\begin{align}
    \hat C_i^{(j)} &= C_i^{(j)} + l^2 C_i^{(j-1)}.
\end{align}
We then have, for $j > 0$,
\begin{align}
    l^2 \hat C_i^{(j)} - a_i^2 \hat C_i^{(j)} &= l^2 \left( C_i^{(j)} +a_i^2 C_i^{(j-1)}\right) - a_i^2 \left( C_i^{(j)} + l^2 C_i^{(j-1)}\right) \nn 
    &= (l^2-a_i^2) C_i^{(j)}.
\end{align}
Thus
\begin{align}
    H^{\bs I}_{\pa_{\psi_j}} &= \f{\mc F}{D-2} \left\{ -\ve \hat C_0^{(j)} + 2\sum_{i=1}^{n-1+\ve} C_i^{(j)}\right\}.
\end{align}
To evaluate $\sum_{i=1}^{n-1+\ve} C_i^{(j)}$, consider a term $a_{k_1}^2 \ldots a_{k_j}^2$, where $1 \leq k_1 < \ldots < k_j \leq n-1+\ve$. Note that this term will appear in each $C_i^{(j)}$ where $i$ is not in the set of $k_l$. Since there are $j$ terms in the set of $k_l$ and $n-1+\ve$ terms in the sum over $i$, the term $a_{k_1}^2 \ldots a_{k_j}^2$ appears  $n-1+\ve-j$ times in total. We thus have
\begin{align}
    \sum_{i=1}^{n-1+\ve} C_i^{(j)} &= (n-1+\ve-j) \sum_{1 \leq k_1 < \ldots < k_j \leq n-1+\ve} a_{k_1}^2 \ldots a_{k_j}^2 \nn 
    &= (n-1+\ve - j) \hat C_0^{(j)}.
\end{align}
Thus,
\begin{align}
    H^{\bs I}_{\pa_{\psi_j}} &= \f{\hat C_0^{(j)} \mc F}{D-2} \left( 2(n-1 -j ) + \ve\right) \nn 
    &= \f{D-2-2j}{D-2} \hat C_0^{(j)} \mc F, \label{Hpapsij}
\end{align}
which also holds for $j = 0$. This has a relatively simple form. 

Of particular interest is the even-dimensional case, where the ``last'' Killing vector $\f{\pa}{\pa \psi_{n-1}}$ has $2(n-1) = D-2$, so that $H^{\bs I}_{\pa_{\psi_{n-1}}} = 0$. It is unclear whether there is any physical significance to the disappearance of the charge associated with $\pa/\pa \psi_{n-1}$ in even dimensions. None of the charges associated with the $\pa/\pa \psi_j$ vanish in odd dimensions (in general).

In $D = 4$, in $(r,\theta,\psi_j)$ coordinates there are only two linearly independent Killing vectors associated with the time--azimuthal sector, $\beta = \pa/\pa \psi_0$ and $\pa/\pa\psi_1$. Of this set, $\bt$ gives a nonzero charge and $\pa/\pa \psi_1$ gives a zero charge. The associated charges for $\xi$ and $\eta$ can be found entirely in terms of the decomposition of $\xi$ and $\eta$ in $\psi_j$ coordinates: $\mc E = \mc F \xi^{\psi_0}$ and $\mc J = - \mc F \eta^{\psi_0}$. 

Some more commentary (which did not appear in the thesis) on $H^{\bs I}_{\psi_j}$ is included in Section \ref{betajsection}. 

\section{Explicit Calculation of Charge Integrals: Four Dimensions} \label{explicitIxi} 

To give a little bit of extra information, I will find the solutions for a ``generalized Kerr--anti-de Sitter'' case, which replaces the constant $m$ with a general function of $r$, $m \to \mu(r)$. The main other physically interesting solution that is included in this form is the four-dimensional Kerr--Newman solution, but it is my hope that solving for an arbitrary function $\mu(r)$ might give some hints as to which aspects of the Kerr--anti-de Sitter solution arise because of the Kerr--Schild form and which arise specifically from the vacuum case. I will report results for the four-dimensional case in this section.

Because the topology is so unusual in the Kerr--NUT--AdS case (let alone a generalized Kerr--NUT--AdS case) I will not be including any NUT parameters. 

In four dimensions, the calculations can be completed using \emph{GRTensorIII} \cite{GrTensor}. In arbitrary dimensions more care is required. The arbitrary-dimensional case will be considered in Section \ref{chargebreakdown}.

\subsection{Quantities Associated with Conserved Charge} 

We will calculate some quantities associated with $\bs I_\chi$ from \eqref{Ixi}, using the form where the covariant derivatives are taken with respect to the background, which in four dimensions reduces to
\begin{align}
    (\bs I_\chi)_{cd} &= \f{1}{16\pi} \left( h^{e[a}\bar \na_e \chi^{b]} - \chi^e \bar \na^{[a}h^{b]}_e + \chi^{[a} \bar \na_e h^{b] e}\right) \bs \ep_{c d ab}.
\end{align}
Unless otherwise indicated, assume that $\bs I_\chi, \bs k_\chi$ are $\bs I_\chi^{EH},\bs k^{EH}_\chi$. The form of the metric is \eqref{KS4D}. Consider KS coordinates, $(t,r,\tht,\phi)$. Choose an integration surface $C$ of constant $(t,r)$ and integrate over $0 \leq \tht \leq \pi, 0 \leq \phi < 2\pi$. Then we have
\begin{align}
    \oint_C \bs I_\chi &= \int_0^{2\pi} \int_0^\pi (\bs I_\chi)_{\tht \phi} d \tht d \phi.
\end{align}
We will then check the values for $\oint_C \bs I_\chi$ substituting in $\xi = \partial_t, \eta = \partial_\phi$ and $\beta = \xi + (a/l^2) \eta$ for $\chi$. 

It will also be interesting to check the combination $\oint_C (\bs I_\xi + \Omega(r) \bs I_\eta)$ where $\Omega(r) = a(1+r^2/l^2)/(r^2+a^2)$. The combination $\xi + \Om(r) \eta$ is equal to $\z$ on the horizon (with $r = r_+$), and is in general proportional to the canonical orthonormal basis vector $e_{\hat n}$. 

In Kerr--AdS, where $\mu(r)=m$, we find, keeping $m$ as constant, the solutions
\begin{align}
    \mc E &= \oint_C \bs I_\xi = \frac{m}{\Xi^2} \nn
    \mc J &= -\oint_C \bs I_\eta = \frac{m a}{\Xi^2} \nn 
    \mc F &= \oint_C \bs I_\beta = \frac{m}{\Xi} \label{ointCIchi4D}
\end{align}
as is expected. (Recall $\Xi = 1-a^2/l^2$.) These results are all independent of the surface. The combination
\begin{align}
    \oint_C (\bs I_\xi + \Omega(r) \bs I_\eta) &= \frac{m}{\Xi}  \frac{r^2}{r^2+a^2}.
\end{align}

If we now send $m \to \mu(r)$, where $\mu$ is allowed to vary, the integrals now depend on $r$. Let $\tilde{\mc E}(r)$ and $\tilde {\mc J}(r)$ be the $r$-dependent functions expressible as
\begin{align}
    \tilde {\mc E}(r) &\equiv \oint_C \bs I_\xi \nn 
    &= \frac{\mu(r)}{\Xi^2} - \frac{(1+r^2/l^2) \left( (r^2+a^2)\arctan(a/r) - ar\right)}{2 a \Xi^2} \mu'(r)  \nn 
    \tilde {\mc J}(r) &= -\oint_C \bs I_\eta \nn 
    &= \frac{\mu a}{\Xi^2} - \frac{(r^2+a^2)\left((r^2+a^2)\arctan(a/r)-ar\right)}{2 a^2 \Xi^2} \mu'(r), \label{tildeErtildeJr}
\end{align}
where $\mu'(r) = d\mu(r)/dr$. We note that if $\mu(r) = m$ is a constant, then $\tilde{\mc E}(r)$ and $\tilde{\mc J}(r)$ are equal to $\mc E$ and $\mc J$ for Kerr--AdS. 

Similarly, let $\tilde{\mc F}(r) = \oint_C \bs I_\bt$, giving
\begin{align}
    \tilde{\mc F}(r) \equiv \oint_C \bs I_\beta &= \tilde{\mc E}(r) - \frac{a}{l^2} \tilde{\mc J}(r) = \frac{\mu}{\Xi} - \frac{(r^2+a^2)\arctan(a/r)-ar}{2\Xi}\mu'(r).    
\end{align}
We also calculate the combination $\oint_C (\bs I_\xi + \Om(r) \bs I_\eta)$.
\begin{align}
    \oint_C (\bs I_\xi + \Omega(r) \bs I_\eta) &= \tilde {\mc E}(r) - \Omega(r) \tilde{\mc J}(r) = \frac{\mu}{\Xi} \frac{r^2}{r^2+a^2}.
\end{align}
We note that the $\mu'(r)$ terms cancel out exactly in $\tilde{\mc E}(r) - \Omega(r) \tilde{\mc J}(r)$. In fact, the vanishing of the $\mu'(r)$ terms in the combination $\bs I_\xi + \Omega(r) \bs I_\eta$ occurs not just in the integral over the surface at constant $r$, but to $\bs I_\xi + \Omega(r) \bs I_\eta$---all components of which are independent of $\mu'(r)$. The $\tht \phi$ component of $\bs I_\xi + \Omega(r) \bs I_\eta$ is, except for a factor of $\sin \tht$, also a function of $r$ only:
\begin{align}
    (\bs I_\xi + \Omega(r) \bs I_\eta)_{\tht \phi} &= \frac{\mu r^2}{\Xi (r^2+a^2)} \sin \tht.
\end{align}

There are Killing horizons in the full spacetime for values of $r$ satisfying
\begin{align} 
    2\mu(r) r = (r^2+a^2)(1+r^2/l^2) \label{horizonequation4D}
\end{align}
(generalizing the constant-$m$ case). For simplicity I will only work with the outer horizon (if it exists) which I will label $r = r_+$ and is the largest value which satisfies \eqref{horizonequation4D}. 

As discussed in Section \ref{BHHorizonGKAdS}, the horizon null vector is $\z = \xi + \Om \eta$ with $\Om \equiv \Om(r_+)$,
\begin{align}
    \Om &= \f{a(r_+^2+l^2)}{l^2 (r_+^2+a^2)}.
\end{align}
The surface gravity is given by \eqref{kappawithmu} and the area is given by \eqref{AGibbonsLu}. In four dimensions, these are
\begin{align}
    \kappa &= \f{3 r_+^4 + r_+^2(a^2+l^2)-a^2l^2- 2 \mu'(r_+)l^2r^2}{2 r_+(r_+^2+a^2)l^2} \nn
    A &= \f{4\pi (r_+^2+a^2)}{\Xi}. \label{kappaArea4D}
\end{align}
Consider some variation $\de$ of the parameters of the spacetime in which $r_+,a$ and $l$ are allowed to vary ($\de r_+, \de a, \de l \neq 0$). $\de A$ then is
\begin{align}
    \de A &= \frac{8\pi r_+}{\Xi} \de r_+ + \frac{8\pi(1+r_+^2/l^2) a}{\Xi^2} \de a - \frac{8 \pi a^2(r_+^2+a^2)}{l^3 \Xi^2} \de l.
\end{align}

Before continuing I will note the values that $\tilde {\mc E}$ and $\tilde {\mc J}$ take on $r = r_+$. We can replace $\mu(r_+)$ with $V(r_+)/2$. For compactness let 
\begin{align}
    \Psi &= \frac{(r_+^2+a^2)\left((r_+^2+a^2)\arctan(a/r_+)-a r_+\right)}{2 a^2 \Xi^2}.
\end{align}
Then we have
\begin{align}
    \tilde {\mc E}(r_+) &= \frac{V(r_+)}{2 \Xi^2} - \Omega \Psi \mu'(r_+) \nn 
    \tilde {\mc J}(r_+) &= \frac{V(r_+) a}{2\Xi^2} - \Psi \mu'(r_+) \nn 
    \tilde {\mc E}(r_+) - \Omega \tilde{\mc J}(r_+) &= \frac{V(r_+) r_+^2}{2 \Xi(r_+^2+a^2)} = \frac{r_+(1+r_+^2/l^2)}{2 \Xi}. \label{mcEetcatrplus}
\end{align}

First I will consider the particular cases of Kerr--AdS and Kerr--Newman--AdS, and then continue to show some properties that emerge for general $\mu(r)$.

\subsection{Kerr--AdS}

The values for $\mc E, \mc J$ and $\mc F$ are given by \eqref{ointCIchi4D} using $\oint \bs I_\chi$ for the respective values of $\chi$ ($\xi, -\eta,\beta$), and match the values of $\mc E, \mc J$ and $\mc F$ in GPP and CGKP. 

Here $\mu(r) = m$, so that $\mu'(r) = 0$ for all $r$. $\kappa$ takes the form
\begin{align}
    \kappa &= \f{3 r_+^4 + r_+^2(a^2+l^2)-a^2l^2}{2 r_+(r_+^2+a^2)l^2}.
\end{align}

It is easy to verify 
\begin{align}
    \de \mc E &= \f{\kappa \de A}{8\pi} + \Om \de \mc J.
\end{align}
This is best accomplished by rewriting $\mc E, \mc J, A, \kappa, \Om$ in terms of $r_+, l$ and $a$, replacing $m$ with $(1+r_+^2/l^2)(r_+^2+a^2)/2r_+$. This is the expected result, because the conserved charges $\bs I_\chi$ were constructed so that they satisfied the horizon variation law.

For the Smarr relation, it is useful to find
\begin{align}
    \oint_H \bs C_{\z;\g} &= -\oint_H \z \cdot \bs \tht^{EH} \nn 
    &= \f{m}{2\Xi} \nn 
    &= \f{\mc F}{2}.
\end{align}
This is found by direct evaluation. In fact this is the value of $-\oint_S \z \cdot \bs \tht^{EH}$ on any surface $S$, since $d (\z \cdot \bs \tht^{EH}) = 0$. We then have, from \eqref{SmarrF}, 
\begin{align}
    \mc F &= 2 \left( \f{\kappa A}{8\pi} + \Om \mc J\right) - 2 \Th' \tilde \La, \label{FSmarrKAdS}
\end{align}
where $\Th' \propto \mc V_{\xi,\mc B}$, which follows from $-\mc F + 2\oint_H \bs C_{\z;\gamma} = 0$. It is also easy to verify \eqref{FSmarrKAdS} directly, which gives the same result as found by CGKP \cite{Cvetic}. The fact that the Smarr relationship works nicely with $\mc F$ (based on $\bt$) follows from the vanishing of $-\mc F + 2\oint_H \bs C_{\z;\gamma} = -\mc F - 2\oint_H \z \cdot \bs \tht^{EH}$. 

We can also revisit and verify the Smarr relation in terms of $\mc E$ \eqref{Smarrwithextraterm}, \eqref{Smarr}, and write
\begin{align}
    \mc E &= 2 \left( \f{\kappa A}{8\pi} + \Om \mc J\right) - 2 \Th \tilde \La,
\end{align}
with \eqref{ThetaminusThetaPrime}
\begin{align}
    \Th &= \Th' - \f{2 \oint_H \bs C_{\z;\g}- \mc E}{2 \tilde \La} \nn 
    &= \Th' - \f16 a \mc J,
\end{align}
reproducing the results from CGKP \cite{Cvetic}.

\subsection{Kerr--Newman--AdS}

Here I show explicitly the values for Kerr--Newman--AdS, including the variation law and Smarr relation, which are very similar to those for the Kerr--AdS case. The metric is given by \eqref{KS4D} with $m \to \mu(r) = m - Q^2/2r$, as described in Section \ref{GKAdSSection}, and I will use the electromagnetic potential from \eqref{Aproptok}.

We use the conserved charges from Section \ref{KSFormEinsteinMaxwell}. $\bs I_\chi^{EM}$ is given by \eqref{EMI}. 

The electric charge $\mc Q$ is given by \eqref{mcQdef}, which gives
\begin{align}
    \mc Q = \f{Q}{\Xi}.
\end{align}
The electrostatic potential $\Phi^{EM}$ on the horizon is given by \eqref{PhiEM}, which in this case gives
\begin{align}
    \Phi^{EM} &= -\f{Q r_+}{r_+^2+a^2}.
\end{align}

The (outer) horizon location $r=r_+$ is given by the largest solution to \eqref{horizonequation4D} with $\mu(r) = m-Q^2/2r$, or
\begin{align}
    (r_+^2+a^2)(1+r_+^2/l^2)-2mr_+ + Q^2 = 0.
\end{align}

The surface gravity and area are given by \eqref{kappaArea4D} with $\mu(r) = m - Q^2/2r$, so that $\kappa$ takes the form
\begin{align}
    \kappa &= \f{3 r_+^4+(a^2+l^2)r_+^2-a^2l^2-l^2Q^2}{2 r_+(r_+^2+a^2)l^2}. \label{kappaKNewmanAdS}
\end{align}

We have
\begin{align}
    \mc E &= \oint_S \left( \bs I_\xi^{EH} + \bs I^{EM}_\xi\right) = \f{m}{\Xi^2} \nn 
    \mc J &= - \oint_S \left( \bs I_\eta^{EH} + \bs I^{EM}_\eta\right) = \f{m a}{\Xi^2} \nn 
    \mc F &= \oint_S \left( \bs I_\bt^{EH} + \bs I^{EM}_\bt\right) = \f{m}{\Xi}.
\end{align}
In other words, we recover the Kerr--AdS results. The effect of adding the $\bs I_\chi^{EM}$ term to the $\bs I_\chi^{EH}$ terms is to cancel, exactly, the $\mu'(r)$ terms which appear. 

By construction, we expect these conserved charges to satisfy the first-law relation. Indeed, we find
\begin{align}
    \de \mc E = \f{\kappa}{8\pi} \de A + \Om \de \mc J + \Phi^{EM} \de \mc Q.
\end{align}

(I note here that the approach in the next few sections is similar to Hajian and Sheikh-Jabbari \cite{HajianSheikh-Jabbari}, who define ``parametric variations'' in which field perturbations are generated by variations of
the solution parameters---that is, by quantities like $m,a$ and so on.) 

We have,
\begin{align}
    - \oint_H \z \cdot \bs \tht^{EH} &= \f{m}{2 \Xi} \nn 
    &= \f{\mc F}{2} \nn 
    -\oint_H \z \cdot \bs \tht^{EM} &= -\f{Q^2 r_+}{2(r_+^2+a^2)\Xi} \nn 
    &= -\f{\Phi^{EM} \mc Q}{2}.
\end{align}
(In fact, $-\oint_S \z \cdot \bs \tht^{EH}$ is independent of $S$, whereas $-\oint_S \z \cdot \bs \tht^{EM} = - Q^2 r/(\Xi(r^2+a^2))$.) Consequently,
\begin{align}
    \mc F - \Phi^{EM} \mc Q - 2 \oint_H \z \cdot (\bs \tht^{EH}+\bs \tht^{EM}) &= 0,
\end{align}
so that, revisiting \eqref{SmarrFEM}, we find
\begin{align}
    \mc F &= 2 \left( \f{\kappa A}{8\pi} + \om \mc J\right) + \Phi^{EM} \mc Q - 2 \Th' \tilde \La,
\end{align}
so that as with Kerr--AdS, the Smarr relation is satisfied with $\Th' \propto \mc V_{\xi,\mc B}$ provided that $(\mc F, \om)$ are used rather than $(\mc E, \Om)$. Using $(\mc E, \Om)$ instead we have (from \eqref{ThetaThetaprimeEM})

\begin{align}
    \mc E &= 2 \left( \f{\kappa A}{8\pi} + \Om \mc J\right) + \Phi^{EM} \mc Q - 2 \Th \tilde \La 
\end{align}
with
\begin{align}
    \Th &= \Th' + \f{\mc E - \mc F}{2 \tilde \La} \nn 
    &= \Th' + \f{1}{6} a \mc J,
\end{align}
the same $\Th$ as for Kerr--AdS. This follows from the fact that $\mc E$ and $-\oint_H \z \cdot \bs \tht^{EH}$ are unchanged from their Kerr--AdS values and that $-\Phi^{EM} \mc Q - 2 \oint_H \z \cdot \bs \tht^{EM} = 0$. 

These results are similar to the results found by CGKP \cite{Cvetic} for the charged, rotating black holes in gauged super-gravity, which similarly have that results are simpler for $\Th'$, evaluated in the asymptotically-rotating frame. Further, for the rotating pairwise-equal 4-charge black hole in $D = 4$ gauged super-gravity solution, which is quite complicated, CGKP show that $\Th - \Th' = \f16 a \mc J$. It would be interesting to see whether it is possible to extend my analysis to the gauged super-gravity metrics, but this is left for future work.

The thermodynamics of Kerr--Newman--AdS black holes are dealt with in \cite{CaldarelliCognola} and later works (such as \cite{Zhao}), which uses the same quantities for $\mc E$ (called $M$ there), $\Om, \mc J, \Phi^{EM}, \mc Q, T$ and $S$ as here, and recovers the first law as well as the Smarr relation. It is worth noting in passing that their method for calculating $M$ differs from mine, basing it on the Komar integral (with background subtraction) associated with $\beta / \Xi$. 

\section{Variations in Four Dimensions} \label{variationsinfourdimensions}

In this subsection I consider the variations in the various quantities calculated in Section \ref{explicitIxi}, and under what circumstances we recover the horizon variation law $\de \mc E = \Om \de J + \kappa \de A/8\pi$.

This breaks down into a number of checks, coming down to checking:
\begin{itemize}
    \item Under what circumstances is $\de \bs I_\chi = \bs k_{\bar \chi}^{EH}[\de g;g]$?
    \item Under what circumstances is $\oint_H \bs k_{\bar \chi}^{EH}[\de ;g] = \kappa \de A/8\pi$?
\end{itemize}
(In this section, $\bs I_\chi$ is always assumed to mean $\bs I^{EH}_\chi$.) The combination of the two questions of course checks whether or not we can use $\oint_H \de \bs I_\chi = \kappa \de A/8\pi$. The general conclusions of this section are: that we can generally use $\de \bs I_\chi = \bs k^{EH}_{\bar \chi}[\de g;g]$ provided that $\de \chi^a = 0$ and $\de \bar g_{ab} = 0$, and \emph{under specific circumstances} when $\de \bar g_{ab} \neq 0$ as well; and that we can use $\oint_H \bs k_{\bar \chi}^{EH}[\de ;g] = \kappa \de A/8\pi$ under a wide variety of circumstances, so that we can more or less trust this formula to hold under the situations we are considering. 

Along the way, I will put forward the argument that the reason why the variation law fails when it is constructed from $\bt$ rather than $\xi$ has to do with the fact that, in most frames, the components of $\bt^a$ vary when comparing two different values of $a$ (or $l$); and, further, that in an asymptotically rotating frame, $\bar g_{ab}$ changes too much under a variation of $\de a$ to be able to use $\de \bs I_\chi = \bs k^{EH}_{\bar \chi}[\de g;g]$. This gives an answer to the question of why it is preferable to use an asymptotically static vector $\xi$.

The variations are performed in four dimensions for computational and conceptual simplicity. I believe that the conclusions generalize to higher dimensions in a straightforward way. 

(Since the writing of my thesis, I found that Jing and Peng \cite{Jing:2017jxw} examine charges in the off-shell generalized Abbott--Deser--Tekin formalism, which has a similar form to the charges that I integrate with my form, and also examine the consequences of allowing either $m$ only, or $m$ and $a$, to vary, in both a frame which is asymptotically static and a frame which is non-rotating, and they find that charges are indeed integrable when associated with, in my notation, $\xi$ but not with $\bt$, when applied to variations in both $m$ and $a$. On the other hand, the charges for $\bt$ are also integrable when \emph{only} $m$ is varied, but are not integrable of both $m$ and $a$ are varied. These match up with my results here.)

\subsection{Variations Involving the Mass Function Only}  \label{massfunctionvars} 

To begin with, consider a variation in which the mass function $\mu(r)$ changes, but $a$ and $l$ remain constant. Let us call this variation $\hat \de$, so that $\hat \de a = \hat \de l = 0, \hat \de \mu(r) \neq 0$ in general. The meaning of this is that $\mu(r)$ and $(\mu + \hat \de \mu)(r)$ are both functions of $r$, so that at every point there is some nonzero $ \hat \de \mu(r)$, but $\hat \de \mu(r_1) \neq \hat \de \mu (r_2)$ for $r_1 \neq r_2$ in general. We also demand that the variation leave the coordinates themselves unaffected.

For variations involving $\mu(r)$ only, the background metric $\bar g_{ab}$ has a zero variation, since $\bar g_{ab}$ does not depend on $\mu(r)$. Similarly, the null vector $k^a$ (in both covariant and contravariant components) has $\hat \de k^a = \hat \de k_a = 0$. Finally, we have $\hat \de \xi^a = \hat \de \eta^a = \hat \de \bt^a = 0$, since none of these vectors depend on $\mu(r)$. 

As discussed in Section \ref{varyingk}, $\de \bs I_\chi = \bs k_\chi^{EH}[\de g;g]$ for any variation which leaves the background metric and the Killing vector $\chi$ unchanged. If $\de \chi^a = 0$, there is no distinction between $\chi^a$ and $\bar \chi^a$, so I will use them interchangeably in this subsection. Thus we expect $\hat \de \bs I_\chi = \bs k^{EH}_\chi[\hat \de g;g]$, which was the impetus for the definition of $\bs I_\chi$. That means that we expect
\begin{align}
    \hat \de \oint_H \bs I_{\bar \zeta} = \oint_H \bs k^{EH}_{\bar \zeta}[\hat \de g;g] = \f{1}{8\pi} \kappa \hat \de A.
\end{align}
I will verify this in this section for four-dimensional metrics, and along the way make some clarifications about how to perform the variation. 

Applying the variation to \eqref{tildeErtildeJr},
\begin{align}
    \hat \de \tilde{\mc E}{(r)} &= \frac{\hat \de \mu(r)}{\Xi^2} - \frac{(1+r^2/l^2)\left( (r^2+a^2) \arctan(a/r) - ar\right)}{2 a \Xi^2}\hat \de \mu'(r) \nn 
    \hat \de \tilde{\mc J}{(r)} &= \frac{a \hat \de \mu(r)}{a^2 \Xi^2} - \frac{(r^2+a^2)\left((r^2+a^2)\arctan(a/r) - ar\right)}{2 \Xi^2} \hat \de \mu'(r),
\end{align}
leaving the combination
\begin{align}
    \hat \de \tilde{\mc E}(r) - \Omega(r) \hat \de \tilde{\mc J}(r) &= \frac{r^2}{\Xi(r^2+a^2)} \hat \de \mu(r). \label{detildemceEr}
\end{align}
(Here, $\mu'(r) = d\mu/dr$ and so $\hat \de \mu'(r) = \hat \de ( d\mu/dr)$.)

We can be more explicit by letting the metric and the function $\mu$ be functions of $s$, where $s$ is a parameter in ``solution space,'' that is, the space of possible spaces corresponding to different values of $\mu(r)$. (These need not be, in this case, solutions to Einstein's equations, so ``solution space'' is a bit of a misnomer. The point is that this includes variation in the space of spacetimes with this particular form, where only $\mu(r)$ is free to vary.) Then we have, say, $\mu^{(s)}(r)$, where $\mu^{(0)}(r)$ is the value of $\mu$ in the unperturbed spacetime. We can then write $\hat \de$ as an operator $\hat \de s \frac{\pa }{\pa s}$ evaluated at $s = 0$:
\begin{align}
    \hat \de \mu (r) &\equiv \hat \de s \left.\frac{\pa \mu^{(s)}(r)}{\pa s}\right|_{s = 0, r \textrm{ const.}} \nn 
    \hat \de \mu'(r) &\equiv \hat \de s \left. \frac{\pa^2 \mu^{(s)}(r)}{\pa s \pa r}\right|_{s = 0, r\textrm{ const.}} = \hat \de s  \frac{\pa}{\pa s} \left[ \frac{\pa \mu^{(s)}(r)}{\pa r}\right]_{s = 0}.
\end{align}
(The $s$ derivative is taken in solution space and the $r$ derivative is evaluated within the spacetime of a given metric.) We then have that $\tilde {\mc E}(r) \equiv \tilde {\mc E}^{(s)}(r)$ and $\tilde{\mc J}(r) \equiv \tilde{\mc J}^{(s)}(r)$, though I will largely not include the $s$ superscript on $\tilde {\mc E}, \tilde {\mc J}$. 

The horizon quantity $r_+$ now depends on $s$, since varying $\mu(r)$ changes the value at which $V(r_+) = 2 \mu(r_+)$:
\begin{align}
    2 \mu^{(s)}(r_+^{(s)}) = V(r_+^{(s)}). \label{mus}
\end{align}
Because $a$ and $l$ are constant, $\hat \de V(r) = 0$, although $V(r_+^{(s)})$ depends on $s$ through the $r$-dependence. $r = r_+^{(0)}$ is the \emph{unperturbed} horizon location. \eqref{mus} can be expanded for small $s$:
\begin{align}
    2(\mu^{(0)} + \hat \de \mu)(r_+^{(0)} + \hat \de r_+) &= V(r_+^{(0)} + \hat \de r_+) \nn
    2\left(\mu^{(0)}(r_+^{(0)}) + (\mu^{(0)})'(r_+^{(0)}) \hat \de r_+ + \hat \de \mu (r_+^{(0)}) \right) &= V(r_+^{(0)}) + V'(r_+^{(0)}) \hat \de r_+.
\end{align}
Using $2\mu^{(0)}(r_+^{(0)}) = V(r_+^{(0)})$ and rearranging,
\begin{align}
    \hat \de \mu ( r_+^{(0)}) &=\f12 \left(  V'(r_+^{(0)}) - 2(\mu^{(0)})' (r_+^{(0)})\right) \hat \de r_+ \nn 
    &= \frac{V(r_+^{(0)})}{1+r_+^{(0)}l^{-2}} \kappa^{(0)} \hat \de r_+ \nn 
    &= \frac{(r_+^{(0)})^2 + a^2}{r_+^{(0)}} \kappa^{(0)} \hat \de r_+, \label{variationinmuatrplus}
\end{align}
where $\kappa^{(0)}$ is the unperturbed surface gravity.

Let us now evaluate \eqref{detildemceEr} at $r = r_+^{(0)}$, with $\Omega = \Omega^{(0)} \equiv \Omega(r_+^{(0)})$. We note that these values of $r$ and $\Omega(r)$ are \emph{constant} and unchanging with the perturbation. I will include the superscript $(s)$ or $(0)$ to make clear which quantities are unperturbed and which are allowed to vary. We have
\begin{align}
    \hat \de \tilde{\mc E}(r_+^{(0)}) - \Omega^{(0)} \hat \de \tilde{\mc J}(r_+^{(0)}) &= \frac{(r_+^{(0)})^2}{\Xi ( (r_+^{(0)})^2+a^2)} \hat \de \mu(r_+^{(0)})\nn
    &= \kappa^{(0)} \frac{r_+^{(0)} \hat \de r_+}{\Xi} \nn 
    &= \f{1}{8\pi} \kappa^{(0)} \hat \de A, \label{kappa0dA}
\end{align}
noting that the only parameter in $A$ which varies under $\hat \de$ is $r_+$. 

It is important to note that when performing variations we must keep the functions $\tilde {\mc E}$ etc.~evaluated at a constant value of $r$. If we tried to proceed directly and vary the expressions \eqref{mcEetcatrplus}, we would not recover \eqref{kappa0dA} in general. Indeed,
\begin{align}
    \tilde{\mc E}(r_+^{(s)}) - \Om(r_+^{(s)}) \tilde{\mc J}(r_+^{(s)}) &= \f{r_+^{(s)} (1+ r_+^{(s)}/l^2)}{2 \Xi} \nn 
    \hat \de \left[ \tilde{\mc E}(r_+^{(s)}) - \Omega(r_+^{(s)}) \tilde{\mc J}(r_+^{(s)}) \right] &= \frac{1 + 3 (r_+^{(0)})^2/l^2}{2 \Xi} \hat \de r_+ \nn 
    &= \frac{1+3 r_+^2/l^2}{16\pi l^2 r_+} \hat \de A, \label{kappa0dAplusextra1}
\end{align}
which is not equal to $\kappa^{(0)} \hat \de A/8\pi$ in general. (It is equal to $\kappa^{(0)}$ when $a = 0$ and $\mu(r) = m$ is constant.) (We can send $r_+^{(0)} \to r_+$ on the last line because we are only working to first order in the variation and we are already multiplying by $\hat \de A$, so that the distinction between $r_+^{(0)} \hat \de A$ and $r_+^{(s)} \hat \de A$ is not meaningful. Along similar lines, we could have instead just written $\kappa$ instead of $\kappa^{(0)}$.) The slightly modified expression where $\Omega(r_+^{(0)})$ is fixed at the unvaried value but $\tilde{\mc E}$ and $\tilde{\mc J}$ are evaluated at $r_+^{(s)}$ gives
\begin{align}
    \hat \de \left[ \tilde{\mc E}(r_+^{(s)}) - \Omega(r_+^{(0)}) \tilde {\mc J}(r_+^{(s)})\right] &= \hat \de \left[ \tilde{\mc E}(r_+^{(s)})\right] - \Omega(r_+^{(0)}) \left[\hat \de \tilde {\mc J}(r_+^{(s)})\right] \nn 
    &= \hat \de \left[ \tilde {\mc E}(r_+^{(s)}) - \Omega(r_+^{(s)}) \tilde {\mc J}(r_+^{(s)})\right] + \tilde{\mc J}(r_+^{(0)}) \hat \de \left[\Omega(r_+^{(s)})\right],
\end{align}
where the last line is taken to first order. This gives
\begin{align}
    \hat \de \left[ \tilde{\mc E}(r_+^{(s)}) - \Omega(r_+^{(0)}) \tilde {\mc J}(r_+^{(s)})\right] &= \left(\f{1+3 r_+^2/l^2}{2\Xi}- \frac{2\Omega(r_+^{(0)}) \Xi}{V(r_+^{(0)})}\tilde{\mc J}(r_+^{(0)})\right) \hat \de r_+ \nn 
    &= \frac{\kappa}{8\pi} \hat \de A + \frac{r_+ \arctan(a/r_+)\mu'(r_+)}{a} \hat \de r_+, \label{kappa0dAplusextra2}
\end{align}
which matches with $\kappa \hat \de A/8\pi$ only if $\mu'(r_+) = 0$.

The idea then is that if we associate to the horizon the values $\tilde{\mc E}(r_+^{(s)})$ and $\tilde{\mc J}(r_+^{(s)})$, we cannot recover the first law (unless $\mu' = 0$). However, we can recover the first law provided that we evaluate both $\tilde{\mc E}$ and $\tilde{\mc J}$ at the location of the \emph{unperturbed} horizon. 

The fact that $\mu'(r) = 0$ simplifies calculations, so that we can evaluate $\tilde{\mc E}$ and $\tilde{\mc J}$ at either the original or perturbed values of the horizon, has a simple explanation. It is a consequence of the fact that, if $\mu'(r_+) = 0$ (or at least $(\mu^{(0)})'(r_+^{(0)}) = 0$), then $(d/dr)\tilde {\mc E}(r)$ and $(d/dr) \tilde {\mc J}(r)$ are zero when evaluated at $r_+^{(0)}$ (in the unperturbed spacetime), and so the distinction between varying $\tilde {\mc E}(r_+^{(0)})$ and varying $\tilde {\mc E}(r_+^{(s)})$ vanishes (and similarly for $\tilde {\mc J}$). The situation in which $\mu'(r) = 0$ everywhere is Kerr--AdS. 

The result \eqref{kappa0dA} is reassuring and suggests that $\tilde {\mc E}$ and $\tilde {\mc J}$ (with care) satisfy a first law with the area, for specific kinds of variations which only affect the function $\mu(r)$ which enter into the Kerr--Schild function $H$. Of course $\bs I_\chi$ was \emph{constructed} so that its variation would produce $\bs k_\chi^{EH}$ for variations which \emph{only} vary the Kerr--Schild correction. To what extent the quantities $\tilde {\mc E}$ and $\tilde {\mc J}$ are useful when, say, $a$ and $l$ are allowed to vary as well is the next thing to check. I will first consider a variation in $a$ only.

As one last stop before continuing, I will check what happens if we use $\beta$ and $\eta$, instead of $\xi = \partial/\partial t$ and $\eta$, as the vectors for defining the quantities. I will let $\omega(r) = a \Xi/(r^2+a^2)$. Recalling that $\tilde{\mc F}(r) = \oint_C \bs I_\beta$, it turns out 
\begin{align}
    \hat \de \tilde{\mc F}(r_+^{(0)}) - \omega(r_+^{(0)}) \hat \de \tilde {\mc J}(r_+^{(0)}) = \f{1}{8\pi} \kappa^{(0)} \hat \de A,
\end{align}
and, similarly, when the variation is performed on $\de \tilde{\mc F}(r_+^{(s)})$ (with either $\omega(r_+^{(0)})$ being varied or kept constant) there are extra terms analogous to, though not equal to, those in \eqref{kappa0dAplusextra1} and \eqref{kappa0dAplusextra2}. I will note that because $a$ does not vary, $\hat \de \tilde{\mc F}(r) =\hat \de \tilde{\mc E}(r) - al^{-2} \hat \de \tilde{\mc J}(r)$. There is thus no preference for $\xi$ over $\bt$ when considering this particular variation. Note that $\hat \de \xi^a = \hat \de \bt^a = \hat \de \eta^a = 0$.

\subsection{Variations Involving Both Mass Function and rotation parameter} \label{massandspecificangularmomentumvary} 

I will now consider a variation $\de$ with $\de l = 0$, but where $a$ is allowed to vary and $\mu$ is also allowed to vary. ($\hat \de$ is a special case of $\de$ with $\de a = 0$.) The general form will still be that of \eqref{KS4D}, with $m \to \mu(r)$. The variations in $\mu$ are similar to those for $\hat \de$. Assume that $\mu(r)$ has no dependence on $a$.

We now have,
\begin{align}
    \de \tilde{\mc E}(r) =& \frac{\de \mu(r)}{\Xi^2} - \frac{(1+r^2/l^2)\left((r^2+a^2) \arctan(a/r) - ar\right)}{2 a \Xi^2} \de \mu'(r) \nn 
    & +\left[\f{4 a \mu(r)}{\Xi^3 l^2}-\f{(1+r^2/l^2)\left((3 a^4 + (l^2+5 r^2)a^2-l^2r^2)\arctan(a/r)-a(5a^2-l^2)r\right)}{2 l^2 \Xi^3 a^2} \mu'(r)\right]\de a \nn 
    \de \tilde{\mc J}(r) =& \f{a \de \mu(r)}{\Xi^2}-\f{(r^2+a^2)\arctan(a/r)-ar}{2 \Xi^2} \de \mu'(r)  \nn 
    &\qquad +\left[\frac{1+3a^2/l^2}{\Xi^3} \mu(r) - \frac{a \left( (a^2+l^2+2r^2)\arctan(a/r)-2ar\right)}{l^2 \Xi^3} \mu'(r)\right] \de a,
\end{align}
with combination
\begin{align}
    \de \tilde{\mc E}(r) &- \Omega(r) \de \tilde{\mc J}(r) = \frac{r^2}{\Xi (r^2+a^2)} \de \mu(r) \nn
    & - \frac{2a^3(1-3r^2/l^2)\mu(r) + (r^2-a^2)(1+r^2/l^2)\left( (r^2+a^2)\arctan(a/r)-ar\right)\mu'(r)}{2 a^2 (r^2+a^2) \Xi^2} \de a.
\end{align}

As with the $\hat \de$ variation, let $r = r_+^{(0)}$ be the location of the horizon in the unvaried coordinates and let $r = r_+^{(0)} + \de r_+$ be the value of $r$ where the horizon is located in the new coordinates. 

Recalling that the horizon is defined by \eqref{mus}, because $V$ depends on $a$ as well as $r$ we must also take into account the variation in the \emph{function} $V$. Let $a^{(0)}$ be the unvaried value of $a$ and $a^{(0)} + \de a$ its perturbed counterpart. We can, say, let $V^{(0)}(r) = (r^2+(a^{(0)})^2)(1+r^2/l^2)/2r$, and then the perturbed $V$, $(V^{(0)}+\de V)(r)$, is given by
\begin{align}
    (V^{(0)} + \de V)(r) &= \frac{(r^2+(a^{(0)}+\de a)^2)(1+r^2/l^2)}{2 r} \nn 
    &= V^{(0)} (r) + \frac{(1+r^2/l^2)a^{(0)}}{r} \de a.
\end{align}
$(1+r^2/l^2)a/r$ is just $\pa V/\pa a$ at constant $r$ and $l$, treating $a$ as a variable; we can write
\begin{align}
    \de V(r) &= \left. \frac{\pa V(r)}{\pa a}\right|_{r,l} \de a.
\end{align}
The horizon will be located at, letting prime denote differentiation with respect to $r$ only and neglecting terms quadratic in the variation, 
\begin{align}
    2(\mu^{(0)} + \de \mu)( r_+^{(0)}+\de r_+) &= (V^{(0)}+\de V)(r_+^{(0)} + \de r_+) \nn 
    2\left(\mu^{(0)}(r_+^{(0)}) + (\mu^{(0)})'(r_+^{(0)}) \de r_+ + \de \mu (r_+^{(0)}) \right) &= V^{(0)}(r_+^{(0)}) + (V^{(0)})'(r_+^{(0)}) \de r_+ + \de V(r_+^{(0)}) \nn 
    \de \mu(r_+^{(0)}) &= \frac{1}{2} \left( (V^{(0)})'(r_+^{(0)}) - 2 (\mu^{(0)})'(r_+^{(0)}) + \de V(r_+^{(0)})\right) \nn 
    &= \frac{ (r_+^{(0)})^2 + (a^{(0)})^2}{r_+^{(0)}} \kappa^{(0)} \de r_+ + \f12 \left.\frac{\pa V(r_+^{(0)})}{\pa a}\right|_{r_+^{(0)},l} \de a
\end{align}
The ${}^{(0)}$ superscripts are becoming cumbersome and so at this point we will drop them, having already performed the variations we need to, but it is understood that any quantities which appear in the expression unvaried are evaluated in the unvaried spacetime. I will continue to include the superscripts in $\de \tilde {\mc E}(r_+^{(0)})$ and $\de \tilde {\mc J}(r_+^{(0)})$ to make clear that I am considering these variations evaluated at constant $r_+$ and not the difference between the quantities evaluated at the unvaried and varied horizon.  We then have
\begin{align}
    \de \tilde{\mc E}(r_+^{(0)}) &- \Omega^{(0)} \de \tilde{\mc J}(r_+^{(0)}) = \kappa \frac{r_+ \de r_+}{\Xi} + \frac{r_+^2}{2\Xi(r_+^2+a^2)} \frac{\pa V(r_+)}{\pa a} \de a- \nn
    &\qquad \frac{a^3(1-3r_+^2/l^2)V(r_+) + (r_+^2-a^2)(1+r_+^2/l^2)\left( (r_+^2+a^2)\arctan(a/r_+)-ar_+\right)\mu'(r_+)}{2 a^2 (r_+^2+a^2) \Xi^2} \de a \nn
    &= \frac{\kappa}{8\pi} \de A + \frac{\left((r_+^2-a^2)\arctan(a/r_+)+r_+a\right) \mu'(r_+)}{16\pi a^3} \left.\frac{\pa A}{\pa a}\right|_{r_+,l} \de a
\end{align}

The extent to which $\de \tilde{\mc E}(r_+^{(0)}) - \Omega^{(0)} \de \tilde{\mc J}(r_+^{(0)})$ fails to be equal to $\kappa \de A/8\pi$ depends explicitly on the variation in the $a$ parameter. It bears repeating that it is not surprising that this occurs, because in general we do \emph{not} expect $\de \bs I_\chi$ and $\bs k_\chi^{EH}[\de g;g]$ to be equal for variations in which the background metric is also allowed to vary. Variations $\hat \de$ which only change the function $\mu$ will only affect $H$ and so will have $\hat \de \bs I_\chi = \bs k_{\chi}^{EH}[\hat \de g;g]$, but variations which also affect $a$ are a different story.

If $\mu'(r_+) = 0$ then we do recover $\de \tilde{\mc E}(r_+^{(0)}) - \Omega^{(0)} \de \tilde{\mc J}(r_+^{(0)}) = \kappa \de A/8\pi$. This will of course be the case in Kerr--AdS, with constant $\mu(r) = m$, and in which $\tilde{\mc E}(r_+^{(0)}) = \mc E$ and $\tilde {\mc J}(r_+^{(0)}) = \mc J$ (the Kerr--AdS energy and angular momentum), the values of which selected by \cite{GibbonsPerry} to ensure that they satisfy a first-law relation. 

We now turn to the case where we use $\beta$ and $\eta$ instead of $\xi$ and $\eta$. We note that as a consequence of the linearity of $\bs I_\xi$ with $\xi$ (and similarly for $\bs I_\eta, \bs I_\beta)$,
\begin{align}
    \tilde{\mc F}(r) &= \tilde{\mc E}(r) - \frac{a}{l^2} \tilde{\mc J}(r) \nn 
    \de \tilde{\mc F}(r) &= \de \tilde{\mc E}(r) - \frac{a}{l^2} \de \tilde{\mc J}(r) - \frac{1}{l^2} \tilde{\mc J}(r) \de a.
\end{align}
Consequently,
\begin{align}
    \tilde{\mc F}(r) - \omega(r) \tilde{\mc J}(r) &= \tilde{\mc E}(r) - \Omega(r) \tilde{\mc J}(r) \nn 
    \de \tilde{\mc F}(r) - \omega(r) \de \tilde{\mc J}(r) &= \de \tilde{\mc E}(r) - \Omega(r) \de \tilde{\mc J}(r) - \f{1}{l^2} \tilde{\mc J}(r) \de a.
\end{align}
We then have
\begin{align}
    \de \tilde{\mc F}(r_+^{(0)}) &- \omega(r_+^{(0)}) \de \tilde{\mc J}(r) = \frac{\kappa}{8\pi} \de A \nn 
    &\qquad + \left[\frac{\left((r_+^2-a^2)\arctan(a/r_+)+r_+a\right) \mu'(r_+)}{16\pi a^3} \left.\frac{\pa A}{\pa a}\right|_{r_+,l} - \frac{V(r_+)a}{2 \Xi^2 l^2} + \f{\Psi \mu'(r_+)}{l^2}\right]\de a.
\end{align}
In general this is not equal to $\kappa \de A / 8 \pi$. If $\mu(r) = m$ (constant) then this reduces to
\begin{align}
    \de \mc F - \omega \de {\mc J} &= \frac{\kappa}{8\pi} \de A + \frac{V(r_+)a}{2\Xi^2l^2} \de a = \frac{\kappa}{8 \pi} \de A - \frac{\mc J}{l^2} \de a, \label{deFvariationlaw}
\end{align}
where $\mc F = \tilde{\mc F}(r)$ is constant (and similar for $\mc J$). We see that the first law does not hold in this case, as was observed by \cite{GibbonsPerry}. 

There are two effects that we see when $\de a$ changes, both of which contribute to the failure of the variation law. The first is that the background metric varies if $\de a \neq 0$. The second is that the components of $\beta^a = \de^a_t + a l^{-2} \de^a_\phi$ also vary under the variation in $a$, so that $\de \beta^a \neq 0$, though $\de \xi^a = 0$. I aim to show that it is the result of the background metric variation that the first law fails for both $\tilde {\mc E}(r)$ and $\tilde {\mc F}(r)$ in general, but $\mc E = \tilde{\mc E}(r)$ when $\mu'(r)=0$. It is because $\de \beta^a \neq 0$ that the variation law fails for $\mc F = \tilde {\mc F}(r)$ even if $\mu'(r) = 0$. 

The argument for this is subtle and will be teased out in the rest of this section. I will make a few preliminary comments. The first is that while the background metric is changing, the background metric is still the spacetime associated with AdS with constant $l$---it is just that the coordinate system changes, since it depends explicitly on $a$. One way to look at it is that while the $t$ and $\phi$ coordinates retain their meanings, the interpretations of the coordinates $r$ and $\tht$ depend on $a$, when comparing to an $a$-independent background such as the spherical polar coordinate system or pseudo-Cartesian coordinates. The meaning of a ``constant-$r$ surface'' thus, in some senses, changes, as the constant-$r$ surface moves in the spherical polar or pseudo-Cartesian background. One way to account for this is to perform calculations entirely in a coordinate frame where the background metric does not change. This will be the subject of the next subsection: what happens if $a$ varies, but we find a way to keep $\bar g_{ab}$ unchanging? The advantage of this is that it will allow us to disentangle the effect of the changing components of $\beta^a$ from the effect of the changing components of $\bar g_{ab}$. 

\subsection{Constant Background Metric with Variable Rotation Parameter} \label{constantmetricbackgroundvariablea}  

Here I consider the case where the background metric does not change, even as $a$ changes. There are two basic approaches, which I will consider in turn. The first is to use a set of coordinates for the AdS background which do not depend on $a$. The second is to use a set of coordinates for the AdS background which depend only on the \emph{unperturbed} value of $a$. The latter is a little more conceptually involved but turns out to be easier to work with computationally. I will begin with the former case.

Two coordinate systems for AdS I have already introduced which do not depend on $a$ are the spherical polar coordinates \eqref{dbarsspherical} and the pseudo-Cartesian coordinates from Section \ref{pseudoCartesian}. It is possible to write $r$ and $\tht$---or, more conveniently, $u = \cos \tht$---in terms of the spherical polar or pseudo-Cartesian coordinates in closed form, but that closed form involves nested radicals, which become computationally intensive. The workaround here is either to work at large values of $r$ where the expressions for $r$ and $u$ simplify considerably, or, as I will do later on, to use a form of the spheroidal coordinates with the unvaried $a$ parameter $a^{(0)}$. 

To explain my approach more clearly, I will start with some physically uninteresting examples motivated by their computational simplicity. Consider a metric with (four-dimensional) Minkowski background in Cartesian coordinates,
\begin{align}
    \bar g_{a b} = -dt^2 + dx^2 + dy^2 + dz^2,
\end{align}
with affinely parametrized null vector $(k^{(s)})^a$ given by
\begin{align}
    (k^{(s)})^a \f{\pa}{\pa x^a} &= \f{\pa}{\pa t} + \cos(\Phi^{(s)}) \f{\pa}{\pa x} + \sin(\Phi^{(s)}) \f{\pa}{\pa y}.
\end{align}
Here, $\Phi$ is an arbitrary function which is constant in spacetime and depends on $s$, which is a parameter which varies through solution space. For any value of $\Phi$, $(k^{(s)})^a$ is a null geodesic. Then we can write a Kerr--Schild metric as
\begin{align}
    g^{(s)}_{a b} = \bar g_{a b} + h^{(s)}_{a b}
\end{align}
where
\begin{align}
    h^{(s)}_{a b} &= H^{(s)}(t,x,y,z) k^{(s)}_a k^{(s)}_b.
\end{align}

Substituting in a \emph{general} (not necessarily Killing) vector 
\begin{align}
    \chi^a \f{\pa}{\pa x^a} &= \chi^t(t,x,y,z) + \chi^x (t,x,y,z) + \chi^y(t,x,y,z) + \chi^z (t,x,y,z),
\end{align}
satisfying $\de \chi^a = 0$, and considering a variation of the form $\de \equiv d s \frac{d}{d s}$ (where the variation does not affect the coordinates or $\bar g_{a b}$), I used \emph{GRTensorIII} to calculate $\bs I_\chi$ (using the $\bar \na_a$ form), $\de \bs I_\chi$, and $\bs k^{EH}_\chi[\de g;g]$, to confirm that indeed $\de \bs I_\chi = \bs k^{EH}_\chi[\de g;g]$. 

Moving to a more physically interesting case, consider Kerr in Kerr--Schild form, in Cartesian coordinates. Its metric is, following Poisson \cite{Poisson}, 
\begin{align}
    g_{a b} &= \eta_{a b} + \f{2 m r}{r^4+a^2z^2} k_a k_b,
\end{align}
where $\eta_{a b} = \textrm{diag}(-1,1,1,1)$, and
\begin{align}
    k_a dx^a &= - dt - \f{r x + a y}{r^2+a^2} dx - \f{ry-ax}{r^2+a^2}dy - \f{z}{r} d z. \label{Kerrka}
\end{align}
(Note that the sign of $k_a dx^a$ in the convention of \cite{Poisson} is the opposite of that of \eqref{MyersPerry}, but since $k_a dx^a$ appears quadratically this has no effect.) $r$ satisfies 
\begin{align}
    \frac{x^2 + y^2}{r^2+a^2} + \f{z^2}{r^2} = 1. \label{Kerrr}
\end{align}
The explicit solution to \eqref{Kerrr} is
\begin{align}
    r^2 &= \f{1}{2} \left( 2(x^2+y^2+z^2-a^2) + 2 \sqrt{ (x^2+y^2+z^2-a^2)^2+4 a^2 z^2}\right).
\end{align}
Sending $a \to a^{(s)}$ and allowing $H \to H^{(s)}(t,x,y,z)$, I verified that $\de \bs I_\chi = \bs k_\chi^{EH}[\de g;g]$ (for $g_{a b} = \bar g_{a b} + h_{a b}$ and with $\de \bar g_{a b} = 0$) for arbitrary $\chi$ using \emph{GRTensorIII}. 

For a KS perturbation to an AdS background, it is tempting to use the pseudo-Cartesian coordinates, but to simplify the form of the background metric I will use the spherical polar form of the metric $(t,R,U,\phi)$ of the form
\begin{align}
    d\bar s^2 &= \bar g_{a b} dx^a dx^b - (1 + R^2/l^2) dt^2 + \f{dR^2}{1+R^2/l^2} + R^2 \left( \f{dU^2}{1-U^2} + (1-U^2) d \phi^2\right). \label{AdS4Dspherical}
\end{align}
(I am using $R$ and $U$ for the spherical polar expressions and are not to be confused with other meanings of $R$ or $U$. $R$ is the aerial radius and $U = \cos \Th$ where $\Th$ is the equatorial angle: $d U^2/(1-U^2) + (1-U^2) d \phi^2$ is the metric of the 2-sphere.) This metric is equivalent to \eqref{dbarsspherical} with $\hat \mu_1 = \sqrt{1-U^2}, \hat \mu_2 = U, y = R$. These are related to the spheroidal coordinates with background metric given by $\bar g_{a b}$ of \eqref{KS4D} by 
\begin{align}
    R U &= r u \nn 
    1 &= R^2 \left( \frac{ (1-U^2)\Xi}{r^2+a^2} + \frac{U^2}{r^2}\right). \label{sphericalspheroidal}
\end{align}
The explicit solution for $r$ in terms of $(R,U)$ is
\begin{align}
    r^2 &= \f{P + \sqrt{P^2 + 4 R^2 U^2 a^2}}{2} \nn 
    P &= R^2 \Xi - a^2 + R^2 U^2 (1 - \Xi).
\end{align}
We also need the form of $k_a dx^a$. Use \eqref{KS4D}. We can replace the $\cos \tht$ appearing in $k_t$ and $k_\phi$ with $\cos \tht = u = R U/r$, where $r$ is a function of $R$ and $U$ (and similarly for $\sin^2\tht = 1-u^2$). We also have $k_r dr = k_R dR + k_U dU$; by expanding out $dr$, we conclude
\begin{align}
    k_R dR + k_U dU &= k_r \left( \f{\pa r}{\pa R} d R + \f{\pa r}{\pa U} d U\right).
\end{align}
At this point I applied $H = H(R,U)$ and wrote $h_{ab} = H k_a k_b$, as usual. I allowed $H$ to have a particularly general form because writing $H = 2\mu(r)r/(r^2+a^2u^2)$ ends up being more computationally intensive because of the complicated way $r$ and $u$ depend on $R$ and $U$; of course $2\mu(r)r/(r^2+a^2u^2)$ is a specific case of a general function $H(R,U)$. Using \emph{GRTensorIII}, I then calculated $\bs I_\xi$ and $\bs I_{\eta}$ using the form \eqref{Ixi} using the background covariant derivatives $\bar \na_a$. 

To simplify calculations, I set $l = 1$, which amounts to rescaling $R, t, r$ and $a$. I then set $a = a^{(s)}$, $H(R,U) = H^{(s)}(R,U)$, and found expressions for $\de \bs I_\chi = \f{\pa }{\pa s} \bs I_\chi \de s$ (keeping coordinates fixed when calculating $\pa/\pa s$), for both $\chi = \xi$ and $\chi = \eta$. 

For $\bs k_\chi^{EH}[\de g;g]$, I calculate $\bs K^K_\chi$ as, from the decomposition in \eqref{komardifferenceformula}, 
\begin{align}
    (\bs K^K_\chi)_{ab} &= \f{1}{16\pi} \bs \ep_{ab cd} \left(\overline{ \na^c \chi^d} - h^{e[a}\bar \na_e \chi^{b]} + \chi^e \bar \na^{[a} h^{b]}_e \right).
\end{align}
I calculated this for $\chi = \xi,\eta$, and then varied it, using the same method as for the calculation of $\de \bs I_\chi$. I calculated $\bs \Th^{EH}$ using $\de g_{ab} = \de h_{ab} = \f{\pa }{\pa s} h_{ab} \de s$, as well as
\begin{align}
    v_a &= \na^b \de g_{ab} - g^{bc} \na_d \de g_{bc} \nn 
    &= \na^b \de h_{ab} - \na_d (g^{bc} \de g_{bc}).
\end{align}
Of course $g^{bc} \de g_{bc} = \de \ln g = \de \ln \bar g = 0$. I also checked explicitly that this is the case, checking $ g^{ab} \de h_{ab} = 0$. The remaining term is calculated as
\begin{align}
    v_a &= \bar \na^b \de h_{ab} - g^{bc} (\Delta^d_{ca} \de h_{ab} + \Delta^d_{cb} \de h_{ad}),
\end{align}
where $\Delta^a_{bc} = \G^a_{bc} - \bar \G^a_{bc}$ is given by \eqref{Deltaabc}.

I then wrote $\bs k_\chi^{EH} = -\de \bs K^K_\chi + \chi \cdot \bs \Th^{EH}$, and checked
\begin{align}
    \de \bs I_\chi - \bs k_\chi^{EH}[\de g;g].
\end{align}
As expected, for both $\chi = \xi$ and $\chi = \eta$, the result is precisely zero. 

We also expect, letting $\bt = \xi + a^{(s)} \eta$ with $\bar \bt = \xi + a^{(0)} \eta$,
\begin{align}
    \de (\bs I_\bt) &= \de \bs I_{\bar \bt} + \bs I_{\de \bt},
\end{align}
where $\de \bt = \de a \eta$, with $\de a = \f{d a^{(s)}}{ds} \de s$. $\bar \bt$ has constant (contravariant components). We thus expect
\begin{align}
    \de (\bs I_\bt) &= \bs k_{\bar \bt}^{EH}[\de g;g] + \de a \bs I_\eta,
\end{align}
where $\bs k_{\bar \bt}^{EH}[\de g;g] = \bs k_\xi^{EH}[\de g;g] + a^{(0)} \bs k_\eta^{EH}[\de g;g].$ I also confirmed this relationship, which demonstrates that the failure of $\de \bs I_\bt $ to be equal to $\bs k_{\bar \bt}^{EH}[\de g;g]$, in this case where the background is kept fixed, is directly attributable to the fact that $\bt$ has changing contravariant components. 

We can thus interpret $\de \bs I_\chi = \bs k_\chi^{EH}[\de g;g]$ for any variations which keep $\de \bar g_{a b} = 0$ and have confirmed it in a few specific cases. The problem with this line of argument is that it is less convenient to keep the background metric $\bar g_{a b}$ fixed than it is to use the spheroidal coordinates which in general depend on the rotational parameter $a$. This does however point at a significant advantage of using the background metric which is asymptotically static, in the form $t,\phi$, over the form of the background in which the metric is asymptotically non-static. While the definitions of the spheroidal $r$ and $u$ depend implicitly on $a$ in either case, the definitions of $t$ and $\phi$ are fixed with respect to the fundamental symmetries of the spacetime. 

I will note before continuing that this ``workaround'' to show $\de \bs I_\chi = \bs k_\chi^{EH}[\de g;g]$ was doable if $a$ is varied specifically because the spacetime is still AdS, and so it is possible to find coordinates where the background does not depend on $a$. It is simply not true if $l$ is allowed to vary, because $l$ appears not only in the background metric but also in the curvature invariants associated with that background (most obviously in $R = 4\La = -12/l^2$) and so no choice of coordinates, no matter how judicious, can remove the change in the background metric. I want to emphasize that this section is meant as a check---I already showed in Section \ref{varyingk} that $\de \bs I_\chi = \bs k_\chi^{EH}[\de g;g]$ if $\de \chi^a = \de \bar g_{ab} = 0$ for Kerr--Schild form. What is useful about this exercise is to show how the result $\de \bs I_\chi \neq \bs k_\chi^{EH}[\de g;g]$ if $a$ varies and we compare different Kerr--AdS solutions can be avoided if we do use a constant background. The cost is that it is more computationally intensive to write the spheroidal $r$ and $u = \cos \tht$ in terms of the ``fixed'' spherical polar $R, U$ coordinates.

The other option, which I will discuss in Section \ref{infinitesimalcoordinatetransformations}, is to use the background metric associated with unvaried parameter $a^{(0)}$ and find a way to transform the final background metric back to the metric associated with $a^{(0)}$.

We now turn to verify the horizon area rule. 

\subsection{Checking Horizon Area Variation Rule} \label{checkinghorizonareavariationrule}

The previous section considered under what circumstances it is possible to argue $\de \bs I_\chi = \bs k_{\bar \chi}^{EH}[\de g;g]$. This section considers under what circumstances \eqref{kdarule}, $\oint_H \bs k_{\bar \chi}^{EH}[\de g;g] = \kappa \de A/8\pi$, holds, as well as how to confirm it. I consider the ABL, KS and BL forms of the metric for Kerr--Newman--AdS where the parameters $m, a, l, Q$ are all allowed to vary. The choice of Kerr--Newman--AdS is to consider a situation somewhat more general than Kerr--AdS, but collapsing the solution space down to something smaller than for the general $\mu(r)$. 

To begin with, we will take the four-dimensional Kerr--Newman--AdS metric in the ABL form, given by \eqref{ABL4D}, with $\Delta = \Delta(r) = (r^2+a^2)(1+r^2l^{-2}) - 2mr + Q^2$. I will also use $u = \cos \tht$ for one of the coordinates, so that the coordinates are $(\tau,r,u,\hat \vp)$. As a reminder, the key features of the ABL form of the metric is that the metric is asymptotically static (in this case, $\lim_{r\to\infty} g_{\tau \hat \varphi} = 0$), that the $r$ cross-terms in the metric are zero (here $g_{\tau r} = g_{\hat \varphi r} = 0$) and that the metric is irregular on the horizon (with $\lim_{r\to r_+} g_{rr} \to \pm \infty$). The outer horizon $r = r_+$ is the largest root of $\Delta(r) = 0$. The ABL coordinates are in fact not regular on the horizon, which appears to be a problem, but we will proceed in our calculations anyway, and find that it does not affect our calculations. The Killing vector tangent to the null generators of the horizon is $\zeta$ as in \eqref{zeta4D}, with $\xi = \pa_\tau, \eta = \pa_{\hat \vp}$. 

Now I will send $m\to m^{(s)}, a \to a^{(s)}, Q \to Q^{(s)}, l \to l^{(s)}$, where $s$ is a parameter used to parametrize different Kerr--Newman--AdS solutions ``close to each other'' in solution space. As a reminder, $s$ is constant throughout the whole spacetime in any given solution. We write $\delta \Phi \equiv (\pa \Phi^{(s)}/\pa s)\de s$ for quantities $\Phi$ in the variation formulas. (The $\pa/\pa s$ derivative keeps the coordinates constant.) 

Quantities like the metric will depend on $s$ only through the dependence on $m^{(s)}, Q^{(s)}, l^{(s)}$ and $a^{(s)}$, and can be written $g_{ab}^{(s)}$. Let the unvaried spacetime be given by $g^{(0)}_{ab}$. In general this will be a Kerr--Newman--AdS solution, generally not pure AdS, and I will, when there is no confusion, refer to it as simply $g_{ab}$. Similarly when there is no confusion I will let the parameters $m^{(0)}, a^{(0)}$ and so on just be called $m, a$, etc.,~and include the ${}^{(s)}$ superscript when necessary. 

Let us consider only infinitesimal variations, and compare the spacetimes $g_{ab} = g^{(0)}_{ab}$ and the infinitesimally perturbed spacetime $g^{(\de s)}_{ab} = g_{ab} + \f{\pa g_{ab}^{(s)}}{\pa s} \de s$. 

We will keep the integration surface at the same coordinates before and after variation. That integration surface will be $H$ given by constant $\tau$ and $r = r_+^{(0)}$, the location of the horizon in the $s = 0$ spacetime. Importantly, the final horizon will not in general be at $r = r_+^{(0)}$, but at $r = r_+^{(\de s)}$. I want to emphasize this point now because the derivation of $\oint_H \bs k_\chi^{EH}[\de g;g] = \kappa \de A/8\pi$ as in \cite{Compere} assumes that the initial and final horizons are at the same coordinates. As is perhaps surprising, $\oint_H \bs k_\chi^{EH}[\de g;g] = \kappa \de A/8\pi$ will still hold if we keep $H$ at $r = r_+^{(0)}$ even if the final horizon is at $r = r_+^{(\de s)} \neq r_+^{(0)}$. To address this issue I will do two things: first to provide a ``fix'' to the metrics under consideration so that the horizon really does remain at the same coordinates in the unperturbed and perturbed metrics, and second to provide an argument why $\oint_H \bs k_\chi^{EH}[\de g;g] = \kappa \de A/8\pi$ still holds in certain circumstances even if the integration surface does not match the perturbed horizon location. 

Let $\bar \z = \xi + \Om^{(0)} \eta$ be the unperturbed Killing vector tangent to the null generators; $\de \bar \z^a = 0$. 

The metric determinant is $\sqrt{-g} = (r^2+a^2u^2)/\Xi$. (Again, technically, the metric is not regular at $r = r_+$ but we can set $\sqrt{-g}$ to be equal to $(r_+^2+a^2u^2)/\Xi$ in a limiting process at that value.) For $\bs \ep$ we can write
\begin{align}
    \bs \ep_{a b c d} = \sqrt{-g} e_{abcd},
\end{align}
where $e_{abcd}$ is the completely antisymmetric symbol with $e_{\tau r u \hat \vp} = 1$. The Komar form $\bs K^K_{\bar \z}$ has $(\bs K^K_{\bar \z})_{ab} = (16\pi)^{-1} \bs \ep_{abcd} \na^c \bar \z^d$. Recall that $\de \bar \z^a = 0$, but because $\sqrt{-g}$ does have a nonzero variation (since $\de a, \de l \neq 0$, and since $\na^c \equiv (\na^{(s)})^c$ is the covariant derivative associated with the (variable) metric $g^{(s)}_{ab}$, we have
\begin{align}
    (\de \bs K^K_{\bar \zeta})_{a b} &= \f{1}{16\pi} \frac{\pa}{\pa s} (\bs \ep_{a b c d}^{(s)} (\na^{(s)})^c \bar \zeta^d) \de s.
\end{align}

For $v^{EH}$ we have \eqref{ThetaEH}. The covariant derivatives can be evaluated in the $g^{(0)}_{ab}$ spacetime, so I will just write $\na^b$ and so on. We have,
\begin{align}
    v^{EH}_a &= \na^b \de g_{a b} - g^{b c} \na_a \de g_{b c} \nn 
    &= \left( \na^b \frac{\pa g_{a b}^{(s)}}{\pa s} +  g^{bc} \na_a \frac{\pa g_{b c}^{(s)}}{\pa s}\right) \de s,
\end{align}
where any quantities with no ${}^{(s)}$ superscript are evaluated on $s = 0$. Consequently, we have for $\bs k^{EH}_{\bar \z}[\de g; g]$ from \eqref{kEHincludingdeltachi},
\begin{align}
    \bs k^{EH}_{\bar \z}[\de g; g] &= -\de  \bs K^K_{\bar \z} -  \bar \z \cdot \bs \Th^{EH}[\de g; g] \nn 
    (\bs k^{EH}_{\bar \z}[\de g;g])_{cd}
    &= -\f{1}{16\pi} \left( \frac{\pa}{\pa s}\left(\bs \ep^{(s)}_{a b c d} (\na^{(s)})^c \bar \z^d\right) + \bar \z^d g^{c f}\left(  \na^e \frac{\pa g_{ef}^{(s)}}{\pa s} +  g^{eh}  \na_f \frac{\pa g_{eh}^{(s)}}{\pa s}\right){\bs \ep}_{abcd}\right) \de s.
\end{align}
I am writing these terms out explicitly in order to try to be as clear as possible about which quantities are evaluated at what values of $s$. 

We then will want to evaluate $(\bs k^{EH}_{\bar \z}[\de g; g])_{u\hat \varphi}$. In general this expression is very complicated, but it simplifies considerably on $H$. It is easier to work with the set $(r_+,a,Q,l)$ than $(m,a,Q,l)$, so let $r_+^{(s)}$ be a constant associated with each metric (at constant $s$), for which the horizon corresponds to $r = r_+^{(s)}$. The horizon equation implies
\begin{align}
    m^{(s)} &= \frac{((r_+^{(s)})^2 +(a^{(s)})^2)((r_+^{(s)})^2+(l^{(s)})^2)}{2 r_+^{(s)} (l^{(s)})^2} + \frac{(Q^{(s)})^2}{2 r_+^{(s)}}. \label{mofs}
\end{align}
The integration surface is at $r = r_+^{(0)} \equiv r_+$. I want to emphasize again that $\bar \z$ has constant components: $\Omega$ depends on $a = a^{(0)}, l = l^{(0)}, r_+ = r_+^{(0)}$. We then have that at $r = r_+ = r_+^{(0)}$,
\begin{align}
    (\bs k^{EH}_{\bar \z}&[\de g;g])_{u\hat \vp} = \nn
    &\f{(l^2-r_+^2)a^2+(Q^2-r_+^2)l^2-3 r_+^4)}{8\pi (l^2-a^2)^2 (r_+^2+a^2)r_+l} \left( \left[a (r_+^2+a^2) \frac{dl^{(s)}}{ds}-l(l^2+r_+^2) \frac{da^{(s)}}{ds} \right]a -r_+ l^3\Xi \frac{d r_+^{(s)}}{ds}\right) \de s. \label{kzetaEHhorizon4D}
\end{align}
Remarkably, this is constant on $H$, so if we integrate over $dud\hat \varphi$ we get an extra factor of $4 \pi$. 
Since the horizon area as a function of $s$ is $A^{(s)} = 4\pi((r_+^{(s)})^2+(a^{(s)})^2)/(1-(a^{(s)})^2/(l^{(s)})^2)$ and the surface gravity at $s = 0$ is \eqref{kappaKNewmanAdS}, we have
\begin{align}
    \oint_H \bs k^{EH}_{\bar \z}[\de g;\bar g] &= 4 \pi (\bs k^{EH}_{\bar \z}[\de g;\bar g])_{u\hat \varphi} \nn 
    &= \frac{\kappa \de A}{8\pi}, \label{intkzeta}
\end{align}
as expected. 

Note that there are several possible ways to perform this calculation incorrectly. It is essential that we keep $\Omega$ a constant rather than allowing it to vary with $s$ to follow the way $\Omega^{(s)}$ varies with the variation in $a^{(s)},l^{(s)}$ and $r_+^{(s)}$, for instance. It is also important that we allow $r_+^{(s)}$ to vary after setting $m^{(s)}$ to a function of $r_+^{(s)}$ (and other terms) but to leave $r$ fixed otherwise. 

Again the issue remains that the derivation of $\oint_H \bs k^{EH}_{\bar \z}[\de g;g] = \kappa \de A/8\pi$ assumed that the horizon was at the same location in the initial and final spacetime. One thing to address this issue is to modify the ABL form of the metric so that the horizon remains at the same coordinates under the variation with $s$. The fix I suggest is to define the coordinate $p \equiv r/r_+^{(s)}$. This coordinate is $r$ times a quantity which is constant within a given spacetime, but where the constant of proportionality is different for different values of $s$. Then the horizon is always located at $p = 1$. To write the metric in $(\tau, p, u, \hat \vp)$ coordinates for arbitrary $s$, send $m^{(s)}$ to its value in \eqref{mofs} earlier in the process and then to replace $r$ with $p r_+^{(s)}$. The metric then takes the ungainly form (omitting the $^{(s)}$ superscript)
\begin{align}
    ds^2 &= g_{\tau \tau} d\tau^2 +g_{\hat \varphi\hat \varphi} d\hat \varphi^2 \nn 
    &\qquad -\frac{a(((a^2+Q^2+r_+^2)l^2+a^2r_+^2+r_+^4)p-l^2Q^2)(1-u^2)(l^2-a^2u^2)}{(l^2-a^2)^2(p^2r_+^2+a^2u^2)}d\tau d\hat \varphi \nn
    &\qquad + \frac{l^2r_+^2(a^2u^2+p^2r_+^2)}{(p-1)(p^3r_+^4+r_+^4p^2+r_+^2(a^2+l^2+r_+^2)p-l^2(a^2+Q^2)}dr^2 +\frac{l^2(a^2u^2+p^2r_+^2)}{(1-u^2)(l^2-a^2u^2)}du^2, \nn
    g_{\tau \tau} &= -\rho^{-2} \Xi^{-2}l^{-4} (1-a^2u^2/l^2)  \left\{ ((u^2-p)a^2+(p-1)(pr_+^2-Q^2))l^4\right. \nn 
    &\qquad \left.+(u^2(p-1)a^4+((u^2-1)r_+^2p^2+(r_+^2(u^2-1)+Q^2u^2)p-Q^2u^2)a^2+r_+^4p^4-r_+^4p)l^2 \right. \nn 
    &\qquad \left.- a^2r_+^2(u^2(p-1)a^2+r_+^2(p^3-u^2))p\right\}\nn
    g_{\hat \varphi \hat \varphi} &= \rho^{-2} \Xi^{-2} l^{-2} (u^2-1) \left\{ a^6u^2 + ((u^2+1)r_+^2p^2+(u^2-1)(l^2+r_+^2)p-l^2u^2)a^4 -l^2 p^4 r_+^4 \right. \nn 
    &\qquad \left.+(r_+^4p^4-l^2(u^2+1)r_+^2p^2+((Q^2+r_+^2)l^2+r_+^4)(u^2-1)p+Q^2(1-u^2)l^2)a^2 \right\},
\end{align}
with $\rho^2 = r^2+a^2\cos^2\tht = p^2r_+^2 + a^2u^2$ in these coordinates. $g_{ab}^{(s)}$ is the above metric with $r_+^{(s)}, a^{(s)}, l^{(s)}, Q^{(s)}$, as before. For all $s$, the metric represents a solution to Kerr--New6man--AdS, and indeed it is the same solution of Kerr--Newman--AdS as the metric in the form \eqref{ABL4D} with $\Delta = \Delta(r) = (r^2+a^2)(1+r^2l^{-2}) - 2mr + Q^2$, rewritten so that the spheroidal radius is rescaled by the value of $r_+$. 

Using the same arguments as before we calculate $(\bs k_{\bar \z}[\de g;g])_{u\hat \vp}$, this time on constant-$\tau$ and $p = 1$ and recover \eqref{kzetaEHhorizon4D} and thus \eqref{intkzeta}. It is interesting that the initial and final surfaces of the horizon being at the same coordinates is not a strict requirement to find \eqref{kzetaEHhorizon4D}, but it is reassuring that the case where the initial and final horizon are at the same coordinates is consistent with it. 

Next on the agenda is to check the Kerr--Schild form of the spacetime. Again for convenience $u = \cos \tht$, but otherwise the base metric is of the same form as \eqref{KS4D}. Here the coordinates are $(t,r,u,\phi).$ Following the same procedure as described above, we find \eqref{intkzeta}, as well as \eqref{kzetaEHhorizon4D} (with $\hat \varphi$ replaced with $\phi$). Similarly, we can replace $r$ as a coordinate with $p = r/[r_+(s)]$ and again recover \eqref{kzetaEHhorizon4D} (with $\hat \varphi$ replaced with $\phi$) and \eqref{intkzeta}. 

The procedure was also repeated for the BL form of the metric (using coordinates $\tau,\varphi$ where $\hat \varphi = \varphi + a l^{-2} \tau$). The BL case with $p = r/[r_+(s)]$ is also considered, as was the case \eqref{BLKS4D} in Kerr--Schild form with the background metric in BL form (BLKS coordinates). Recall that $\bar \z = \xi + \Om^{(0)} \eta$, which has constant components in the ABL and KS coordinates. In the case of the BL or BLKS coordinates, it is not $\xi$ and $\eta$ but $\bt$ and $\eta$ which have constant components, so the Killing vector which is tangent to the null generators of the horizon in the unperturbed spacetime which has constant contravariant components is, instead, $\bt + \omega^{(0)} \eta$, where $\omega^{(0)}$ is the value of $\omega$ with the initial values of $a,l,r_+$. I will call this vector $\hat \z$ instead of $\bar \z$:
\begin{align}
    \hat \z &= \bt + \om^{(0)}\eta.
\end{align}
We once again recover \eqref{kzetaEHhorizon4D} (with $\hat \varphi$ replaced with $\varphi$ and $\bar \z$ replaced with $\hat \z$) and \eqref{intkzeta}, in all three cases. This tells us that, once we have chosen the Killing vector which has constant components, it is not a requirement that we be in an asymptotically-static frame for $\oint_H \bs k_{\bar \z}^{EH} [\de g;g] = \ka \de A/8\pi$. 

If we instead use $\bar \zeta = \xi + \Om^{(0)} \eta$, which has contravariant component $\bar \zeta^\varphi$ changing as $a$ changing, the result is different. Note,
\begin{align}
    \xi &= \bt - \f{a}{l^2} \eta \nn 
    \de \xi^a &= - \f{1}{l^2} \eta^a \de a.
\end{align}
This means we have, in BL or BLKS coordinates,
\begin{align}
    \de \bar \z^a &= \de \xi^a \nn 
    &= -\f{1}{l^2} \eta^a \de a.
\end{align}
The distinction between $\hat \z$ and $\bar \z$ is important and will be revisited in Section \ref{asymptoticallyrotatingframe}.

In general my finding was that \eqref{kdarule} is very robust and applies for all the forms of the metric under consideration. To explain this robustness, when the derivation for $\oint_H \bs k^{EH}_{\bar \z} = \ka \de A/8\pi$ was only derived for the case where the horizon remains at constant location, I consider the effect of infinitesimal coordinate transformations in the following section.

\section{Representing Infinitesimal Coordinate Transformations with the Lie Derivative} \label{infinitesimalcoordinatetransformations}

A gap in our argument so far is that $\oint_H \bs k^{EH}_{\bar \z}[\de g;g] = \kappa \de A/8\pi$ was proven by \cite{Compere}, following \cite{BCH}, to hold provided the horizon has the same coordinates in the initial and final spacetime, but appears to hold for $\oint_H \bs k^{EH}_{\bar \z} [\de g;g] = \kappa \de A/8\pi$ even when the initial and final horizon locations $r = r_+^{(s)}$ are different. Further, the statement $\de \bs I_{\bar \z} = \bs k^{EH}_{\bar \z}[\de g;g]$ relies on the background metric being unchanging, but we still seem to be able to use the \emph{integral} equation $\oint_H \de \bs I_{\bar \z} = \oint_H  \bs k^{EH}_{\bar \z}[\de g;g]$ even if $\de \bar g_{ab} \neq 0$, albeit only in vacuum rather than in full generality. The former issue can be addressed by moving to coordinates where the radial coordinate is rescaled, so that the horizon remains at the same location, as in Section \ref{checkinghorizonareavariationrule}, and the latter issue can be addressed by moving to coordinates where the background metric remains constant, as in Section \ref{constantmetricbackgroundvariablea}, but generally we cannot satisfy both requirements at once. 

In this section I give an explanation for what currently seems to be very fortuitous, that both equalities $\oint \de \bs I_{\bar \z} = \oint_H \bs k^{EH}_{\bar \z}[\de g;g] = \kappa \de A/8\pi$ appear to hold in a more general setting than one might expect.

To do so I use the Lie derivative and its connection to infinitesimal diffeomorphisms/coordinate transformations. (As pointed out by, e.g.,~Wald \cite{Wald84}, we can either take the active view of diffeomorphisms, in which a diffeomorphism $f$ mapping points in $M$ to other points in $M$ also defines a map acting on tensors, or the passive view, in which the points and tensors are unchanged, but $f$ is associated with a coordinate transformation, affecting the coordinate representations of the points and tensor components. These points of view are philosophically different but functionally work out pretty similarly, so I won't dwell on the distinction.) 

To begin, when considering a metric variation $\de g_{ab}$, there are two parts to consider: the part of the change in $g_{ab}$ which results from moving between two distinct spacetimes, and the part of $\de g_{ab}$ which is ``pure gauge,'' and only the result of a diffeomorphism/coordinate transformation. The latter can be connected to the Lie derivative (see, for instance, \cite{Wald84,Poisson,Yano}) and will be the focus of this section.

I will, roughly following the above references, lay out the argument in the following way, taking the passive (coordinate transformation) point of view. Let $M$ be a spacetime manifold and let $x^a$ be a set of coordinates for it, with metric tensor field whose components depend on the coordinates: $g_{ab} = g_{ab}(x^c)$. Let $\tilde x^a = \phi^a (x^b)$ be a new set of coordinates, with functions $\phi^a$ each functions of $x^a$. Schematically we can represent this as $\tilde x = \phi(x)$, and so $\phi$ represents a map. We demand that the coordinate transformation be invertible, so that $x = \phi^{-1}(\tilde x)$, or $x^a = (\phi^{-1})^a(\tilde x^b).$ $\tilde g_{ab}$ are the components for the metric tensor resulting from the coordinate transformation from $x$ to $\tilde x$. 

$\tilde g_{ab}$ is a function of the $\tilde x^c$ coordinates, so that we can write $\tilde g_{ab} = \tilde g_{ab}(\tilde x^c)$. Using the usual tensor transformation equation:
\begin{align}
    \left.\tilde g_{ab}\right|_{\tilde x = \phi(x)} &=  g_{cd} \f{\pa x^c}{\pa \tilde x^a} \f{\pa x^d}{\pa \tilde x^b},
\end{align}
where both sides are evaluated at some particular value for the original coordinates $x$.

Now consider the case where we have some specific set of values $X^a$ and wish to compare the values of the components of $\tilde g_{ab}$ evaluated at $\tilde x^c = X^c$ and the components of $g_{ab}$ evaluated at $x^c = X^c$. These will have a difference
\begin{align}
    \tilde g_{ab}|_{\tilde x = X} - g_{ab}|_{x = X}.
\end{align}
In general these represent different points in the manifold. Specifically, the point $x = X$ corresponds to $\tilde x = \phi(X)$, and the point $\tilde x = X$ corresponds to $x = \phi^{-1}(X)$ (so that $\tilde x = \phi ( \phi^{-1}(X)) = X$). Consequently,
\begin{align}
    \tilde g_{ab}|_{\tilde x = X} - g_{ab}|_{x=X} &= \tilde g_{ab}|_{x = \phi^{-1}(X)} - g_{ab}|_{x=X} \nn 
    &=\left. \left( g_{cd} \f{\pa x^c}{\pa \tilde x^a} \f{\pa x^d}{\pa \tilde x^b}\right)\right|_{x = \phi^{-1}(X)} - g_{ab}|_{x=X}. \label{expandedLieDerivativeMetric}
\end{align}
Now consider a set of maps $\phi^{(\la)}$ parametrized smoothly by $\la$, such that $\phi^{(0)}$ is the identity map. Then we can let
\begin{align}
    \tilde x^{(\la)} = \phi^{(\la)} (x),
\end{align}
or, including indices,
\begin{align}
    (\tilde x^{(\la)})^a = (\phi^{(\la)})^a ( x^b).
\end{align}
For small values of $\la$ this expands to
\begin{align}
    \tilde x^{(\la)} &= x + \f{d \phi^{(\la)}}{d \la} \la \nn 
    (\tilde x^{(\la)})^a &= x^a + \f{d (\phi^{(\la)})^a}{d \la} \la,
\end{align}
where the derivative is evaluated at $x^a$ and at $\la = 0$. Define $w^a$ as
\begin{align}
    w^a &\equiv \left. \f{d (\phi^{(\la)})^a}{d \la} \right|_{\la = 0}.
\end{align}
$w^a$ is a vector field and so is a function of $x$. Let $\tilde g^{(\la)}_{ab}$ be the pull-back of $g_{ab}$ under the map $\phi^{(\la)}$. 

We then have,
\begin{align}
    \left.\f{d}{d\la} \left( \tilde g_{ab}^{(\la)}|_{\tilde x^{(\la)} = X} - g_{ab}|_{x=X}\right)\right|_{\la = 0} &= -\lie_w g_{ab}|_{x=X}. \label{ddlatildeg}
\end{align}
This is (in an informal way) the definition of the Lie derivative for a covariant tensor and so the right-hand side follows automatically. To convince the reader, I will expand this out and compare with the generally-known expression. From \eqref{expandedLieDerivativeMetric}, 
\begin{align}
    \left(\tilde g^{(\la)}_{ab}|_{\tilde x^{(\la)} = X} - g_{ab}|_{x = X}\right) &=   \left(\left.\left( g_{cd} \f{\pa x^c}{\pa (\tilde x^{(\la)})^a} \f{\pa x^d}{\pa (\tilde x^{(\la)})^b}\right)\right|_{x = (\phi^{(\la)})^{-1}(X)} - g_{ab}|_{x = X}\right).
\end{align}
I will expand this out and keep only terms zeroth and first-order in $\la$. $x^a = (\tilde x^{(\la)})^a -  w^a \la$ for small values of $\la$. Any difference between $w^a$ and its push-forward $(\tilde w^{(\la)})^a$ will be first-order or higher in $\la$, and so can be neglected when considering only first-order differences. Similarly the partial derivative with respect to $(\tilde x^{(\la)})^a$ can be treated as being the same as a partial derivative with respect to $x^a$, to zeroth order in $\la$. We can then write, to first-order, 
\begin{align}
    \f{\pa x^c}{\pa (\tilde x^{(\la)})^a} &= \de^c_a - \la \f{\pa w^c}{\pa x^a},
\end{align}
and thus, to first order,
\begin{align}
    \tilde g_{ab}^{(\la)}|_{\tilde x^{(\la)} = \phi^{(\la)}(x)} &= g_{ab} - \la \left( g_{cb} \f{\pa w^c}{\pa x^a} + g_{ac} \f{\pa w^c}{\pa x^b}\right),
\end{align}
with the right-hand side evaluated at $x$. We then have
\begin{align}
    \tilde g^{(\la)}_{ab}|_{\tilde x^{(\la)} = X} - g_{ab}|_{x = X} &=  \left.\left( g_{ab} - \la \left( g_{cb} \f{\pa w^c}{\pa x^a} + g_{ac} \f{\pa w^c}{\pa x^b}\right)\right)\right|_{x = (\phi^{(\la)})^{-1}(X)} - g_{ab}|_{x=X}
\end{align}
We can now apply $\tilde x = X \implies x^a = \tilde x^a - w^a \la = X^a - w^a \la$, with $w^a$ evaluated at $x = X$. 
\begin{align}
    \tilde g^{(\la)}_{ab}|_{\tilde x^{(\la)} = X} - g_{ab}|_{x = X} &= \left.\left( g_{ab} - \la \left( g_{cb} \f{\pa w^c}{\pa x^a} + g_{ac} \f{\pa w^c}{\pa x^b}\right)\right)\right|_{x = X - w \la} - g_{ab}|_{x=X} \nn 
    &= \left.\left(g_{ab} - \la w^c \f{\pa g_{ab}}{\pa x^c}\right)\right|_{x=X} - \left.\la \left( g_{cb} \f{\pa w^c}{\pa x^a} + g_{ac} \f{\pa w^c}{\pa x^b}\right)\right|_{x=X} - g_{ab}|_{x=X} \nn 
    &= \la \left.\left( - w^c \f{\pa g_{ab}}{\pa x^c} - g_{cb} \f{\pa w^c}{\pa x^a} - g_{ac} \f{\pa w^c}{\pa x^b}\right)\right|_{x=X} \nn 
    &= - \la \lie_w g_{ab}|_{x=X}. 
\end{align}
\eqref{ddlatildeg} follows.

Recall also that $\lie_w g_{ab} = 2 \na_{(a}w_{b)}$ for any vector $w$.

The above argument applies to any arbitrary tensor, not just the metric tensor. Letting ${T^{a_1 \ldots a_k}}_{b_1 \ldots b_l}$ be a tensor and letting its pull-back under the coordinate map $\tilde x^{(\la)} = \phi^{(\la)}(x)$ be ${(\tilde T^{(\la)})^{a_1 \ldots a_k}}_{b_1 \ldots b_l}$, then 
\begin{align}
    \left.\f{d}{d\la} \left({(\tilde T^{(\la)})^{a_1 \ldots a_k}}_{b_1 \ldots b_l}|_{\tilde x = X} - {T^{a_1 \ldots a_k}}_{b_1 \ldots b_l}|_{x=X}\right)\right|_{\la = 0} = - \lie_w {T^{a_1 \ldots a_k}}_{b_1 \ldots b_l}|_{x=X}.
\end{align}

As a consequence of the above argument, we can also write that for infinitesimal variations in coordinates, $\tilde x^a = x^a + w^a \de \la$,
\begin{align}
    \left({(\tilde T^{(\la)})^{a_1 \ldots a_k}}_{b_1 \ldots b_l}|_{\tilde x = X} - {T^{a_1 \ldots a_k}}_{b_1 \ldots b_l}|_{x=X}\right) = - \de \la \lie_w {T^{a_1 \ldots a_k}}_{b_1 \ldots b_l}|_{x=X}.
\end{align}

To make connection with BC notation I will use ${}^{(s)}$ rather than ${}^{(\la)}$ in the following subsections.

Now I want to consider two important applications. The first is to consider what happens when the radius is allowed to change for a black hole horizon. The second is to consider a way to use the infinitesimal coordinate transformation approach to keep $\bar g_{ab}$ constant while $a$ varies when discussing the GKAdS solutions.

\subsection{Varying Horizon Radius} \label{varyinghorizonradius}

We have that, under a variation in the spacetime where the metric changes but the horizon location $H$ is at the same coordinates $r = r_+$ in the initial and final spacetime, $\oint_H \bs k_{\bar \z}^{EH}[\de g;g] = \kappa \de A/8\pi$. I will now show what happens when the horizon radius is also allowed to vary. I will initially work in arbitrary dimension, under the assumption that the horizon remains at some value $r = r_+^{(s)}$, which is constant within each spacetime. 

Consider, as in Section \ref{massandspecificangularmomentumvary}, a metric which depends on a parameter $s$ in solution space, $g_{ab}^{(s)}$. We have $\de = \de s \f{d}{ds}$, evaluated at $s = 0$. The event horizon is located at $r = r_+^{(s)}$, which depends on $s$ directly. This much is true for all forms of the metric (ABL, BL, KS) in Section \ref{massandspecificangularmomentumvary}. 
It is useful to find a way to keep the horizon radius fixed, so let us perform the following coordinate transformation. Let (for this section) $\tilde g^{(s)}_{ab}$ be the result of applying the transformation
\begin{align}
    r \to \f{r_+^{(0)}}{r_+^{(s)}} r
\end{align}
(keeping the other coordinates unchanged) to $g^{(s)}_{ab}$. To be more precise, define the set of coordinates $(\tilde x^{(s)})^a$ to be $(\tilde x^{(s)})^a = x^a$ for all coordinates $x^a$ except for $r$, and
\begin{align}
    \tilde r^{(s)} = \f{r_+^{(0)}}{r_+^{(s)}} r.
\end{align}

Of course $\tilde g^{(0)}_{ab} = g^{(0)}_{ab}$. The idea then is that $\tilde g^{(s)}_{ab}$ represents the same spacetime as $g^{(s)}_{ab}$, but has the horizon located at $r = r_+^{(0)}$ throughout, by rescaling of the coordinates.

We can now define variations $\de$, $\de_1$ and $\de_2$ as follows. Let $X^a$ be some set of values that coordinates can take. Then define $\de g_{ab}, \de_1 g_{ab}$ and $\de_2 g_{ab}$ to be functions of $x$ given by
\begin{align}
    \de g_{ab}|_{x=X} &\equiv \de s \left.\f{d g^{(s)}_{ab}}{ds}\right|_{x = X,s=0} \nn 
    \de_1 g_{ab}|_{x=X} &\equiv \de s \left. \f{d \tilde g^{(s)}_{ab}}{ds}\right|_{\tilde x^{(s)}=X,s=0} \nn 
    \de_2 g_{ab}|_{x=X} &\equiv \de s \left. \f{d}{ds}\left( g_{ab}^{(s)}|_{x=X} - \tilde g^{(s)}_{ab}|_{\tilde x^{(s)}=X}\right)\right|_{s=0}. \label{definitionsofdeltag}
\end{align}
We also define expressions for other tensors analogously: if ${(T^{(s)})^{a_1 \ldots a_k}}_{b_1 \ldots b_l}$ is a tensor (in general dependent on $s$) and ${(\tilde T^{(s)})^{a_1 \ldots a_k}}_{b_1 \ldots b_l}$ is the result of the coordinate transformation from $x^a$ to $(\tilde x^{(s)})^a$, then define
\begin{align}
    \de {T^{a_1 \ldots a_k}}_{b_1 \ldots b_l} |_{x=X} &\equiv \de s \left. \f{d {(T^{(s)})^{a_1 \ldots a_k}}_{b_1 \ldots b_l}}{ds}\right|_{x=X,s=0} \nn 
    \de_1 {T^{a_1 \ldots a_k}}_{b_1 \ldots b_l} |_{x=X} &\equiv \de s \left. \f{d {(\tilde T^{(s)})^{a_1 \ldots a_k}}_{b_1 \ldots b_l}}{ds}\right|_{\tilde x^{(s)}=X,s=0} \nn 
    \de_2 {T^{a_1 \ldots a_k}}_{b_1 \ldots b_l} |_{x=X} &\equiv \de s \left. \f{d}{ds} \left( {(T^{(s)})^{a_1 \ldots a_k}}_{b_1 \ldots b_l}|_{x=X} - {(\tilde T^{(s)})^{a_1 \ldots a_k}}_{b_1 \ldots b_l}|_{\tilde x^{(s)}=X}\right) \right|_{s=0} \label{tensorvariations}
\end{align}
I will, for now, focus on the metric variations but the results generalize readily to arbitrary tensors.

While $\de g_{ab}$ is straightforward, for $\de_1$ and $\de_2$ there is ambiguity in whether it would be $x$ or $\tilde x^{(s)}$ which would be evaluated at $X^a$, hence the specificity. $\de$ represents the variation in $g_{ab}^{(s)}$, which includes a horizon changing location, $\de_1$ represents a variation in $\tilde g_{ab}^{(s)}$ and so represents the variation in the physical spacetime but keeping the horizon location fixed (which we can represent by writing $\de_1 r_+ = 0$), and $\de_2$ represents the variation between $g_{ab}^{(s)}$ and $\tilde g_{ab}^{(s)}$ resulting from the ``pure gauge'' coordinate transformation between them. We note that $\de g_{ab} = \de_1 g_{ab} + \de_2 g_{ab}$.

We can use the infinitesimal coordinate transformation approach for $\de_2 g_{ab}$. We can expand the definition of $\de_2 g_{ab}|_{x=X}$ to be
\begin{align}
    \de_2 g_{ab}|_{x=X} = \de s \f{d}{ds} \left( g^{(0)}_{ab} |_{\tilde x^{(s)}=X} - \tilde g^{(0)}_{ab} |_{x=X}\right) + \de s \f{d}{ds} \left( g^{(s)}_{ab}|_{x=X} - \tilde g^{(s)}_{ab}|_{\tilde x^{(0)}= X}\right),
\end{align}
splitting the $s$-dependence into the $s$-dependence of the metric and the $s$-dependence of the coordinates. For the first term, we are considering only the components of the metric $g_{ab}^{(0)}$, that is, associated with the original spacetime. This first term is $+ \de s \lie_w g_{ab}|_{x=X}$ where $\tilde x^{(s)} = x^a + s w^a$, from \eqref{ddlatildeg}.  The second term is zero, because $\tilde x^{(0)} = x$ and so the term being differentiated is zero. We conclude,
\begin{align}
    \de_2 g_{ab} = \de s \lie_w g_{ab},\label{delta2gab}
\end{align}
where it is understood that the evaluation of the right-hand side is at $s = 0$. The left- and right-hand sides are both evaluated at the same value of $x^c$. 

The effect of $\de_2$ acting on any tensor $T$ which is in general a function of $s$ is analogous:
\begin{align}
    \de_2 {T^{a_1 \ldots a_k}}_{b_1 \ldots b_l} &= \de s \lie_w {T^{a_1 \ldots a_k}}_{b_1 \ldots b_l}. \label{delta2T}
\end{align} 
In particular, 
\begin{align}
    \de_2 \bs K^K_{\bar \z} = \de s \lie_w \bs K^K_{\bar \z}
\end{align}
(where the last term is evaluated at $s =0$).

$w^a$ is given by $d (\tilde x^{(s)})/ds$, so 
\begin{align}
    w^a &= \f{d \tilde x^{(s)}}{ds} \nn 
    &= \de^a_r \f{d \tilde r^{(s)}}{ds} \nn 
    &= -\de^a_r \f{d r_+^{(s)}}{ds} \f{r_+^{(0)} r}{(r_+^{(s)})^2},
\end{align}
which we can write compactly as $w^a = f(r) \de^a_r$ (with $f(r)$ implicitly including the $s$ dependence). When evaluated at $s = 0$,
\begin{align}
    w^a &= -\de^a_r \f{r}{r_+^{(0)}} \left. \f{d r_+^{(s)}}{d s}\right|_{s=0}.
\end{align}

The utility of this breakdown is the following. Because $\bs k_{\bar \z}^{EH}[\de g;g]$ is linear in the variation, we can write
\begin{align}
    \bs k_{\bar \z}^{EH}[\de g;g] &= \bs k_{\bar \z}^{EH}[\de_1 g;g] + \bs k_{\bar \z}^{EH}[\de_2 g;g].
\end{align}
We can then treat the ``purely physical with unchanging horizon'' and ``pure gauge'' parts of the variation in the metric independently. The formula \eqref{kappadaBC2} applies when the horizon radius is unchanging, so that we can write
\begin{align}
    \oint_H \bs k_{\bar \z}^{EH}[\de_1 g;g] &= \f{\kappa \de A}{8\pi},
\end{align}
where $\de A = \de s \f{dA^{(s)}}{d s}$ and $\kappa = \kappa^{(0)}$ (the value of the horizon associated with $g^{(0)}_{ab}$). Applying \eqref{delta2gab} as well, we have 
\begin{align}
    \oint_H \bs k_{\bar \z}^{EH}[\de g;g] &= \f{\kappa \de A}{8\pi} + \de s \oint_H \bs k_{\bar \z}^{EH}[\lie_w g;g].
\end{align}
Here $H$ is at constant $t$ and with the unvaried horizon location $r = r_+^{(0)}$. 

I will now show that $\oint_H \bs k_{\bar \z}^{EH}[\lie_w g;g] = 0$ in the cases under consideration in Section \ref{massandspecificangularmomentumvary}, which tells us that we need not keep the horizon strictly at constant radius in order to apply \eqref{kappadaBC2}. 

We can break down $\bs k_{\bar \z}^{EH}[\lie_w g;g]$ using \eqref{kEHincludingdeltachi}. Let $ v_w^a \de s$ be the $v^a$ associated with $\de g_{ab} = \de s \lie_w g_{ab}$. Then we have, from $\lie_w g_{ab} = 2 \na_{(a}w_{b)}$,
\begin{align}
    (v_w)_a &= 2 \na^b \na_{(a}w_{b)} - 2 \na_a \na^b w_b.
\end{align}
As in, for instance, \cite{Rossi}, this can be simplified to
\begin{align}
    (v_w)^a &= 2 R^a_b w^b - 2 \na_b \na^{[a}w^{b]}.
\end{align}
Meanwhile, the variation in $\bs K_{\bar \z}^K$ can be shown (using Cartan's identity) to be
\begin{align}
    \de_2 \bs K_{\bar \z}^K &= \de s \lie_w \bs K_{\bar \z}^K \nn 
    &= \de s \left( w \cdot d \bs K_{\bar \z}^K + d (w \cdot \bs K_{\bar \z}^K)\right).
\end{align}
Using \eqref{dKKxigeneral} the first term can be written
\begin{align}
    (w \cdot d \bs K^K_{\bar \z})_{a_1 \ldots a_{D-2}} &= \f{1}{8\pi} R^b_c \bar \z^c w^d \bs \ep_{b d a_1 \ldots a_{D-2}}.
\end{align}
The second term can be rewritten using \eqref{dKxiKdefinition}, \eqref{Vdotstaromega} and \eqref{dstaromega} as
\begin{align}
    d (w \cdot \bs K^K_{\bar \z}) &= \f{1}{96\pi} d * (d \bar \z^\flat \wedge w^\flat) \nn 
    &= \f{1}{32\pi} * \mathrm{div} ( d \bar \z^\flat \wedge w^\flat) \nn 
    (d (w \cdot \bs K_{\bar \z}^K))_{a_1 \ldots a_{D-2}} &= \f{3}{16\pi} \na_b (\na^{[b} \bar \z^c w^{d]}) \bs \ep_{c d a_1 \ldots a_{D-2}},
\end{align}
giving (slightly rearranging)
\begin{align}
    (\lie_w \bs K_{\bar \z}^K)_{a_1 \ldots a_{D-2}} &= \f{1}{16\pi} \left(2 R^b_e \bar \z^e w^c + 3 \na_e (w^{[b} \na^c \bar \z^{e]}\right) \bs \ep_{b c a_1 \ldots a_{D-2}}.
\end{align}
This results in, after some further rearranging,
\begin{align}
    (\bs k^{EH}_{\bar \z}[\lie_w g;g])_{a_1 \ldots a_{D-2}} &= \f{1}{16 \pi} \left( - 4 R^b_e (\bar \z^{(e}w^{c)} ) + \left( 2 \bar \z^c \na_e \na^{[b} w^{e]}- 3 \na_e(w^{[b} \na^c \bar \z^{e]}\right) \right) \bs \ep_{b c a_1 \ldots a_{D-2}}.
\end{align}

On the horizon itself, $H$, we can replace terms like $\bs B^{ab} \bs \ep_{ab c_1 \ldots c_{D-2}} dx^{c_1} \ldots d x^{c_{D-2}}$ with $\bs B^{ab} d S_{ab}$, with $dS_{ab}$ given by \eqref{horizonbinormal}. $\z^a$ is tangent to the null generators of the horizon and $n_a$ is the auxiliary null vector satisfying $\z_a n^a = -1$. Because all quantities are evaluated at $s = 0$, we can also drop the bar on $\bar \z$ on the right-hand side. On $H$, we can write
\begin{align}
    \bs k_{\bar \z}^{EH}[\lie_w g;g] &= \f{dS}{8\pi} \left( - 4 R^b_e ( \z^{(e}w^{c)} ) + \left( 2  \z^c \na_e \na^{[b} w^{e]}- 3 \na_e(w^{[b} \na^c \z^{e]}\right) \right)  \z_{[b} n_{c]}.
\end{align}
We can use $ \z_a  \z^a = 0$ and $ \z_a n^a = -1$. As proven in \cite{BCH}, using Raychaudhuri's equation and assuming the dominant energy condition, $R_{a b}  \z^a$ must be parallel to $ \z_b$ (or zero) on the horizon. Let us write $R_{a b}  \z^a|_{H} \equiv \Psi \z_b$ where $\Psi$ is some scalar. Expanding out the Ricci term gives
\begin{align}
    -4 R^b_e  \z^{(e} w^{c)}  \z_{[b} n_{c]} &= -R^b_e \left(  \z^e w^c  \z_b n_c -  \z^e w^c \z_c n_b +  \z^c w^e  \z_b n_c -  \z^c w^e  \z_c n_b\right) \nn 
    &= - R^b_e \left( \z^e w^c  \z_b n_c -  \z^e w^c  \z_c n_b +  \z^c w^e  \z_b n_c\right) \nn 
    &= - \Psi \left(  \z^b w^c  \z_b n_c -  \z^b w^c  \z_c n_b +  \z^c w^e  \z_e n_c\right) \nn 
    &= 0.
\end{align}

After some straightforward simplification, the remaining terms can be written as
\begin{align}
    \bs k^{EH}_{\bar \z} [\lie_w g;g] &= -\f{dS}{8\pi} \left( \z_b \na_e \na^{[b}w^{e]} + 3 \na_e (w^{[b}\na^c \z^{e]}) \z_{b}n_{c}\right) \nn
    &= -\f{dS}{8\pi} \f{\z_b }{\sqrt{-g}} \left( \f{\pa}{\pa x^e} \left( \sqrt{-g} \na^{[b} w^{e]}\right) + 3 n_c \f{\pa}{\pa x^e} \left(\sqrt{-g} w^{[b} \na^{c} \z^{e]}\right)\right).
\end{align}

Now let us specify to the situation where $\z = \partial_t + \sum_{i} \Omega_i \partial_{\phi_i}$ in some set of coordinates, and also where the horizon is located at constant $r$, $r = r_+$; the horizon can be represented as a function $F(r) = 0$. On the horizon, $ \z_a$ will be proportional to $\pa_a F$ and so we have $ \z_a dx^a = \z_r d r$---there is only an $r$ component. If we wish to take the spatial surface $H$ at constant $(t,r)$, then the binormal $ \z_{[a} n_{b]}$ should only have $tr$ terms, so that $n_a$ must only have $t$ and $r$ components. Then $n_t = -1$ by $n_a \z^a = -1$. ($n_r$ is found by the condition that $n$ is null.) Using this information, we can further simplify the expression for $\bs k^{EH}_{\bar \z}[\lie_w g;g]$ to
\begin{align}
    \bs k_{\bar \z}^{EH} [\lie_w g;g]|_{H} &= - \f{dS}{8\pi} \f{\z_r}{\sqrt{-g}} \f{\pa}{\pa x^e} \left( \sqrt{-g} \left( \na^{[r} w^{e]} - 3 w^{[r} \na^t \z^{e]}\right)\right). \label{kEHzetaliewg}
\end{align}
The terms where $e = r$ disappear due to antisymmetrization (as well as $t$ for the $w^{[r}\na^t \z^{e]}$ term). 

It is possible further general simplifications are possible, but at this point I will specify to the situations we were considering: the GKAdS metric in four dimensions. For definiteness consider first the ABL form of the metric in ($\tau,r,u,\hat \vp)$ coordinates (assuming that the metric being singular on the horizon does not cause problems), and let $w^a$ be, as yet, a vector of the form
\begin{align}
    w &= w^r(r,u) \pa_r + w^u(r,u) \pa_u. \label{wru}
\end{align}
The reason for demanding that $w^a$ has no $\tau, \hat \vp$ dependence is to keep things simple by ensuring that $\lie_\xi w = \lie_\eta w = 0$, so that $w$ shares these symmetries with the metric overall. Having $w$ also have no components in the $\tau$ or $\hat \vp$ directions is another choice made to keep things simple. On the horizon, in these coordinates, $dS = \Xi^{-1} (r_+^2+a^2) d u d \hat \vp, \z_r = (r_+^2+a^2u^2)/(r_+^2+a^2)$ and $\sqrt{-g} = \Xi^{-1}(r_+^2 +a^2u^2)$, so that $dS \z_r/\sqrt{-g} = du d \hat \vp$. 

Because the metric and $w^e$ have no dependence on $\tau, \hat \vp$, the only nonzero derivative terms in \eqref{kEHzetaliewg} are the $x^e = u$ terms. We find, evaluating on $r = r_+$,
\begin{align}
    \left.\left(\sqrt{-g} \left(\na^{[r} w^{u]} - 3 w^{[r} \na^t \bar \z^{u]}\right) \right)\right|_{ H} &= \f{-(1-u^2)(1-a^2u^2/l^2) \f{\pa w^r}{\pa u} - 2(r_+^2+a^2)\kappa w^u}{2\Xi} \label{dwmwtz}
\end{align}
Thus \eqref{kEHzetaliewg} becomes 
\begin{align}
    \oint_{H} \bs k_{\bar \z}^{EH}[\lie_w g;g] &= -\f{1}{8\pi} \oint_H du d \hat \varphi \f{\pa}{\pa x^e}\left( \sqrt{-g} \left( \na^{[r}w^{e]} - 3 w^{[r}\na^t \bar \z^{e]}\right) \right) \nn 
    &= -\f{1}{4} \int_{-1}^1 du \f{\pa}{\pa u} \left( \sqrt{-g} \left( \na^{[r}w^{u]} - 3 w^{[r}\na^t \bar \z^{u]}\right)\right) \nn 
    &= -\f{1}{4} \left[ \sqrt{-g} \left( \na^{[r}w^{u]} - 3 w^{[r}\na^t \bar \z^{u]}\right)\right]^{u=1}_{u=-1} \nn 
    &= -\f{1}{4} \left[\f{-(1-u^2)(1-a^2u^2/l^2) \f{\pa w^r}{\pa u} - 2(r_+^2+a^2)\kappa w^u}{2\Xi}\right]^{u=1}_{u=-1}. \label{horizonkliewgsimplification}
\end{align}
If $\pa w^r/\pa u$ is finite at $u = \pm 1$, the first term will disappear, and we are left with a term proportional to $w^u|_{u=-1}^{u=1}$.

Now consider $w^a$---still not necessarily a function of $r$ only---resulting from a coordinate variation $\de x^a = w^a \de s$. The values $u = \pm 1$ are the poles (the location where $g_{\hat \vp \hat \vp}$ vanishes). If we demand that the poles of the coordinate system remain unchanged under the coordinate mapping, which is to say $u=1 \to \tilde u^{(s)} =1$ and $u = -1 \to \tilde u^{(s)} = -1$ or $\de \tilde u^{(s)}|_{u=\pm1} = 0$, then we have that $w^u|_{u=\pm 1} = 0$, so that in this case $\oint_H \bs k^{EH}_{\bar \z}[\lie_w g;g] = 0$. 

(In fact, a more general case might be true; preliminary calculations show that even off the horizon, $\oint \bs k^{EH}_{\bar \z} [\lie_w g;g]$ on a surface of constant $r,t$ is zero provided that $w^u = 0$ at $u = \pm 1$ and that $w^r$ is regular, even if $r \neq r_+$. I want to investigate further.)

The case that we are specifically considering in this section is the $w^a = f(r) \de^a_r$ resulting in the coordinate transformation required to rescale $r$ to keep the horizon at constant radius, then automatically $\bs k^{EH}_{\bar \z}[\lie_w g;g]|_H = 0$, even before applying the derivative. This is what we intended to show, and so we have given an explanation why we can use \eqref{kappadaBC2} even when the horizon does not remain at constant radius.

\subsubsection{KS Form of Metric} \label{varyinghorizonradiusKSMetricForm}

The same basic argument applies in a natural way to the KS and BL cases, and so it is nothing particular about the ABL form of the metric that results in the fact that we can set $\oint_H \bs k_{\bar \z}^{EH}[\lie_w g;g] = 0$. The argument in the KS form of the metric is particularly neat and is given below. While the focus in this section is on four dimensions, the argument for the KS form of the metric also generalizes readily to higher dimensions. Here I focus exclusively on the case \eqref{wru}.

Consider the GKAdS metric in KS form, in coordinates $(t,r,u,\phi)$ ($u = \cos \tht$). Here I remind the reader of some pertinent facts. As usual $h_{ab} = H k_a k_b$, with $k^r = 1, k^u = 0$, $k^t, k^\phi$ functions of $r$ only. The background metric $\bar g_{ab}$ is diagonal. On the horizon, $H = \bar g^{rr}$, and $H \bar g_{rr} = 2 \mu(r)/V$ is a function of $r$ everywhere. 

Because the metric and $w^a$ have no $t$ or $\phi$ dependence, \eqref{kEHzetaliewg} can be written
\begin{align}
    \bs k_{\bar \z}^{EH}[\lie_w g;g]|_H &= - \f{dS}{8\pi} \f{\z_r}{\sqrt{-g}} \f{\pa}{\pa u} \left( \sqrt{-g} \left( \na^{[r}w^{e]} - 3 w^{[r} \na^t \z^{e]}\right)\right). \label{kEHliewintermsofdA}
\end{align}
Analogously to the case in ABL coordinates, $dS \z_r / \sqrt{-g} $ simplifies to just $d u d \phi$ and so
\begin{align}
    \oint_H \bs k_{\bar \z}^{EH}[\lie_w g;g] &= - \f{1}{4} \left.\left[\sqrt{-g} \left( \na^{[r} w^{u]} - 3 w^{[r} \na^t \z^{u]}\right)\right]\right|^{u=1}_{u=-1},
\end{align}
by analogous simplifications to those in \eqref{horizonkliewgsimplification}. We have,
\begin{align}
    \na^{[r} w^{u]} &= g^{a [r} g^{u] b} \na_a w_b \nn 
    &= g^{ra} g^{ub} \na_{[a} w_{b]} \nn 
    &= g^{ra} g^{ub} \pa_{[a} w_{b]}.
\end{align}
$g^{ub} = g^{uu} \de^b_u$ (there are no cross-terms involving $u$), so
\begin{align}
    \na^{[r} w^{u]} &= g^{ra} g^{uu} \pa_{[a} w_{u]} \nn 
    &= \f12 g^{uu} g^{ra} ( \pa_a w_u - \pa_u w_a).
\end{align}
$g^{ar} \pa_a w_u = g^{rr} \pa_r w_u$, since $\pa_t w_u = \pa_\phi w_u = 0$. Additionally,
\begin{align}
    g^{ra} \pa_u w_a &= g^{ra} \pa_u (g_{ra} w^r) \nn 
    &= g^{ar} g_{ra} \pa_u w^r + w^r g^{ra} \pa_u g_{ar} \nn 
    &= \pa_u w^r + w^r g^{ra} \pa_u g_{ra}.
\end{align}
We can evaluate $g^{ra} \pa_u g_{ra}$ as follows,
\begin{align}
    g^{r a} \pa_u g_{r a} &= (\bar g^{r a} - h^{ra}) \pa_u (\bar g_{r a} + h_{r a}) \nn 
    &= \bar g^{ra} \pa_u \bar g_{r a} - h^{r a} \pa_u \bar g_{ra} + \bar g^{ra} \pa_u h_{r a} - h^{ra} \pa_u h_{ra} \nn 
    &= \bar g^{rr} \pa_u \bar g_{rr} - h^{rr} \pa_u \bar g_{rr} + \bar g^{rr} \pa_u h_{rr} - h^{ra} \pa_u h_{ra}.
\end{align}
Since $h^{a b}$ can be raised and lowered with either metric, and since $h^{rr} = H k^r k^r = H$, the middle two terms simplify to
\begin{align}
    -h^{rr} \pa_u \bar g_{rr} + \bar g^{rr} \pa_u h_{rr} &= -h^{rr} \pa_u \bar g_{rr} + \bar g^{rr} \pa_u ( (\bar g_{rr})^2 h^{rr}) \nn 
    &= - H \pa_u \bar g_{rr} + (\bar g_{rr})^{-1} \pa_u ((\bar g_{rr})^2 H) \nn 
    &= \pa_u (\bar g_{rr} H) \nn 
    &= 0,
\end{align}
since $\bar g_{rr} H$ is a function of $r$ only. 

Since $H k_r = H \bar g_{rr}$, $\pa_u (H k_r) = 0$. Then $\pa_u h_{r a} = H k_r \pa_u k_a$. The final term becomes
\begin{align}
    -h^{r a} \pa_u h_{ra} &= - (H k^r k^a) H k_r \pa_u k_a \nn 
    &= -H^2 k^r k_r \left( \pa_u (k^a k_a) - k_a \pa_u k^a\right) \nn 
    &= 0,
\end{align}
since $k_ak^a = \pa_u k^a = 0$. Thus we are left with
\begin{align}
    g^{r a} \pa_u g_{a r} &= \pa_u \ln \bar g_{rr}
\end{align}
Thus,
\begin{align}
    \na^{[r} w^{u]} &= \f12 g^{uu} \left( g^{rr} \pa_r w_u - \pa_u w^r - w^r \pa_u \ln \bar g_{rr}\right) \nn 
    &= \f12 g^{uu} \left( g^{rr} \pa_r w_u - \bar g^{rr} \pa_u (\bar g_{rr} w^r)\right) \nn 
    \na^{[r}w^{u]}|_H &= - \f12 g^{uu} \bar g^{rr} \pa_u \bar g_{rr} w^r - \f12 g^{uu} \pa_u w^r,
\end{align}
using $g^{rr} = 0$ on the horizon on the last line. 

We also have under the restriction that $w^t = 0$,
\begin{align}
    3 w^{[r} \na^t \z^{u]} &= w^r \na^t \z^u - w^u \na^t \z^r .
\end{align}
On the horizon, $\z_a = \z_r \de_a^r$ (no components except $\z_r$), and $\z^a \na_a \z^b = \kappa \z^b$. Finally, $\z^a$ is Killing with $\z^t = 1$ (everywhere). This means that, on the horizon,
\begin{align}
    \na^t \z^r &= - \na^r \z^t \nn 
    &= - \f{1}{\z_r} \left( \z_r \na^r \z^t \right) \nn 
    &= -\f{1}{\z_r} \left( \z_a \na^a \z^t\right)\nn
    &= - \f{1}{\z_r} \kappa \z^t \nn 
    &= - \f{\kappa}{\z_r}.
\end{align}
Additionally, again using the fact that $\z$ is Killing, that $\z_u = 0$, and that there are no $u$ cross-terms in the metric,
\begin{align}
    \na^t \z^u &= g^{ta} g^{uu} \na_a \z_u \nn 
    &= g^{ta} g^{uu} \na_{[a} \z_{u]} \nn 
    &=-\f12 g^{ta} g^{uu} \pa_u \z_a.
\end{align}
Since only $\z_r \neq 0$ on the horizon (and since the direction $\pa_u$ is tangent to the horizon), $\pa_u \z_a = \pa_u \z_r \de_a^r$. We then have, 
\begin{align}
    \na^t \z^u &= - \f12 g^{tr} g^{uu} \pa_u \z_r \nn 
    &= -\f12 g^{uu} \left( \pa_u (g^{tr} \z_r) - \z_r \pa_u g^{tr}\right) \nn 
    &= - \f12 g^{uu} \left( \pa_u(g^{ta} \z_a) - \z_r \pa_u g^{tr}\right) \nn 
    &= -\f12 g^{uu} \left( \pa_u \z^t + \z_r \pa_u h^{tr}\right) \nn 
    &= -\f12 g^{uu} \z_r \pa_u h^{tr}.
\end{align}
$\pa_u k^a = 0$, so
\begin{align}
    \na^t \z^u &= -\f12 g^{uu} \z_r \pa_u (k^t k^r H) \nn 
    &= -\f12 g^{uu} \z_r k^t k^r \pa_u H \nn 
    &= -\f12 g^{uu} \z_r h^{tr} \f{\pa_u H}{H} \nn 
    &= -\f12 g^{uu} \z_r g^{tr} \f{\pa_u H}{H} \nn 
    &= \f12 g^{uu} \z^t \f{\pa_u H}{H} \nn 
    &= \f12 g^{uu} \f{\pa_u H}{H}.
\end{align}
We then have
\begin{align}
    3 w^{[r}\na^t \z^{u]} &= \f12 w^r g^{uu} \f{\pa_u H}{H} + \f{w^u \kappa}{\z_r}.
\end{align}

The combination $\na^{[r} w^{u]} - 3 w^{[r}\na^t \z^{u]}$ is then (evaluated on the horizon)
\begin{align}
    \na^{[r} w^{u]} - 3 w^{[r}\na^t \z^{u]} &= -\f12 w^r g^{uu} \bar g^{rr} \pa_u \bar g_{rr} - \f12 g^{uu} \pa_u w^r - \f12 w^r g^{uu} \f{\pa_u H}{H} + \f{w^u \kappa}{\z_r} \nn 
    &= -\f12 w^r g^{uu} \f{\pa_u (\bar g_{rr}H)}{\bar g_{rr} H } - \f12  g^{uu} \pa_u w^r - \f{w^u \kappa}{\z_r} \nn 
    &= -\f12  g^{uu} \pa_u w^r - \f{w^u \kappa}{\z_r},
\end{align}
since $\bar g_{rr}H$ is a function of $r$ only everywhere (and is, specifically, 1 on the horizon). 

We then have
\begin{align}
    \oint_H \bs k_{\bar \z}^{EH}[\lie_w g;g] &= \f14 \sqrt{-g}  \left. \left(-\f12  g^{uu} \pa_u w^r - \f{w^u \kappa}{\z_r} \right) \right|^{u=1}_{u=-1}.
\end{align}
$g^{uu} \propto 1-u^2$, so $g^{uu}|_{u=\pm 1 = 0}$. Assume $w^u|_{u=\pm 1} = 0$, for the same reasons as for the ABL metric form. Then, provided regularity of $\pa_u w^r$, we conclude 
\begin{align}
    \oint_H \bs k_{\bar \z}^{EH} [\lie_w g;g] = 0. \label{intHkEHzetaliewgKS}
\end{align}

One advantage working through this in detail is that the above argument generalizes to higher dimensions. Specifically, in the KS form of the GKAdS metric in coordinates $(t,r,y_\alpha ,\phi_i)$, consider the case
\begin{align}
    w = w^r(r,y_\alpha) + \sum_{\alpha=1}^{n-1} w^{y_\alpha}(r,y_\beta),
\end{align}
that is, $w$ has no components in the $t$ or $\phi_i$ directions and $w^a$ has no dependence on $t$ or $\phi_i$. Then we have 
\begin{align}
    \left.\na^{[r}w^{y_\alpha]} - 3 w^{[r} \na^t \z^{y_\alpha]}\right|_H &= -\f12 g^{y_\alpha y_\alpha} \pa_{y_\alpha} w^r - \f{w^{y_\alpha} \kappa}{\z_r}, \label{nablarwyterms}
\end{align}
for any $y_\alpha$, by following the same argument as the four-dimensional case with $u = y_1/a$ and noting that there are no cross-terms in the metric with each $y_\alpha$. We then have, generalizing \eqref{kEHliewintermsofdA},
\begin{align}
    \left.\bs k_{\bar \z}^{EH}[\lie_w g;g]\right|_H &= \left.-\f{dS}{8\pi} \f{\z_r}{\sqrt{-g}} \sum_{\alpha=1}^{n-1} \f{\pa}{\pa y_\alpha} \left( \sqrt{-g} \left( \na^{[r}w^{y_\alpha]} - 3 w^{[r} \na^t \z^{y_\alpha]}\right)\right)\right|_H.
\end{align}

In higher dimensions, $dS \z_r/\sqrt{-g} = d^{D-2}x \sqrt{\sigma} \z_r/\sqrt{-g} = d^{D-2}x$ from \eqref{gsigmazeta}, where $d^{D-2}x$ is the product of all coordinate differentials except $dt$ and $dr$. Consequently, 
\begin{align}
    \left. \bs k_{\bar \z}^{EH} [\lie_w g;g]\right|_H &= - \f{d^{D-2} x}{8\pi} \left.\sum_{\alpha=1}^{n-1} \f{\pa}{\pa y_\alpha} \left( \sqrt{-g} \left( \na^{[r}w^{y_\alpha]} - 3 w^{[r} \na^t \z^{y_\alpha]}\right)\right)\right|_H.
\end{align}
This suggests that we may have a similar argument for $\oint \bs k_{\bar \z}^{EH}[\lie_w g;g]$ vanishing under certain circumstances. The argument is complicated, however, by the fact that unlike $u = \cos \tht = y/a$, the limits on $y_\alpha$ are $a_i$ dependent. This complicates matters sufficiently that I leave the case where $w^{y_\alpha}$ are nonzero to future work. (One workaround is to use $\mu_i$ coordinates instead of $y_\alpha$, with one written in terms of the others to satisfy the constraint. The $\mu_i$ will then have limits which are independent of $a_i$. The disadvantage of this method is the presence of $\mu_i \mu_j$ cross-terms in the metric.)

In the specific case where $w = f(r) \pa_r$, then the right-hand side of \eqref{nablarwyterms} is zero. This implies then that 
\begin{align}
    \left.\bs k_{\bar \z}^{EH}[\lie_w g;g]\right|_H = 0.
\end{align}
As a consequence, we have
\begin{align}
    \oint_H \bs k_{\bar \z}^{EH}[\lie_w g;g] = 0
\end{align}
on the $(D-2)$-surface $H$ of constant $t,r = r_+^{(0)}$.  As in the four-dimensional case, this tells us that, in arbitrary dimension, for the GKAdS metrics we can apply $\oint_H \bs k_{\bar \z}^{EH}[\de g;g] = \kappa \de A/8\pi$ \emph{even if} the horizon radius changes between the two versions of the spacetime under comparison, because the infinitesimal coordinate transformation required to change the radial coordinate gives a zero contribution to the $\oint_H \bs k^{EH}_{\bar \z}[\de g;g]$ integral. 

\subsection{Using Infinitesimal Coordinate Transformations to Keep the Kerr--Schild Background Fixed} \label{keepingKSbackgroundfixed}

In Section \ref{constantmetricbackgroundvariablea}, I considered the case where the background metric was independent of rotation parameter $a$, in order to assure $\de \bar g_{ab} = 0$. One way to address the GKAdS case where $a$ is allowed to vary is to perform a coordinate transformation so that the background metric always depends on the initial rotational parameter $a^{(0)}$. 

Consider the KS form of the metric, and let $g^{(s)}_{ab}$ be a family of GKAdS spacetimes, parametrized by $s$, given by
\begin{align}
    g^{(s)}_{ab} = \bar g^{(s)}_{ab} + H^{(s)} k^{(s)}_a k^{(s)}_b.
\end{align}
The values for the background metric and $k^{(s)}_a$ are given by \eqref{KS4D}, with $a = a^{(s)}$, in general dependent on $s$. $H^{(s)} = 2 \mu^{(s)}(r)/U$. For convenience in calculation I will work in coordinates $(t,r,u,\phi)$, where $u = \cos \tht$. In this section, $l$ is kept fixed.

Let $\de \equiv \de s \f{d}{d s}$. Then we have 
\begin{align}
    \de g^{(s)}_{ab} \neq 0,
\end{align}
since $g^{(s)}_{ab}$ depends on $a^{(s)}$. Significantly this is true not just of the full metric but also the background metric:
\begin{align}
    \de \bar g^{(s)}_{ab} \neq 0.
\end{align}
I will now give an argument for constructing a coordinate system in which the background metric does not depend on $a$. I will refer to this as the background metric having ``constant components'' with respect to $s$, by which I mean not that the background metric components are constant \emph{in space} but that they do not depend, at a particular point, on $s$. 

Similarly to Section \ref{varyinghorizonradius}, I will now perform a coordinate transformation to coordinates $(\tilde x^{(s)})^a$, which depend on $s$. $\tilde g^{(s)}_{ab}$ will then be the metric tensor resulting from the coordinate transformation from $x^a$ coordinates to $(\tilde x^{(s)})^a$ coordinates, $\tilde{\bar g}^{(s)}_{ab}$ will be the \emph{background} metric written in $(\tilde x^{(s)})^a$ coordinates, and $\tilde k^{(s)}_a$ will be $k^{(s)}_a$ written in $(\tilde x^{(s)})^a$ coordinates. Specifically let us choose coordinates $(\tilde x^{(s)})^a$ such that the background metric has constant components:
\begin{align}
    \tilde{\bar g}_{ab}^{(s)} = \bar g_{ab}^{(0)}.
\end{align}
This also implies 
\begin{align}
    \de \tilde {\bar g}^{(s)}_{ab} = \de s \f{d \tilde {\bar g}^{(s)}_{ab}}{d s} = 0.
\end{align}
In Section \ref{varyinghorizonradius}, we wanted to find a transformation which allowed the horizon radius to be at the same location. Here, we wish to find a coordinate transformation so that the background metric is unchanged. Again we specify that $(\tilde x^{(0)})^a = x^a$.

How can we fix the coordinates $(\tilde x^{(s)})^a$ so that $\tilde {\bar g}^{(s)}_{ab} = \bar g^{(0)}_{ab}$? One way to do this is to note that the $x^a$ coordinates that automatically are associated with the background AdS metric can be anchored to the spherical polar AdS background coordinates, according to \eqref{sphericalspheroidal}, along with $t = t, \phi = \phi$. I will rearrange \eqref{sphericalspheroidal} somewhat, writing $RU = ru$ as well as 
\begin{align}
    R^2(1-U^2) &= \f{(r^2+a^2)(1-u^2)}{\Xi}.
\end{align}
To make things clearer, I will write out an $(s)$-dependence for the (non-tilde) coordinates $x^a$ explicitly, $(x^{(s)})^a$ here. The coordinates $t$ and $\phi$ are the same as those for the spherical polar case, and so we can just write $t^{(s)} = \tilde t^{(s)} =t, \phi^{(s)} = \tilde \phi^{(s)} = \phi$ without confusion. For $r^{(s)}$ and $u^{(s)}$, allowing $a = a^{(s)}$, we can write,
\begin{align}
    RU &= r^{(s)} u^{(s)} \nn 
    R^2(1-U^2) &=\f{((r^{(s)})^2+(a^{(s)})^2)(1-(u^{(s)})^2)}{1 - (a^{(s)})^2/l^2} \label{rsus}
\end{align}
The coordinate transformation required to get $\tilde {\bar g}_{ab}^{(s)} = \bar g^{(0)}_{ab}$ is then just to have $\tilde r^{(s)}, \tilde u^{(s)}$ satisfy \eqref{sphericalspheroidal} with $a = a^{(0)}$, or
\begin{align}
    RU &= \tilde r^{(s)} \tilde u^{(s)} \nn 
    R^2(1-U^2) &=\f{((\tilde r^{(s)})^2+(a^{(0)})^2)(1-(\tilde u^{(s)})^2)}{1 - (a^{(0)})^2/l^2} 
 \label{tilderu}
\end{align}
Combining \eqref{rsus} and \eqref{tilderu},
\begin{align}
    r^{(s)} u^{(s)} &= \tilde r^{(s)} \tilde u^{(s)} \nn 
    \f{((r^{(s)})^2+(a^{(s)})^2)(1-(u^{(s)})^2)}{1 - (a^{(s)})^2/l^2} &= \f{((\tilde r^{(s)})^2+(a^{(0)})^2)(1-(\tilde u^{(s)})^2)}{1 - (a^{(0)})^2/l^2}.
\end{align}
Now we can drop the explicit $s$ dependence on $x^a$, because we wish to interpret $\de x^a = 0$ (and so to view the $x^a$ coordinates as unchanging under the transformation). We then have
\begin{align}
    r u &= \tilde r^{(s)} \tilde u^{(s)} \nn 
    \f{(r^2+(a^{(s)})^2)(1-u^2)}{1 - (a^{(s)})^2/l^2} &= \f{((\tilde r^{(s)})^2+(a^{(0)})^2)(1-(\tilde u^{(s)})^2)}{1 - (a^{(0)})^2/l^2}. \label{rutildertildeu}
\end{align}
This can be solved explicitly for $\tilde r^{(s)}, \tilde u^{(s)}$, but involves radicals. We can solve for $d \tilde r^{(s)}/ds, d \tilde u^{(s)}/ds$ directly and find
\begin{align}
    \f{d \tilde r^{(s)}}{ds} &= \f{\tilde r^{(s)} a^{(s)} (1-u^2) (1+r^2/l^2) \Xi^{(0)}}{(\Xi^{(s)})^2 \left( (\tilde r^{(s)})^2 + (a^{(s)})^2 (\tilde u^{(s)})^2\right)} \f{d a^{(s)}}{ds} \nn 
    \f{d \tilde u^{(s)}}{ds} &= -\f{d \tilde r^{(s)}}{d s} \f{\tilde u^{(s)}}{\tilde r^{(s)}}. \label{rutildertildeu2}
\end{align}
$\Xi^{(s)} = 1 - (a^{(s)})^2/l^2$. For sufficiently small $s$, then, we can write $\tilde r^{(\de s)}$, for infinitesimal $\de s$ (neglecting terms quadratic in $\de s$) as
\begin{align}
    \tilde r^{(\de s)} &= \tilde r^{(0)} + \left.\f{d \tilde r^{(s)}}{ds}\right|_{s =0} \de s \nn
    &= r + \f{a^{(0)} (1-u^2)(1+r^2/l^2)}{ \Xi^{(0)} ( r^2+ (a^{(0)})^2 u^2)} \f{d a^{(s)}}{ds} r \de s.
\end{align}
Similarly for $\tilde u^{(\de s)}$. Note again $\tilde t^{(s)} = t, \tilde \phi^{(s)} = \phi$. We can then write $(\tilde x^{(\de s)})^a - x^a$ as
\begin{align}
    (\tilde x^{(\de s)})^a - x^a &= w^a \de s, \label{tildexdeltas}
\end{align}
expressed in $x^a$ coordinates, with
\begin{align}
    w^a &=  \f{a^{(0)} (1-u^2)(1+r^2/l^2)}{ \Xi^{(0)} ( r^2+ (a^{(0)})^2 u^2)} \f{d a^{(s)}}{ds}\left( r \de^a_r - u \de^a_u\right).
\end{align}
Of course we expect $(\tilde x^{(s)})^a - x^a = w^a s + \mc O(s^2)$ for a more general value of $s$.

We repeat the definitions for \eqref{definitionsofdeltag} and \eqref{tensorvariations} and recover \eqref{delta2gab} and \eqref{delta2T}.

Of particular interest is the tensor $\bar g_{ab}$. We have $\tilde{\bar g}_{ab}^{(s)} = \bar g^{(0)}_{ab}$, so that we expect $\de_1 \bar g_{ab} = 0$. We also have $\de_2 \bar g_{ab} = \de s \lie_w \bar g_{ab}^{(0)}$. As a check, to confirm $\de_1 \bar g_{ab} = 0$ we can evaluate
\begin{align}
    0 &= \de_1 \bar g_{ab} \nn 
    &= \de \bar g_{ab} - \de_2 \bar g_{ab},
\end{align}
or
\begin{align}
    \left.\f{d \bar g^{(s)}_{ab}}{ds}\right|_{s=0} - \lie_w \bar g^{(0)}_{ab} = 0. \label{dbargdsliewbarg}
\end{align}
I did indeed confirm this using \emph{GRTensorIII}. 

I will now revisit the form $\bs I_{\bar \z}$ associated with the conserved charge. As usual $\bar \z$ is the Killing vector tangent to the null generators associated with the unvaried horizon, which is here $\bar \z = \pa_t + \Om^{(0)} \pa_\phi, \Om^{(0)} = a^{(0)} (1 + r_+^{(0)}/l^2)/((r_+^{(0)})^2 + (a^{(0)})^2)$. Because the $t,\phi$ coordinates are unchanged when going from $x^a$ to $(\tilde x^{(s)})^a$ coordinates, and because $\bar \z^a$ has components which are independent of $s$, the coordinate expression for $\tilde {\bar \z}^a = \bar \z^a$, and is independent of $s$. 

Because $\de_1 \bar g_{ab} = 0$, we expect $\de_1 \bs I_{\bar \z} = \bs k^{EH}_{\bar \z}[\de_1 g;g]$ from \eqref{varyingkcheck} (verified in Section \ref{varyingk}). ($\bs k^{EH}_{\bar \z}[\de_1 g;g]$ is evaluated at $s = 0$.) I also verified directly that $\de_1 \bs I_{\bar \z} = \bs k_{\bar \z}^{EH}[\de_1 g;g]$, by checking
\begin{align}
    \de \bs I_{\bar \z} - \de s \lie_w \bs I_{\bar \z} = \bs k_{\bar \z}^{EH}[\de g;g] - \de s \bs k^{EH}_{\bar \z}[\lie_w g;g],
\end{align}
or, more specifically, that
\begin{align}
    \left.\f{d}{ds} \bs I_{\bar \z}^{(s)}\right|_{s=0} - \lie_w \bs I_{\bar \z}^{(0)} &= \bs k_{\bar \z}^{EH} \left[ \left.\f{d g^{(s)}}{ds}\right|_{s=0},g\right] - \bs k^{EH}_{\bar \z}[\lie_w g;g],
\end{align}
using \emph{GRTensorIII}. 

We then also have
\begin{align}
    \oint_H \de_1 \bs I_{\bar \z} &= \oint_H \bs k^{EH}_{\bar \z} [\de_1 g;g].
\end{align}
Is the right-hand side equal to $\kappa \de A/8\pi$? It turns out yes. I verified this directly using \emph{GRTensorIII}. To explain why, here I use the results of Section \ref{varyinghorizonradius}, as follows. Expand the right-hand side as
\begin{align}
    \oint_H \bs k^{EH}_{\bar \z}[\de_1 g;g] &= \oint_H \bs k^{EH}_{\bar \z}[\de g;g] - \oint_H \bs k^{EH}_{\bar \z}[\de_2 g;g] \nn 
    &= \oint_H \bs k^{EH}_{\bar \z} [\de g;g] - \de s \oint_H \bs k^{EH}_{\bar \z}[\lie_w g;g].
\end{align}
$\oint_H \bs k^{EH}_{\bar \z}[\de g;g]$ is just $\kappa \de A/8\pi$, as we verified directly in Section \ref{checkinghorizonareavariationrule} and explained in Section \ref{varyinghorizonradius}. From Section \ref{varyinghorizonradiusKSMetricForm}, specifically \eqref{intHkEHzetaliewgKS}, $\oint_H \bs k_{\bar \z}^{EH}[\lie_w g;g] = 0$. (This follows from the fact that $w$ has the form where $w^a = w^r(r,u)\pa_r + w^u(r,u)\pa_u$ and that all components of $w$ vanish at $u = \pm1$.) Thus we conclude 
\begin{align}
    \oint_H \de_1 \bs I_{\bar \z} &= \oint_H \bs k_{\bar \z}^{EH}[\de_1 g;g] \nn 
    &= \f{\kappa \de A}{8\pi}.
\end{align}
I also verified this directly.

We also have,
\begin{align}
    \de_2 \bs I_{\bar \z} &= \lie_w \bs I_{\bar \z} \nn 
    &= w \cdot d \bs I_{\bar \z} + d (w \cdot \bs I_{\bar \z}). 
\end{align}
The integral of any exact form over a surface without a boundary is zero \cite{Lee}. Since $H$ has no boundary, $\oint_H d (w \cdot \bs I_{\bar \z}) = 0$. For $d \bs I_{\bar \z}$ we have \eqref{dIchi}. We then have
\begin{align}
    \oint_H \de_2 \bs I_{\bar \z} &=-\f{1}{8\pi} \oint_H w^a (G^b_c - \bar G^b_c) \bar \z^c \bs \ep_{b a a_1 \ldots a_{D-2}}. \label{de2Iexp}
\end{align}
Using $\de\bs I_{\bar \z} = \de_1 \bs I_{\bar \z} + \de_2 \bs I_{\bar \z}$, 
\begin{align}
    \oint_H \de \bs I_{\bar \z} &= \f{\kappa \de A}{8\pi} -\f{1}{8\pi} \oint_H w^a (G^b_c - \bar G^b_c) \bar \z^c \bs \ep_{b a a_1 \ldots a_{D-2}},
\end{align}
and, in particular, if $G^a_b - \bar G^a_b = 0$, which is the case in vacuum,
\begin{align}
    \oint_H \de \bs I_{\bar \z} &= \f{\kappa \de A}{8\pi}.
\end{align}
This tells us that, in vacuum, we can reliably use $\bs I_\chi$ to give conserved charges whose variations recover $\kappa \de A/8\pi$. Conceptually, we can think of the reason that the variation law using $\bs I_\chi$ might fail away from vacuum as resulting from $d \bs I_{\chi} \neq 0$ (in particular, $d \bs I_{\bar \z} \neq 0$), which means that the value of $\oint \bs I_\chi$ is sensitive to integration surface. The expression \eqref{de2Iexp} is a consequence of the fact that the infinitesimal coordinate transformation means that the varied value of $\bs I_\chi$ is evaluated at different points in the manifold (both represented by the same coordinates $t = const., r = r_+$) depending on whether the $\de$ or $\de_1$ variation is used. 

So far I have focused on performing variations in coordinates which are asymptotically nonrotating. It is useful to examine the case where the coordinates are asymptotically rotating: that is to say, the BL coordinates and the accompanying/related Kerr--Schild form \eqref{BLKS4D}. Here I will demonstrate why, if $a$ is allowed to vary, $\oint_H \de \bs I_{\bar \z} \neq \kappa \de A/8\pi$ in this case even in vacuum, again using the infinitesimal coordinate transformation argument.

\subsection{Asymptotically Rotating Frame} \label{asymptoticallyrotatingframe}

In Section \ref{checkinghorizonareavariationrule}, I checked that $\oint_H \bs k^{EH}_{\hat \z} = \kappa \de A/8\pi$ holds in the BL and BLKS coordinates (including the Kerr--Schild form thereof). Here $\hat \z$ is the vector which has constant contravariant components in the BL and BLKS coordinates. The argument given before is that we expect that $\de \bs I_{\chi} = \bs k^{EH}_{\chi}$ if the background metric and the vector $\chi$ are both unchanging. 

Consider the case where we have Kerr--Newman--AdS in the BLKS form (\eqref{BLKS4D} with $m \to m-Q^2/2r$, and also using $u = \cos \tht$ as a coordinate instead of $\tht$) so again we can apply the infinitesimal coordinate transformation method. Let $g^{(s)}_{ab}$ and $\bar g^{(s)}_{ab}$ be the full and background metric of the form \eqref{BLKS4D} with $m = m^{(s)}, a = a^{(s)}, Q = Q^{(s)}$, and $l$ fixed as in Section \ref{keepingKSbackgroundfixed}. Let $\tilde g^{(s)}_{ab}, \tilde{\bar g}^{(s)}_{ab}$ be the full and background metric under a coordinate transformation from $x^a$ to $(\tilde x^{(s)})^a$, and, again as in Section \ref{keepingKSbackgroundfixed}, we construct the new coordinates so that $\tilde{\bar g}^{(s)}_{ab} = \bar g^{(0)}_{ab}$, so that $\de \tilde{\bar g}^{(s)}_{ab} = 0$ where $\de = \de s \f{d}{ds}$. As in \eqref{tildexdeltas}, we can write $(\tilde x^{(\de s)})^a - x^a = w^a \de s$. The $x^a$ coordinates are $(t,r,u,\breve \phi)$.

Let $\hat \z$ be given by
\begin{align}
    \hat \z &= \bt + \om^{(0)} \eta \nn 
    &= \f{\pa}{\pa t}+ \om^{(0)} \f{\pa}{\pa \breve \phi},
\end{align}
where 
\begin{align}
    \om^{(0)} &= \f{a^{(0)} \Xi^{(0)}}{(r_+^{(0)})^2 + (a^{(0)})^2}
\end{align}
is the angular velocity (in these coordinates) for the $s = 0$ spacetime. 

The relationships between $r, u$ and $\tilde r^{(s)}, \tilde u^{(s)}$ are automatically the same as those given by \eqref{rutildertildeu} and \eqref{rutildertildeu2}. The crucial difference from the asymptotically static KS coordinates is that now the $\breve \phi$ coordinate changes. Because we had $\tilde t^{(s)} = t, \tilde \phi^{(s)} = \phi$ in Section \ref{keepingKSbackgroundfixed}, and because $\breve \phi = \phi - a l^{-2} t$ from \eqref{BLKS4D}, we can write 
\begin{align}
    \breve \phi &= \phi - a^{(s)}l^{-2} t.
\end{align}
Meanwhile, we wish to have $\tilde{\breve \phi}^{(s)}$ defined in such a way so that $\tilde{\bar g}_{ab}$ has constant components. This requires that $\tilde{\breve \phi}^{(s)}$ be coordinate transformed from $\phi,t$ as from $a^{(0)}$, rather than $a^{(s)}$, giving
\begin{align}
    \tilde{\breve \phi}^{(s)} &= \phi - a^{(0)}l^{-2} t.
\end{align}
Combining the two expressions we have
\begin{align}
    \tilde{\breve \phi}^{(s)} &= \breve \phi + (a^{(s)} - a^{(0)}) l^{-2} t,
\end{align}
or
\begin{align}
    \f{d \tilde{\breve \phi}^{(s)}}{d s} &= \f{t}{l^2} \f{d a^{(s)}}{ds}.
\end{align}
If the argument is unclear to the reader I will emphasize that the argument here is the same as the one for $\tilde r^{(s)},\tilde u^{(s)}$ in the previous section, with the new feature being that $\breve \phi$ and thus $\tilde{\breve \phi}$ depends in an $a$-related way on the $t,\phi$ appearing in \eqref{AdS4Dspherical}. 

Now again writing \eqref{tildexdeltas} we have 
\begin{align}
    w^a &= \left[\f{a^{(0)} (1-u^2)(1+r^2/l^2)}{ \Xi^{(0)} ( r^2+ (a^{(0)})^2 u^2)} \left( r \de^a_r - u \de^a_u\right) + \f{t}{l^2} \de^a_{\breve \phi} \right]\f{d a^{(s)}}{ds}.
\end{align}

Another wrinkle to consider before proceeding is that, unlike in the case of Section \ref{keepingKSbackgroundfixed}, where $\lie_w \bar \z = 0$, we have $\lie_w \hat \z \neq 0$---specifically,
\begin{align}
    \lie_w \hat \z^a &= -\f{1}{l^2} \f{d a^{(s)}}{ds} \de^a_{\breve \phi},
\end{align}
or $\lie_w \hat \z = - l^{-2} (d a^{(s)}/ds) \eta$. This means that if we treat $\hat \z^a$ as a vector under a coordinate transformation in the usual way, then its components $(\tilde{\hat \z}^{(s)})^a$ in the $\tilde x^{(s)}$ coordinate system will vary under the coordinate transformation. As established, it is important for the equation $\oint_H \bs k_{\hat \z}^{EH}[\de g;g] = \ka \de A/8\pi$ that the components of $\hat \z$ do not change. We must be careful. I will still maintain the definitions for $\de, \de_1, \de_2$ as given by \eqref{definitionsofdeltag} and \eqref{tensorvariations}, recovering \eqref{delta2gab} and \eqref{delta2T}.

Let $(\tilde{\hat \z}^{(s)})^a$ be the components of $\hat \z^a$ in the $(\ti x^{(s)})^a$ coordinates.  Its coordinate expression is
\begin{align}
    (\tilde {\hat \z}^{(s)})^a &= \hat \z^b \f{\pa (\tilde x^{(s)})^a}{\pa x^b} \nn 
    &= \f{\pa (\tilde x^{(s)})^a}{\pa t} + \om^{(0)} \f{\pa (\tilde x^{(s)})^a}{\pa \breve \phi} \nn 
    &= \de^a_t + \left( \om^{(0)} + \f{a^{(s)}-a^{(0)}}{l^2} \right) \de^a_{\tilde{\breve \phi}^{(s)}}. \label{tildehatzeta}
\end{align}

Now we wish to define an alternate vector which is equal to $\hat \z^a$ when $s = 0$ and has constant components in the $(\tilde x^{(s)})^a$ coordinates. I will call this vector $\bar \z^{(s)}$ and want to construct it so that
\begin{align}
    (\tilde{\bar \z}^{(s)})^t &= 1 \nn 
    (\tilde{\bar \z}^{(s)})^{\tilde{\breve \phi}^{(s)}} &= \om ^{(0)},
\end{align}
or, assigning $(\tilde x^{(s)})^1 = t, (\tilde x^{(s)})^4 = \tilde{\breve \phi}^{(s)}$, 
\begin{align}
    (\tilde{\bar \z}^{(s)})^a &= \de^a_1 + \om^{(0)} \de^a_4
\end{align}
so that
\begin{align}
    \f{d \tilde{\bar \z}^{(s)}}{ds} &= 0.
\end{align}
The reason I use the name $\bar \z$ for this vector is because it is the one that corresponds to the vector $\bar \z$ with unvarying components in the ABL coordinate system.

The components of $\bar \z^{(s)}$ in the $x^a$ coordinates are
\begin{align}
    (\bar \z^{(s)})^a &= (\tilde{\bar \z}^{(s)})^b \f{\pa x^a}{\pa (\tilde x^{(s)})^b} \nn 
    &= \f{\pa x^a}{\pa t} + \om^{(0)} \f{\pa x^a}{\pa \tilde{\breve \phi}^{(s)}} \nn 
    &= \de^a_t + \left( \om^{(0)} - \f{a^{(s)}-a^{(0)}}{l^2}\right) \de^a_{\breve \phi}.
\end{align}
Of course this is equal to
\begin{align}
    \bar \z^{(s)} &= \bt + \left( \om^{(0)} - \f{a^{(s)} - a^{(0)}}{l^2}\right) \eta.
\end{align}
Using $\xi^{(s)} = \bt - l^{-2} a^{(s)} \eta$ from the usual $\bt = \xi + l^{-2} a \eta$ relationship, we get $\bar \z^{(s)} = \xi^{(s)} + (\om^{(0)} + l^{-2} a^{(0)}) \eta$ or
\begin{align}
    \bar \z^{(s)} &= \xi^{(s)} + \Om^{(0)} \eta,
\end{align}
which corresponds to the usual definition of $\bar \z$, this time where we think of the vector $\xi^{(s)}$ as changing in its contravariant components as we change spacetimes. The relationship between $\bar \z^{(s)}$ and $\hat \z$ is
\begin{align}
    \bar \z^{(s)} &= \hat \z - \f{a^{(s)}-a^{(0)}}{l^2} \eta,
\end{align}
giving $\bar \z^{(0)} = \hat \z$ as desired. Again, to make absolutely clear, by construction, $\hat \z^a$ has constant components in $x^a$ coordinates and $\tilde{\bar \z}^a$ has constant components in $(\tilde x^{(s)})^a$ coordinates.

Because $\tilde{\bar \z}^{(s)}$ has constant contravariant components we can drop the superscript. Since $d \tilde{\bar g}^{(s)}_{ab}/ds = 0$ and $d \tilde{\bar \z}^{(s)}/ds = 0$, we expect, from Section \ref{varyingk},
\begin{align}
    \left.\f{d}{ds} \tilde{\bs I}^{(s)}_{\tilde{\bar \z}}\right|_{s=0} &= \tilde{\bs k}^{EH}_{\tilde{\bar \z}}\left[ \left.\f{d \tilde g^{(s)}}{ds}\right|_{s=0};g^{(0)}\right],
\end{align}
noting that $\tilde g^{(0)}_{ab} = g^{(0)}_{ab}$. This means we expect,
\begin{align}
    \de_1 \bs I_{\bar \z} &= \bs k^{EH}_{\bar \z} [\de_1 g;g], \label{de1Ikde1g}
\end{align}
or, writing each side in terms of $\de$ and $\lie_w$,
\begin{align}
    \left.\f{d}{ds} \bs I^{(s)}_{\bar \z^{(s)}}\right|_{s=0} - \lie_w \bs I^{(0)}_{\bar \z^{(0)}} &= \bs k^{EH}_{\bar \z}\left[\left. \f{d g^{(s)}}{ds}\right|_{s=0};g^{(0)}\right] -  \bs k^{EH}_{\bar \z^{(0)}}[\lie_w g^{(0)};g^{(0)}]. \label{de1Ikde1gexpanded}
\end{align}
To be clear, here I mean, since $\de \bar \z^a \neq 0$ in $x^a$ coordinates,
\begin{align}
    \bs k^{EH}_{\bar \z}[\de g;g] &= -\de \bs K^K_{\bar \z} + \bs K^K_{\de \bar \z} - \bar \z \cdot \bs \Th^{EH}[\de g;g] \nn 
    \bs k^{EH}_{\bar \z} [\lie_w g;g] &= - \lie_w \bs K^K_{\bar \z} + \bs K^K_{\lie_w \bar \z} - \bar \z \cdot \bs \Th^{EH}[\lie_w g;g].
\end{align}
These can be somewhat simplified by noting that evaluation happens at $s = 0$, where $\hat \z = \bar \z$; in particular, $\bs k^{EH}_{\hat \z}[\lie_w g;g] = \bs k^{EH}_{\bar \z}[\lie_w g;g]$. 

First of all, I verified, using \emph{GRTensorIII}, that 
\begin{align}
    \oint_H \bs k^{EH}_{\bar \z}[\de_1 g;g] = \f{\kappa \de A}{8\pi},
\end{align}
which does hold in this case, as expected, since $\tilde{\bar \z}$ has constant contravariant components. I then verified \eqref{de1Ikde1gexpanded}, showing that indeed $\de_1 \bs I_{\bar \z} = \bs k^{EH}_{\bar \z}[\de_1 g;g]$. This finally tells us that we have
\begin{align}
    \oint_H \de_1 \bs I_{\bar \z} &= \f{\kappa \de A}{8\pi}.
\end{align}

What happens if we instead try to evaluate $\de_1 \bs I_{\hat \z}$, where $\hat \z$ is the vector with constant components in the $x^a$ coordinates but not $(\tilde x^{(s)})^a$ coordinates? We find immediately,
\begin{align}
    \de_1 \bs I_{\hat \z} &= \de_1 \bs I_{\bar \z} + \bs I_{\de_1 \hat \z},
\end{align}
since $\de_1 \bar \z^a = \de s (d \tilde{\bar \z}^a/ds) = 0$. From \eqref{tildehatzeta}, 
\begin{align}
    \f{d}{ds} \left( \tilde{\hat \z}^{(s)}\right)^a &= \f{1}{l^2} \f{d a^{(s)}}{ds} \de^a_{\tilde{\breve \phi}^{(s)}},
\end{align}
or
\begin{align}
    \de_1 \hat \z &= \f{1}{l^2} \de a \eta.
\end{align}
Consequently we have
\begin{align}
    \de_1 \bs I_{\hat \z} &= \de_1 \bs I_{\bar \z} + \f{1}{l^2} \de a \bs I_\eta,
\end{align}
giving
\begin{align}
    \oint_H \de_1 \bs I_{\hat \z} &= \oint_H\de_1 \bs I_{\bar \z} + \f{1}{l^2} \de a \oint_H \bs I_\eta \nn 
    &= \f{\kappa \de A}{8\pi} - \f{1}{l^2} \de a \mc J_K
\end{align}
where $\mc J_K$ is the Komar angular momentum evaluated at the horizon. Similarly, if we try to evaluate $\de \bs I_{\hat \z}$, we can use $\de \bs I_{\hat \z} = \de_1\bs I_{\hat \z} + \de_2 \bs I_{\hat \z} = \de \bs I_{\hat \z} + \lie_w \bs I_{\hat \z}$. The result is
\begin{align}
    \de \bs I_{\hat \z} &= \bs k^{EH}_{\bar \z}[\de_1 g,g]+ \f{1}{l^2} \de a \bs I_\eta  + \de s \lie_w \bs I_{\hat \z} ,
\end{align}
where all quantities are evaluated at $s = 0$. Integrating over the horizon gives
\begin{align}
    \oint_H \de \bs I_{\hat \z} &= \f{\kappa \de A}{8\pi} - \f{\de a}{l^2} \mc J_K + \de s \oint_H \lie_w \bs I_{\hat \z}.
\end{align}
As in Section \ref{keepingKSbackgroundfixed}, we can use Cartan's identity to expand the last term,
\begin{align}
    \oint_H \lie_w \bs I_{\hat \z} &= \oint_H \left( d (w \cdot \bs I_{\hat \z} ) + w \cdot (d \bs I_{\hat \z})\right),
\end{align}
where again $\oint_H d (w \cdot \bs I_{\hat \z}) = 0$ and $w \cdot (d \bs I_{\hat \z}) = 0$ if $G^a_b - \bar G^a_b = 0$, since $d \bs I_{\hat \z}$ is proportional to $G^a_b - \bar G^a_b$. Assuming for the moment this is the case---that $G^a_b - \bar G^a_b = 0$, then we find
\begin{align}
    \oint_H \de \bs I_{\hat \z} &= \f{\kappa \de A}{8\pi} - \f{\de a}{l^2} \mc J_K.
\end{align}
Consider specifically Kerr--AdS vacuum. Then $\mc J_K = \mc J$. $\bs I_{\hat \z} = \bs I_{\beta} + \om^{(0)} \bs I_{\eta}$, and so
\begin{align}
    \oint_H \de \bs I_{\hat \z} &= \de \oint_H \bs I_\bt + \om^{(0)} \de \oint_H \bs I_\eta \nn 
    &= \de \mc F - \om^{(0)} \de \mc J,
\end{align}
giving, finally,
\begin{align}
    \de \mc F - \om \de J = \f{\kappa \de A}{8\pi} - \f{\de a}{l^2} \mc J.
\end{align}
This is of course the result already obtained through other methods. This means that an extra term arrives in the variation law if we try to base our conserved charge on a fixed $\bt, \eta$ rather than a fixed $\xi, \eta$.

What is the takeaway from this section? $\hat \z = \bt + \om^{(0)} \eta$ is the Killing vector tangent to the null generators in the unperturbed spacetime whose components are constant in a frame in which $\bt, \eta$ have constant components. It turns out that $\oint_H \bs k_{\hat \z}^{EH}[\de g;g] = \kappa \de A/8\pi$ still holds if we use coordinates $x^a$ such that $\hat \z^a$ has the same contravariant components throughout the variation, which is to say a set of asymptotically rotating (BL or BLKS) coordinates. Nothing breaks down about the $\oint_H \bs k^{EH}_{\hat \z}[\de g;g] = \kappa \de A/8\pi$ formula. However, the other part of the equivalences, namely the requirement that the variation in $\bs I_{\hat \z}$ produce $\bs k_{\hat \z}^{EH}[\de g;g]$, fails, even in vacuum. This is because even though we can perform an infinitesimal coordinate transformation to $(\tilde x^{(s)})^a$ coordinates so that $\bar g_{ab}^{(s)}$ is unchanging, doing so automatically causes $(\tilde{\hat \z}^{(s)})^a$ to have varying contravariant components. 

This all comes down to the following: when comparing spacetimes with different $a^{(s)}$ parameters, we can, through a judicious choice of coordinates, either find coordinates where $\hat \z$ has constant contravariant components \emph{or} where the background metric $\bar g_{ab}$ has constant components, but not both. Because the statement $\de \bs I_{\chi} = \bs k^{EH}_{\chi}[\de g;g]$ requires that $\de \bar g_{ab} = 0$ \emph{and} that $\de \chi^a = 0$, we cannot use $\hat \z$ as the vector on which to base a variation---or, in other words, we cannot use $\bt$ as one of our ``fixed'' vectors, if we are varying $a$.

This gives a slightly different perspective on the failure of the variation law. In this perspective, it is not so much that using an asymptotically rotating frame is itself a problem, because we can still use $\oint_H \bs k^{EH}_{\hat \z}[\de g;g] = \kappa \de A/8\pi$ in an asymptotically rotating frame. However, the conserved charges resulting from integrating through solution space do not work when they are based on components which are fixed in the asymptotically rotating frame, because the vector $\bt = \xi + l^{-2} a^{(s)} \eta$ points in different directions, with respect to the background metric $\bar g_{ab}$, as a function of $a^{(s)}$. If we choose coordinates where $\bt$ has constant components, the background metric changes its components, and if we choose coordinates where the background metric components do not change as we compare solutions, then $\bt$ changes its components. By contrast, $\xi$ (and vectors based around $\xi$, such as $\bar \z = \xi + \Om^{(0)} \eta$) can have fixed contravariant components in coordinates in which the metric does not depend on $s$. This means that we can find coordinates such that $\de \bs I_{\xi} = \bs k^{EH}_\xi[\de g;g]$, but not $\de \bs I_{\bt} = \bs k^{EH}_{\bt}[\de g;g]$. This addresses the question of why it is necessary to use $\xi$ rather than $\bt$ in the construction of conserved charges.

Next I turn to a breakdown of the conserved charge in arbitrary dimension.

\section{Breakdown of Conserved Charge in Arbitrary Dimension} \label{chargebreakdown}

At this point I will calculate the terms associated with ${\bs i}_\chi^{ab}$ for a surface of constant $r$ given an arbitrary $\mu(r)$. While $\oint_S \bs I_\chi$ is only a conserved charge associated with the Hamiltonian when the Einstein--Hilbert equations of motion are satisfied, that is to say, in vacuum, it is interesting to find the general solution for arbitrary $\mu(r)$, so that we can see what results are specifically due to vacuum and which remain true for generic $\mu(r)$.

It will be useful to break ${\bs{i}}_\chi^{ab}$ down into the Komar terms and the non-Komar terms. Throughout this section I use the KS form of the metric, with coordinates $(t,r,\phi_i)$, and, depending on convenience, either $y_\alpha$ or $\mu_i$.

The simplest calculations will be ones taken at large radius $r$, which are similar to those performed in \cite{DeruelleKatz06}. In fact, $\bs I_\chi$ is very close in form and definition to the KBL superpotential used in \cite{DeruelleKatz06}, as discussed in Section \ref{KBLSection}. The key difference is that the KBL superpotential is generally considered only at $r \to \infty$.  

Then I will write down the result of $\oint \bs I_{\chi}$ on a surface of constant $t,r$. I will break this down into the ``Komar only'' and ``non-Komar only'' versions, and also include the results for specific values of $\chi$. There are some interesting results along the way. 

I begin by breaking down ${\bs{i}}_\chi^{a b}$ into two terms, one associated with the Komar terms, which I will call $({\bs{i}}^K_\chi)^{a b}$, and one associated with $\bs \tht$---throughout this section I will assume that $\bs \tht = \bs\tht^{EH}$ and ignore any other possible contributions---which I will call $({\bs{i}}^{\tht}_\chi)^{a b}$. The Komar contribution is $-1/16\pi$ times the difference of the covariant derivatives of $\chi$ in the full and background metric:
\begin{align}
    ({\bs{i}}^K_\chi)^{a b} &\equiv -\f{1}{16\pi} \left( \na^a \chi^b - \overline{\na^a \chi^b}\right) \nn 
    &= \f{1}{16\pi} \left( h^{d[a}\na_d \chi^{b]} - \chi^d \na^{[a} h^{b]}_d\right) \nn 
    &= \f{1}{16\pi} \left( h^{d[a}\bar \na_d \chi^{b]} - \chi^d \bar \na^{[a} h^{b]}_d\right). \label{iKchiab}
\end{align}
Then we have
\begin{align}
    -(\bs K_\chi^K - \overline{\bs K_\chi^K}) &=  *{\bs{i}}^K_\chi.
\end{align}
(The negative sign is because $\bs I_\chi = - (\bs K^K_\chi - \overline{\bs K^K_\chi}) - \chi \cdot \bs \tht$.) 

(While not mentioned in the thesis, I think it's worth pointing out here that, since $\lie_\chi k^a = 0$ for all of the Killing vectors we are interested in, such as linear combinations of $\xi$ and $\eta_i$, we have $k^d \bar \na_d \chi^b = \chi^d \bar \na_d k^b$. Then we can rewrite \eqref{iKchiab} as
\begin{align}
    (\bs i^K_\chi)^{ab} &= \f{1}{16\pi} \left( H k^d k^{[a} \bar \na_d \chi^{b]} - \chi^d \bar \na^{[a} h^{b]}_d\right) \nn 
    &= \f{1}{16\pi} \left( H \chi^d k^{[a} \bar \na_d k^{b]} - \chi^d \bar \na^{[a} h^{b]}_d\right) \nn 
    &= \f{1}{16\pi} \chi^d \left( H k^{[a} \bar \na_d k^{b]} - \bar \na^{[a} h^{b]}_d\right).
\end{align}
This form is nice because it shows the linearity of $\bs i_\chi^K$ with respect to $\chi$ very directly.) 

The non-Komar term can be defined as
\begin{align}
    ({\bs{i}}^{\tht}_\chi)^{a b} &\equiv \f{1}{16\pi} \chi^{[a}\na_d h^{b]d} \nn 
    &= \f{1}{16\pi} \chi^{[a} \bar \na_d h^{b]d} \nn 
    &= \f{1}{16\pi} \chi^{[a} V^{b]},
\end{align}
using $V^a = \na_b h^{a b} = \bar \na_b h^{a b}$ from \eqref{VEHdef} (dropping the EH superscript). This corresponds to the contribution from the $\bs \tht$ term:
\begin{align}
    -\chi \cdot \bs \tht &= *\bs{i}^{ \tht}_\chi.
\end{align}

The breakdown of ${\bs{i}}_\chi^{a b}$ into the Komar and $\bs \tht$ constituent elements is important, particularly because it is the Komar terms which contribute to the area and volume terms ($\kappa A$ and $\Lambda \mc V_{\xi,\mc B}$) in the Smarr relation on the horizon, as discussed in Section \ref{smarrrevisited}. The $\bs \tht$-related term fixes the value of $\mc E$, as well as the form of the Killing potential, as discussed in detail in Section \ref{relationshiptokillingpotential}. 

I will examine the conserved charges which are found by taking surfaces of constant $(t,r)$ in asymptotically nonrotating Kerr--Schild coordinates. We will thus be interested only in the ${\bs{i}}_\chi^{t r}$ (and $({\bs{i}}^K_\chi)^{t r}, ({\bs{i}}_\chi^{\tht})^{tr}$) components. This allows us to make a few observations immediately. 

\subsection{Non-Komar Term} \label{nonKomarterm}

If we consider the conserved charges associated with the rotational Killing vectors $\eta_i$, then we have
\begin{align}
    ({\bs{i}}^{\tht}_{\eta_i})^{t r} &= \f{1}{16\pi} \eta_i^{[t} V^{r]}.
\end{align}
Since $\eta_i^t = \eta_i^r = 0$, this is exactly zero. We can also use the fact that the background metric $\bar g_{a b}$ is diagonal to write
\begin{align}
    \overline{\na^t \eta_i^r} &= \overline{ g^{tt} g^{rr} \na_t (\eta_i)_r} \nn 
    &= \overline{ g^{tt} g^{rr} \pa_{[t}(\eta_i)_{r]}} \nn 
    &= 0,
\end{align}
since $\overline{(\eta_i)_t} = \overline{(\eta_i)_r} = 0$. The only remaining term corresponds to the Komar integrand for $\eta_i$ associated with the full metric.

Let 
\begin{align}
    \upsilon \equiv \pa_t + \sum_i c_i \eta_i
\end{align}
be a Killing vector, with as yet unspecified constants $c_i$. This includes three vectors we are interested in: $\upsilon = \xi = \pa_t$, $\upsilon = \beta = \xi + \sum_i a_i l^{-2} \eta_i$ and $\zeta = \xi + \sum_i \Omega_i \eta_i$. Because the only nonzero contribution to ${\bs{i}}_{\eta_i}^{t r}$ is through the Komar integrand for the full spacetime, the only dependence on $c_i$ in the expressions ${\bs{i}}^{t r}_\upsilon$ will come through the $({\bs{i}}^K_\upsilon)^{t r}$. 

For all $c_i$, $({\bs{i}}^{\tht}_\upsilon)^{tr}$ is
\begin{align}
    ({\bs{i}}^{\tht}_\upsilon)^{t r} &= \f{1}{16\pi} \upsilon^{[t} V^{r]} \nn 
    &= \f{1}{32\pi} \upsilon^t V^r \nn 
    &= \f{1}{32 \pi} V^r. \label{KthetaupsisVr}
\end{align}
$V^r$ is given by \eqref{VEHproptok} 
\begin{align}
    V^r &= \bar \na_b (H k^b) k^r \nn 
    &= \bar \na_b (H k^b) \nn 
    &= \f{1}{\sqrt{-\bar g}} \f{\pa}{\pa x^b} (\sqrt{-\bar g} H k^b).
\end{align}
$k^{y_\alpha} = 0$ for all $\alpha$ (or, using the $\mu_i$ coordinates, $k^{\mu_i} = 0$ for all $i$). Since $H$ and $k^b$ are independent of $t$ and $\phi_i$, the only nonzero derivative in the above sum is with respect to $r$, giving
\begin{align}
    V^r &= \f{1}{\sqrt{-\bar g}} \f{\pa}{\pa r} (\sqrt{-\bar g} H). \label{Vr}
\end{align}
Because the $\mu_i$ are functions of the $y_\alpha$ only, whether the above is calculated in $(t,r,\mu_i,\phi_i)$ (with the largest $\mu_i$ suppressed using the constraint) or $(t,r,y_\alpha,\phi_i)$ will not make a difference. Use the $\mu_i$ form and use \eqref{Gibbonssqrtg} for $\sqrt{-\bar g}$ (which is equal to $\sqrt{-g}$). Then using $H = 2\mu(r)/U$, $\sqrt{-\bar g} H = 2 r \mu(r) \prod_{i = 1}^{n-1+\ve} \mu_i/(\mu_{n}\prod_{j=1}^{n-1+\ve} \Xi_j)$ and so
\begin{align}
    \f{1}{\sqrt{-\bar g} H} \f{\pa}{\pa r} (\sqrt{-\bar g} H) &= \f{1}{\mu(r)r} \f{d}{dr} (\mu(r) r)\nn 
    &= \f{\mu'(r) r + \mu(r)}{\mu(r) r} \nn
    &= \f{1}{r} + \f{\mu'(r)}{\mu(r)}.
\end{align}
Of course if $\mu(r) = m$ (constant) this reduces to $1/r$. We then have
\begin{align}
    V^r &= \f{H}{\sqrt{-\bar g} H} \f{\pa}{\pa r} (\sqrt{-\bar g} H) \nn 
    &= \left(\f{1}{r} + \f{\mu'(r)}{\mu(r)}\right) H.
\end{align}
We further have
\begin{align}
    \sqrt{-\bar g} ({\bs{i}}^{\tht}_\ups)^{t r} &= \f{1}{32\pi} \sqrt{-\bar g} H \left(\f{1}{r} + \f{\mu'(r)}{\mu(r)}\right) \nn 
    &= \f{2( \mu(r) + r \mu'(r)) \prod_{i=1}^{n-1+\ve} \mu_i}{32\pi\mu_{n}\prod_{j = 1}^{n} \Xi_j}. \label{sqrtgmcKthetaups}
\end{align}
The combination of the $\mu_i$ terms appearing here is the one that appears in \eqref{sqrtgmuonly}, representing the metric of the unit $(D-2)$-sphere, and so integrating over the $\mu_i$ and $\phi_i$ gives the area $\mc A_{D-2}$ of the $(D-2)$-sphere (see \eqref{A2N} and \eqref{A2Nm1}). Consequently,
\begin{align}
    \int d \mu_1 \ldots d \mu_{n-1} d \phi_1 \ldots d \phi_{n-1+\ve} \sqrt{-\bar g} ({\bs{i}}^{\tht}_\ups)^{t r} &= \f{2 (\mu(r) + r \mu'(r)) \mc A_{D-2}}{32\pi \prod_{j=1}^{n} \Xi_j}. \label{ithetaupsilonintegral}
\end{align}
We conclude,
\begin{align}
    -\oint_C \upsilon \cdot \bs \tht &= \f{\mc A_{D-2}}{8\pi \prod_{j=1}^{n} \Xi_j} (\mu(r) + r \mu'(r)), \label{upsilonthetagenericmu}
\end{align}
where $C$ corresponds to a surface of constant ($t,r$). Note that if $\mu(r) = m$ (a constant) this reduces to
\begin{align}
    -\oint_C \upsilon \cdot \bs \tht &= \f{\mc A_{D-2} m}{8 \pi \prod_{j = 1}^{n} \Xi_j}.
\end{align}
(The extra factor of 2 appears because 
\begin{align}
    -(\upsilon \cdot \bs \tht)_{\mu_1 \ldots \mu_{n-1} \phi_1 \ldots \phi_{n-1+\ve}} &= ({\bs{i}}^{\tht}_\ups)^{t r} \bs \ep_{\mu_1 \ldots \mu_{n-1} \phi_1 \ldots \phi_{n-1+\ve} t r} + ({\bs{i}}^{\tht}_\ups)^{r t} \bs \ep_{\mu_1 \ldots \mu_{n-1} \phi_1 \ldots \phi_{n-1+\ve} r t} \nn
    &= 2 ({\bs{i}}^{\tht}_\ups)^{t r} \sqrt{-\bar g}.)
\end{align}

We will further consider the limit as $r \to \infty$. Assume that $\mu(r)$ can be Taylor expanded in powers of $1/r$ about $r \to \infty$, so takes the form 
\begin{align}
    \mu(r) = m_\infty + \sum_{j = 1}^\infty \f{d_j}{r^j} \label{murexpansion}
\end{align} 
for constants $d_j$. In this case, 
\begin{align}
    \lim_{r\to\infty} (\mu(r) + r \mu'(r)) &= \lim_{r\to\infty} \left( m_\infty + \sum_{j = 1}^\infty \left(\f{d_j}{r^j} - \f{j d_j}{r^{j}}\right)\right) \nn 
    &= m_\infty.
\end{align}

Consequently, from \eqref{sqrtgmcKthetaups},
\begin{align}
    \lim_{r\to\infty} \sqrt{-\bar g} (\bs{i}^\tht_\ups)^{tr} &= \f{m_\infty \prod_{i=1}^{n-1+\ve}\mu_i}{16\pi (\prod_{j=1}^n \Xi_j) \mu_n}. \label{Kthetalarger}
\end{align}
Letting $C_\infty$ be the surface $t$-constant with $r \to \infty$, then
\begin{align}
    -\oint_{C_\infty} \upsilon \cdot \bs \tht &= \f{\mc A_{D-2} m_\infty}{8\pi \prod_{j = 1}^{n} \Xi_j}. \label{upsilonthetalimit}
\end{align}
If $\mu(r)$ has a different functional form than \eqref{murexpansion}, then the above will not necessarily hold. If $\mu(r)$ has terms which diverge as $r \to \infty$, the integral will in general be divergent.

I will note before continuing that the form here for $\bs \tht$ is not coincidental. For the vacuum case, we require $R-\bar R = 0$ which means $\bar \na_a \bar \na_b h^{ab}=0$ from \eqref{RequalsRbarplusdoublederivatives}. This means that if $R - \bar R = 0$, $d (\ups \cdot \bs \tht) = 0$, so that $\oint_S \ups \cdot \bs \tht$ is independent of $S$ (provided it encloses the central singularity!). This is the case for Kerr--AdS, but also for (four-dimensional) Kerr--Newman--AdS, despite not being vacuum. 

We can also show that the form of $V^r$ is also determined by $R - \bar R$. We have, since $\bar \na_b h^{ab} = V^a$,
\begin{align}
    \bar \na_a V^a &= 0 \nn 
    \pa_a (\sqrt{-\bar g} V^a) &= 0.
\end{align}
Since $V^a \propto k^a$, $k^a$ has no $y_\alpha$ components, and $V^a$ and $\sqrt{-\bar g}$ have no $t$ or $\phi_i$ derivatives, this reduces to
\begin{align}
    \f{\pa}{\pa r} (\sqrt{-\bar g} V^r) &= 0,
\end{align}
which implies that 
\begin{align}
    \f{\pa}{\pa r} \int d^{D-2} x \sqrt{-\bar g} (\bs{i}^\tht_\ups)^{tr} &= \f{1}{32\pi} \int d^{D-2} x \f{\pa}{\pa r} (\sqrt{-\bar g} V^r) \nn 
    &= 0,
\end{align}
matching the idea that $\oint_s \ups \cdot \bs \tht = 0$ from above.

Using \eqref{Vr}, this implies
\begin{align}
    \f{\pa^2}{\pa r^2} (\sqrt{-\bar g} H) &= 0. \label{d2rgH}
\end{align}
Again, this follows from the fact that $k^r = 1$ (that $r$ is the affine parameter for $k^a$), that $k^{y_\alpha} = 0$ and that the metric, $k^a$ and $H$ are all independent of $t$ and $\phi_i$.  The solution to \eqref{d2rgH} (with $\sqrt{-\bar g} H$ being independent of $t,\phi_i$) is
\begin{align}
    H &= \f{f_1 r + f_2}{\sqrt{-\bar g}},
\end{align}
where $f_1, f_2$ are functions of $y_\alpha$ (or $\mu_i$). This implies
\begin{align}
    \f{\pa}{\pa r} (\sqrt{-\bar g} H) &= f_1,
\end{align}
irrespective of $r$, so that from \eqref{KthetaupsisVr},
\begin{align}
    \oint \sqrt{-\bar g} ( \bs{i}^\tht_\ups)^{tr} d^{D-2} x &= \f1{32\pi} \int d^{D-2} x f_1,
\end{align}
so that not only the integral but the integrand is obviously independent of $r$. 

Using \eqref{Gibbonssqrtg}, 
\begin{align}
    H &= \f{F_1 + F_2/r}{U},
\end{align}
where $F_{1,2} = f_{1,2} \mu_n \prod_{i=1}^n \Xi_i/\prod_{i=1}^{n-1+\ve} \mu_i$. It requires more information to fix the functional forms of $F_1$ and $F_2$, but I will note that $F_1 = 2m,F_2=0$ is the Kerr--AdS solution and $F_1=2m,F_2=-Q^2$ is the (four-dimensional) Kerr--Newman--AdS solution.

\subsection{Terms Associated with Principal Vector} \label{conservedchargebetasection}

For the ${\bs{i}}^K$ terms, it will be useful to start by looking at $({\bs{i}}^K_\bt)^{a b}$, where several terms will cancel or simplify very nicely due to some of the particular properties of the $\beta$ vector. I will remind the reader of some of the properties that $\beta$, the Killing vector associated with the Principal Conformal Killing--Yano tensor, has:
\begin{itemize}
    \item It is a Killing vector, with components only in the ``time-and-azimuthal'' sector of the metric.
    \item It has a time component of 1 ($\beta^t = 1$).
    \item It has $\beta^ak_a = 1$ \eqref{kdotbeta}.
    \item It has $\nabla_k \beta \propto k$ \eqref{nabetak} and $\bar \na_k \bt \propto k$ \eqref{kdbeta}.
\end{itemize}
Of course the first two properties are shared by many of the Killing vectors under consideration, but the latter two are particular to $\beta$, and, as shown in Section \ref{KSNullVector}, follow from the fact that $k^a$ is one of the null eigenvectors of $\bs h$, affinely parametrized. These properties will turn out to simplify the expression for the Komar term associated with $\beta$ considerably. 

Two other properties of $\bt$ which are related to the above which I think are also suggestive of its key role, but which I do not use directly below, are
\begin{itemize}
    \item It is proportional to the leading term in the ``time--angular'' part of $k$, from \eqref{kpsicoords}, so that at large $r$ (using the large-$r$ form of $X$), $k \simeq \pa_r - \f{l^2}{r^2} \bt $; and
    \item Using \eqref{nabetacurvatureresult} $\na^a \bt^b$ and thus the Komar integrand can be related directly to the PCKY tensor and to the Riemann tensor and its decomposition. In particular, for an AdS background with a vacuum Kerr--AdS full spacetime, $R^a_b = \bar R^a_b = -(D-1)l^{-2} \de^a_b$ so that 
    \begin{align}
        \na^a \bt^b - \overline{\na^a \bt^b} &= \f{1}{2(D-2)} C^{abcd} \bs h_{cd}, \label{KomardifferenceWeyl}
    \end{align}
    so that the ``full minus background'' Komar integrand is related to the Weyl tensor and the PCKY tensor, for Kerr--AdS specifically. This is not true for just any Killing vector $\chi$ but is specifically due to the symmetries of $\bt$ following from $\bs h$. I elaborate on this in Chapter \ref{additionalNote} (which did not appear in the thesis). 
\end{itemize}
It is not so much that one couldn't use these last two points to draw conclusions about what makes $\bt$ special as that I emphasized the first few points instead. There are so many reasons that $\bt$ is special that it is hard to tease out which are the most important.

I will begin with simplifying the expression for $({\bs i}^K_\beta)^{ab}$.

\subsubsection{Simplification of Expansion of Komar Term}

Expand out $({\bs{i}}^K_\bt)^{a b}$ as follows:
\begin{align}
    ({\bs{i}}_\bt^K)^{a b} &= \f{1}{16\pi} \left(H k^d k^{[a}\bar \na_d \beta^{b]} - \beta^d \bar \na^{[a} (H k^{b]} k_d)\right)
\end{align}
First of all, $k^d \bar \na_d \beta^b \propto k^b$ \eqref{kdbeta}. This means that the first term disappears entirely, since 
\begin{align}
    k^d k^{[a}\bar \na_d \beta^{b]} \propto k^{[a}k^{b]} = 0.
\end{align}

Expand what remains as
\begin{align}
    ({\bs{i}}^K)^{a b}_{\beta} &= -\f{1}{16\pi} \beta^d \bar \na^{[a} (H k^{b]} k_d) \nn 
    &= -\f{1}{16\pi} \beta^d \left( k_d \bar \na^{[a} (H k^{b]})  + H k^{[b} \bar \na^{a]} k_d\right).
\end{align}
The first term simplifies because $\beta^a k_a = 1$. As for the second term, 
\begin{align}
    \beta^d k^{[b} \bar \na^{a]} k_d &= k^{[b} \left( \bar \na^{a]}(k_d \beta^d) - k_d \bar \na^{a]} \beta^d\right).
\end{align}
$\bar \na^a (k_d \beta^d) = 0$ since $k_d \beta^d$ is constant. $k_d \bar \na^a \beta^d = - k_d \bar \na^d \beta^a$ (since $\beta$ is Killing), which is proportional to $k^a$. Thus $\beta^d k^{[b}\bar \na^{a]} k_d \propto k^{[b}k^{a]} = 0$. Thus $H k^{[b} \bar \na^{a]}k_d = 0$. We have
\begin{align}
    ({\bs{i}}^K_\bt)^{a b} &= -\f{1}{16\pi} \bar \na^{[a} ( H k^{b]}).
\end{align}
I want to emphasize that these simplifications follow from the two key features which are particular to $\beta$, as opposed to a generic Killing vector: that $k^a \bar \na_a \beta^b \propto k^b$ and that $k_a \beta^a = 1$.  Recalling that the background metric is diagonal, $({\bs{i}}^K_\bt)^{t r}$ can be expanded to give,
\begin{align}
    ({\bs{i}}^K_\bt)^{t r} &= -\f{1}{16\pi} \bar g^{tt} \bar g^{rr} \bar \na_{[t}(H k_{r]}) \nn 
    &= \f{1}{32 \pi} \bar g^{tt} \bar g^{rr} \pa_r (H k_t) .
\end{align}
$k_t$ is $r$-independent, so this reduces to
\begin{align}
    ({\bs{i}}^{K}_\bt)^{t r} &= \f{1}{32\pi} \bar g^{tt} \bar g^{rr} \pa_r H \nn 
    &= -\f{1}{16\pi} \bar g^{tt} \bar g^{rr} \pa_{[t}(Hk_{r]}) \nn
    &= \f{1}{32 \pi} k^t \bar g^{rr} \f{\pa}{\pa r} H.
\end{align}
On the last line we used that $\pa_t(H k_r) = 0$. Somewhat anticipating later simplifications, I will note that 
\begin{align}
    \lim_{r\to\infty} k^t \bar g^{rr} = -1.
\end{align}

Using \eqref{backgroundds2rytphi}, $({\bs{i}}^K_\bt)^{t r}$ can be re-expressed as 
\begin{align}
    ({\bs{i}}^K_\bt)^{t r} &= -\f{1}{32 \pi} \f{1}{1+r^2/l^2} \f{\bar X}{U_n} \f{\pa}{\pa r} \left(\frac{2 \mu(r) r^{1-\varepsilon}}{U_n}\right) \nn 
    &= -\f{1}{32\pi} \f{\prod_{i = 1}^{n-1+\ve}(r^2+a_i^2)}{r^{2 \varepsilon} U_n} \f{\pa}{\pa r} \left( \f{2\mu(r) r^{1-\varepsilon}}{U_n}\right) \label{mcKKtrbeta}
\end{align}

We will ultimately want to integrate over constant $r$. This expression can then be broken down into
\begin{align}
    ({\bs{i}}^K_\bt)^{t r} &= -\f{1}{16\pi} \f{\prod_{i=1}^{n-1+\ve} (r^2+a_i^2)}{r^{2\varepsilon} U_n} \left[ \mu(r) \f{\pa}{\pa r} \f{r^{1-\varepsilon}}{U_n} + \f{\mu'(r) r^{1-\varepsilon}}{U_n}\right] 
\end{align}

We then have
\begin{align}
    -\oint (\bs K^K_\beta - \overline{\bs K^K_\beta}) &= \oint d^{D-2} x\f{\sqrt{-\bar g}}{8\pi} \f{\prod_{i=1}^{n-1+\ve} (r^2+a_i^2)}{r^{2\varepsilon} U_n} \f{\pa}{\pa r} \left( \f{2 \mu(r) r^{1-\varepsilon}}{U_n}\right). 
\end{align}
(The extra factor of 2 comes from the fact that summation includes both $({\bs{i}}^K_\bt)^{tr} \bs \ep_{t r \ldots}$ and $(\bs{i}^K)^{rt}_\beta \bs \ep_{r t \ldots}$.)

To evaluate this, first I will consider the case where either $\mu(r) = m$ is constant, or where we consider the integral at infinity where $\mu(r) \simeq m_\infty$.

\subsubsection{Asymptotic or Vacuum Value of Komar Term}

The most sensible coordinates to use are either $y_\alpha$ coordinates for the latitude angles or $\mu_i$ (with one suppressed using the constraint). The $y_\alpha$ have the advantage of elegance, that they diagonalize the metric, and that the integration limits on the $y_\alpha$ are constants, whereas the $\mu_i$ have the advantage that their global geometric interpretations are clearer. We have $U_n$ expressed in terms of $y_\alpha$ but also have $U$ expressed in terms of $\mu_i$, and so can switch back and forth between the two, and we have the metric determinant expressed in both systems. 

In the $\mu_i$ coordinates (with $\mu_{n}$ suppressed), we have $U_n = r^{1-\varepsilon} U$. Using \eqref{Gibbonssqrtg} for $\sqrt{-\bar g}$, 
\begin{align}
    \sqrt{-\bar g}(\bs{i}^K_\bt)^{tr} &= -\f{r U \prod_{i = 1}^{n-1+\ve} \mu_i}{\mu_{n} \prod_{j = 1}^{n} \Xi_j} \f{1}{32\pi} \f{\prod_{i = 1}^{n-1+\ve} (r^2+a_i^2)}{r^{1+\varepsilon} U} \f{\pa}{\pa r} \left( \f{2 \mu(r)}{U}\right) \nn 
    &= -\f{1}{32\pi \prod_{j = 1}^{n} \Xi_j} \f{\prod_{i=1}^{n-1+\ve} \mu_i}{\mu_{n}} r^{-\varepsilon} \left(\prod_{i=1}^{n-1+\ve} (r^2+a_i^2) \right) \f{\pa}{\pa r} \left(\f{2 \mu(r)}{U}\right).
\end{align}

To begin with, consider the limiting case of large $r$. Assume again that $\mu(r)$ has the functional form of \eqref{murexpansion} near $r \to \infty$. Then we need only take the leading term in $r$ in the following. From \eqref{Udefinition}, the leading order terms in $U$ are
\begin{align}
    U &\simeq r^{1-\ve} \sum_{i = 1}^{n} \mu_i^2 r^{2(n-2+\ve)} \nn 
    &\simeq r^{D-3},
\end{align}
applying the constraint $\sum_{i = 1}^{n} \mu_i^2 = 1$. So we have
\begin{align}
    \f{\pa}{\pa r} \left(\f{2\mu(r)}{U}\right) &\simeq -(D-3) \f{2 m_\infty}{r^{D-2}},
\end{align}
where $\simeq$ again indicates taking only the leading term in $r$. Consequently, 

\begin{align}
    \lim_{r\to\infty} \sqrt{-\bar g} (\bs{i}^K_\bt)^{tr} &= \lim_{r\to\infty} -\f{1}{32\pi \prod_{j=1}^{n} \Xi_j} \f{\prod_{i = 1}^{n-1+\ve} \mu_i}{\mu_{n}} r^{-\varepsilon} r^{2(n-1+\ve)} \f{\pa}{\pa r}\left( \f{2\mu(r)}{U}\right) \nn 
    &= \f{(D-3) m_\infty}{16\pi \prod_{j = 1}^{n}\Xi_j} \f{\prod_{i = 1}^{n-1+\ve} \mu_i}{\mu_{n}}. \label{asymptoticsqrtgibetaKtr}
\end{align}
This implies
\begin{align}
    \lim_{r\to\infty}\oint d\mu_1 \ldots d \mu_{n-1} d\phi_1 \ldots d \phi_{n-1+\ve} \sqrt{-\bar g} (\bs{i}^K_\bt)^{t r} &= \f{(D-3) m_\infty \mc A_{D-2}}{16\pi \prod_{j = 1}^{n} \Xi_j}
\end{align}
and so, letting $C_\infty$ be the surface of constant $t$ and constant $r \to \infty$,
\begin{align}
    -\oint_{C_\infty}(\bs K^K_\beta - \overline{\bs K^K_\beta}) &= \f{(D-3) m_\infty \mc A_{D-2}}{8 \pi \prod_{j = 1}^{n} \Xi_j}. \label{KomarIntegralTermforbetaasymptotic}
\end{align}
As stated, in vacuum the result for the Komar integral is independent of integration surface (provided it encloses the singularity), since $d(\bs K_\bt^K - \overline{\bs K_\bt^K})=0$ in this case. Let $C$ be any such surface. Then 
\begin{align}
    -\oint_C (\bs K^K_\beta - \overline{\bs K^K_\beta}) &= \f{(D-3) m \mc A_{D-2}}{8\pi \prod_{j = 1}^{n} \Xi_j}, \label{KomarIntegralTermforbeta}
\end{align}
again \emph{specifically for Kerr--anti-de Sitter}, where $\mu(r) = m_\infty = m$ (constant). 

Note that while the result was calculated in using the $(t,r,\mu_i,\phi_i)$ coordinates it will also be true in the $(t,r,y_\alpha,\phi_i)$ coordinates since the $y_\alpha$ are a function of the $\mu_i$ only. The equality (up to a factor of $(D-3)$) of the integrals for the Komar and non-Komar parts of the conserved charge, at infinity (or anywhere for Kerr--AdS), associated with the $\beta$ is a very important result. 

Now we turn to the calculation where $\mu(r)$ is an arbitrary function of $r$ and we evaluate the integral at an arbitrary value of $r$.

\subsubsection{Komar Term for Arbitrary Mass Function and Radius}

Now again we recall that for Kerr--anti-de Sitter ($\mu(r) = m$, constant), the integration is independent of surface. This implies that necessarily, taking $m = 1$,
\begin{align}
    \oint d^{D-2} x \f{\prod_{i=1}^{n-1+\ve} \mu_i}{\mu_{n}} r^{-\varepsilon} \left(\prod_{i=1}^{n-1+\ve} (r^2+a_i^2)\right) \f{\pa}{\pa r} \f{1}{U} &= -(D-3) \mc A_{D-2}, \label{m=1}
\end{align}
for any value of $r$. $(d^{D-2}x$ means the integration over all the $\mu_i$, with one suppressed, and $\phi_i$. It implicitly includes the factor of 2 in even dimensions to account for the fact that the suppressed $\mu_n$ can take on positive or negative values.) 

(It is perhaps surprising that the result \eqref{m=1} is independent of $r$. The argument presented in this section appeals to the vanishing of $R^a_b - \bar R^a_b$. A supplementary argument is included in Appendix Section \ref{explicitJacobiTransformed} which shows the $r$-independence of $\oint\sqrt{-\bar g} (\bs i_\bt^K)^{tr}d^{D-2}x$  explicitly by making use of the $y_\alpha$ coordinates.)

Consequently, we must have
\begin{align}
    \oint d^{D-2} x \f{\prod_{i=1}^{n-1+\ve} \mu_i}{\mu_{n}}\f{\pa}{\pa r} \f{1}{U} &= -\f{(D-3) \mc A_{D-2} r^{\varepsilon}}{\prod_{i=1}^{n-1+\ve} (r^2+a_i^2)} \nn 
    \f{\pa}{\pa r} \oint d^{D-2} x \f{\prod_{i=1}^{n-1+\ve} \mu_i}{\mu_n U} &= -\f{(D-3) \mc A_{D-2} r^\varepsilon}{\prod_{i=1}^{n-1+\ve} (r^2+a_i^2)}. \label{betaintegral}
\end{align}
The differentiation by $r$ can be taken all the way to left of the integral on the left-hand side because the $\mu_i$ are $r$-independent. 
Integrating both sides with respect to $r$, 
\begin{align}
    \oint d^{D-2} x \f{\prod_{i=1}^{n-1+\ve} \mu_i}{\mu_n U} &= -(D-3) \mc A_{D-2} \int \f{r^\varepsilon d r}{\prod_{i=1}^{n-1+\ve} (r^2+a_i^2)}. \label{muibymunUintegral}
\end{align}
The additive constant in the indefinite integral is chosen to ensure the indefinite integral vanishes in the large $r$ limit. (This is because $1/U$ vanishes in the large-$r$ limit, so that its integral over the $\mu_i$ must vanish for large $r$.) 

The integral over $r$ can be calculated by rewriting the term using the $\Ups_i$. As stated in \cite{KrtousKubiznak}, $\sum_{\mu=1}^n \f{1}{U_\mu} = 0$. This is provided $n \geq 2$. This implies we have, analogously using both the $\Ups_i$ and $\hat \Ups_i$ sets,
\begin{align}
    \sum_{i=1}^{n-1+\ve} \f{1}{\Ups_i} &= 0, \label{sumUpsi}
\end{align}
provided $n+\ve \geq 1$, true for $D \geq 5$, and
\begin{align}
    \sum_{i=0}^{n-1+\ve} \f{1}{\hat \Ups_i} &= 0. \label{sumhatUpsi}
\end{align}
provided $n+\ve \geq 2$, true for $D \geq 4$ (and even $D=3$, though we do not consider that case). The $\hat \Ups_i$ can be expanded using \eqref{hatnonthatrelationsforCetc} to give 
\begin{align}
    \f{1}{\prod_{i=1}^{n-1+\ve} (a_i^2-l^2)} + \sum_{i=1}^{n-1+\ve} \f{1}{(l^2-a_i^2) \Ups_i} &= 0.
\end{align}
This is just a numerical relation and so it will be satisfied under the substitution $l^2 \to -r^2$, provided $-r^2$ is not equal to any of the $a_i^2$. Then we have
\begin{align}
    \f{1}{\prod_{i=1}^{n-1+\ve} (r^2+a_i^2)} &= \sum_{i=1}^{n-1+\ve} \f{1}{(r^2+a_i^2) \Ups_i}. \label{raproductintermsofUpsilon}
\end{align}

This allows for integration. We then have,
\begin{align}
    \int \f{r^\varepsilon d r}{\prod_{i=1}^{n-1+\ve} (r^2+a_i^2)} &= \sum_{i =1}^{n-1+\ve} \int \f{r^\varepsilon dr}{(r^2+a_i^2) \Upsilon_i}.
\end{align}

In even dimension, we have $\varepsilon = 0$, and so
\begin{align}
    \int \f{dr}{r^2+a_i^2} &= \f{1}{a_i} \arctan\f{r}{a_i} + c,
\end{align}
for some constant $c$. Choosing $c = -\pi/(2a_i)$ and noting $\arctan(r/a_i) = \pi/2-\arctan(a_i/r)$, we can write
\begin{align}
    \int \f{dr}{r^2+a_i^2} &= -\f1{a_i}\arctan\f{a_i}{r},
\end{align}
which is zero as $r \to \infty$. Consequently in even dimension,
\begin{align}
    \int \f{dr}{\prod_{i=1}^{n-1} (r^2+a_i^2)} &= -\sum_{i=1}^{n-1} \f{\arctan(a_i/r)}{a_i \Upsilon_i}. \label{evendimensiononebyr2a2integral}
\end{align}

In odd dimension ($\ve = 1$),
\begin{align}
    \int \f{r dr}{r^2+a_i^2} &= \f{1}{2} \ln (r^2+a_i^2) \nn 
    &= \ln r + \f12 \ln\left( 1 + \f{a_i^2}{r^2}\right),
\end{align}
possibly plus a constant. We then have,
\begin{align}
    \int \f{r dr}{\prod_{i=1}^{n}(r^2+a_i^2)} &=\sum_{i=1}^{n}\f{1}{\Ups_i}\left[ \ln r + \f12 \ln\left(1 + \f{a_i^2}{r^2}\right)\right] \nn 
    &= \ln r \sum_{i=1}^n \f{1}{\Ups_i} +  \sum_{i=1}^n \f{1}{2\Ups_i} \ln\left(1 + \f{a_i^2}{r^2}\right),
\end{align}
again possibly plus an integration constant. From \eqref{sumUpsi}, $\sum_{i=1}^n \Ups_i^{-1}=0$, so that this reduces to
\begin{align}
    \int \f{rdr}{\prod_{i=1}^n (r^2+a_i^2)} &= \sum_{i=1}^n \f{\ln(1+a_i^2/r^2)}{2\Ups_i}, \label{odddimintrbyr2a2}
\end{align}
which is already 0 when $r \to \infty$ (so we need add no integration constant). 

Putting it all together, we can then write, taking $C$ to be a surface of constant $(t,r)$, with arbitrary $r$ and with arbitrary $\mu(r)$, in \emph{even} dimensions
\begin{align}
    -\oint_C (\bs K^K_\beta - \overline{\bs K^K_\beta}) &= \f{(D-3) \mc A_{D-2}}{8 \pi \prod_{j=1}^{n} \Xi_j} \left( \mu(r) - \mu'(r) \sum_{i=1}^{n-1} \f{\arctan(a_i/r)}{a_i \Upsilon_i} \prod_{j=1}^{n-1} (r^2+a_j^2)\right) \label{evendimKomarbeta}
\end{align}
and, in \emph{odd} dimensions,
\begin{align}
    -\oint_C (\bs K^K_\beta - \overline{\bs K^K_\beta}) &= \f{(D-3) \mc A_{D-2}}{8\pi \prod_{j=1}^n \Xi_j} \left( \mu(r) + \mu'(r)\sum_{i=1}^{n} \f{\ln(1+a_i^2/r^2)}{2 \Upsilon_i} \f{\prod_{j=1}^n (r^2+a_j^2)}{r}\right) \label{odddimKomarbeta}
\end{align}

Having found these explicit expressions, I now turn to comparing the Komar and non-Komar terms.

\subsubsection{Comparison to Non-Komar Term and Total Conserved Charge} \label{ComparisonBetweenKomarAndNonKomarTerms}

It is worth comparing the Komar term and non-Komar term associated with $\bt$, again specifically in the case where either we have vacuum or are only working asymptotically. 

We then note how close (though not identical) the forms of the Komar and $\bs \tht$ constituents of the expression for ${\bs{i}}_\beta$ are, when focusing on the $tr$ components:
\begin{align}
    ({\bs{i}}^K_\bt)^{tr} &= -\f{1}{16\pi} \bar \na^{[t} (H k^{r]}) \nn 
    ({\bs{i}}^{\tht}_\bt)^{tr} &= \f{1}{32\pi} \bar \na_b (H k^b). \label{KomarnonKomarcomparison1}
\end{align}
These expressions are very close to each other, though of course they are not the same.

We also note the comparison
\begin{align}
    ({\bs{i}}^K_\bt)^{tr} &= \f{1}{32 \pi} k^t \bar g^{rr} \f{\pa}{\pa r} H \nn
    ({\bs{i}}^{ \tht}_\bt)^{t r} &= \f{1}{32\pi} \f{1}{\sqrt{-\bar g}} \f{\pa}{\pa r} (\sqrt{-\bar g} H),
\end{align}
and the expressions look very similar. In the large $r$ limit, since $\lim_{r\to\infty} k^t \bar g^{rr} = -1$, then we have
\begin{align}
    \lim_{r\to\infty} \sqrt{-\bar g} (\bs{i}^K_\bt)^{tr} &= -\f{1}{32\pi} \lim_{r\to \infty} \sqrt{-\bar g} \f{\pa}{\pa r} H \nn 
    \lim_{r\to\infty} \sqrt{-\bar g} (\bs {i}^\tht_\bt)^{tr} &= \f{1}{32\pi} \lim_{r\to\infty} \f{\pa}{\pa r} (\sqrt{-\bar g} H).
\end{align}
Not only do these expressions look similar, but in the large $r$ limit the key feature that will contribute to the limit is the leading power in $r$ for $H$ and $\sqrt{-\bar g}H$. This is easier to see in the related form
\begin{align}
    \lim_{r\to\infty} \sqrt{-\bar g} (\bs{i}^K_\bt)^{tr} &= -\f1{32\pi} \lim_{r\to\infty} \sqrt{-\bar g} H \f{\pa}{\pa r} \ln H \nn 
    \lim_{r\to\infty} \sqrt{-\bar g} (\bs{i}^\tht_\bt)^{tr} &= \f{1}{32\pi} \lim_{r\to\infty} \sqrt{-\bar g} H \f{\pa}{\pa r} \ln (\sqrt{-\bar g}H).
\end{align}
Assuming that $\mu(r) = m_\infty + \sum_{j=1}^\infty d_j/r^j$, the leading terms in $H$ and $\sqrt{-\bar g} H$ are proportional to $r^{-(D-3)}$ and $r$, respectively. Thus $\pa \ln H/\pa r \simeq -(D-3)/r$ and $\pa \ln (\sqrt{-\bar g} H)/\pa r \simeq 1/r$, where $\simeq$ means is equal to at large $r$ (discarding everything but the leading-order terms). Thus we find 
\begin{align}
    \lim_{r\to\infty} \sqrt{-\bar g} (\bs{i}^K_\bt)^{tr} &= (D-3) \lim_{r\to\infty} \sqrt{-\bar g} (\bs{i}^\tht_\bt)^{tr}. \label{asymptoticintegrand}
\end{align}
We note immediately that this \emph{asymptotic form} gives 
\begin{align}
    -\oint_{C_\infty} \left( \bs K^K_{\beta} - \overline{\bs K^K_\beta}\right) &= (D-3) \left(-\oint_{C_\infty} \beta \cdot \bs \tht\right). \label{asymptoticintegral}
\end{align}
(This follows from the fact that \eqref{upsilonthetalimit} holds for $\upsilon = \beta$.) If $D = 4$, $D-3=1$ and so 
\begin{align}
    -\oint_{C_\infty}( \bs K^K_\beta - \overline{\bs K^K_\beta}) = - \oint_{C_\infty} \beta \cdot \bs \tht \qquad (D=4).
\end{align}
Thus the integrals, and even the integrands, of the Komar and non-Komar terms are equal in the large-$r$ limit (up to a factor of $D-3$). This is a remarkable result! It follows from the specific features of the PCKY tensor $\beta$. 

In vacuum (Kerr--AdS), both $-\oint_C (\bs K^K_\bt - \overline{\bs K^K_\bt})$ and $-\oint_C \bt \cdot \bs \tht$ are independent of integration surface, so we also have, specifically for vacuum,
\begin{align}
    -\oint_C (\bs K^K_\beta - \overline{\bs K^K_\beta}) &= -(D-3) \oint_C \beta \cdot \bs \tht. \label{KerrAdSKomarnonKomarequality}
\end{align}
(Note that whereas, in the vacuum case, the integrand $\sqrt{-\bar g} (\bs{i}^\tht_\bt)^{tr}$ 
is independent of $r$, the integrand $\sqrt{-\bar g} (\bs {i}^K_\beta)^{tr}$ is not constant, even if its integral is constant in vacuum.)

This allows us to write
\begin{align}
    \oint_{C_\infty} \bs I_\bt &= (D-2) \left( - \oint_{C_\infty} \bt \cdot \bs \tht\right) \nn 
    &= \f{(D-2) m_\infty \mc A_{D-2}}{8\pi \prod_{j=1}^n \Xi_j}, \label{IbetaintCinfty}
\end{align}
provided that $\mu(r) = m_\infty + \sum_{j=1}^\infty d_j r^{-j}$. The same statements are true on an arbitrary surface $C$ in Kerr--AdS, 
\begin{align}
    \mc F &= \oint_{C} \bs I_\bt \nn 
    &= \f{(D-2) m \mc A_{D-2}}{8\pi \prod_{j=1}^n \Xi_j} \nn
    &=(D-2) \left( - \oint_{C} \bt \cdot \bs \tht\right) \qquad\textrm{(vacuum)}.\label{KAdSFRelationship}
\end{align}
Thus we recover \eqref{D-2C=F}! That the Komar and non-Komar terms are equal, up to a numerical factor, when the charge related to $\beta$ is used, is the basic reason why the volume appears in a natural way in the Smarr relationship for $\bt$ and thus $\mc F$, as discussed in Section \ref{SmarrPrincipal}. This equality follows from the simplifications that result from the properties of $\beta$, listed at the beginning of this section. This is one of the main results of this thesis.

It might seem surprising that the asymptotically-rotating Killing vector is the one for which the Komar and non-Komar terms have such a simple relationship. Note however that Kerr--AdS permits a smooth limit to Kerr (or Myers--Perry, in higher dimensions) by taking $l \to \infty$. In this case, we still have \eqref{KAdSFRelationship}, but here $\bt = \xi = \f{\pa}{\pa t}$, since all the terms $a_i/l^2 \to 0$. The idea then is that the Principal Vector is the same as the asymptotically-static vector in Kerr/Myers--Perry.

It is plain to see that, for generic $\mu(r)$ (even with the asymptotic form $\mu(r) = m_\infty + \sum_j d_{j=1}^\infty r^{-j}$), the equality \eqref{KerrAdSKomarnonKomarequality} does not hold on an arbitrary surface $C$, in general. One need only compare \eqref{evendimKomarbeta} or \eqref{odddimKomarbeta} to \eqref{upsilonthetagenericmu}. One way to put it is that the relationship at infinity is guaranteed (provided regularity of $\mu(r)$), and that it is only in vacuum when the Komar and non-Komar terms maintain their values ``from infinity'' to a generic integration surface. 

If $\mu(r)$ is not constant and we evaluate at finite radius, the contributions from the Komar and non-Komar terms associated with $\mu'(r)$ do not combine or cancel in any simple way. The total charge in this case is, from \eqref{upsilonthetagenericmu}, \eqref{evendimKomarbeta}, and \eqref{odddimKomarbeta}
\begin{align}
    \oint_C \bs I_\beta &= \f{\mc A_{D-2}}{8\pi \prod_{j=1}^n \Xi_j} \left[ (D-2) \mu(r) + \mu'(r)  \left( r - (D-3) \sum_{i=1}^{n-1} \f{\arctan(a_i/r)}{a_i \Ups_i} \prod_{j=1}^{n-1}(r^2+a_j^2)\right)\right] \label{Ibetaeven}
\end{align}
in even dimensions and
\begin{align}
    \oint_C \bs I_\bt &= \f{\mc A_{D-2}}{8\pi \prod_{j=1}^n \Xi_j} \left[ (D-2) \mu(r) + \mu'(r) \left( r + (D-3) \sum_{i=1}^n \f{\ln(1+a_i^2/r^2)}{2 \Ups_i} \f{\prod_{j=1}^n (r^2+a_j^2)}{r}\right)\right] \label{Ibetaodd}
\end{align}
in odd dimensions.

\subsection{Komar Term Associated with More General Killing Vector}

Having examined $(\bs{i}^K_\bt)^{tr}$, it is worth considering the Komar term $(\bs{i}^K_\chi)^{tr}$ associated with a more general Killing vector $\chi$ that has components $\chi^r = \chi^{\mu_i} = 0$ (or $\chi^{y_\alpha} = 0$) and $\chi^t, \chi^{\phi_i}$ constant. The term
\begin{align}
    H k^d k^{[a}\bar \na_d \chi^{b]}
\end{align}
can be rewritten as
\begin{align}
    H k^d k^{[a}\bar \na_d \chi^{b]} &= \f12 H k^d (k^a \bar \na_d \chi^b - k^b \bar \na_d \chi^a) \nn 
    &= \f12 H k^d (k^a \bar g^{be} \bar \na_d \bar \chi_e - k^b \bar g^{ae} \bar \na_d \bar \chi_e) \nn 
    &= \f12 H k^d (k^a \bar g^{be} \pa_{[d}\bar \chi_{e]} - k^b \bar g^{ae} \pa_{[d}\bar \chi_{e]}) \nn 
    &= \f14 H k^d (k^a \bar g^{be} \pa_d \bar \chi_e -k^a \bar g^{be}\pa_e \bar \chi_d - k^b \bar g^{ae} \pa_d \bar \chi_e + k^b \bar g^{ae} \pa_e \bar \chi_d)
\end{align}
Specifically considering $a = t, b = r$ and recalling that the background metric is diagonal, that $\chi^r = 0$ and that the metric is independent of $t$ or $\phi_i$:
\begin{align}
    H k^d k^{[t}\bar \na_d \chi^{r]} &= \f14 H k^d (k^t \bar g^{re} \pa_d \bar \chi_e - k^t \bar g^{r e} \pa_e \bar \chi_d - k^r \bar g^{t e} \pa_d \bar \chi_e + k^r \bar g^{t e} \pa_e \bar \chi_d) \nn 
    &= \f14 H k^d (k^t \bar g^{rr} \pa_d \bar \chi_r - k^t \bar g^{rr} \pa_r \bar \chi_d - k^r \bar g^{tt} \pa_d \bar \chi_t + k^r \bar g^{tt} \pa_t \bar \chi_d) \nn 
    &= -\f14 H k^d (k^t \bar g^{rr} \pa_r \bar \chi_d + k^r \bar g^{tt} \pa_d \bar \chi_t) \nn 
    &= -\f14 H( k^t k^d \bar g^{rr} \pa_r \bar \chi_d + \bar g^{tt} \pa_r \bar g_{tt} \chi^t).  
    \label{firstKomarterm}
\end{align}
We also used $k^r = 1$ on the final line.

Consider the term $\chi^d \bar \na^{[a}h^{b]}_d.$ This can be expanded into
\begin{align}
    \chi^d \bar \na^{[a}h^{b]}_d &= \chi^d \bar \na^{[a} (H k^{b]} k_d) \nn 
    &= \chi^d k_d \bar \na^{[a} (H k^{b]}) + \chi^d H k^{[b} \bar \na^{a]} k_d \nn 
    &= \chi^d k_d \bar \na^{[a} (H k^{b]}) + H k^{[b} \bar \na^{a]} ( \chi^d k_d) - H k_d k^{[b} \bar \na^{a]} \chi^d
\end{align}
When we specifically consider $a=t,b=r$ this reduces further to:
\begin{align}
    \chi^d \bar \na^{[t} h^{r]}_d &= \chi^d k_d \bar \na^{[t} (H k^{r]}) + H k^{[r} \bar g^{t]c} \pa_c (\chi^d k_d) - H k^d k^{[r} \bar g^{t]c} \bar \na_c \bar \chi_d.
\end{align}
Of course $\bar \chi_d = \bar g_{d e} \chi^e$. Because the background metric is diagonal, the middle term has nonvanishing $c = t$ or $r$. However, $k_t$ and $k_{\phi_i}$ do not depend on either $t$ or $r$, so the middle term vanishes. The others can then be rewritten in terms of antisymmetrized partial derivatives as follows:
\begin{align}
    \chi^d \bar \na^{[t}h^{r]}_d &= \chi^d k_d \bar g^{tt} \bar g^{rr} \bar \na_{[t}(H k_{r]}) - H k^d k^{[r}\bar g^{t]c} \pa_{[c}\bar \chi_{d]} \nn 
    &= \chi^d k_d \bar g^{tt} \bar g^{rr} \pa_{[t}( H k_{r]}) - H k^d k^{[r} \bar g^{t] c} \pa_{[c} \bar \chi_{d]} \nn 
    &= - \f{1}{2} \chi^d k_d \bar g^{tt} \bar g^{rr} \pa_r (H k_t) - \f{1}{4} H k^d \left(k^r \bar g^{tt} (\pa_t \bar \chi_d - \pa_d \bar \chi_t) - k^t \bar g^{rr} (\pa_r \bar \chi_d - \pa_d \bar \chi_r)\right).
\end{align}
Recall $\bar \chi_r = 0$ (since $\chi^r = 0$), $k^r = 1$ and $k_t$ is $r$-independent. 
\begin{align}
    \chi^d \bar \na^{[t} h^{r]}_d &= -\f12 \chi^d k_d k^t \bar g^{rr} \pa_r H - \f{1}{4} H k^d (-\bar g^{tt} \pa_d \bar \chi_t - k^t \bar g^{rr} \pa_r \bar \chi_d) \nn
    &= -\f12 \chi^d k_d k^t \bar g^{rr} \pa_r H + \f14 H ( \bar g^{tt} \pa_r \bar \chi_t + k^t \bar g^{rr} k^d \pa_r \bar \chi_d) \nn 
    &= -\f12 \chi^d k_d k^t \bar g^{rr} \pa_r H - H k^d k^{[t} \bar \na_d \chi^{r]} \label{secondKomarterm},
\end{align}
noting on the last line the interesting effect that much of this term matches that of \eqref{firstKomarterm}. The combination is therefore 
\begin{align}
    (\bs{i}^K_\chi)^{t r} &= \f{1}{16\pi} \left(h^{d[t}\bar \na_d \chi^{r]} - \chi^d \bar \na^{[t}h^{r]}_d\right) \nn &= \f1{32\pi} (\chi^d k_d k^t \bar g^{rr} \pa_r H + 4 H k^d k^{[t}\bar \na_d \chi^{r]}) \nn 
    &= \f1{32\pi} \left[\chi^d k_d k^t \bar g^{rr} \pa_r H - H( k^t k^d \bar g^{rr} \pa_r \bar \chi_d + \bar g^{tt} \pa_r \bar g_{tt} \chi^t)\right].   \label{Komarbreakdown}
\end{align}

\subsection{Angular Momentum Komar Term} \label{AngularMomentumKomarTerm}

In this section we discuss the Komar terms associated with the angular momentum Killing vectors $\eta_i$. 

As stated in \eqref{angularmomentumPureAdS}, provided that we choose a surface $C$ of constant $t$ in the KS coordinates, the Komar integral associated with $\eta_i$ for the \emph{background} spacetime is zero. This means that the only contribution to Komar terms come from the full, and not the background, spacetime. (Explicitly, $\oint_C(\bs K_{\eta_i}^K - \overline{\bs K_{\eta_i}^K}) = \oint_C \bs K_{\eta_i}^K$.) With $\chi = \eta_i$, \eqref{Komarbreakdown} reduces to
\begin{align}
    (\bs{i}^K_{\eta_i})^{tr} &= \f{1}{32\pi} \left[ k_{\phi_i} k^t \bar g^{rr} \pa_r H - H k^t k^{\phi_i}\bar g^{rr} \pa_r \bar g_{\phi_i\phi_i}\right]
\end{align}
Because $\pa_r k_{\phi_i} = \pa_r \bar k_{\phi_i} = 0, k^{\phi_i}\pa_r \bar g_{\phi_i\phi_i} = - \bar g_{\phi_i\phi_i} \pa_r k^{\phi_i}$. 
\begin{align}
    (\bs{i}^K_{\eta_i})^{tr} &= \f{1}{32\pi} \left( H k^t \bar g_{\phi_i\phi_i} \bar g^{rr} \pa_r k^{\phi_i} + \bar g_{\phi_i\phi_i} k^{\phi_i} k^t \bar g^{rr} \pa_r H\right) \nn 
    &= \f{1}{32\pi} \bar g_{\phi_i\phi_i} k^t \bar g^{r r} \pa_r (H k^{\phi_i}).
\end{align}

In $\mu_i$ coordinates (with one suppressed), using \eqref{GibbonsXiWF} and \eqref{GibbonsV} for the definitions of $F$ and $V$,
\begin{align}
    (\bs{i}^K_{\eta_i})^{tr} &= \f{1}{32\pi} \f{\mu_i^2(r^2+a_i^2)}{\Xi_i} \f{l^2}{(r^2+l^2)F} \f{\pa}{\pa r} \left( \f{2\mu(r)}{U} \f{a_i}{r^2+a_i^2}\right) \label{KKetamuform} \\ 
    \sqrt{-\bar g} (\bs{i}^K_{\eta_i})^{tr} &= \f{1}{32\pi} \f{r U}{\mu_{n}} \left(\prod_{j =1}^{n-1+\ve} \f{\mu_j}{\Xi_j}\right) \f{\mu_i^2(r^2+a_i^2)}{\Xi_i} \f{l^2}{(r^2+l^2)F} \f{\pa}{\pa r} \left( \f{2\mu(r)}{U} \f{a_i}{r^2+a_i^2}\right) \nn 
    &= \f{\prod_{j=1}^{n-1+\ve} \mu_j}{32\pi \mu_n \prod_{j=1}^{n} \Xi_j} \f{\mu_i^2}{\Xi_i} \f{r (r^2+a_i^2) l^2 V}{(r^2+l^2) } \f{\pa}{\pa r} \left( \f{2\mu(r)}{U} \f{a_i}{r^2+a_i^2}\right) \nn 
    &= \f{\prod_{j=1}^{n-1+\ve} \mu_j}{32\pi \mu_n \prod_{j=1}^{n} \Xi_j} \f{\mu_i^2}{\Xi_i} (r^2+a_i^2) r^{-\ve} \left( \prod_{j=1}^{n-1+\ve} (r^2+a_j^2)\right) \f{\pa}{\pa r} \left( \f{2\mu(r)}{U} \f{a_i}{r^2+a_i^2}\right).  \label{Komaretaarbitraryr}
\end{align}

(We note in passing,
\begin{align}
    (\bs{i}^K_{\eta_i})^{tr} &= - \f{\mu_i^2 (r^2+a_i^2)}{\Xi_i} (\bs{i}_\bt^K)^{tr} \f{\f{\pa}{\pa r} (2\mu(r) a_i/[U(r^2+a_i^2)])}{\f{\pa}{\pa r} (2 \mu(r)/U)}. \label{KKetabetarelation}
\end{align}
A recurring theme throughout this subsection is that the calculation for the Komar term associated with $\eta_i$ is similar to the calculation for the Komar term associated with $\beta$, and it follows from the similarities between the expressions for the two.)

As in Section \ref{conservedchargebetasection}, we will start by considering the large-$r$ limit in the case where $\mu(r)$ has the asymptotic form \eqref{murexpansion}. Then, taking the leading terms in $r$, and noting that $V \simeq r^{D-1}/l^2$ for large $r$, we have
\begin{align}
    \lim_{r\to\infty} \sqrt{-\bar g} (\bs{i}^K_{\eta_i})^{t r} &= \f{\prod_{j=1}^{n-1+\ve} \mu_j}{32 \pi \mu_{n} \prod_{j=1}^{n-1+\ve} \Xi_j} \f{\mu_i^2}{\Xi_i} \f{r^3 r^{D-1}}{r^2} \f{\pa}{\pa r}\left( \f{2 m_\infty}{r^{D-1}}\right) \nn
    &= \f{\prod_{j=1}^{n-1+\ve} \mu_j}{32 \pi \mu_n \prod_{j=1}^n \Xi_j} \f{\mu_i^2}{\Xi_i} r^D \f{\pa}{\pa r}\left( \f{2 m_\infty a_i}{r^{D-1}}\right) \nn 
    &= -\f{m_\infty a_i (D-1) \prod_{j = 1}^{n-1+\ve} \mu_j}{16\pi \mu_n \prod_{j=1}^n \Xi_j} \f{\mu_i^2}{\Xi_i}. \label{etaintegrallarger}
\end{align}

The calculation of the angular momentum Komar integral for Kerr--AdS (the above integral with $m_\infty \to m$) was performed in \cite{GibbonsPerry} explicitly for $D \leq 7$ (and, it appears, the results extrapolated from there to higher dimension). The higher-dimensional cases were also calculated in \cite{DeruelleKatz05} in the large-$r$ limit; the authors of that paper get to \eqref{etaintegrallarger} and then state that they use spherical angular coordinates to calculate the integral. Not all the intermediate steps are shown explicitly there and so I think it is worth stating my method for calculating \eqref{etaintegrallarger} explicitly. Afterward I will return to the general-$r$ case. 

\subsubsection{Asymptotic or Vacuum Value}

$(\prod_{j=1}^{n-1+\ve} \mu_j)/\mu_n$ is the integrand associated with $\mc A_{D-2}$, but there is an extra multiplicative term $\mu_i^2$ which alters the calculation. To calculate the impact of this $\mu_i^2$, I will split into even- and odd-dimensional cases.

Consider first even dimensions. Integrate over $\mu_1,\ldots, \mu_{n-1}$ as well as $\phi_1,\ldots,\phi_{n-1}$. The $\phi_1,\ldots,\phi_{n-1}$ bring a factor of $(2\pi)^{n-1}$, and the limits on $-1 \leq \mu_{n} \leq 1$ (only appearing via the constraint) lead to an extra factor of 2, so that we can write
\begin{align}
    \int d^{2(n-1)}x \f{\prod_{j =1}^{n-1} \mu_j}{\mu_{n}} \mu_i^2 &= 2(2\pi)^{n-1} \int_0^1 d \mu_{n-1} \int_0^{\sqrt{1-\mu_{n-1}^2}} d \mu_{n-2} \ldots \int_0^{\sqrt{1-\sum_{j=2}^{n-1} \mu_j^2}} d \mu_1 \f{\prod_{j=1}^{n-1} \mu_j^2}{\mu_n} \mu_i^2.
\end{align}
(Technically perhaps the 2 from the fact that there are two solutions for $\mu_n$ for every $\{\mu_1, \ldots, \mu_{n-1}\}$ except the trivial $\mu_n = 0$ case should be included on the left hand side separately, but hopefully the meaning is clear.) 

Consider $i = n-1$. Making the substitution $M_i = \mu_i^2$ as in Section \ref{volumeandareaformulas}, and using the general argument from that section, this becomes
\begin{align}
    \int d^{2(n-1)}x \f{\prod_{j=1}^{n-1} \mu_j}{\mu_{n}} \mu_{n-1}^2 &= 2 \pi^{n-1} \int_0^1 M_{n-1} d M_{n-1} \int_0^{1-M_{n-1}} d M_{n-1} \ldots \int_0^{1-\sum_{j=2}^{n-1} M_j} d M_1 \f{1}{\sqrt{1-\sum_{j=1}^{n-1} M_j}}.
\end{align}
The only difference from the calculation for $\mc A_{2N}$ is the extra factor of $M_{n-1}$ ($= \mu_{n-1}^2$). All the integrals except the final one take the same general form as in the arguments leading up to \eqref{evendimmucalc}, so that we get
\begin{align}
    \int d^{2(n-1)}x \f{\prod_{j=1}^{n-1} \mu_j}{\mu_{n}} \mu_{n-1}^2 &= 2 \pi^{n-1} \int_0^1 M_{n-1} d M_{n-1} \left(1 - M_{n-1}\right)^{(2(n-1)-3)/2} \f{2}{1} \cdot \f{2}{3} \ldots \f{2}{2 (n-1)-3} \nn 
    &= \f{(2\pi)^{n-1}}{(2n-5)!!} \int_0^1 dM_{n-1} M_{n-1}(1-M_{n-1})^{(2n-5)/2} \nn 
    &=\f{4 (2\pi)^{n-1}}{(2n-1)!!} \nn 
    &= \f {2 \mc A_{D-2}}{D-1}.
\end{align}

Geometrically there is nothing special about $\mu_{n-1}$, so the above integral holds for any value of $i$ being singled out to have the extra factor of $\mu_i^2$. (If $i < n-1$, all that is needed is to perform the integration in a different order for the above argument to hold. The case $\mu_n$ might seem to be distinct, since $\mu_n$ has different limits than the other $\mu_n$ terms, but this issue can be averted by letting $0 \leq \mu_n \leq 1$ and then multiplying the integral by an overall factor of 2. Doing so, $\mu_n$ can be chosen as one of the latitude variables to be integrated over and one of the others, say $\mu_{n-1}$, can be the latitude variable eliminated via the constraint.) 

In odd dimensions, integrate over $\mu_1, \ldots, \mu_{n-1}, \phi_1 ,\ldots, \phi_n$. 
\begin{align}
    \int d^{2n-1} x \left(\prod_{j=1}^{n-1} \mu_j\right) \mu_i &= (2\pi)^{n} \int_0^1 d \mu_{n-1} \mu_{n-1} \int_0^{\sqrt{1-\mu_{n-1}^2}} d \mu_{n-2} \mu_{n-2} \ldots \int_0^{\sqrt{1-\sum_{j=2}^{n-1} \mu_j^2}} d \mu_1 \mu_1 \mu_i^2.
\end{align}
Making the substitution $\mu_j^2 = M_j$, this becomes
\begin{align}
    \int d^{2n-1} x \left(\prod_{j=1}^{n-1} \mu_j\right) \mu_i^2 &= 2 \pi^n \int_0^1 d M_{n-1} \int_0^{1 - M_{n-1}} d M_{n-2} \ldots \int_0^{1-\sum_{j=2}^{n-1} M_j}  d M_1 M_i.
\end{align}
Consider $i = n-1$. 
\begin{align}
    \int d^{2n-1} x \left( \prod_{j=1}^{n-1} \mu_j\right) \mu_{n-1}^2 &= 2 \pi^n \int_0^1 {M_{n-1}} d M_{n-1} \int_0^{1-M_{n-1}} d M_{N-2} \ldots \int_0^{1-\sum_{j=2}^{n-1} M_j} d M_1.
\end{align}
By the steps leading up to \eqref{odddimensionMiintegrals},
\begin{align}
    \int_0^{1-M_{n-1}} d M_{n-2} \ldots \int_0^{1-\sum_{j=2}^{n-1} M_j} d M_1 &= \f{(1-M_{n-1})^{n-2}}{(n-2)!},
\end{align}
so
\begin{align}
    \int d^{2n-1} x \left( \prod_{j=1}^{n-1} \mu_j\right) \mu_{n-1}^2 &= 2 \pi^n \int_0^1 M_{n-1} d M_{n-1} \f{(1-M_{n-1})^{n-2}}{(n-2)!} \nn 
    &= \f{2 \pi^n}{n!} \nn 
    &= \f{2 \mc A_{D-2}}{D-1}.
\end{align}
Again, there is nothing special about $i = n-1$ and so the result will be the same regardless of the value of $i$.

Consequently, in either odd or even dimension, 
\begin{align}
    -\oint_{C_\infty} (\bs K^K_{\eta_i} - \overline{\bs K^K_{\eta_i}}) &= 2 \lim_{r\to\infty} \int d^{D-2} x \sqrt{-\bar g} (\bs{i}^K)_{\eta_i}^{tr} \nn 
    &= -\f{m_\infty a_i (D-1)}{8\pi \Xi_i \prod_{j=1}^n \Xi_j} \int d^{D-2} x \f{\mu_i^2 \prod_{j=1}^{n-1+\ve} \mu_j}{\mu_n} \nn 
    &= - \f{m_\infty a_i \mc A_{D-2}}{4\pi \Xi_i \prod_{j=1}^n \Xi_j}. \label{asymptoticetaKomar}
\end{align}

As with the $\beta$ calculation, for Kerr--AdS itself, where $\mu(r) = m$ (constant), the fact that $ d(\bs K_{\eta_i}^K - \overline{\bs K_{\eta_i}^K}) = 0$ implies that the result will be independent of integration surface (provided it encloses the singularity), so the result for this case is
\begin{align}
    -\oint_C \left( \bs K_{\eta_i}^K - \overline{\bs K_{\eta_i}^K}\right) &= -\f{m a_i \mc A_{D-2}}{4\pi \Xi_i \prod_{j=1}^n \Xi_j} \qquad \textrm{(Kerr--anti-de Sitter)}.
\end{align}
This is of course the result given by \cite{GibbonsPerry,DeruelleKatz05}.

Now I will use this result to find the Komar integral for arbitrary $r$ for the Generalized Kerr--AdS solution.

\subsubsection{Arbitrary Mass Function and Radius}

Using the expression for Kerr--AdS and substituting $\mu(r) = m$ into \eqref{Komaretaarbitraryr}, 
\begin{align}
    2\int d^{D-2}x \f{\prod_{j=1}^{n-1+\ve} \mu_j}{32 \pi \mu_n \prod_{j=1}^{n} \Xi_j} \f{\mu_i^2}{\Xi_i} \f{r(r^2+a_i^2) l^2 V}{(r^2+l^2)} \f{\pa}{\pa r} \left( \f{2 m a_i}{U (r^2+a_i^2)}\right) &= -\oint_C (\bs K^K_{\eta_i} - \overline{\bs K^K_{\eta_i}}) \qquad \textrm{(Kerr--AdS)} \nn 
    &= -\f{m a_i \mc A_{D-2}}{4\pi \Xi_i \prod_{j=1}^n \Xi_j} \nn 
    \int d^{D-2} x \f{\mu_i^2\prod_{j=1}^{n-1+\ve} \mu_j}{\mu_n} \f{\pa}{\pa r} \left(\f{1}{U(r^2+a_i^2)}\right) &= -\f{2 \mc A_{D-2} (r^2+l^2)}{r(r^2+a_i^2)l^2 V}\nn 
    &= -\f{2 r^{\ve} \mc A_{D-2}}{(r^2+a_i^2)\prod_{j=1}^{n-1+\ve} (r^2+a_j^2)} \label{muismdividedbym}
\end{align}

Integrating both sides with respect to $r$,
\begin{align}
    \int d^{D-2} x \f{\mu_i^2 \prod_{j=1}^{n-1+\ve} \mu_j}{\mu_n} \f{1}{U (r^2+a_i^2)} &= - 2 \mc A_{D-2} \int \f{r^\ve dr}{(r^2+a_i^2) \prod_{j=1}^{n-1+\ve}(r^2+a_j^2)}.
\end{align}

Purely as a mathematical trick, we note that this expression can be expressed in terms of a partial derivative over $a_i$, keeping the $a_j, j\neq i$ as well as $r$ constant:
\begin{align}
    \f{r^\ve}{(r^2+a_i^2)\prod_{j=1}^{n-1+\ve} (r^2+a_j^2)} &= -\f{1}{2a_i} \f{\pa}{\pa a_i} \f{r^\ve}{\prod_{j=1}^{n-1+\ve}(r^2+a_j^2)}.
\end{align}
Thus we have
\begin{align}
    \int d^{D-2} x \f{\mu_i^2 \prod_{j=1}^{n-1+\ve} \mu_j}{\mu_n} \f{1}{U(r^2+a_i^2)} &= \f{\mc A_{D-2} }{a_i} \f{\pa}{\pa a_i} \int \f{r^\ve dr}{\prod_{j=1}^{n-1+\ve} (r^2+a_j^2)}.
\end{align}
This is a very similar integral to that for the $\beta$-associated Komar term. As in that case I will choose the constant in the definite integral to be zero as $r \to \infty$. 

In even dimensions, using \eqref{evendimensiononebyr2a2integral},
\begin{align}
    &\f{\pa}{\pa a_i} \int \f{dr}{\prod_{j=1}^{n-1} (r^2+a_j^2)} = -\f{\pa}{\pa a_i} \sum_{j=1}^{n-1} \f{\arctan(a_j/r)}{a_j \Ups_j} \nn 
    &= -\sum_{j=1,j\neq i}^{n-1} \f{\arctan(a_j/r)}{a_j} \f{\pa}{\pa a_i} \f{1}{\Ups_j} - \arctan(a_i/r)\f{\pa}{\pa a_i} \left(\f{1}{a_i \Ups_i}\right) - \f{1}{a_i \Ups_i} \f{\pa}{\pa a_i}\arctan(a_i/r).
\end{align}
For $j \neq i$,
\begin{align}
    \f{\pa}{\pa a_i} \f{1}{\Upsilon_j} &= \f{\pa}{\pa a_i} \left(\f{1}{a_i^2-a_j^2} \prod_{k=1, k \neq i, j}^{n-1} \f{1}{a_k^2-a_j^2}\right) \nn 
    &= - \f{2 a_i}{(a_i^2-a_j^2)^2} \prod_{k=1,k\neq i,j}^{n-1} \f{1}{a_k^2-a_j^2} \nn 
    &= - \f{2 a_i}{(a_i^2-a_j^2) \Upsilon_j}. \label{dUpsjinvdai}
\end{align}

For $\Ups_i$, we have, similarly,
\begin{align}
    \f{\pa}{\pa a_i} \f{1}{a_i \Ups_i} &= -\f{1}{a_i^2 \Ups_i^2} \f{\pa}{\pa a_i} (a_i \Ups_i) \nn 
    &= -\f{1}{a_i^2 \Ups_i^2} \left( \Ups_i + a_i \f{\pa \Ups_i}{\pa a_i}\right).
\end{align}
Meanwhile, using $\Ups_i = \prod_{j=1,j\neq i}^{n-1} (a_j^2 - a_i^2)$,
\begin{align}
    \f{\pa \Ups_i}{\pa a_i} &= \sum_{j=1,j\neq i}^{n-1} \f{-2 a_i}{a_j^2-a_i^2} \Ups_i.
\end{align}
So we have
\begin{align}
    \f{\pa}{\pa a_i} \f{1}{a_i \Ups_i} &= -\f{1}{\Ups_i} \left( \f{1}{a_i^2} - \sum_{j=1,j\neq i}^{n-1} \f{2}{a_j^2-a_i^2}\right).
\end{align}

Also using $\pa \arctan(a_i/r)/\pa a_i = r/(r^2+a_i^2)$, we get
\begin{align}
    &\f{\pa}{\pa a_i} \int \f{dr}{\prod_{j=1}^{n-1} (r^2+a_j^2)} \nn 
    &= \sum_{j=1,j\neq i}^{n-1} \f{2 a_i}{a_j(a_i^2-a_j^2) \Ups_j} \arctan(a_j/r) + \f{\arctan(a_i/r)}{\Ups_i} \left( \f{1}{a_i^2} - \sum_{j=1,j\neq i}^{n-1} \f{2}{a_j^2-a_i^2}\right) - \f{r}{a_i \Ups_i(r^2+a_i^2)} \nn 
    &= \sum_{j=1,j\neq i}^{n-1} \f{ 2 a_i}{a_i^2-a_j^2} \left( \f{\arctan(a_j/r)}{a_j \Ups_j} + \f{\arctan(a_i/r)}{a_i \Ups_i}\right) + \f{\arctan(a_i/r)}{a_i^2 \Ups_i} - \f{r}{a_i \Ups_i(r^2+a_i^2)},
\end{align}
so
\begin{align}
    &\int d^{D-2}x \f{\mu_i^2 \prod_{j=1}^{n-1+\ve} \mu_j}{\mu_n} \f{1}{U(r^2+a_i^2)}\nn &=  \left(  \sum_{j=1,j\neq i}^{n-1} \f{ 2 }{a_i^2-a_j^2} \left( \f{\arctan(a_j/r)}{a_j \Ups_j} + \f{\arctan(a_i/r)}{a_i \Ups_i}\right) + \f{\arctan(a_i/r)}{a_i^3 \Ups_i} - \f{r}{a_i^2 \Ups_i(r^2+a_i^2)} \right) \mc A_{D-2}.
\end{align}

In odd dimensions, from \eqref{odddimintrbyr2a2} we have
\begin{align}
    \f{\pa}{\pa a_i} \int \f{r dr}{\prod_{j=1}^n (r^2+a_j^2)} &= \f{\pa}{\pa a_i} \sum_{j=1}^n \f{\ln(1+a_j^2/r^2)}{2 \Ups_j} \nn 
    &= -\sum_{j=1,j\neq i}^n \f{\ln (1 + a_j^2/r^2) }{2} \f{\pa}{\pa a_i}\f{1}{\Ups_j} - \f{\ln (1+a_i^2/r^2)}{2} \f{\pa}{\pa a_i}\f{1}{\Ups_i} - \f{1}{2 \Ups_i} \f{\pa}{\pa a_i} \ln (1+a_i^2/r^2).
\end{align}
We can use \eqref{dUpsjinvdai}. $\pa \ln (1+a_i^2/r^2)/\pa a_i = 2 a_i/(r^2+a_i^2)$. Additionally, using $\Ups_i = \prod_{j=1,j\neq i}^n (a_j^2-a_i^2)$ (in odd dimensions),
\begin{align}
    \f{\pa}{\pa a_i} \f{1}{\Ups_i} &= -\f{1}{\Ups_i^2} \f{\pa \Ups_i}{\pa a_i} \nn 
    &= \f{2 a_i}{\Ups_i} \sum_{j=1,j\neq i}^n \f{1}{a_j^2-a_i^2}.
\end{align}
This gives
\begin{align}
    &\f{\pa}{\pa a_i} \int \f{r dr}{\prod_{j=1}^n (r^2+a_j^2)} \nn 
    &= \sum_{j=1,j\neq i}^n \f{a_i}{(a_i^2-a_j^2)\Ups_j} \ln (1+a_j^2/r^2) - \f{a_i}{\Ups_i} \sum_{j=1,j\neq i}^n \f{1}{a_j^2-a_i^2}\ln(1+a_i^2/r^2) -\f{a_i}{(r^2+a_i^2) \Ups_i} \nn 
    &=  \sum_{j=1,j\neq i}^n \f{a_i}{a_i^2-a_j^2} \left( \f{\ln(1+a_j^2/r^2)}{\Ups_j} + \f{\ln(1+a_i^2/r^2)}{\Ups_i}\right) - \f{a_i}{(r^2+a_i^2)\Ups_i}.
\end{align}
so that the odd-dimensional case is
\begin{align}
    &\int d^{D-2}x \f{\mu_i^2 \prod_{j=1}^n \mu_j}{\mu_n} \f{1}{U (r^2+a_i^2)} \nn 
    &=- \left( \sum_{j=1,j\neq i}^n \f{1}{a_i^2-a_j^2} \left( \f{\ln(1+a_j^2/r^2)}{\Ups_j} + \f{\ln(1+a_i^2/r^2)}{\Ups_i}\right) - \f{1}{(r^2+a_i^2)\Ups_i}\right) \mc A_{D-2}
\end{align}

The Komar term for the generalized Kerr--AdS solution is
\begin{align}
    -\oint_C (\bs K^K_{\eta_i} - \overline{\bs K^K_{\eta_i}}) &= -\oint_C \bs K^K_{\eta_i} \nn 
    &= 2\int d^{D-2}x \f{\prod_{j=1}^{n-1+\ve} \mu_j}{32 \pi \mu_n \prod_{j=1}^{n} \Xi_j} \f{\mu_i^2}{\Xi_i} \f{r(r^2+a_i^2) l^2 V}{(r^2+l^2)} \f{\pa}{\pa r} \left( \f{2 \mu(r) a_i}{U (r^2+a_i^2)}\right)  \nn 
    &= \f{4 a_i r (r^2+a_i^2) l^2 V}{32 \pi \Xi_i (\prod_{j=1}^n \Xi_j) (r^2+l^2)} \int d^{D-2} x \f{\mu_i^2 \prod_{j=1}^{n-1+\ve} \mu_j}{\mu_n}\left( \mu(r) \f{\pa}{\pa r}\f{1}{U(r^2+a_i^2)} + \f{\mu'(r)}{U(r^2+a_i^2)}\right),
\end{align}
giving
\begin{align}
    &-\oint_C (\bs K_{\eta_i}^K - \overline{\bs K^K_{\eta_i}}) = \f{-a_i \mc A_{D-2}}{4 \pi \Xi_i \prod_{j=1}^{n-1} \Xi_j} \left\{ \mu(r) - \f12 \mu'(r)(r^2+a_i^2) \left( \prod_{j=1}^{n-1}(r^2+a_j^2)\right) \right.\times \nn
    &\qquad \left.\left(  \sum_{j=1,j\neq i}^{n-1} \f{ 2 }{a_i^2-a_j^2} \left( \f{\arctan(a_j/r)}{a_j \Ups_j} + \f{\arctan(a_i/r)}{a_i \Ups_i}\right) + \f{\arctan(a_i/r)}{a_i^3 \Ups_i} - \f{r}{a_i^2 \Ups_i(r^2+a_i^2)} \right) \right\} \label{Komaretaeven}
\end{align}
in even dimensions and
\begin{align}
    &-\oint_C (\bs K_{\eta_i}^K - \overline{\bs K^K_{\eta_i}}) = \f{-a_i \mc A_{D-2}}{4 \pi \Xi_i \prod_{j=1}^n \Xi_j}\left\{ \mu(r) + \f{\mu'(r) (r^2+a_i^2)\prod_{j=1}^n(r^2+a_j^2)}{2 r} \times \right. \nn 
    &\qquad \left. \left( \sum_{j=1,j\neq i}^n \f{1}{a_i^2-a_j^2} \left( \f{\ln(1+a_j^2/r^2)}{\Ups_j} + \f{\ln(1+a_i^2/r^2)}{\Ups_i}\right) - \f{1}{(r^2+a_i^2)\Ups_i}\right)\right\} \label{Komaretaodd}
\end{align}
in odd dimensions. 

This is also equal to simply $\oint_C \bs I_{\eta_i}$, since $\oint_C \eta_i \cdot \bs \tht = 0$.

\subsection{Asymptotically Static Vector}

Here I relate the charges associated with $\xi = \f{\pa}{\pa t}$. Since $\xi^t = 1$, for the non-Komar term we can just use the expressions involving $\ups$. For the Komar terms, we have two options. One is to attempt to evaluate the Komar terms directly. The other is to write the conserved charges associated with $\xi$ by using the breakdown $\xi = \beta - \sum_{i=1}^{n-1+\ve} \f{a_i}{l^2} \eta_i$ and the expressions associated with $\beta$ and $\eta_i$.

If we start form \eqref{Komarbreakdown}, with $\chi = \xi$, a few terms simplify and we have
\begin{align}
    (\bs i^K_\xi)^{tr} &= \f1{32\pi} \left[ k_t k^t \bar g^{rr} \pa_r H - H (k^t k^t \bar g^{rr} \pa_r \bar g_{tt} + \bar g^{tt} \pa_r \bar g_{tt})\right] \nn 
    &= \f{1}{32\pi} \left[ k_t k^t \bar g^{rr} \pa_r H - H \pa_r \bar g_{tt} ( (k^t)^2 \bar g^{rr} + \bar g^{tt})\right].
\end{align}
We can rewrite $(k^t)^2 \bar g^{rr} + \bar g^{tt} = \bar g^{rr} \bar g^{tt} (k^t k_t + k^r k_r)$, but this does not significantly help. While this is a little simpler than the general expression \eqref{Komarbreakdown}, it is not simpler than the ones for $(\bs i^K_\beta)^{tr}$ or $(\bs i^K_{\eta_i})^{tr}$. Despite the privileged status $\xi$ holds in the spacetime globally, and its role in black hole mechanics, the conserved charges associated with $\xi$ are easier to think of as being a linear combination of the ones associated with $\beta$, which simplifies due to the properties of the principal vector, and $\eta_i$, which simplify due to the fact that $\eta_i$ is tangent to the integration surfaces $C$. 

Putting in explicit values,
\begin{align}
    (\bs i^K_\xi)^{tr} &= \f1{32\pi} \left[ \f{(k^t)^2 \bar g_{tt}}{\bar g_{rr}} \pa_r H - H \pa_r \bar g_{tt} \left(\f{(k^t)^2}{\bar g_{rr}} + \f{1}{\bar g_{tt}} \right) \right] \nn 
    &= \f1{32\pi \bar g_{tt} \bar g_{rr}} \left[ (k_t)^2 \pa_r H - H \pa_r \bar g_{tt} ( \bar g_{tt} (k^t)^2 + \bar g_{rr})\right] \nn 
    &= -\f{1}{32 \pi (l^2+r^2)^2 F} \left( W l^2 (r^2+l^2) \pa_r H + 2 r H (F(r^2+l^2) - l^2 W)\right) \nn 
    \sqrt{-\bar g} (\bs i_\xi^K)^{tr} &= - \frac{r U \prod_{i=1}^{n-1+\ve} \mu_i}{\mu_{n} \prod_{j = 1}^{n} \Xi_j} \f{1}{32 \pi (l^2+r^2)^2 F} \left( W l^2 (r^2+l^2) \pa_r H + 2 r H (F(r^2+l^2) - l^2 W)\right) \nn 
    &= - \f{r V \prod_{i=1}^{n-1+\ve} \mu_i}{32 \pi (l^2+r^2)^2 \mu_n \prod_{j=1}^n \Xi_j} (W l^2 (r^2+l^2) \pa_r H + 2 r H (F (r^2+l^2)-l^2W)).
\end{align}
It is not immediately clear that integrating this will be a simple matter, so I will turn to evaluating the Komar integral by summing the terms for $\beta$ and $\eta_i$. 

The linear combination is
\begin{align}
    -\oint_C (\bs K^K_\xi - \overline{\bs K^K_\xi}) &= -\oint_C (\bs K^K_\bt - \overline{\bs K^K_\bt}) + \sum_{i=1}^{n-1+\ve}\f{a_i}{r^2+a_i^2} \oint_C \bs K^K_{\eta_i}.
\end{align}

The asymptotic solutions are from \eqref{KomarIntegralTermforbetaasymptotic} and \eqref{asymptoticetaKomar}, and so give
\begin{align}
    -\oint_{C_\infty} (\bs K^K_\xi - \overline{\bs K^K_\xi}) &= \f{m_\infty \mc A_{D-2}}{8\pi \prod_{j=1}^n \Xi_j} \left( \sum_{i=1}^{n-1+\ve} \Xi_i^{-1} -\ve-1\right).
\end{align}
This implies, using \eqref{upsilonthetalimit}
\begin{align}
    \oint_{C_\infty} \bs I_\xi &= \oint_{C_\infty} \left( - \bs K^K_\xi - \overline{\bs K^K_\xi} - \xi \cdot \bs \tht\right) \nn 
    &= \f{m_\infty \mc A_{D-2}}{4\pi \prod_{j=1}^n \Xi_j} \left(  \sum_{i=1}^{n-1+\ve} \Xi_i^{-1} - \f\ve2\right).
\end{align}
In the case where $\mu(r) = m = const.$, we get, on arbitrary surface $C$,
\begin{align}
    -\oint_{C} (\bs K^K_\xi - \overline{\bs K^K_\xi}) &= \f{m \mc A_{D-2}}{8\pi \prod_{j=1}^n \Xi_j} \left( 2\sum_{i=1}^{n-1+\ve} \Xi_i^{-1} -\ve - 1\right) \nn
    \oint_C \bs I_\xi &=\f{m \mc A_{D-2}}{4\pi \prod_{j=1}^n \Xi_j} \left(  \sum_{i=1}^{n-1+\ve} \Xi_i^{-1} - \f\ve2\right),
\end{align}
matching the results for $\mc E = \oint_C \bs I_\xi$.

With a general $\mu(r)$ and at arbitrary radius, we use \eqref{evendimKomarbeta}, \eqref{odddimKomarbeta}, \eqref{Komaretaeven} and \eqref{Komaretaodd}. 

In even dimensions, 
\begin{align}
    &-\oint_C (\bs K_\xi^K - \overline{\bs K_\xi^K}) = \nn
    &\f{\mc A_{D-2}}{8\pi \prod_{j=1}^n \Xi_j} \left[  \mu(r) \left( D-3 + \sum_{i=1}^{n-1} \f{2 a_i^2}{l^2 \Xi_i}\right) - \mu'(r) \left(\prod_{j=1}^{n-1} (r^2+a_j^2) \right) \sum_{i=1}^{n-1} \left\{ \f{\arctan(a_i/r)}{a_i \Ups_i}(D-3) + \f{a_i^2(r^2+a_i^2)}{l^2-a_i^2}\times \right.\right. \nn 
    & \left. \left. \qquad\qquad\qquad\left(  \sum_{j=1,j\neq i}^{n-1} \f{ 2 }{a_i^2-a_j^2} \left( \f{\arctan(a_j/r)}{a_j \Ups_j} + \f{\arctan(a_i/r)}{a_i \Ups_i}\right) + \f{\arctan(a_i/r)}{a_i^3 \Ups_i} - \f{r}{a_i^2 \Ups_i(r^2+a_i^2)} \right)\right\} \right]
\end{align}
As in the $\mu(r) = m$ case, in even dimension, $D-3+\sum_{i=1}^{n-1} 2 a_i^2/(l^2 \Xi_i) = 2 \sum_{i=1}^{n-1} \Xi_i^{-1} - 1$. The terms multiplying $\mu'(r)$ either involve $\arctan(a_i/r)$ for one of the $a_i$ or do not. The non-arctan terms simplify:
\begin{align}
    \sum_{i=1}^{n-1} \f{a_i^2(r^2+a_i^2)}{l^2-a_i^2} \f{r}{a_i^2 \Ups_i(r^2+a_i^2)} &= r\sum_{i=1}^{n-1}\f{1}{(l^2-a_i^2)\Ups_i} \nn 
    &= r \sum_{i=1}^{n-1} \f{1}{\hat \Ups_i} \nn 
    &= -\f{r}{\hat \Ups_0} \nn 
    &= -\f{r}{l^{2(n-1)} \prod_{j=1}^{n-1} \Xi_j},
\end{align}
using \eqref{sumhatUpsi}.  For the arctan terms, we have,
\begin{align}
    &\sum_{i=1}^{n-1} \f{a_i^2(r^2+a_i^2)}{l^2-a_i^2} \sum_{j=1,j\neq i}^{n-1} \f{2}{a_i^2-a_j^2} \left( \f{\arctan(a_j/r)}{a_j \Ups_j} + \f{\arctan(a_i/r)}{a_i \Ups_i}\right)  \nn 
    &= \sum_{i=1}^{n-1} \sum_{j=1,j\neq i}^{n-1} \f{2 a_i^2(r^2+a_i^2) \arctan(a_j/r)}{(l^2-a_i^2) (a_i^2-a_j^2) a_j \Ups_j} + \sum_{i=1}^{n-1} \sum_{j=1,j\neq i}^{n-1} \f{2 a_i^2(r^2+a_i^2) \arctan(a_i/r)}{(l^2-a_i^2) a_i \Ups_i (a_i^2-a_j^2)}
\end{align}
Relabel the first sum $i \leftrightarrow j$. 
\begin{align}
&= \sum_{i=1}^{n-1} \f{2 \arctan (a_i/r)}{a_i \Ups_i} \sum_{j=1,j\neq i}^{n-1} \f{a_j^2(r^2+a_j^2)(l^2-a_i^2)-a_i^2(r^2+a_i^2)(l^2-a_j^2)}{(a_j^2-a_i^2)(l^2-a_i^2)(l^2-a_j^2)} \nn 
&= \sum_{i=1}^{n-1} \f{2 \arctan(a_i/r)}{ a_i \Ups_i} \sum_{j=1,j\neq i}^{n-1} \f{(a_i^2+a_j^2+r^2)l^2-a_i^2a_j^2}{l^4 \Xi_i \Xi_j}\nn 
&= \sum_{i=1}^{n-1} \f{2 \arctan(a_i/r)}{ l^2 a_i \Ups_i \Xi_i} \sum_{j=1,j\neq i}^{n-1} \f{a_i^2 (l^2-a_j^2) + l^2 (r^2 + a_j^2) }{l^2 \Xi_j} \nn 
&= \sum_{i=1}^{n-1} \f{2 \arctan(a_i/r)}{l^2 a_i \Ups_i \Xi_i} \sum_{j=1,j\neq i}^{n-1} \left( a_i^2 + \f{r^2+a_j^2}{\Xi_j}\right) \nn 
&= \sum_{i=1}^{n-1} \f{2 \arctan(a_i/r)}{l^2 a_i \Ups_i \Xi_i} \left( (n-2)a_i^2 + \sum_{j=1,j\neq i}^{n-1} \f{r^2+a_j^2}{\Xi_j}\right) \nn 
&= \sum_{i=1}^{n-1} \f{2 \arctan(a_i/r)}{l^2 a_i \Ups_i \Xi_i} \left( (n-2) a_i^2 + \sum_{j=1,j\neq i}^{n-1} \f{r^2 + l^2 - (l^2-a_i^2)}{\Xi_j}\right) \nn 
&= \sum_{i=1}^{n-1} \f{2 \arctan(a_i/r)}{l^2 a_i \Ups_i \Xi_i} \left( (n-2) a_i^2 + (r^2+l^2) \sum_{j=1,j\neq i}\Xi_j^{-1} - l^2 (n-2)\right) \nn 
&= \sum_{i=1}^{n-1} \f{2 \arctan(a_i/r)}{l^2 a_i \Ups_i \Xi_i} \left( -(n-2) l^2 \Xi_i + (r^2 +l^2) \sum_{j=1,j\neq i}^{n-1} \Xi_j^{-1}\right) \nn 
&= \sum_{i=1}^{n-1} \f{2 \arctan(a_i/r)}{a_i \Ups_i \Xi_i} \left( (1+r^2/l^2) \sum_{j=1,j\neq i}^{n-1} \Xi_j^{-1} - (n-2) \Xi_i\right).
\end{align}
We then have,
\begin{align}
    &\sum_{i=1}^{n-1} \left\{ \f{\arctan(a_i/r)}{a_i \Ups_i}(D-3) + \f{a_i^2(r^2+a_i^2)}{l^2-a_i^2}\times \right. \nn 
    &\left.\left(  \sum_{j=1,j\neq i}^{n-1} \f{ 2 }{a_i^2-a_j^2} \left( \f{\arctan(a_j/r)}{a_j \Ups_j} + \f{\arctan(a_i/r)}{a_i \Ups_i}\right) + \f{\arctan(a_i/r)}{a_i^3 \Ups_i} - \f{r}{a_i^2 \Ups_i(r^2+a_i^2)} \right)\right\} \nn 
    &= \sum_{i=1}^{n-1} \f{\arctan(a_i/r)}{a_i \Ups_i} \left\{ D-3 + \f{2}{\Xi_i}\left((1+r^2/l^2) \sum_{j=1,j\neq i}^{n-1} \Xi_j^{-1} - (n-2) \Xi_i \right) + \f{r^2+a_i^2}{l^2-a_i^2}\right\} + \f{ r}{l^{2(n-1)} \prod_{j=1}^{n-1} \Xi_j} \nn 
    &= \sum_{i=1}^{n-1} \f{2\arctan(a_i/r)(1+r^2/l^2)}{a_i \Ups_i \Xi_i}\left( \f12 + \sum_{j=1,j\neq i}^{n-1} \Xi_j^{-1}\right) + \f{r}{l^{2(n-1)} \prod_{j=1}^{n-1} \Xi_j}
\end{align}
using $D = 2n$ (and some other simplifications) on the last line. We then have,
\begin{align}
    &-\oint_C (\bs K_\xi^K - \overline{\bs K^K_\xi}) = \f{\mc A_{D-2}}{8\pi \prod_{j=1}^n \Xi_j}\left[ \mu(r)\left( 2 \sum_{i=1}^{n-1} \Xi_i^{-1} -1\right) \right. \nn 
    &\qquad \left.- \mu'(r) \left(\prod_{j=1}^{n-1} (r^2+a_j^2)\right) \left\{\sum_{i=1}^{n-1} \f{\arctan(a_i/r)(1+r^2/l^2)}{a_i \Ups_i \Xi_i}\left( 1 + 2\sum_{j=1,j\neq i}^{n-1} \Xi_j^{-1}\right)  + \f{r}{l^{2(n-1)} \prod_{j=1}^{n} \Xi_j} \right\}\right]
\end{align}
(for even dimensions). (Recall $\Xi_n = 1$.) 

In odd dimensions, 
\begin{align}
    -\oint_C (\bs K^K_\xi - \overline{\bs K^K_\xi}) &= \f{(D-3) \mc A_{D-2}}{8\pi \prod_{j=1}^n \Xi_j} \left( \mu(r) + \mu'(r)\sum_{i=1}^{n} \f{\ln(1+a_i^2/r^2)}{2 \Upsilon_i} \f{\prod_{j=1}^n (r^2+a_j^2)}{r}\right) \nn 
    &\qquad + \sum_{i=1}^{n} \f{a_i^2 \mc A_{D-2}}{4 \pi l^2 \Xi_i \prod_{j=1}^n \Xi_j}\left\{ \mu(r) + \f{\mu'(r) (r^2+a_i^2)\prod_{j=1}^n(r^2+a_j^2)}{2 r} \times \right. \nn 
    &\qquad \left. \left( \sum_{j=1,j\neq i}^n \f{1}{a_i^2-a_j^2} \left( \f{\ln(1+a_j^2/r^2)}{\Ups_j} + \f{\ln(1+a_i^2/r^2)}{\Ups_i}\right) - \f{1}{(r^2+a_i^2)\Ups_i}\right)\right\} 
\end{align}

We have,
\begin{align}
    \sum_{i=1}^n \f{a_i^2 (r^2+a_i^2)}{l^2 \Xi_i} \f{1}{(r^2+a_i^2)\Ups_i} &= \sum_{i=1}^n \f{a_i^2}{(l^2-a_i^2)\Ups_i} \nn 
    &= \sum_{i=1}^n \f{l^2 - (l^2-a_i^2)}{(l^2-a_i^2)\Ups_i} \nn
    &= \sum_{i=1}^n \left( \f{l^2}{(l^2-a_i^2)\Ups_i} - \f{1}{\Ups_i}\right) \nn 
    &= \sum_{i=1}^n \left( \f{l^2}{\hat \Ups_i} - \f{1}{\Ups_i}\right) \nn 
    &= - \f{l^2}{\hat \Ups_0} \nn 
    &= - \f{1}{l^{2(n-1)} \prod_{j=1}^n \Xi_j},
\end{align}
using \eqref{sumUpsi} and \eqref{sumhatUpsi}.

We also have, relabelling $i \leftrightarrow j$ for the $\ln (1+a_j^2/r^2)$ terms,
\begin{align}
    &\sum_{i=1}^n \f{a_i^2 (r^2+a_i^2)}{l^2 \Xi_i} \sum_{j=1,j\neq i}^n \f{1}{a_i^2-a_j^2} \left( \f{\ln(1+a_j^2/r^2)}{\Ups_j} + \f{\ln(1+a_i^2/r^2)}{\Ups_i}\right) \nn 
    &= \sum_{i=1}^n \f{\ln(1+a_i^2/r^2)}{\Ups_i} \sum_{j=1,j\neq i}^n \f{1}{a_i^2-a_j^2} \left( \f{a_i^2(r^2+a_i^2)}{l^2-a_i^2} - \f{a_j^2(r^2+a_j^2)}{l^2-a_j^2}\right) \nn 
    &= \sum_{i=1}^n \f{\ln(1+a_i^2/r^2)}{\Ups_i} \sum_{j=1,j\neq i}^n \f{l^2 a_i^2 + l^2 a_j^2 + l^2 r^2 - a_i^2 a_j^2}{l^4 \Xi_i \Xi_j} \nn
    &= \sum_{i=1}^n \f{\ln(1+a_i^2/r^2)}{\Ups_i} \sum_{j=1,j\neq i}^n \left( \f{1+r^2/l^2}{\Xi_i\Xi_j} - 1\right) \nn 
    &= \sum_{i=1}^n \f{\ln (1+a_i^2/r^2)}{\Ups_i} \left( \f{1+r^2/l^2}{\Xi_i} \sum_{j=1,j\neq i}^n \Xi_j^{-1} - (n-1)\right)
\end{align}

We then have
\begin{align}
    -\oint_C (\bs K^K_\xi - \overline{\bs K_\xi^K}) &= \f{\mc A_{D-2}}{4 \pi \prod_{j=1}^n \Xi_j} \left[ \left( \sum_{i=1}^n \Xi_i^{-1} - 1\right) \mu(r) + \f{\mu'(r) \prod_{j=1}^n (r^2+a_j^2)}{2 r}\times   \right.  \nn 
    &\qquad \left. \left( \f{1}{l^{2(n-1)}\prod_{j=1}^n \Xi_j} + (1+r^2/l^2)\sum_{i=1}^n \f{\ln(1+a_i^2/r^2)}{\Ups_i \Xi_i}  \sum_{j=1,j\neq i}^n \Xi_j^{-1} \right) \right].
\end{align}

The non-Komar term is given by the right hand side of \eqref{upsilonthetagenericmu}, since $\xi^t = 1$. The total charge is
\begin{align}
    \oint_C \bs I_\xi &= -\oint_C( \bs K^K_\xi - \overline{\bs K^K_\xi}) - \oint_C \xi \cdot \bs \tht.
\end{align}
In even dimensions, this is 
\begin{align}
    \oint_C \bs I_\xi &= \f{\mc A_{D-2}}{8\pi \prod_{j=1}^n \Xi_j}\left[ 2\mu(r) \sum_{i=1}^{n-1} \Xi_i^{-1} +\mu'(r) \left\{ r - \left( \prod_{j=1}^{n-1} (r^2+a_j^2)\right) \times \right. \right.  \nn 
    &\qquad \left. \left. \left\{\sum_{i=1}^{n-1} \f{\arctan(a_i/r)(1+r^2/l^2)}{a_i \Ups_i \Xi_i}\left( 1 + 2\sum_{j=1,j\neq i}^{n-1} \Xi_j^{-1}\right)  + \f{r}{l^{2(n-1)} \prod_{j=1}^{n} \Xi_j} \right\} \right\} \right]
\end{align}
In odd dimensions,
\begin{align}
    \oint_C \bs I_\xi &= \f{\mc A_{D-2}}{8 \pi \prod_{j=1}^n \Xi_j} \left[ \left( 2\sum_{i=1}^n \Xi_i^{-1} - 1\right) \mu(r) + \mu'(r) \left \{ r + \f{\prod_{j=1}^n (r^2+a_j^2)}{r}\right.\times   \right.  \nn 
    &\qquad \left. \left. \left( \f{1}{l^{2(n-1)}\prod_{j=1}^n \Xi_j} + (1+r^2/l^2)\sum_{i=1}^n \f{\ln(1+a_i^2/r^2)}{\Ups_i \Xi_i}  \sum_{j=1,j\neq i}^n \Xi_j^{-1} \right)\right\} \right].
\end{align}
These are the most complicated expressions yet! Interestingly, the conserved charge $\oint_C \bs I_\chi$ simplifies if we use a ``co-rotating'' Killing vector, which I specify in the following subsection. 

\subsection{``Co-rotating'' Killing Vector Charge} \label{corotating} 

Here I will state some results for the full conserved charge associated with the vector $\zeta$, and a more general form of it. As I will show, the term $\mu'(r)$ (the \emph{derivative} of the scalar mass function) drops out of these expressions entirely, due to a fortuitous cancellation of terms. An obvious generalization of $\z$ is a vector proportional to $e_{\hat n}$, that is to say, in the direction of the orthonormal frame vector which points in the direction of $\z$ on the horizon. 

In arbitrary dimension, we have $(\bs I_\chi)_{c_1 \ldots c_{D-2}} = \bs{i}^{a b}_\chi \bs \ep_{a b c_1 \ldots c_{D-2}}$. Using $h_{a b} = H k_a k_b$, we can rewrite $\bs{i}_\chi$ from \eqref{mcKab} as
\begin{align}
    \bs{i}^{a b}_\chi &= \f{1}{16\pi}\left[ H \left( k^d k^{[a} \bar \na_d \chi^{b]} - \chi^d \bar \na^{[a}\left( k^{b]} k_d\right) + \chi^{[a}\bar \na_d\left( k^{b]} k^d\right) \right) - \chi^d k_d k^{[b} \bar \na^{a]}H +  \chi^{[a}k^{b]} k^d \bar \na_d H \right]
\end{align}
We can now make some observations. Assume that we are in the generalized Kerr--AdS situation. Define $Z$ to be the vector field
\begin{align}
    Z &\equiv \partial_t + \sum_i \Omega_i(r) \partial_{\phi_i},
\end{align}
where $\Omega_i(r) = a_i(r^2+l^2)/[l^2(r^2+a_i^2)]$. We have that \emph{on the horizon}, $\zeta = Z$ (since there $\Omega_i(r_+) = \Omega_i$). We also have $Z \propto e_{\hat n}$, so that it is associated with the natural orthonormal frame. Then we have
\begin{align}
    k &= -\frac{l^2}{l^2+r^2} Z + \f{\pa}{\pa r}.
\end{align}
Let $w \equiv \pa_r$ (for this section only), so that we can rewrite this as
\begin{align}
    Z &= - \left(1 + \f{r^2}{l^2}\right) ( k - w).
\end{align}

While $\bs{i}^{a b}_\chi$ was defined specifically for Killing vectors $\chi$, it will be useful for intermediate steps to show how it works for an arbitrary vector in place of $\chi$. For example, we find immediately that $\bs{i}_k^{a b} = 0$. Explicitly,
\begin{align}
    \bs{i}^{a b}_k &= \f{1}{16\pi}\left[ H \left( k^d k^{[a} \bar \na_d k^{b]} - k^d \bar \na^{[a}\left( k^{b]} k_d\right) + k^{[a}\bar \na_d\left( k^{b]} k^d\right) \right) - k^d k_d k^{[b} \bar \na^{a]}H +  k^{[a}k^{b]} k^d \bar \na_d H \right].
\end{align}
$k^{[a} k^{b]} = k^{[a}\bar \na_d k^{b]} = k^{[a} \bar \na_d k^{b]} k^d = k^{[a}k^{b]} \bar \na_d k^d = 0$ due to the antisymmetrization over symmetric expressions. The other terms vanish because $k$ is null. (For example, $k^d \bar \na^a k_d = \f12 \bar \na^a (k^d k_d) = 0$.)

This can actually be extended to a more general case. For a scalar function $f$ which depends on the coordinates in general,
\begin{align}
    \bs{i}_{f k}^{a b} &= f \bs{i}^{a b}_k + \f{1}{16\pi} H \left(k^d k^{[b}k^{a]} \bar \na_d f - k^d k_d k^{[b}\bar\na^{a]} f + k^d k^{[a}k^{b]} \bar \na_d f\right) \nn 
    &= 0.
\end{align}

At this point I will show that the combination $\bs{i}^{a b}_\xi + \sum_i \Omega_i(r) \bs{i}^{a b}_{\eta_i}$ does not depend on $\mu'(r)$, in arbitrary dimension. The limits on the sum over $i$ are from 1 to $n-1+\ve$.

This combination is \emph{almost} equal to $\bs{i}^{ab}_Z$, but because the $\Omega_i(r)$ terms are taken ``out front'' all the terms which involve derivatives of the $\Omega_i(r)$ must be subtracted off:
\begin{align}
    \bs{i}^{ab}_\xi + \sum_i \Omega_i(r) \bs{i}^{ab}_{\eta_i} &= \bs{i}^{ab}_Z - \f1{16\pi}\sum_{i} h^{d[a} \left(\bar \na_d \Omega_i(r)\right) \eta_i^{b]} \nn 
    &= \bs{i}_Z^{ab} - \f1{16\pi} H k^{[a}\eta_i^{b]} k^d \bar \na_d \Omega_i(r) \nn 
    &= \bs{i}_Z^{ab} - \f{1}{16\pi} H k^{[a} \eta_i^{b]} \Omega_i'(r),
\end{align}
where $\Omega_i'(r) = d \Omega_i(r)/dr$. Note that when taking $\bs{i}^{tr}_\xi + \sum_i \Omega_i(r) \bs{i}^{tr}_{\eta_i}$, the last term disappears because $k^{[t}\eta_i^{r]} = 0$ (since $\eta_i^t = \eta_i^r = 0$).

Return then to $\bs i_Z^{ab}$. The only terms that might depend on $\mu'(r)$ are terms which involve the $H$ derivative. They will be (up to the $1/16\pi$ factor)
\begin{align}
    A^{ab} &\equiv -Z^d k_d k^{[b}\bar \na^{a]}H + Z^{[a} k^{b]} k^d \bar \na_d H,
\end{align}
where $A^{ab}$ is just a temporary name for the next few lines. I will show that $A^{ab}$ never has terms of the form $\mu'(r)$. 

It is convenient to rewrite this in terms of $w$. Note,
\begin{align}
    k_d Z^d &= -\left(1+\f{r^2}{l^2}\right) k_d (k^d - w^d) \nn 
    &= \left(1+\f{r^2}{l^2}\right) k_d w^d \nn 
    &= \left(1+\f{r^2}{l^2}\right) k_r \nn 
    k^{[a} Z^{b]} &= -\left(1 + \f{r^2}{l^2}\right) k^{[a}\left( k^{b]} - w^{b]}\right) \nn 
    &= \left(1 + \f{r^2}{l^2}\right) k^{[a} w^{b]}.
\end{align}
Then,
\begin{align}
    A^{ab} &= \left(1+\f{r^2}{l^2}\right) \left( -k_r k^{[b} \bar \na^{a]} H + w^{[a} k^{b]} k^d \bar \na_d H\right).
\end{align}
$k^d \bar \na_d H = k^d \pa_d H = k^r \pa_r H$, since $H$ does not depend on $t$ or $\phi_i$. We can rewrite this as $k_r \bar \na^r H$, using the diagonality of the metric, to have
\begin{align}
    A^{ab} &= \left(1 + \f{r^2}{l^2}\right) k_r \left( -k^{[b} \bar \na^{a]} H + w^{[a} k^{b]} \bar \na^r H\right) \nn 
    &= \left(1 + \f{r^2}{l^2}\right) k_r  k^{[b} \left(\de^{a]}_r \bar \na^r H - \bar \na^{a]} H \right).
\end{align}
Using coordinates $(t,r,y_\alpha,\phi_i)$, we note that $H$ only has nonzero derivatives $\bar \na^r H$ and $\bar \na^{y_\alpha}H$ and that $k$ only has nonzero components $k^r,k^t,k^{\phi_i}$. Any terms of the form where $a$ or $b$ are $r$ will have contributions from both $\de^{a]}_r \bar \na^r H$ and $-\bar \na^{a]} H$, which cancel, so that the $r$ derivative terms exactly cancel! The only terms which survive are ones for which the $y_\alpha$ terms are nonzero; they have the form
\begin{align}
    A^{y_\alpha b} &= -\f12 \left(1 + \f{r^2}{l^2}\right) k_r k^{[b} \bar \na^{y_\alpha]} H.
\end{align}
Since $\mu(r)$ is a function of $r$ only, there are no terms in $\bs i_Z^{ab}$ which depend on $\mu'(r)$. Further, $\bs i_Z^{tr}$ has no terms which depend on derivatives of $H$ at all.

What remains then is
\begin{align}
    \bs i_Z^{tr} &= \f{H}{16\pi} \left( k^d k^{[t} \bar \na_d Z^{r]} - Z^d \bar \na^{[t}(k^{r]} k_d) + Z^{[t} \bar \na_d (k^{r]}k^d)\right) \nn 
    &= \f{H}{16\pi} \left( k^d k^{[t}\bar \na_d Z^{r]} - Z^d k_d \bar \na^{[t}k^{r]} - Z^d k^{[r}\bar \na^{t]} k_d + Z^{[t}k^{r]} \bar \na_d k^d\right),
\end{align}
using $k^d \bar \na_d k^b = 0$. We can further rearrange to
\begin{align}
    \bs i_Z^{tr} &= \f{H}{16\pi} \left( k^d k^{[t}\bar \na_d Z^{r]} - Z^d k_d \bar \na^{[t}k^{r]} - k^{[r}\bar \na^{t]} (k_d Z^d) + k_d k^{[r}\bar \na^{t]} Z^d + Z^{[t}k^{r]} \bar \na_d k^d\right).
\end{align}
Let (for this section only) $v^a = (1+r^2/l^2)\de^a_r$. Then we can replace $Z^a$ with $v^a$ in all the above terms, noting that $Z^a = v^a - (1+r^2/l^2)k^a$ and using $k^d \bar \na_d k^a = k^{[a}k^{b]} = 0$. Then,
\begin{align}
    \bs i^{tr}_Z &=\f{H}{16\pi} \left( k^d k^{[t}\bar \na_d v^{r]} - v^d k_d \bar \na^{[t}k^{r]} - k^{[r}\bar \na^{t]} (k_d v^d) + k_d k^{[r}\bar \na^{t]} v^d + v^{[t}k^{r]} \bar \na_d k^d\right).
\end{align}
We also have, letting $\bar v_a = \bar g_{ab} v^b$,
\begin{align}
    k^d k^{[t}\bar \na_d v^{r]} + k_d k^{[r}\bar \na^{t]}v^d &= k_d \left( \bar \na^d v^{[r}k^{t]} + k^{[r}\bar \na^{t]} v^d\right) \nn 
    &= \f12 k_d \left( \bar \na^d v^{r}k^{t} - \bar \na^d v^t k^r + k^{r}\bar \na^{t} v^d - k^t \bar \na^r v^d \right) \nn 
    &= \f12 k_d ( k^t \bar \na^{[d}v^{r]} + k^r \bar \na^{[t}v^{d]}) \nn 
    &= \f12 \bar g^{rr} \bar g^{tt} k^d ( k_t \bar \na_{[d} \bar v_{r]} + k_r \bar \na_{[t} \bar v_{d]}) \nn 
    &= \f12 \bar g^{rr} \bar g^{tt} k^d (k_t \pa_{[d} \bar v_{r]} + k_r \pa_{[t}\bar v_{d]}).
\end{align}
Because $\bar g_{ab}$ is diagonal, the only nonzero component of $\bar v_a$ is $\bar v_r$, which has no $t$ or $\phi_i$ derivatives. Consequently this expression is zero. 

We also have that $k_d v^d$ has no $t$ derivative, so $\bar \na^{t} (k_d v^d) = 0$. We have, remaining,
\begin{align}
    \bs i^{tr}_Z &= \f{H}{16\pi} \left( -v^r k_r \bar \na^{[t} k^{r]} + \f12  k^{t} \bar \na^{r} (k_d v^d) - \f12 k^t v^r \bar \na_d k^d\right) \nn 
    &= \f{H}{16\pi} \left( - v^r k_r \bar g^{tt} \bar g^{rr} \bar \na_{[t} k_{r]} + \f12 k^t \bar g^{rr} \pa_r (k_d v^d) - \f12 k^t v^r \bar \na_d k^d\right) \nn 
    &= \f{H}{16\pi} \left( - v^rk_r \bar g^{tt} \bar g^{rr} \pa_{[t}k_{r]} + \f12 k^t \bar g^{rr} \pa_r (k_d v^d) - \f12 k^t v^r \bar \na_d k^d\right) \nn 
    &= \f{H}{32\pi} \left( v^r k_r \bar g^{tt} \bar g^{rr} \pa_r k_t + k^t \bar g^{rr} \pa_r (v^rk_r) - k^t v^r \bar \na_d k^d\right).
\end{align}
$\bar \na_d k^d = \f{1}{\sqrt{-\bar g}} \pa_d (\sqrt{-\bar g} k^d) = \f{1}{\sqrt{-\bar g}} \pa_r (\sqrt{-\bar g} k^r))$, since $\pa_t(\sqrt{-\bar g} k^t) = \pa_{\phi_i} (\sqrt{-\bar g} k^{\phi_i}) = 0$. Of course $k^r = 1$, so this reduces to $\bar \na_d k^d = \f{1}{\sqrt{-\bar g}}\pa_r \sqrt{-\bar g}.$ Furthermore, $k_t$ is $r$-independent so $\pa_r k_t = 0$. We can rearrange,
\begin{align}
    \bs i_Z^{tr} &= \f{H \bar g^{rr}\bar g^{tt}}{32 \pi \sqrt{-\bar g}} \left( \sqrt{-\bar g} k_t \pa_r (v^r k_r) - k_t \bar g_{rr} v^r \pa_r \sqrt{-\bar g} \right) \nn 
    &= \f{H k^t \bar g^{rr}}{32\pi \sqrt{-\bar g}} \left(\sqrt{-\bar g}  \pa_r (v^r k_r) - k_r v^r \pa_r \sqrt{-\bar g}\right) \nn 
    &= \f{H k^t \bar g^{rr} \sqrt{-\bar g}}{32 \pi} \f{\pa}{\pa r} \left(\f{v^rk_r}{\sqrt{-\bar g}}\right) \nn 
    &= -\f{H}{32\pi} \f{r V}{1+r^2/l^2} \f{d}{dr} \left( \f{1+r^2/l^2}{r V}\right) \nn 
    &= - \f{H}{32\pi} \f{d}{dr} \ln \left( \f{1+r^2/l^2}{r V}\right) \nn 
    &= \f{H}{32\pi} \f{d}{dr} \ln \left(\f{r V}{1 + r^2/l^2}\right).
\end{align}
Recall that $V$ is given by \eqref{GibbonsV}. This reduces to $H$ times a function of $r$ only. We then further have,
\begin{align}
    \sqrt{-\bar g} \bs i^{tr}_Z &= \f{2\mu(r) r \prod_{i=1}^{n-1+\ve} \mu_i}{32\pi \mu_n \prod_{j=1}^n \Xi_j} \f{d}{dr} \ln \left( \f{rV}{1+r^2/l^2}\right),
\end{align}
so that the combination $\sqrt{-\bar g} \bs i^{tr}_Z$ is a function of $r$ only times $\mu_n^{-1} \prod_{i=1}^{n-1+\ve} \mu_i$---a simple result. The only angular dependence comes in through the factor which integrates to give exactly $\mc A_{D-2}$, as in \eqref{ithetaupsilonintegral}. The integral of $\oint_C \bs I_Z = 2 \int \bs i^{tr}_Z \sqrt{-\bar g} d^{D-2}x$, so 
\begin{align}
    \oint_C \left(\bs I_\xi + \sum_{i=1}^{n-1+\ve} \Om_i(r) \bs I_{\eta_i}\right) &= \oint_C \bs I_Z \nn 
    &= \f{\mu(r)r \mc A_{D-2}}{8 \pi \prod_{j=1}^{n} \Xi_j} \f{d}{dr} \ln \left( \f{r V}{1+r^2/l^2}\right). \label{IzC}
\end{align}
Applying \eqref{GibbonsV} for $V$, we get, for arbitrary $r$,
\begin{align}
    \oint_C \left(\bs I_\xi + \sum_{i=1}^{n-1+\ve} \Om_i(r) \bs I_{\eta_i}\right) &= \f{\mu(r) r \mc A_{D-2}}{8 \pi \prod_{j=1}^n \Xi_j} \f{d}{dr} \ln \left(r^{-\ve} \prod_{i=1}^{n-1+\ve} (r^2+a_i^2)\right) \nn 
    &= \f{\mu(r) r \mc A_{D-2}}{8\pi \prod_{j=1}^n \Xi_j} \left( \sum_{i=1}^{n-1+\ve} \f{2r}{r^2+a_i^2} -\f{\ve}{r}\right). \label{intIxiplusOmegaIeta}
\end{align}

In the limit $C\to C_\infty$, $rV/(1+r^2/l^2) \simeq rV = r^{D-2}$, so that, letting $\lim_{r\to\infty} \mu(r) = m_\infty$ as usual,
\begin{align}
    \oint_{C_\infty} \left( \bs I_\xi + \sum_{i=1}^{n-1+\ve} \Om_i(r) \bs I_{\eta_i}\right) &= \f{(D-2) m_\infty \mc A_{D-2}}{8\pi \prod_{j=1}^n \Xi_j}.
\end{align}
This is the same as the expression for $\oint_{C_\infty} \bs I_\bt$ from \eqref{IbetaintCinfty}, which is unsurprising because $\lim_{r \to \infty} \Om_i(r) = \f{a_i}{l^2}$, so that $\lim_{r\to\infty} (\xi + \sum_{i=1}^{n-1+\ve} \Om_i(r) \eta_i) = \xi + \sum_{i=1}^{n-1+\ve} \f{a_i}{l^2} \eta_i = \bt$.

The combination $\oint_C(\bs I_\xi + \sum_{i=1}^{n-1+\ve} \Om_i(r) \bs I_{\eta_i})$ is thus special in that the contribution from $\mu'(r)$ drops out entirely. We can then think of another reason that $\bt$ is special is that it is equal to this combination, at infinity. Note however that while $\mu'(r)$ drops out of the expressions for the combination $\oint_C (\bs I_\xi + \sum_{i=1}^{n-1+\ve} \Om_i(r) \bs I_{\eta_i})$, it does not drop out of the \emph{individual} contributions from the Komar and non-Komar parts, each of which has some (in general nonzero) $\mu'(r)$ term. The term multiplying $\mu'(r)$ for the Komar and non-Komar terms is just (plus or minus) the one which appears in \eqref{upsilonthetagenericmu}, and so the arctan or ln terms are avoided in \eqref{intIxiplusOmegaIeta}. 

Of particular interest is the combination $\z = \xi + \sum_i \Om_i \eta_i$, which corresponds to $r = r_+$ and the horizon. We have,
\begin{align}
    \oint_H \bs I_\z &= \f{\mu(r_+)r_+ \mc A_{D-2}}{8 \pi \prod_{j=1}^{n} \Xi_j} \f{d}{dr} \left. \ln \left( \f{r V}{1+r^2/l^2}\right)\right|_{r= r_+}.
\end{align}
On the horizon, $V = 2\mu(r)$, so we have
\begin{align}
    \oint_H \bs I_\z &= \f{V(r_+) r_+ \mc A_{D-2}}{16 \pi \prod_{j=1}^n \Xi_j} \left( \f{V'(r_+)}{V(r_+)} + \f{1}{r_+} - \f{2 r_+}{l^2+r^2}\right).
\end{align}

\subsection{Terms Associated with \texorpdfstring{$\bt^{(j)}$}{beta (j)}} \label{betajsection} 

This subsection is not in the thesis. One interesting fact is that it is not just $\bt$, but also the $\bt^{(j)} = \pa/\pa \psi_j$, which give results where the Komar and non-Komar terms are related by a simple numerical factor, as shown in \eqref{Hpapsij}. Note that since $(\bt^{(j)})^t = \hat C_0^{(j)}$, we have (if $\mu(r) = m$ is constant)
\begin{align}
    H^{\bs I}_{\pa_{\psi_j}} &= \f{D-2-2j}{D-2} \mc F (\bt^{(j)})^t \nn
    &= \f{(D-2-2j) m (\bt^{(j)})^t \mc A_{D-2}}{8 \pi \prod_j \Xi_j}.
\end{align}
This only holds either in the large-$r$ limit or if $\mu(r) = m$ is constant; I'll focus exclusively on the large-$r$ limit in this section. So we have
\begin{align}
    \oint_{C_{\infty}} \bs I_{\bt^{(j)}} &= \f{(D-2-2j) m_\infty (\bt^{(j)})^t \mc A_{D-2}}{8\pi \prod_j \Xi_j},
\end{align}
where $C_\infty$ is the surface of constant $t$ with $r \to \infty$. This can be broken down, using \eqref{upsilonthetalimit}, as
\begin{align}
    \oint_{C_\infty} (-\bt^{(j)} \cdot \bs \tht) &= \f {\mc A_{D-2} m_\infty (\bt^{(j)})^t}{8 \pi \prod_j \Xi_j} \nn 
    \oint_{C_\infty} \left(-\left(\bs K^K_{\bt^{(j)}} - \overline{\bs K^K_{\bt^{(j)}}}\right)\right) &= \f{(D-3-2j) \mc A_{D-2} m_\infty (\bt^{(j)})^t}{8\pi \prod_j \Xi_j},
\end{align}
from which we have
\begin{align}
    -\oint_{C_\infty} \left(\bs K^K_{\bt^{(j)}} - \overline{\bs K^K_{\bt^{(j)}}}\right) &= (D-3-2j) \left( -\oint_{C_\infty} \bt^{(j)} \cdot \bs \tht\right), \label{betajKomarnonKomarrelationship}
\end{align}
generalizing \eqref{asymptoticintegral}. It is interesting that this simple numerical relationship occurs only for the $\bt^{(j)}$ (and constant multiples thereof!), but it is only for $\bt$ that we recover the constant of proportionality $D-3$, which is what we really want for the Smarr relation.

What makes the $\bt^{(j)}$ special? One way to proceed is the following. Any of the $\chi$ can be constructed from $\bt$ and the $\eta_i$. We note that the asymptotic form of $\sqrt{-\bar g} (\bs i^K_\bt)^{tr}$ from \eqref{asymptoticsqrtgibetaKtr} and $\sqrt{-\bar g} (\bs i^K_{\eta_i})^{tr}$ from \eqref{etaintegrallarger} can be rewritten using $\bt \cdot k = 1$ and $\eta_i \cdot k = k_{\phi_i} = - \mu_i^2 a_i/\Xi_i$ from \eqref{kmu} as
\begin{align}
    \lim_{r\to\infty} \sqrt{-\bar g} ( \bs i_{\eta_i}^K)^{tr}  &= \f{m_\infty  \prod_{j = 1}^{n-1+\ve} \mu_j}{16\pi \mu_n \prod_{j=1}^n \Xi_j}  (D-1) \eta_i \cdot k \nn 
    \lim_{r\to\infty} \sqrt{-\bar g} (\bs{i}^K_\bt)^{tr} &= \f{m_\infty  \prod_{j = 1}^{n-1+\ve} \mu_j}{16\pi \mu_n \prod_{j=1}^n \Xi_j} (D-3) \nn 
    &= \f{m_\infty  \prod_{j = 1}^{n-1+\ve} \mu_j}{16\pi \mu_n \prod_{j=1}^n \Xi_j} ((D-1) \bt \cdot k - 2 \bt^t),
\end{align}
so that for generic $\chi$, which is a linear combination of the $\bt$ and $\eta_i$, we can conclude
\begin{align}
    \lim_{r\to\infty} \sqrt{-\bar g} (\bs i_\chi^K)^{tr} &= \f{m_\infty  \prod_{j = 1}^{n-1+\ve} \mu_j}{16\pi \mu_n \prod_{j=1}^n \Xi_j} ((D-1) \chi \cdot k - 2 \chi^t)
\end{align}

To interpret this, we can switch from $\mu_i$ coordinates to $y_\al$ coordinates, as is done in Appendix \ref{explicitJacobiTransformed}. From \eqref{sqrtgKtrby}, taking the large-$r$ limit,
\begin{align}
    \lim_{r\to\infty} \sqrt{-\bar g} (\bs i_\bt^K)^{tr} &= \f{(D-3) m_\infty \ti P (\prod_{\al=1}^{n-1} y_\al^\ve)}{16\pi C (\prod_j \Xi_j) (\prod_{j=1}^{n-1+\ve} a_i^{1-\ve})}.
\end{align}
Here, $\ti P = P/U_n = \prod_{\al < \bt} (y_\al^2 - y_\bt^2)$. We then have, for the general case $\chi$,
\begin{align}
    \lim_{r\to\infty} \sqrt{-\bar g} (\bs i_\chi^K)^{tr} &= \f{ m_\infty \ti P (\prod_{\al=1}^{n-1} y_\al^\ve)}{16\pi C (\prod_j \Xi_j) (\prod_{j=1}^{n-1+\ve} a_i^{1-\ve})} ((D-1) \chi \cdot k - 2 \chi^t).
\end{align}
Now consider the case $\bt^{(j)}$. $\bt^{(j)} \cdot k = k_{\psi_j} = A_n^{(j)}$ from \eqref{kexp} and \eqref{emuehatmuehat0}. So then, also using $(\bt^{(j)})^t = \hat C_0^{(j)}$,
\begin{align}
    \lim_{r\to\infty} \sqrt{-\bar g} (\bs i_{\bt^{(j)}}^K)^{tr} &= \f{ m_\infty \ti P (\prod_{\al=1}^{n-1} y_\al^\ve)}{16\pi C (\prod_j \Xi_j) (\prod_{j=1}^{n-1+\ve} a_i^{1-\ve})} ((D-1) A_n^{(j)} - 2 \hat C_0^{(j)}).
\end{align}
Then to take the integral, we integrate over $d^{n-1}y d^{n-1+\ve} \phi$, where the limits on the $y_\al$ involve the $a_i$. This integral, when $j = 0$, is considered in Appendix \ref{AreaCalculationAppendix}, where $\int \tilde P (\prod y_\al)^\ve d^{n-1} y$ is given by \eqref{tildePintegral}, or 
\begin{align}
    \int dy_1 \ldots dy_{n-1} \tilde P \left( \prod_{\alpha=1}^{n-1} y_\alpha\right)^\ve = \f{2^{1-\ve} C \prod_{i=1}^{n-1} a_i^{1-\ve}}{(D-3)!!}.
\end{align}
I won't go through the full derivation, but checking values for small $D$ gives the related relation
\begin{align}
    \int d y_1 \ldots d y_{n-1} \tilde P A_n^{(j)} \left( \prod_{\al=1}^{n-1} y_\al\right)^\ve &= \f{2^{1-\ve} C \prod_{i=1}^{n-1} a_i^{1-\ve} (D-1-2j) \hat C^{(j)}_0}{(D-1)!!} \nn 
    &= \f{D-1-2j}{D-1} \hat C_0^{(j)} \int dy_1 \ldots dy_{n-1} \ti P \left(\prod_{\al=1}^{n-1} y_\al \right)^\ve.
\end{align}
The key is that the integral, including the $A_n^{(j)}$ term, is proportional to $\hat C^{(j)}_0$, and the constant of proportionality depends on $j$ ($\propto D-1-2j$). We then have, comparing to the $\bt$ value,
\begin{align}
    \lim_{r\to\infty} \oint \sqrt{-\bar g} (\bs i^K_{\bt^{(j)}})^{tr} d^{D-2} x &= \f{(D-3-2j) \hat C_0^{(j)}}{D-3}  \lim_{r\to\infty} \oint \sqrt{-\bar g} (\bs i^K_{\bt})^{tr} d^{D-2} x \nn 
    &= (D-3-2j) \hat C_0^{(j)}\lim_{r\to\infty} \oint \sqrt{-\bar g} (\bs i_\bt^{\bs \tht})^{tr} d^{D-2}x,
\end{align}
the last line from \eqref{asymptoticintegrand}. Since $(\bs i^{\bs \tht}_{\bt^{(j)}})^{tr} = (\bt^{(j)})^t (\bs i^{\bs \tht}_{\bt})^{tr}$ (as $(\bs i_\chi^{\bs \tht})^{tr}$ depends only on $\chi^t$ and not on the other components of $\chi$), we then conclude
\begin{align}
    \lim_{r\to\infty} \sqrt{-\bar g} (\bs i_{\bt^{(j)}}^K)^{tr} d^{D-2} x &= (D-3-2j) \lim_{r\to\infty} \sqrt{-\bar g} (\bs i_{\bt^{(j)})}^{\bs \tht})^{tr},
\end{align}
from which \eqref{betajKomarnonKomarrelationship} follows. 

To summarize, then, for all the $\bt^{(j)}$, the asymptotic form of the Komar and non-Komar terms are equal up to a multiplicative factor of $D-3-2j$, and that particular factor arises due to the presence of the $A_n^{(j)}$ and $\hat C_0^{(j)}$ terms in the expression for the Komar integral, corresponding respectively to the contributions from $\bt^{(j)} \cdot k$ and from $(\bt^{(j)})^t$. For a more general vector $\chi$, the relationship between the Komar and non-Komar terms will not be as straightforward. If we express $\chi$ as
\begin{align}
\chi &= \sum_j \chi^{\psi_j} \f{\pa}{\pa \psi_j},
\end{align}
then we have
\begin{align}
    -\oint_{C_\infty} \left(\bs K^K_{\chi} - \overline{\bs K^K_{\chi}}\right) &= \sum_{j} (D-3-2j) \chi^{\psi_j} \left( -\oint_{C_\infty} \bt^{(j)} \cdot \bs \tht\right).
\end{align}

\subsection{Variation}

As stated in Section \ref{explicitIxi}, we can use the expressions $\oint \bs I_\chi$ to find $\kappa \de A$ for the horizon, on the condition that the background metric does not change. The calculation will get complicated quickly if I try to repeat the results of Section \ref{explicitIxi} for arbitrary dimensions. One thing I can do though is to consider the case where the $a_i$, and thus the background metric, do not change, and the only change is in $\mu$, with $\mu(r) \to \mu(r)+\de \mu(r)$. In this case, we would expect that if we calculate
\begin{align}
    \de \oint_H\left( \bs I_\xi+ \sum_i \Om_i \bs I_{\eta_i}\right),
\end{align}
where $H$ is the value $r = r_+$ of the horizon for the original $\mu(r)$, and where $\xi^a, \eta_i^a$ and $\Om_i$ are constant, the result will be $\kappa \de A$. I will proceed and demonstrate this. Fortunately, despite the complicated form for the expressions for $\oint_C \bs I_\xi$ and so on for an arbitrary $\mu(r)$, there are several simplifications.

As in Section \ref{massfunctionvars}, but generalized to higher dimensions, let $\hat \de$ represent a variation which does not affect $a_i$ or $l$, and consider a variation where $r$ is kept constant. We require that the function $\mu(r)$ varies, so $\hat \de \mu(r) \neq 0$. We then have, starting from a general constant-$r$ surface $C$, using \eqref{IzC},
\begin{align}
    \hat \de \oint_C \left( \bs I_\xi + \sum_{i=1}^{n-1+\ve} \Om_i(r) \bs I_{\eta_i}\right) &=  \hat \de \left(\f{\mu(r)r \mc A_{D-2}}{8 \pi \prod_{j=1}^{n} \Xi_j} \f{d}{dr} \ln \left( \f{r V}{1+r^2/l^2}\right)\right) \nn 
    &=  \f{\hat \de \mu(r)r \mc A_{D-2}}{8 \pi \prod_{j=1}^{n} \Xi_j} \f{d}{dr} \ln \left( \f{r V}{1+r^2/l^2}\right),
\end{align}
that is, the only term which has a nonzero variation is $\hat \de \mu(r)$. 

The same argument leading to \eqref{variationinmuatrplus} (except the last line) in Section \ref{massfunctionvars} applies here. Whereas $r$ itself is unchanged by the variation, the value of the horizon location, $r_+$, does depend on the functional form $\mu$, which varies under $\hat \de$, satisfying $2 (\mu+\hat \de \mu)(r_+ + \hat \de r_+) = V(r_+ + \hat \de r_+)$. We can then write
\begin{align}
    \hat \de \mu(r_+) &= \f{\kappa V(r_+)}{1 + r_+^2/l^2} \hat \de r_+,
\end{align}
where $\hat \de \mu(r_+)$ is the evaluation of $\hat \de \mu (r)$ as the unperturbed $r_+$, and similarly all the quantities on the right-hand side except for $\hat \de r_+$ are the unperturbed values. 

We can also write, from \eqref{AGibbonsLu},
\begin{align}
    A &= \f{\mc A_{D-2}}{\prod_{j=1}^n \Xi_j} \left(\f{r_+ V(r_+)}{1+r_+^2/l^2}\right).
\end{align}
Because $\hat \de a_i = \hat \de l = 0, \hat \de \Xi_j = 0$. The functional form $V(r)$ does not vary either. Thus the only contribution to $\hat \de A$ will be from $\hat \de r_+$, and so
\begin{align}
    \hat \de A &= \f{\pa A}{\pa r_+} \de \hat r_+ \nn 
    &= \f{\mc A_{D-2}}{\prod_{j=1}^n \Xi_j} \left. \f{d}{dr} \left( \f{r V}{1+r^2/l^2}\right)\right|_{r=r_+} \hat \de r_+.
\end{align}
We then have,
\begin{align}
    \hat \de \oint_H \left( \bs I_\xi + \sum_{i=1}^{n-1+\ve} \Om_i(r) \bs I_{\eta_i}\right) &=\left.\f{\hat \de \mu(r_+)r_+ \mc A_{D-2}}{8 \pi \prod_{j=1}^{n} \Xi_j} \f{d}{dr} \ln \left( \f{r V}{1+r^2/l^2}\right)\right|_{r=r_+} \nn 
    &= \f{\mc A_{D-2} r_+}{8 \pi \prod_{j=1}^n \Xi_j} \hat \de \mu(r_+) \f{1+r_+^2/l^2}{r_+V(r_+)} \left. \f{d}{dr} \left( \f{r V}{1+r^2/l^2}\right) \right|_{r=r_+} \nn 
    &= \f{\mc A_{D-2}}{8\pi \prod_{j=1}^n \Xi_j} \kappa \hat \de r_+ \left. \f{d}{dr} \left( \f{rV}{1+r^2/l^2}\right)\right|_{r=r_+} \nn 
    &= \f{\kappa \hat \de A}{8\pi}.
\end{align}

If we let $\ti{\mc E}(r), \ti{\mc J}_i(r)$ be given by $\ti{\mc E}(r) = \oint_C \bs I_\xi, \ti{\mc J}_i(r) = - \oint_C \bs I_{\eta_i}$ as in Section \ref{massfunctionvars}, then we recover
\begin{align}
    \hat \de \ti{\mc E}(r_+^{(0)}) - \sum_{i=1}^{n-1+\ve} \Om_i \hat \de \ti{\mc J}_i(r_+^{(0)})  &= \f{\kappa \hat \de A}{8\pi},
\end{align}
where I specify here that the evaluation at the left-hand side is evaluated at the value of $r_+$ in the unperturbed spacetime. This does not require that $\mu(r)$ be a constant, or that $\mu(r)$ be a function so that the resulting spacetime satisfies Einstein's equations!

For a more general variation, in which $a_i, l$ also vary, more care needs to be taken for the same reasons as in Section \ref{explicitIxi} in general. Of course in variations from Kerr--AdS to another Kerr--AdS solution (with the same $l$), the first law with $\mc E, \mc J_i$, which are constants, is automatically satisfied. The case of variation with generic function $\mu(r)$ and where $a_i$ are allowed to vary would follow along similar lines to Section \ref{massandspecificangularmomentumvary} and following sections. 

\subsection{Summary}

The main takeaway of this section is that the expressions for $\oint_C \bs I_\chi$ are fairly complicated in general, and that applying either $\chi = \bt$ or $\chi = Z$ simplifies the expressions considerably. It is due to the special properties of $\bt$ that the Komar and non-Komar terms associated with $\oint_C \bs I_\bt$ end up being equal in the $\mu(r)=m$ case, and it is due to the fact that $Z$ shares its $t,\phi_i$ components with $k$ (up to a constant multiple) that the expressions in $\oint_C \bs I_Z$ simplify so readily. The latter has the application that it simplifies the horizon-variation calculation considerably and allows for a demonstration of the utility of the $\ti{\mc E}(r), \ti{\mc J}_i(r)$ ``energy and angular momentum'' expressions for discussing the horizon variation law even away from a solution to Einstein's equations. The $\bt$ statement answers why it is that the geometric volume appears naturally in the Smarr law and thus variation law including $\La$ for the charge related to $\bt$ rather than for the charge related to $\xi$. 

\section{Physical Process Version} \label{physicalprocessversion}

As pointed out in, e.g.,~\cite{GaoWald, Poisson}, there is also a ``physical process'' version of the first law of BH mechanics. In this case, instead of comparing distinct spacetimes, the same spacetime is allowed to vary with time, in that a small amount of matter is allowed to pass by the BH horizon. This is done as a quasi-stationary process, so that the black hole remains nearly stationary throughout. Following the explanation in \cite{Poisson}, the argument is (briefly) as follows. 

Let a four-dimensional black hole, initially stationary (not necessarily vacuum), be perturbed by a small amount of matter flowing through the horizon. Assume that after the small amount of matter flow, the horizon settles down to a new stationary state. The initial and final black holes will both have a stationarity Killing vector $\xi$ and azimuthal symmetry Killing vector $\eta$. Let the event horizon be given by $\mc H$. Let $v$ be a parameter along the null generators of the horizon, with $v = -\infty$ in the distant past and $v = +\infty$ in the distant future. (In general $v$ will not be affine.) Assume that the black hole is quasi-stationary for both $v = \pm \infty$. 

Let, following Poisson (modifying notation), let the change in black hole mass $\de \mc E_{\mc H}$ and change in black hole angular momentum $\de \mc J_{\mc H}$ be given by, respectively,
\begin{align}
    \de \mc E_{\mc H} &= -\int_\mc H T^a_b \xi^b d \Si_a \nn 
    \de \mc J_{\mc H} &= + \int_\mc H T^a_b \eta^b d \Si_a. \label{deEdeJmcH}
\end{align}
Poisson works in first order in $T^a_b$ and assumes that $\xi^a,\eta^a$ and $d \Si_a$ can be taken as their unperturbed values. 

The argument that the transfer of energy can be written in the above form is roughly the following. Let $\varepsilon^a = -T^a_b \xi^b$ and $\ell^a = T^a_b \eta^b$ be vector fields where $T^{ab}$ is a test stress--energy tensor that does not influence the spacetime. We can interpret $\ve^a$ as an energy-density flux vector and $\ell^a$ as an angular momentum-flux vector, based on the decomposition of the stress--energy tensor $T^{ab}$. If $\xi^a$ and $\eta^a$ are Killing vectors, then we find that $\ve$ and $\ell$ are divergence-free:
\begin{align}
    \na_a \ve^a &= - \na_a (T^a_b \xi^b) \nn 
    &= - \na_a (T^{ab} \xi_b) \nn 
    &= - \na_a T^{ab} \xi_b - T^{ab} \na_a \xi_b.
\end{align}
Since $\na_a T^{ab} = 0$, $T^{ab}$ is symmetric, and $\na_{(a}\xi_{b)} = 0$ is the Killing condition, this vanishes. The argument for $\na_a \ell^a= 0$ is analogous. Then from Gauss' law, the amount of flux of $\ve^a$ flowing into any region must be equal to the amount flowing out, so it can be interpreted as the flux of energy (similarly for $\ell$ and angular momentum). It is not necessarily the case that this interpretation corresponds to the same interpretation as energy and angular momentum for the black hole, but we will accept this idea for the moment. I include the subscripts $\mc H$ to try to emphasize that these are the energy and angular momentum transfers specifically across the horizon according to \eqref{deEdeJmcH}. 

Combining the expressions for energy and angular momentum transfer from \eqref{deEdeJmcH} along with the (initial) angular velocity $\Om$ of the black hole gives
\begin{align}
    \de \mc E_{\mc H} - \Om \de \mc J_{\mc H} &= -\oint_{\mc H} T^a_b (\xi^b + \Om \eta^b) d \Si_a \nn 
    &= -\oint_{\mc H} T^a_b \bar \z^b d \Si_a,
\end{align}
where as usual $\bar \z = \xi + \Om \eta$ is the Killing vector tangent to the initial black hole null generators. The initial value of $d \Si_a$ can be written 
\begin{align}
    d \Si_a &= -\bar \z_a d S d v,
\end{align}
where $dS$ is the area element and $v$ is the parameter along $\bar \z^a$, $\bar \z = \pa/\pa v$. Assuming that $d \Si_a$ does not change under the incoming matter, we have
\begin{align}
    \de \mc E_{\mc H} - \Om \de \mc J_{\mc H} &= \oint_{\mc H} T^a_b \bar \z^b \bar \z_a d S d v \nn 
    &= \oint_{\mc H} T_{ab} \bar \z^a \bar \z^b d S d v.
\end{align}
Raychaudhuri's equation, when neglecting terms quadratic in $T^a_b$, gives
\begin{align}
    \f{d \tht}{d v} &= \kappa \tht - 8 \pi T_{ab} \bar \z^a \bar \z^b,
\end{align}
where $\tht$ is the expansion of the null geodesics with tangent vector $\bar \z$. $\tht$ is also equal to the fractional rate of change of cross-sectional area, $\tht = \f{1}{dS} \f{d}{dv} dS$. We can then write, letting $H$ be the 2-surface on the horizon at a constant $v$,
\begin{align}
    \de \mc E_{\mc H} - \Om \de \mc J_{\mc H} &= - \f{1}{8\pi} \int dv \oint_{H} \left( \f{d \tht}{dv} - \kappa \tht \right) d S \nn 
    &= - \left.\f{1}{8\pi} \oint_{\mc H} \tht d S\right|_{-\infty}^\infty + \f{\kappa}{8\pi} \int dv \oint_{ H} \tht d S.
\end{align}
$\tht = 0$ at $v = \pm \infty$ because, as the black hole is stationary at both of these points, the expansion of the null geodesics are zero. The remaining term gives
\begin{align}
    \de \mc E_{\mc H} - \Om \de \mc J_{\mc H} &= \f{\kappa}{8\pi} \int dv \oint_{H} \left( \f{1}{dS} \f{d}{dv} dS\right) d S \nn 
    &= \left.\f{\kappa}{8\pi} \oint_{H} d S \right|_{v=-\infty}^{v=+\infty}\nn 
    &= \f{\kappa}{8\pi} \de A.
\end{align}
The derivation, at least in this form, relies on the assumption that the Killing vector tangent to the null generators after the matter falling into the BH, $\bar \z + \de \z$, is sufficiently close to $\bar \z$ that Raychaudhuri's equation can still be used. 

This argument is very similar to the argument that $\kappa \de A/8\pi = \oint_H \bs k^{EH}_{\bar \z} [\de g;g]$, comparing different nearby spacetimes, with the integral over $\oint_{\mc H} T^a_b \chi^b d \Si_a$ having similarities to $\oint_H \bs k^{EH}_{\chi} [\de g;g]$. 

Another note I want to make, before continuing, is that because $\bar \z^a \bar \z_a = 0$ (and remains so throughout the spacetime, because we presume that the deviation of $\bar \z_a$ from being null is only first order in the change in the stress--energy tensor), if $G_{ab} + \La g_{ab} = 8\pi T_{ab}$, this implies that $G_{ab} \bar \z^a \bar \z^b = 8\pi T_{ab} \bar \z^a \bar \z^b$, as $\La g_{ab} \bar \z^a \bar \z^b = 0$.

I wanted to check how the $\de \mc E_{\mc H}, \de \mc J_{\mc H}$ related to the $ \mc E,  \mc J$ quantities associated with Kerr--AdS. My thinking is that since we have $\de \mc E_{\mc H} - \Om \de \mc J_{\mc H} = \kappa \de A/8\pi$, if we could show that $\mc E_{\mc H} = \mc E, \mc J_{\mc H} = \mc J$ for some values $\mc E, \mc J$, that we would have values of $\mc E, \mc J$ that would automatically satisfy the first law. This is another approach that amounts to integrating the first law, in a manner similar to GPP---and also in a manner similar to integrating $\bs k^{EH}_\chi$ to get $\de \mc E, \de \mc J_i$. 

The idea of interpreting the energy and angular momentum in a spacetime according to the amount that must be transferred across some boundary surface when comparing two different spacetimes has a long history. To give a key example, much of the argument behind the AMD formalism \cite{AshtekarMagnon,AshtekarDas,AshtekarPawlowski} relies on showing how the change in AMD integral values corresponds to the stress--energy tensor flux through a hypersurface at infinity. I have not seen someone make the precise argument that I am laying out here, but both the results and my ability to position it within the broader literature (as well as its overall uniqueness) are best considered tentative, hence its inclusion as a final section in this chapter.

Here is a specific method I considered. First I will consider the Kerr spacetime, which is simpler and more familiar, and then Kerr--AdS spacetime, and then some other spacetimes of interest which I did not consider elsewhere in the thesis.

The calculations in the following subsections are performed/checked using \emph{GRTensorIII}.

\subsection{Kerr} \label{physicalprocesskerr}

The Kerr metric in ingoing Eddington--Finkelstein coordinates is given by \cite{Poisson} (for ease of computation using $u = \cos \tht$)
\begin{align}
    ds^2 &= - \left(1 - \f{2 m r}{\rho^2}\right) dv^2 + 2 dv dr - \f{2 m(1-u^2) a r}{\rho^2} dv d \psi - 2 a (1-u^2) d r d \psi + \nn
    &\qquad \f{\rho^2}{1-u^2} du^2 + (1-u^2) \left( r^2+a^2 - \f{2 m r a^2 (1-u^2)}{\rho^2}\right) d\psi^2, \label{kerringoing}
\end{align}
where as usual $\rho^2 = r^2+a^2u^2$. The horizon is as usual at $r^2+a^2-2mr = 0$. Trajectories with constant $(v,u,\psi)$ are ingoing null geodesics, with $r$ decreasing to the future. 

As a toy model, consider the case where $m$ and $a$ are allowed to vary as functions of $v$. That means that every constant-$v$ surface corresponds to a constant-$v$ surface in a Kerr spacetime, but for different Kerr spacetimes. I then let $m(-\infty) = m_0, a(-\infty) = a_0$ and $m(+\infty) = m_1, a(+\infty) = a_1$, both constant. Eventually I will take $\de m = m_1 - m_0, \de a = a_1 - a_0$ to be very small. The approach of taking the Eddington--Finkelstein ingoing coordinates and replacing $m$ and, here, $a$ with functions of $v$ is similar to the way the Vaidya spacetime generalizes Schwarzschild by writing Schwarzschild in ingoing Eddington--Finkelstein coordinates and sending $m \to m(v)$. I call this a toy model because I am not claiming there is a specific physical mechanism which would give rise to this metric, but we can consider it a sketch of a scenario in which ingoing radiation transforms a Kerr solution with parameters $m_0, a_0$ into ones with $m_1, a_1$. 

Consider, for simplicity, the surface $r = r_0$. I do not yet specify that it is given by $r_0 = r_{+,0}$ (where $r_{+,0}$ is the value of $r_+$ in Kerr with $m = m_0, a = a_0$), denoted by $\Si$. Then I will calculate the quantities
\begin{align}
    T^a_b \chi^b d \Si_a
\end{align}
on the surface $\Si$, where $\chi^a = \xi^a = \de^a_v$ or $\chi^a = \eta^a = \de^a_\psi$. In a scenario where $m,a$ are allowed to vary with $v$ but not $\psi$, $\eta$ is still a Killing vector, but $\xi$ is not---it is only ``nearly'' a Killing vector, for sufficiently small values of $m' = dm/dv, a' = da/dv$. Because $\La = 0$ we have $G^a_b = 8\pi T^a_b$.

Up to choices of orientation---and I will state here that I chose the orientation that gives a sensible result for the energy transfer later on---these are, respectively,
\begin{align}
    T^a_b \xi^b d\Si_a &= \f{1}{8\pi} G^r_v \bs \epsilon_{r v u \psi} d v d u d \psi \nn 
    &= -\f{1}{8\pi} \sqrt{-g} G^r_v d v d u d \psi \nn 
    T^a_b \chi^b d\Si_a &= -\f{1}{8\pi} \sqrt{-g} G^r_\psi d v d u d \psi.
\end{align}
The energy transfer $\de \mc E_\Si$ across $\Si$ then is denoted by
\begin{align}
    \de \mc E_\Si &= \int_\Si T^a_b \xi^b d \Si_a \nn 
    &= \f{1}{8\pi} \int_\Si \sqrt{-g} G^r_v d v d u d \psi \nn 
    &= \f{1}{8\pi} \int_{-\infty}^\infty d v \oint \sqrt{-g} G^r_v d u d \psi,
\end{align}
where the last integral is on a constant $r = r_0$, constant $v$ surface. We can then, in perhaps somewhat an abuse of notation, represent the rate of energy transfer $d \mc E_\Si/dv$ at a particular value of $v$ by taking the $v$-derivative of both sides and so write
\begin{align}
    \f{d \mc E_\Si}{d v} &= \f{1}{8\pi} \oint \sqrt{-g} G^r_v du d \psi. \label{dEsigmadv}
\end{align}
Here and above, $\sqrt{-g}$ refers to the determinant of the full four-dimensional metric $g_{ab}$. The reason this is perhaps an abuse of notation is that we do not in general know that there is some quantity $\mc E_\Si$ that is having its derivative taken. The justification is that $(d \mc E_\Si/d v) d v $ evaluated at $v$ represents the contribution from between $v$ and $v + dv$ to $\de \mc E_\Si$ overall.

Analogously the rate of angular momentum transfer along $\Si$ can be represented by 
\begin{align}
    \f{d \mc J_\Si}{dv} &= - \f{1}{8\pi} \oint \sqrt{-g} G^r_\psi du d \psi. \label{dJsigmadv}
\end{align}

I then evaluated these, using the Kerr--Vaidya-type toy metric \eqref{kerringoing} with $m \to m(v), a \to a(v)$. The exact results for $d \mc E_\Si/d v$ and $d \mc J_\Si/dv$ are quite involved if we include all terms, which include several terms involving $d^2m/dv^2, d^2 a/dv^2$, $(dm/dv)^2, (da/dv)^2$ and $(dm/dv)(da/dv)$. If we treat all terms like this which involve either second derivatives or the products of two first-derivative terms (or combinations involving further derivatives!) as being ``of second order'' then we can write (just writing $r$, rather than $r_0$, but it being understood that $\Si$ is on some particular value $r = r_0$
\begin{align}
    \f{d \mc E_\Si}{dv} &= \f{d m}{dv} + \textrm{(second order terms)} \nn 
    \f{d \mc J_\Si}{dv} &= \f{d(ma)}{dv} + \textrm{(second order terms)},
\end{align}
irrespective of the value of $r_0$. 

(To evaluate these straightforwardly, I set $m(v) = m_0 + \la m'_0 v + \la^2 f_1(v), a(v) = a_0 + \la a'_0 v + \la^2 f_2(v) $ where $m_0, m'_0, a_0, a'_0$ are constants, and then expanded out $G^a_b$ in terms of $\la$, so that all terms that involve second derivatives or a combination of first derivatives appear as powers of $\la^2$ or higher, and then took only the zeroth and first order terms in $\la$, setting $\la = 1$.) 

This means that if we assume that the functions $m(v), a(v)$ are slowly-varying and that $dm/dv, da/dv$ dominate over all ``second-order terms'' (that is, either the second- or higher-derivative terms or any products of the first-order derivatives), then we have $d \mc E_\Si/dv = dm/dv, d \mc J_\Si/dv = d(ma)/dv$. If we further assume that the horizon remains close enough to $r = r_{+,0}$ that we can treat $\mc H$ as being at nearly constant radius, then we find 
\begin{align}
    \f{d \mc E_{\mc H}}{d v} &= \f{d m}{dv} \nn 
    \f{d \mc J_{\mc H}}{dv} &= \f{d (ma)}{dv}.
\end{align}
Of course, in Kerr, $\mc E = m, \mc J = ma$, and so we have, through this method, very readily recovered the mass and energy. 

This result, that the $d \mc E_{\mc H}/dv, d \mc J_{\mc H}/dv$, gives $d \mc E/dv, d \mc J/dv$, is remarkable. It is perhaps not entirely surprising. In Vaidya proper, which is what the above solution is with $a = 0$, then the stress--energy tensor is simply given by \cite{Poisson}
\begin{align}
    T_{a b} &= \f{dm/dv}{4\pi r^2} l_a l_b,
\end{align}
where $l_a = -\pa_a v$ is tangent to the ingoing null geodesics. In this case we have $\sqrt{-g} G^r_v = 2 dm/dv$ (exactly). 

My initial thought was that it made the most sense to have a treatment in the ingoing Eddington--Finkelstein coordinates both because the horizon is regular in these coordinates and that the layout with $v$ labeling ingoing coordinates meant that this toy model seemed a good generalization of Vaidya. However I also thought it might be worth checking other forms of the Kerr spacetime. Using the standard Boyer--Lindquist coordinates, which are not regular on the horizon, and again for ease of computation using $u  = \cos \tht$ as one of the coordinates, the metric is written
\begin{align}
    ds^2 &= -\left(1-\f{2mr}{\rho^2}\right) dt^2 - \f{2 m a r (1-u^2)}{\rho^2} dt d\phi + \f{\Si}{\rho^2} (1-u^2) d \phi^2 + \f{\rho^2}{\Delta} dr^2 + \f{\rho^2}{1-u^2} du^2
\end{align}
with $\rho^2$ as above, $\Si = (r^2+a^2)-a^2 \Delta (1-u^2)$ and $\Delta = r^2+a^2-2mr$. Define, without much physical justification, a ``time-dependent Kerr'' where $m \to m(t), a \to a(t)$. Again take $\eta = \pa_\phi, \xi = \pa_t$. We can, by the same process as above, define $d \mc E_\Si/dt, d \mc J_\Si/dt$ on a surface $\Si$ at $r = r_0$ (constant) as in \eqref{dEsigmadv} and \eqref{dJsigmadv}, with $v \to t, \psi \to \phi$:
\begin{align}
    \f{d \mc E_\Si}{d t} &= \f{1}{8\pi} \oint \sqrt{-g} G^r_t d u d \phi \nn 
    \f{d \mc J_\Si}{d t} &= -\f{1}{8\pi} \oint \sqrt{-g} G^r_\phi d u d \phi. 
\end{align}
If we use these expressions with this ``time-dependent Kerr'' spacetime, we find,
\begin{align}
    \f{d \mc E_\Si}{d t} &= \f{dm}{dt} \nn
    \f{d \mc J_\Si}{d t} &= \f{d (ma)}{dt},
\end{align}
\emph{exactly}---that is, any higher-order derivatives or products of first-order derivatives vanish! This result is independent of $r$. This means once again we can take it on $\mc H$, in the case where we assume that the radius does not change too much on $\mc H$. Also, in both the ``Kerr--Vaidya'' and ``time-dependent Kerr'' solutions we can also take the constant-$r$ surface out to infinity.

Finding the exact location of the event horizon is nontrivial. Using the argument involving Raychaudhuri's equation suggests that we do expect that the combination of $\de \mc E_{\mc H} - \Om \de \mc J_{\mc H} = \kappa \de A/8\pi$ where $\de A$ is the overall change in the horizon area and we take $\Om, \kappa$ at their unperturbed values, provided the amount of energy and angular momentum flowing through the horizon is sufficiently small. It would be nice if we could find a way to bring this into a derivative form, to combine 
\begin{align}
    \f{d \mc E_{\mc H}}{dv} - \Om \f{d \mc J_{\mc H}}{dv} &= \kappa \f{dS}{dv}, \label{firstlawdvversion}
\end{align}
or, equivalently, with the $t$ derivative for the ``time-dependent Kerr'' spacetime. A fully rigorous treatment of this is left for future work. What I will say at the moment is the following. We expect from the Raychaudhuri's equation argument that \eqref{firstlawdvversion} should hold, and we can also check the validity of it by evaluating the left-hand side. Using $\Om$ given by
\begin{align}
    \Om &= \f{a(v)}{r_+^2(v) + a^2(v)},
\end{align}
with $r_+^2(v) -2 m(v) r_+(v) + a^2(v)$, we find
\begin{align}
    \f{d \mc E_{\mc H}}{dv} - \Om \f{d \mc J_{\mc H}}{dv} &= \f{1}{16\pi} \f{(r_+^2-a^2)}{r_+(r_+^2+a^2)}\f{d}{dv} (4 \pi (r_+^2+a^2)) + \textrm{(second-order terms)},
\end{align}
as expected from the right-hand side being interpreted as $\kappa dS/dv$ using the usual values in Kerr, where $r_+$ and $a$ are allowed to vary with $v$. In the ``time-dependent Kerr'' solution we get exactly
\begin{align}
    \f{d \mc E_{\mc H}}{dt} - \Om \f{d \mc J_{\mc H}}{dt} &= \f{1}{16\pi} \f{(r_+^2-a^2)}{r_+(r_+^2+a^2)}\f{d}{dt} (4 \pi (r_+^2+a^2)),
\end{align}
with no higher-order terms, where $r_+$ and $a$ are allowed to vary with $t$.

To what extent can we link these usual values of $\kappa$ and $A$ to an actual $\kappa$ and $A$ in the dynamic spacetime? As a preliminary measure, here are a few observations. Let $\z^a$ in this spacetime be the vector given by
\begin{align}
    \z^a &= \de^a_v + \Om(v) \de^a_\psi \nn 
    \Om(v) &= \f{a(v)}{r_+^2(v) + a^2(v)}.
\end{align}
Then $\z^a \z_a = 0$ at the value $r = r_+(v)$. Of course if $m,a$ are actually constants, then $r = r_+$ is the usual Kerr Killing horizon. In a variable-$v$ spacetime we do not actually expect $r = r_+(v)$ to be a horizon. 

We proceed as follows. Single out a particular value of $v$, which without loss of generality we set to be $v = 0$. Then consider the hypersurface $\Si$ given by $r = r_+(0)$ (that is, the value of $r_+(v)$ where $v = 0$). Then on $v = 0$, the hypersurface intersects with the vanishing of $\z^a \z_a$. On a surface of constant $r$, constant $v$, the area element is given by $\sqrt{ g_{uu} g_{\psi\psi}} du d \psi$. On \emph{particular} surface $v, r = r_+(v)$ for some value of $v$, this is
\begin{align}
    dS &= (r_+^2(v)+a^2(v)) du d \psi,
\end{align}
with of course $dS = (r_+^2(0)+a^2(0)) d u d \psi$ at $v = 0$. This means that $\Si$ is null \emph{specifically} at $v=0$, with area
\begin{align}
    A = 4\pi (r_+^2(0)+a^2(0)).
\end{align}

$\z^a$ is null on $\Si$ at $v = 0$, and additionally is a null normal to $\Si$, because $\z_v = \z_u = \z_\psi = 0$ on $r = r_+(0),v=0$ (specifically, though $\z_u = 0$ everywhere). What we can say is that there is a null hypersurface which is tangent to $\Si$ at the 2-surface $r=r_+(0),v=0$, and that $\z$ is a null normal to $\Si$ (and to the null hypersurface) at that point. It would take more work to find the null geodesic field that $\z^a$ is tangent to at $v = 0, r = r_+(0)$, and the $\z^a$ that we have defined is not geodesic in general, even at $v=0,r=r_+(0)$. 

Nevertheless, it is possible to check that we have
\begin{align}
    \z^b \na_b \z^a &= \kappa(v) \z^a + \textrm{(first-order derivative terms)},
\end{align}
when evaluated at $v = 0, r= r_+(0)$, so that to zeroth order $\z^a$ is a null geodesic, and where $\kappa$ takes its usual form with the usual constants now allowed to vary with $v$, here
\begin{align}
    \kappa(v) &= \f{r_+^2(v)-a^2(v)}{2 r_+(v)(r_+^2(v)+a^2(v))},
\end{align}
with both $r_+$ and $a$ evaluated at $v = 0$. This is not a wholly satisfying argument for the value of $\kappa$ and $A$ but shows some progress in this direction. What we can say then is that the combination
\begin{align}
    \kappa(0) \left. \f{d A(v)}{dv}\right|_{v=0}
\end{align}
can be identified with the hypersurface $\Si$ at $v = 0, r = r_+(0)$ and the vector $\z^a$ in the case where changes to the metric functions $a,m$ are very slow so that the first-order derivative terms do not significantly impact $\z^b \na_b \z^a$. It is to be expected too that in the limit where changes happen extremely slowly, the location of the event horizon at $v = 0$ will be \emph{close to} $\Si$, though the exact location will be slightly outside $\Si$ (since with incoming matter the null geodesics tangent to $\z^a$ at this point will fall inside the black hole as the area grows). We can thus write
\begin{align}
    \left.\left(\f{d \mc E_\Si}{dv} - \Om(v) \f{d \mc J_\Si}{dv}\right)\right|_{v=0} = \f{1}{8\pi} \kappa(0) \left. \f{d A}{dv}\right|_{v=0}, \label{firstlawdbydv}
\end{align}
where terms which constitute ``second-order derivatives'' or higher are dropped, and then generalize to note that this holds for any value of $v$, with $\Si$ being defined by $r = r_+(v)$ for that particular value of $v$. 

I believe that similar arguments apply for the ``time-dependent Kerr'' case, though because the horizon is not regular Kerr in Boyer--Lindquist coordinates a little more care might need to be taken and this is left for future work.

I will note before continuing that in the ``time-dependent Kerr'' spacetime, the Komar integrals associated with $\xi$ and $\eta$ also give $m$ and $ma$. On a 2-surface $C$ of constant $r,t$ are given by 
\begin{align}
    -\oint_C \bs K^K_\xi &= \f{m}{2} \nn 
    +\oint_C \bs K^K_\eta &= m a,
\end{align}
which holds not just for the usual Kerr spacetime but also for the time-dependent ones as well, where here $m, a$ are functions of $t$. Here $\xi$ is no longer a Killing vector, because the metric depends on $t$, but I will still use the expression $\bs K^K_\xi = * d \xi^\flat / 32\pi$. I don't believe that this need be on a constant-$r$ surface specifically (thought it need be constant-$t$), but used this for simplicity of computation.

The equivalent expressions for the ``Kerr--Vaidya'' spacetime are much more complicated; the terms ``zeroth order'' in the derivatives on a surface of constant $v,t$ are $m(v)/2, m(v)a(v)$ respectively for $-\oint_C \bs K^K_\xi$ and $\oint_C \bs K^K_\eta$. 

Another interesting wrinkle is the following. We actually have, exactly, in the ``Kerr--Vaidya'' spacetime,
\begin{align}
    \f{\pa}{\pa v} (\bs K^K_\eta)_{u \psi} &=- \f{1}{8\pi} \sqrt{-g} G^r_\psi, \label{dKetadv}
\end{align}
and similarly in the ``time-dependent Kerr'' spacetime,
\begin{align}
    \f{\pa}{\pa t} (\bs K^K_\eta)_{u \phi} &= -\f{1}{8\pi} \sqrt{-g} G^r_\phi. \label{dKetadt}
\end{align}
Thus not only do the \emph{integral} $d \mc J_\Si/dv$, $d \mc J_\Si/dt$ match with the Komar integral associated with $\eta$, but the \emph{integrands} match on both sides as well! On the other hand, in general
\begin{align}
    \f{\pa }{\pa v} (\bs K^K_\xi)_{u\psi} &\neq -\f{1}{16\pi} \sqrt{-g} G^r_\psi
\end{align}
in ``Kerr--Vaidya'' and similarly
\begin{align}
    \f{\pa}{\pa t} (\bs K^K_\xi)_{u \phi} &\neq - \f{1}{16\pi} \sqrt{-g} G^r_\phi
\end{align}
in ``time-dependent Kerr,'' even though equality holds after integrating over $du d\psi$ or $du d \phi$ (respectively).

\subsection{Other Vacuum Spacetimes with Zero Cosmological Constant}

I was interested by how effective this strategy of checking the stress--energy tensor was at other spacetimes which are in $\La = 0$ vacuum. Here I show two examples: the Curzon--Chazy metric in Stachel coordinates, and the five dimensional black ring spacetime. I will not attempt to describe the black hole horizon for these spacetimes and am only evaluating the energy and angular momentum transfer terms associated with the stress--energy tensor, and how they relate to the ADM and Komar quantities.

\subsubsection{Curzon--Chazy}

The Curzon--Chazy metric is a $\La = 0$ vacuum form of the Weyl metric. It can be written in Stachel coordinates (from \cite{abdelqaderlake} with the substitution $u = \cos \tht$)
\begin{align}
    ds^2 = - e^{-2m/r} dt^2 + e^{2m/r}\left( e^{-m^2 (1-u^2)/r^2} (dr^2 + r^2 du^2/(1-u^2)) + r^2(1-u^2) d \phi^2\right).
\end{align}
The sole parameter is $m$. I will define the ``time-dependent Curzon--Chazy'' metric by the substitution $m \to m(t)$.

The ADM energy of Curzon--Chazy is $m$, and its Komar energy is also $m$ as can be verified very easily---in fact, even with variable $m$, 
\begin{align}
    \sqrt{-g} \nabla^r \xi^t = m,
\end{align}
exactly. The angular momentum (either ADM or Komar) is zero. 

Then, allowing $m = m(t)$ in the metric, we get
\begin{align}
    G^r_t &= e^{\frac{(1-u^2)m^2 - 2 m r}{r^2}} \frac{(3 m u^2 - m + 2r)}{r^3} \frac{d m}{dt} \nn 
    G^r_\phi &= 0.
\end{align}
We then find, on any surface $C$ of constant $r,t$, using $\Si$ again to represent a constant-$r$ surface,
\begin{align}
    \f{d \mc E_\Si}{d t} &= \frac{1}{8\pi} \oint G^r_t \sqrt{-g} du d \phi \nn &= \frac{dm}{dt} \nn 
    \f{d \mc J_\Si}{dt} &= - \frac{1}{8\pi} \oint G^r_\phi \sqrt{-g} du d\phi \nn 
    &= 0,
\end{align}
exactly. Thus we have recovered the ADM energy (and angular momentum) using the ``stress--energy tensor'' method. Note that we are not necessarily considering the black hole region---just looking at the energy flow approach as a way to define energy and angular momentum.

\subsubsection{Black Ring}

The five-dimensional black ring was discovered in \cite{EmparanReall}. I use coordinates from Elvang and Emparan \cite{ElvangEmparan}. 

The black ring spacetime is a vacuum ($\Lambda = 0$) solution with a horizon with a $S^2 \times S^1$ geometry. In coordinates $(t,x,y,\phi,\psi)$, the black ring takes the form
\begin{align}
    ds^2 &= -\frac{F(x)}{F(y)} \left( dt + R \sqrt{\lambda \nu} (1+y) d \psi\right)^2 + \nonumber \\
    &\qquad \frac{R^2}{(x-y)^2} \left[ - F(x) \left( G(y) d \psi^2 + \frac{F(y)}{G(y)} d y^2\right) + F(y)^2 \left( \frac{dx^2}{G(x)} + \frac{G(x)}{F(x)} d \phi^2\right)\right],
\end{align}
where
\begin{align}
    F(\xi) = 1 - \lambda \xi, \qquad G(\xi) = (1 - \xi^2)(1 - \nu \xi).
\end{align}
$R$ is a parameter of dimension length and $\lambda$ and $\nu$ are dimensionless parameters. The variables $x$ and $y$ take the values
\begin{align}
    -1 \leq x \leq 1, \qquad - \infty < y \leq -1, \lambda^{-1} < y < \infty.
\end{align}
$\psi$ and $\phi$ are coordinates with period
\begin{align}
    \Delta \phi = \Delta \psi = \frac{4 \pi \sqrt{F(-1)}}{|G'(-1)|} = \frac{2 \pi \sqrt{1+\lambda}}{1 +\nu},
\end{align}
a period which is introduced in order to avoid conical singularities at $x = -1$ and $y = -1$. The conical singularities at $x=+1$ can be avoided either by setting $\lambda = 1$ which produces a black hole, or $\lambda = 2 \nu/(1+\nu^2)$, which produces the black ring solution. Then the black ring horizon is located at $y = \nu^{-1}$ and has geometry $S^2 \times S^1$. $x$ and $\phi$ are coordinates which, on $y = \nu^{-1}$, parametrize the 2-sphere, and $\psi$ parametrizes the circle. (If $\lambda = 1$, then $x,\phi,\psi$ parametrize the 3-sphere.) There is no angular momentum associated with $\phi$, but there is one associated with $\psi$. The parameter $\nu$ varies as $0 \leq \nu < 1$. 

I'll take the $\lambda = 2\nu/(1+\nu^2)$ and also introduce coordinates
\begin{align}
    \bar \phi &= \frac{2 \pi \phi}{\Delta \phi} = \frac{1 + \nu}{\sqrt{1 + \lambda}} \phi \nonumber \\
    \bar \psi &= \frac{2 \pi \psi}{\Delta \psi} = \frac{1 + \nu}{\sqrt{1 + \lambda}} \psi
\end{align}
which have the canonical period $2 \pi$. 

We can now take \emph{this} metric, in terms of coordinates $(t,x,y,\bar \phi,\bar \psi)$, and allow $R = R(t), \nu = \nu(t)$ (and $\lambda = 2\nu(t)/(1+\nu(t)^2)$). 

Using this metric, we have, on a constant-$y$, constant-$t$ surface, integrating from $-1 \leq x \leq 1$ and over $2\pi$ for both $\phi$ and $\psi$,
\begin{align}
    \f{d \mc E_\Si}{dt} = \frac{1}{8\pi} \oint G^y_t \sqrt{-g} dx d \phi d \psi &= \frac{d}{d t} \left(\frac{3 \pi R^2 \nu (1 +\nu)}{2 (1 + \nu^2)^2}\right) \nonumber \\
    \f{d \mc J_\Si}{dt} = -\frac{1}{8\pi} \oint G^y_{\bar \psi} \sqrt{-g} dx d \phi d \psi &= \frac{d}{dt} \left(-\frac{ \pi R^3 \nu (1 + \nu)^3}{\sqrt 2 (1 +\nu^2)^3}\right) \label{blackringderivatives} 
\end{align}
exactly (in both cases). Here I am using $\mc J$ to refer to the angular momentum associated with the $\upsilon \equiv \pa_{\bar \psi}$ direction. (Note that $\sqrt{-g} = (1+\lambda)(1-\lambda x)(1-\lambda y)^2 R^4/[(1+\nu)^2(x-y)^4]$; for this range of $\nu$ and $x$, $1 - \lambda x > 0$.) The expressions within the parentheses match with the expressions for ADM energy and angular momentum respectively as reported in \cite{ElvangEmparan}, once $\lambda = 2 \nu/(\nu^2+1)$ is substituted, and with an additional negative sign for the angular momentum. The latter might not be physically meaningful. Note that these results are independent of $y$. The metric is not regular at the horizon $y = \nu^{-1}$, since $G(y) = 0$ and $G(y)$ appears in the numerator and denominator of different metric components, but the metric is regular as $y \to \nu^{-1}$ (from above). 

The angular momentum associated with the $\phi$ direction is zero, and indeed $G^y_{\bar \phi} = 0$ identically.

It is further worth noting that the negative of the Komar integral, on a constant $t,y$ surface, associated with $\xi = \pa_t$ is given by, with orientation given by $\bs\ep_{t y x \bar \phi \bar \psi} > 0$,
\begin{align}
    -\oint_C \bs K^K_\xi &= \f{\pi R^2 \nu(1+\nu)}{(1+\nu^2)^2},
\end{align}
that is, $2/3$ the ADM energy. This holds if $R,\nu$ are constant and also holds even if they are allowed to vary with $t$.  Using $\ups = \pa_{\bar \psi}$ we similarly have
\begin{align}
    \oint_C \bs K^K_\ups &=- \f{\pi R^3 \nu(1+\nu)^3}{\sqrt 2 (1+\nu^2)^3},
\end{align}
matching with the ADM angular momentum again up to a sign. As with the $\xi$ case, the right-hand side has the $t$-dependent values of $R, \nu$ in the $t$-dependent spacetime. The Komar integrals then once again produce (up to a multiplicative constant in the $t$ case) the quantities appearing inside the derivative in \eqref{blackringderivatives}. 

\subsection{Kerr--AdS}

I work in four dimensions only for simplicity.

Here I follow the basic pattern laid out in Section \ref{physicalprocesskerr} for the Kerr case, beginning with the case where $\La$ is held constant, where we find many analogous results. In particular I will show that when we use ABL coordinates, we find exact time derivatives associated with $d \mc E_\Si/d \tau, d \mc J_\Si/d\tau$ which correspond to the expected $\mc E, \mc J$ values, and when we use a similar ``ingoing Kerr--AdS--Vaidya'' type spacetime we get similar expressions with extra terms. I also show that if we use the BL form of the metric we do not get the same clean results, further bolstering the argument that the ABL asymptotically static frame is better suited to describing the conserved quantities. The main difference, which we expect, is that unlike with Kerr, Curzon--Chazy or the Black Ring spacetime, the Komar integral associated with $\xi$ does not reproduce the conserved energy.

Having gone through the case where $\La$ does not vary, I consider the case where it is allowed to vary, via $l$. 

\subsubsection{$\La$ Unvarying}

For a ``Kerr--AdS--Vaidya'' metric, take the form of the metric from \eqref{kadsIngoing4D} and allow $m$ and $a$ (at first) to vary with $v$. In the next subsection, I will allow $l$ to vary with $v$ as well. The $l \to \infty$ limit of this spacetime is the ``Kerr--Vaidya'' from Section \ref{physicalprocesskerr}. The justification for this ``Kerr--AdS--Vaidya'' spacetime is the same as for the ``Kerr--Vaidya'' one (with the same caveats).

With a constant background $\La$, we have $G^a_b = 8\pi T^a_b - \La \de^a_b$. That means we have
\begin{align}
    T^r_v &= \f{1}{8\pi} G^r_v \nn 
    T^r_\psi &= \f{1}{8\pi} G^r_\psi,
\end{align}
i.e.~the $\La$ term does not appear in the expressions for the (mixed-component) stress--energy terms which are not on the diagonal. That means we can still define $d \mc E_\Si/dv, d \mc J_\Si/dv$ as in \eqref{dEsigmadv} and \eqref{dJsigmadv}. The evaluation is, again denoting any terms involving either second- or higher-derivatives of $m,a$ or products of at least two first-derivatives thereof (including the squares of first-derivatives) as second-order terms,
\begin{align}
    \f{d \mc E_\Si}{d v} &= \f{1}{8\pi} \oint \sqrt{-g} G^r_v d u d \psi = \f{d}{dv} \left(\f{m}{(1-a^2/l^2)^2}\right) + \textrm{(second-order terms)} \nn 
    \f{d \mc J_\Si}{dv} &= -\f{1}{8\pi} \oint \sqrt{-g}G^r_\psi du d \psi = \f{d}{dv}\left(\f{m a}{(1-a^2/l^2)^2}\right) + \textrm{(second-order terms)}.
\end{align}
$\Si$ is a constant-$r$ hypersurface and the integral is on a 2-surface of constant $r,v$. 

We find that this method has given a way to recover $\mc E, \mc J$ for Kerr--AdS, with which we are very familiar by this point. The first-order results are independent of $r$. As with the Kerr case we find that the $v$-derivative of Komar integrand associated with $\eta = \pa_\psi$ matches with the $d \mc J_\Si/dv$, so that we recover \eqref{dKetadv}. 

As stated previously, the Komar integral associated with $\xi$ does not give the ADM energy (even up to a constant multiple). 

As in the Kerr case, define $r_+(v)$ to be the outer solution to
\begin{align}
    (r_+^2(v)+a^2(v))(1+ r_+^2(v)/l^2) - 2m(v) r_+(v)=0,
\end{align}
and let $\Si$ correspond to $r = r_+(0)$, the value where $v = 0$. We can define the $v$-dependent vector $\z^a$ to be
\begin{align}
    \z^a &= \de^a_v + \Om(v) \de^a_\psi \nn 
    \Om(v) &= \f{a(v) (1 + r_+^2(v)/l^2)}{a^2(v)+r_+^2(v)}.
\end{align}
Then $\z^a \z_a = 0$ when $r = r_+(v)$. We thus have as in the Kerr case that the hypersurface $\Si$ is null specifically at $v = 0$, and at specifically $v = 0, r = r_+(0)$, $\z^a$ is a null vector tangent and normal to $\Si$, with $\z_v = \z_u = \z_\psi = 0$. Again we find on $r = r_+(v)$
\begin{align}
    \z^b \na_v \z^a &= \kappa(v) \z^a + \textrm{(first-order terms)}
\end{align}
with
\begin{align}
    \kappa(v) &= \f{3 r_+^4(v) + (l^2 + a^2(v)) r_+^2(v) - a^2(v)l^2}{2 l^2 r_+(v) (a^2(v)+r_+^2(v))}.
\end{align}
The area element of the 2-surface of constant $v, r = r_+(v)$ is given by 
\begin{align}
    dS &= \f{ r_+^2(v) + a^2(v)}{1-a^2(v)/l^2} d u d \psi,
\end{align}
so that the (null) 2-surface of constant $v, r = r_+(v)$ has area $A(v) = 4 \pi (r_+^2(v) + a^2(v))/(1-a^2(v)/l^2)$. Similarly to the Kerr case we can identify the combination
\begin{align}
    \kappa(0) \left.\f{d A(v)}{d v}\right|_{v=0}
\end{align}
with the surface $\Si$ given by $r =r(0)$ and the vector $\z^a$, recovering, to first order in the $v$-derivatives, \eqref{firstlawdbydv}. 

Next we can check the ABL form of the metric given by \eqref{ABL4D} with additionally $u=\cos\tht$, setting $m \to m(\tau), a \to a(\tau)$ to create a ``time-dependent ABL'' form of the metric. In this case we find
\begin{align}
    \f{d\mc E_\Si}{d \tau} &= \f{1}{8\pi}\oint \sqrt{-g} G^r_\tau du d\hat\vp = \f{d}{d\tau} \left(\f{m}{(1-a^2/l^2)^2}\right) \nn
    \f{d\mc J_\Si}{d\tau} &= -\f{1}{8\pi} \oint \sqrt{-g} G^r_{\hat\vp} du d\hat \vp = \f{d}{d\tau} \left( \f{m a}{(1-a^2/l^2)^2}\right),
\end{align}
exactly in both cases. Thus using the ABL form of the metric we exactly recover $\mc E,\mc J$ as the expressions whose $\tau$ derivatives are being taken. We also have
\begin{align}
    \f{d}{d\tau}(\bs K^K_\eta)_{u\hat\vp} = -\f{1}{8\pi} \sqrt{-g} G^r_{\hat \vp}.
\end{align}

It is also worth checking the BL form of the metric, with $u = \cos \tht$ as usual, creating a ``time-varying BL'' by allowing $m,a$ to depend on $\tau$. In this case, $\bt = \pa/\pa \tau$, which is a Killing vector if $m,a$ are constant with respect to $\tau$. Letting $d \tilde{\mc E}_\Si/d\tau$ be defined analogously to $d \mc E_\Si/d\tau$ by
\begin{align}
    \f{d  \tilde{\mc E}_\Si}{d\tau} &= \f{1}{8\pi} \oint \sqrt{-g} G^r_\tau d u d \varphi
\end{align}
on a 2-surface of constant $r,\tau$, we have
\begin{align}
     \f{d \tilde{\mc E}_\Si}{d\tau} &= \f{1}{8\pi} \oint \sqrt{-g} G^r_\tau d u d \varphi = \f{1}{(1-a^2/l^2)} \f{dm}{d\tau} + \f{3 a m}{(1-a^2/l^2)^2} \f{da }{d\tau},
\end{align}
which is not a total $\tau$-derivative for generic functions $m(\tau),a(\tau)$. By contrast we have
\begin{align}
    \f{d \mc J_\Si}{d \tau} &= -\f{1}{8\pi} \oint \sqrt{-g} G^r_\vp du d \vp = \f{d}{d\tau} \left( \f{m a}{(1-a^2/l^2)}\right).
\end{align}
As expected setting 
\begin{align}
    \om(\tau) &= \f{a(\tau) (1 - a^2(\tau)/l^2)}{r_+^2(\tau) + a^2(\tau)}
\end{align}
with $(r_+^2(\tau)+a^2(\tau))(1+r_+^2(\tau)/l^2) - 2 m(\tau) r_+(\tau)$, the combination $d \tilde{\mc E}_\Si/d\tau - \om(\tau) d \mc J_\Si/d\tau$ gives
\begin{align}
    \f{d \tilde{\mc E}_\Si}{d\tau} - \om(\tau) \f{d \mc J_\Si}{d\tau} &=  \f{1}{8\pi} \f{3 r_+^4 + (l^2+a^2)r_+^2 - a^2l^2}{2 l^2 r_+(r_+^2+a^2)} \f{d}{d\tau} \left( \f{4\pi (r_+^2 + a^2)}{1-a^2/l^2}\right),
\end{align}
so that the combination still matches the expected $\kappa d A/d\tau$-type term. The key result from using the BL coordinates is that the ``energy transfer'' with respect to the $\bt$ direction is not a total $\tau$ derivative, unlike in the ABL coordinates.

Because of the connection between the ``integration of $\sqrt{-g} G^r_\tau d u d \hat \vp$'' argument and the ``integration of $\bs k^{EH}_\chi [\de g;g]$'' argument, which I will elaborate on in Section \ref{relationshiptokEHchi}, the observation that the ``energy'' associated with $\bt$ is not integrable is related to the observation in \cite{HajianSheikh-Jabbari} that the ``energy'' associated with $\bt$ is not integrable by evaluating the integral of $\oint \bs k^{EH}_\bt[\de g;g]$ through solution space. Other authors who laid out the non-integrability of the ``energy'' associated with $\bt$ in BL coordinates (again, through other methods than the one I mentioned here) include Blagojevi\'c and Cvetkovi\'c \cite{Blagojevic:2020edq} and Jing and Peng \cite{Jing:2017jxw}.

\subsubsection{$\La$ Varying}

The situation is more complicated if we also allow $l$ to be a function of $v$ or $\tau$ as well. This corresponds to $\La$ effectively being time-dependent. Without a specific physical theory which would give rise to the changing value of $\La$, it is difficult to say what the relationship between the Einstein tensor and stress--energy tensor should be. We may consider writing
\begin{align}
    G^a_b &= 8\pi T^a_b - \La(v) \de^a_b,
\end{align}
but it may be that if the change in $\La$ results from some physical field it might be that should be included in some way in $T^a_b$; or it might be that some terms involving $d\La/dv$ should be moved ``outside'' $T^a_b$, in addition to $\La(v)$. (Similarly for when $\La = \La(\tau)$.) To avoid dealing with these issues right now I will continue writing quantities in terms of $G^a_b$ and not include direct reference to $T^a_b$. I will not distinguish, as a result, between the effect of matter moving across a boundary surface or the result of $\La$ changing.

The ``Kerr--AdS--Vaidya,'' ``time-varying ABL'' and ``time-varying BL'' spacetimes are all the same as in the $\La$ unvarying case except that now $l = l(v)$ (``Kerr--AdS--Vaidya'') or $l = l(\tau)$ (the other two cases). I take $\La(v) = -3 / l^2(v)$ or $\La(\tau) = -3/l^2(\tau)$.

For the ``Kerr--AdS--Vaidya'' spacetime we now have
\begin{align}
    \f{d \mc E_\Si}{d v} &= \f{1}{8\pi} \oint \sqrt{-g} G^r_v d u d \psi = \f{d}{dv} \left(\f{m}{(1-a^2/l^2)^2}\right) - \Th(r,v) \f{d\La}{d v}+ \textrm{(second-order terms)} \nn 
    \f{d \mc J_\Si}{dv} &= -\f{1}{8\pi} \oint \sqrt{-g}G^r_\psi du d \psi = \f{d}{dv}\left(\f{m a}{(1-a^2/l^2)^2}\right) + \textrm{(second-order terms)},
\end{align}
where
\begin{align}
    \Th(r,v) &= -\f{r(r^2+a^2)}{6(1-a^2/l^2)} - \f{m a^2}{6 (1-a^2/l^2)^2} \nn 
    &= -\f{1}{8\pi} \mc V(r,v) - \f{ma^2}{6(1-a^2/l^2)^2}. \label{Thetarv}
\end{align}
The term $\mc V(r_0,v)$
\begin{align}
    \mc V(r_0,v) &= \int_{0}^{r_0} dr \oint \sqrt{-g} du d \phi
\end{align}
is a generalization of the vector volume or geometric volume of the constant-$v$ hypersurface between $r =0$ and $r =r_0$. ($r_0$ here is the outer radius in question and is not related to the $r_0$ satisfying $r_0^2+a_n^2 = 0$.) In \eqref{Thetarv}, $a$ and $m$ are functions of $v$, and the $r$ appearing in $\Th$ is the value of (constant) $r$ on the hypersurface $\Si$. Of course if we evaluate on $r = r_+$ then we get 
\begin{align}
    \Th &= -\f{r_+(r_+^2+a^2)}{6(1-a^2/l^2)} - \f{m a^2}{6 (1-a^2/l^2)^2} \nn 
    &= \f{1}{8\pi} \mc V(r_+,v) - \f{m a^2}{6 (1-a^2/l^2)^2}.
\end{align}
This is just the $\Th$ which appears in CGKP and is described in Section \ref{CveticSection}, modified so that now $r_+,a,m$ are all functions of $v$.   

It is interesting then that the $\Th$ term arises entirely from the $\pa_v$-based terms, and nothing from the $\pa_\psi$-based terms. It is also interesting that the ``geometric volume'' part is modified if instead the integration surface is taken to be a different value of $r$, in which case the equivalent volume term up to that value of $r$ appears instead.

The relationship between the angular momentum variation term and the Komar integral \eqref{dKetadv} still holds.

Generalizing $r_+(v),\Om(v)$ and $\z$ from the $\La$ unvarying case only requires allowing $l = l(v)$, and otherwise the arguments are identical, so we can, evaluating everything at $v = 0$, state that to first-order in the derivatives,
\begin{align}
    \f{d\mc E_\Si}{dv} - \Om(0) \f{d\mc J_\Si}{dv} &= \f{\kappa(0)}{8\pi} \f{d A}{dv}.
\end{align}
This gives
\begin{align}\f{d}{dv}\left( \f{m}{(1-a^2/l^2)^2}\right) - \Th(r_+(0),0) \f{d \La}{dv} - \Om(0) \f{d}{dv} \left( \f{m a}{(1-a^2/l^2)}\right) &= \f{\kappa(0)}{8\pi} \f{d A}{dv},
\end{align}
where $\Si$ is specifically the surface $r = r_+(0)$. It is noteworthy that now we must take the surface $r=r_+(0)$ on the left-hand side, since $d \mc E_\Si/dv$ now depends not just on $v$ but also $r$. This naturally leads to the identification of $m/(1-a^2/l^2)^2$ with the $\mc E$ in the first-law (enthalpy, when $\La$ is allowed to change).

Once again the second-order terms drop out if we use the ``time-dependent ABL'' form of the metric, here also using $l = l(\tau)$. We then have on a constant-$r$ hypersurface $\Si$
\begin{align}
    \f{d\mc E_\Si}{d \tau} &= \f{1}{8\pi}\oint \sqrt{-g} G^r_\tau du d\hat\vp = \f{d}{d\tau} \left( \f{m}{(1-a^2/l^2)^2} \right) - \Th(r,\tau) \f{d \La}{d\tau} \nn 
    \f{d \mc J_\Si}{d\tau} &= -\f{1}{8\pi} \oint \sqrt{-g} G^r_{\hat \vp} du d\hat\vp = \f{d}{d\tau} \left( \f{ma}{(1-a^2/l^2)^2}\right),
\end{align}
exactly. $\Theta(r,\tau)$ is the same as in \eqref{Thetarv} with $a,l$ functions of $\tau$ instead of $v$; 
\begin{align}
    \Th(r,\tau) &= -\f{1}{8\pi} \mc V(r,\tau) -\f{m a^2}{6(1-a^2/l^2)^2} \nn 
    \mc V(r_0,\tau) &= \int_0^{r_0} dr \oint \sqrt{-g} du d \hat\vp.
\end{align}

\subsection{Angular Momentum}

Here I explain why we keep recovering the relationship between the ``physical process'' angular momentum variation and the Komar term. I will consider coordinates $(v,r,u,\psi)$ for this but it applies equally well to $(v,r,u,\psi)$, $(\tau,r,u,\hat \vp)$ or $(\tau,r,u,\vp)$ coordinate systems. Let $\eta = \pa/\pa \phi$ and let it be a Killing vector. We can rewrite the result \eqref{dKetadv} as stating
\begin{align}
    \sqrt{-g} G^r_\psi &= -\f{\pa}{\pa v} (\sqrt{-g} \na^v \eta^r). \label{Gnavetarexp}
\end{align}
Since $\eta^a$ is a Killing vector, \eqref{Boxchi} gives $\na_b \na^b \eta^a = - R^a_b \eta^b$ and $\na^v \eta^r = \na^{[v}\eta^{r]}$. Thus we have
\begin{align}
    \na_b \na^b \eta^a &= \na_b \na^{[b} \eta^{a]} \nn 
    &= \f{1}{\sqrt{-g}} \pa_b \left( \sqrt{-g} \na^{[b} \eta^{a]}\right) \nn 
    \na_b \na^b \eta^r &= \f{1}{\sqrt{-g}} \pa_v \left( \sqrt{-g} \na^{[v} \eta^{r]}\right) + \pa_u \left( \sqrt{-g} \na^{[u} \eta^{r]}\right),
\end{align}
on the last line using the fact that the spacetime and the derivatives of $\eta^a$ are independent of $\psi$. It turns out that $\na^{[u} \eta^{r]} = 0$, for the spacetimes under consideration, so that we have
\begin{align}
    \na_b \na^b \eta^r &= \f{1}{\sqrt{-g}} \pa_v \left( \sqrt{-g} \na^{[v} \eta^{r]}\right),
\end{align}
which, combining with \eqref{Boxchi} and the fact that $G^r_\psi = R^r_\psi$, gives \eqref{Gnavetarexp}. 

\subsection{Why Are Results Exact with ``Time Varying'' But Not ``Vaidya'' Spacetimes?} \label{timevaryingbutnotvaidya}

Why is it that the second-order derivative terms drop out when we use the ``time-varying'' spacetimes and not the ``Vaidya'' ones? 

To check, I defined a general metric of the form
\begin{align}
    ds^2 = A dt^2 + B dr^2 + C du^2 + D d\phi^2 + 2 E dt d\phi, \label{ABCDE}
\end{align}
where each of the metric functions are functions of $t,r,u$. Then (using \emph{GRTensorIII}), I checked and verified that all terms are of 0th or 1st order in $t$ derivative in $G^r_t$ and $G^r_\phi$. In fact, using $\eta = \partial_\phi$,
\begin{align}
    G^r_\phi &= \frac{\partial}{\partial t}( \nabla^t \eta^r) + \nabla^t \eta^r \Gamma^\alpha_{t \alpha} \textrm{ (exactly)},
\end{align}
or, writing it another way,
\begin{align}
    \sqrt{-g} G^r_\phi &= \frac{\partial}{\partial t}\left(\sqrt{-g} \nabla^t \eta^r\right),\label{GrphiId}
\end{align}
so that the relationship between the angular momentum transfer and Komar formula for angular momentum is reconstructed. (And again, we do not get an equivalent formula for $\nabla^t\xi^r$ where $\xi = \partial/\partial t$.)  Specifically here,
\begin{align}
    \sqrt{-g} G^r_\phi &= \f{\pa}{\pa t} \left( -\f{C}{2 \sqrt{-g}} \left( D \f{\pa E}{\pa r} - E \f{\pa D}{\pa r}\right)\right).
\end{align}

By contrast, in general $\sqrt{-g} G^r_t$ is not the $t$ derivative of some quantity. This is to be expected, because in the ``time-varying Kerr--AdS ABL'' spacetime the $d \mc E_\Si/d\tau$ was not a $\tau$ derivative of some quantity when $l$ was allowed to vary. Nevertheless, $ G^r_t$ is only first-order in $t$-derivatives, reassuringly. It is given by, using the convention $A_t = \pa A/\pa t, A_{tr} = \pa^2 A/\pa t \pa r$ and so on,
\begin{align}
    G^r_t &= \f{1}{\sqrt{-g}} \f{\pa}{\pa t}\left( \sqrt{-g} \na^{[t}\xi^{r]}\right) -\f{C D A_{rt}}{2 g} + \f{BC^2A_t}{4g^2}\left( D^2 A_r+E^2 D_r - 2 D E E_r\right) +\f{B_t}{8g^2}\f{\pa}{\pa r}\left(\f{g^2}{B^2}\right) + \f{C_rC_t}{4 B C^2} \nn
    &\qquad - \f{C_{rt}}{2 B C} - \f{ACD_{rt}}{2g} + \f{BC^2D_t}{4g^2}\left( A^2 D_r + E^2 A_r-2AEE_r\right) \nn 
    &\qquad + \f{CE}{g} E_{rt} - \f{BC^2}{2g^2}\left( -(AD+E^2)E_r + ED A_r +AE D_r\right) E_t. \label{GrtABCDE}
\end{align}
$g$ is the metric determinant. This is not very illuminating, but it is noteworthy that the $u$ derivatives all drop out and that the expression is completely linear in the $t$ derivatives (note that $\na^{[t}\xi^{r]}$ does not contain any $t$ derivatives of the metric components).

Interestingly, even changing the metric to 
\begin{align}
    ds^2 = A dt^2 + B dr^2 + C du^2 + D d\phi^2 + 2 E dt d \phi + 2 F dr du + 2 G dr d \phi + 2 H du d \phi, \label{FGH}
\end{align}
including other cross-terms, introduces terms which are either quadratic in the first $t$ derivative or which are second derivatives in $t$ in $G^r_t$ and $G^r_\phi$. Further investigation reveals that it is the $G$ and $H$ terms above which are the culprit, and there are no terms quadratic in the first $t$ derivative or linear in the second $t$ derivative in the metric
\begin{align}
    ds^2 = A dt^2 + B dr^2 + C du^2 + D d\phi^2 + 2 E dt d \phi + 2 F dr du.
\end{align}
(Specifically, $R_{r t}$ contains only terms to the first order in $t$ derivatives even if $G = 0$ but $H \neq 0$, but then $R^r_t$ is related to $R_{u t}$, which contains terms which are higher order if $H \neq 0$, since $g^{r u} \neq 0$, so it requires that both $G$ and $H$ be zero, or for $F = 0$ so that $g^{r u} = 0$, and so on.)

I think the next step would be to break the Ricci tensor down into its constituent components to show that second-order terms drop out when the metric is in the form \eqref{ABCDE} but not the form \eqref{FGH}, but this is left for future work.

\subsection{Relationship to \texorpdfstring{$\bs k^{EH}_\chi$}{kEHchi}} \label{relationshiptokEHchi}

Again consider the \eqref{ABCDE} general metric, now modified so that the functions are $t$-independent, but also that the metric as a whole depends on $s$ in solution space, so that $A = A^{(s)}(r,u)$ and so on. Then we can calculate $\bs k^{EH}_\chi$. Specifically let 
\begin{align}
    \bs k^{EH}_\chi [\de g;g] &= \de s \bs k^{EH}_\chi \left[ \f{d g^{(s)}}{ds} ; g\right] \nn
    \bs k^{EH}_\chi \left[ \f{d g^{(s)}}{ds} ; g\right] &= -\f{d}{ds} (\bs K^K_\chi)^{(s)} - \chi \cdot \bs \Th\left[ \f{d g^{(s)}}{ds};g\right],
\end{align}
where all terms are evaluated at $s = 0$ and it is assumed that $\de \chi^a = 0$. Here $\xi = \pa_t$ is a Killing vector (with constant components) and so we can apply the above to $\chi = \xi$. Letting $A_{s}$ represent $\pa A^{(s)}/\pa s$, $A_{r} = \pa A^{(s)}/\pa r$, $A_{rs} = \pa^2 A^{(s)}/\pa r \pa s$ and so on, we find that the $u\phi$ component of $\bs k^{EH}_\xi [dg^{(s)}/ds;g^{(s)}]$ is
\begin{align}
    &\f{8\pi}{\sqrt{-g}} \left(\bs k^{EH}_\xi \left[ \f{d g^{(s)}}{ds}; g^{(s)}\right]\right)_{u\phi}  \nn 
    &= \f{1}{\sqrt{-g}} \f{\pa}{\pa s}\left( \sqrt{-g} \na^{[t}\xi^{r]}\right) -\f{C D A_{rs}}{2 g} + \f{BC^2A_s}{4g^2}\left( D^2 A_r+E^2 D_r - 2 D E E_r\right) +\f{B_s}{8g^2}\f{\pa}{\pa r}\left(\f{g^2}{B^2}\right) + \f{C_rC_s}{4 B C^2} \nn
    &\qquad - \f{C_{rs}}{2 B C} - \f{ACD_{rs}}{2g} + \f{BC^2D_s}{4g^2}\left( A^2 D_r + E^2 A_r-2AEE_r\right) \nn 
    &\qquad + \f{CE}{g} E_{rs} - \f{BC^2}{2g^2}\left( -(AD+E^2)E_r + ED A_r +AE D_r\right) E_s.
\end{align}
That is, comparing to \eqref{GrtABCDE}, we have that, \emph{exactly}, the result for $(\bs k^{EH}_\xi [dg^{(s)}/ds;g^{(s)}])_{u\phi}$ is the same as $\sqrt{-g}G^r_t / 8\pi$ with every $t$ partial derivative associated with the metric component functions in the latter being replaced with an $s$ partial derivative! 

This tells us that in a metric of the form \eqref{ABCDE}, the $\bs k^{EH}_\xi [\de g;g]$ and $G^a_b\xi^b$ forms correspond to each other, in a one-to-one correspondence! That is to say, if we can use 
\begin{align}
    \f{d \mc E}{d s} &= \oint \bs k^{EH}_\xi\left[\f{dg^{(s)}}{ds} ; g^{(s)}\right],
\end{align}
on a 2-surface of constant $r,t$, then this will yield the same $\mc E$ which appears in the time derivative in 
\begin{align}
    \f{d \mc E_\Si}{dt} &= \f{1}{8\pi} \oint \sqrt{-g} G^r_t du d\phi
\end{align}
on the same 2-surface! This tells us that methods derived to find conserved quantities by comparing nearby solutions in solution space using $\bs k^{EH}_\xi[\de g;g]$ and ones using $G^a_b\xi^b$ on a hypersurface in one space which is physically varying are equivalent to each other, in the specific case of a metric of the simple form \eqref{ABCDE}. 

For $\chi = \eta$ we also have that on a $u\phi$ surface, the $\eta\cdot \bs\Th$ term automatically drops out, so we have
\begin{align}
    \left(\bs k_\eta^{EH}\left[\f{d g^{(s)}}{ds};g^{(s)}\right]\right)_{u\phi} = -\f{d}{ds}( (\bs K^K_\eta)^{(s)})_{u \phi},
\end{align}
whereas the relationship \eqref{GrphiId} implies
\begin{align}
    \f{1}{8\pi}\sqrt{-g} G^r_\phi &= -\f{\pa}{\pa t}( (\bs K^K_\eta)^{(s)})_{u \phi}.
\end{align}
Thus the relationship between $\left(\bs k^{EH}_\eta[d g^{(s)}/ds,g^{(s)}]\right)_{u\phi}$ and $\sqrt{-g} G^r_\phi/8\pi$ matches the one for $\xi,t$.

I think with some further work that this ``equivalence'' can be formulated in a more precise way. What is nice about this is not just that it provides some justification for the physical process method in terms of methods we have used elsewhere but also uses the analogy with the physical process method to lend additional weight to the use of $\bs k^{EH}_\chi[\de g;g]$. Again, this reinforces the idea that this ``physical process version'' is similar to integrating $\bs k^{EH}_\chi$ directly to get the conserved charges (as in the BC method, or as in \cite{HajianSheikh-Jabbari}). 

Of course the ``time-varying ABL metric'' form of Kerr--AdS, as well as ``time-varying Kerr'' and ``time-varying Curzon--Chazy,'' are of the general form \eqref{ABCDE}. 

\subsection{Physical Process Version Summary}

Results here are tentative but this suggests that we can use the approach of looking at the flux of the stress--energy tensor for slow variations which transform one solution in a class of solutions to another (e.g.~one Kerr--AdS solution to another) in order to integrate out conserved quantities. I find it interesting that we can recover $\Th$ by writing $d \mc E/dv = d\mc E/dv - \Th d \La/dv$ for the Kerr--AdS case with varying $\La$, and also that the close relationship between the flux of angular momentum and angular momentum Komar integral gives further justification for the use of the Komar integral for angular momentum. Finally by showing how, in metrics of the form \eqref{ABCDE}, the $\sqrt{-g}G^r_b \chi^b/8\pi$ terms match up with the $\bs k^{EH}_\chi[dg^{(s)}/ds;g]$ terms, I connected this section to the rest of the thesis.

An obvious next step would be to consider the Kerr--Newman and Kerr--Newman--AdS solution, in this case hopefully separating out the ``matter'' and electromagnetic stress--energy tensor components in some way. I have done some early work on the Kerr--NUT spacetime as well. I would like to understand better the relationship between the above quantities and a more physically motivated model of a dynamic horizon, particularly one with varying $\La$. 

\chapter{Conclusion} \label{conclusionchapter}

The central concern of this thesis is the role of the volume in black hole thermodynamics. In Chapter \ref{PRDPaper} I defined the vector volume for a region $\mc R$ and a divergence-free vector $v^a$ such that $v^a$ is tangent to the boundary $\pa \mc R$ as the rate of growth of the $D$-dimensional volume of $\mc R$ along the flow of $v$. I showed that this is a constant, and that it applies particularly if $v$ is a Killing vector. I showed that in coordinates adapted to the vector $v$ such that $v^a = \de^a_0$, the vector volume can be written as $\int \sqrt{-g} d^{D-1}x$, where $g$ is the determinant of the metric and $d^{D-1}x$ is the product of the coordinate differentials with $dx^0$ omitted. For a Kerr--Schild spacetime, the background and full metric determinants are the same, showing that the vector volume for the full and background spacetimes are equal. For spacetimes where the background is Minkowski space, such as the Kerr spacetime, and where the vector in question is the staticity Killing vector in the Minkowski background, this implies the interpretation of the vector volume as the Euclidean volume as calculated in the Cartesian coordinates in the Minkowski background. I showed that the vector volume is linear with respect to the vector and that the vector volume associated with an azimuthal symmetry vector vanishes for an azimuthally symmetric region. 

I showed that for the Kerr--Newman--AdS or Kerr--Newman--de Sitter black holes, a canonical black hole volume can be defined by using the black hole region between the singularity and the (outer) event horizon and using the canonically normalized stationarity Killing vector. I showed that this canonical black hole volume, denoted by $\mc V_C$ or $\mc V_{\xi,\mc B}$, is equal to volumes which had previously appeared in the literature, especially the Parikh volume \cite{Parikh} and the geometric volume appearing in \cite{Cvetic}. 

In Chapter \ref{GKAdSChapter}, I stated some important facts about the higher-dimensional Kerr--AdS solutions and their place in the larger set of Generalized Kerr--NUT--AdS spacetimes. In addition to stating some important properties of the PCKY tensor $\bs h$ and associated principal vector $\bt$ needed in future chapters, I showed how the Jacobi transformations by \cite{Chen} applied to the $\mu_i$ from the original higher-dimensional solutions from \cite{GibbonsLu2} to the $y_\alpha$ parameters worked because they were in effect ``finishing the job'' of the Jacobi transformation as applied to the $\nu_i$, counting $x_n = i r$ as one of the $x_\mu$ parameters, and showed how the Generalized Kerr--AdS solutions can be written in pseudo-Cartesian coordinates. 

In Chapter \ref{NoetherChapter}, I reviewed how the papers by Gibbons, Perry and Pope \cite{GibbonsPerry} and Cveti\v{c}, Gibbons, Kubiz\v{n}\'ak and Pope \cite{Cvetic} laid out an argument for a first law variation. The GPP paper proposed that a definition of energy $\mc E$ for Kerr--AdS black holes must satisfy the first law $\de \mc E = \sum_i \Om_i \de \mc J_i + T \de S$ where $T \de S = \kappa \de A/8\pi$, with black hole angular momenta $\mc J_i$, angular velocities $\Om_i$, temperature $T$, entropy $S$, surface gravity $\kappa$ and area $A$. They obtained values of $\mc E$ by integrating the first law and found that it was necessary to use an asymptotically static frame in order to get a working first law. The CGKP paper then allowed $\tilde \La$ to vary, interpreting $\mc E$ as an enthalpy, and showed that this resulted in the presence of a thermodynamic volume $V_{th}$, with (in the presence of charges $\mc Q_\alpha$ and potentials $\Phi_\alpha$) $\de \mc E = \sum_i \Om_i \de \mc J_i + \sum_\alpha \Phi_\alpha \de \mc Q_\alpha + T \de S + V_{th} \de P$, where the pressure $P$ is proportional to $\tilde \La$. They point out that this follows directly from a scaling relation and the existence of a Smarr relation, written, for dimension $D$, $(D-3)\mc E = (D-2) \left( \sum_i \Om_i \mc J_i + T S\right) + (D-3) \sum_\alpha \Phi_\alpha \mc Q_\alpha - 2 V_{th} P$. They also showed that if, instead (using my notation) $\mc F$, the ``energy/enthalpy'' associated with the principal vector $\bt$ rather than the asymptotically static vector $\xi$, with corresponding angular velocity $\om_i$, is used, the Smarr relation is $(D-3) \mc F = (D-2) \left(\sum_i \Om_i \mc J_i + TS\right) +(D-3) \sum_\alpha \Phi_\alpha \mc Q_\alpha - 2 V_{geo} P$, where $V_{geo}$ is the geometric volume, which is the one which corresponds to $\mc V_{\xi,\mc B}$. They also pointed out the relation for Kerr--AdS $V_{geo} = r_+A/(D-1)$, where $r_+$ is the value of the spheroidal radius on the horizon.

To address why it was necessary to use the conserved charge associated with the asymptotically static Killing vector $\xi$ rather than the principal vector $\bt$ to construct a first law relation and why $\mc V_{\xi,\mc B}$ appears more naturally in the Smarr relation associated with $\bt$ rather than $\xi$, I gave a new definition for a conserved charge, making use of the covariant phase space methods of Wald and the idea from Barnich and Comp\`ere to integrate through solution space, in concert with the Kerr--Schild decomposition of the Kerr--AdS spacetimes. It is possible that the Relativity community was not clamouring for another definition of a conserved charge associated with a Killing vector, one only applicable to Kerr--Schild spacetimes, but it possesses many advantages. Because of its derivation via the covariant phase space formulation, it is justified to think of it as a conserved quantity associated with the Hamiltonian. It is defined ``in the bulk'' and not only at infinity. It corresponds to the ``first-law-integrated'' quantities. It has a simple form that matches expressions for the KBL superpotential but is exact even away from infinity. It is closely related to the stress--energy tensor. Finally, it is closely related to the first-law area relation at the horizon. 

This conserved quantity is equal to the difference in the Komar integrals associated with a given Killing vector in the background and full spacetimes, plus a non-Komar term. I showed how, when the conserved charge is a stationarity Killing vector (plus some azimuthal symmetry Killing vectors), the Komar integral associated with the background spacetime is (up to a multiplicative constant) equal to $\La$ times the canonical black hole volume $\mc V_{\xi,\mc B}$, explaining the black hole volume's natural appearance in the Smarr relation paired with $\La$ (or $\tilde \La$). 

In Chapter \ref{VolumeAreaChapter}, having previously shown that the geometric volume which appears in CGKP is the same as the canonical black hole vector volume $\mc V_{\xi,\mc B}$, I also showed how to interpret this as the Euclidean volume of the ellipsoid represented by the spatial part of the black hole region in the pseudo-Cartesian coordinates for the background AdS metric. I showed how the canonical volume appears naturally in the Euclidean approach to defining conserved charges. I showed how the method from \cite{KastorEtal:2009,Cvetic} 
of defining a Killing potential leads to the natural emergence of the vector volume. I explained the $\mc V_{\xi,\mc B} = r_+ A/(D-1)$ relationship by making use of the PCKY tensor $\bs h$, the equivalence of the volumes associated with $\xi$ and $\bt$, the Gauss--Stokes theorem, and the fact that the null generator of the horizon is an eigenvector of the PCKY tensor with eigenvalue $r_+$. 

In Chapter \ref{ExplicitGKAdSChapter}, by making use of the ``infinitesimal coordinate transformation'' method I demonstrated in four dimensions how, when $l$ is kept constant but the angular parameter $a$ varies, the variation of integral of $\bs I_{\bar \z}$ is equal to $\kappa \de A/8\pi$ in a set of coordinates where the AdS background metric does not vary, and how the difference between $\de \bs I_{\bar \z}$ and $\kappa \de A/8\pi$ can be expressed in terms of the Lie derivative associated with the coordinate transformation resulting from the variation in $a$. I showed too how in vacuum, this Lie derivative term vanishes, demonstrating the suitability of $\oint_H \bs I_{\bar \z}$ as a quantity which, when varied, gives $\kappa \de A/8\pi$. I showed that the formula $\oint_H \bs k^{EH}_{\bar \z}[\de g;g] = \kappa \de A/8\pi$ could be used even when the horizon radius changes, using the Lie derivative method. 

I argued that a fundamental reason that the charge $\mc F$ associated with $\bt$ cannot be used to construct a first-law relation is not so much that it is based on an asymptotically rotating frame---or, relatedly, that the vector $\bt$ is asymptotically rotating, rather than asymptotically static---but that its degree of rotation is proportional to the rotation parameter $a$ (parameters $a_i$ in higher dimensions), so that varying $a$ ($a_i$) causes the $\bt$ vector to point ``in different directions,'' with respect to the AdS background. This makes it impossible to find a set of coordinates in which both components of the background metric $\bar g_{ab}$ and the vector $\bt^a$ are unvarying. By contrast there are coordinates where $\bar g_{ab}$ and the vectors $\xi^a$ and $\eta^a$ ($\eta_i^a$ in higher dimensions) are unvarying even as $a$ varies ($a_i$ vary). This gives my interpretation for why it is important to use the asymptotically static Killing vector $\xi$ rather than the principal vector $\bt$ when constructing conserved charges. This amounts to a particular explanation for the observation, made elsewhere in the literature, that the Hamiltonian associated with $\bt$ is not integrable, but the one associated with $\xi$ is. 

In showing how the expression for $\bs I_\chi$ simplifies considerably for $\chi = \bt$, and in demonstrating the equivalence, up to a numerical constant, of the Komar-difference and non-Komar terms associated with $\bs I_\bt$, I gave an explanation for the canonical black hole volume appearing naturally when the Smarr relation is constructed using the quantity $\mc F$ associated with $\bt$, rather than in the energy term $\mc E$ associated with $\xi$. I found expressions for the conserved charges for an arbitrary $\mu(r)$ in order to highlight the simplifications that occur in vacuum.

\section{Future Work} \label{futurework}

To follow up on the Addendum for Chapter \ref{PRDPaper}, Section \ref{addendum}, I would like to look more carefully at what we can say about the vector volume's dependence on choice of hypersurface in situations where adapted coordinates do not exist covering the full manifold. For spacetimes where the vector field $v^a$ vanishes at some point in the manifold, the statement that the hypersurface must intersect each of the integral curves of $v$ once does not make sense when talking about the whole manifold rather than the portion of the manifold covered by a particular adapted coordinate chart. There may be a generalization of that requirement that does make sense globally.

I find the embedding spacetime introduced in Section \ref{embedding} and continuing in Appendix \ref{embeddingcalcs} to be intriguing and would like to study it further.

In Section \ref{varyingk}, I mentioned that I wanted to expand out $\de \bs \tht^{EH}$ when the background metric is allowed to vary.

The explanation for the equality, up to a numerical constant, of the Komar difference and non-Komar terms associated $\bt$ comes down to the observation \eqref{asymptoticintegral}, resulting from \eqref{asymptoticintegrand}, which are
\begin{align}
     \lim_{r\to\infty} \sqrt{-\bar g} (\bs{i}^K_\bt)^{tr} &= (D-3) \lim_{r\to\infty} \sqrt{-\bar g} (\bs{i}^\tht_\bt)^{tr} \nn 
     -\oint_{C_\infty} \left( \bs K^K_{\beta} - \overline{\bs K^K_\beta}\right) &= (D-3) \left(-\oint_{C_\infty} \beta \cdot \bs \tht\right). 
\end{align}
It is, on the one hand, satisfying to get to this point and to find this equality. On the other hand, it is unclear whether this is something that ``had to happen'' or something that ``just worked out.'' Because the only Kerr--Schild solution to vacuum Einstein's equations which are globally asymptotically AdS is Kerr--AdS, and because the Kerr/Myers--Perry solutions exist as a smooth limit to $l \to \infty$ so that they too can essentially be treated the same way, we do not have other spacetimes to compare this to. It could be that the result \eqref{asymptoticintegral} is the consequence of some more fundamental property, or it could be a fortuitous result that happens only in Kerr--AdS and Kerr/Myers--Perry. It would be interesting to broaden the class of spacetimes under consideration, particularly to other spacetimes with a PCKY tensor and thus a principal vector, to see whether there is a more general rule. (There is some discussion of another approach to this in Chapter \ref{additionalNote}.) 

I mentioned some future work in Section \ref{varyinghorizonradius} regarding the $\bs k^{EH}_{\chi}[\lie_w g;g]$ terms: that $\oint \bs k_{\bar \z}^{EH}[\lie_w g;g] = 0$ seems to hold on a surface of constant $t,r$ under more general circumstances than I considered explicitly in that section, and the consideration of the higher-dimensional case.

Throughout this work I mostly omitted the treatment of Kerr--de Sitter, with positive $\La$. In fact mathematically there are many similarities, but the existence of a cosmological horizon means that most results having to do with infinite spatial radius are complicated. Some treatments of the asymptotically de Sitter case which shows the similarity to the Kerr--AdS case are given in \cite{DolanKastor,KubiznakSimovic}. Because Kerr--de Sitter is so mathematically similar to Kerr--AdS, it also possesses a PCKY tensor and can be written in Kerr--Schild form, and so it would be interesting to see whether the $\bs I_\chi$ forms could be useful in interpreting the Kerr--de Sitter solutions in the framework given in \cite{DolanKastor,KubiznakSimovic}. I noted already that the relationship between the horizon area and geometric volume $\mc V_C = r \mc A/(D-1)$ observed in these papers can be explained simply in terms of the PCKY tensor (in Section \ref{AreaVolumeRelationship}). Additionally, the value for the thermodynamic volume associated with the cosmological (outer) horizon is given by \cite{DolanKastor} as
\begin{align}
    \mc V_{c} &= \f{r_c \mc A_c}{D-1} + \f{8\pi}{(D-1)(D-2)} \sum_i a_i \mc J_i,
\end{align}
where $r_c, \mc A_c$ are the values of the $r$ radius parameter and the area of the cosmological horizon, respectively. This is the same form as \eqref{VgeominusVth}.  Because the Kerr--de Sitter solutions and Kerr--AdS solutions are related by sending $l \to \pm il$, the arguments given in this thesis---that the geometric volume will appear in the Smarr relationship if a conserved ``energy'' $\mc F$ is defined associated with the principal Killing vector $\bt$, and the as a consequence the thermodynamic volume associated with $\mc E$ will have extra terms related to the difference between $\xi$ and $\bt$---will still hold, although in this case $\xi$ is no longer an asymptotically \emph{static} Killing vector (since it is not timelike beyond the cosmological horizon), though it remains asymptotically hypersurface-orthogonal and it continues to have the important property $\xi$ does for the asymptotically AdS black holes, which is that its direction remains fixed with respect to the de Sitter background (as opposed to $\bt$). 

In CGKP and future papers (for example, see \cite{Altamirano} and references therein), the isoperimetric inequality is introduced, which states that the ratio of the thermodynamic volume and area always satisfies a specific inequality. This interesting result is something I would like to examine further.

Even in CGKP's original paper there are several other solutions which are examined which I have not addressed here at all. There are of course many other papers dealing with thermodynamic volumes in the many years since CGKP's paper came out, a description of which is beyond my scope. Two interesting examples from CGKP are the rotating pairwise-equal 4-charge black hole in $D=4$ gauged supergravity and the charged rotating black hole in minimal $D = 5$ gauged supergravity. What is interesting is that these solutions do not have Kerr--Schild form (with an anti-de Sitter background), and so my approach here would have to be modified (or even might be inapplicable). I would like to try to apply the BC method of integrating through solution space for these cases too, so as to get a conserved charge at the horizon. It is interesting to me, too, that the pattern is still present where an asymptotically static Killing vector must be used for the first law but a geometric volume term appears when an asymptotically rotating vector is used. Because, at large distance from the central charges, the solution approaches Kerr--AdS, it might be that many of the methods used here to discuss Kerr--AdS could be used asymptotically and then ``transferred to the horizon'' in some way.  

Another example of a spacetime with interesting properties are the black ring spacetimes, the thermodynamics of which are described in \cite{Altamirano}. These solutions are vacuum and are rotating, but have a non-spherical horizon topology. They do not a admit Kerr--Schild decomposition. An exact solution is known for the $D=5$ black ring with $\La = 0$ (see \cite{EmparanReall}) but the method in \cite{Altamirano} was to take an approximation of an extremely thin ring. One interesting feature is that the thermodynamic volume differs by a large amount from the expected geometric volume. Because the thermodynamic volume can be taken even for the $\La = 0$ case by taking a $l \to \infty$ limit, it would be interesting to compare in more detail the thermodynamic volume to the vector volume for the $\La = 0$ exact solution.

Another natural next step would be to consider solutions with NUT parameters. These complicate the topology very considerably, but the advantage of the Kerr--NUT and Kerr--AdS--NUT spacetimes is that they still possess the PCKY tensor $\bs h$. A treatment of the Kerr--NUT case appears in \cite{Bordo} and a treatment of Kerr--AdS--NUT appears in \cite{Rodriguez}. I argued in Section \ref{multiplekerrschild} that the Kerr--NUT--AdS spacetimes could be interpreted as the result of a ``multiple Kerr--Schild'' perturbation to a pure AdS solution, with the significant caveat that the Kerr--Schild vectors for the NUT parameters would be complex. It might be possible to use $\bs I_\chi$ constructed from the real null vector to account for the difference between Kerr--NUT--AdS and simply NUT--AdS, or it might be possible to construct a $\bs I_\chi$ using the complex null vector to account for the NUT parameters. I also note that the existence of $\bs h$ means that we might be able to find some sort of equivalent for the volume--area relationship.

There is interesting work on dynamic black holes, such as \cite{Hayward04} (and references therein). For spherically symmetric dynamic black holes, I already pointed out in Chapter \ref{PRDPaper} that the black hole volume, introduced by Hayward, associated with the Kodama vector, is a generalization of the vector volume. There might be a similar generalization in dynamic black holes away from spherical symmetry.  

I touched on the physical process version of the first law of black hole mechanics in Section \ref{physicalprocessversion}. I would like to examine the stress--energy tensor for a more realistic model of energy and angular momentum flow into a black hole, perhaps in the presence of an accretion disk.

My general approach was to interpret the thermodynamic volume $ V_{th}$ more by its appearance in the Smarr relation than by its appearance in the first law. This is because the scaling relation means that its appearance in the Smarr relation automatically implies its appearance in the first law, given that the first law is otherwise satisfied with $\La$ (or $\tilde \La)$ kept constant. This meant my focus was somewhat narrow, in trying to interpret the appearance of $ V_{th}$ in the Smarr relation, rather than in the first law. Along those lines, I did not spend time in this writing on the physical mechanisms which could give rise to a varying $\La$. 

My focus has generally also been more on black hole \emph{mechanics} rather than thermodynamics, so that I have not discussed, for example, Hawking--Page phase transitions, and only touched on Euclidean methods in Section \ref{EuclideanAction} to show how the volume enters into them. It would be worthwhile to take a closer look at thermodynamic methods and how my work related to black hole mechanics and the geometric interpretation of volume fits in. This general field has been dubbed ``black hole chemistry,'' and is an extremely active area of research: see \cite{KubiznakMannTeo,Mann} for reviews. 

In Section \ref{timevaryingbutnotvaidya}, I mentioned that I would like to try to break the Ricci tensor down into its constituent components to show why the second-order terms drop out when the metric is of the form \eqref{ABCDE} but not of the form \eqref{FGH}. 

The work of \cite{Chernyavsky,HajianOzsahin} considering $\La$ as a type of ``electric charge'' resulting from a gauge field is intriguing and may be worth pursuing, particularly in terms of whether there is some way to use the BC method to construct the Smarr relation in this case. 

\section{Final Words}

With the increased number of observations of black holes with the Event Horizon Telescope and via gravitational wave detection of black hole mergers, black holes and their properties are a constant, ongoing subject of interest. I hope that my analysis of the role that the black hole volume plays in black hole thermodynamics enriches this developing and fascinating field. 

\chapter{Additional Note} \label{additionalNote} 

Since submitting the thesis, I noticed the following (not appearing in the thesis proper). 

One of the questions addressed in my thesis is why the conserved quantity $\mc F$ associated with the vector $\bt$ is equal, up to a simple numerical factor, to the difference between the Komar integrals associated with $\bt$ in the full and background spacetime, when this is not true for the asymptotically-static vector $\xi$. In the thesis I addressed this by showing that the covariant phase space methods of Wald and the integration through solution space method of Barnich and Comp\`ere could be adapted in Kerr--Schild spacetimes to produce a conserved quantity which is equal to a Komar difference term plus an extra term, related to the symplectic potential form $\bs \Th$, and then showed how this last term is equal (up to a numerical factor) to the Komar difference term associated with $\bt$. 

Another, in some ways simpler, explanation is the following. $\mc F$ can also be defined in terms of the Ashtekar--Magnon--Das (AMD) conserved charge. Here I will show that the AMD conserved charge associated with $\bt$ is proportional to the Komar difference term. First I will rewrite the AMD conserved quantity $\mc F$ in a simple form, having to do with the canonical orthonormal frame components. (The canonical orthonormal frame, which I will refer to henceforth as the orthonormal frame, is the one introduced in Section \ref{GKNAdSSection}.) Then I will show the Komar difference term and show how the PCKY tensor and the Weyl tensor appear naturally in the Komar difference, in a way that shows the equivalence of the Komar difference term and the AMD term (up to a multiplicative factor of $(D-2)/(D-3)$). I then revisit the Smarr relationship explicitly with this equivalence.

\section{AMD Conserved Charge}

Following Ashtekar and Das \cite{AshtekarDas}, let $g_{ab}$ be the physical spacetime for an asymptotically anti-de Sitter spacetime, which for this document will be Kerr--AdS. Let $\hat g_{ab}$ be the conformally-related spacetime
    \begin{align}
        \hat g_{ab} = \Om^2 g_{ab}
    \end{align}
    where $\Om$ is a conformal factor. Quantities with carets are quantities in the conformally-related spacetime, and quantities without are quantities in the physical spacetime. The contravariant metrics $g^{ab}, \hat g^{ab}$ are the inverses of $g_{ab}$ and $\hat g_{ab}$ respectively and satisfy
    \begin{align}
        \hat g^{ab} = \Om^{-2} g^{ab}.
    \end{align}
    The idea then is that $\Om$ is chosen so that $\Om = 0$ on $\mc I$, representing spatial infinity ($r \to \infty$).

    One such choice is $\Om = 1/r$, where $r$ is the spheroidal Boyer--Lindquist $r$. (The spherical--polar radius could also be used, but it will turn out that the Boyer--Lindquist $r$ will be more useful.) I will use the KS form of the metric.

    Rearranging the formula from Ashtekar and Das \cite{AshtekarDas}, the expression for the conserved charge associated with Killing vector $\chi$ in the conformally-related spacetime is
        \begin{align}
        Q_C[\chi] &= -\f{l^3}{8\pi (D-3)} \oint_{ \mc I} \Om^{3-D} \hat C_{acbd} \hat n^c \hat n^d \chi^a \hat N^b d \hat S, \label{conformalspacetimeversion}
    \end{align}
    where $\mc I$ is a constant-$t$ surface within $\Om = 0$. Here $\hat N_a$ is a unit timelike normal within $\mc I$, $\hat C_{acbd}$ is the Weyl tensor in the conformally-related spacetime, $\hat n_a = \hat \na_a \Om = \pa_a \Om$, and where $d \hat S$ is the $(D-2)$-area element on $\mc I$. Using $\hat g_{ab} = \Om^2 g_{ab}$, the above integral can be expressed in terms of quantities in the physical spacetime. The integral is calculated at constant $(r,t)$, and then $r$ is taken to infinity, to match with $\mc I$. The result is
    \begin{align}
        Q_C[\chi] &= -\f{l^3}{8\pi (D-3)} \lim_{r\to \infty} \oint \Om^{-2} C_{acbd} n^c n^d \chi^a N^b d S,
    \end{align}
    where $n_a = \na_a \Om = \pa_a \Om = \hat n_a$,  $N^b$ is the is the unit future-directed timelike normal to a 2-surface of constant $(r,t)$ (satisfying $N^a n_a = 0$), and $C_{acbd}$ and $dS$ are the Weyl tensor and $(D-2)$-area element for the physical spacetime respectively. This can be found from \eqref{conformalspacetimeversion} by using ${\hat{C}^a}_{\ \ bcd} = {C^a}_{bcd}, d\hat S = \Om^{D-2} d S$, and so on, following from the conformal relation. I have chosen the sign so that $Q_C[\beta] > 0$ for future-oriented $N^b$. 

            To make full comparison with the Komar integral with background subtraction for $\bt$ it will be useful to rearrange. Defining $q_a$ to be the unit spacelike outward-pointing normal to constant $r$ surfaces $(q_a \propto \pa_a r$, $q_r > 0, q_a q^a = 1$),
    \begin{align}
        q_a  \simeq - \f{ l n_a}{\Om}, \label{qa}
    \end{align}
    where $\simeq$ means is equal to, to leading order in $r$, as $r \to \infty$. In fact, $q_a = e^n_a$ exactly, where $e^n_a$ is the orthonormal frame one-form.

    We then have, up to choice of orientation,
    \begin{align}
        dS_{ab} &= 2 q_{[a} N_{b]} dS. \label{dSab}
    \end{align}
    The choice here matches that in Poisson \cite{Poisson} equation (3.2.8). Using \eqref{qa} and rearranging, we have
    \begin{align}
        Q_C[\chi] &= -\f{l}{16\pi (D-3)} \lim_{r\to\infty} \oint C^{abcd} q_a \chi_b d S_{cd}. \label{QCchi}
    \end{align}
    When applied to, specifically, $\chi = \bt$, this is
    \begin{align}
        \mc F &\equiv Q_C[\bt] \nn 
        &= -\f{l}{16\pi(D-3)} \lim_{r\to\infty} \oint C^{abcd} q_a \bt_b d S_{cd}.
    \end{align}

    It is useful at this point to use the orthonormal basis introduced in Section \ref{GKNAdSSubsection}. Capital Roman letters are used for frame components. Now significantly, for large $r$, $\bt$ is \emph{almost} given by $ir l^{-1} e^{\hat n}$, in the sense that if $\bt$ is written in frame components with $\bt^a = \bt^A e_A^a$, 
    \begin{align}
        \bt^{\hat n} &\simeq \f{i r}{l},
    \end{align}
    where $\simeq$ means equal to to leading order in $r$. Meanwhile, the only other nonzero frame components of $\bt$ are $\bt_{\hat \al}$ for $\al \neq n$ and $\bt_{\hat 0}$ in odd dimensions, which are $\mc O(r^{-1})$. 
    The frame components of $\bt$ are calculated in Section \ref{betaframecalculation}.  The reason that $\bt$ points primarily in the $e_{\hat n}$ direction comes down to \eqref{betacanonicalbasis}, and that $Q_\al, S$ are $\mc O(r^{-2})$ but $Q_n$ is $\mc O(r^2)$. The simple form for $\bt$ in terms of the frame components seems to be because of $\bt$'s role as Principal Vector, tied up with the ``hidden symmetries'' of the spacetime, along with the fact that $r$ is large and so dominates over the $y_\al$ terms. 
    
    We can then write
    \begin{align}
        Q_C[\bt] &= -\f{1}{16\pi(D-3)} \lim_{r\to\infty} \oint C^{abcd} i r e^n_a e^{\hat n}_b d S_{cd} - \f{l}{16\pi(D-3)} \lim_{r\to\infty} \oint C^{abcd} e_a^n \left( \sum_{\al=1}^{n-1} \bt^{\hat \al} e_b^{\hat \al} + \ve \bt^{\hat 0} e_b^{\hat 0}\right) d S_{cd}.
    \end{align}
    
    In terms of the orthonormal frame components, this gives
    \begin{align}
        Q_C[\bt] &= -\f{1}{16\pi(D-3)} \lim_{r\to\infty} \oint i r C^{n \hat n AB} dS_{AB} - \f{l}{16\pi(D-3)} \lim_{r\to\infty} \oint \left( \sum_{\al=1}^{n-1} C^{n\hat \al CD}\bt_{\hat \al} + \ve C^{n \hat 0 CD} \bt_{\hat 0}\right) dS_{CD}.
    \end{align}

    From \eqref{dSab}, 
    \begin{align}
        dS_{AB} = 2 \de_{[A}^n N_{B]} d S. \label{dSAB} 
    \end{align}
    The frame components of $N_A$ are calculated in Section \ref{dtcalculation}, and $N_{\hat \mu}$ and $N_{\hat 0}$ are $\mc O(r^0)$ (with $N_{n} = \mc O(r^{-D-1})$ and $N_\al = 0$ for $\al \neq n$). The $\hat \al, \hat 0$ terms give
    \begin{align}
        & \f{l}{16\pi(D-3)} \lim_{r\to\infty} \oint \left( \sum_{\al=1}^{n-1} C^{n\hat \al CD}\bt_{\hat \al} + \ve C^{n \hat 0 CD} \bt_{\hat 0}\right) dS_{CD} \nn 
        &=  \f{l}{16\pi(D-3)} \lim_{r\to\infty} \oint \left( \sum_{\al=1}^{n-1} C^{n\hat \al n D} \bt_{\hat \al} + \ve C^{n \hat 0 n D} \bt_{\hat 0} \right) N_D dS.
    \end{align}
    The only nonzero contributions to the Weyl tensor of the form $C^{nBnD}$, from Hamamoto \cite{Hamamoto}, are for $B = D \neq n$. The Weyl frame components are calculated in Section \ref{WeylFrameComponents}. For $B=D = \hat \al$ or $B= D = \hat 0$, $C^{n \hat \al n \hat \al}, C^{n \hat 0 n \hat 0}$ are $\mc O(r^{1-D})$, and $\bt_{\hat \al}$ and $ \bt_{\hat 0}$ are $\mc O(r^{-1})$. Along with $dS = \mc O(r^{D-2})$, we conclude that the $\hat \al, \hat 0$ terms do not contribute. What remains is
    \begin{align}
        \mc F &= Q_C[\bt] = - \f{1}{16\pi(D-3)} \lim_{r\to\infty} \oint i r C^{n \hat n AB} dS_{AB}.
    \end{align}
    This can also be written
    \begin{align}
        \mc F &= Q_C[\bt] = -\f{1}{8\pi(D-3)} \lim_{r\to\infty} \oint i r C^{n \hat n n \hat n} N_{\hat n} d S. \label{FQCbetaformula}
    \end{align}
    This simple form comes down to the fact that $\bt$ points ``almost entirely'' in the $e_{\hat n}$ direction for large $r$, in the sense described above. 

    Now consider the Komar difference term.

    \section{Komar Difference Term} \label{komardifference}

    The Komar difference term is given by $- \lim_{r\to\infty} \oint (\bs K^K_\bt - \overline{\bs K^K_\bt})$. To evaluate it, we begin by evaluating
    \begin{align}
        -\lim_{r\to\infty} \oint \left(\bs{K}^K_\bt - \overline{ \bs{K}^K_\bt}\right) &= -\f1{16\pi} \lim_{r\to\infty} \oint \left(\na^a \bt^b dS_{ab} - \overline{\na^a \bt^b d S_{ab}}\right) \nn 
        &= -\f{1}{16\pi} \lim_{r\to\infty} \oint \left( \na^a \bt^b - \overline{\na^a \bt^b}\right) dS_{ab}.
    \end{align}
    The last equality comes down to the fact that $\bs \ep = \overline{\bs \ep}$; choosing the integration surface to be a constant-$r$, constant-$t$ surface means that $dS_{ab} = \overline{dS_{ab}}$, as a result. I will evaluate $dS_{ab}$ in the full spacetime; it is given by \eqref{dSab}. 
    
    \eqref{nabetacurvatureresult}, raised, gives
    \begin{align}
        \na^a \bt^b &= \f{1}{2(D-2)} \left[ C^{abcd} \bs h_{cd} + \f{2(D-4)}{D-2} R^{[a}_d \bs h^{b]d} - \f{2}{(D-1)(D-2)} R \bs h^{ab}\right].  \label{nablabeta}
    \end{align}
    Recall too that $\bs h^{ab} = \bar {\bs h}^{ab}$. For Kerr--AdS, $R^a_b = \bar R^a_b = -(D-1)l^{-2} \de^a_b$, with $\bar C^{abcd} = 0$ in the background, implying \eqref{KomardifferenceWeyl}. This means that the Weyl tensor appears immediately in the Komar difference term. We then have
    \begin{align}
        -\lim_{r\to\infty} \oint \left( \bs K^K_\bt - \overline{\bs K^K_\bt}\right) &= - \f{1}{32\pi(D-2)} \lim_{r\to\infty} \oint C^{abcd} \bs h_{cd} d S_{ab}. \label{Komardifferencebetaadditionalnote}
    \end{align}
    (An integral of the Weyl tensor contracted with a CKY tensor for Kerr--de Sitter is considered in \cite{Chrusciel}.)  Using $C^{abcd} = C^{cdab}$ and the decomposition $\bs h = \sum_{\mu=1}^n x_\mu e^\mu \wedge e^{\hat \mu}$ as well as $x_n = ir, x_\al = y_\al$,
    \begin{align}
        -\lim_{r\to\infty} \oint \left( \bs K^K_\bt - \overline{\bs K^K_\bt}\right) &= - \f{1}{16\pi(D-2)} \lim_{r\to\infty} \oint \sum_{\mu=1}^n C^{\mu \hat \mu AB} \bs h_{\mu \hat \mu} d S_{ab} \nn 
        &= - \f{1}{16\pi(D-2)} \lim_{r\to\infty} \oint i r C^{n \hat n AB} dS_{AB} - \f{1}{16\pi(D-2)} \lim_{r\to\infty} \oint \sum_{\al=1}^{n-1} y_\al C^{\al \hat \al AB} dS_{AB}.
    \end{align}
    Using \eqref{dSAB}, the second term becomes
    \begin{align}
        -\f{1}{16\pi(D-2)} \lim_{r\to\infty} \oint \sum_{\al=1}^{n-1} y_\al C^{\al \hat \al n B} N_{\hat B} d S.
    \end{align}
    The only nonzero term $C^{\al \hat \al n B}$ is $C^{\al \hat \al n \hat n}$, which, as shown in Section \ref{Cnhatnalhatal}, goes as $\mc O(r^{-D})$, so that the combination $C^{\al \hat \al n \hat n} d S$ vanishes in the large-$r$ limit. Thus the only contribution from $\bs h$ which remains is the $\mu = n$ portion. We are left with
    \begin{align}
        -\lim_{r\to\infty} \oint \left( \bs K^K_\bt - \overline{\bs K^K_\bt}\right) &= -\f{1}{16\pi(D-2)} \lim_{r\to\infty} \oint i r C^{n \hat n AB} dS_{AB} \nn 
        &= \f{D-3}{D-2} Q_C[\bt] \nn
        &= \f{D-3}{D-2} \mc F. \label{KomarDifferenceIsConformalChargeBeta}        
    \end{align}

    This equality is the central result of this note. (Note that because the orthonormal frame has Euclidean signature, any terms written in the orthonormal frame can be readily raised and lowered.)
    
    Thus we establish immediately that the Komar difference term gives a term proportional to $\mc F = Q_C[\bt]$. The argument can be summarized by noting that the combination $\na^a \bt^b - \overline{\na^a \bt^b} \propto C^{abcd}\bs h_{cd}$, and that, given the form of $dS_{ab}$, the only part of $\bs h$ which contributes is the $x_n \bs \om^n$ part of $\bs h$, so that $r = -ix_n$ also arises naturally. 

    One reason for why the Komar difference term is proportional to the AMD charge for $\bt$ but not for $\xi$ is that it is $\bt$ which permits the decomposition \eqref{nablabeta}, which follows from the properties of the Conformal Killing--Yano tensor $\bs h$, so that $(\na^a \bt^b - \overline{\na^a \bt^b})dS_{ab} \simeq -2 l C^{abcd}q_c \bt_d dS_{ab}$. It is the ``hidden symmetries'' of the spacetime, once again, that lead to nice results!

    We can now revisit the Smarr relation.

    \section{Smarr Relation Revisited}

    As established in the thesis, the \emph{Komar difference} for Kerr--AdS is independent of integration surface, so we can move the integration from near infinity to the horizon, $H$, so that \eqref{KomarDifferenceIsConformalChargeBeta} can be rearranged to
    \begin{align}
        \mc F &= -\f{D-2}{D-3} \oint_H \left(\bs K^K_\bt - \overline{\bs K^K_\bt} \right) \label{mcFeqn}
    \end{align}

    Recall from the thesis that $-\oint_H \bs K^K_\z = \f{\ka A}{8\pi}$ where $\z = \xi + \sum_i \Om_i \eta_i = \bt + \sum_i \om_i \eta_i$ is the Killing vector tangent to the null generators of the horizon with $\z^t = 1$, and $\ka$ and $A$ are the black hole surface gravity and area. $\Om_i$ are the angular velocities with respect to a frame adapted to $\xi$ and $\om_i$ are the angular velocities with respect to a frame adapted to $\bt$. The angular momenta $\mc J_i$ are given by a Komar integral (performed anywhere enclosing the black hole),
    \begin{align}
        \mc J_i &= \oint \bs K^K_{\eta_i}.
    \end{align}
    Thus we have, by linearity,
    \begin{align}
        -\oint_H \bs K^K_\bt &= -\oint_H \bs K^K_\z + \sum_i \om_i \oint_H \bs K^K_{\eta_i} \nn 
        &= \f{\ka A}{8\pi} + \sum_i \om_i \mc J_i.
    \end{align}
    
    The integral $\oint_H \overline{\bs K^K_\bt}$, as shown in the thesis, gives 
    \begin{align}
        \oint_H \overline{\bs K^K_\bt} &= \f{1}{8\pi} \tilde \La \mc V_{\bt,\mc B} = \f{1}{8\pi} \tilde \La \mc V_{\bt, \mc B},
    \end{align}
    where $\tilde \La$ is the cosmological constant normalized to $R_{ab} = \tilde \La g_{ab}$ in vacuum (given by $\tilde \La = 2\La/(D-2)$ where $G_{ab} + \La g_{ab} = 0$) and  $\mc V_{\bt,\mc B} = \mc V_{\xi,\mc B}$ is the black hole vector volume, also known as the geometric volume $V_{geo}$. Also using the pressure definition $P = - \f{D-2}{16\pi} \tilde \La,$ we can write \eqref{mcFeqn} as
    \begin{align}
         \mc F &= \f{D-2}{D-3} \left( \f{\ka A}{8\pi} + \sum_i \om_i \mc J_i\right) - \f{2}{D-3} P V_{geo},
    \end{align}
    which is one of the Smarr relations found by CGKP \cite{Cvetic}.

    This is a somewhat different argument than the one presented in the thesis. This one makes a closer link between the conformally-calculated charge and the Komar integrals, and does not (directly) use Wald's conformal phase space methods. 
    \section{Connection to Peng}

    Peng et al.~\cite{Peng} introduce a definition of conserved charge which is a modification of the Komar integral, given by the following. (Note that they use $\ell = l^{-1}$ with $R_{\mu \nu} = - (D-1) \ell^2 g_{\mu \nu}$ for the Kerr--AdS solutions; I maintain the use of $l$ here to be consistent with my notation.) A generalized Komar potential associated with a Killing vector $\chi$ is given by
    \begin{align}
        \mc K[\chi] &= \f{1}{2(D-3)}\left( d \chi^\flat  - \f{l^2 \Box d \chi^\flat}{2 }\right),
    \end{align}
    where $\Box = \na_a \na^a$. Then a conserved charge $\mc Q$ is given by
    \begin{align}
        \mc Q[\chi] &= \f{1}{16\pi} \oint_{\pa \Si} * \mc K[\chi],
    \end{align}
    where $*$ is the Hodge dual and $\pa \Si$ denotes the $(D-2)$-dimensional $t = const.$ surface at spatial infinity $r \to \infty$. (Peng has $1/8\pi$, due to a difference in Hodge dual convention.) Peng et al.~also express $\mc K$ (I will omit the $[\chi]$ here) using Killing vector identities as
    \begin{align}
        \mc K^{ab} &= \f{1}{D-3} \left( \na^{[a} \chi^{b]} + \f{l^2}{2 } {R^{a b}}_{cd} \na^c \chi^d\right). \label{Peng2.17}
    \end{align}
    Peng et al.~then show that when $\mc Q$ is calculated (again, specifically at infinity) for the Kerr--AdS black holes, $\mc E$ is recovered for $\mc Q$ associated with $-\xi$ and $\mc J_i$ is recovered for $\mc Q$ associated with $+\eta_i$, so that we have $\mc Q[\chi] = - Q_C[\chi]$. 

    My observation is that, using the fact that Kerr--AdS is an Einstein space, \eqref{Peng2.17} can be rewritten in terms of the Weyl tensor. Using
    \begin{align}
        C_{abcd} &= R_{abcd} - \f{2}{D-2} \left( g_{a[c} R_{d]b} - g_{b[c} R_{d]a}\right) + \f{2}{(D-1)(D-2)} R g_{a[c} g_{d]b} \nn 
        &= R_{abcd} + \f{2}{l^2} g_{a[c}g_{d]b},
    \end{align}
    so that (also using $\na^{[a}\chi^{b]} = \na^a \chi^b$)
    \begin{align}
        \mc K^{a b} &= \f{1}{D-3} \left( \na^a \chi^b + \f{l^2}{2} \left( {C^{ab}}_{cd} - \f{2}{l^2} \de^a_{[c} \de^b_{d]}\right) \na^c \chi^d\right) \nn 
        &= \f{l^2}{2  (D-3)} C^{abcd} \na_c \chi_d,\label{PengPotentialWeylForm}
    \end{align}
    implying
    \begin{align}
        \mc Q[\chi] &= \f{l^2}{32\pi(D-3)} \oint C^{abcd} \na_c \chi_d dS_{ab}.
    \end{align}
    When applied to $\chi = \bt$, this form is intriguingly similar to the integrand in \eqref{Komardifferencebetaadditionalnote}. Using \eqref{KomardifferenceWeyl} and the fact that $\overline{\na_a \bt_b} = \f{1}{l^2} \bs h_{ab}$, we can write
    \begin{align}
        \na_a \bt_b &= \f{1}{l^2} \bs h_{ab} + \f{1}{2(D-2)} C_{abcd} \bs h^{cd},
    \end{align}
    so that
    \begin{align}
    \mc K^{a b} &= \f{l^2}{2 (D-3)} C^{abcd} \left(l^{-2} \bs h_{cd} + \f{1}{2(D-2)} C_{cdef} \bs h^{ef}\right) \nn 
    &\simeq \f{1}{2  (D-3)} C^{abcd} \bs h_{cd},
    \end{align}
    where I make the assumption that when evaluated at large radius, the Weyl tensor is small enough that terms quadratic in the Weyl tensor vanish. Thus the integrand matches that in \eqref{Komardifferencebetaadditionalnote} precisely. 

    More generally it appears that Peng et al.'s~conserved charge is closely connected with the AMD conserved charge, given the appearance of the Weyl tensor. To tease out why they give the same result, by comparing \eqref{QCchi} and \eqref{PengPotentialWeylForm} and noting $\mc Q[\chi] = - Q_C[\chi]$, it is only necessary to show the equality of the following integrals in the large $r$ limit for arbitrary Killing vector $\chi$: 
    \begin{align}
        \f{l}{2} \lim_{r\to\infty} \oint C^{abcd} \na_c \chi_d dS_{ab} &= \lim_{r\to\infty} \oint C^{abcd} q_c \chi_d d S_{ab}. \label{PengtoAMDintegrandcomparison}
    \end{align}
    It appears to hold not just for $\bt$ but for arbitrary $\chi$ that we have that the contribution to the integral from $(l/2) \na_c \chi_d$ is simply $q_{[c} \chi_{d]}$. 

    To show that this holds, not just for $\bt$ but for arbitrary $\chi$, I'll consider the cases $\chi = \xi = \pa_t$ and $\chi = \eta_i = \pa_{\phi_i}$. I will assume that, just as with $\bt$, 
    \begin{align}
        \lim_{r\to\infty} \oint C^{abcd} \na_c \chi_d d S_{ab} = \lim_{r \to \infty} \oint C^{abcd} \overline{\na_c \chi_d} d S_{ab}, \label{PengAMDcomparison}
    \end{align}
    that is, we need only consider $\overline{\na_a \chi_b}$ in the background spacetime. Using $\xi$ as an example, we have
    \begin{align}
        \overline{\na_a \xi_b} &= \overline{ \pa_{[a} (g_{tt} \de_{b]}^t) }\nn
        &=\overline{ \pa_{[a} \ln |g_{tt}| \xi_{b]}}.
    \end{align}
    The relatively simple form is because the background metric is diagonal. Then letting $w_a = \pa_a \ln|\bar g_{tt}|$, $\overline{\na_a \xi_b} = w_{[a}\bar \xi_{b]},$ so the left-hand side of \eqref{PengAMDcomparison} is
    \begin{align}
        \lim_{r\to\infty} \oint C^{abcd} w_c \bar \xi_d d S_{ab}.
    \end{align}

    We have, from \eqref{backgroundds2rytphi}, 
    \begin{align}
        \pa_r \ln |\bar g_{tt}| &= \f{2r}{r^2+l^2} \simeq \f{2}{r} \nn 
        \pa_{y_\al} \ln |\bar g_{tt}| &= -\f{2 y_\al}{l^2-y_\al^2},
    \end{align}
    taking the large-$r$ limit in the first expression. This means we can write
    \begin{align}
        w^\flat &\simeq \f{2}{r} dr - \sum_\al \f{2 y_\al}{l^2-y_\al^2} dy_\al \nn 
        &\simeq \f{2}{l} q^\flat - \sum_\al \f{2 y_\al \sqrt{Q_\al}}{l^2-y_\al^2} e^\al.
    \end{align}
    From \eqref{Qalphafalloff}, $Q_\al$ falls off as $\mc O(r^{-2})$ for large $r$. We can then expand out the contribution from $e^\al$ as
    \begin{align}
        C^{abcd} \left( - \f{2 y_\al \sqrt{Q_\al}}{l^2-y_\al^2}\right) e^\al_c \bar \xi_d d S_{ab} &= \left( - \f{4 y_\al \sqrt{Q_\al}}{l^2-y_\al^2}\right) C^{AB\al D} \bar \xi_D d S_{AB} \nn 
        &= \left( - \f{4 y_\al \sqrt{Q_\al}}{l^2-y_\al^2} \right) 2 C^{n B \al D} N_B \bar \xi_D d S,
    \end{align}
    using \eqref{dSAB}. The only nonzero contribution to $C^{n B \al D}$ are from $B = n, D = \al$ and $B = \al, D=n$. We then have 
    \begin{align}
        C^{abcd} \left( - \f{2 y_\al \sqrt{Q_\al}}{l^2-y_\al^2}\right) e^\al_c \bar \xi_d d S_{ab} &= \left( - \f{4 y_\al \sqrt{Q_\al}}{l^2-y_\al^2}\right) 2 C^{n \hat n \al \hat \al} ( N_{\hat n} \bar \xi_{\hat \al} - N_{\hat \al} \bar \xi_{\hat n}) d S.
    \end{align}
    We then have $N_{\hat n}, N_{\hat \al} \simeq \mc O(r^0)$, $\bar \xi_{\hat n} \simeq \mc O(r^1), \bar \xi_{\hat \al} \simeq \mc O(r^{-1})$. The dominant contribution then will be from $N_{\hat \al} \bar \xi_{\hat n}$ which will be $\mc O(r)$. Meanwhile, $\sqrt{Q_\al} \simeq \mc O(r^{-1}), C^{n \hat n \al \hat \al} \simeq \mc O(r^{-D})$ from \eqref{Cnnhatalphaalphahat2} and $d S \simeq \mc O(r^{D-2})$, so that this term overall is $\mc O(r^{-1})$, and so vanishes in the large $r$ limit. We are left with
    \begin{align}
        \lim_{r\to\infty} \oint C^{abcd} w_c \bar \xi_d d S_{ab} &= \f{2}{l} \lim_{r\to\infty} \oint C^{abcd} q_c \bar \xi_d d S_{ab},
    \end{align}
    so that, given the smallness of $\xi_d - \bar \xi_d$ compared to $\bar \xi_d$ itself, confirms \eqref{PengtoAMDintegrandcomparison} for $\chi^a = \xi^a$. The argument for $\chi^a = \eta_i^a$ is essentially identical; in this case
    \begin{align}
        \overline{\na_a (\eta_i)_b} &= \overline{\pa_{[a} (g_{\phi_i \phi_i} \de^{\phi_i}_{b]})} \nn 
        &= \overline{\pa_{[a} \ln |g_{\phi_i\phi_i} (\eta_i)_{b]}}
    \end{align}
    Since, in Kerr--AdS (except for special cases like Schwarzschild--AdS which have even more symmetries), $\chi^a$ can be constructed as a linear combination of $\xi^a$ and $\eta_i^a$, we conclude that \eqref{PengtoAMDintegrandcomparison} holds for all $\chi^a$. And so we have the interesting result that Peng's charge is closely connected with the AMD conserved charge.
    
    \section{Additional Note Conclusion}

    In this additional note chapter, I showed that the particular properties of $\bt$ related to the Principal Conformal Killing--Yano tensor $\bs h$ lead to the Komar difference term being $(D-3)/(D-2)$ times the conformal charge $\mc F$ associated with $\bt$. In doing so, the following properties of $\bt$ and of the spacetime were were used:
    \begin{itemize}
        \item That $\bt$ points in the direction of the orthonormal frame unit vector $e_{\hat n}$, for large $r$, according to $\bt \simeq i r l^{-2} e_{\hat n}$, with all other frame components $\bt_A$ vanishing;
        \item That the combination
        $\na^a \bt^b - \overline{\na^a \bt^b}$ can be written as the contraction of the Weyl tensor and the PCKY tensor $\bs h$;
        \item That $\bs h$ can be decomposed into $r e^n \wedge e^{\hat n}$ (as well as other terms which do not end up contributing to the large-$r$ calculation for the Komar difference term);
        \item That $\Om = r^{-1}$ can be used as the conformal factor for defining the AMD charge;
        \item That the Weyl tensor components $C_{AnBn}$ are zero unless $A = B$, and, similarly, that $C_{\mu \hat \mu n B}$ is zero unless $B = \hat n$.
    \end{itemize}

    These all come down to the central role that the orthonormal frame has in describing these spacetimes as well as how closely $\bs h$ and $\bt$ are tied up with the structure of the spacetime. This further justifies the appearance of $\bt$ in the Smarr relation in which the volume comes out naturally.

    As a final note, recall that, from the thesis, $\overline{\na^a \bt^b} = l^{-2} \bs h^{ab}$. The reader uncomfortable with the comparison between the full spacetime and the background spacetime could instead replace $\overline{\na^a \bt^b}$ with $l^{-2} \bs h^{ab}$ in the above analysis and the results would be unaffected.

    In Appendix \ref{AdditionalNoteAppendix}, I calculate the frame components and also evaluate the integral \eqref{FQCbetaformula}.

\appendix

\addtocontents{toc}{\protect\renewcommand{\protect\cftchappresnum}{\appendixname\space}} 

\chapter{Calculation of Horizon Area for GKAdS Spacetimes} \label{AreaCalculationAppendix} 

This Appendix fills in the details for Section \ref{AreaRevisited}. This originally appeared in the main body of the thesis, but I thought it was sufficiently involved to be moved to an appendix. 

I will use the KS form of the metric, because the metric is regular across the horizon in those coordinates.

Consider the $(D-2)$-surface $t = const., r = r_+$ and call it $\Gamma$. Let coordinates be $z^I = (y_\alpha, \phi_i)$ where $\alpha=1, \ldots, n-1$ and $i =1, \ldots, n-1+\ve$; $I$ runs from (say) $3$ to $D$. Then let the metric intrinsic to this surface be $\sigma_{IJ}$:
\begin{align}
    ds^2|_{\Gamma} &= \sigma_{IJ} d z^I dz^J \nn 
    &= \sum_{\alpha=1}^{n-1} g_{y_{\alpha} y_\alpha} dy_\alpha^2 + \sum_{i=1}^{n-1+\ve} g_{\phi_i\phi_i} d \phi_i^2 \nn 
    &= \sum_{\alpha=1}^{n-1} \bar g_{y_\alpha y_\alpha} dy_\alpha^2 + \sum_{i=1}^{n-1+\ve} \bar g_{\phi_i \phi_i}d\phi_i^2  + H (k_I d z^I)^2.
\end{align}
Let 
\begin{align}
    \bar \sigma_{IJ} dz^I dz^J &= \sum_{\alpha=1}^{n-1} \bar g_{y_\alpha y_\alpha} dy_\alpha^2 + \sum_{i=1}^{n-1+\ve} \bar g_{\phi_i \phi_i} d \phi_i^2,
\end{align}
so that $\bar \sigma_{IJ}$ is the part of $\sigma_{IJ}$ corresponding to the background spacetime. Of course all quantities are evaluated at $r = r_+$. $\sigma_{IJ}$ has a Kerr--Schild-like form, though $k^I$ is not a null vector \emph{within} $\Gamma$. Because $k^a$ can be raised and lowered using $\bar g_{ab}$, and because $\bar g_{ab}$ is diagonal, write $k_a = \bar g_{ab} k^b$, or, restricting to the $z^I$ coordinates, $k_I = \bar \sigma_{I K} k^K$. This implies
\begin{align}
    \sigma_{IJ} &= \bar \sigma_{I K} ( \de^K_J + H k^K k_J),
\end{align}
where $K$ also runs over $3,\ldots, D$. Sylvester's determinant theorem and the fact that the determinant of the product of matrices is the product of the determinants of those matrices implies
\begin{align}
    \mathrm{det} (\sigma_{IJ}) = \mathrm{det}(\bar \sigma_{IK}) \left(1 + H k_J k^J\right)
\end{align}
For $1 + H k_J k^J$, we have $k^J k_J = \sum_{a=1}^D k_ak^a - k_t k^t - k_r k^r = -k_t k^t - k_r k^r$. Also using $k^r = 1$ as well as $H = \bar g^{rr}$ on the horizon we have $H k_r k^r = 1$, so that 
\begin{align}
    1 + H k_J k^J &= 1 - H k_r k^r - H k_t k^t \nn 
    &= - 1 - H k_t k^t \nn 
    &= -H k_t k^t \nn 
    &= - H \bar g_{tt} (k^t)^2 \nn 
    &= - \f{1}{(1+r_+^2/l^2)^2} \f{\bar g_{tt}}{\bar g_{rr}},
\end{align}
using \eqref{kmu} for $k^t$. For $\mathrm{det}(\bar \sigma_{IK})$, we have
\begin{align}
    \mathrm{det} (\bar \sigma_{IJ}) &= \prod_{I=3}^{D} \bar g_{II} \nn 
    &= \f{\prod_{a=1}^D \bar g_{aa}}{\bar g_{tt} \bar g_{rr} }\nn 
    &= \f{\mathrm{det}( \bar g_{ab})}{\bar g_{tt} \bar g_{rr}}.
\end{align}
Letting $\sigma = \mathrm{det}(\sigma_{IJ})$ and again using $-g = -\bar g$ be the negative of the determinant of $g_{ab}$ or $\bar g_{ab}$ (which are equal), we have
\begin{align}
    \sqrt{\sigma} &= \f{\sqrt{-\bar g}}{(1+r_+^2/l^2) \bar g_{rr}} \nn 
    &= \f{P \bar X \left(r_+ \prod_{\alpha=1}^{n-1} y_\alpha\right)^\ve}{ C U_n \left(\prod_{i=1}^n \Xi_i\right) \left(\prod_{i=1}^{n-1} a_i\right)^{1-\ve} (1+r_+^2/l^2)}. \label{sqrtsigma1}
\end{align}
Note that this also implies, from \eqref{zetaflatH},
\begin{align}
    \sqrt{-\bar g} = \sqrt{\sigma} \z_r \label{gsigmazeta}
\end{align}
(on the horizon).

Of course $r = r_+$ also in $P, \bar X$ and $U_n$. Defining $\tilde P$ to be
\begin{align}
    \tilde P &\equiv \prod_{1 \leq \alpha < \beta \leq n-1} (y_\alpha^2 - y_\beta^2), \label{tildePdef}
\end{align}
we note
\begin{align}
    \tilde P = \f{P}{U_n}
\end{align}
and so have
\begin{align}
    \sqrt{\sigma} &= \f{\tilde P \bar X \left(r_+ \prod_{\alpha=1}^{n-1} y_\alpha\right)^\ve}{ C \left(\prod_{i=1}^n \Xi_i\right) \left(\prod_{i=1}^{n-1} a_i\right)^{1-\ve} (1+r_+^2/l^2)}
\end{align}
The area is then
\begin{align}
    A &= \int dy_1 \ldots dy_{n-1} d\phi_1 \ldots d\phi_{n-1+\ve} \sqrt{\sigma} \nn 
    &= (2\pi)^{n-1+\ve} \f{\bar X r_+^{\ve}}{C \left(\prod_{i=1}^n \Xi_i\right) \left(\prod_{i=1}^{n-1} a_i\right)^{1-\ve} (1+r_+^2/l^2) } \int dy_1 \ldots dy_{n-1} \tilde P \left( \prod_{\alpha=1}^{n-1} y_\alpha\right)^\ve.  \label{Aintermediateexpression}
\end{align}

Integrals of the form 
\begin{align}
    \int dy_1 \ldots dy_{n-1} \tilde P \left( \prod_{\alpha=1}^{n-1} y_\alpha\right)^\ve
\end{align}
are considered in Appendix \ref{ValueOfVandermondeDeterminant}. We can invert \eqref{Ptildetimesyalphapderivative} with $p-1 = \ve$ to give the indefinite integral
\begin{align}
    \int dy_1 \ldots dy_{n-1} \tilde P \left( \prod_{\alpha=1}^{n-1} y_\alpha\right)^\ve &= \f{(\ve-1)!!}{(2(n-2)+\ve-1)!!} \tilde P \prod_{\alpha=1}^{n-1} y_\alpha^{1+\ve} \nn
    &= \f{1}{(D - 3)!!} \tilde P \prod_{\alpha=1}^{n-1} y_\alpha^{1+\ve}. \label{areaintegral}
\end{align}
Here $!!$ is the double factorial function defined by \eqref{doublefactorial}. The definite integral is found by applying the integration limits from Section \ref{limitsonjacobitransformedcoordinates}. In both odd and even dimensions, $a_\alpha \geq y_\alpha \geq a_{\alpha+1}$ for $\alpha < n-1$, and then in odd dimensions $a_{n-1} \geq y_{n-1} \geq a_n$, whereas in even dimension $a_{n-1} \geq y_{n-1} \geq -a_{n-1}$. 

In even dimensions, we can rewrite \eqref{areaintegral} as
\begin{align}
    \int dy_1 \ldots dy_{n-1} \tilde P &= \f{1}{(D - 3)!!} \tilde P \prod_{\alpha=1}^{n-1} y_\alpha \nn 
    &= \f{1}{(D-3)!!}\left( \prod_{\gamma < \delta < n-1} (y_\gamma^2 - y_\delta^2)\right)  \left( \prod_{\beta=1}^{n-2} (y_\beta^2-y_{n-1}^2)\right) \left( \prod_{\beta=1}^{n-2} y_\beta\right) y_{n-1}.
\end{align}

Let (for the next few lines) $f(y_1,\ldots, y_{n-1}) = \tilde P \prod_{\alpha=1}^{n-1} y_\alpha$. By antisymmetry, if any two entries are the same in $f(y_1,\ldots, y_{n-1})$ (if any $y_{\alpha} = y_\beta$ for $\alpha \neq \bt$), $f = 0$. The definite integral, found by applying the limits on $y_\alpha$, is
\begin{align}
    \int d y_1 \ldots dy_{n-1} \tilde P &= \f{1}{(D-3)!!} \left.\left.\left. f(y_1,\ldots, y_{n-1})\right|_{y_{n-1}=-a_{n-1}}^{y_{n-1}=a_{n-1}}\right|_{y_{n-2}=a_{n-1}}^{y_{n-2}=a_{n-2}} \ldots \right|_{y_1 = a_2}^{y_1 = a_1}.
\end{align}
Applying first the limit on $y_{n-1}$, we have
\begin{align}
    \int dy_1 \ldots dy_{n-1} \tilde P &= \left. \left.\f{1}{(D-3)!!} \left(f(y_1 ,\ldots, y_{n-2},a_{n-1}) - f(y_1,\ldots, y_{n-2},-a_{n-1}) \right) \right|^{y_{n-2}=a_{n-2}}_{y_{n-2}=a_{n-1}} \ldots \right|^{y_1 = a_1}_{y_1=a_2}.
\end{align}
$y_{n-1}$ always appears in $\tilde P$ as $y_{n-1}^2$, so $\tilde P$ is even in $y_{n-1}$; since $f(y_1,\ldots, y_{n-1}) = \tilde P y_{n-1} \prod_{\beta=1}^{n-2} y_\alpha$, $f$ is odd in $y_{n-1}$. Thus $f(y_1, \ldots, y_{n-2},-a_{n-1}) = -f(y_1,\ldots, y_{n-2},a_{n-1})$. 
\begin{align}
    \int dy_1 \ldots dy_{n-1} \tilde P &= \left. \left.\f{2}{(D-3)!!} f(y_1 ,\ldots, y_{n-2},a_{n-1})\right|^{y_{n-2}=a_{n-2}}_{y_{n-2}=a_{n-1}} \ldots \right|^{y_1 = a_1}_{y_1=a_2}.    
\end{align}
If we apply the bottom limit on $y_{n-2}$, we get a term $f(y_1, \ldots, a_{n-1},a_{n-1}) = 0$, since $f$ is antisymmetric in its arguments. Thus only the top limit on $y_{n-2}$ is nonvanishing, yielding $f(y_1, \ldots, y_{n-3}, a_{n-2},a_{n-1})$. Similarly, if we take the bottom limit on $y_{n-3}$, $a_{n-2}$, we will have $f(y_1, \ldots, a_{n-2},a_{n-2},a_{n-1}) = 0$, and so only the top limit on $y_{n-3}$ survives. The argument continues all the way to $y_1$, where the bottom integration limit on \emph{any} of the $y_\alpha$ (except $y_{n-1}$) will always vanish, so that the only nonvanishing term is $f(a_1,\ldots, a_{n-1})$:
\begin{align}
    \int dy_1 \ldots dy_{n-1} \tilde P &= \f{2}{(D-3)!!} f(a_1, \ldots, a_{n-1}) \nn 
    &= \f{2 C \prod_{i=1}^{n-1} a_i}{(D-3)!!} .
\end{align}
$C$ is from \eqref{Cdefinition}. 

In odd dimensions, similarly let (for the next few lines) $f(y_1,\ldots, y_{n-1})$ be
\begin{align}
    f(y_1,\ldots, y_{n-1}) &= \tilde P \prod_{\alpha=1}^{n-1} y_\alpha^2.
\end{align}
As in the even-dimensional case, $f$ is antisymmetric in $y_\alpha$ so again if any two $y_\alpha$ are equal to each other, $f$ will be zero. 

Apply the integration limits to $f$. Then we will have
\begin{align}
    \left.\left. f(y_1,\ldots, y_{n-1})\right|_{y_{n-1}=a_n}^{y_{n-1}=a_{n-1}}\ldots \right|_{y_1=a_2}^{y_1=a_1}.
\end{align}
If at any point, we take the lower limit of $y_\alpha$ but the upper limit of $y_{\alpha+1}$, we will have $f(\ldots, a_{\alpha+1},a_{\alpha+1}, \ldots) = 0$. This means that the only nonzero entries when taking these integration limits are the ones where the upper limit is taken for the first several entries and then the lower limits are taken for the rest of them (including the cases where all upper or all lower limits are taken). These are entries for which $y_\beta = a_\beta$ for $\beta < \alpha$ and $y_\beta = a_{\beta+1}$ for $\beta \geq \alpha$. Thus all the $a_i$ will appear, except for $a_\alpha$.

Let $f(\hat a_i)$ be the value of $f$ where all the $a_j$ appear, in order, except $a_i$,
\begin{align}
    f(\hat a_i) = f(\ldots, a_{i-1}, a_{i+1},\ldots).
\end{align}
For example, $f(\hat a_1) = f(a_2,\ldots, a_n)$, $f(\hat a_n) = f(a_1, \ldots, a_{n-1})$ and $f(\hat a_2) = f(a_1, a_3, \ldots, a_n)$. The appearance of $f(\hat a_i)$ comes from taking the upper limit for the first $i-1$ entries and the lower limit for the last $n-i$ entries, so that it will pick up a sign of $(-1)^{n-i}$. Thus
\begin{align}
    \left.\left. f(y_1,\ldots, y_{n-1})\right|_{y_{n-1}=a_n}^{y_{n-1}=a_{n-1}}\ldots \right|_{y_1=a_2}^{y_1=a_1} &= \sum_{i=1}^n (-1)^{n-i} f(\hat a_i).    
\end{align}

$f(\hat a_i)$ is given by
\begin{align}
    f(\hat a_i) &= \left(\prod_{\substack{1 \leq j < k \leq n \\ j,k \neq i}} (a_j^2-a_k^2)\right) \prod_{j=1, j \neq i}^n a_j^2 \nn 
    &= \f{\left(\prod_{1 \leq j < k \leq n} (a_j^2-a_k^2)\right) \prod_{j=1}^n a_j^2}{\left( \prod_{j=1}^{i-1} (a_j^2-a_i^2)\right) \left(\prod_{j=i+1}^n (a_i^2-a_{j}^2)\right) a_i^2} \nn 
    &= (-1)^{n-i} \f{C \prod_{j=1}^n a_j^2}{\Ups_i a_i^2}.
\end{align}
Consequently,
\begin{align}
    \left.\left. f(y_1,\ldots, y_{n-1})\right|_{y_{n-1}=a_n}^{y_{n-1}=a_{n-1}}\ldots \right|_{y_1=a_2}^{y_1=a_1} &= \left(C \prod_{j=1}^n a_j^2\right) \sum_{i=1}^n \f{1}{\Ups_i a_i^2}.
\end{align}
In \cite{KrtousKubiznak}, the $A_\mu^{(l)}$ and $U_\mu$ are stated to satisfy
\begin{align}
    \sum_{\mu=1}^n \f{A_\mu^{(l)}}{x_\mu^2 U_\mu} &= \f{A^{(l)}}{A^{(n)}},
\end{align}
which, using $l = 0$, gives $\sum_{\mu=1}^n 1/(x_\mu^2 U_\mu) = \f{1}{A^{(n)}} = 1/\prod_{\mu=1}^n x_\mu^2$. The analogous expression for $\Ups_i$ and $a_i^2$ is
\begin{align}
    \sum_{i=1}^n \f{1}{a_i^2 \Ups_i} &= \f{1}{\prod_{j=1}^n a_j^2}.
\end{align}
Thus we have
\begin{align}
    \left.\left. f(y_1,\ldots, y_{n-1})\right|_{y_{n-1}=a_n}^{y_{n-1}=a_{n-1}}\ldots \right|_{y_1=a_2}^{y_1=a_1} = C \f{\prod_{j=1}^n a_j}{\prod_{j=1}^n a_j} &= C.
\end{align}

Thus we have, in arbitrary dimension,
\begin{align}
    \int dy_1 \ldots dy_{n-1} \tilde P \left( \prod_{\alpha=1}^{n-1} y_\alpha\right)^\ve &= \f{2^{1-\ve} C \prod_{i=1}^{n-1} a_i^{1-\ve}}{(D-3)!!} \label{tildePintegral}
\end{align}
\eqref{Aintermediateexpression} becomes
\begin{align}
    A &= (2\pi)^{n-1+\ve} 2^{1-\ve} \f{\bar X r_+^{\ve}}{(D-3)!! \left(\prod_{i=1}^n \Xi_i\right)  (1+r_+^2/l^2) } \nn 
    &= \f{\bar X r_+^\ve}{\left(\prod_{i=1}^n \Xi_i\right) (1+r_+^2/l^2)} \mc A_{D-2},
\end{align}
using \eqref{ADm2}. Using the value of $\bar X$ from \eqref{Xbar} with $r = r_+$, we recover \eqref{AGibbonsLu}. 

\chapter{Explicit Calculations for Constancy of Komar Difference Term for Principal and Azimuthal Vectors for Arbitrary Radius in Vacuum} \label{ExplicitCalculationsAppendixChapter}

In this appendix chapter, I revisit the Komar difference terms associated with $\beta$ (Section \ref{explicitJacobiTransformed}) and $\eta_i$ (Section \ref{explicitetaintegral}) in the Kerr--AdS vacuum. In the main text, I appeal to the vanishing of $R^a_b - \overline{R^a_b}$ to show that these terms must be constant. In this chapter I take an alternate approach, calculating the Komar integral at arbitrary $r$ in both cases (of course, recovering the constancy). The impetus for doing this was to 
illuminate what it is about the structure of the quantity being integrated gives rise to this constancy. The arguments contained herein are interesting, if complicated, so they are included here in the appendix. 

Additionally, integrals similar to those in this chapter appear in the calculation of the horizon area in Section \ref{AreaRevisited}, itself moved to Appendix \ref{AreaCalculationAppendix}.

\section{Explicit Calculation of Komar Term Associated with Principal Vector in Jacobi Transformed Coordinates} \label{explicitJacobiTransformed}

Here I calculate the value of $\sqrt{-\bar g} (\bs i^K_\bt)^{tr}$ for a constant mass function, $\mu(r) = m$. Note again that the constancy of $\oint_C \sqrt{-\bar g} (\bs i^K_\bt)^{tr}$ with respect to surface $C$ follows directly from the vanishing of $R^a_b-\overline{R^a_b}$ and so this appendix section is an independent check/direct verification. 

It is easier to work with the integrals in $y_\alpha$ rather than $\mu_i$ coordinates. In coordinates $(t,r,y_\alpha,\phi_i)$, the combination $\sqrt{-\bar g} (\bs i^K_\bt)^{t r}$ is, from \eqref{mcKKtrbeta} and \eqref{sqrtgrytphi},
\begin{align}
    \sqrt{-\bar g} (\bs i^K_\bt)^{t r} &= -\f{P}{C \prod_{i = 1}^{n} \Xi_i} \f{(r \prod_{\alpha=1}^{n-1} y_\alpha)^\varepsilon}{(\prod_{i = 1}^{n-1+\ve} a_i)^{1-\varepsilon}} \f{1}{32 \pi} \f{\prod_{i = 1}^{n-1+\ve} (r^2+a_i^2)}{r^{2 \varepsilon} U_n} \f{\pa}{\pa r} \left(\f{2 m r^{1-\varepsilon}}{U_n}\right). \label{sqrtgKtrby}
\end{align}
Define $\tilde P$ to be $P/U_n$:
\begin{align}
    \tilde P = \f{P}{U_n} &= \prod_{\alpha < \beta} (y_\alpha^2-y_\beta^2).
\end{align}
This is $r$-independent. 
We then have
\begin{align}
    \sqrt{-\bar g} (\bs i_\bt^K)^{tr} &= -\f{\prod_{i=1}^{n-1+\ve}(r^2+a_i^2)}{32 \pi C (\prod_{i=1}^{n} \Xi_i )(\prod_{i=1}^{n-1+\ve} a_i)^{1-\varepsilon} r^{\varepsilon}} \tilde P \left(\prod_{\alpha=1}^{n-1} y_\alpha\right)^\varepsilon \f{\pa}{\pa r} \left(\f{2 m r^{1-\varepsilon}}{U_n}\right) \nn 
    &= -\f{2m \prod_{i=1}^{n-1+\ve} (r^2+a_i^2)}{32 \pi C (\prod_{i=1}^{n} \Xi_i) (\prod_{i=1}^{n-1+\ve} a_i)^{1-\ve} r^\ve} \tilde P \left( \prod_{\alpha=1}^{n-1} y_\alpha^\ve\right) \f{\pa}{\pa r} \f{r^{1-\ve}}{U_n}. \label{sqrtgiKbetatrycoords}
\end{align}

The integrals over $\phi_i$ are straightforward (each bringing in a factor of $2\pi$). For the integrals over $y_\alpha$ we will need the integration limits from \eqref{oddyalphalimits} and \eqref{evenyalphalimits}, and we will want the values of 
\begin{align}
    \int dy_1 \ldots dy_{n-1} \tilde P \left(\prod_{\alpha=1}^{n-1} y_\alpha^\ve\right) \f{1}{U_n} 
\end{align}
as well as its $r$ (partial) derivative. 

It is not obvious at first glance how to calculate this, but it turns out that the indefinite integral has a relatively simple form. As shown in Appendix \ref{ValueOfVandermondeDeterminant} in the steps leading up to \eqref{tildePypbyUnderivatives}, for arbitrary $p > 0$,
\begin{align}
    \f{\pa^{n-1}}{\pa y_1 \ldots \pa y_{n-1}} \left( \f{\tilde P \prod_{\alpha=1}^{n-1} y_\alpha^p}{U_n}\right) &= - \f{(2(n-3)+p)!!}{(p-2)!!} \left( r \prod_{\alpha=1}^{n-1} y_\alpha\right)^{p-1}  \f{\pa}{\pa r} \left( \f{\tilde P r^{2-p}}{U_n}\right).
\end{align}
Here $x!!$ is the double factorial of $x$ from \eqref{doublefactorial}. We can substitute in $p = 1+ \ve$. The $(p-2)!!$ term becomes $(\ve-1)!! = 1$ since $0!! = (-1)!! = 1$. After a bit of rearranging,
\begin{align}
    \int dy_1 \ldots dy_{n-1} \tilde P \left( \prod_{\alpha=1}^{n-1} y_\alpha^\ve\right) \f{\pa}{\pa r} \f{r^{1-\ve}}{U_n} &= -\f{1}{(D-5)!!} \f{\tilde P \prod_{\alpha=1}^{n-1} y_\alpha^{1+\ve}}{U_n r^\ve}. \label{indefiniteintegral}
\end{align}
($D = 4$ gives $(D-5)!! = (-1)!!= 1$.)

This is the indefinite integral. In both odd and even dimensions, $a_\alpha \geq y_\alpha \geq a_{\alpha+1}$ for $\alpha < n-1$, and then in odd dimensions $a_{n-1} \geq y_{n-1} \geq a_n$, whereas in even dimension $a_{n-1} \geq y_{n-1} \geq -a_{n-1}$. 

The arguments for how to apply the integration limits in even and odd dimensions are essentially the same as those in Appendix \ref{AreaCalculationAppendix}. I will state the argument in detail here as well.

\subsection{Even-Dimensional Calculation}

Consider first the even-dimensional case. We can expand the right-hand side of \eqref{indefiniteintegral} to isolate the $y_{n-1}$ terms as follows.
\begin{align}
    -\f{1}{(D-5)!!} \f{\tilde P \prod_{\alpha=1}^{n-1} y_\alpha^{1+\ve}}{U_n r^\ve} &= -\f{1}{(D-5)!!} \f{\left( \prod_{\gamma<\delta < n-1} (y_\gamma^2-y_\delta^2)\right) \left( \prod_{\beta=1}^{n-2} (y_\beta^2-y_{n-1}^2)\right) \left(\prod_{\beta=1}^{n-2} y_\beta\right) y_{n-1}}{\left( \prod_{\beta=1}^{n-2}(r^2+y_\beta^2)\right) (r^2+y_{n-1}^2)}.
\end{align}
This is an odd function of $y_{n-1}$. 
Evaluating this for $y_{n-1} = + a_{n-1}$ minus $y_{n-1}=-a_{n-1}$ gives
\begin{align}
    &\left[-\f{1}{(D-5)!!} \f{\tilde P \prod_{\alpha=1}^{n-1} y_\alpha^{1+\ve}}{U_n r^\ve}\right]^{y_{n-1}=a_{n-1}}_{y_{n-1}=-a_{n-1}}\nn &=   -\f{2}{(D-5)!!} \f{\left( \prod_{\gamma<\delta < n-1} (y_\gamma^2-y_\delta^2)\right) \left( \prod_{\beta=1}^{n-2} (y_\beta^2-a_{n-1}^2)\right) \left(\prod_{\beta=1}^{n-2} y_\beta\right) a_{n-1}}{\left( \prod_{\beta=1}^{n-2}(r^2+y_\beta^2)\right) (r^2+a_{n-1}^2)}. 
\end{align}
(The 2 comes from the fact that the expression is an odd function of $y_{n-1}$, so its value when $y_{n-1}=-a_{n-1}$ is the negative of its value when $y_{n-1} = a_{n-1}$.) Now we can set the limits on $y_{n-2}$, which are from $a_{n-1}$ to $a_{n-2}$. Note however that the presence of the term $y_{n-2}^2-a_{n-1}^2$ in the numerator means that the value of this expression evaluated at $y_{n-2}=a_{n-1}$ is simply zero. Consequently, applying the limits on $y_{n-2}$ means substituting in $y_{n-2} = a_{n-2}$ in the expression. Similarly, the presence of the term $y_{n-3}^2 - a_{n-2}^2$ in the numerator means that substituting in $y_{n-3} = a_{n-2}$ will give zero, so again applying the limits on $y_{n-3}$ amounts to setting $y_{n-3} = a_{n-3}$ in the expression. This continues all the way down to $y_1$, so that once the limits are applied we have the right-hand side of \eqref{indefiniteintegral} with $y_\alpha = a_\alpha$, times 2 to account for the limits on $y_{n-1}$ including $-a_{n-1}$:
\begin{align}
    \int_{a_2}^{a_1} dy_1 \ldots \int_{-a_{n-1}}^{a_{n-1}} dy_{n-1} \tilde P  \f{\pa}{\pa r} \f{r}{U_n} &= -\f{2}{(D-5)!!}\f{\left(\prod_{1 \leq i < j \leq n-1} (a_i^2-a_j^2)\right) \prod_{i=1}^{n-1} a_i}{\prod_{i = 1}^{n-1} (r^2+a_i^2)} \nn 
    &= -\f{2 C \prod_{i=1}^{n-1} a_i}{(D-5)!! \prod_{i=1}^{n-1} (r^2+a_i^2)}. \label{evendimbetaintegral}
\end{align}

\subsection{Odd-Dimensional Calculation} \label{betaintegraloddsection}
In odd dimensions, the argument proceeds along different lines. Let the right-hand side of \eqref{indefiniteintegral} be given by $f(y_1, \ldots, y_{n-1})$. $f$ is completely antisymmetric in the $y_\alpha$. The integration limits are such that $f$ is evaluated between $y_\alpha = a_{\alpha+1}$ and $a_\alpha$. We can expand this out as
\begin{align}
    \left[ \left[\left[f(y_1, \ldots, y_{n-1})\right]_{y_1=a_2}^{y_1=a_1}\right]_{y_2 = a_3}^{y_2 = a_2} \ldots \right]_{y_{n-1}=a_{n}}^{y_{n-1} = a_{n-1}} &= f(a_1, a_2, \ldots, a_{n-2},a_{n-1}) - f(a_1, a_2, \ldots, a_{n-2},a_n) + \ldots.
\end{align}
There are, before cancellations, $2^{n-1}$ terms in the expansion, since there are two choices (upper and lower limit) for each variable $y_\alpha$ in the set. Because of the antisymmetry of $f$, however, all $f$ terms vanish if two of the entries are equal. Since $a_{\alpha+1}$ is the lower limit for $y_\alpha$ and the upper limit for $y_{\alpha+1}$, any terms where the lower limit is chosen for $y_\alpha$ and the upper limit is chosen for $y_{\alpha+1}$ will lead to $f$ being zero. This means that if the lower limit is taken for $y_\alpha$, the lower limit must also be taken for $y_{\alpha+1}$ in order to have a nonvanishing term.

This means that only terms which survive will be the ones for which the upper limit is taken for the first $i-1$ terms and the lower limit taken for the final $n-i$ terms, where $i$ can vary from $1$ to $n$. In the slots with $1 \leq j < i$, $a_j$ will appear, and in the slots $i \leq j \leq n-1$ $a_{j+1}$ will appear, so that $a_{i}$ itself does not appear. Since the lower limit is chosen $n-i$ times, this term will be accompanied by a sign of $(-1)^{n-i}$. Let $f(\hat a_i)$ be equal to the value of $f$ in this case---where $a_i$ is omitted:
\begin{align}
    f(\hat a_i) &= f(\ldots, a_{i-1},a_{i+1}, \ldots). \label{fhatai}
\end{align}
Then we can expand 
\begin{align}
    \left[ \left[\left[f(y_1, \ldots, y_{n-1})\right]_{y_1=a_2}^{y_1=a_1}\right]_{y_2 = a_3}^{y_2 = a_2} \ldots \right]_{y_{n-1}=a_{n}}^{y_{n-1} = a_{n-1}} &= \sum_{i = 1}^n (-1)^{n-i}f(\hat a_i).
\end{align}

Explicitly,
\begin{align}
    f(\hat a_i) &= -\f{1}{(D-5)!!} \f{ \left(\prod_{\substack{1 \leq j < k \leq n \\ j,k \neq i}} (a_j^2-a_k^2) \right) \prod_{j=1,j\neq i}^n a_j^2}{r \prod_{j=1,j\neq i}^n (r^2+a_j^2)}. 
\end{align}
This can be simplified by noting
\begin{align}
    C &= \prod_{1 \leq j < k \leq n} (a_j^2 - a_k^2) \nn 
    &=\left( \prod_{\substack{1 \leq j < k \leq n \\ j,k\neq i}}(a_j^2-a_k^2)\right) \left(\prod_{1 \leq j < i} (a_j^2-a_i^2)\right) \prod_{i < k \leq n} (a_i^2-a_k^2) \nn 
    &= \left( \prod_{\substack{1 \leq j < k \leq n \\ j,k\neq i}}(a_j^2-a_k^2)\right) \left(\prod_{1 \leq j < i} (a_j^2-a_i^2)\right) (-1)^{n-i} \prod_{i < k \leq n} (a_k^2-a_i^2) \nn 
    &= (-1)^{n-i} \Upsilon_i \left( \prod_{\substack{1 \leq j < k \leq n \\ j,k\neq i}}(a_j^2-a_k^2)\right).
\end{align}
Thus
\begin{align}
    f(\hat a_i) &= -\f{1}{r(D-5)!!} (-1)^{n-i} \f{C}{\Upsilon_i} \f{\prod_{j=1}^n a_j^2}{a_i^2} \f{r^2+a_i^2}{\prod_{j=1}^n (r^2+a_j^2)}. \label{fahatvalue}
\end{align}
Consequently,
\begin{align}
    \sum_{i=1}^n (-1)^{n-i} f(\hat a_i) &= -\f{C \prod_{j=1}^n a_j^2}{(D-5)!! r \prod_{j=1}^n (r^2+a_j^2)}\sum_{i=1}^n \f{r^2+a_i^2}{a_i^2 \Upsilon_i}.
\end{align}
As stated in \eqref{sumUpsi}, $\sum_{i=1}^n \f{1}{\Upsilon_i} = 0$. It is stated in \cite{KrtousKubiznak} that $\sum_{\mu=1}^n \f{A_\mu^{(l)}}{x_\mu^2 U_\mu} = \f{A^{(l)}}{A^{(n)}}$, which taking $l = 0$ gives $\sum_{\mu=1}^n \f{1}{x_\mu^2 U_\mu} = \f{1}{A^{(n)}}$. Applying the same principle with $U_\mu \to \Upsilon_i, x_\mu \to a_i$ and $A^{(n)} \to \prod_{j=1}^n a_j^2$ gives
\begin{align}
    \sum_{i=1}^n \f{1}{a_i^2 \Upsilon_i} &= \f{1}{\prod_{j=1}^n a_j^2}. \label{sumofa2Upsilonreciprocal}
\end{align}
Consequently,
\begin{align}
    \int_{a_2}^{a_1} dy_1 \ldots \int_{a_n}^{a_{n-1}} dy_{n-1} \tilde P \left( \prod_{\alpha=1}^{n-1} y_\alpha\right) \f{\pa}{\pa r} \f{1}{U_n} &=
    \sum_{i=1}^n (-1)^{n-i} f(\hat a_i) \nn 
    &= -\f{C r}{(D-5)!! \prod_{j=1}^n (r^2+a_j^2)}. \label{odddimensionalbetaintegral}
\end{align}

\subsection{Summary}

In arbitrary dimension, the definite integral with the proper limits, combined with the integral over the $n-1+\ve$ azimuthal angles $\phi_i$, is
\begin{align}
    \int dy_1 \ldots dy_{n-1} d\phi_1 \ldots d \phi_{n-1+\ve} \tilde P \left( \prod_{\alpha=1}^{n-1} y_\alpha^\ve\right) \f{\pa}{\pa r} \f{r^{1-\ve}}{U_n} &= -\f{(2\pi)^{n-1+\ve} (2 \prod_{i=1}^{n-1+\ve}a_i)^{1-\ve} r^\ve C}{(D-5)!! \prod_{i=1}^{n-1+\ve} (r^2+a_i^2)}. 
\end{align}

Consequently, substituting into \eqref{sqrtgKtrby}, 
\begin{align}
    \int d^{D-2} x \sqrt{-\bar g} (\bs i_\bt^K)^{tr} &= \f{2m (2\pi)^{n-1+\ve} 2^{1-\ve}}{32\pi (D-5)!! \prod_{i=1}^n \Xi_i} \nn 
    &= \f{m (D-3) \mc A_{D-2}}{16 \pi \prod_{i=1}^n \Xi_i} ,
\end{align}
using \eqref{ADm2} for $\mc A_{D-2}$.  Consequently, we reproduce \eqref{KomarIntegralTermforbeta} without using the Gauss--Stokes theorem.

\section{Explicit Calculation of Angular Momentum Komar Integral in Jacobi Transformed Coordinates} \label{explicitetaintegral}

The value of $(\bs i_{\eta_i}^K)^{tr}$ for generalized Kerr--AdS is given in terms of $(t,r,\mu_i,\phi_i)$ coordinates in \eqref{KKetamuform}. Converting from $\mu_i$ to $y_\alpha$ coordinates and using the value of $\sqrt{-\bar g}$ from \eqref{sqrtgKtrby}, in $(t,r,y_\alpha,\phi_i)$ coordinates, 
\begin{align}
    &\sqrt{-\bar g} (\bs i^K_{\eta_i})^{tr} = \f{1}{32\pi} \f{P (r\prod_{\alpha=1}^{n-1} y_\alpha)^\ve}{C (\prod_{j=1}^n \Xi_j)(\prod_{j=1}^{n-1} a_j)^{1-\ve}} \f{\mu_i^2(r^2+a_i^2)}{\Xi_i} \f{l^2}{(r^2+l^2) F} \f{\pa}{\pa r} \left( \f{2\mu(r)}{U} \f{a_i}{r^2+a_i^2}\right) \nn 
    &= \f{1}{32\pi} \f{P (r \prod_{\alpha=1}^{n-1} y_\alpha)^\ve}{C (\prod_{j=1}^n \Xi_j) (\prod_{j=1}^{n-1} a_j)^{1-\ve}} \f{(r^2+a_i^2)}{\Xi_i} \f{\prod_{\alpha=1}^{n-1} (a_i^2-y_\alpha^2)}{\prod_{j=1,j\neq i}^{n} (a_i^2-a_j^2)} \f{r^{1-\ve} l^2 V}{(r^2+l^2) U_n} \f{\pa}{\pa r} \left(\f{2 \mu(r) r^{1-\ve}}{U_n} \f{a_i}{r^2+a_i^2}\right) \nn 
    &= \f{a_i}{32\pi \Xi_i} \f{\tilde P (r \prod_{\alpha=1}^{n-1} y_\alpha)^\ve}{C (\prod_{j=1}^n \Xi_j) (\prod_{j=1}^{n-1} a_j)^{1-\ve}}  \f{(r^2+a_i^2)\prod_{\alpha=1}^{n-1}(y_\alpha^2-a_i^2)}{(-a_i^2)^{1-\ve} \Upsilon_i} \f{\prod_{j=1}^n(r^2+a_j^2)}{r^2} \f{\pa}{\pa r} \left( \f{2 \mu(r) r^{1-\ve}}{U_n (r^2+a_i^2)} \right) 
    \label{sqrtgKKtretayform}
\end{align}

I will calculate the integral of this over $d^{D-2}x = dy_1 \dots dy_{n-1} d\phi_1 \ldots d \phi_{n-1+\ve}$ explicitly for the Kerr--AdS case $(\mu(r) = m)$. 

We will need to calculate integrals of the form
\begin{align}
    \int dy_1 \ldots dy_{n-1} \tilde P \left(\prod_{\alpha=1}^{n-1} y_\alpha\right)^\ve \left( \prod_{\alpha=1}^{n-1} (y_\alpha^2-a_i^2)\right) \f{\pa}{\pa r} \f{r^{1-\ve}}{U_n(r^2+a_i^2)}.
\end{align}
To approach this integral it should first be noted how similar it is to \eqref{indefiniteintegral} in form. Indeed, \eqref{indefiniteintegral} is really a statement about an arbitrary set of $n-1$ distinct variables $y_\alpha$. Consider the case where we add an extra variable, say $z$, to the set of $y_\alpha$ to give the set $\{y_1, \ldots, y_{n-1}, z\}$. $z$ is not labelled like the $y_\alpha$, and is not physically meaningful; rather, this is a mathematical trick that will help us perform a calculation. Let $D = 2n+\ve$ still. Then, keeping the same meaning of $\tilde P$ and $U_n$ (that is, $\tilde P = \prod_{1 \leq \alpha < \beta \leq n-1} (y_\alpha^2-y_\beta^2)$ and $U_n = \prod_{\alpha=1}^{n-1} (r^2+y_\alpha^2)$, neither of which involve $z$), we can write a similar equation to \eqref{indefiniteintegral} by replacing the integral over $y_\alpha$ with one over $y_\alpha$ \emph{and} $z$, replacing $\tilde P$ with $\tilde P \prod_{\alpha=1}^{n-1} (y^2_\alpha - z^2)$, replacing $U_n$ with $U_n (r^2+z^2)$, and replacing $\prod_{\alpha=1}^{n-1} y_\alpha^\ve$ with $\left(\prod_{\alpha=1}^{n-1} y_\alpha^\ve\right) z$. The right-hand side will be similar but the same substitutions must be made for $\tilde P$ and $U_n$, the numerical factor will be $-1/(D-3)!!$ instead of $-1/(D-5)!!$ (since the $(D-5)!!$ is really $(2(n-1-2)+1+\ve)!!$, which becomes $(2(n-2)+1+\ve)!!$ under $n-1\to n$ for the size of the set) and $\prod_{\alpha=1}^{n-1} y_\alpha^{1+\ve}$ becomes $\left( \prod_{\alpha=1}^{n-1} y_\alpha^{1+\ve}\right) z^{1+\ve}$. The result is
\begin{align}
    &\int dy_1 \ldots dy_{n-1} dz \tilde P \left( \prod_{\alpha=1}^{n-1} (y_\alpha^2-z^2)\right) \left( \prod_{\alpha=1}^{n-1} y_\alpha^\ve\right) z^\ve \f{\pa}{\pa r} \f{r^{1-\ve}}{U_n(r^2+z^2)} \nn 
    &= - \f{1}{(D-3)!!} \f{\tilde P \left( \prod_{\alpha=1}^{n-1}(y_\alpha^2-z^2)\right) \left(\prod_{\alpha=1}^{n-1} y_\alpha^{1+\ve}\right) z^{1+\ve}}{U_n (r^2+z^2) r^\ve}.
\end{align}
Differentiating both sides with respect to $z$,
\begin{align}
    &\int dy_1 \ldots dy_{n-1}\tilde P \left( \prod_{\alpha=1}^{n-1} (y_\alpha^2-z^2)\right) \left( \prod_{\alpha=1}^{n-1} y_\alpha^\ve\right) z^\ve \f{\pa}{\pa r} \f{r^{1-\ve}}{U_n(r^2+z^2)} \nn 
    &= - \f{1}{(D-3)!!} \f{\tilde P  \left(\prod_{\alpha=1}^{n-1} y_\alpha^{1+\ve}\right) }{U_n r^\ve} \f{\pa}{\pa z} \f{\left( \prod_{\alpha=1}^{n-1}(y_\alpha^2-z^2)\right) z^{1+\ve}}{r^2+z^2} \nn 
    &= -\f{1}{(D-3)!!} \f{\tilde P \prod_{\alpha=1}^{n-1} y_\alpha^{1+\ve}}{U_n r^\ve} \left( \f{z^{\ve}( (\ve-1) z^2 + (1+\ve) r^2}{(r^2+z^2)^2} \prod_{\alpha=1}^{n-1} (y_\alpha^2-z^2) - \f{2 z^{2+\ve}}{r^2+z^2}\sum_{\alpha=1}^{n-1} \prod_{\beta=1,\beta\neq\alpha}^{n-1}(y_\beta^2-z^2)\right)
\end{align}
We can now substitute in $z = a_i$, to give,
\begin{align}
    &\int dy_1 \ldots dy_{n-1}\tilde P \left( \prod_{\alpha=1}^{n-1} (y_\alpha^2-a_i^2)\right) \left( \prod_{\alpha=1}^{n-1} y_\alpha^\ve\right) a_i^\ve \f{\pa}{\pa r} \f{r^{1-\ve}}{U_n(r^2+a_i^2)} \nn 
    &= -\f{1}{(D-3)!!} \f{\tilde P \prod_{\alpha=1}^{n-1} y_\alpha^{1+\ve}}{U_n r^\ve} \left( \f{a_i^{\ve}( (\ve-1) a_i^2 + (1+\ve) r^2}{(r^2+a_i^2)^2} \prod_{\alpha=1}^{n-1} (y_\alpha^2-a_i^2) - \f{2 a_i^{2+\ve}}{r^2+a_i^2}\sum_{\alpha=1}^{n-1} \prod_{\beta=1,\beta\neq\alpha}^{n-1}(y_\beta^2-a_i^2)\right). \label{indefiniteetaintegral}
\end{align}
(The attentive reader may note that it was assumed that $z$ is not equal to the $y_\alpha$, but $y_i$ and $y_{i-1}$, if they exist, are equal to $a_i$ at the limits. Because it is only on the limiting cases that this is the case I don't believe it affects the integral. Additionally, no terms of the form $(y_\alpha^2-a_i^2)$ appear in the denominator the way I have written the expression.) 

The right-hand side can be simplified by noting that if $y_\alpha^2 = a_i^2$, $y_\alpha^2-a_i^2 = 0$ (for some particular $y_\alpha$). 

\subsection{Even-Dimensional Case}

In even dimensions, by the same argument as that leading up to \eqref{evendimbetaintegral}, the evaluation of the definite integral form of \eqref{indefiniteetaintegral} with the limits $a_1 \geq y_1 \geq a_2, a_2 \geq y_2 \geq a_3$ and so on, up to $a_{n-1} \geq y_{n-1} \geq -a_{n-1}$, will just be twice the value of the right-hand side of \eqref{indefiniteetaintegral} with $y_\alpha = a_\alpha$. This being the case, any terms with $y_i^2 - a_i^2$ will be zero, leaving only
\begin{align}
    &\int_{a_2}^{a_1} dy_1 \ldots \int_{-a_{n-1}}^{a_{n-1}} dy_{n-1}\tilde P \left( \prod_{\alpha=1}^{n-1} (y_\alpha^2-a_i^2)\right)  \f{\pa}{\pa r} \f{r}{U_n(r^2+a_i^2)} \nn 
    &= \left.-\f{2}{(D-3)!!} \f{\tilde P \prod_{\alpha=1}^{n-1} y_\alpha}{U_n} \left(  - \f{2 a_i^{2}}{r^2+a_i^2} \prod_{\beta=1,\beta\neq i}^{n-1}(y_\beta^2-a_i^2)\right)\right|_{y_\alpha = a_\alpha} \nn 
    &= +\f{4}{(D-3)!!} \f{C \prod_{j=1}^{n-1} a_j}{\prod_{j=1}^{n-1} (r^2+a_j^2)}  \f{a_i^2 \Upsilon_i}{(r^2+a_i^2)}.
\end{align}
(In the sum of $\alpha$, only $\alpha = i$ survives since otherwise a term $y_i^2-a_i^2$ will appear in the product $\prod_{\beta = 1, \beta \neq \alpha}^{n-1} (y_\beta^2-a_i^2)$.) 

\subsection{Odd-Dimensional Case}

As with the calculation for the $\beta$ vector, the odd-dimensional case requires more care. Let $\tilde f(y_1, \ldots, y_{n-1})$ be the right-hand side of \eqref{indefiniteetaintegral}. Recalling $f(y_1, \ldots, y_{n-1})$ from Section \ref{betaintegraloddsection}, we have
\begin{align}
    \tilde f(y_1, \ldots, y_{n-1}) &= f(y_1, \ldots, y_{n-1}) \f{1}{D-3} \left( \f{2 a_i r^2}{(r^2+a_i^2)^2} \prod_{\alpha=1}^{n-1} (y_\alpha^2-a_i^2) - \f{2 a_i^3}{r^2+a_i^2} \sum_{\alpha=1}^{n-1} \prod_{\beta = 1, \beta \neq \alpha}^{n-1} (y_\beta^2-a_i^2)\right).
\end{align}

By the same argument as in the steps leading up to \eqref{odddimensionalbetaintegral}, let $\tilde f(\hat a_j)$ be $\tilde f$ evaluated for $y_\alpha = a_\alpha$ for $\alpha < j$ and $y_\alpha = a_{\alpha+1}$ for $\alpha > j$:
\begin{align}
    \tilde f(\hat a_j) &= \tilde f( \ldots, a_{j-1}, a_{j+1}, \ldots).
\end{align}
This is analogous to $f(\hat a_i)$ from \eqref{fhatai}.

We will need separate expressions for $\tilde f(\hat a_j)$ for $j \neq i$ and $j =i$. If $j < i$, then $y_{i-1} = a_i$; $\prod_{\alpha=1}^{n-1} (y_\alpha^2-a_i^2) = 0$ and the only term in the $\alpha$ sum which survives is $\alpha = i-1$. If $j > i$, then $y_i = a_i$; $\prod_{\alpha=1}^{n-1} (y_\alpha^2-a_i^2) = 0$ and the only term in the $\alpha$ sum which survives is $\alpha = i$. In both these cases, the product $\prod_{\beta=1,\beta\neq \alpha}^{n-1} (y_\beta^2-a_i^2) = \prod_{k=1, k \neq i,j}^{n} (a_k^2-a_i^2) = \Upsilon_i/(a_j^2-a_i^2)$. For $j \neq i$ then, using \eqref{fahatvalue},
\begin{align}
    \tilde f(\hat a_j) &= - \f{2 a_i^3 \Upsilon_i}{(D-3)(r^2+a_i^2)(a_j^2-a_i^2)}f(\hat a_j) \nn 
    &= \f{2a_i^3 C \Upsilon_i \prod_{k=1}^n a_k^2}{r (r^2+a_i^2) (D-3)!! \prod_{k=1}^n (r^2+a_k^2)} \f{(-1)^{n-j} (r^2+a_j^2)}{\Upsilon_j a_j^2 (a_j^2-a_i^2)}.
\end{align}

If $j = i$, then none of the $y_\alpha$ are equal to $i$ and so no terms vanish.
\begin{align}
    \tilde f(\hat a_i) &= \f{1}{D-3} \left( \f{2 a_i r^2}{(r^2+a_i^2)^2} \Upsilon_i - \f{2 a_i^3}{r^2+a_i^2} \sum_{j=1, j \neq i}^n \f{\Upsilon_i}{a_j^2-a_i^2}\right) f(\hat a_i) \nn 
    &= \f{2 a_i \Upsilon_i}{(r^2+a_i^2)(D-3)} \left( \f{r^2}{r^2+a_i^2} - \sum_{j=1,j\neq i}^n\f{a_i^2}{a_j^2-a_i^2}\right) f (\hat a_i) \nn 
    &= -\f{2 (-1)^{n-i} C \prod_{k=1}^n a_k^2}{r (D-3)!! a_i \prod_{k=1}^n (r^2+a_k^2)} \left( \f{r^2}{r^2+a_i^2} - \sum_{j=1,j\neq i}^{n} \f{a_i^2}{a_j^2-a_i^2}\right)
\end{align}

Analogously to Section \ref{betaintegraloddsection}, 
\begin{align}
    &\int_{a_2}^{a_1} dy_1 \ldots \int_{a_n}^{a_{n-1}} dy_{n-1} \tilde P \left( \prod_{\alpha=1}^{n-1} (y_\alpha^2-a_i^2)\right) \left( \prod_{\alpha=1}^{n-1} y_\alpha\right) a_i \f{\pa}{\pa r} \f{1}{U_n(r^2+a_i^2)} \nn
    &= \sum_{j=1}^n (-1)^{n-j} \tilde f(\hat a_j) \nn 
    &= \f{2 C \prod_{k=1}^n a_k^2}{r(D-3)!!\prod_{k=1}^n (r^2+a_k^2)} \f{a_i^3 \Upsilon_i}{r^2+a_i^2} \left[ -\f{r^2}{a_i^4 \Upsilon_i} + \sum_{j=1, j\neq i}^n \f{1}{a_j^2-a_i^2} \left( \f{r^2+a_j^2}{\Upsilon_j a_j^2} + \f{r^2+a_i^2}{\Upsilon_i a_i^2 }\right)\right] \nn 
    &= \f{2 C \prod_{k=1}^n a_k^2}{r(D-3)!!\prod_{k=1}^n (r^2+a_k^2)} \f{a_i^3 \Ups_i}{r^2+a_i^2} \times \nn
    &\qquad \left[ r^2 \left( - \f{1}{a_i^4 \Ups_i} + \sum_{j=1,j\neq i} \f{1}{a_j^2-a_i^2}\left( \f{1}{a_j^2 \Ups_j}+\f{1}{a_i^2\Ups_i}\right)\right) + \sum_{j=1,j\neq i}^n \f{1}{a_j^2-a_i^2} \left(\f{1}{\Ups_j}+\f{1}{\Ups_i}\right)\right]
    \label{definiteintegraletasection}
\end{align}

To evaluate the expression in brackets, note that \eqref{sumUpsi} and \eqref{sumofa2Upsilonreciprocal} hold regardless of the values of $a_j$ (provided they are not equal to each other). We can differentiate those expressions with respect to $a_i$ keeping the other $a_j$ constant (again, as a mathematical trick).
\begin{align}
    \f{\pa}{\pa a_i} \f{1}{\Ups_j} &= \f{a_i^2-a_j^2}{\Ups_j}\f{\pa}{\pa a_i}\f{1}{a_i^2-a_j^2} \qquad (\textrm{if }j \neq i) \nn 
    &= -\f{2 a_i}{(a_i^2-a_j^2) \Ups_j} \nn 
    &= +\f{2 a_i}{(a_j^2-a_i^2)\Ups_j} \nn 
    \f{\pa}{\pa a_i} \f{1}{\Ups_i} &= \sum_{j=1,j\neq i}^n \f{a_j^2-a_i^2}{\Ups_i} \f{\pa}{\pa a_i} \f{1}{a_j^2-a_i^2} \nn 
    &= \f{2a_i}{\Upsilon_i} \sum_{j=1,j\neq i}^n \f{1}{a_j^2-a_i^2}.
\end{align}
Consequently, we have from \eqref{sumUpsi}
\begin{align}
    0 &= \f{\pa}{\pa a_i} \sum_{j=1}^n \f{1}{\Ups_j} \nn 
    &= 2 a_i \sum_{j=1,j\neq i}^n \f{1}{a_j^2-a_i^2}\left( \f{1}{\Ups_i}+\f{1}{\Ups_j}\right), \label{aiderivativeofsuminverseUpsilon}
\end{align}
and from \eqref{sumofa2Upsilonreciprocal},
\begin{align}
    \f{\pa}{\pa a_i} \f{1}{\prod_{j=1}^n a_j^2} &= \f{\pa}{\pa a_i} \sum_{j=1}^n \f{1}{a_j^2 \Ups_j} \nn 
    -\f{2}{a_i \prod_{j=1}^n a_j^2} &= -\f{2}{a_i^3 \Ups_i} + \sum_{j=1}^n \f{1}{a_j^2} \f{\pa}{\pa a_i} \f{1}{\Ups_j} \nn 
    &= -\f{2}{a_i^3 \Ups_i} + \sum_{j=1,j\neq i}^n \f{1}{a_j^2-a_i^2} \left( \f{1}{a_i^2 \Ups_i} + \f{1}{a_j^2 \Ups_j} \right) \nn 
    -\f{1}{a_i^2 \prod_{j=1}^n a_j^2} &= -\f{1}{a_i^4 \Ups_i} + \sum_{j=1,j\neq i}^n \f{1}{a_j^2-a_i^2} \left( \f{1}{a_i^2 \Ups_i} + \f{1}{a_j^2 \Ups_j}\right).
\end{align}

Returning to \eqref{definiteintegraletasection}, the coefficient of $r^0$ (inside the square brackets) vanishes from \eqref{aiderivativeofsuminverseUpsilon} and the coefficient of $r^2$ (inside the square brackets) becomes $-1/(a_i^2 \prod_{j=1}^n a_j^2)$:
\begin{align}
    &\int_{a_2}^{a_1} dy_1 \ldots \int_{a_n}^{a_{n-1}} dy_{n-1} \tilde P \left( \prod_{\alpha=1}^{n-1} (y_\alpha^2-a_i^2)\right) \left( \prod_{\alpha=1}^{n-1} y_\alpha\right) a_i \f{\pa}{\pa r} \f{1}{U_n(r^2+a_i^2)} \nn
    &= -\f{2 C \Ups_i r a_i}{ (r^2+a_i^2) (D-3)!! \prod_{k=1}^n (r^2+a_k^2)}.
\end{align}

\subsection{Summary}

In arbitrary dimension $(D \geq 4)$, applying the correct limits,
\begin{align}
    \int d^{n-2}y \tilde P \left(\prod_{\alpha=1}^{n-1} (y_\alpha^2 - a_i^2) y_\alpha^\ve\right)a_i^\ve \f{\pa}{\pa r} \f{r^{1-\ve}}{U_n(r^2+a_i^2)} &= \f{(-2 a_i)^{2-\ve} C \Ups_i r^\ve \left(\prod_{j=1}^{n-1} a_j\right)^{1-\ve}}{(D-3)!! (r^2+a_i^2) \prod_{j=1}^{n-1+\ve} (r^2+a_j^2)}.
\end{align}
Integrating over the $n-1+\ve$ azimuthal angles brings an additional factor of $(2\pi)^{n-1+\ve}$, so that
\begin{align}
    \int d^{D-2} x \tilde P \left(\prod_{\alpha=1}^{n-1} (y_\alpha^2 - a_i^2) y_\alpha^\ve\right)a_i^\ve \f{\pa}{\pa r} \f{r^{1-\ve}}{U_n(r^2+a_i^2)} &= \f{(2\pi)^{n-1+\ve}(-2 a_i)^{2-\ve} \Ups_i r^\ve C \left(\prod_{j=1}^{n-1} a_j\right)^{1-\ve}}{(D-3)!! (r^2+a_i^2) \prod_{j=1}^{n-1+\ve} (r^2+a_j^2)} \nn 
    &= \f{(-1)^\ve 2 \mc A_{D-2} \Ups_i r^\ve C a_i^{2-\ve} \left( \prod_{j=1}^{n-1} a_j\right)^{1-\ve}}{(r^2+a_i^2) \prod_{j=1}^{n-1+\ve}(r^2+a_j^2)}.
\end{align}

Thus we find from \eqref{sqrtgKKtretayform}
\begin{align}
    &\int d^{D-2} x \sqrt{-\bar g} (\bs i^K_{\eta_i})^{tr} \nn
    &= \int d^{D-2} x\f{a_i}{32\pi \Xi_i} \f{\tilde P (r \prod_{\alpha=1}^{n-1} y_\alpha)^\ve}{C (\prod_{j=1}^n \Xi_j) (\prod_{j=1}^{n-1} a_j)^{1-\ve}}  \f{(r^2+a_i^2)\prod_{\alpha=1}^{n-1}(y_\alpha^2-a_i^2)}{(-a_i^2)^{1-\ve} \Upsilon_i} \f{\prod_{j=1}^n(r^2+a_j^2)}{r^2} \f{\pa}{\pa r} \left( \f{2 m r^{1-\ve}}{U_n (r^2+a_i^2)} \right) \nn 
    &= \f{2 m r^\ve a_i^{1-\ve} (r^2+a_i^2) \prod_{j=1}^n (r^2+a_j^2)}{32 \pi \Xi_i C (\prod_{j=1}^n \Xi_j) (\prod_{j=1}^{n-1} a_j^{1-\ve}) (-a_i^2)^{1-\ve} \Ups_i r^2} \int d^{D-2} x \tilde P \left( \prod_{\alpha=1}^{n-1} (y_\alpha^2-a_i^2) y_\alpha^\ve\right) a_i^\ve \f{\pa}{\pa r} \f{r^{1-\ve}}{U_n(r^2+a_i^2)} \nn 
    &= \f{ (-1)^{1-\ve}2 m (r^2+a_i^2) \prod_{j=1}^{n-1+\ve} (r^2+a_j^2)}{32 \pi \Xi_i C (\prod_{j=1}^{n} \Xi_j) (\prod_{j=1}^{n-1} a_j^{1-\ve}) a_i^{1-\ve} \Ups_i r^{\ve}} \int d^{D-2} \tilde P \left( \prod_{\alpha=1}^{n-1} (y_\alpha^2-a_i^2) y_\alpha^\ve\right) a_i^\ve \f{\pa}{\pa r} \f{r^{1-\ve}}{U_n (r^2+a_i^2)} \nn 
    &= \f{ (-1)^{1-\ve}2 m (r^2+a_i^2) \prod_{j=1}^{n-1+\ve} (r^2+a_j^2)}{32 \pi \Xi_i C (\prod_{j=1}^{n} \Xi_j) (\prod_{j=1}^{n-1} a_j^{1-\ve}) a_i^{1-\ve} \Ups_i r^{\ve}} \f{(-1)^\ve 2 \mc A_{D-2} \Ups_i r^\ve C a_i^{2-\ve} \prod_{j=1}^{n-1} a_j^{1-\ve}}{(r^2+a_i^2) \prod_{j=1}^{n-1+\ve} (r^2+a_j^2)} \nn 
    &= -\f{m a_i \mc A_{D-2}}{4 \pi \Xi_i \prod_{j=1}^n \Xi_j}.
\end{align}

\section{Value of Derivatives Related to \texorpdfstring{$\tilde P$}{tilde P}}
\label{ValueOfVandermondeDeterminant}

\subsection{Monomial Expansion of \texorpdfstring{$\tilde P$}{tilde P}}

It will be useful to expand $\tilde P$ in terms of a sum of monomials. In what follows we treat the quantities $y_\alpha$ as generic and do not require, for the moment, that they obey the implicit ordering $y_1 > y_2 > \ldots > y_{n-1}$. 

Let $c_1, \ldots, c_{n-1}$ be nonnegative integers and let $B_{c_1 \ldots c_{n-1}}$ be an associated coefficient. Since $\tilde P$ consists of terms where $y_\alpha$ appear entirely in the numerator, and only appear in squares, $\tilde P$ will necessarily be a sum of monomials which consist of products of nonnegative even powers of $y_\alpha$. Thus we can write it as
\begin{align}
    \tilde P &= \sum_{c_1=0}^\infty \sum_{c_2 = 0}^\infty \ldots \sum_{c_{n-1} = 0}^\infty B_{c_1 \ldots c_{n-1}} y_1^{2 c_1} y_2^{2 c_2} \ldots y_{n-1}^{2 c_{n-1}}. \label{tildePexpansion}
\end{align}
The $B_{c_1 \ldots c_{n-1}}$ are coefficients for the corresponding monomial. Because the monomial $y_1^{2 c_1} \ldots y_{n-1}^{2 c_{n-1}}$ occurs only once in this expansion, the coefficient $B_{c_1 \ldots c_{n-1}}$ is unique for each set $\{c_1 ,\ldots, c_{n-1}\}$.

Singling out two values, $\alpha$ and $\beta$ (say $\alpha < \beta$), $\tilde P$ can be expanded as
\begin{align}
    \tilde P &= \left(\prod_{\gamma< \alpha} (y_\gamma^2 - y_\alpha^2)(y_\gamma^2-y_\beta^2)\right) \left( \prod_{\alpha < \gamma < \beta} (y_\alpha^2 - y_\gamma^2)(y_\gamma^2-y_\beta^2)\right) \left(\prod_{\gamma > \beta} (y_\alpha^2-y_\gamma^2)(y_\beta^2-y_\gamma^2)\right) (y_\alpha^2-y_\beta^2).
\end{align}
We can rewrite (in the first product) $(y_\gamma^2-y_\alpha^2)(y_\gamma^2-y_\beta^2) = (y_\alpha^2-y_\gamma^2)(y_\beta^2-y_\gamma^2)$, and in the second product, there will be $\beta-\alpha-1$ possible $\gamma$ terms between $\alpha$ and $\beta$, so that $\prod_{\alpha < \gamma < \beta} (y_\alpha^2-y_\gamma^2)(y_\gamma^2-y_\beta^2) = (-1)^{\beta-\alpha-1} \prod_{\alpha < \gamma < \beta} (y_\alpha^2-y_\gamma^2)(y_\beta^2-y_\gamma^2)$. Then we can write $\tilde P$ as
\begin{align}
    \tilde P &= (y_\alpha^2-y_\beta^2) (-1)^{\beta-\alpha-1}\prod_{\gamma \neq \alpha,\beta} (y_\alpha^2-y_\gamma^2)(y_\beta^2-y_\gamma^2).
\end{align}
Treating $\tilde P$ as a function of $\{y_1, \ldots, y_{n-1}\}$, $\tilde P$ is thus antisymmetric under the interchange of the values $y_\alpha$ and $y_\beta$ (keeping the other $y_\gamma$ constant), since $(-1)^{\beta-\alpha-1}\prod_{\gamma\neq \alpha,\beta}(y_\alpha^2-y_\gamma^2)(y_\beta^2-y_\gamma^2)$ is symmetric under the interchange and $y_\alpha^2-y_\beta^2$ is antisymmetric. Taking the example of $\alpha=1,\beta=2$, $\tilde P$ is equal to the negative of the result if $y_1$ and $y_2$ are interchanged in $\tilde P$. Consequently, interchanging $y_1$ and $y_2$ in each term in the expansion \eqref{tildePexpansion},
\begin{align}
    \tilde P &= -\sum_{c_1=0}^\infty \sum_{c_2 = 0}^\infty \ldots \sum_{c_{n-1}=0}^\infty B_{c_1 c_2 c_3 \ldots c_{n-1}} y_2^{2 c_1} y_1^{2 c_2} \ldots y_{n-1}^{2 c_{n-1}}.
\end{align}
We can now relabel $c_1 \leftrightarrow c_2$ (since both are just summation indices), and so find
\begin{align}
    \tilde P &= -\sum_{c_2 = 0}^\infty \sum_{c_1 = 0}^\infty \sum_{c_3 = 0}^\infty \ldots \sum_{c_{n-1}=0}^\infty B_{c_2 c_1 c_3 \ldots c_{n-1}} y_1^{2c_1} y_2^{2 c_2} y_3^{2c_3} \ldots y_{n-1}^{2 c_{n-1}} \nn
    &= -\sum_{c_1 = 0}^\infty \sum_{c_2 = 0}^\infty \ldots \sum_{c_{n-1}=0}^\infty B_{c_2 c_1 c_3 \ldots c_{n-1}}  y_1^{2 c_1} y_2^{2 c_2} y_3^{2 c_3} \ldots y_{n-1}^{2 c_{n-1}}.
\end{align}
Comparing to \eqref{tildePexpansion} and noting that the $B$ coefficients are unique, we conclude
\begin{align}
    B_{c_1 c_2 c_3 \ldots c_{n-1}} = - B_{c_2 c_1 c_3 \ldots c_{n-1}}.
\end{align}
This means that coefficients $B$ are antisymmetric under interchange of the values in the first and second slots. This also tells us that for any set of values $c_3, \ldots, c_{n-1}$, if $c_1 = c_2 = c$, the result will be zero, since $B_{cc c_3 \ldots c_{n-1}} = - B_{cc c_3 \ldots c_{n-1}}$, which can only be the case of $B_{cc c_3 \ldots c_{n-1}}$ vanishes: 
\begin{align}
    B_{c c c_3 \ldots c_{n-1}} = 0.
\end{align}
This followed from $\tilde P$ being antisymmetric under interchange of $y_1$ and $y_2$. Since $\tilde P$ is antisymmetric under any interchange of $y_\alpha, y_\beta$ for $\alpha \neq \beta$, we can conclude through the same line of argument that $B_{c_1 \ldots c_{n-1}}$ is antisymmetric under exchange of any two of its index values and that the coefficient $B_{c_1 \ldots c_{n-1}}$ where any of the $c_i$ are equal to each other must be zero. 

There are $\binom{n-1}{2} = \f{(n-1)(n-2)}{2}$ possible ways to choose two distinct values from a set of size $n-1$. Since $\tilde P$ is the product of terms of the form $y_\alpha^2 - y_\beta^2$ with $1 \leq \alpha < \beta \leq n-1$, each of which has degree 2, the total degree of $\tilde P$ is necessarily $(n-1)(n-2)$ (counting any $y_\alpha$ as having degree 1). This means that the only terms in $\tilde P$ which are nonzero the expansion in terms of the $c_i$ are ones for which the degree, equal to twice the sum over $c_i$, is also $(n-1)(n-2)$. These have
\begin{align}
    2\sum_{i=1}^{n-1} c_i &= (n-1)(n-2).
\end{align}
If we additionally consider that the nonzero $B$ terms all have distinct $c_i$, then we note that the smallest possible power is from one of the $c_i$ having value 0, one having value 1, and so on, up to $n-2$. Since $\sum_{i=0}^{n-2} i = \f{(n-1)(n-2)}{2} = \sum_{i=1}^{n-1} c_i$, we conclude that the only nonzero $B$ terms such that all the $c_i$ are distinct and such that they sum to $(n-1)(n-2)/2$ are the ones where the $c_i$ are a permutation of the integers from 0 to $n-2$. We can now determine the value of the coefficients $B_{c_1 \ldots c_{n-1}}$ by taking considering a representative example, say $B_{n-2, n-3, \ldots, 1, 0}$ and then recalling that $B_{c_1 \ldots c_{n-1}}$ is antisymmetric under permuting any of its indices.

$B_{n-2, n-3, \ldots, 1, 0}$ is the coefficient of the term $y_1^{2(n-2)} y_2^{2(n-3)} \ldots y_{n-2}^{2} y_{n-1}^0$. This can be found by expanding $\tilde P$, in binomial form, directly. $\tilde P$ can be expressed as
\begin{align}
    \tilde P &= (y_1^2 - y_2^2) \ldots (y_1^2 - y_{n-1}^2) (y_2^2 - y_3^2) \ldots (y_2^2-y_{n-1}^2) \ldots (y_{n-2}^2-y_{n-1}^2)  \nn 
    &= \left[\prod_{\beta=2}^{n-1} (y_1^2 - y_\beta^2)\right] \left[ \prod_{\beta=3}^{n-1} (y_2^2 - y_\beta^2)\right] \ldots \left[y_{n-2}^2-y_{n-1}^2\right]. \label{Ptildeexpansion}
\end{align}
There are $(n-1)(n-2)/2$ binomials in the product, and so when expanding this out, before cancellation, there will be $2^{(n-1)(n-2)/2}$ monomials being summed. Each binomial consists of a left-hand term with a positive sign and a right-hand term with a negative sign. So each of the monomials in the expansion will have a sign of $\pm1$ where the sign is positive if an even number of negative signs appear in the product---that is, if that monomial corresponds to an expansion where the right-hand term was selected in the binomials an even number of time. Of all the $2^{(n-1)(n-2)/2}$ monomials in the expansion (before cancellation), only one will be of the form $y_1^{2(n-2)} y_2^{2(n-3)} \ldots y_{n-2}^{2} y_{n-1}^0$, and it will have a coefficient of $+1$. This is because it is the monomial resulting from choosing the left-hand side in every binomial in the product \eqref{Ptildeexpansion}. 

To see this, note that $y_1^2$ appears in the first $n-2$ binomial terms on the left-hand side, and only in those terms, which means that the left-hand term must be chosen for each of those terms in order to have a total power of $y_1$ of $2(n-2)$. Of the remaining terms, $y_2^2$ only appears in the next $n-3$ terms, each on the left-hand side, so the left-hand term must be chosen in each of those terms as well. There remain $n-4$ terms which include $y_3^2$, all on the left-hand side, and so on. Consequently the left-hand side must be taken in every binomial when expanding to get a $y_1^{2(n-2)} y_2^{2(n-3)}\ldots y_{n-2}^2 y_{n-1}^0$ term. 

Consequently, we can write
\begin{align}
    B_{n-2,n-3,\ldots, 0} &= 1.
\end{align}
$B_{c_1 \ldots c_{n-1}}$ is $\pm1$, if $\{c_1, \ldots, c_{n-1}\}$ is a permutation of $\{n-2, n-3 \ldots, 0\}$ and $0$ otherwise, so that we can write
\begin{align}
    B_{c_1 \ldots c_{n-1}} &= (n-1)! \delta_{c_1 \ldots c_{n-1}}^{[n-2,\ldots, 0]}.
\end{align}

We then have,
\begin{align}
    \tilde P &= \sum_{c_i} (n-1)! \de^{[n-2, \ldots, 0]}_{c_1 \ldots c_{n-1}} y_1^{2c_1} \ldots y_{n-1}^{2 c_{n-1}},
\end{align}
where the sum over $c_i$ includes all integer values for all the $c_i$ terms. Another, perhaps more convenient way to express this is to note that all the terms will necessarily have one of the $y_\alpha$ to the power of $2(n-2)$, one to a power of $2(n-3)$, and so on down to 0, with signs $\pm1$. Using the $S_n,\sigma$ permutation notation (see Nomenclature Section), we find 
\begin{align}
    \tilde P &= \sum_{\sigma \in S_{n-1}} (-1)^{|\sigma|} \prod_{\alpha=1}^{n-1} y_{\sigma(\alpha)}^{2(n-1-\alpha)} \nn 
    &= \sum_{\sigma \in S_{n-1}} (-1)^{|\sigma|}y_{\sigma(1)}^{2 (n-2)} \ldots y_{\sigma(n-1)}^{0}. \label{tildePexpandedform}
\end{align}

\subsection{Derivatives Related to \texorpdfstring{$\tilde P$}{tilde P}}

Now we turn to the derivatives and integrals. The convenience of breaking $\tilde P$ down into monomials is that it makes it easier to calculate the multivariable derivatives and integrals. For example, the derivative of a monomial $y_1^{p_1} y_2^{p_2} \ldots y_{n-1}^{p_{n-1}} \equiv \prod_{\al=1}^{n-1} y_\al^{p_\al}$ with respect to each of the $y_\alpha$ is
\begin{align}
    \f{\pa^{n-1}}{\pa y_1 \ldots \pa y_{n-1}}( y_1^{p_1} y_2^{p_2} \ldots y_{n-1}^{p_{n-1}}) &= \f{\pa y_1^{p_1}}{\pa y_1} \f{\pa y_2^{p_2}}{\pa y_2} \ldots \f{\pa y_{n-1}^{p_{n-1}}}{\pa y_{n-1}} \nn 
    &= (p_1 y_1^{p_1-1})(p_2 y_2^{p_2-1}) \ldots (p_{n-1} y_{n-1}^{p_{n-1}-1}) \nn 
    &= \prod_{\alpha=1}^{n-1} p_\alpha y_{\alpha}^{p_\alpha-1}.
\end{align}

It will be convenient to find the derivative of $\tilde P \prod_{\alpha=1}^{n-1} y_\alpha^p$, where $p$ is a constant power, with respect to all the $y_\alpha$. This is
\begin{align}
    \f{\pa^{n-1}}{\pa y_1 \ldots \pa y_{n-1}} \left( \tilde P \prod_{\alpha=1}^{n-1} y_\alpha^p\right) &= \f{\pa^{n-1}}{\pa y_1 \ldots \pa y_{n-1}} \sum_{\sigma \in S_{n-1}}(-1)^{|\sigma|} \prod_{\alpha=1}^{n-1} y_{\sigma(\alpha)}^{2 (n-2-\alpha)+p} \nn 
    &= \sum_{\sigma \in S_{n-1}} (-1)^{|\sigma|} \prod_{\alpha=1}^{n-1} \f{\pa}{\pa y_{\sigma(\alpha)}} y_{\sigma(\alpha)}^{2(n-2-\alpha)+p} \nn 
    &= \sum_{\sigma \in S_{n-1}} (-1)^{|\sigma|} \prod_{\alpha=1}^{n-1} (2(n-2-\alpha)+p) y_{\sigma(\alpha)}^{2(n-2-\alpha)+p-1}.
\end{align}
The powers $2(n-2-\alpha)+p$ will be, in some order, $2(n-2)+p, 2(n-3)+p, \ldots, 2+p, p$. 
\begin{align}
    \f{\pa^{n-1}}{\pa y_1 \ldots \pa y_{n-1}} \left( \tilde P \prod_{\alpha=1}^{n-1} y_\alpha^p\right) &= (2(n-2)+p)(2(n-3)+p) \ldots (2+p) p \sum_{\sigma \in S_{n-1}} (-1)^{|\sigma|} \prod_{\alpha = 1}^{n-1} y_{\sigma(\alpha)}^{2(n-2-\alpha)+p-1} \nn 
    &= (2(n-2)+p)(2(n-3)+p) \ldots (2+p) p \tilde P \prod_{\alpha=1}^{n-1} y_\alpha^{p-1}.
\end{align}
If $p = 0$ this is exactly zero, so that $\pa^{n-1} \tilde P / \pa y_1 \ldots \pa y_{n-1} = 0$. If $p > 0$ the product can be written compactly as
\begin{align}
    \f{\pa^{n-1}}{\pa y_1 \ldots \pa y_{n-1}} \left( \tilde P \prod_{\alpha=1}^{n-1} y_\alpha^p\right) &= \f{(2(n-2)+p)!!}{(p-2)!!} \tilde P \prod_{\alpha=1}^{n-1} y_\alpha^{p-1}. \label{Ptildetimesyalphapderivative}
\end{align}
(Note that $(-1)!!$ and $0!!$ are defined to be 1.) 

\subsection{Derivatives Related to \texorpdfstring{$\tilde P/U_n$}{tilde P/Un}}

We want to calculate derivatives that are related to $\tilde P/U_n$. As in \cite{KrtousKubiznak} we have 
\begin{align}
    \sum_{\mu = 1}^n \f{1}{U_\mu} &= 0.
\end{align}
This means we can rewrite
\begin{align}
    \f{1}{U_n} &= -\sum_{\alpha=1}^{n-1} \f{1}{U_\alpha} \nn 
    &= \sum_{\alpha=1}^{n-1}\f{1}{(r^2+y_\alpha^2) \prod_{\beta=1,\beta \neq \alpha}^{n-1} (y_\beta^2 - y_\alpha^2)}.
\end{align}
Further,
\begin{align}
    \f{\tilde P}{U_n} &= \sum_{\alpha} \f{1}{r^2+y_\alpha^2} \f{\tilde P}{\prod_{\beta=1,\beta \neq \alpha}^{n-1}(y_\beta^2-y_\alpha^2)}.
\end{align}
Define $\tilde P_\alpha$ as
\begin{align}
    \tilde P_\alpha &\equiv \f{\tilde P}{\prod_{\beta =1, \beta \neq \alpha}^{n-1} (y_\beta^2 - y_\alpha^2)} \nn 
    &= \f{\prod_{\gamma < \delta} (y_\gamma^2 - y_\delta^2)}{\prod_{\beta=1,\beta \neq \alpha}^{n-1}(y_\beta^2-y_\alpha^2)}.
\end{align}
All the terms with $y_\alpha^2$ in the numerator will cancel out. For all the terms $y_\gamma^2-y_\alpha^2$ with $\gamma < \alpha$, the sign will match that in the denominator, and for those with $y_\alpha^2 - y_\delta^2$ with $\delta > \alpha$ the signs will not match, leading to an overall factor of $(-1)^{n-1-\alpha}$:
\begin{align}
    \tilde P_\alpha &= (-1)^{n-1-\alpha} \prod_{\substack{\beta < \gamma\\  \beta,\gamma \neq \alpha}} (y_\beta^2 - y_\gamma^2).
\end{align}

We now note that $\tilde P_\alpha$ is, up to a possible sign difference, the equivalent of $\tilde P$ for the set of $n-2$ $y_\beta$ instead of the full set of $n-1$ $y_\beta$ terms (since $y_\alpha$ is suppressed). This means that we can adapt \eqref{Ptildetimesyalphapderivative}. Taking $\alpha = n-1$ as an example, 
\begin{align}
    \f{\pa^{n-2}}{\pa y_1 \ldots \pa y_{n-2}} \left( \tilde P_{n-1} \prod_{\beta = 1}^{n-2}y_\beta^p\right) &= \f{(2(n-3)+p)!!}{(p-2)!!} \tilde P_{n-1} \prod_{\beta = 1}^{n-2} y_\beta^{p-1}.
\end{align}
(The term in the numerator is $(2(n-3)+p)!!$ instead of $(2(n-2)+p)!!$ because there are $n-2$ $y_\beta$ terms instead of $n-1$.) Let $\left[ \f{\pa^{n-2}}{\pa^{n-2}y}\right]_\alpha$ be the operator which takes all the $\pa / \pa y_\beta$ derivatives except for $\pa/\pa y_\alpha$. (For example, $\left[ \f{\pa^{n-2}}{\pa^{n-2}y}\right]_{n-1} = \f{\pa^{n-2}}{\pa y_1 \ldots \pa y_{n-2}}$. Then,
\begin{align}
    \left[ \f{\pa^{n-2}}{\pa^{n-2} y}\right]_\alpha \left( \tilde P_\alpha \prod_{\beta = 1, \beta \neq \alpha}^{n-1} y_\beta^p\right) &= \f{(2(n-3)+p)!!}{(p-2)!!} \tilde P_\alpha \prod_{\beta=1, \beta \neq \alpha}^{n-1} y_\beta^{p-1}. \label{ddnm2}
\end{align}

We can now find derivatives for $\tilde P \prod_{\alpha=1}^{n-1} y_\alpha^p/U_n$. We have,
\begin{align}
    \f{\tilde P \prod_{\alpha=1}^{n-1} y_\alpha^p}{U_n} &= \sum_{\alpha =1 }^{n-1} \f{y_\alpha^p}{r^2+y_\alpha^2} \tilde P_\alpha \prod_{\beta \neq \alpha} y_\beta^p \nn 
    \f{\pa^{n-1}}{\pa y_1 \ldots \pa y_{n-1}} \left( \f{\tilde P \prod_{\alpha=1}^{n-1} y_\alpha^p}{U_n}\right) &= \f{\pa^{n-1}}{\pa y_1 \ldots \pa y_{n-1}} \sum_{\alpha=1}^{n-1} \f{y_\alpha^p}{r^2+y_\alpha^2} \tilde P_\alpha \prod_{\beta \neq \alpha} y_\beta^p \nn 
    &= \sum_{\alpha=1}^{n-1} \f{\pa}{\pa y_\alpha} \f{y_\alpha^p}{r^2+y_\alpha^2} \left[\f{\pa^{n-2}}{\pa^{n-2} y}\right]_\alpha \left( \tilde P_\alpha \prod_{\beta \neq \alpha} y_\beta^p\right). \label{PybyU}
\end{align}
We can write
\begin{align}
    \f{\pa}{\pa y_\alpha} \f{y_\alpha^p}{r^2+y_\alpha^2} &= \f{ ((p-2)y_\alpha^2 + p r^2)y_\alpha^{p-1}}{(r^2+y_\alpha^2)^2}.
\end{align}
Similarly,
\begin{align}
    \f{\pa}{\pa r} \f{ r^q}{r^2+y_\alpha^2} &= \f{(( q-2) r^2 + q y_\alpha^2)r^{q-1}}{(r^2+y_\alpha^2)^2}.
\end{align}
Setting $q=2-p$ we find
\begin{align}
    \f{\pa}{\pa y_\alpha} \f{y_\alpha^p}{r^2+y_\alpha^2} &= \f{((p-2)y_\alpha^2 + p r^2)y_{\alpha}^{p-1}}{(r^2+y_\alpha^2)^2} = -\left(y_\alpha r\right)^{p-1} \f{\pa}{\pa r} \f{r^{2-p}}{r^2+y_\alpha^2}. \label{ddyddr}
\end{align}
Plugging \eqref{ddyddr} and \eqref{ddnm2} into \eqref{PybyU},
\begin{align}
    \f{\pa^{n-1}}{\pa y_1 \ldots \pa y_{n-1}} \left( \f{\tilde P \prod_{\alpha=1}^{n-1} y_\alpha^p}{U_n}\right) &=- \sum_{\alpha=1}^{n-1} (y_\alpha r)^{p-1} \f{\pa}{\pa r} \f{r^{2-p}}{r^2+y_\alpha^2} \f{(2(n-3)+p)!!}{(p-2)!!} \tilde P_\alpha \prod_{\beta \neq \alpha} y_\beta^{p-1} \nn 
    &= - \f{(2(n-3)+p)!!}{(p-2)!!} \left( r \prod_{\alpha=1}^{n-1} y_\alpha\right)^{p-1} \sum_{\alpha=1}^{ n-1} \f{\pa}{\pa r} \f{r^{2-p}}{r^2+y_\alpha^2} \tilde P_\alpha \nn 
    &= - \f{(2(n-3)+p)!!}{(p-2)!!} \left( r \prod_{\alpha=1}^{n-1} y_\alpha\right)^{p-1}  \f{\pa}{\pa r} \sum_{\alpha=1}^{n-1} \f{r^{2-p} \tilde P_\alpha}{r^2+y_\alpha^2} \nn 
    &= - \f{(2(n-3)+p)!!}{(p-2)!!} \left( r \prod_{\alpha=1}^{n-1} y_\alpha\right)^{p-1}  \f{\pa}{\pa r} \left( \f{\tilde P r^{2-p}}{U_n}\right). \label{tildePypbyUnderivatives}
\end{align}
Again this is provided $p > 0$. If $p = 0$, the left-hand side of \eqref{tildePypbyUnderivatives} equals 0.

\chapter{Embedding} \label{embeddingcalcs} 

This follows from Section \ref{embedding}. This appendix is expanded from the version of the appendix appearing in the thesis.

We will explicitly write out the metric for the ``embedding'' spacetime. This is the spacetime of dimension $(D+1)$ which is associated with the $D$-dimensional GKAdS spacetime in a natural way. 

The metric satisfies
\begin{align}
    ds^2 &= d \bar s^2 + \f{2\mu(r)}{U} (k_a dx^a)^2, \label{embeddingsimpleform}
\end{align}
where
\begin{align}
    d\bar s^2 &= - du^2 -dv^2 + \sum_{i = 1}^{n-1+\ve} (dx_i^2 + dy_i^2) + (1-\ve) dz^2
\end{align}
(a flat metric of signature $(D-3)$, or $(- - + \ldots +)$ in $(D+1)$-dimensional spacetime).

$U$ and $k_a$ are exactly the expressions \eqref{Udemocratic} and \eqref{kmudxmuuvxyz} (respectively), where $r$ is defined implicitly by \eqref{requationuvxyz}. 
The condition that $k_a$ is null is simply the same as the definition of $r$, so $k_a$ is automatically null and so the metric is of Kerr--Schild form.

If the constraint \eqref{UVXYZConstraint} is applied to this spacetime we recover GKAdS.

I will now specify to the case $\mu(r) = m$, so that it is the metric associated with the embedding of Kerr--AdS. I will show that this metric is Ricci-flat ($R_{ab} = 0$). To do so I will introduce alternate coordinates which are easier to work with for this purpose. Let 
\begin{align}
    L^2 &\equiv u^2 + v^2 - \sum_{i = 1}^{n-1+\ve} (x_i^2+y_i^2) - (1-\ve) z^2. \label{L2embedding}
\end{align}
Now, assuming that the $a_i$ (including $a_0 = l$) are distinct, we will, essentially, make the substitutions from \eqref{uvxyzjacobi} multiplied by $L/l$. Explicitly,
\begin{align}
    u &= L \left( \f{(1+r^2/l^2) \prod_{\alpha = 1}^{n-1} (1-y_\alpha^2/l^2)}{\prod_{i = 1}^n \Xi_i}\right)^{1/2} \cos(t/l) \nn
    v &= L \left( \f{(1+r^2/l^2)\prod_{\alpha=1}^{n-1}(1-y_\alpha^2/l^2)}{\prod_{i=1}^n \Xi_i}\right)^{1/2} \sin(t/l) \nn 
    x_i &= L\left( \f{ (r^2+a_i^2)\prod_{\alpha = 1}^{n-1} (a_i^2 - y_\alpha^2)}{\Xi_i {\prod}_{j=1, j \neq i}^{n} (a_i^2-a_j^2)}\right)^{1/2} \cos \phi_i \nn 
    y_i &= L \left( \f{(r^2+a_i^2)\prod_{\alpha=1}^{n-1}(a_i^2-y_\alpha^2)}{\Xi_i \prod_{j=1,j\neq i}^n (a_i^2-a_j^2)}\right)^{1/2} \sin \phi_i \nn 
    z &= L \f{r \prod_{\alpha=1}^{n-1} y_\alpha}{l\prod_{i = 1}^{n-1} a_i}, \label{uvxyzjacobiwithL}
\end{align}
or (equivalently)
\begin{align}
    u &= L \nu_0 \cos \phi_0 \nn 
    v &= L \nu_0 \sin \phi_0 \nn 
    x_i &= -i L \nu_i \cos \phi_i \nn 
    y_i &= -i L \nu_i \sin \phi_i \nn 
    z &= -i L \nu_n.\label{pseudoCartesiansnuembedding}
\end{align}

It is now useful to distinguish expressions in the $(D+1)$-dimensional space in coordinates $(L,t,r,y_\alpha,\phi_i)$ and expressions in the $D$-dimensional spacetime in coordinates $(t,r,y_\alpha,\phi_i)$. I will use subscripts $D+1$ and $D$ (respectively) to distinguish them. The coordinates $(t,r,y_\alpha,\phi_i)$ are common to both (so need no subscripts). The constraint leading from $(D+1)$ dimensions to $D$ is $L_D = l$. The pseudo-Cartesian coordinates satisfy $u_{D+1} = (L/l) u_D, v_{D+1} = (L/l) v_D$ and so on.

We have
\begin{align}
    d u_{D+1} = \f{u_D dL + L d u_D}{l},
\end{align}
which means we have
\begin{align}
    (d\bar s^2)_{D+1} &= \left[ -du^2 -dv^2 + \sum_{i=1}^{n-1+\ve} (dx_i^2+dy_i^2) + (1-\ve) dz^2\right]_{D+1} \nn 
    &=  \f{dL^2}{l^2} \left[ -u^2 - v^2 + \sum_{i=1}^{n-1+\ve} (x_i^2+y_i^2) + (1-\ve) z^2\right]_D \nn
    &\qquad + \f{L^2}{l^2} \left[ -du^2 -dv^2 + \sum_{i=1}^{n-1+\ve} (dx_i^2+dy_i^2) + (1-\ve) dz^2\right]_{D} \nn 
    &\qquad +\f{2L dL}{l^2} \left[ -u du - v dv + \sum_{i=1}^{n-1+\ve} (x_i dx_i + y_i dy_i) + (1-\ve) z dz \right]_D.
\end{align}
Applying the constraint for the $D$-dimensional coordinates, $[-u^2-v^2+\sum_{i=1}^{n-1+\ve} (x_i^2+y_i^2) + (1-\ve) z^2]_D = -l^2$ and $\left[ -u du - v dv + \sum_{i=1}^{n-1+\ve} (x_i dx_i + y_i dy_i) + (1-\ve) z dz \right]_D = 0$. Further the AdS background metric in pseudo-Cartesian coordinates is $(d\bar s^2)_D = \left[ -u^2 - v^2 + \sum_{i=1}^{n-1+\ve} (x_i^2+y_i^2) + (1-\ve) z^2\right]_D$.
\begin{align}
    (d\bar s^2)_{D+1} =  -dL^2 + \f{L^2}{l^2} (d \bar s^2)_D.
\end{align}

For $U$ we note that $U$ is proportional to the square of the pseudo-Cartesian coordinates, so that $U_{D+1} = (L^2/l^2) U_D$. 

To find $(k_a d x^a)_{D+1}$ we write
\begin{align}
    (k_a dx^a)_{D+1} &= \left[-\frac{(ru+lv) du + (rv-lu) dv}{r^2+l^2} + \sum_{i = 1}^{n-1+\ve}\frac{(rx_i + a_i y_i) d{x_i}+ (ry_i-a_ix_i) d{y_i}}{r^2+a_i^2} + (1-\ve) \frac{z dz}{r}\right]_{D+1} \nn 
    &= \f{L^2}{l^2} \left[-\frac{(ru+lv) du + (rv-lu) dv}{r^2+l^2} + \sum_{i = 1}^{n-1+\ve}\frac{(rx_i + a_i y_i) d{x_i}+ (ry_i-a_ix_i) d{y_i}}{r^2+a_i^2} + (1-\ve) \frac{z dz}{r}\right]_{D} \nn 
    &\qquad + \f{L dL}{l^2} \left[  -\f{(r u + l v)u + (rv-lu) v}{r^2+l^2} + \sum_{i=1}^{n-1+\ve} \f{(rx_i+a_i y_i) x_i + (r y_i-a_ix_i) y_i}{r^2+a_i^2} + (1-\ve) \f{z^2}{r}\right]_{D}.
\end{align}
The first line is simply $(L^2/l^2) (k_a dx^a)_D$ and the second line expands to 
\begin{align}
    \f{r L dL}{l^2} \left[ -\f{u^2+v^2}{r^2+l^2} + \sum_{i=1}^{n-1+\ve} \f{x_i^2+y_i^2}{r^2+a_i^2} + (1-\ve) \f{z^2}{r^2}\right]_D,
\end{align}
which is equal to zero by the implicit definition of $r$ in the pseudo-Cartesian coordinates. Thus we have
\begin{align}
    (k_a dx^a)_{D+1} &= \f{L^2}{l^2} (k_a dx^a)_D.
\end{align}

Combining with $U$ (and since $m$ is a constant, $m_D = m_{D+1}$), we have
\begin{align}
    (ds^2)_{D+1} &= \left[d\bar s^2 + \f{2m}{U} (k_a dx^a)^2\right]_{D+1} \nn 
    &= -dL^2 + \f{L^2}{l^2} (d s^2)_D, \label{embeddingdp1tod}
\end{align}
where $(ds^2)_D$ is the metric for Kerr--AdS in $(t,r,y_\alpha,\phi_i)$ coordinates. 

In the form \eqref{embeddingdp1tod}, it becomes clear that $(D+1)$-dimensional embedding spacetime is a Brinkmann warped product spacetime of the form introduced by \cite{Brinkmann}, and examined in detail, for instance, by Ortaggio et al.~\cite{OrtaggioBrinkmann}. (I did not mention this connection explicitly in the thesis.) Following the latter, if $ds^2$ is the metric for an $n$-dimensional Einstein space with $R = 2 n \La/(n-2)$ which can be written in the form
\begin{align}
    ds^2 = \f{dz^2}{f(z)} + f(z) d \ti s^2
\end{align}
with
\begin{align}
f(z) &= -\la z^2 + (2  d) z + b, \qquad \la = \f{\La}{(n-1)(n-2)}
\end{align}
and where $d \ti s^2$ is the metric for an $(n-1)$-dimensional space, then the $(n-1)$-dimensional space is also an Einstein space with $\ti R = (n-1)(n-2)(\la b + d^2)$. The particular case $\la = b = 0$ then is
\begin{align}
    ds^2 &= \f{dz^2}{(2d) z} + (2d)z d \ti s^2,
\end{align}
where I've inserted the parentheses around the constant $d$ to distinguish $(d) z$ from $dz$. Taking $y = \int dz/\sqrt{ (d) z} = 2 \sqrt{z/d}$, 
\begin{align}
    ds^2 &= dy^2 + (d^2) y^2 d \ti s^2. \label{dsdtisOrtaggio}
\end{align}
In this case the metric associated with $dy^2$ is Ricci-flat and the one associated with $d \ti s^2$ has $\ti R = d^2(n-1)(n-2)$. This is almost the same as the form \eqref{embeddingdp1tod}, except for the leading negative sign. In a footnote, Ortaggio et al.~also point out that they could include a timelike $z$ by performing a Wick rotation, which is even closer to the form I have here, but they still assume that the full spacetime with $ds^2$ is Lorentzian, so that $d \ti s^2$ would be Euclidean in signature. Indeed, taking $y = i L, d = i/l$ and applying to \eqref{dsdtisOrtaggio} gives
\begin{align}
    ds^2 &= -dL^2 + \f{L^2}{l^2} d \ti s^2,
\end{align}
reproducing \eqref{embeddingdp1tod}. As pointed out by Ortaggio, the results of Brinkmann are essentially signature-independent, and so indeed we expect the $(D+1)$-dimensional spacetime given by \eqref{embeddingdp1tod} to be Ricci-flat. 

Embedding Kerr--de Sitter solutions into Ricci-flat spacetimes according to something like \eqref{dsdtisOrtaggio} has been performed in examining black strings, as in \cite{Wesson} (for the Schwarzschild-de Sitter case), and Ortaggio include an embedding of Kerr--AdS black holes into a negative-curvature spacetime of dimension one higher in \cite{OrtaggioBrinkmann}. Embeddings having to do with Kerr--NUT--AdS were examined in \cite{Krtous16}, which includes several useful proofs, including the relationship between the existence of a closed conformal Killing--Yano tensor in the ``seed'' metric and in the embedding metric. (I will elaborate on this point in a few paragraphs.) I have not encountered the particular embedding I describe here summarized by \eqref{embeddingdp1tod}, as applied to Kerr--AdS specifically embedded into a Ricci--flat spacetime (as opposed to Kerr--dS) in the literature. One possible reason this (to my knowledge) has not been examined in the literature is suggested by  \cite{YangWarpedEmbeddings}, who state, regarding that the case where an Einstein space with negative curvature is embedded into Ricci-flat space of dimension one higher, ``Real embeddings exist, but may involve bulk geometries with multiple timelike directions. Not knowing of any physical interpretations of spacetimes with multiple timelike directions, we regard this class of embeddings as containing some physical illness, however if Wick rotations are taken into account, such embeddings may still yield interesting results.'' 

The point I make here is that this ``embedding spacetime'' should not be a big surprise, and the existence of two timelike dimensions is one reason it may not have been examined before. It is nevertheless interesting that the embedding spacetime has such an attractive form, \eqref{embeddingsimpleform}, in terms of the pseudo-Cartesian coordinates.

While the vanishing of the Ricci tensor for the $(D+1)$-dimensional metric follows directly from the Brinkmann warped product argument, I perform the calculation via the Christoffel symbols, following a similar calculation to many papers such as \cite{Romero,CaldarelliCamps}, in Appendix \ref{RicciTensorCalculation}. It is also noteworthy that $k^a$ is not only null, but also an affinely parametrized geodesic, which I demonstrate in Section \ref{embeddingnullgeodesic}. This is not surprising: it was pointed out by \cite{OrtaggioBrinkmann} that a WAND in the seed spacetime is also a WAND in the full spacetime, and that \cite{Malek} considers Brinkmann warped products of Kerr--Schild spacetimes, pointing out that the results are still generalized Kerr--Schild spacetimes. 

An open question is whether this embedding spacetime, which has a very natural-seeming form and is Ricci flat, can be used to gain some deeper insight about the Kerr--AdS spacetimes. One first step is to consider the equivalent of the PCKY tensor. As stated, Krtou\v{s} et al.~\cite{Krtous16} consider warped products involving the Kerr--NUT--AdS class of solutions. Their general setting involves a seed spacetimes of dimensions $\ti D$ and $\bar D$, with respective metrics $\ti g$ and $\bar g$, along with ``warp factor'' $\ti w$, to produce a full $(\ti D + \bar D)$-dimensional spacetime with metric
\begin{align}
    g = \ti g + \ti w^2 g.
\end{align}
In our case here, $\ti g = -dL^2, \ti w = L/l$ and $g$ is the metric for Kerr--AdS. One of the results in \cite{Krtous16} is that if there exists a closed conformal Killing--Yano tensor $q$-form $\bar{\bs h}$ associated with $\ti g$, then there will be a closed CKY tensor associated with the full metric $g$ given by
\begin{align}
    \bs h = \ti w^{q+1} \ti{\bs \ep} \wedge \bar{\bs h},
\end{align}
where $\ti {\bs \ep}$ is the Levi-Civita tensor associated with $\ti g$. Applied to our case here, using the PCKY tensor 2-form, we expect that the embedding spacetime will have a closed CKY tensor 3-form $\bs h_3$
\begin{align}
    \bs h_3 &= \f{L^3}{l^3} d L \wedge \bs h_2
\end{align}
where $\bs h_2$ is the PCKY tensor for the $D$-dimensional spacetime. This can be written using the form \eqref{hnuform} as
\begin{align}
    \bs h_3 = - \f{L^3}{l} d L \wedge \sum_{i=0}^{n-1+\ve} a_i \nu_i d \nu_i \wedge d \phi_i,
\end{align}
which can be rewritten
\begin{align}
    \bs h_3 &= -\f{L^3}{2l} d L \wedge \sum_{i=0}^{n-1+\ve} a_i d (\nu_i^2) \wedge d \phi_i \nn 
    &= -\f{L^3}{2l} d L \wedge \sum_{i=0}^{n-1+\ve} \f{1}{L^2} a_i d ( L^2 \nu_i^2) \wedge d \phi_i \nn 
    &= -\f{1}{4 l} d (L^2) \wedge \sum_{i=0}^{n-1+\ve} a_i d (L^2 \nu_i^2) \wedge d \phi_i
\end{align}
(since the terms proportional to $dL$ in $d (L^2 \nu_i^2)$ get wiped out by the wedge product with $dL$), which be rewritten using \eqref{pseudoCartesiansnuembedding} as
\begin{align}
    \bs h_3 &= -\f{1}{4 l} d \left(u^2+v^2-\sum_{i=1}^{n-1+\ve} (x_i^2+y_i^2) - (1-\ve) z^2\right) \wedge \left( l du \wedge dv - \sum_{i=1}^{n-1+\ve} a_i dx_i \wedge dy_i\right). \label{h3embedding}
\end{align}
It would be interesting to examine the implications of this. 

\section{Ricci Tensor Calculation} \label{RicciTensorCalculation}

To simplify the notation, let $\xi^i$ represent the coordinates $(t,r,y_\alpha,\phi_i)$, where $i$ (and $j,k,p,q)$ ranges from $1$ to $D$. Let $x^a$ represent $(L,\xi^i) = (L,t,r,y_\alpha,\phi_i)$ where $x^0 = L$; $a$, along with $b,c,d$, ranges from $0$ to $D$. Let $(ds^2)_{D+1} = g_{ab} dx^a dx^b$ and let $\G^a_{bc}$ and $R_{ab}$ represent the associated Christoffel symbols and Ricci tensor. Let $(ds^2)_D = \gamma_{ij} d\xi^i d\xi^j$ and let $c^i_{jk}$ and $r_{ij}$ be the associated Christoffel symbols and Ricci tensor. We then have
\begin{align}
    g_{00} &= -1 \nn 
    g_{0 i} &= 0 \nn 
    g_{ij} &= \f{L^2}{l^2} \g_{ij}.
\end{align}
$\gamma_{ij}$ is a function of $\xi^k$ only (not $L$). The metric inverse is 
\begin{align}
    g^{00} &= -1 \nn 
    g^{0 i} &= 0 \nn 
    g^{i j} &= \f{l^2}{L^2} \g^{ij}
\end{align}
where $\g^{ij}$ is the inverse of $\g_{ij}$. 

The Christoffel symbols satisfy
\begin{align}
    \G^a_{bc} &= \f12 g^{a d} (\pa_b g_{dc} + \pa_c g_{bd} - \pa_d g_{bc}) \nn 
    c^i_{jk} &= \f12 \gamma^{ip}(\pa_j \gamma_{pk} + \pa_k \gamma_{jp} - \pa_p \gamma_{jk}).
\end{align}
Considering individual terms,
\begin{align}
    \G^i_{jk} &= \f12 g^{i d} (\pa_j g_{d k} + \pa_k g_{j d} - \pa_d g_{jk}) \nn 
    &= \f12 g^{i p} (\pa_j g_{pk} + \pa_k g_{jp} - \pa_p g_{jk}) \nn 
    &= \f12 l^2 L^{-2} \gamma^{i p} (\pa_j (L^2 \gamma_{pk}) + \pa_k (l^{-2} L^2 \gamma_{jp}) - \pa_p (l^{-2} L^2 \gamma_{jk})) \nn 
    &= \f12 \gamma^{ip}(\pa_j \gamma_{pk} + \pa_k \gamma_{jp} - \pa_p \gamma_{jk}) \nn 
    &= c^i_{jk}.
\end{align}
\begin{align}
    \G^i_{j 0} &= \f12 g^{i d} (\pa_j g_{d0} + \pa_0 g_{jd} - \pa_d g_{j0}) \nn 
    &= \f12 g^{i k} (\pa_j g_{k0} + \pa_0 g_{jk} - \pa_d g_{j0}) \nn 
    &= \f12 g^{ik} \pa_0 g_{jk} \nn 
    &= \f12 l^2 L^{-2} \gamma^{ik} \pa_L(l^{-2} L^2 \gamma_{jk}) \nn 
    &= L^{-1} \gamma^{ik} \gamma_{jk} \nn 
    &= L^{-1} \de^i_j.
\end{align}

\begin{align}
    \G^i_{00} &= \f12 g^{id} (2 \pa_0 g_{d0} - \pa_d g_{00}) \nn 
    &= g^{ij}(2 \pa_0 g_{j0} - \pa_j g_{00}) \nn 
    &= 0.
\end{align}

\begin{align}
    \G^0_{ij} &= \f12 g^{0d}(\pa_i g_{dj} + \pa_j g_{id} - \pa_d g_{ij}) \nn 
    &= \f12 g^{00} (\pa_i g_{0j} + \pa_j g_{i0} - \pa_0 g_{ij}) \nn 
    &= \f12 \pa_L (l^{-2} L^2 \gamma_{ij}) \nn 
    &= \f{L}{l^2} \gamma_{ij}.
\end{align}

\begin{align}
    \G^0_{0i} &= \f12 g^{0 d}(\pa_0 g_{di} + \pa_i g_{0d} - \pa_d g_{0i}) \nn 
    &= \f12 g^{00} (\pa_0 g_{0i} + \pa_i g_{00} - \pa_0 g_{0i}) \nn 
    &= 0.
\end{align}

\begin{align}
    \G^0_{00} &= \f12 g^{0d}(2 \pa_0 g_{0d} - \pa_d g_{00} ) \nn 
    &= \f12 g^{00} \pa_0 g_{00} \nn 
    &= 0.
\end{align}

Consequently, the only nonzero $\G^a_{b c}$ have at most one of $(a,b,c)$ equal to $0$. 

I will now show how the Ricci tensors $R_{ab}$ and $r_{ij}$ are related. In terms of Christoffel symbols, \cite{Poisson}
\begin{align}
    R_{a b} &= \pa_c \G^c_{ab} - \pa_b \G^c_{ac} + \G^c_{dc} \G^d_{ab} - \G^c_{db} \G^d_{ac} \nn 
    r_{ij} &= \pa_k c^k_{ij} - \pa_j c^k_{ik} + c^k_{pk} c^p_{ij} - c^k_{pj} c^p_{ik}.
\end{align}

We then have,
\begin{align}
    R_{00} &= \pa_c \G^c_{00} - \pa_0 \G^c_{0c} + \G^c_{dc} \G^d_{00} - \G^c_{d 0} \G^d_{0 c} \nn 
    &= 0 - \pa_0 \G^i_{0i} + 0 - \G^i_{j 0} \G^j_{0 i} \nn 
    &= - \pa_L (L^{-1} \de^i_i) - L^{-2} \de^i_j \de^j_i \nn 
    &= L^{-2} (\de^i_i - \de^i_i) \nn 
    &= 0.
\end{align}

\begin{align}
    R_{0 i} &= \pa_c \G^c_{0i} - \pa_i \G^c_{0 c} + \G^c_{d c} \G^d_{0i} - \G^c_{d i} \G^d_{0 c} \nn 
    &= \pa_j \G^j_{0i} - \pa_i \G^j_{0j} + \G^j_{kj} \G^k_{0i} - \G^j_{ki} \G^k_{0 j} \nn 
    &= \pa_j (L^{-1} \de^j_i) - \pa_i (L^{-1} \de^j_j) + c^j_{kj} L^{-1} \de^k_i - c^j_{ki} L^{-1} \de^k_j \nn 
    &= 0 - 0 + c^j_{ij} L^{-1} - c^k_{i k} L^{-1} \nn 
    &= 0.
\end{align}

\begin{align}
    R_{ij} &= \pa_c \G^c_{ij} - \pa_j \G^c_{i c} + \G^c_{dc} \G^d_{ij} - \G^c_{d j} \G^d_{i c} \nn 
    &= (\pa_0 \G^0_{ij} + \pa_k \G^k_{ij}) - \pa_j \G^k_{ik} + (\G^k_{0k} \G^0_{ij} + \G^k_{pk} \G^p_{ij}) - (\G^0_{p j} \G^p_{i 0} + \G^k_{0 j} \G^0_{i k} + \G^k_{p j} \G^p_{i k}) \nn 
    &= \pa_L (l^{-2} L\gamma_{ij}) + \pa_k c^k_{ij} - \pa_j c^k_{ik} + L^{-1} \de^k_k l^{-2} L \gamma_{ij} + c^k_{pk} c^p_{ij} - l^{-2} L\gamma_{pj}L^{-1} \de^p_i - L^{-1} \de^k_j l^{-2} L \gamma_{ik} - c^k_{pj} c^p_{ik} \nn 
    &= (\pa_k c^k_{ij} - \pa_j c^k_{ik} + c^k_{pk} c^p_{ij} - c^k_{pj} c^p_{ik}) + l^{-2} \left(\gamma_{ij} + D \gamma_{ij} - \gamma_{ij} - \gamma_{ij}\right) \nn 
    &= r_{ij} + \f{D-1}{l^2} \gamma_{ij}.
\end{align}

This means if $r_{ij} = -\f{D-1}{l^2} \gamma_{ij}$, which is the case for the case where $(ds^2)_{D}$ is the metric for Kerr--AdS with radius of curvature $l$, then we have $R_{ij} = 0$ and thus $R_{ab} = 0$. 

\section{Direct Calculation of Geodesicity of Null Vector} \label{embeddingnullgeodesic}

As an additional interesting observation, we can show that $k^a$ is geodesic in the flat $(D+1)$-dimensional spacetime. We could do so by appealing to the Christoffel symbols (and showing that if $k^a$ is geodesic in the $D$-dimensional spacetime, it follows that it will be geodesic in the $(D+1)$-dimensional spacetime). Another approach is to proceed directly, using the fact that the background spacetime is flat, so that, in the pseudo-Cartesian coordinates, the covariant derivative with respect to the background spacetime is just the partial derivative. 

To simplify expressions, let $x_0 \equiv i u, y_0 \equiv i v$. Then we have
\begin{align}
    ds^2 &= \sum_{i = 0}^{n-1+\ve} (dx_i^2 + dy_i^2) + (1-\ve) dz^2,
\end{align}
so the metric is the Euclidean flat metric (though $x_0$ and $y_0$ take on imaginary values). Let $a_0 \equiv l$. Then $k^a \pa_a$ takes the simple form
\begin{align}
    k^a \pa_a &= \sum_{i = 0}^{n-1+\ve} \frac{(r x_i + a_i y_i) \pa_{x_i} + (r y_i - a_i x_i) \pa_{y_i}}{r^2+a_i^2} + (1-\ve)\f{z}{r} \pa_z.
\end{align}
The covariant derivatives for the background metric are just partial derivatives and we can also raise and lower using the background metric just by using the identity matrix. 

One advantage of proceeding in the pseudo-Cartesian coordinates over the $(r,y_\alpha)$ coordinates is that it is not necessary that the $a_i$ be distinct when using the pseudo-Cartesian coordinates.

We can confirm $k^a \bar \na_a k^b = 0$ in the following way. Let a prime denote differentiation by $r$ keeping all explicit appearances of the coordinates $x_i, y_i, z$ fixed. Further let $\tilde \pa_{x_i}$ be the partial derivative of an expression with respect to $x_i$, keeping the explicit $r$ terms appearing in an expression fixed. As a result we have
\begin{align}
    \pa_{x_i} f &= f' \pa_{x_i} r + \tilde \pa_{x_i} f,
\end{align}
for any $f$. 

$r$ is defined by \eqref{requationuvxyz}, which can be rewritten as
\begin{align}
    \sum_{i = 0}^{n-1+\ve} \frac{x_i^2+y_i^2}{r^2+a_i^2} + (1-\ve) \f{z^2}{r^2} & =0.
\end{align}
Let $J_n$ for any integer $n$ be defined by
\begin{align}
    J_n &\equiv \sum_{i = 0}^{n-1+\ve} \frac{x_i^2+y_i^2}{(r^2+a_i^2)^n} + (1-\ve) \f{z^2}{r^{2n}}.
\end{align}
It is easy to verify that $J'_n = - 2 n r J_{n+1}$.
Then the $r$ equation is $J_1 = 0$. In Kerr--AdS we also have the constraint $J_0 = -l^2$, but we are dropping this. Taking $d J_1 = 0$ (from $J_1 = 0$) we have
\begin{align}
    0 &= J'_1 dr + \sum_{i = 0}^{n-1+\ve} \f{2 (x_i dx_i + y_i dy_i)}{r^2+a_i^2} + (1-\ve)\f{2 z dz}{r^{2}},
\end{align}
and since $J'_1 = -2 r J_2$, we have
\begin{align}
    \f{\pa r}{\pa x_i} &= \f{x_i}{r (r^2+a_i^2) J_2} \nn 
    \f{\pa r}{\pa y_i} &= \f{y_i}{r (r^2+a_i^2) J_2} \nn 
    \f{\pa r}{\pa z} &= \f{z}{r^3 J_2}.
\end{align}

We also have,
\begin{align}
    k^a \pa_a r &= \sum_{i = 0}^{n-1+\ve} (k^{x_i} \pa_{x_i} r + k^{y_i} \pa_{y_i} r) + \ep k^z \pa_z r \nn 
    &= \f{1}{rJ_2} \left[ \sum_{i = 0}^{n-1+\ve} \left(\f{x_i(r x_i+a_i y_i) + y_i(r y_i - a_i x_i)}{(r^2+a_i^2)^2} \right) + (1-\ve) \f{z^2}{r^3} \right] \nn 
    &= \f{1}{J_2} \left[ \sum_{i=0}^{n-1+\ve} \f{x_i^2+y_i^2}{(r^2+a_i^2)^2} + (1-\ve) \f{z^2}{r^4}\right] \nn 
    &= 1.
\end{align}

We then find
\begin{align}
    k^a \bar \na_a k^b &= k^a \pa_a k^b \nn 
    &= (k^b)' k^a \pa_a r + k^a \tilde \pa_a k^b \nn 
    &= (k^b)' + k^a \tilde \pa_a k^b.
\end{align}
Consider $k^{x_i}$. The only nonzero terms $\tilde \pa_a k^{x_i}$ are from $a = x_i$ and $a = y_i$. So we have
\begin{align}
    k^a \bar \na_a k^{x_i} &= (k^{x_i})' + k^{x_i} \tilde \pa_{x_i} k^{x_i} + k^{y_i} \tilde \pa_{y_i} k^{x_i} \nn
    &= \f{(a_i^2-r^2)x_i - 2 r a_i y_i}{(r^2+a_i^2)^2} + \f{(rx_i+a_iy_i) r + (r y_i - a_i x_i) a_i}{(r^2+a_i^2)^2} \nn 
    &= 0.
\end{align}
The argument for $k^a \bar \na_a k^{y_i}$ is essentially the same. For $k^a \bar \na_a k^z$, similarly $\tilde \pa_a k^z$ is zero for all terms but $a = z$, so that
\begin{align}
    k^a \bar \na_a k^z &= (k^z)' + k^z \tilde \pa_z k^z \nn 
    &= -\f{z}{r^2} + \f{z}{r} \f{1}{r} \nn 
    &= 0.
\end{align}
This confirms that $k^a$ is tangent to an affinely parametrized geodesic.

\chapter{Hodge Dual Identity} \label{HodgeAppendix}

We have
\begin{align}
    ( d * \bs \om)_{a_p \ldots a_D} &= (D-p+1) \pa_{[a_p} (*\bs \om)_{a_{p+1} \ldots a_D]} \nn 
    &= (D-p+1) \pa_{[a_p} ( \bs \om^{b_1 \ldots b_p} \bs \ep_{b_1 \ldots b_p| a_{p+1} \ldots a_D]}),
\end{align}
where the antisymmetrization is only over $\{a_p, a_{p+1}, \ldots, a_D\}$ and not over $\{b_1, \ldots, b_p\}$. 
\begin{align}
    (d * \bs \om)_{a_p \ldots a_D} &= (D-p+1) \pa_{[a_p} ( \bs \om^{b_1 \ldots b_p} \sqrt{-g} e_{b_1 \ldots b_p|a_{p+1} \ldots a_D]}) \nn 
    &= (D-p+1) e_{b_1\ldots b_p [a_{p+1} \ldots a_D} \pa_{a_p]} (\sqrt{-g} \bs \om^{b_1 \ldots b_p}).
\end{align}
As a representative example consider the component $(d*\bs \om)_{p \ldots D}$. 
\begin{align}
    (d*\bs \om)_{p \ldots D} &= (D-p+1) e_{b_1\ldots b_p [p+1 \ldots D} \pa_{p]} (\sqrt{-g} \bs \om^{b_1 \ldots b_p}) \nn 
    &=  e_{b_1 \ldots b_p, p+1 \ldots D} \pa_p (\sqrt{-g} \bs \om^{b_1 \ldots b_p}) - e_{b_1 \ldots b_p, p, p+2 \ldots D} \pa_{p+1} (\sqrt{-g} \bs \om^{b_1\ldots b_p}) \nn 
    &\qquad + e_{b_1 \ldots b_p, p, p+1, p+3 \ldots D} \pa_{p+2} (\sqrt{-g} \bs \om^{b_1 \ldots b_p}) + \ldots \nn 
    &= e_{b_1 \ldots b_p, p+1 \ldots D} \pa_p (\sqrt{-g} \bs \om^{b_1 \ldots b_p}) + e_{b_1 \ldots b_{p-1}, p, b_{p+1}, p+2 \ldots D} \pa_{p+1} (\sqrt{-g} \bs \om^{b_1\ldots b_{p-1} b_{p+1}}) \nn 
    &\qquad + e_{b_1 \ldots b_{p-2}, p, p+1, b_{p+2}, p+3 \ldots D} \pa_{p+2} (\sqrt{-g} \bs \om^{b_1 \ldots b_{p-2}, b_{p+2}}) + \ldots,
\end{align}
permuting indices and relabelling on the last line. The expression $e_{b_1 \ldots b_p, p+1 \ldots D} \pa_p (\sqrt{-g} \bs \om^{b_1 \ldots b_p})$ is equal to $p! \pa_p (\sqrt{-g} \bs \om^{1 \ldots p})$, which follows from the fact that the nonzero terms are ones for which $\{b_1 , \ldots, b_p\}$ are permutations of $\{1, \ldots, p\}$, as well as the antisymmetry of $\bs \om$. Similar expressions exist for other terms, so we have
\begin{align}
    (d*\bs \om)_{p \ldots D} &= p!(\pa_p (\sqrt{-g} \bs \om^{1 \ldots p}) + \pa_{p+1} (\sqrt{-g} \bs \om^{1 \ldots p-1, p+1}) + \pa_{p+2} (\sqrt{-g} \bs \om^{1 \ldots p-1, p+2}) + \ldots ) \nn 
    &= p! \pa_a (\sqrt{-g} \bs \om^{1 \ldots p-1, a}).
\end{align}
($a$ here is an index, whereas $p$ represents a number, the dimensionality of $\bs \om$.) From \cite{Poisson} and using the antisymmetry of $\bs \om$,
\begin{align}
    \pa_a (\sqrt{-g} \bs \om^{1 \ldots p-1, a}) &= p! \sqrt{-g} \na_a \bs \om^{1 \ldots p-1, a},
\end{align}
from which
\begin{align}
    (d * \bs \om)_{p \ldots D} &= p! \sqrt{-g} \na_a \bs \om^{1 \ldots p-1, a} \nn 
    &= p \na_a \bs \om^{b_1 \ldots b_{p-1} a} \bs \ep_{b_1 \ldots b_{p-1} p \ldots D},
\end{align}
where we divided through by $(p-1)!$ on the last line to account for the multiple values of $b_1 \ldots b_p$. Since there is nothing special about choosing the components $p \ldots D$ we conclude
\begin{align}
    (d * \bs \om)_{a_p \ldots a_D} &= p \bs \ep_{b_1 \ldots b_{p-1} a_p \ldots a_D} \na_a \bs \om^{b_1 \ldots b_{p-1} a}, \label{dstaromegaAppendix}
\end{align}
giving \eqref{dstaromega} after $\bs \om^{a b_1 \ldots b_{p-1}} = (-1)^{p-1} \bs \om^{b_1 \ldots b_{p-1} a}$.

\chapter{Appendices Related to Generalized Kerr--AdS Chapter}

In this Appendix chapter, I collect various calculations related to Chapter \ref{GKAdSChapter}.

\section{Generalized Kerr--NUT--AdS Curvature} \label{curvature}

Let $\bs R_{AB}$ be the curvature 2-forms for the GKNAdS spacetimes in the canonical orthonormal basis. These satisfy (with two vectors $u,v$)
\begin{align}
    (\bs R_{A B})_{cd} &= R_{abcd} e_A^a e_B^b \nn 
    \bs R_{A B} (u,v) &= (\bs R_{AB})_{c d} u^c v^d = R_{abcd} e_A^a e_B^b u^c v^d.
\end{align}

These were calculated by \cite{Hamamoto} and can be written compactly in terms of $\beta^2 = \sum_{\mu=1}^n Q_\mu + \ve S$ from \eqref{beta2}. (\cite{Hamamoto} recognize the importance of $\sum_{\mu=1}^n Q_\mu + \ve S$ in simplifying their expressions, but do not explicitly connect it to the vector $\beta$.) I will define (for here and the rest of this section assume $\mu \neq \nu$)
\begin{align}
    p_{\mu \nu} &= \frac{1}{2(x_\mu^2-x_\nu^2)} \left( x_\mu \frac{\partial \beta^2}{\partial x_\mu} - x_\nu \frac{\partial \beta^2}{\partial x_\nu}\right) \nonumber \\
    q_{\mu \nu} &= \frac{1}{2(x_\mu^2-x_\nu^2)}\left( x_\nu \frac{\partial \beta^2}{\partial x_\mu} - x_\mu \frac{\partial \beta^2}{\partial x_\nu}\right). \label{pmnqmn}
\end{align}
(To be clear, $\pa \beta^2 / \pa x_\mu = \pa(\beta^2)/\pa x_\mu$.)

Then the nonvanishing forms are
\begin{align}
    \bs R_{\mu \nu} &= -p_{\mu \nu} e^\mu \wedge e^\nu - q_{\mu \nu} e^{\hat \mu} \wedge e^{\hat \nu} \nonumber \\
    \bs R_{\mu \hat \mu} &= -\frac{1}{2} \frac{\partial^2 \beta^2}{\partial x_\mu^2} \bs \omega^\mu - 2\sum_{\rho \neq \mu} q_{\mu \rho} \bs \omega^\rho \nonumber \\
    \bs R_{\mu \hat \nu} &= -p_{\mu \nu} e^\mu \wedge e^{\hat \nu} - q_{\mu \nu} e^{\nu} \wedge e^{\hat \mu} \nonumber \\
    \bs R_{\hat \mu \hat \nu} &= - q_{\mu \nu} e^\mu \wedge e^\nu - p_{\mu \nu} e^{\hat \mu} \wedge e^{\hat \nu} \nn 
    \bs R_{\mu \hat 0} &= -\f{1}{2x_\mu} \f{\pa \beta^2}{\pa x_\mu} e^\mu \wedge e^{\hat 0} \nn 
    \bs R_{\hat \mu \hat 0} &= -\f{1}{2 x_\mu} \f{\pa \beta^2}{\pa x_\mu} e^{\hat \mu} \wedge e^{\hat 0}. \label{Rmn}
\end{align}
(Recall $\bs \omega^\mu = e^\mu \wedge e^{\hat \mu}$.)

Let $I_\mu$ be the differential operator 
\begin{align}
    I_\mu &= \f12 \f{\pa^2}{\pa x_\mu^2} + \sum_{\rho \neq \mu} \f{1}{x_\rho^2-x_\mu^2} \left( x_\rho \f{\pa}{\pa x_\rho} - x_\mu \f{\pa}{\pa x_\mu}\right).
\end{align}

Let the frame components of the Ricci tensor be given by $\mc R_{A B}$, so that
\begin{align}
    \mc R_{A B} &= R_{a b} e^a_A e^b_B.
\end{align}
Then \cite{Hamamoto} find that in the $e^A$ basis, the Ricci tensor is diagonal:
\begin{align}
    \mc R_{A B} &= 0\textrm{ if } A \neq B \nn
    \mc R_{\mu \mu} = \mc R_{\hat \mu \hat \mu} &= - I_\mu \beta^2 - \varepsilon \f{1}{2 x_\mu} \f{\pa \beta^2}{\pa x_\mu} \nn 
    \mc R_{\hat 0 \hat 0} &= - \sum_{\rho=1}^n \f{1}{x_\rho} \f{\pa \beta^2}{\pa x_\rho}.
\end{align}
Finally the Ricci scalar $R$ is calculated as
\begin{align}
    R &= -\sum_{\mu = 1}^n \f{X''_\mu}{U_\mu} - \varepsilon \sum_{\mu=1}^n \left( \f{2}{x_\mu} \f{X'_\mu}{U_\mu} + \f{2(-1)^n c}{x_\mu^4 U_\mu}\right)
\end{align}
where $X'_\mu = d X_\mu/dx_\mu, X''_\mu = d^2 X_\mu/dx_\mu^2$. (Note that the $c$ here is the negative of the $c$ from \cite{Hamamoto}.)

The Weyl curvature frame components can also be calculated as follows. Let $R_{ABCD}$ be the Riemann frame components, 
\begin{align}
    R_{ABCD} &= \bs R_{AB}(e_C,e_D) \nn 
    &= R_{abcd} e^a_A e^b_B e^c_C e^d_D.
\end{align}
Then similarly let the Weyl frame components be
\begin{align}
    C_{ABCD} &= C_{abcd} e^a_A e^b_B e^c_C e^d_D.
\end{align}
The Weyl decomposition \eqref{Weyldefinition} implies 
\begin{align}
    C_{ABCE} &= R_{ABCE} - \f{2}{D-2} \left( \de_{A[C} \mc R_{E]B} - \de_{B[C}\mc R_{E]A}\right) + \f{2}{(D-1)(D-2)} R \de_{A[C}\de_{E]B}.
\end{align}
Since $\mc R_{AB}$ is diagonal, as is $\de_{AB}$, the only terms where $C_{ABCE}-R_{ABCE} \neq 0$ will be ones for $C=A, E= B$ (or $C=B,E=A$). In these cases we have
\begin{align}
    C_{ABAB} &= R_{ABAB} - \f{\mc R_{AA} + \mc R_{BB}}{D-2} + \f{ R}{(D-1)(D-2)}.
\end{align}

\subsection{The Vanishing of a Traceless Ricci Invariant in Four-Dimensional Generalized Kerr--NUT--AdS} \label{r2zeroinfourdimensions} 

Let $S_{ab}$ be the (well-known) traceless Ricci curvature tensor,
\begin{align}
    S_{ab} = R_{ab} - \f{1}{D} R g_{ab},
\end{align}
with $S^a_a = 0$. As with the Ricci tensor let $\mc S_{AB}$ be the frame components of $S_{ab}$ in the canonical orthonormal frame,
\begin{align}
    \mc S_{AB} &= S_{ab} e^a_A e^b_B \nn 
    &= \mc R_{AB} - \f{1}{D} R \de_{AB}.
\end{align}
Because $\mc R_{AB}$ and $\de_{AB}$ are diagonal and additionally have $\mc R_{\mu \mu} = \mc R_{\hat \mu \hat \mu}$ (as well as $\de_{\mu \mu} = \de_{\hat \mu \hat \mu} = 1$), $\mc S_{\mu \mu} = \mc S_{\hat \mu \hat \mu}$. 

Following Carminati and McLenaghan \cite{CM}, let $r_2$ be the scalar curvature invariant
\begin{align}
    r_2 &= -\f18 S^a_b S^b_c S^c_a.
\end{align}
While Carminati and McLenaghan's definition was for $D = 4$ I will allow the definition to hold in arbitrary dimension $D \geq 4$. I will now show that for the Generalized Kerr--NUT--AdS spacetimes, $r_2 = 0$ if $D = 4$ but not necessarily if $D > 4$.

Expand out $r_2$ in the canonical basis. To begin with I consider arbitrary $D \geq 4$. We then have, writing the summation explicitly,
\begin{align}
    -8 r_2 &= \sum_{A=1}^D \sum_{B = 1}^D \sum_{C=1}^D \mc S^A_B \mc S^B_C \mc S^C_A \nn 
    &= \sum_{A=1}^D \sum_{B=1}^D \sum_{C=1}^D \mc S_{AB} \mc S_{BC} \mc S_{CA},
\end{align}
using the orthonormality of the frame. We can then write $\mc S_{AB} = \mc S_{AA} \de_{AB}$ (no sum), so that $\mc S_{AB} \mc S_{BC} \mc S_{CA} = \mc S_{AA} \mc S_{BB} \mc S_{CC} \de_{AB} \de_{BC} \de_{CA} = (\mc S_{AA})^3 \de_{AB} \de_{BC}$ (no sum), or
\begin{align}
    -8r_2 &= \sum_{A=1}^D \sum_{B=1}^D \sum_{C=1}^D (\mc S_{AA})^3 \de_{AB} \de_{BC} \nn 
    &= \sum_{A=1}^D (\mc S_{AA})^3 \nn 
    &= \sum_{\mu=1}^n \left( (\mc S_{\mu\mu})^3 + (\mc S_{\hat \mu \hat \mu})^3\right) + \ve (\mc S_{\hat 0 \hat 0})^3 \nn 
    &= 2 \sum_{\mu=1}^n (\mc S_{\mu \mu})^3 + \ve (\mc S_{\hat 0 \hat 0})^3.
\end{align}
Here we used $\mc S_{\hat \mu \hat \mu} = \mc S_{\mu \mu}$. 

If $D = 4$, this reduces to
\begin{align}
    -8r_2 &= 2 \left( (\mc S_{11})^3 + (\mc S_{22})^3 \right) \nn 
    &= 2 ( \mc S_{11} + \mc S_{22}) \left( (\mc S_{11})^2 - \mc S_{11} \mc S_{22} + (\mc S_{22})^2\right).
\end{align}
Since $0 = S^a_a = \sum_{A=1}^4 \mc S_{AA} = 2 \sum_{\mu=1}^2 \mc S_{11} = 2 (\mc S_{11} + \mc S_{22})$, we conclude that $r_2=0$. 

$r_2$ is not zero for arbitrary functions $X_\mu$ in higher dimensions. For example, consider $D = 6$. Then we have 
\begin{align}
    -8r_2 &= 2 \left((\mc S_{11})^3 + (\mc S_{22})^3 + (\mc S_{33})^3\right).
\end{align}
Since $\mc S_{11}+\mc S_{22}+\mc S_{33} = 0$, we can rewrite this as
\begin{align}
    -8r_2 &= 2 \left( (\mc S_{11})^3 + (\mc S_{22})^3 + (-\mc S_{11} - \mc S_{22})^3\right) \nn 
    &= -6 \mc S_{11} \mc S_{22} (\mc S_{11} + \mc S_{22}) \nn 
    &= 6 \mc S_{11}\mc S_{22}\mc S_{33}.
\end{align}
Thus it is required that at least one of $\mc S_{11}, \mc S_{22}$ or $\mc S_{33}$ is zero, which restricts the possible solutions.

\section{Volume Form in \texorpdfstring{$(x_\mu, \psi_j)$}{(x-mu, psi-j)} Coordinates} \label{Determinantinxmupsijcoords}

The determinant of the metric in the $(x_\mu, \psi_j)$ form is given in \cite{KubiznakFrolov} as 
\begin{align}
    g &= (-c A^{(n)})^\ve \left[\mathrm{det} A_\mu^{(j)} \right]^2,
\end{align}
interpreting $A_\mu^{(j)}$ as an $n \times n$ matrix. (This considers the Euclidean signature, hence the sign.) 

Here I perform my own calculation showing this (up to a sign), and in so doing calculate the volume element $*1$, in $(x_\mu, \psi_j)$ coordinates. In $x_\mu, \psi_j$ coordinates, the metric has a positive-definite signature---it is only once we use the $r$ coordinate instead of $x_n$ that the metric again has Lorentzian signature. 

I will begin by calculating $\bigwedge_{A=1}^D e^A$, as follows. Note that because $e^{\hat n}$ is imaginary and the other $e^A$ are real, $\bigwedge_{A=1}^D e^A$ is an imaginary form, so that the real form $*1$ will be $\pm i \bigwedge_{A=1}^D e^A$, with the sign depending on choice of orientation.

We now proceed:
\begin{align}
    \bigwedge_{A=1}^D e^A &= \left(\bigwedge_{\mu = 1}^n e^\mu\right) \wedge \left(\bigwedge e^{\hat \mu} \right) \wedge (e^{\hat 0})^\varepsilon \nonumber \\
    &=\left( \bigwedge_{\mu = 1}^n dx_\mu\right) \wedge \left(\bigwedge_{\mu = 1}^n A_\mu^{(j)} d\psi_j\right) \wedge  \left(\sqrt{S} \sum_{j= 0}^n A^{(j)} d \psi_j\right)^{\varepsilon} \nonumber \\
    &= \left(\bigwedge_{\mu = 1}^n d x_\mu\right)\left( \wedge \bigwedge_{\mu = 1}^n A_\mu^{(j)} d \psi_j\right) \wedge \left(\sqrt{S} A^{(n)} d \psi_n\right)^{\varepsilon}
\end{align}
The last simplification is because the only appearance of $d \psi_n$ is in the $e^{\hat 0}$ term. 
$\sqrt{S} A^{(n)} = \sqrt{-c A^{(n)}} = \sqrt{-c} \prod_{\mu=1}^n x_\mu$. (This is real, because $x_n$ is imaginary.)

For 
\begin{align}
    \bigwedge_{\mu = 1}^n A_\mu^{(j)} d \psi_j,
\end{align}
the antisymmetry of the wedge product means we have to calculate the determinant of $A_\mu^{(j)}$ treated as a matrix. Its determinant can be calculated by noting that it is related to the Vandermonde matrix (see, e.g.,~\cite{Turner, ProofWiki}). The simplest argument (adapted from the above references) is as follows. Let $x$ be the order of any term $x_\mu$ (in the sense of, $x_\mu^2$ is of order $x^2$). If the $\mu$ label rows and the $(j)$ label columns, the $j$th column is of order $x^{2j}$. Thus the total order of the determinant must be $x^{\left(\sum_{j = 0}^{n-1} (2j)\right)} = x^{(n-1)n}$. The determinant must be a polynomial in the $x_\mu^2$ since all the entries in the matrix $A_\mu^{(j)}$ are polynomials in $x_\mu^2$. We also require that exchanging the order of entries $\mu$ and $\nu$ will reverse the overall sign. This means that $x_\mu^2 - x_\nu^2$ must divide the overall determinant. (It cannot simply be $x_\mu - x_\nu$ because the $x_\mu$ and $x_\nu$ terms only appear quadratically in $A_\mu^{(j)}$, never linearly.) This implies that the expression must be
\begin{align}
    \textrm{det} A_\mu^{(j)} \propto P,
\end{align}
where $P$ is given by \eqref{Pdefinition}.

Since the product is itself of order $x^{n(n-1)}$ in $x$ (where the exponent of $x$ is $2 \binom{n}{2}$), and the expression is a polynomial, the ``constant of proportionality'' must be independent of $x_\mu$, and thus be a constant. Let us call this constant $\mc C$ (for this section only). So we have
\begin{align}
    \textrm{det} A_\mu^{(j)} &= \mc C P. \label{detprod}
\end{align}
To determine the value of $\mc C$, we need only find one term in the expansion and check its coefficient. We know that the entry associated with 
\begin{align}
    \prod_{\mu=1}^n A_\mu^{(\mu-1)} = A_1^{(0)} A_2^{(1)} \ldots A_n^{(n-1)}
\end{align}
should have coefficient $+1$. This is equal to
\begin{align}
    \prod_{\mu = 1}^n x_\mu^{2(n-1-\mu)} = x_1^{2(n-1)} \ldots x_\mu^{2(n-1-\mu)} \ldots x_n^0.
\end{align}

We now show that this term only appears in the final result due to the term $A_1^{(0)} \ldots A_n^{(n-1)}$. For $x_1$ to have exponent $2(n-1)$, it must appear once in all terms $A_\mu^{(j)}$ except for $j = 0$, so the $j = 0$ term must have $\mu = 1$. Then we require that $x_1^2$ appears in every other term. That means that the contribution to this sum from $A_\mu^{(1)}$ must only be $x_1^2$. Then since $x_2$ has exponent $2(n-2)$, it must appear in every term except two---the $j = 0$ term, and, since the only term allowed from the $j = 1$ term is $x_1$, also the $j = 1$ term. This means that $\mu = 2$ for $j = 2$. The same argument follows through, and so we conclude that the only product of $A_\mu^{(j)}$ which includes the term $\prod_{\mu = 1}^n x_\mu^{2(n-1-\mu)}$ is $A_1^{(0)} \ldots A_n^{(n-1)}$. Thus we require that the term $\prod_{\mu = 1}^n x_\mu^{2(n-1-\mu)}$ have coefficient $+1$. But this is just the term that results in the expansion of \eqref{detprod} from choosing the left entry in every difference in the product, and so will have coefficient $\mc C$. Thus $\mc C = 1$. 

Then we have
\begin{align}
    \bigwedge_{A=1}^D e^A &= P \left(\sqrt{-c} \prod_{\nu=1}^n x_\nu\right)^\ve \left(\bigwedge_{\mu = 1}^n d x_\mu \right)\wedge \bigwedge_{j = 0}^{n-1+\ve} d \psi_j.
\end{align}
We then have
\begin{align}
    *1 &= \pm i P \left(\sqrt{-c} \prod_{\nu=1}^n x_\nu\right)^\ve\left( \bigwedge_{\mu=1}^n d x_\mu \right) \wedge \bigwedge_{j=0}^{n-1+\ve} d \psi_j \nn 
    &= \pm P \left( \sqrt{c} r \prod_{\alpha=1}^{n-1} y_\alpha\right)^\ve \left(\bigwedge_{\beta=1}^{n-1} d y_\beta\right) \wedge dr \wedge \bigwedge_{j=0}^{n-1+\ve} d \psi_j.
\end{align}
Comparing to $*1 = \sqrt{-g} dx^1 \wedge \ldots \wedge dx^D$, we find in $(y_\alpha, r, \psi_j)$ coordinates $\sqrt{-g} = P (\sqrt c r \prod_{\alpha=1}^{n-1} y_\alpha)^\ve$ (up to orientation).

\section{Calculation of Covariant Derivative of Principal Vector} \label{dbetaSection} 

In this section I calculate $d \beta^\flat $. 
\begin{align}
    d \beta^\flat &= \sum_{\mu = 1}^n \left( d \sqrt{Q_\mu} \wedge e^{\hat \mu} + \sqrt{Q_\mu} d e^{\hat \mu}\right) + \varepsilon \left( d \sqrt{S} \wedge e^{\hat 0} + \sqrt{S} d e^{\hat 0}\right)
\end{align}

Using the definition \eqref{QmuUmuSdefinition} for $U_\mu$, we have for $\nu \neq \mu$
\begin{align}
    \frac{\partial U_\mu}{\partial x_\nu} &=- \frac{2 x_\nu}{x_\mu^2 - x_\nu^2} U_\mu.
\end{align}
We also have, using the connection forms from \cite{Hamamoto},
\begin{align}
    d e^{\hat \mu} &= \f{\pa \sqrt{Q_\mu}}{\pa x_\mu} e^\mu \wedge e^{\hat \mu} + \sum_{\nu \neq \mu} \left( \f{2 x_\nu \sqrt{Q_\mu}}{x_\mu^2-x_\nu^2} e^\nu \wedge e^{\hat \nu} - \f{x_\nu \sqrt{Q_\nu}}{x_\mu^2-x_\nu^2} e^\nu \wedge e^{\hat \mu}\right).
\end{align}

If $\varepsilon = 0$, we get
\begin{align}    
    d \beta^\flat &= \sum_{\mu = 1}^n \left( \frac{\partial \sqrt{Q_\mu}}{\partial x_\mu} d x_\mu \wedge e^{\hat \mu} - \frac{1}{2} \sum_{\nu \neq \mu} \frac{\sqrt{Q_\mu}}{U_\mu} \frac{\partial U_\mu}{\partial x_\nu}   d x_\nu \wedge e^{\hat \mu} + \sqrt{Q_\mu} d e^{\hat \mu}\right) \nonumber \\
    &=  \sum_{\mu = 1}^n \sqrt{Q_\mu} \left( \frac{\partial \sqrt{Q_\mu}}{\partial x_\mu} e^\mu \wedge e^{\hat \mu} + \sum_{\nu \neq \mu}  \frac{\sqrt{Q_\nu} x_\nu}{x_\mu^2-x_\nu^2} e^\nu \wedge e^{\hat \mu} + d e^{\hat \mu} \right) \nonumber \\
    &= \sum_{\mu = 1}^n \sqrt{Q_\mu} \left( 2 \frac{\partial \sqrt{Q_\mu}}{\partial x_\mu} e^\mu \wedge e^{\hat \mu} + \sum_{\nu \neq \mu} \frac{2 x_\nu \sqrt{Q_\mu}}{x_\mu^2-x_\nu^2} e^\nu \wedge e^{\hat \nu}\right) \nonumber \\
    &= \sum_{\mu = 1}^n \left( \frac{\partial Q_\mu}{\partial x_\mu} - \sum_{\nu \neq \mu} \frac{2 x_\mu Q_\nu}{x_\mu^2-x_\nu^2}\right) e^{\mu} \wedge e^{\hat \mu} \nonumber \\
    &= \sum_{\mu = 1}^n \frac{\partial Q_T}{\partial x_\mu} e^\mu \wedge e^{\hat \mu}. \label{dbeta}
\end{align}
relabelling somewhat on the final two lines and using $\pa Q_T/\pa x_\mu = \pa Q_\mu/\pa x_\mu + \sum_{\nu \neq \mu} \pa Q_\nu/\pa x_\mu = \pa Q_\mu/\pa x_\mu - \sum_{\nu \neq \mu} Q_\nu U_\nu^{-1} \pa U_\nu/\pa x_\mu$.

The $\varepsilon = 1$ case also has the terms involving $\sqrt{S}$ and $e^{\hat 0}$ to contend with. Also using the connection forms from \cite{Hamamoto},
\begin{align}
    d e^{\hat 0} &= \sum_{\mu=1}^n \left( -\f{2 \sqrt{S}}{x_\mu} e^\mu \wedge e^{\hat \mu} + \f{\sqrt{ Q_\mu}}{x_\mu} e^\mu \wedge e^{\hat 0}\right).
\end{align}
We then have,
\begin{align}
    d \sqrt{S} \wedge e^{\hat 0} + \sqrt{S} d e^{\hat 0} &= -\frac{1}{2} \sqrt S d \ln S \wedge e^{\hat 0} + \sqrt{S} d e^{\hat 0} \nonumber \\
    &= -\frac{1}{2} \sqrt S \sum_{\mu = 1}^n \frac{2 d x_\mu}{x_\mu} \wedge e^{\hat 0} + \sqrt S d e^{\hat 0} \nonumber \\
    &= \sqrt S \left( -\sum_{\mu = 1}^n \frac{\sqrt{Q_\mu}}{x_\mu} e^\mu \wedge e^{\hat 0} + d e^{\hat 0}\right) \nonumber \\
    &= \sqrt{S} \sum_{\mu = 1}^n \left( -\frac{\sqrt{Q_\mu}}{x_\mu} e^\mu \wedge e^{\hat 0} - 2 \frac{\sqrt{S}}{x_\mu} e^\mu \wedge e^{\hat \mu} + \frac{\sqrt{Q_\mu}}{x_\mu} e^\mu \wedge e^{\hat 0}\right) \nonumber \\
    &=- \sum_{\mu = 1}^n \frac{2 S}{x_\mu} e^\mu \wedge e^{\hat \mu} \nonumber \\
    &= \sum_{\mu = 1}^n \frac{\partial S}{\partial x_\mu} e^\mu \wedge e^{\hat \mu}.
\end{align}

Using $\beta^2 = Q_T + \ve S$, \eqref{dbetaflat} follows.

\section{Explicit Derivation of Kerr--Schild Null Vector--Principal Vector Relation} \label{kdbetaSection}

Here I will show directly that $k^a \bar \na_a \beta^b \propto k^b$. 

Use the Kerr--Schild coordinates for which the background metric is asymptotically static. If we use $(t,r,y_\alpha,\phi_i)$, the background (pure AdS) metric is diagonal (and even if we use $\mu_i$ instead of $y_\alpha$, the $t,r,\phi_i$ part of the metric is still diagonal). 

Use $\phi_0 \equiv t/a_0$ where $a_0 \equiv l$. We have that $\bar g_{\phi_i \phi_i} \propto (r^2+a_i^2)$ and otherwise has no $r$ dependence, so
\begin{align}
    \f{\pa}{\pa r} \bar g_{\phi_i\phi_i} &= \f{2 r}{r^2+a_i^2} \bar g_{\phi_i\phi_i}.
\end{align}
We also have $\beta^{\phi_i} = a_i/l^2$, that $k^{\phi_i} = -a_i/(r^2+a_i^2)$, that $k^r = 1$, and that all other contravariant components of $\beta$ and $k$ are zero, recalling that $t$ is treated as one of the $\phi$ coordinates. (This holds taking either the $\mu_i$ or $y_\alpha$ coordinates.) 

Let $\bar \beta_a = \bar g_{a b} \beta^b$. (Since $k$ can be raised and lowered using either metric, there is no ambiguity in writing $k_a$.)
\begin{align}
    k^a \bar \na_a \bar \beta_b &= k^a \pa_{[a}\bar \beta_{b]} \nn 
    &= \f12 k^a (\pa_a \bar \beta_b - \pa_b \bar \beta_a).
\end{align}
The first line follows from $\beta^a$ being a Killing vector. Because both the metric and $\beta^a$ do not depend explicitly on the $\phi_i$, the only value of the index on the derivative which will be nonvanishing is $r$.
\begin{align}
    k^a \bar \na_a \bar \beta_b &= \f{1}{2} (k^r \pa_r \bar \beta_b - k^a \de^r_b \pa_r \bar \beta_a).
\end{align}
Explicitly, and using $k^r = 1$,
\begin{align}
    k^a \bar \na_a \bar \beta_{\phi_i} &= \f12 \pa_r \bar \beta_{\phi_i} \nn 
    &= \f{a_i}{2l^2} \f{\pa}{\pa r} \bar g_{\phi_i\phi_i} \nn 
    &= \f{r a_i}{l^2(r^2+a_i^2)} \bar g_{\phi_i\phi_i} \nn 
    &= -\f{r}{l^2} k^{\phi_i} \bar g_{\phi_i\phi_i} \nn 
    &= -\f{r}{l^2} \bar k_{\phi_i}
\end{align}
\begin{align}
    k^a \bar \na_a \bar \beta_r &= -\f12 k^a \pa_r\bar\beta_a \nn 
    &= -\f12 \sum_{i} k^{\phi_i} \pa_r \bar \beta_{\phi_i} \nn 
    &= -\f12 \sum_i k^{\phi_i} \pa_r \bar g_{\phi_i\phi_i} \f{a_i}{l^2} \nn 
    &= -\f{r}{l^2} \sum_{i} \f{a_i}{r^2+a_i^2} \bar g_{\phi_i\phi_i} k^{\phi_i} \nn 
    &= \f{r}{l^2} \sum_i \bar k_{\phi_i} k^{\phi_i} \nn 
    &= -\f{r}{l^2} \bar k_r k^r \nn 
    &= -\f{r}{l^2} \bar k_r
\end{align}
\begin{align}
    k^a \bar \na_a \bar \beta_{y_\alpha} &= \f12 \pa_r \bar \beta_{y_\alpha} \nn 
    &= 0.
\end{align}
Thus we conclude 
\begin{align}
    k^a \bar \na_a \bar \beta_b &= -\f{r}{l^2} \bar k_b. 
\end{align}

\section{\texorpdfstring{$U,U_n$}{U, Un} Relation} \label{UUnrelation}

Define $\tht_i$ to be
\begin{align}
    \tht_i &\equiv -(r^2+a_i^2) \prod_{j=1,j\neq i}^n(a_j^2-a_i^2) \nn 
    &= -(r^2+a_i^2) \Upsilon_i (-a_i^2)^{1-\ve}.
\end{align} 
Then, using \eqref{Jacobi},
\begin{align}
    \f{\mu_i^2}{r^2+a_i^2} &= -\f{\prod_{\alpha=1}^{n-1} (y_\alpha^2 - a_i^2)}{\tht_i} \nn 
    &= -\f{1}{\tht_i} \sum_{j=0}^{n-1} A_n^{(j)} (-a_i^2)^{n-1-j}.
\end{align}
If $a_n = 0$ (in even dimension, say), we interpret $(-a_n^2)^0$ as being 1 in this expression. 

The sum, then, is
\begin{align}
    \sum_{i=1}^{n} \f{\mu_i^2}{r^2+a_i^2} &= -\sum_{j=0}^{n-1} A_n^{(j)} \sum_{i=1}^n \f{(-a_i^2)^{n-1-j}}{\tht_i}.
\end{align}
We further have,
\begin{align}
    \sum_{i=1}^n \f{(-a_i^2)^{k}}{\tht_i} + \f{r^{2k}}{\prod_{j=1}^n (r^2+a_i^2)} &= \de^{n}_0,
\end{align}
where $0 \leq k \leq n$. This follows from adapting $\sum_{\mu=1}^n (-x_\mu^2)^k/U_\mu = \de^{n-1}_0$ \cite{KrtousKubiznak} to the set $\{a_1^2, \ldots, a_n^2, -r^2\}$ instead of $\{x_1^2, \ldots, x_n^2\}$. This result allows us to write
\begin{align}
    \sum_{i=1}^n \f{\mu_i^2}{r^2+a_i^2} &= \sum_{j=0}^{n-1} A_n^{(j)} \f{r^{2(n-1-j)}}{\prod_{j=1}^n (r^2+a_j^2)} \nn 
    &= \f{\prod_{\alpha=1}^n (r^2+y_\alpha^2)}{\prod_{j=1}^n (r^2+a_j^2)} \nn 
    &= \f{U_n}{\prod_{j=1}^n (r^2+a_j^2)}.
\end{align}
\eqref{UUn} follows.

\section{Calculations Associated with Jacobi Transformations} \label{CalculationsAssociatedWithJacobiTransformations}

In this Appendix section, I calculate explicitly some of the consequences of the transformation from $\mu_i$ to $x_\mu$ coordinates. These calculations are verifying the results found by CLP \cite{Chen}. 

It is convenient to define, in even dimensions, $C^{(l)}$ by
\begin{align}
    \prod_{j=1}^{n-1} (1 + \alpha a_j^2) &\equiv \sum_{l=0}^{n-1} \alpha^l C^{(l)} \nn 
    C^{(l)} &= \sum_{1 \leq j_1 < \ldots < j_l \leq n-1, j_k \neq i} \prod_{k=1}^l a^2_{j_k}.
\end{align}

Another relation given in \cite{KrtousKubiznak} is
\begin{align}
    \sum_{\mu = 1}^n \f{A_\mu^{(l)}}{x_\mu^2 U_\mu} &= \f{A^{(l)}}{A^{(n)}},
\end{align}
so that, again in even dimensions,
\begin{align}
    \sum_{i=1}^{n-1} \f{C_i^{(l)}}{a_i^2 \Ups_i} &= \f{C^{(l)}}{C^{(n-1)}}. \label{sumoverCa2Ups}
\end{align}

Using these and \eqref{Jacobi}, $\mu_i^2$ can be written as
\begin{align}
    \mu_i^2 &= \f{(-1)^{n+1} \G_i}{(r^2+a_i^2)(-a_i^2)^{1-\ve} \Ups_i},
\end{align}
for $i \leq n-1+\ve$. If $\ve = 0$ then $\mu_n$ is special (since $a_n = 0$) and can be compactly written
\begin{align}
    \mu_n^2 &= \f{\prod_{\alpha=1}^{n-1} y_\alpha^2}{\prod_{j = 1}^{n-1} a_j^2} = \f{A_n^{(n-1)}}{C^{(n-1)}}.
\end{align}

We also have, for $i \leq n-1+\ve$,
\begin{align}
    \mu_i^2 &= \f{(-1)^{n+1} \prod_{\alpha=1}^{n-1} (a_i^2 - y_\alpha^2)}{(-a_i^2)^{1-\ve} \Ups_i} \nn 
    \prod_{\alpha=1}^{n-1} (a_i^2 - y_\alpha^2) &= a_i^{2(n-1)} \prod_{ \alpha = 1}^{n-1} (1 - a_i^{-2} y_\alpha^2) \nn 
    &= a_i^{2(n-1)} \sum_{j = 0}^{n-1} (-a_i^2)^{-j} A_n^{(j)} \nn 
    &= (-1)^{n-1} \sum_{j=0}^{n-1} (-a_i^2)^{n-1-j} A_n^{(j)} \nn 
    \mu_i^2 &= \sum_{j = 0}^{n-1} A_n^{(j)} D^i_{(j)}.
\end{align}

Split the sum over $\mu_i^2$ into odd and even dimensional cases. In odd dimension,
\begin{align}
    \sum_{i=1}^n \mu_i^2 &= \sum_{j=0}^{n-1} \sum_{i=1}^n A_n^{(j)} D^i_{(j)} \nn 
    &= \sum_{j=0}^{n-1} A_n^{(j)} \sum_{i=1}^n D^i_{(j)} C_i^{(0)} \nn 
    &= \sum_{j=0}^{n-1} A_n^{(j)} \de_j^0 \nn 
    &= A_n^{(0)} \nn 
    &= 1.
\end{align}
In even dimension,
\begin{align}
    \sum_{i=1}^n \mu_i^2 &= \sum_{i=1}^{n-1} \mu_i^2 + \mu_n^2 \nn 
    &= \sum_{j=0}^{n-1} \sum_{i=1}^{n-1} A_n^{(j)} D^i_{(j)} + \f{A_n^{(n-1)}}{C^{(n)}} \nn 
    &= \sum_{j=0}^{n-2} A_n^{(j)} \sum_{i=1}^{n-1} D^i_{(j)} C_i^{(0)} + A_n^{(n-1)} \sum_{i=1}^{n-1} D^i_{(n-1)} + \f{A_n^{(n-1)}}{C^{(n)}} \nn 
    &= \sum_{j=0}^{n-2} A_n^{(j)} \de^0_{j} + A_n^{(n-1)} \left[\sum_{i=1}^{n-1}\f{1}{-a_i^2 \Ups_i} + \f{1}{C^{(n)}}\right] \nn 
    &= A_n^{(0)} + 0 \nn 
    &= 1.
\end{align}

\begin{align}
    \G_i &= a_i^{2n} \prod_{\nu = 1}^n \left(1 - a_i^{-2} x_\nu^2\right) \nn 
    &= a_i^{2n} \sum_{j = 0}^{n} (-a_i^{2})^{-j} A^{(j)} \nn 
    &= (-1)^n \sum_{j=0}^{n} (-a_i^2)^{n-j} A^{(j)}
\end{align}

Write $W$ as follows. Using $l = a_0$,
\begin{align}
    W &= \sum_{i=1}^n \f{\mu_i^2}{\Xi_i} \nn 
    &= l^2 \sum_{i=1}^n \f{\mu_i^2}{l^2-a_i^2} \nn 
    &= l^2 \sum_{i=1}^n \f{ \prod_{\alpha=1}^{n-1} (y_\alpha^2 - a_i^2)}{\prod_{j=0,j\neq i}^n (a_j^2-a_i^2)} \nn 
    &= l^2 \sum_{k=0}^{n-1} A_n^{(k)}\sum_{i=1}^n   \f{(-a_i^2)^{n-1-k}}{\prod_{j=0,j\neq i}^n (a_j^2-a_i^2)} \nn 
    &= l^2 \sum_{k=0}^{n-1} A_n^{(k)} \left[\sum_{i=0}^n  \f{(-a_i^2)^{n-1-k}}{\prod_{j=0,j\neq i}^n (a_j^2-a_i^2)} - \f{(-l^2)^{n-1-k}}{\prod_{j=1}^n (a_i^2-l^2)}\right].
\end{align}
Because $\sum_{\mu = 1}^n B^\mu_{(j)} = 0$ for $1 \leq j \leq n-1$, we can conclude that, replacing the set of $n$ quantities $x_\mu$ with the set of $n+1$ quantities $a_i$ ($0 \leq i \leq n$), 
\begin{align}
    \sum_{i=0}^n  \f{(-a_i^2)^{n-p}}{\prod_{j=0,j\neq i}^n (a_j^2-a_i^2)} = 0
\end{align}
for $1 \leq p \leq n$. Plugging in $p = k+1$ we find this is the case for all values of $k$, and we are left with
\begin{align}
    W &= l^2 \sum_{k=0}^{n-1} A_n^{(k)} (-1)  \f{(-l^2)^{n-1-k}}{\prod_{j=1}^n (a_i^2-l^2)} \nn 
    &= \sum_{k=0}^{n-1} A_n^{(k)} \f{(-l^2)^{-k}}{\prod_{j=1}^n \Xi_j} \nn 
    &= \f{\prod_{\alpha=1}^{n-1} (1-y_\alpha^2/l^2)}{\prod_{j=1}^n \Xi_j}.
\end{align}

$U$ can be related to $U_n$ according to \eqref{UUn}, or $U = U_n/r^{1-\ve}$.

This allows us to compute the time--azimuthal parts of the metric. We have, from the ABL form of the metric \eqref{ABL}, 
\begin{align}
    ds^2_{time-azimuthal} &= -\f{\prod_{\mu = 1}^n (1-x_\mu^2/l^2)}{\prod_{j=1}^n \Xi_j} d \tau^2 + \sum_{i=1}^{n-1+\ve} \f{(-1)^{n+1} \G_i}{(-a_i^2)^{1-\ve} \Ups_i \Xi_i} d \hat \vp_i^2 \nn 
    &\qquad + \f{2m r^{1-\ve}}{U_n} \left( \f{\prod_{\alpha=1}^{n-1} (1-y_\alpha^2/l^2) }{\prod_{j=1}^n \Xi_j} d \tau - \sum_{i=1}^{n-1+\ve} \f{a_i(-1)^{n+1} \G_i d \hat \vp_i}{(r^2+a_i^2)(-a_i^2)^{1-\ve} \Ups_i \Xi_i}\right)^2.
\end{align}

\section{Calculations Related to Canonical Frame Components}

\subsection{Canonical One-Forms in KS Coordinates} \label{canonicalbasisKScoordssection}

Here I provide the derivation of \eqref{canonicalbasisKScoords}.

Use \eqref{dphidphihat} (which also applies for $i = 0$ if we use $\phi_0 = t/l, \tau = \hat \vp_0/l$ and $l = a_0$) and plug them into \eqref{ehatmuhatphi}. Let $d Z = 2m dr/(V-2m)$ (for more compact writing). We have,
\begin{align}
    e^{\hat \mu} &= (-1)^{n-1+\ve} \sqrt{Q_\mu}\sum_{i=0}^{n-1+\ve} \f{l^2 a_i^{2\ve-1} \G_i}{\hat \Ups_i(a_i^2-x_\mu^2)} \left( d \phi_i - \f{a_i}{r^2+a_i^2} d Z\right) \nn 
    &= (-1)^{n-1+\ve}\sqrt{Q_\mu} \sum_{i=0}^{n-1+\ve} \f{l^2 a_i^{2\ve-1}\G_i}{\hat \Ups_i (a_i^2-x_\mu^2)}d \phi_i - (-1)^{n-1+\ve} l^2 \sqrt{Q_\mu} d Z \sum_{i=0}^{n-1+\ve} \f{a_i^{2\ve} \G_i}{\hat \Ups_i (a_i^2-x_\mu^2)(a_i^2+r^2)}.
\end{align}
The sum associated with $dZ$ takes on different values depending on the cases $\mu = n$ and $\mu \neq n$. Consider first $\mu \neq n$. Define $A_{\mu n}^{(j)}$ to be the value of $A_\mu^{(j)}$ with $x_n^2 = -r^2 \to 0$, or
\begin{align}
    A_{\mu n}^{(j)} &= \sum_{\substack{\nu_1 < ... < \nu_j \\ \nu_i \neq \mu,n}} \prod_{i = 1}^j x^2_{\nu_i} \nonumber \\
    \prod_{\nu \neq \mu,n} (1 + \alpha x_\nu^2) &= \sum_{j = 0}^{n-2} A_{\mu n}^{(j)} \alpha^j. 
\end{align}
Then
\begin{align}
    \f{\G_i}{(a_i^2-x_\mu^2)(a_i^2+r^2)} &= \prod_{\nu \neq \mu,n} (a_i^2-x_\mu^2) \nn 
    &= (-1)^{n-2} \sum_{j=0}^{n-2} A_{\mu n}^{(j)} (-a_i^2)^{n-2-j} 
\end{align}
from which
\begin{align}
    \sum_{i=0}^{n-1+\ve} \f{a_i^{2\ve} \G_i}{\hat \Ups_i (a_i^2-x_\mu^2)(a_i^2+r^2)} &= (-1)^{n-2} \sum_{j=0}^{n-2} A_{\mu n}^{(j)} \sum_{i=0}^{n-1+\ve} \f{a_i^{2 \ve}}{\hat \Ups_i}(-a_i^2)^{n-2-j} \nn 
    &= (-1)^{n-2+\ve} \sum_{j=0}^{n-2+\ve} A_{\mu n}^{(j)} \sum_{i=0}^{n-1+\ve} \f{(-a_i^2)^{n-2-j+\ve}}{\hat \Ups_i} \nn 
    &= (-1)^{n-2+\ve} \sum_{j=0}^{n-2+\ve} A_{\mu n}^{(j)} \sum_{i=0}^{n-1+\ve} \hat D^i_{(j+1)} \nn 
    &= (-1)^{n-2+\ve} \sum_{j=0}^{n-2+\ve} A_{\mu n}^{(j)} \sum_{i=0}^{n-1+\ve} \hat D^i_{(j+1)} \hat C_i^{(0)} \nn 
    &= 0.
\end{align}
Thus for all $\mu \neq n$ the $dZ$ term exactly cancels, and $e^{\hat \mu}$ is still a function of the azimuthal coordinates \emph{only} with the functional form precisely that resulting from sending $\hat \varphi_i \to \phi_i$. 

For $\mu = n$ we have instead, using \eqref{Gammabya2mx2},
\begin{align}
    \f{\G_i}{(r^2+a_i^2)(a_i^2-x_n^2)} &= \f{1}{r^2+a_i^2} \f{\G_i}{r^2+a_i^2} \nn 
    &= \f{1}{r^2+a_i^2} (-1)^{n-1} \sum_{j=0}^{n-1} A_n^{(j)} (-a_i^2)^{n-1-j}
\end{align}
from which
\begin{align}
    \sum_{i=0}^{n-1+\ve} \f{a_i^{2\ve} \G_i}{\hat \Ups_i (a_i^2+r^2)^2} &= \sum_{j=0}^{n-1} A_n^{(j)} (-1)^{n-1} \sum_{i=0}^{n-1+\ve} \f{a_i^{2\ve} (-a_i^2)^{n-1-j}}{\hat \Ups_i(r^2+a_i^2)} \nn 
    &=(-1)^{n-1+\ve} \sum_{j=0}^{n-1} A_n^{(j)} \sum_{i=0}^{n-1+\ve} \f{(-a_i^2)^{n-1+\ve-j}}{\hat \Ups_i (r^2+a_i^2)}.
\end{align}
To interpret the sum over $i$, consider the set of $n+1+\ve$ variables $\{l^2, a_1^2, \ldots, a_{n-1+\ve}^2, -r^2\}$. We then have, adapting $\sum_{i = 0}^{n-1+\ve} \hat D^i_{(j)} = 0$ for $1 \leq j \leq n-1+\ve$ to the case where the set also includes $-r^2$,
\begin{align}
    \sum_{i=0}^{n-1+\ve} \f{(-a_i^2)^{n+\ve-l}}{(-r^2-a_i^2)\prod_{k=0,k\neq i}^{n-1+\ve} (a_k^2-a_i^2)} + \f{r^{2(n+\ve-l)}}{\prod_{k=0}^{n-1+\ve} (a_k^2+r^2)} &= 0, \qquad 1 \leq l \leq n+\ve \nn 
    \sum_{i=0}^{n-1+\ve} \f{(-a_i^2)^{n+\ve-l}}{(r^2+a_i^2) \hat \Ups_i} &= \f{r^{2(n+\ve-l)}}{\prod_{k=0}^{n-1+\ve} (r^2+a_k^2)}, \qquad 1 \leq l \leq n+\ve \nn 
    \sum_{i=0}^{n-1+\ve} \f{(-a_i^2)^{n-1+\ve-j}}{(r^2+a_i^2)\hat \Ups_i} &= \f{r^{2(n-1+\ve-j)}}{\prod_{k=0}^{n-1+\ve}(r^2+a_k^2)},\qquad 0 \leq j \leq n-1+\ve\nn 
    \sum_{i=0}^{n-1+\ve} \f{\hat D^i_{(j)}}{r^2+a_i^2} &= \f{r^{2(n-1-j)}}{l^2 \bar X}, \qquad 0 \leq j \leq n+\ve -1. \label{sumDhatbyr2pa2}
\end{align}

Consequently
\begin{align}
    \sum_{i=0}^{n-1+\ve} \f{a_i^{2\ve} \G_i}{\hat \Ups_i (a_i^2+r^2)^2} &= (-1)^{n-1+\ve} \sum_{j=0}^{n-1} A_n^{(j)} \f{r^{2(n-1-j)}}{l^2 \bar X} \nn 
    &= \f{(-1)^{n-1+\ve}}{l^2 \bar X} \prod_{\mu=1}^{n-1} (x_\mu^2+r^2) \nn 
    &= \f{(-1)^{n+\ve} U_n}{l^2 \bar X_n},
\end{align}
using $\bar X_n = -\bar X$ on the last line. 

We also have, in odd dimension,
\begin{align}
    e^{\hat 0} &= (-1)^n \sqrt{S} \sum_{i=0}^n \f{l^2 \G_i}{a_i \hat \Ups_i} \left( d \phi_i - \f{a_i}{r^2+a_i^2} d Z\right) \nn 
    &= (-1)^n \sqrt{S} \sum_{i=0}^n \f{l^2 \G_i}{a_i \hat \Ups_i} d \phi_i - (-1)^n \sqrt{S} l^2 d Z \sum_{i=0}^n \f{\G_i}{(a_i^2+r^2)\hat \Ups_i}
\end{align}
Using \eqref{Gammabya2mx2},
\begin{align}
    \sum_{i=0}^n \f{\G_i}{(a_i^2+r^2)\hat \Ups_i} &= (-1)^{n-1} \sum_{j=0}^{n-1} A_n^{(j)} \sum_{i=0}^n \f{(-a_i^2)^{n-1-j} }{\hat \Ups_j} \nn 
    &= (-1)^{n-1} \sum_{j=0}^{n-1} A_n^{(j)} \sum_{i=0}^n \hat D^i_{(j+1)} \nn 
    &= 0.
\end{align}

Using \eqref{Vm2mX} and $X = -X_n$, \eqref{canonicalbasisKScoords} follows.

\subsection{Canonical Basis Vectors in KS Coordinates} \label{basisvectorsKS}

First we find the canonical basis vectors in $(x_\mu, \hat \vp_i)$ coordinates. The $e_\mu$ are unchanged from those for $(x_\mu, \psi_j)$ coordinates. For the $e_{\hat \mu}$,
\begin{align}
    e_{\hat \mu} &= \f{1}{\sqrt{Q_\mu}} \sum_{j=0}^{n-1+\ve} B^\mu_{(j)} \sum_{i=0}^{n-1+\ve} \hat C^{(j)}_i \f{a_i}{l^2} \f{\pa}{\pa \hat \vp_i}.
\end{align}
Specifically,
\begin{align}
    \sum_{j=0}^{n-1+\ve} B^\mu_{(j)} \hat C^{(j)}_i &= \f{1}{U_\mu} \sum_{j=0}^{n-1+\ve} (-x_\mu^2)^{n-1-j} \hat C^{(j)}_i \nn 
    &= \f{1}{(-x_\mu^2)^\ve U_\mu} \sum_{j=0}^{n-1+\ve} (-x_\mu^2)^{n-1+\ve-j} \hat C_i^{(j)} \nn 
    &= \f{1}{(-x_\mu^2)^\ve U_\mu} \prod_{k=0, k \neq i}^{n-1+\ve} (a_k^2-x_\mu^2) \nn 
    &= \f{1}{(-x_\mu^2)^\ve (a_i^2-x_\mu^2) U_\mu} \prod_{k=0}^{n-1+\ve} (a_k^2-x_\mu^2) \nn 
    &= -\f{l^2 \bar X_\mu}{(a_i^2-x_\mu^2) U_\mu} \nn 
    &= -\f{ l^2 \bar Q_\mu}{a_i^2-x_\mu^2},
\end{align}
using $\bar Q_\mu = \bar X_\mu/U_\mu$. Thus,
\begin{align}
    e_{\hat \mu} &= -\f{1}{\sqrt{Q_\mu}} \bar Q_\mu \sum_{i=0}^{n-1+\ve} \f{a_i}{a_i^2-x_\mu^2} \f{\pa}{\pa \hat \vp_i}. \label{ehatmudihatvarphii}
\end{align}
For $\mu = \alpha \leq n-1$ recall $X_\alpha = \bar X_\alpha$ so that $Q_\alpha = \bar Q_\alpha$. For $\mu = n$ recall $\bar X = -\bar X_n$. Finally, $e_{\hat 0}$ can be calculated using \eqref{psivarphiequations} and that $\hat C_0^{(n)} = c$ (odd dimension). 
\begin{align}
    e_{\hat \alpha} &= -\sqrt{Q_\alpha}  \sum_{i=0}^{n-1+\ve} \f{a_i}{a_i^2-y_\alpha^2} \f{\pa}{\pa \hat \vp_i} \nn 
    e_{\hat n} &= -\f{\bar Q_n}{\sqrt{Q_n}}  \sum_{i=0}^{n-1+\ve} \f{a_i}{a_i^2+r^2} \f{\pa}{\pa \hat \vp_i} \nn
    e_{\hat 0} &=-\sqrt{S} \sum_{i=0}^n \f{1}{a_i} \f{\pa}{\pa \hat \vp_i}.
\end{align}
It is now straightforward to convert from $(r,y_\alpha,\hat \vp_i)$ to $(r,y_\alpha, \phi_i)$ coordinates. Under the coordinate transformation from $\hat \vp_i$ to $\phi_i$ coordinates, the vectors $\pa/\pa \hat \vp_i$ (at constant $(r,y_\alpha, \hat \vp_j)$ where $j \neq i$) are equal to the vectors $\pa/\pa \phi_i$ (at constant $(r,y_\alpha, \phi_j)$ where $j \neq i$). The $\pa/\pa y_\alpha$ vectors similarly point in the same direction because $y_\alpha$ remain unaltered by the coordinate transformation. We have,
\begin{align}
    \left.\f{\pa}{\pa r}\right|_{y_\alpha,\hat \vp_i} &=  \f{\pa}{\pa r} + \f{2 \mu(r)}{V-2 \mu(r)} \sum_{i=0}^{n-1+\ve} \f{a_i}{r^2+a_i^2} \f{\pa}{\pa \phi_i},
\end{align}
where the right-hand side is in terms of coordinates $(r,y_\alpha, \phi_i)$. We thus have, for $(r,y_\alpha,\phi_i)$ coordinates,
\begin{align}
    e_{\alpha} &= \sqrt{Q_\alpha} \f{\pa}{\pa y_\alpha} \nn 
    e_{n} &= \f{1}{\sqrt{ X U_n}} \left( X \f{\pa}{\pa r} - (X-\bar X) \sum_{i=0}^{n-1+\ve} \f{a_i}{r^2+a_i^2} \f{\pa}{\pa \phi_i}\right) \nn 
    e_{\hat \alpha} &= -\sqrt{Q_\alpha} \sum_{i=0}^{n-1+\ve} \f{a_i}{a_i^2-y_\alpha^2} \f{\pa}{\pa \phi_i} \nn 
    e_{\hat n} &= -\f{i \bar X}{\sqrt{X U_n}} \sum_{i=0}^{n-1+\ve} \f{a_i}{r^2+a_i^2} \f{\pa}{\pa \phi_i} \nn 
        e_{\hat 0} &= -\sqrt{S} \sum_{i=0}^n \f{1}{a_i} \f{\pa}{\pa \phi_i}. \label{canonicalformsinryphicoords}
\end{align}

\subsection{Canonical Frame Components of Kerr--Schild Null Vector} \label{kCanonicalFrame}

First we shall transform $k^a$ into $(r,y_\alpha, \psi_j)$ coordinates. 

The expression for $d \psi_j$ in terms of $d \phi_i$ and $dr$ is as follows. Use \eqref{psivarphiequations} and \eqref{dphidphihat} (with $t = l \phi_0, \tau = l \hat \vp_0$). We also use \eqref{sumDhatbyr2pa2}. Note further that $2mr^{1-\ve} = -(X - \bar X)$. (We can send $m \to \mu(r)$ at any point without problems.)

We have
\begin{align}
    d \psi_j &= \sum_{i=0}^{n-1+\ve} \hat D^i_{(j)} \f{l^2 d \hat \vp_i}{a_i} \nn 
    &= \sum_{i=0}^{n-1+\ve} \hat D^i_{(j)} \f{l^2}{a_i} \left( d \phi_i - \f{a_i}{r^2+a_i^2} \f{2mdr}{V-2m}\right) \nn 
    &= \sum_{i=0}^{n-1+\ve} \hat D^i_{(j)} \f{l^2}{a_i} d \phi_i - \f{2m l^2 dr}{V-2m} \sum_{i=0}^{n-1+\ve} \f{\hat D^i_{(j)}}{r^2+a_i^2} \nn 
    &= \sum_{i=0}^{n-1+\ve} \hat D^i_{(j)} \f{l^2}{a_i} d \phi_i - \f{2m dr}{V-2m} \f{r^{2(n-1-j)}}{\bar X} \nn 
    &= \sum_{i=0}^{n-1+\ve} \hat D^i_{(j)} \f{l^2}{a_i} d \phi_i - \f{(2 m r^{1-\ve}) r^{2(n-1-j)} dr}{ X \bar X} \nn 
    &= \sum_{i=0}^{n-1+\ve} \hat D^i_{(j)} \f{l^2}{a_i} d \phi_i + \f{X-\bar X}{X \bar X} r^{2(n-1-j)} dr.
\end{align}

We can then transform from $(r,\mu_i,\phi_i)$ to $(r,y_\alpha,\psi_j)$ coordinates as follows. We note that the $\mu_i$ and $y_\alpha$ are functions of each other only, so that the expression for $k^a\pa_a$ in $(r,y_\alpha,\phi_i)$ coordinates is just the same as \eqref{kmu} (with $t = a \phi_0$). For the other coordinate transformations, note
\begin{align}
    \f{\pa}{\pa \phi_i} &= \sum_{j=0}^{n-1+\ve} \f{\pa \psi_j}{\pa \phi_i} \f{\pa}{\pa \psi_j} \nn 
    &= \sum_{j=0}^{n-1+\ve} \hat D^i_{(j)} \f{l^2}{a_i} \f{\pa}{\pa \psi_j}.
\end{align}
Consequently,
\begin{align}
    \sum_{i=0}^{n-1+\ve} \f{a_i}{r^2+a_i^2} \f{\pa}{\pa \phi_i} &= \sum_{i=0}^{n-1+\ve} \sum_{j=0}^{n-1+\ve}\f{a_i}{r^2+a_i^2} \hat D^i_{(j)} \f{l^2}{a_i} \f{\pa}{\pa \psi_j} \nn 
    &= l^2 \sum_{j=0}^{n-1+\ve} \f{\hat D^i_{(j)}}{r^2+a_i^2} \f{\pa}{\pa \psi_j} \nn 
    &= \f{1}{\bar X} \sum_{j=0}^{n-1+\ve} r^{2(n-1-j)} \f{\pa}{\pa \psi_j},
\end{align}
using \eqref{sumDhatbyr2pa2} again.

For $\pa_r$ we have
\begin{align}
    \left.\f{\pa}{\pa r}\right|_{\phi_i} &= \left.\f{\pa r}{\pa r}\right|_{\psi_j} \f{\pa}{\pa r} + \sum_{j=0}^{n-1+\ve} \f{\pa \psi_j}{\pa r} \f{\pa}{\pa \psi_j},
\end{align}
where it is understood that on the left-hand side the expressions are taken with respect to the coordinate system $(r,y_\alpha,\phi_i)$ and on the right they are taken with respect to the coordinate system $(r,y_\alpha,\psi_j)$, and so I will drop the explicit statements on the right-hand-side. We have
\begin{align}
    \left. \f{\pa}{\pa r}\right|_{\phi_i} &= \f{\pa}{\pa r} + \sum_{j=0}^{n-1+\ve} \f{X-\bar X}{X \bar X} r^{2(n-1-j)} \f{\pa}{\pa \psi_j}.
\end{align}
Consequently, writing in $(r,y_\alpha,\psi_j)$ coordinates,
\begin{align}
    k^a\pa_a &= \f{\pa}{\pa r} + \sum_{j=0}^{n-1+\ve} \f{X-\bar X}{X \bar X} r^{2(n-1-j)} \f{\pa}{\pa \psi_j}- \sum_{j=0}^{n-1+\ve} \f{1}{\bar X} r^{2(n-1-j)} \f{\pa}{\pa \psi_j} \nn 
    &= \f{\pa}{\pa r} - \f{1}{X} \sum_{j=0}^{n-1+\ve} r^{2(n-1-j)} \f{\pa}{\pa \psi_j}. \label{kpsij}
\end{align}

Comparing to the definitions of the orthonormal frame components \eqref{emuehatmuehat0} 
\begin{align}
    k &= \f{e_{\hat n} + i e_{n}}{\sqrt{Q_n}}.
\end{align}
$k$ is real-valued ($k^a$ and $k_a$ are real), since $e_n$ is real-valued, $e_{\hat n}$ is imaginary-valued, and $\sqrt{Q_n}$ is imaginary. 

 \section{Calculations Related to PCKY Tensor}
\subsection{PCKY Vector Exterior Derivative in Terms of Curvature Tensors and PCKY Tensor} \label{dbetacurvature}

$\na_a \bt_b$ is written in terms of $\bs h$ and the Ricci and Riemann tensors in various places in the literature (see, for instance, \cite{FrolovReview}). I show a derivation for interest, then rewrite the result in terms of the Weyl tensor and make some additional comments. 

We can write, using the definition of $\beta$, the CKY condition on $\bs h$, and the relationship between the second covariant derivatives and the Riemann tensor (this is essentially the same as in \cite{FrolovReview}), 
\begin{align}
    (D-1)\na_a \beta^b &= \na_a \na_c \bs h^{c b} \nn 
    &= \na_c \na_a \bs h^{c b} - {R^c}_{d c a} \bs h^{d b} - {R^b}_{dca} \bs h^{cd} \nn 
    &= \na_c ( \de^c_a \beta^b - \de^b_a \beta^c) - R_{da} \bs h^{d b} - {R^b}_{dca} \bs h^{cd} \nn 
    &= \na_a \beta^b - R_{da} \bs h^{d b} - {R^b}_{dca} \bs h^{cd} \nn 
    \na_a \beta^b &= \f{-R_{da} \bs h^{d b} - {R^b}_{dca} \bs h^{cd}}{D-2} \nn 
    \na_a \beta_b &= \f{R^d_a \bs h_{b d} - R_{b d c a} \bs h^{c d}}{D-2}.
\end{align}

The Riemann term can be rewritten a little more conveniently by using the cyclic identity and the antisymmetry of $\bs h^{cd}$:
\begin{align}
    R_{bdca}\bs h^{c d} + R_{b c a d} \bs h^{c d} + R_{badc} \bs h^{c d} &= 0 \nn 
    R_{b d c a}\bs h^{c d} - R_{b c d a} \bs h^{c d} + R_{badc} \bs h^{c d} &= 0 \nn 
    2 R_{b d c a} \bs h^{c d} &= -R_{a bcd} \bs h^{c d}.
\end{align}
Thus,
\begin{align}
    \na_a \beta_b &= \f{2 R^d_a \bs h_{b d} + R_{abcd} \bs h^{c d}}{2(D-2)}.
\end{align}

Because $e^\mu$ and $e^{\hat \mu}$ are eigenvectors of the Ricci tensor with eigenvalue $\mc R_{\mu \mu}$, and because $\bs h = \sum_{\mu=1}^n x_\mu \bs \om^\mu$,
\begin{align}
    R^d_a \bs h_{bd} = -\sum_{\mu=1}^n \mc R_{\mu \mu} x_\mu \bs \om^\mu.
\end{align}
We can rewrite $R_{abcd} \bs h^{cd}$ as
\begin{align}
    R_{abcd}\bs h^{cd} &=\sum_{A,B=1}^D e^A_a e^B_b R_{AB cd} \bs h^{cd} \nn 
    &= \sum_{A,B=1}^D e_A^a e_B^b \sum_{\mu=1}^n 2 x_\mu \bs R_{AB}(e_\mu,e_{\hat \mu}).
\end{align}
Checking the curvature forms from Section \ref{curvature}, the only nonzero $\bs R_{AB}(e_\mu, e_{\hat \mu})$ are $\bs R_{\nu \hat \nu}(e_\mu,e_{\hat \mu})$ (where $\nu$ is not necessarily equal to $\mu$); the result will be
\begin{align}
    R_{abcd} \bs h^{cd} &= R_{cd ab} \bs h^{cd} \nn 
    &= \sum_{\mu=1}^n R_{cd ab} x_\mu (e_\mu^c e_{\hat \mu}^d - e_{\hat \mu}^c e_\mu^d) \nn 
    &= 2 \sum_{\mu=1}^n x_\mu R_{cdab} e_\mu^c e_{\hat \mu}^d \nn 
    &= 2 \sum_{\mu=1}^n x_\mu (\bs R_{\mu \hat \mu})_{ab}.
\end{align}
Because all the $\bs R_{\mu \hat \mu}$ can be expanded in terms of the $\bs \om^\nu$ (not necessarily $\mu \neq \nu$) as per \eqref{Rmn}, this shows that $\na_a \beta_b$ can be expanded entirely in terms of the $\bs \om^\mu$. 

The Riemann tensor can be split up into Weyl and Ricci terms, giving (keeping in mind the antisymmetry of $\bs h$)
\begin{align}
    R_{abcd} \bs h^{cd} &= C_{abcd} \bs h^{cd} - \f{4}{D-2}R^d_{[a} \bs h_{b]d}  - \f{2}{(D-1)(D-2)} R \bs h_{a b}.
\end{align}
The result is
\begin{align}
    \na_a \beta_b &= \f{1}{2(D-2)}\left[ C_{abcd} \bs h^{c d} + 2 R^d_a \bs h_{b d} - \f{4}{D-2} R^d_{[a}\bs h_{b]d} - \f{2}{(D-1)(D-2)} R \bs h_{ab}\right].
\end{align}

The results so far (except the ones that used the explicit values for the curvature forms) apply to a generic CKY tensor (which does not need to be closed) and its associated vector. In the GKNAdS spacetimes, where $\beta$ is a Killing vector, we can use the Killing condition, $\na_a \beta_b = \na_{[a} \beta_{b]}$ to simplify the expression to \eqref{nabetacurvatureresult}.

The fact that $\na_a \beta_b$ can be expanded in terms of the $\bs \om^\mu$ itself guarantees that $k$ will be an eigenvector of $\na_a \beta_b$. However it is worth teasing this out a little further. We have,
\begin{align}
    k^a \na_a \beta_b &= \f{1}{2(D-2)} C_{abcd} \bs h^{c d} k^a + \f{D-3}{(D-2)^2} R^c_a \bs h_{b c} k^a + \f{1}{(D-2)^2} R^c_b \bs h_{a c} k^c - \f{1}{(D-1)(D-2)^2} R \bs h_{a b} k^a.
\end{align}
We now have three types of terms, the Weyl term, the Ricci tensor terms, and the Ricci scalar term. I will show that each of these is proportional to $k_b$.

For the Ricci scalar term, $k$ is an eigenvector of $\bs h_{a b}$ so that $\bs h_{a b} k^a \propto k_b$.

For the Ricci tensor terms, $k$ is also an eigenvector of the Ricci tensor, since $k = \tilde m_n$ and $\tilde m_n$ is an eigenvector of the Ricci tensor as stated in Section \ref{PCKYNull}. It is also stated in \cite{Ortaggio,Malek} that if $k^a$ is geodetic, it is automatically an eigenvector of the Ricci tensor and a WAND, if the background is Minkowski or anti-de Sitter. (\cite{Malek} also shows that vacuum solutions with $\na_a k^a \neq 0$ require that $H$ essentially take the form it does for Kerr--AdS, though the form it takes in \cite{Malek} is slightly more general because it includes the case where some of the eigenvalues $y_\alpha$ are degenerate.) 

Ricci tensor terms here are proportional to $k$ because $k$ is an eigenvector of both $R_{a b}$ and $\bs h$: $R^c_a \bs h_{b c} k^a \propto k^c \bs h_{b c} \propto k_b$ and $R^c_b \bs h_{a c} k^a \propto R^c_b k_c \propto k_b$. 

For the Weyl term, we note first that $C_{abcd} \bs h^{c d} k^a k^b = 0$ by symmetry--antisymmetry of $a,b$. Next, we will use the Petrov Type D condition. Following \cite{Hamamoto}, use the convention that $C_{a b c d} p^a u^b v^c w^c = W(p,u,v,w)$ for arbitrary vectors $p,u,v,w$. Using the decomposition of $\bs h$ given by \eqref{hkl}, for an as-yet arbitrary vector $v$ we have
\begin{align}
    C_{abcd} \bs h^{c d} k^a v^b &= \sum_{\nu = 1}^{n-1} y_\nu W(k,v,e_\nu, e_{\hat \nu}) + r W(k,v,k,l).
\end{align}
We can now let $v$ equal various basis components. We already established that $C_{abcd} \bs h^{cd} k^a k^b = 0$. If $v = e_\alpha$ (which is to say that $v$ is orthogonal to both $k$ and $\ell$) then \begin{align}
    C_{abcd} \bs h^{cd} k^a e_\alpha^b &= \sum_{\nu = 1}^{n-1} y_\nu W(k,e_\alpha,e_\nu, e_{\hat \nu}) + r W(k,e_\alpha,k,\ell). 
\end{align} 
$W(k,e_\alpha,e_\nu,e_{\hat \nu})$ is of the form $W(k,e_\alpha,e_\beta,e_\gamma)$ (since $e_\nu,e_{\hat \nu}$ are both orthogonal to $k,\ell)$ and $W(k,e_\alpha,k,\ell) = W(k,\ell,k,e_\alpha)$. Both of these are zero by the Petrov Type D condition \eqref{TypeD}. This implies that $C_{abcd} \bs h^{c d} k^a$ is orthogonal to $k^b$ and $e_\alpha^b$. The only basis vector in the set $(k,\ell, e_\alpha)$ which is orthogonal to $k$ and $e_\alpha$ is $k$, and so we conclude that $C_{abcd} \bs h^{c d} k^a \propto k_b$. 

Thus we conclude that $k^a \na_a \beta_b \propto k_b$ and we can relate it to $k$ being an eigenvector of both $\bs h$ and the Ricci tensor as well as being a WAND. 

\subsection{Alternate Forms of \texorpdfstring{$\bs b, \bs h$}{b,h}}  
\label{alternatebh}

Using \eqref{bpotentialdefinition} and \eqref{psivarphiequations}, substituting $k = j+1$ partway through,
\begin{align}
    2 \bs b &= \sum_{j=0}^{n-1} A^{(j+1)} \sum_{i=0}^{n-1+\ve} \hat D^i_{(j)} \f{l^2 d \hat \vp_i}{a_i} \nn 
    &= \sum_{i=0}^{n-1+\ve} \f{l^2 d \hat \vp_i}{a_i} \sum_{j=0}^{n-1} A^{(j+1)} \f{(-a_i^2)^{n-1+\ve-j}}{\hat \Ups_i} \nn 
    &= \sum_{i=0}^{n-1+\ve} \f{l^2 d \hat \vp_i}{a_i \hat \Ups_i} \sum_{k=1}^n A^{(k)} (-a_i^2)^{n+\ve-k} \nn 
    &= l^2 \sum_{i=0}^{n-1+\ve} \f{(-1)^\ve a_i^{2\ve - 1} d \hat \vp_i}{\hat \Ups_i} \left[ \sum_{k=0}^n A^{(k)} (-a_i^2)^{n-k} - (-a_i^2)^n\right] \nn 
    &= l^2 \sum_{i=0}^{n-1+\ve} \f{(-1)^\ve a_i^{2\ve - 1} d \hat \vp_i}{\hat \Ups_i} \left[ \prod_{\mu=1}^n (x_\mu^2-a_i^2) - (-a_i^2)^n \right] \nn 
    &= l^2 \sum_{i=0}^{n-1+\ve} \f{(-1)^{n+\ve} a_i^{2\ve-1} d \hat \vp_i}{\hat \Ups_i} \left[ \G_i - a_i^{2n}\right].
\end{align}
We recognize the term involving $\Gamma_i$ to be related to the Jacobi transformed terms $\nu_i^2$, from \eqref{nui2} and \eqref{nu02v1}. We can then rewrite as
\begin{align}
    2 \bs b &= \sum_{i=0}^{n-1+\ve} \left( (-l^2 \nu_i^2) a_i d \hat \vp_i + \f{l^2 (-1)^{n-1+\ve} a_i^{2(n+\ve) - 1} d \hat \vp_i}{\hat \Ups_i}\right).
\end{align}
The term $(-1)^{n+\ve} \hat \Ups_i^{-1} a_i^{2(n+\ve)-1} d \hat \vp_i$ is closed, and so does not contribute to $\bs h = d \bs b$.

It is worth collecting these terms. Let $\bs \sigma$ be defined by
\begin{align}
    2 \bs \sigma &= l^2 (-1)^{n-1+\ve} \sum_{i=0}^{n-1+\ve} \f{a_i^{2(n+\ve)-1} d \hat \vp_i}{\hat \Ups_i} \nn 
    &= l^2 (-1)^{n-1+\ve} \sum_{i=0}^{n-1+\ve} \f{a_i^{2(n-1+\ve) + 1}}{\hat \Ups_i} d \hat \vp_i \nn 
    &= l^2 \sum_{i=0}^{n-1+\ve} \hat D^i_{(0)} a_i d \hat \vp_i.
\end{align}
Let 
\begin{align}
    \bs b'' &\equiv \bs b - \bs \sigma, 
\end{align}
which is another form with $d \bs b'' = \bs h$. This turns out to be (up to a sign) the form that is used in \cite{Cvetic}. More on this later in the section.

The pseudo-Cartesian coordinates are written in terms of $\phi_i$ (including $t = l \phi_0$) rather than in terms of $\vp_i$. At this point we will apply the transformation between $\hat \vp_i$ and $\phi_i$ \eqref{phivarphi}, using $dZ = 2m dr/(V-2m)$ for compactness. We then have
\begin{align}
    2 \bs b &= \sum_{i=0}^{n-1+\ve} \left( (-l^2 \nu_i^2) a_i + \f{l^2 (-1)^{n-1+\ve} a_i^{2(n+\ve)-1}}{\hat \Ups_i}\right) \left( d \phi_i + \f{a_i}{r^2+a_i^2} d Z\right) \nn 
    &= \sum_{i=0}^{n-1+\ve} \left( (-l^2 \nu_i^2) a_i + \f{l^2 (-1)^{n-1+\ve} a_i^{2(n+\ve)-1}}{\hat \Ups_i}\right) d \phi_i + \left[ \sum_{i=0}^{n-1+\ve} \left( \f{-l^2 a_i^2 \nu_i^2}{r^2+a_i^2}  + \f{l^2 (-1)^{n-1+\ve} a_i^{2(n+\ve)-1}}{(r^2+a_i^2) \hat \Ups_i} \right)\right] d Z
\end{align}
Noting that $a_n = 0$ if $\ve = 0$,
\begin{align}
    \sum_{i=0}^{n-1+\ve} \f{a_i^2 \nu_i^2}{r^2+a_i^2} &= \sum_{i=0}^n \f{a_i^2 \nu_i^2}{r^2+a_i^2}\nn 
    &= \sum_{i=0}^n \left( \nu_i^2 - \f{r^2 \nu_i^2}{r^2+a_i^2}\right) \nn 
    &= 1,
\end{align}
using \eqref{nusum1} and \eqref{nu02v1}. (Note that \eqref{nu02v1} implies $\sum_{i=0}^{n} \f{\nu_i^2}{r^2+a_i^2} = 0$.)

We then have that the $dZ$ term is
\begin{align}
    -l^2 \left[1 + \f{(-1)^{n+\ve} a_i^{2(n+\ve)-1}}{(r^2+a_i^2) \hat \Ups_i}\right] d Z.
\end{align}
This is of the form $f(r) d r$ and so is closed, and thus also does not contribute to $\bs h$. Let $\bs b'$ be equal to $\bs b$ up to a closed form, so that $\bs h = d \bs b'$, and write
\begin{align}
    2 \bs b' &= \sum_{i=0}^{n-1+\ve} (-l^2 \nu_i^2) a_i d \phi_i. \label{2bprimenuform}
\end{align}
Using the definitions \eqref{pseudoCartesiansnu} this becomes
\begin{align}
    2 \bs b' &= l (v du - u dv) + \sum_{i=1}^{n-1+\ve} a_i (x_i dy_i - y_i d x_i).
\end{align}
We then have
\begin{align}
    \bs h &= d \bs b \nn 
    &= d \bs b' \nn
    &= -l du \wedge dv + \sum_{i=1}^{n-1+\ve} a_i dx_i \wedge dy_i,
\end{align}
as required. 

Incidentally, going from \eqref{2bprimenuform}, we also get the form for $\bs h$
\begin{align}
    \bs h &= \sum_{i=0}^{n-1+\ve} (-l^2) a_i \nu_i d \nu_i \wedge d \phi_i. \label{hnuform}
\end{align} 

We can also show explicitly that $k \cdot \bs h = -r k^\flat$ fairly easily in these coordinates. Using \eqref{kmudxmuuvxyz},
\begin{align}
    k \cdot \bs h &= \f{-l(ru+lv) dv + l(rv-lu) du}{r^2+l^2} + \sum_{i=1}^{n-1+\ve} \f{a_i(rx_i+a_iy_i) dy_i - a_i (ry_i - a_ix_i) dx_i}{r^2+a_i^2} \nn 
    &= \f{-lr(udv-vdu) - l^2(udu+vdv)}{r^2+l^2} + \sum_{i=1}^{n-1+\ve} \f{-ra_i(y_i dx_i - x_idy_i) +a_i^2 (x_idx_i+y_idy_i)}{r^2+a_i^2} \nn 
    &= \f{-lr(udv-vdu) + r^2(udu+vdv)}{r^2+l^2} - (udu+vdv) + \nn
    &\qquad \sum_{i=1}^{n-1+\ve} \left(\f{-r a_i (y_i dx_i-x_idy_i)-r^2(x_idx_i+y_idy_i)}{r^2+a_i^2} + (x_idx_i+y_idy_i) \right) \nn 
    &= -r \left[-\f{(ru+lv)du + (rv-lu)dv}{r^2+l^2} + \f{(r x_i + a_iy_i) dx_i + (r y_i - a_i x_i) dy_i}{r^2+a_i^2} + (1-\ve) \f{z dz}{r}\right]  \nn 
    &\qquad + \left(-udu-vdv+\sum_{i=1}^{n-1+\ve} (x_idx_i + y_i dy_i) + (1-\ve) zdz\right) \nn 
    &= -r k^\flat,
\end{align}
using that $-u^2 - v^2 + \sum_{i=1}^{n-1+\ve} (x_i^2+y_i^2) + (1-\ve)z^2 = -l^2$ so that $-udu - vdv + \sum_{i=1}^{n-1+\ve} (x_idx_i+y_idy_i) + (1-\ve) z dz = 0$. 

It is also useful to have an expression for $\bs b$ in terms of the canonical basis one-forms. Write,
\begin{align}
    2 \bs b &= \sum_{j=0}^{n-1} A^{(j+1)} d \psi_j \nn 
    &= \sum_{j=0}^{n-1} A^{(j+1)} \sum_{\mu=1}^n \f{B^\mu_{(j)}}{\sqrt{Q_\mu}} e^{\hat \mu} \nn 
    &= \sum_{\mu=1}^n \sum_{j=0}^{n-1} A^{(j+1)} \f{(-x_\mu^2)^{n-1-j}}{\sqrt{Q_\mu} U_\mu} e^{\hat \mu} \nn 
    &= \sum_{\mu=1}^n \f{1}{\sqrt{Q_\mu} U_\mu} e^{\hat \mu} \sum_{j=0}^{n-1} A^{(j+1)} (-x_\mu^2)^{n-1-j} \nn 
    &= \sum_{\mu=1}^n \f{1}{\sqrt{Q_\mu} U_\mu} e^{\hat \mu} \sum_{k = 1}^n A^{(k)} (-x_\mu^2)^{n-k} \nn 
    &= \sum_{\mu=1}^n \f{1}{\sqrt{Q_\mu} U_\mu} e^{\hat \mu} \left[ \sum_{k=0}^n A^{(k)} (-x_\mu^2)^{n-k} - A^{(0)} (-x_\mu^2)^n\right] \nn 
    &= (-1)^{n+1} \sum_{\mu=1}^n \f{x_\mu^{2n}}{\sqrt{ Q_\mu} U_\mu} e^{\hat \mu}, \label{bincanonicalbasis}
\end{align}
using the fact that
\begin{align}
    \sum_{k=0}^n A^{(k)} (-x_\mu^2)^{n-k} &= \prod_{\nu=1}^n (x_\nu^2 - x_\mu^2) \nn 
    &= 0.
\end{align}

We will now find an expression for $\bs b''$. Using the expression for $d \hat \vp_i$ in terms of $d \psi_j$, $\bs \sigma$ can be rewritten
\begin{align}
    2 \bs \sigma &= -\sum_{j=0}^{n-1+\ve} \sum_{i=0}^{n-1+\ve}  \f{(-a_i^2)^{n+\ve} \hat C_i^{(j)}}{\hat \Ups_i} d \psi_j.
\end{align}
This sum has a simple form. Let $\hat C^{(j)}$ be the elementary symmetric polynomials in the $a_i^2$ for $0 \leq i \leq n-1+\ve$, with none omitted. Then $\hat C^{(j+1)} = a_i^2 \hat C_i^{(j)} + \hat C_i^{(j+1)}$. We can then write
\begin{align}
    \sum_{i=0}^{n-1+\ve} \f{(-a_i^2)^{n+\ve} \hat C_i^{(j)}}{\hat \Ups_i} &= \sum_{i=0}^{n-1+\ve}\f{ (-a_i^2)^{n-1+\ve} (-a_i^2) \hat C_i^{(j)}}{\hat \Ups_i} \nn 
    &= \sum_{i=0}^{n-1+\ve} \hat D^i_{(0)}\left( \hat C_i^{(j+1)} - \hat C^{(j+1)}\right) \nn 
    &= - \hat C^{(j+1)} \sum_{i=0}^{n-1+\ve} \hat D^i_{(0)} \hat C_i^{(0)} + \sum_{i=0}^{n-1+\ve} \hat D^i_{(0)} \hat C_i^{(j+1)} \nn 
    &= - \hat C^{(j+1)}.
\end{align}
We therefore conclude
\begin{align}
    2 \bs \sigma &= \sum_{j=0}^{n-1+\ve} \hat C^{(j+1)} d \psi_j.
\end{align}
This tells us that we have
\begin{align}
    2 \bs b'' &= \sum_{j=0}^{n-1+\ve} \left( A^{(j+1)} - \hat C^{(j+1)}\right) d \psi_j,
\end{align}
with $A^{(n+1)} = 0$ if it appears in the sum.

We can also write $\bs b''$ in terms of the canonical basis. Again I begin with $\bs \sigma$, in terms of $d \hat \vp_i$. We have, using \eqref{ehatmudihatvarphii},
\begin{align}
    d \hat \vp_i &= \sum_{A} (e_A \cdot d \hat \vp_i ) e^A \nn 
    &= \sum_{\mu=1}^n (e_{\hat \mu} \cdot d \hat \vp_i ) e^{\hat \mu} + \vp ( e_{\hat 0} \cdot d \hat \vp_i) e^{\hat 0} \nn 
    &= -\sum_{\mu=1}^n \f{\bar Q_\mu a_i}{\sqrt{Q_\mu}(a_i^2-x_\mu^2)} e^{\hat \mu} - \ve \sqrt{S} \f{1}{a_i} e^{\hat 0}.
\end{align}
Then we have,
\begin{align}
    2 \bs \sigma &= l^2 \sum_{i=0}^{n-1+\ve} \hat D^i_{(0)} \left( -\sum_{\mu=1}^n \f{\bar Q_\mu}{\sqrt{Q_\mu}} \f{a_i^2}{a_i^2-x_\mu^2} e^{\hat \mu} - \ve \sqrt{S} e^{\hat 0}\right).
\end{align}
Individually, and using the results of \eqref{sumDhatbyr2pa2}, generalized from $-r^2$ to $x_\mu^2$,
\begin{align}
    \sum_{i=0}^{n-1+\ve} \hat D^i_{(0)} &= \sum_{i=0}^{n-1+\ve} \hat D^i_{(0)} \hat C_i^{(0)} \nn 
    &= 1 \nn 
    \sum_{i=0}^{n-1+\ve} \hat D^i_{(0)} \f{a_i^2}{a_i^2-x_\mu^2} &= \sum_{i=0}^{n-1+\ve} \hat D^i_{(0)} \left[ 1 - \f{x_\mu^2}{a_i^2-x_\mu^2}\right] \nn 
    &= 1 + \f{(-x_\mu^2)^n}{l^2 \bar X_\mu}.
\end{align}
Consequently,
\begin{align}
    2 \bs \sigma &= -\sum_{\mu=1}^n \left(l^2 + \f{(-x_\mu^2)^n}{\bar X_\mu}\right) \f{\bar Q_\mu}{\sqrt{Q_\mu}} e^{\hat \mu} - \ve l^2 \sqrt{S} e^{\hat 0} \nn 
    &= -l^2 \left( \sum_{\mu=1}^n \f{\bar Q_\mu}{\sqrt{Q_\mu}} e^{\hat \mu} + \ve \sqrt{S} e^{\hat 0}\right) - \sum_{\mu=1}^n \f{(-x_\mu^2)^n}{\sqrt{Q_\mu} U_\mu} e^{\hat \mu}.
\end{align}
In the GKAdS solutions, we have $X_\alpha = \bar X_\alpha$ for $1 \leq \alpha \leq n-1$. We then can decompose $\bs \sigma$ as
\begin{align}
    2 \bs \sigma &= -l^2 \left( \sum_{\mu=1}^n \sqrt{Q_\mu} e^{\hat \mu} + \ve \sqrt{S} e^{\hat 0}\right) + l^2 \f{Q_n - \bar Q_n}{\sqrt{Q_n}} e^{\hat n} - \sum_{\mu=1}^n \f{(-x_\mu^2)^n}{\sqrt{Q_\mu} U_\mu} e^{\hat \mu} \nn 
    &= - l^2 \beta^\flat + \f{l^2 H}{\sqrt{Q_n}} e^{\hat n} - \sum_{\mu=1}^n \f{(-x_\mu^2)^n}{\sqrt{Q_\mu} U_\mu} e^{\hat \mu},
\end{align}
using the decomposition of $\beta$ from \eqref{betacanonicalbasis} and using the expression for the Kerr--Schild scalar $H$ from \eqref{HintermsofQn}. Thus,
\begin{align}
    2 \bs b'' &= 2 (\bs b - \bs \sigma) \nn 
    &= l^2 \left( \sum_{\mu=1}^n \sqrt{Q_\mu} e^{\hat \mu} + \ve \sqrt{S} e^{\hat 0}\right) + \f{l^2 H}{\sqrt{Q_n}} e^{\hat n} \nn
    &= l^2 \beta^\flat + \f{ l^2 H}{\sqrt {Q_n}} e^{\hat n}. \label{bprimeprime}
\end{align}

\subsection{Raising and Lowering \texorpdfstring{$\bs{h}$}{h}} \label{RaisingAndLoweringh}

We show that if $\bs h_{ab}$ has the same coordinate expression in the full and background spacetime, that so does $\bs h^{ab}$. We have,
\begin{align}
    g^{ac} g^{bd} \bs h_{cd} &= (\bar g^{ac} - H k^a k^c) (\bar g^{bd} - H k^b k^d) \bs h_{cd} \nn 
    &= (\bar g^{ac} - H k^a k^c) (\bar g^{bd} \bs h_{cd} + H k^b k^d \bs h_{dc}) \nn 
    &= (\bar g^{ac} - H k^a k^c) (\bar g^{bd} \bs h_{cd} + H k^b (-r k_c)) \nn 
    &= \bar g^{ac} \bar g^{bd} \bs h_{cd} - \bar g^{bd} H k^a (k^c \bs h_{cd}) - H r k^b k^a \nn 
    &= \bar g^{ac} \bar g^{bd} \bs h_{cd} - \bar g^{bd} H k^a (-r k_d) - H r k^b k^a \nn 
    &= \bar g^{ac} \bar g^{bd} \bs h_{cd} + H r k^a k^b - H r k^a k^b \nn 
    &= \bar g^{ac} \bar g^{bd} \bs h_{cd}.
\end{align}
Thus the raised form of $\bs h$ is the same for both metrics so we can write $\bs h^{ab}$ without ambiguity. Note however that there is an ambiguity in the mixed form.

\subsection{Potential \texorpdfstring{$\bs \om^{(n)}$}{omega(n)}} \label{potentialomegan}

Let $\beta^{(j)}$ be the Killing vector $\pa/\pa \psi_j$. We wish to find the normalization for the Killing potential $\bs \om^{(n)}$ such that $(\beta^{(n)})^\flat = \mathrm{div} \bs \om^{(n)}$. 

Using \eqref{dstaromega},
\begin{align}
    d * \bs \om^{(j)} &= -2 * \mathrm{div} \bs \om^{(j)} \nn 
    &= -2 * (\beta^{(j)})^\flat .\label{dstaromegaj}
\end{align}

Consider odd dimensions. We have that $\bs h^{(n)}$ is given by
\begin{align}
    \bs h^{(n)} &= n! \sqrt{A^{(n)}} \bigwedge_{\mu=1}^n \bs \om^\mu.
\end{align}
In the canonical basis, this gives
\begin{align}
    (\bs h^{(n)})_{1 \hat 1 \ldots n \hat n} &= n! \sqrt{A^{(n)}}.
\end{align}
In the canonical basis, we have $*1 = \pm i e^1 \wedge \ldots \wedge e^D$, where the factor of $i$ accounts for the fact that one of the one-forms is imaginary-valued and we wish $*1$ to be real-valued. Choose orientation so that $*1 = f i \bs \om^1 \wedge \ldots \bs \om^n \wedge e^D$, where $f$ is either $+1$ or $-1$. Then $\bs \ep_{1 \hat 1 \ldots n \hat n \hat 0} = f i$ again in the canonical basis. Then we have, recalling the convention $\hat 0 = 2n+1$,
\begin{align}
    (* \bs h^{(n)})_{\hat 0} &= (\bs h^{(n)})^{A_1 \ldots A_{2n}} \bs \ep_{A_1 \ldots A_{2n} \hat 0}  \nn 
    &= (2n)! (\bs h^{(n)})^{1 \hat 1 \ldots n \hat n} \bs \ep_{1 \hat 1 \ldots n \hat n \hat 0} \nn 
    &= (2n)! n! \sqrt{A^{(n)}} f i \nn 
    &= f (2n)! n! \sqrt{-A^{(n)}}.
\end{align}
We also have that $(*\bs h^{(n)})_A = 0$ in the canonical basis for any value of $A$ not equal to $\hat 0$, so that we have
\begin{align}
    * \bs h^{(n)} &= f (2n)! n! \sqrt{-A^{(n)}} e^{\hat 0}.
\end{align}
Comparing to $\beta^{(n)} = \sqrt{-c A^{(n)}} e_{\hat 0}$, we then have
\begin{align}
    (\beta^{(n)})^\flat &= f \f{\sqrt c}{(2n)!n!} * \bs h^{(n)}.
\end{align}
Its Hodge dual is
\begin{align}
    * (\beta^{(n)})^\flat &=f  \f{\sqrt c}{(2n)!n!} ** \bs h^{(n)} \nn 
    &= - f \f{\sqrt c}{n!} \bs h^{(n)}
\end{align}
using \eqref{starstaromega}. Since $\bs h^{(n)} = d \bs b^{(n)}$ we have
\begin{align}
    * (\beta^{(n)})^\flat &= -f \f{\sqrt c}{n!} d \bs b^{(n)}.
\end{align}
From \eqref{dstaromegaj},
\begin{align}
    d * \bs \om^{(n)}&= f \f{2 \sqrt c}{n!} d \bs b^{(n)} \nn 
    {*} {\bs \om^{(n)}} &= f \f{2 \sqrt c}{n!} \bs b^{(n)},
\end{align}
up to a closed form. Taking the Hodge dual of both sides,
\begin{align}
    -2 (2n-1)! \bs \om^{(n)} &= f \f{2 \sqrt c}{n!} * \bs b^{(n)} \nn 
    \bs \om^{(n)} &=  -\f{\sqrt c}{n! (2n-1)!} f * \bs b^{(n)}.
\end{align}
(Using my Hodge dual convention/ordering I found $f = +1$, at least for $D = 5$, and I assume it holds for all odd $D \geq 7$.)

Using $\bs b^{(n)} = \bs b'' \wedge \bs h^{(n-1)}$,
\begin{align}
    2 \bs b^{(n)} &= l^2 \left\{ \sum_{\mu=1}^n \f{\bar Q_\mu}{\sqrt{Q_\mu}} e^{\hat \mu} + \sqrt S e^{\hat 0}\right\} \wedge (n-1)! \sum_{\nu=1}^n \left( \prod_{\sigma \neq \nu} x_\nu\right) \bigwedge_{\sigma \neq \nu} \bs \om^\sigma \nn 
    &= l^2 \sqrt{A^{(n)}} (n-1)! \left\{ \sum_{\mu=1}^n \f{\bar Q_\mu}{\sqrt{Q_\mu}} e^{\hat \mu} + \sqrt S e^{\hat 0}\right\} \wedge \sum_{\nu=1}^n \f{1}{x_\nu} \bigwedge_{\sigma \neq \nu} \bs \om^\sigma.
\end{align}
We also have that $e^{\hat \mu} \wedge \bs \om^\mu = 0$, so this can be simplified to
\begin{align}
    2 \bs b^{(n)} &= l^2 \sqrt{A^{(n))}} (n-1)! \sum_{\mu=1}^n \left(\f{\bar Q_\mu}{ x_\mu \sqrt{Q_\mu}} e^{\hat \mu} + \f{\sqrt S}{x_\mu} e^{\hat 0}\right) \wedge \bigwedge_{\nu \neq \mu} \bs \om^\nu. 
\end{align}
The Hodge dual of $e^{\hat \mu} \wedge \bigwedge_{\nu \neq \mu} \bs \om^\nu$ is proportional to $e^\mu \wedge e^{\hat 0}$, since $e^\mu$ and $e^{\hat 0}$ are the only forms absent in the wedge product. To find the constant, we check:
\begin{align}
    *\left(e^{\hat \mu} \wedge \bigwedge_{\nu \neq \mu} \bs \om^\nu\right)_{\mu \hat 0} &= \left(e^{\hat \mu} \wedge \bigwedge_{\nu \neq \mu} \bs \om^\nu\right)^{A_1 \ldots A_{D-2}} \bs \ep_{A_1 \ldots A_{D-2} \mu \hat 0} \nn 
    &= (D-2)! \left(e^{\hat \mu} \wedge \bigwedge_{\nu \neq \mu} \bs \om^\nu\right)^{\hat \mu \nu_1 \hat \nu_1 \ldots \nu_{n-1} \hat \nu_{n-1}} \bs \ep_{\hat \mu \nu_1 \hat \nu_1 \ldots \nu_{n-1} \hat \nu_{n-1} \mu \hat 0},
\end{align}
where the set $\{ \nu_1, \ldots, \nu_{n-1}\}$ is the set $\{1, \ldots, n\}$ with $\mu$ omitted (in that order). $\mu$ must be moved $2n-1$ spots to the right to give $\bs \ep_{\mu \hat \mu \nu_1 \hat \nu_1 \ldots \nu_{n-1} \hat \nu_{n-1} \hat 0}$, and then $\mu \hat \mu$ can be moved together to their proper place with no change in sign, so we have $\bs \ep_{\hat \mu \nu_1 \hat \nu_1 \ldots \nu_{n-1} \hat \nu_{n-1} \mu \hat 0} = - \bs \ep_{1 \hat 1 \ldots n \hat n \hat 0} = f i$. We conclude
\begin{align}
    * \left( e^{\hat \mu} \wedge \bigwedge_{\nu \neq \mu} \bs \om^\nu\right)_{\mu \hat 0} &= - (D-2)! i f \nn 
    * \left( e^{\hat \mu} \wedge \bigwedge_{\nu \neq \mu} \bs \om^\nu\right) &= - (D-2)! if e^{\mu} \wedge e^{\hat 0}.
\end{align}
Similarly, we must have $* \left( e^{\hat 0} \wedge \bigwedge_{\nu \neq \mu} \bs \om^\nu\right) \propto \bs \om^\mu$, so we write
\begin{align}
    * \left( e^{\hat 0} \wedge \bigwedge_{\nu \neq \mu} \bs \om^\nu\right)^{\mu \hat \mu} &= (D-2)! \left( e^{\hat 0} \wedge \bigwedge_{\nu \neq \mu} \bs \om^\nu\right)^{\hat 0 \nu_1 \hat \nu_1 \ldots \nu_{n-1} \hat \nu_{n-1}} \bs \ep_{\hat 0 \nu_1 \hat \nu_1 \ldots \hat \nu_{n-1} \hat \nu_{n-1} \mu \hat \mu}.
\end{align}
In this case, $\bs \ep_{\hat 0 \nu_1 \hat \nu_1 \ldots \nu_{n-1} \hat \nu_{n-1} \mu \hat \mu} = + \bs \ep_{1 \hat 1 \ldots n \hat n \hat 0} = + f i$ and so
\begin{align}
    * \left( e^{\hat 0} \wedge \bigwedge_{\nu \neq \mu} \bs \om^\nu\right) &= (D-2)! f i \bs \om^\mu.
\end{align}
With $D-2 = 2n-1$, we conclude,
\begin{align}
    2 * \bs b^{(n)} &= f i l^2 \sqrt{A^{(n)}}(n-1)! (2n-1)! \sum_{\mu=1}^n \left( \f{\sqrt S}{x_\mu} e^\mu \wedge e^{\hat \mu} - \f{\bar Q_\mu}{x_\mu \sqrt{Q_\mu}} e^\mu \wedge e^{\hat 0} \right) \nn 
    \bs \om^{(n)} &= -\f{ l^2 \sqrt{-c A^{(n)}}}{2 n} \sum_{\mu=1}^n \f{1}{x_\mu} e^\mu \wedge \left( \sqrt {S} e^{\hat \mu} - \f{\bar Q_\mu}{\sqrt{Q_\mu}} e^{\hat 0}\right) \nn 
    &= \f{l^2}{2 n} \sum_{\mu=1}^n \f{e^\mu}{x_\mu \sqrt{Q_\mu}} \wedge \left(  c \sqrt{Q_\mu} e^{\hat \mu} + \sqrt{-c A^{(n)}}\bar Q_\mu e^{\hat 0}\right) \nn 
    &= \f{l^2}{2n} \sum_{\mu=1}^n \f{1}{x_\mu} \left(  c \bs \om^\mu + \f{\sqrt{-cA^{(n)}} \bar Q_\mu}{\sqrt{Q_\mu}} e^\mu \wedge e^{\hat 0}\right). \label{omeganexpression}
\end{align}
(The sign of the orientation, encoded in $f$, drops out, as expected since the orientation will not affect $\mathrm{div} \bs \om^{(n)} = (\beta^{(n)})^\flat$.) 

We already have that the 2-form $\bs \om^\mu$ has, in KS coordinates, $\sqrt{-g} (\bs \om^\mu)^{t r} = 0$ on $r^2+a_n^2 = 0$ for all $\mu$. We are then interested in the $r \psi_j$ component of the remaining terms which are:
\begin{align}
    \f{l^2}{2n} \sum_{\mu=1}^n \f{\bar Q_{\mu}}{x_\mu} \f{e_\mu^r}{\sqrt{Q_\mu}}  \sqrt{-c A^{(n)}} e_{\hat 0}^{\psi_j} &= \f{l^2}{2n} \f{\bar Q_n}{x_n} \f{e_n^r}{\sqrt{Q_n}} \sqrt{-c A^{(n)}} e_{\hat 0}^{\psi_j} \nn 
    &= \f{l^2 \bar Q_n}{2 n r} \de^n_j.
\end{align}
We then have that the $t r$ component is
\begin{align}
    \f{l^2}{2n} \sum_{\mu=1}^n \f{\bar Q_\mu}{x_\mu} \left(\f{e^\mu}{\sqrt {Q_\mu}} \wedge \sqrt{-c A^{(n)}} e^{\hat 0}\right)^{t r} &= - \f{l^2 \bar Q_n}{2 n r} \f{\pa t}{\pa \psi_n} \nn
    &= - \f{l^2 c \bar Q_n}{2nr}.
\end{align}
In $(t,r,y_\alpha, \phi_i)$ components,
\begin{align}
    \sqrt{-g} \f{l^2}{2n} \sum_{\mu=1}^n \f{\bar Q_\mu}{x_\mu} \left(\f{e^\mu}{\sqrt {Q_\mu}} \wedge \sqrt{-c A^{(n)}} e^{\hat 0}\right)^{t r} &= -\f{P r \prod_{\alpha=1}^{n-1} y_\alpha}{C \prod_{i=1}^n \Xi_i} \f{l^2 c \bar Q_n}{2 n r} \nn 
    &= + \f{l^2 c \prod_{\alpha=1}^{n-1} y_\alpha}{2 n C \prod_{i=1}^n \Xi_i} \f{P}{U_n} \bar X.
\end{align}
$P/U_n$ is finite, as are the other terms. $\bar X$ is equal to zero when $r^2+a_n^2$. Consequently we conclude $\sqrt{-g} (\bs \om^{(n)})^{t r} = 0$ on $r^2+a_n^2 = 0$, when constructed from $\bs b^{(n)}$ constructed from $\bs b''$.

\chapter{Appendices Related to Covariant Phase Space Formalism Chapter}

In this Appendix chapter, I collect various calculations related to Chapter \ref{NoetherChapter}.

\section{Lagrangian Variation Calculations }
\subsection{Einstein--Hilbert Lagrangian} \label{EHterms}

The Einstein--Hilbert action is 
\begin{align}
    L_{EH} &= \frac{1}{16\pi} (R - 2 \Lambda).
\end{align}
I will show the calculation of $\bs E$ and $\bs \Th$ under a variation in this case. Many aspects of this calculation appear in, for instance, Poisson \cite{Poisson} or Wald \cite{Wald84}. 

We have
\begin{align}
    \delta ( \delta^a_b) &= 0 \nonumber \\
    \delta (g^{a c} g_{b c} ) &= 0 \nonumber \\
    g_{b c} \delta g^{a c} + g^{a c} \delta g_{b c} &= 0\nonumber \\
    \delta g^{a c} &= -g^{a b} g^{c d} \delta g_{b d}.
\end{align}
Similarly,
\begin{align}
    \delta \sqrt{-g} &= \frac{1}{2} \sqrt{-g} g^{a b} \delta g_{a b} = -\frac{1}{2} \sqrt{-g} g_{a b} \delta g^{a b},
\end{align}
which follows from the properties of the determinant. 

The variation $\delta$ commutes with partial derivatives, but not necessarily with covariant derivatives. 

The Christoffel symbols are $\Gamma^a_{b c}$. Let $\Gamma_{a b c} \equiv g_{a d} \Gamma^d_{b c}$. The Levi-Civita connection is torsion-free, so that $\G^a_{b c} = \G^a_{(bc)}, \G_{a bc} = \G_{a(bc)}$. We have, allowing a comma to represent a partial derivative, 
\begin{align}
    \Gamma_{a b c} &= \frac{1}{2} (g_{b a,c} + g_{a c,b} - g_{b c,a}),
\end{align}
so that
\begin{align}
    \delta \Gamma_{a b c} &= \frac{1}{2} ( \delta g_{b a,c} + \delta g_{a c,b} - \delta g_{b c,a}),
\end{align}
using the convention where $\de g_{ba,c} = \pa_c \de g_{ba} = \de \pa_c g_{ba}$. Then,
\begin{align}
    \delta \Gamma^d_{b c} &= g^{a d} \delta \Gamma_{a b c} + \delta g^{a d} \Gamma_{a b c} \nonumber \\
    g_{a e} \delta \Gamma^e_{b c} &= \delta \Gamma_{a b c} + g_{a e} \de g^{e d} \G_{d b c} \nonumber \\
    &= \de \G_{a b c} - g^{e d} \de g_{a e} \G_{d b c} \nonumber \\
    &= \de \G_{a b c} - \de g_{a e} \G^e_{b c}.
\end{align}
The statement $g_{ae} \de g^{ed} = - g^{e d} \de g_{a e}$ is only valid in the linearized regime (or in a Kerr--Schild case). 

Treating $\de g_{a b}$ as a tensor, we have
\begin{align}
    \nabla_a \de g_{b c} &=\de g_{a b,c} - \G^d_{a b} \de g_{cd} - \G^d_{ac} \de g_{b d}
\end{align}
so that
\begin{align}
    \de \G_{a b c} &= \frac{1}{2} \left( \na_c \de g_{a b} + \G^d_{ca} \de g_{bd} + \G^d_{c b} \de g_{ad} + \na_b \de g_{ac} + \G^d_{ba} \de g_{cd} \right. \nn 
    &\qquad \left. + \G^d_{b c} \de g_{ad} - \na_a\de g_{bc}- \G^d_{a b} \de g_{dc} - \G^d_{ac} \de g_{d b} \right) \nn
    &= \frac{1}{2} \left(\na_c\de g_{a b} + \na_b \de g_{ac} - \na_a \de g_{bc} + 2\G^d_{cb}\de g_{a d} \right) \nn 
    \therefore g_{a d} \de \G^d_{b c} &= \de \G_{a b c} - \de g_{ad} \G^d_{b c} \nonumber \\
    &= \frac{1}{2} \left(\na_c\de g_{a b} + \na_b \de g_{ac} - \na_a \de g_{bc} \right) \nn 
    \therefore \de \G^a_{b c} &= \frac{1}{2} g^{a d} \left(\na_c\de g_{d b} + \na_b \de g_{dc} - \na_d \de g_{bc} \right). \label{deltaGamma}
\end{align}

($\Delta^a_{bc} = \G^a_{bc} - \overline{\G^a_{bc}}$ for the Kerr--Schild metric with $g_{ab} = \bar g_{ab} + h_{ab}, g^{ab} = \bar g^{ab} - h^{ab}$ is given by \cite{Taub81} as
\begin{align}
    \Delta^a_{bc} &= \f12 (\bar \na_c h^a_b + \bar \na_b h^a_c - \bar \na^a h_{bc}) + \f12 k^d \pa_d H k^a h_{bc}, \label{Deltaabc}
\end{align}
where the last term is quadratic in the Kerr--Schild correction. In a previous version of this document, I erroneously did not include that term. This term does not match the otherwise similar expression for $\de \G^a_{bc}$ because $\de G^a_{bc}$ is only included to linear order.

$\Delta^a_{bc}$ can also be written as
\begin{align}
    \Delta^a_{bc} &= \f12 (\na_c h^a_b + \na_b h^a_c - \na^a h_{bc}) - \f12 k^d \pa_d H k^a h_{bc},\label{Deltaabcfull}
\end{align}
where the covariant derivatives are expressed in terms of the full metric.)

Specifically,
\begin{align}
    \de \G^c_{ca} &= \frac{1}{2} g^{b c} \nabla_a \de g_{b c} \nonumber \\
    &= \de \ln (\sqrt{-g}).
\end{align}

If we treat $\de g_{a b}$ as a tensor, then $\de \G^a_{b c}$ is a tensor (though $\de \G_{a b c}$ is not), in the sense that its covariant derivatives are well-defined in the usual tensorial way. 

For the Ricci tensor we have
\begin{align}
    R_{a b} &= \Gamma^c_{a b,c} - \Gamma^c_{a c,b} + \Gamma^c_{c d} \Gamma^d_{a b} - \Gamma^c_{d a} \Gamma^d_{c b} \nonumber \\
    \delta R_{a b} &= \delta \Gamma^c_{a b, c} - \delta \Gamma^c_{a c ,b} + \delta \Gamma^c_{c d} \Gamma^d_{a b} + \Gamma^c_{c d} \delta \Gamma^d_{a b} - \delta \Gamma^c_{d a} \Gamma^d_{c b} - \Gamma^c_{d a} \delta \Gamma^d_{c b}.
\end{align}

Note,
\begin{align}
    \na_a \de \G^b_{c d} = \de \G^b_{c d,a} + \G^b_{a e} \de \G^e_{cd}-\G^e_{a c} \de \G^b_{ed} - \G^e_{a d} \de \G^b_{c e}
\end{align}

This allows us to replace:
\begin{align}
    \de \G^c_{a b,c} &= \na_c \de \G^c_{ab} - \G^c_{cd} \de \G^d_{ab} + \G^d_{ca} \de \G^c_{bd} + \G^d_{cb} \de \G^c_{ad} \nonumber \\
    \de \G^c_{ac,b} &= \na_b \de \G^c_{ac} - \G^c_{bd} \de \G^d_{ca} + \G^d_{bc} \de \G^c_{da} + \G^d_{ba} \de \G^c_{dc} \nonumber \\
    &= \na_b \de \G^c_{ac} + \G^d_{ba} \de \G^c_{dc}
\end{align}
so that
\begin{align}
    \de R_{ab} &= \left(\na_c \de \G^c_{ab} - \G^c_{cd} \de \G^d_{ab} + \G^d_{ca} \de \G^c_{bd} - \G^d_{cb} \de \G^c_{ad}\right) \nonumber \\
    &\qquad - \left(\na_b \de \G^c_{ac} + \G^d_{ba} \de \G^c_{dc}\right) \nonumber \\
    &\qquad + \delta \Gamma^c_{c d} \Gamma^d_{a b} + \Gamma^c_{c d} \delta \Gamma^d_{a b} - \delta \Gamma^c_{d a} \Gamma^d_{c b} - \Gamma^c_{d a} \delta \Gamma^d_{c b} \nonumber \\
    &= \na_c \de \G^c_{ab} - \na_b \de \G^c_{ac} \nonumber \\
    &= \f{1}{2}\left[ \na^d \left( \na_a \de g_{bd} + \na_b \de g_{ad} - \na_d \de g_{ab}\right) - \na_b \na_a \left( g^{cd} \de g_{cd}\right)\right] \nn 
    \therefore g^{a b} \de R_{a b} &= \na^b \na^a \de g_{ab} -\na_c \na^c (g^{ab} \de g_{a b}) \nn 
    &= \nabla_a v^a,
\end{align}
where
\begin{align}
    v_a &= \na^b \de g_{a b} - \na_a (g^{bc} \de g_{bc}).
\end{align}

For the variation in $R$, we have
\begin{align}
    R &= g^{ab} R_{ab} \nn 
    \de R&= g^{ab} \de R_{ab} + \de g^{ab} R_{ab} \nn 
    &= g^{a b} \de R_{ab} - g^{ac} g^{bd} \de g_{cd} R_{ab} \nn
    &= g^{ab} \de R_{ab} - R^{ab} \de g_{ab} \nn 
    &= \na_a v^a - R^{a b} \de g_{a b}.
\end{align}

This implies the variation in $\de(\sqrt{-g} R)$ is
\begin{align}
    \de (\sqrt{-g} R) &= \sqrt{-g} \de R + R \de \sqrt{-g} \nn 
    &= \sqrt{-g} \left( \na_a v^a - R^{a b} \de g_{ab} + \f12 R g^{ab}\de g_{ab}\right) \nonumber \\
    &= \sqrt{-g} \left( \na_a v^a - G^{a b} \de g_{ab}\right).
\end{align}

We thus have
\begin{align}
    \bs L_{EH} = L_{EH} \bs \epsilon = \frac{1}{16\pi} (R - 2 \La) \bs \ep,
\end{align}
with
\begin{align}
    \de \bs \epsilon = \f12 (g^{ab} \de g_{ab}) \bs \ep,
\end{align}
so that
\begin{align}
    \de \bs L_{EH} &= \f1{16\pi} \left( \delta R \bs \ep + (R-2\La) \de \bs \ep \right) \nn 
    &= \f1{16\pi} \left( \na_a v^a - (G^{ab} + \La g^{ab} )\de g_{ab}\right) \bs \ep \nn 
    &= -\f1{16\pi} (G^{ab} + \La g^{ab}) \de g_{ab} \bs \ep + d \bs \Th (g_{ab},\de g_{ab}),
\end{align}
where 
\begin{align}
    \bs \Th_{abc} &= \frac{1}{16\pi}  v^d \bs \ep_{dabc}.
\end{align}

This can be written as
\begin{align}
    \de \bs L_{EH} &= \bs E^{a b} \de g_{ab} + d \bs \Th,
\end{align}
where
\begin{align}
    {\bs E^{a b}}_{c d e f} &= -\f{1}{16\pi} (G^{ab} + \La g^{ab}) \bs \ep_{c d ef}.
\end{align}

\subsection{Maxwell Field} \label{MaxwellTerms}

The Maxwell field can also be considered. Notation, conventions etc.~largely follow Gao and Wald \cite{GaoWald}. I will focus on the four-dimensional case. The field $\bs A$ is a one-form (``vector potential'') with associated physical Maxwell field $\bs F = d \bs A$. The electromagnetic Lagrangian term $\bs L_{EM}$ is 
\begin{align}
    \bs L_{EM} &= - \frac{1}{16\pi} (g^{a c} g^{b d} \bs F_{a b} \bs F_{cd}) \bs \epsilon.
\end{align}
Its variation is
\begin{align}
    \de \bs L_{EM} &= \left(-\f1{16\pi} \de (g^{ac} g^{b d} \bs \epsilon) \right) \bs F_{ab} \bs F_{cd} - 2 \left(\f1{16\pi} g^{ac} g^{bd} \bs \epsilon\right) \bs F_{a b} \de \bs F_{cd}, 
\end{align}
where the factor of 2 accounts for the fact that $g^{a c} g^{b d} \de (\bs F_{a b} \bs F_{cd}) = 2 g^{a c} g^{bd} \bs F_{ab} \de \bs F_{cd}$.

The variation commutes with the exterior derivative so $\de \bs F_{cd} = (d \de \bs A)_{c d} = 2 \na_{[c}(\de \bs A_{d]})$. For the second term,
\begin{align}
    \bs F^{c d} \na_{[c} \de \bs A_{d]} &= \bs F^{c d} \na_c \de \bs A_{d},
\end{align}
by antisymmetry of $\bs F$. 

The second term becomes
\begin{align}
    -2 \f{1}{16\pi} \bs \ep \left( \bs F^{cd} \de \bs F_{cd}\right)  &- \f1{4\pi} \bs \epsilon \left( \bs F^{c d} \na_c \de \bs A_d\right) \nn
    &= -\f1{4\pi} \bs \ep \left( \na_c (\bs F^{cd} \de \bs A_d) - \de \bs A_d \na_c \bs F^{cd} \right) \nn
    &=  \f1{4\pi} (\de \bs A_d \na_c \bs F^{c d}) \bs \ep + d\bs \Th^{EM},
\end{align}
where
\begin{align}
    \bs \Th^{EM}_{abc} &= -\f 1{4\pi} \bs F^{d e} \de \bs A_e \bs \ep_{d a b c}.
\end{align}
The first term is
\begin{align}
    -\f1{16\pi} \de (g^{a c} g^{bd} \bs \ep) \bs F_{a b} \bs F_{cd} &= -\f{2}{16\pi} g^{ac} \de g^{bd} \bs \ep \bs F_{ab} \bs F_{cd} - \f{1}{16\pi} g^{ac} g^{bd} \de \bs \ep \bs F_{ab} \bs F_{cd} \nn
    &= +\f1{8\pi} {\bs F^c}_b \bs F_{c d} g^{be} g^{df} \de g_{e f} \bs \ep - \f{1}{32\pi} \bs F^{cd} \bs F_{cd} g^{ef} \de g_{ef} \bs \ep \nn
    &= \f1{8\pi} \bs F^{ce} {\bs F_c}^f \de g_{ef} \bs \ep - \f1{32\pi} \bs F^{cd} \bs F_{cd} g^{ef} \de g_{ef} \bs \ep \nn
    &= \f1{8\pi} \left[\left( \bs F^{c a} {\bs F_c}^b - \f1{4} \bs F^{cd} \bs F_{c d} g^{ab}\right) \de g_{ab} \right] \bs \epsilon.
\end{align}
The variation overall is
\begin{align}
    \de \bs L_{EM} &= \f1{8\pi} \left[\left( \bs F^{c a} {\bs F_c}^b - \f1{4} \bs F^{cd} \bs F_{c d} g^{ab}\right) \de g_{ab} \right] \bs \epsilon + \f1{4\pi} (\de \bs A_d \na_c \bs F^{c d}) \bs \ep + d \bs \Th^{EM}.
\end{align}

Letting $T^{EM}_{ab}$ be the stress-energy tensor of the electromagnetic field, 
\begin{align}
    T^{EM}_{a b} = \f1{4\pi} \left(\bs F_{a c} {\bs F_b}^c - \f 1 4 g_{ab} \bs F_{de} \bs F^{de}\right), \label{TEM}
\end{align}
we have
\begin{align}
    \de \bs L_{EM} &= \f{8\pi}{16\pi} T^{a b}_{EM} \de g_{ab} \bs \ep + \f1{4\pi} \na_c  \bs F^{cd} \de \bs A_d \bs \ep + d \bs \Th^{EM},
\end{align}
so that for Einstein--Maxwell theory (with no other fields)
\begin{align}
    \de \bs L &= \f{1}{16\pi} \bs \ep \left( -G^{ab} - \La g^{ab} + 8 \pi T^{ab}_{EM} \right) \de g_{ab} + \f1{4\pi} \bs \ep \na_a \bs F^{ab} \de \bs A_b + d \bs \Th,
\end{align}
where
\begin{align}
    \bs \Th &= \f1{16\pi} (v_{EH} + v_{EM}) \cdot \bs \ep
\end{align}
where 
\begin{align}
    (v_{EH})_a &= \na^b \de g_{a b} - \na_a (g^{b c} \de g_{b c}) \nn
    (v_{EM})_a &= - 4 {\bs F_a}^b \de \bs A_b. \label{vehvem}
\end{align}

The Noether charge differential form becomes \eqref{NoetherChargeEM}.

\section{Other Calculations Related to the Difference in Christoffel Symbols in the Kerr--Schild Case} \label{DeltaAppendix}

In the Kerr--Schild case, $\Delta^a_{bc} = \G^a_{bc} - \overline \G^a_{bc}$ is given by \eqref{Deltaabc}. I make a few calculations here, for use in Section \ref{KBLSection}.

$\Delta^a_{ba}$ is given by
\begin{align}
    \Delta^a_{ba} &= \f12 (\bar \na_a h^a_b + \bar \na_b h^a_a - \bar \na^a h_{ab}) - \f12 k^d \pa_d H k^a h_{ba} \nn 
    &= 0. \label{Deltaaba}
\end{align}

$\bar g^{bc} \Delta^a_{bc}$ is given by \cite{Taub81} $\bar g^{bc} \Delta^a_{bc} = 2 k^a \bar \na_b (H k^b)$, which implies
\begin{align}
    \bar g^{bc} \Delta^a_{bc} &=  (V^{EH})^a \label{gbcDeltaabc}
\end{align}
from \eqref{VEHdef}.

We note,
\begin{align}
    k^a \bar \na_a h_{bc} &= k^a \bar \na_a (H k_b k_c) \nn 
    &= k_b k_c k^a \bar \na_a H,
\end{align}
using the affinely parametrized geodesic condition. Also,
\begin{align}
    k^a \bar \na_b h_{ac} &= k^a \bar \na_b (H k_a k_c) \nn 
    &= k^a ( H k_c \bar \na_b k_a + k_a \bar \na_b (H k_c)).
\end{align}
Of course $k^a k_a = 0$. Also, $k^a \bar \na_b k_a = \f12 \bar \na_b (k^a k_b) = 0$. Thus we have 
\begin{align}
    k^a \bar \na_b h_{ac} = 0.
\end{align}
From these we conclude (results which appear in \cite{Taub81})
\begin{align}
    k^b \Delta^a_{bc} &= \f12 k^a k_c k^b \bar \na_b H \nn 
    k_a \Delta^a_{bc} &= -\f12 k_bk_c k^a \bar \na_a H. \label{kDelta}
\end{align}
In particular, this implies
\begin{align}
    h^{bc} \Delta^a_{bc} &= (H k^b k^c) \f12 k^a k_c \bar \na_b H \nn 
    &= 0. \label{hbcDeltaabc}
\end{align}

Using the ``full metric covariant derivative'' form of $\Delta^a_{bc}$, \eqref{Deltaabcfull}, we can calculate $t^a_b$ from \eqref{tab}, which can be rewritten as
\begin{align}
    16\pi t^a_b &= g^{cd} \Delta^f_{ce} \Delta^e_{d h} (-2 \delta^a_f \delta^h_b + \delta^a_b \delta^h_f).
\end{align}
We will now show $g^{cd} \Delta^f_{ce} \Delta^e_{dh} = 0$. 

Recall $k_ak^a = 0, k_a \na^a k^b = 0, k_a \na^b k^a = \f12 \na^b (k^a k_a) = 0$. Recall that $\Delta^f_{ce} = \Delta^f_{(ce)}$. Let us calculate,
\begin{align}
    g^{cd} \Delta^e_{dh} &= \f12 (\na^c h^e_h + \na_h h^{ec} - \na^e h^c_h) - \f12 k^f \pa_f H k^e h^c_h \nn 
    \therefore g^{d(c}\Delta^{e)}_{dh} &= \f12 \na_h h^{ec} - \f12 k^f \pa_f H k^e h^c_h \nn 
    g^{cd} \Delta^a_{ce} \Delta^e_{dh} &= \Delta^f_{(ce)} g^{d(c} \Delta^{e)}_{dh} \nn 
    &= \f12 \Delta^f_{ce} (\na_h h^{ec} - k^g \pa_g H k^e h^c_h) \nn 
    &= \f12 \Delta^f_{ce} ( \de^g_h - h^g_h )\na_g h^{ec} \nn
    &= \f12 \Delta^f_{ce}(\de^g_h - h^g_h) ( \na_g \ln H h^{ec} + H \na_g k^e k^c + H k^e \na_g k^c ).
\end{align}
$h^{ec}\Delta^f_{ce} = 0$ from \eqref{hbcDeltaabc}. The other terms give, using \eqref{kDelta}.
\begin{align}
    g^{cd} \Delta^f_{ce} \Delta^e_{dh} &=  \f12 H (\de^g_h - h^g_h) (\na_g k^e k^c k^f k_e \na_c H +\na_g k^e k^c k^f k_c \na_e H )\nn 
    &= 0
\end{align}
since $k^e k_e = k^c k_c = 0$. Thus we verify that $t^a_b = 0$ in the Kerr--Schild case. 

\section{Expression for \texorpdfstring{$\bs k^{EH}_{\bar \chi}$}{kEH bar chi}} \label{kxicomparison}

Here I show that $\bs k^{EH}_{\bar \chi}[h,\bar g]$ from \eqref{kEHincludingdeltachi} (setting $\bar g_{ab}$ for the background metric) is equal to its value in \eqref{kbarxiterm}, plugging in $ h_{ab}$ and $\bar g_{ab}$ for $\de g_{ab}$ and $g_{ab}$ respectively into \eqref{kEHincludingdeltachi}, as well as using $\bar \chi^a$ instead of $\chi^a$ throughout and and using the Einstein--Hilbert gravity terms throughout.  

To start with, let $\bar \chi^a$ be a Killing vector with $\de \bar \chi^a = 0$ (unvarying components) and $\chi^a = \bar \chi^a$ in the unvaried spacetime. Then \eqref{kEHincludingdeltachi} implies 
\begin{align}
    \bs k_\chi[\de g;g] &= \bs k_{\bar \chi}[\de g;g] \nn 
    &= -\de \bs K^K_{\bar \chi} - \bar \chi \cdot \bs \Th[\de g;g].
\end{align}

Using $D=4$ as a representative example, using the expression for $\bs \Th^{EH}(\phi,\de \phi)$ from \eqref{ThetaEH},
\begin{align}
    -\de \bs K^K_{\bar \chi} - \bar \chi \cdot \bs \Th^{EH}(\phi,\de \phi) &= \f1{16\pi} \left( - \de (\bs \ep_{cdab} \na^c \bar \chi^d) - v^c \bar \chi^d \bs \ep_{cdab})\right) \nn
    &= \f1{16\pi} \bs \ep_{cdab} \left( -\f1{\sqrt{-g}} \de (\sqrt{-g} \na^{[c} \bar \chi^{d]}) - v^{[c} \bar \chi^{d]}\right),
\end{align}
using the fact that $\bs \ep_{abcd} = \sqrt{-g} e_{abcd}$. $v^d$ is as given in \eqref{ThetaEH}; I drop the $EH$ superscript for compactness of writing. Throughout this Appendix section I am using $h_{ab}$ to represent $h^{BC}_{ab}$ (not the Kerr--Schild correction), to simplify the notation. Writing $h_{ab} \equiv \de g_{ab}$ with $h = \bar g^{ab} h_{ab}, h^a_b = \bar g^{ac} h_{cb}$ and so on, and calculating covariant derivatives with respect to the background $\bar g_{ab}$,
\begin{align}
    v^a &= \bar \na_b h^{ab} - \bar \na^a h.
\end{align}
For the $-\de \bs K^K_{\bar\chi}$ term, note that assuming $\de \bar \chi^a = 0$ and $\de g_{a b} = h_{a b}$, we can write
\begin{align}
    \de \na^a \bar \chi^b &= \de ( g^{a c} \na_c \bar \chi^b) \nn
    &= \de g^{ac} \bar \na_c \bar \chi^b + \bar g^{ac} \de (\na_c \bar \chi^b) \nn 
    &= -h^{ac} \bar \na_c \bar \chi^b + \bar g^{ac} \de (\pa_c \bar \chi^b + \G^b_{cd} \bar \chi^d).
\end{align}
Because the variation and the partial derivative commute, $\de \pa_c \bar \chi^b = \pa_c (\de \bar \chi^b) = 0$. We can use \eqref{deltaGamma} to give $\de \G^b_{cd}$, modified so that $\de g_{ab} = h_{ab}$ and the metric and covariant derivative are associated with the background,
\begin{align}
    \de \G^b_{cd} &= \f12 (\bar \na_d h^b_c + \bar \na_c h^b_d - \bar \na^b h_{cd}).
\end{align}
Thus we have
\begin{align}
    \de \na^a \bar \chi^b &= - h^{ac} \bar \na_c \bar \chi^b + \bar g^{ac} \de \G^b_{cd} \bar \chi^d \nn
    &= -h^{ac} \bar \na_c \bar \chi^b + \bar g^{ac} \f12 (\bar \na_c h^b_d + \bar \na_d h^b_c - \bar \na^b h_{cd}) \bar \chi^d \nn 
    &= -h^{ac} \bar \na_c \bar \chi^b + \f12 (\bar \na^a h^b_d + \bar \na_d h^{ab} - \bar \na^b h^a_d) \bar \chi^d \nn 
    &= -h^{ac} \bar \na_c \bar \chi^b + \bar \na^{[a} h^{b]}_c \bar \chi^c + \f12 \bar \na_c h^{ab} \bar \chi^c.
\end{align}
Antisymmetrizing,
\begin{align}
    \de \na^{[a}\bar \chi^{b]} &= -h^{c[a}\bar \na_c\bar \chi^{b]} + \bar\na^{[a} h^{b]}_c \bar \chi^c,
\end{align}
using $h^{[ab]} = 0$ ($h^{ab}$ is symmetric). To bring my expression into agreement with BC's I will apply the Killing identity for $\bar \chi$, $\na^a \bar \chi^b = -\na^b \bar \chi^a$. Letting $\bar \chi_c = \bar g_{cd} \bar \chi^d$, 
\begin{align}
    -h^{c[a} \bar \na_c \bar \chi^{b]}&= \f12 h^{c[b} \bar \na_c \bar \chi^{a]} + \f12 h^{c[a} \bar \na^{b]} \bar \chi_c.
\end{align}
Thus
\begin{align}
    \de \na^{[a} \bar\chi^{b]} &= \bar \na^{[a} h^{b]}_c \bar \chi^c + \f12 h^{c[b} \bar \na_c \bar \chi^{a]} + \f12 h^{c[a} \bar \na^{b]}\bar\chi_c.
\end{align}

The variation in $\sqrt{-g}$ is $\de \sqrt{-g} = \f12 \sqrt{-\bar g} \bar g^{ab} \de g_{ab} =\f12 \sqrt{-\bar g} h$, so
\begin{align}
    -\f1{\sqrt{-g}} \de ( \sqrt{-g} \na^{[a}\chi^{b]}) &= -\f{\de \sqrt{-g}}{\sqrt{-\bar g}} \bar\na^{[a}\bar\chi^{b]} - \de \na^{[a}\bar\chi^{b]} \nn 
    &= \f12 h \bar\na^{[b}\bar\chi^{a]} + \bar\na^{[b} h^{a]}_c \bar\chi^c + \f12 h^{c[a} \bar\na_c \bar\chi^{b]} + \f12 h^{c[b} \bar\na^{a]}\bar\chi_c.
\end{align}
Combining with $v^{[a}\bar\chi^{b]}$,
\begin{align}
    -\f1{\sqrt{-\bar g}} \de (\sqrt{-g} \na^{[a}\bar\chi^{b]}) - v^{[a}\bar\chi^{b]} &= -\f{\de \sqrt{-\bar g}}{\sqrt{-\bar g}} \bar\na^{[a}\bar\chi^{b]} - \de \na^{[a}\chi^{b]} \nn 
    &= \f12 h \bar\na^{[b}\bar \chi^{a]} + \bar\na^{[b} h^{a]}_c \bar\chi^c + \f12 h^{c[a} \bar\na_c \bar\chi^{b]} + \f12 h^{c[b} \bar\na^{a]}\bar\chi_c\nn
    &\qquad + \bar\chi^{[a} \bar\na_c h^{b] c} + \bar\chi^{[b}\bar\na^{a]} h,
\end{align}
switching the $a,b$ order when necessary to keep every term positive. Thus we recover \eqref{kbarxiterm}.

\section{Schwarzschild Interior Calculation} \label{schwint}

Here are collected some of the calculations regarding the Schwarzschild Interior metric as appearing in Section \ref{TxiGauss}.

\begin{align}
    - \int T^a_b \xi^b d \Sigma_a &= - \int T^t_t \sqrt{-g} dr d \tht d \phi \nn 
    &= 4 \pi \rho \int_0^a dr \: r^2 \sqrt{-g_{tt} g_{rr}} \nn 
    &= 4 \pi \rho \int_0^a dr \: r^2 \left(\frac{3}{2} \sqrt{\frac{1-2m/a}{1-2mr^2/a^3}} - \f12\right) \nn 
    &= 4 \pi \rho \left[ \frac{3 a^4}{8m} \left( \sqrt{\f{a}{2m}-1} \arcsin{\sqrt{\f{2m}{a}}} - 1 + \f{2m}{a}\right) - \f{a^3}{6}\right] \nn 
    &= \f74 m - \f 9 8 a \left(1 - \sqrt{\f{a}{2m}-1} \arcsin\sqrt{\f{2m}{a}} \right).
\end{align}

To calculate the limit as $a$ becomes very large compared to $m$, let $x = 2m/a$:
\begin{align}
    -\int T^a_b \xi^b d\Si_a &= \f74 m - \f9 8 \f{2m}{x} \left(1 - \f1{\sqrt{x}} \sqrt{1-x} \arcsin \sqrt{x}\right) \nn 
    &= \f74 m - \f{9 m ( \sqrt{x} - \sqrt{1-x} \arcsin \sqrt{x})}{4 x^{3/2}} .
\end{align}
We now take the limit as $x \to 0$. Let $x = \sqrt y$. The second term is
\begin{align}
    -\frac{9 m (y - \sqrt{1-y^2} \arcsin y)}{4 y^3}.
\end{align}
We can Taylor expand the numerator. The well-known Taylor expansion of $\arcsin y$ is $\arcsin y = y + y^3/6 + \mc O(y^5)$; we also have $\sqrt{1-y^2} = 1 - y^2/2 + \mc O(y^4)$. Thus we have
\begin{align}
    y - \sqrt{1-y^2} \arcsin y = y^3/3 + \mc O(y^5),
\end{align}
so that
\begin{align}
    \lim_{y \to 0} -\frac{9 m(y-\sqrt{1-y^2} \arcsin y)}{4 y^3} = -\f{3m}{4},
\end{align}
giving
\begin{align}
    \lim_{(2m/a)\to 0} T^a_b \xi^b d \Si_a &= \f74 m - \f34 m = m.
\end{align}

\section{Komar Difference Term Calculation} \label{KomarDifference}

Let us proceed as follows. Assume that $\chi$ is a Killing vector field in both metrics and that its contravariant components are constant. We can write,
\begin{align}
    \na^a \chi^b &= g^{a c} g^{b d} \na_c \chi_d \nn 
    &= g^{ac} g^{bd} \na_{[c} \chi_{d]} \nn 
    &= g^{a c} g^{b d} \pa_{[c} \chi_{d]},
\end{align}
using Killing's equation. We can write,
\begin{align}
    \pa_{[c} \chi_{d]} &= \pa_{[c} (g_{d] e} \chi^e) \nn 
    &= \pa_{[c} (\bar g_{d] e}\chi^e + h_{d] e}\chi^e) \nn 
    &= \overline{\na_c \chi_d} + \pa_{[c} (h_{d] e} \chi^e),
\end{align}
where $\overline{\na_c \chi_d} = \bar \na_c (\bar g_{de} \chi^e) = \pa_{[c}(\bar g_{d]e}\chi^e)$ is $\na_c \chi_d$ calculated in the background metric. Since $h_{d e} \chi^e$ are the components of a one-form, we have $\pa_{[c} (h_{d] e} \chi^e) = \bar \na_{[c}(h_{d] e} \chi^e)$. We can also replace $\overline{\na_c \chi_d}$ with the lowered version of $\overline{\na^e \chi^f}$, which is equal to $\bar \na^e \chi^f$ (since $\bar \chi^f = \chi^f$). Then,
\begin{align}
    \pa_{[c}\chi_{d]} &= \bar g_{c e} \bar g_{d f} \bar \na^e \chi^f + \bar \na_{[c} (h_{d]e}\chi^e).
\end{align}
Here $\bar \na^e = \bar g^{e h} \bar \na_h$. 

Consequently,
\begin{align}
    \na^a \chi^b &= g^{ac} g^{b d} \left( \overline{g_{c e} g_{d f} \na^e \chi^f} + \bar \na_{[c} (h_{d] e}\chi^e)\right) \nn 
    &= (\bar g^{ac} - h^{ac})(\bar g^{bd} - h^{bd})\left( \overline{g_{c e} g_{d f} \na^e \chi^f} + \bar \na_{[c} (h_{d] e}\chi^e)\right).
\end{align}
The first set of terms give
\begin{align}
    g^{ac} g^{b d}  \overline{g_{c e} g_{d f} \na^e \chi^f}
    = \overline{\na^a \chi^b} - h^a_e \overline{\na^e \chi^b} - h^b_f \overline{\na^a \chi^f} + h^a_e h^b_f \overline{\na^e \chi^f}.
\end{align}
$h^a_e h^b_f = H^2 k^a k^b k_e k_f$ is symmetric in $e,f$, whereas $\overline{\na^e \chi^f}$ is antisymmetric (since $\chi$ is a Killing vector of the background), so the final term vanishes. We also have
\begin{align}
    -h^a_e \overline{\na^e \chi^b} - h^b_f \overline{\na^a \chi^f} &= - h^a_e \overline{\na^e \chi^b} + h^b_f \overline{\na^f \chi^a} \nn 
    &= -h^{ae} \overline{\na_e \chi^b} + h^{b e} \overline {\na_e \chi^a} \nn 
    &= -2 h^{e[a} \overline{\na_e \chi^{b]}}.
\end{align}
So the first set of terms is
\begin{align}
    g^{a c} g^{b d} \overline{ g_{ce} g_{df} \na^e \chi^f} = \overline{ \na^a \chi^b} - 2 h^{e[a} \overline{\na_e \chi^{b]}}.
\end{align}

We also have
\begin{align}
    g^{ac} g^{bd} \bar \na_{[c} (h_{d]e} \chi^e) &= g^{ac} g^{bd} (\bar \na_{[c} h_{d]e}) \chi^e + g^{ac} g^{bd} h_{e[d} \bar \na_{c]}\chi^e
\end{align}

\begin{align}
    g^{ac} g^{bd} \bar \na_c h_{de} &= g^{a c} (\bar g^{bd} \bar \na_c h_{de} - h^{bd} \bar \na_c h_{de}) \nn 
    &= g^{ac} (\bar \na_c h^b_e - h^{bd} \bar \na_c h_{de}).
\end{align}
\begin{align}
    h^{bd} \bar \na_c h_{de} &= H k^b k^d \bar \na_c (H k_d k_e) \nn 
    &= H k^b k^d (\bar \na_c(H k_e) k_d + H k_e \bar \na_c k_d) \nn 
    &= 0,
\end{align}
since $k^d k_d = k^d \bar \na_c k_d = 0$. Thus
\begin{align}
    g^{ac} g^{bd} \bar \na_{[c} h_{d]e} &= g^{c[a} g^{b] d} \bar \na_c h_{de}.
\end{align}
Consider first the term before antisymmetrization.
\begin{align}
    g^{ac} g^{bd} \bar \na_c h_{de} &= g^{ac} \bar \na_c h^b_e \nn 
    &= \bar \na^a h^b_e - h^{ac} \bar \na_c h^b_e.
\end{align}
For the latter term,
\begin{align}
    h^{ac} \bar \na_c h^b_e &= H k^a k^c \bar \na_c h^b_e \nn 
    &= H k^a k^c k^b k_e \bar \na_c H \nn 
    &= h^{ac} h^b_e \bar \na_c \ln|H|,
\end{align}
using the null geodesic condition. We conclude,
\begin{align}
    g^{ac} g^{bd} \bar \na_c h_{de} &= \bar \na^a h^b_e - h^{ac} h^b_e \bar \na_c \ln |H|.
\end{align}
Under antisymmetrization, $h^{c[a} h^{b]}_e = k^ck^{[a}k^{b]}k_e H^2 = 0$. So we have
\begin{align}
    g^{a c} g^{bd} \bar \na_{[c}h_{d]e} &= \bar \na^{[a}h^{b]}_e.
\end{align}

Further,
\begin{align}
    g^{ac} g^{bd} h_{e[d} \bar \na_{c]} \chi^e &= g^{c[a}g^{b]d} h_{d e} \bar \na_c \chi^e \nn 
    &= (\bar g^{c[a} - h^{c[a} ) h^{b]}_e \bar \na_c \chi^e \nn 
    &= h^{[b}_e \bar \na^{a]} \chi^e,
\end{align}
using $h^{c[a}h^{b]}_e = H^2 k^c k^{[a}k^{b]} k_e = 0$. So
\begin{align}
    g^{ac} g^{bd} \bar \na_{[c} (h_{d]e} \chi^e) &= \bar \na^{[a}h^{b]}_e\chi^e + h^{[b}_e \bar \na^{a]}\chi^e 
\end{align}

Putting it all together,
\begin{align}
    \na^a \chi^b &= \overline{\na^a \chi^b} - 2 h^{e[a} \bar \na_e \chi^{b]} + \bar \na^{[a}h^{b]}_e\chi^e + h^{[b}_e \bar \na^{a]}\chi^e \nn 
    &= \overline{\na^a \chi^b} - 2 \bar \na^e \chi^{[b} h^{a]}_e + \bar \na^{[a} h^{b]}_e \chi^e + h^{[b}_e \bar \na^{a]} \chi^e.
\end{align}
Using $\bar \na^e \chi^b = -\bar \na^b \chi^e$, the second term on the right-hand side can be rewritten as
\begin{align}
    -2 \bar \na^e \chi^{[b} h^{a]}_e = 2 h^{[a}_e\bar \na^{b]}\chi^e
\end{align}
so we have
\begin{align}
    \na^a \chi^b &= \overline{\na^a \chi^b} + h^{[a}_e \bar \na^{b]} \chi^e + \bar \na^{[a} h^{b]}_e \chi^e.
\end{align}
The second term on the right-hand side can now be expanded as, using the Killing condition,
\begin{align}
    h^{[a}_e \bar \na^{b]} \chi^e &= \f12 \left( h^a_e \bar \na^b \chi^e - h^b_e \bar \na^a \chi^e\right) \nn 
    &= \f12 \left( - h^a_e \bar \na^e \chi^b + h^b_e \bar \na^e \chi^a\right) \nn 
    &= \f12 \left( - h^{e a} \bar \na_e \chi^b + h^{e b} \bar \na_e \chi^a \right) \nn 
    &= - h^{e[a} \bar \na_e \chi^{b]}, 
\end{align}
giving
\begin{align}
    \na^a \chi^b &= \overline{\na^a \chi^b} - h^{e[a} \bar \na_e \chi^{b]} + \bar \na^{[a} h^{b]}_e \chi^e. \label{Komardifferenceresult}
\end{align}

Because the proofs did not rely on the particular form of the background or full spacetime, except that they obey the Kerr--Schild form and that $\chi^a$ is a Killing vector of both, we can also run the same proof ``in reverse'' substituting $\bar g_{ab} \to g_{ab}, g_{ab} \to \bar g_{ab}, h_{ab} \to -h_{ab}$, and so get
\begin{align}
    \overline{\na^a \chi^b} &= \na^a \chi^b + h^{e[a}\na_e \chi^{b]} - \na^{[a}h^{b]}_e \chi^e,
\end{align}
where the covariant derivatives are with respect to the full metric. Thus we have
\begin{align}
    \na^a \chi^b - \overline{\na^a \chi^b} &= - h^{e[a} \bar \na_e \chi^{b]} + \bar \na^{[a} h^{b]}_e \chi^e = - h^{e[a}  \na_e \chi^{b]} + \na^{[a} h^{b]}_e \chi^e.
\end{align}

An alternate, shorter demonstration (not appearing in the thesis) is the following. We have
   \begin{align}
       \na^a \chi^b - \overline{\na^a \chi^b} &= g^{ac} \na_c \chi^b - \overline{g^{ac} \na_c \chi^b} \nn 
       &= (\bar g^{ac} - h^{ac}) (\bar \na_c \chi^b + \Delta^b_{cd} \chi^d) - \overline{g^{ac} \na_c \chi^b} \nn 
       &= -h^{ac} \bar \na_c \chi^b + \bar g^{ac} \Delta^b_{cd} \chi^d - h^{ac} \Delta^b_{cd} \chi^d.
   \end{align}
   We also have that since both $\na^a \chi^b = \na^{[a}\chi^{b]}, \overline{\na^a \chi^b} = \overline{\na^{[a}\chi^{b]}}$, we also have $\na^a \chi^b - \overline{\na^a \chi^b} = \na^{[a}\chi^{b]} - \overline{\na^{[a} \chi^{b]}}$. This means,
   \begin{align}
       \na^a \chi^b - \overline{\na^a \chi^b} &= - h^{c[a} \bar \na_c \chi^{b]} + \bar g^{c[a} \Delta^{b]}_{cd} \chi^d - h^{c[a} \Delta^{b]}_{cd} \chi^d.
   \end{align}
   Using \eqref{Deltaabc}, we can write
   \begin{align}
       \bar g^{c[a}\Delta^{b]}_{cd} &= \f12 \bar g^{c[a}\bar g^{b]e} (\bar \na_d h_{ec} + \bar \na_c h_{ed} - \bar \na_e h_{cd} +k^f \pa_f H k_e h_{cd}) \nn 
       &= \f12 \left( \bar \na_d h^{[ba]} + \bar \na^{[a} h^{b]}_d - \bar \na^{[b} h^{a]}_d + k^f \pa_f H h^{[a}_d k^{b]}\right) \nn 
       &= \bar \na^{[a} h^{b]}_d.
   \end{align}
   Using \eqref{kDelta},
   \begin{align}
       h^{c[a} \Delta^{b]}_{cd} &= H k^c k^{[a}\Delta^{b]}_{cd} \nn 
       &= \f12 H k^{[a} k^{b]} k_d k^c \na_c H \nn 
       &= 0.
   \end{align}
   We thus very quickly recover \eqref{Komardifferenceresult}.

\chapter{Appendices Related to Additional Note} \label{AdditionalNoteAppendix}

    This appendix chapter collects information related to the Additional Note, Chapter \ref{additionalNote}. 
    
    \section{ Frame Component Calculations}

    \subsection{Frame Components of \texorpdfstring{$\bt$}{beta}} \label{betaframecalculation}

    $\bt$ satisfies \eqref{betacanonicalbasis}. We then need the large-$r$ expansions of $Q_\mu$ and $S$. As I will show, except for $\mu = n$, the $Q_\al$ and $S$ all go as $\mc O(r^{-2})$, which results from the fact that $r$ does not appear in $X_\al$ and appears as $r^2$ in $U_\al$ and in $A^{(n)}$; $r^2$ is so much larger than the other $y_\al^2$ values in magnitude that it dominates the denominators (appearing in terms like $r^2 + y_\bt^2$). In $Q_n$, $X_n$ is of order $r^{2n}$ and $U_n$ is of order $r^{2(n-1)}$, so that $Q_n$ is of order $r^2$. This means that the $\bt_{\hat n}$ component is enhanced and the other $\bt$ frame components are suppressed as $r \to \infty$. Explicitly,
    \begin{align}
        Q_\mu &= \f{X_\mu}{U_\mu}.
    \end{align}
    For $\alpha \neq n$, using \eqref{Yofy} and \eqref{Xbaralpha}, 
    we have
    \begin{align}
        Q_\al &= \f{X_\al}{U_\al} \nn 
        &= (-1)^{1-\ve} \f{1-y_\al^2/l^2}{y_\al^{2\ve}}\f{\prod_{j=1}^{n-1+\ve} (a_j^2-y_\al^2) }{-(r^2+y_\al^2)\prod_{\bt\neq \al,\bt=1}^{n-1} (y_\bt^2-y_\al^2)} \nn &= \mc O(r^{-2}). \label{Qalphafalloff}
    \end{align}
    (This neglects the case $y_\bt^2 = y_\al^2$, or $y_\al = 0$ if $\ve = 1$; I believe that these contributions will disappear when integrating over all values of $y_\al,y_\bt$.) $S$ is
    \begin{align}
        S &= -\f{c}{A^{(n)}} \nn 
        &= \mc O(r^{-2}).
    \end{align}

    By contrast,
    \begin{align}
        Q_n &= -\f{X}{U_n} \nn 
        &\simeq -\f{r^2}{l^2}.
    \end{align}

    Consequently, we have
    \begin{align}
        \bt_{\hat n} & \simeq \f{i r}{l} \nn 
        \bt_{\hat \al} &\simeq \bt_{\hat 0} \simeq 0.
    \end{align}
    The sign choice is because $\bt$ is future-pointing, and so, from \eqref{canonicalformsinryphicoords}, is $i e_{\hat n}$ (which has $i e_{\hat n}^t > 0$).
    
    We also have $\bt_{\hat \al}, \bt_{\hat 0}$ both $\mc O(r^{-1})$.

    \subsection{Frame Components of \texorpdfstring{$dt$}{dt}, \texorpdfstring{$N_a$}{Na}} \label{dtcalculation}

    Consider $dt$, which can be expanded as
    \begin{align}
        dt &= \sum_A e^A e_A(t) \nn 
        &= - \f{(X-\bar X)}{\sqrt{X U_n}} \f{l^2}{r^2+l^2} e^n - \f{i \bar X}{\sqrt{X U_n}} \f{l^2}{r^2+l^2} e^{\hat n} - \sum_{\alpha=1}^{n-1} \sqrt{Q_\alpha} \f{l^2}{l^2-y_\alpha^2} e^{\hat \alpha} - \ve \sqrt S e^{\hat 0}, \label{dt}
    \end{align}
    using \eqref{canonicalformsinryphicoords}. The frame components $(dt)_A = e_A(t)$. 
    
        We have
    \begin{align}
        (dt)_{\hat \al} &= -\sqrt{Q_\al} \f{l^2}{l^2-y_\al^2} \nn 
        &= - \f{\sqrt{Q_\al (r^2+y_\al^2)}}{\sqrt{r^2+y_\al^2}} \f{l^2}{l^2-y_\al^2} \nn 
        &\simeq -\f{\sqrt{Q_\al(r^2+y_\al^2)}}{r} \f{l^2}{l^2-y_\al^2}.
    \end{align}
    Note that $Q_\al(r^2+y_\al^2)$ is $r$-independent (since $r$ appears only via $U_\al$). This means that $(dt)_{\hat \al} = \mc O(r^{-1})$. 

    $(dt)_{\hat 0}$ is
    \begin{align}
        (dt)_{\hat 0} &= -\sqrt S \nn 
        &= -\f{\sqrt c}{\prod_{\al=1}^{n-1} y_\al} \f{1}{r}.
    \end{align}

    $X-\bar X= -2 \mu(r) r^{1-\ve}$, which is $-2mr^{1-\ve}$ for Kerr--AdS. For Kerr--AdS, $(dt)_n$ is 
    \begin{align}
        (dt)_n &= - \f{(X-\bar X)}{\sqrt{X U_n}} \f{l^2}{r^2+l^2} \nn 
        &= \f{2mr^{1-\ve}}{\sqrt{X U_n}} \f{l^2}{r^2+l^2}.
    \end{align}
    In the large $r$ limit, $U_n \simeq r^{2(n-1)}$ and $X \simeq r^{2n}/l^2$, so that
    \begin{align}
        (dt)_n &\simeq \f{2 m l^3}{r^{D}}.
    \end{align}
    $(dt)_{\hat n}$ is
    \begin{align}
        (dt)_{\hat n} &= - \f{i \bar X}{\sqrt{ X U_n}} \f{l^2}{r^2+l^2} \nn 
        &\simeq -i \sqrt{Q_n} \f{l^2}{r^2} \nn 
        &\simeq -i \f{r}{l} \f{l^2}{r^2} \nn 
        &\simeq - \f{i l}{r}.
    \end{align}

    For the components of $N_a$, we should first calculate the normalization parameter. 
    From \eqref{backgroundds2rytphi}, $\bar g_{tt}$ is
    \begin{align}
        \bar g_{tt} &= -\f{(1+r^2/l^2)\prod_{\al=1}^{n-1}(1-y_\al^2/l^2)}{\prod_{j=1}^n \Xi_j}.
    \end{align}
    Thus $\bar g^{tt}$ is the inverse of this, which is $\mc O(r^{-2})$. We have
    \begin{align}
        g^{tt} = \bar g^{tt} - H (k^t)^2.
    \end{align}
    $H$ is $\mc O(r^{3-D})$ and $k^t$ is $\mc O(r^{-2})$, so that the combination $H (k^t)^2 = \mc O(r^{-1-D})$ which vanishes compared to $\bar g^{tt}$ for large $r$ (considering $D \geq 4$).

    In order to ensure that $N_a$ is a future-directed timelike unit vector when evaluated on a large-$r$ surface, we can set $N_a = -(dt)_a / \sqrt{-g^{tt}}$, which for large $r$ gives
    \begin{align}
        N_a &\simeq -\sqrt{-\bar g_{tt}} (dt)_a \nn 
        &\simeq -\f{r}{l} \sqrt{ \f{\prod_{\al=1}^{n-1} (1-y_\al^2/l^2)}{\prod_{j=1}^n \Xi_j}} (dt)_a
    \end{align}
    We then have frame components
    \begin{align}
        N_{\hat n} &\simeq + i\sqrt{ \f{\prod_{\al=1}^{n-1} (1-y_\al^2/l^2)}{\prod_{j=1}^n \Xi_j}} \nn 
        N_{\hat \al} &\simeq \f{l\sqrt{Q_\al(r^2+y_\al^2)}}{l^2-y_\al^2} \sqrt{ \f{\prod_{\al=1}^{n-1} (1-y_\al^2/l^2)}{\prod_{j=1}^n \Xi_j}} \nn 
        N_{\hat 0} &\simeq \f{\sqrt c}{ l \prod_{\al=1}^{n-1} y_\al} \sqrt{ \f{\prod_{\al=1}^{n-1} (1-y_\al^2/l^2)}{\prod_{j=1}^n \Xi_j}} \nn 
        N_{n} &\simeq -\f{2m l^2}{r^{D+1}} \sqrt{ \f{\prod_{\al=1}^{n-1} (1-y_\al^2/l^2)}{\prod_{j=1}^n \Xi_j}}.
    \end{align}
    $N_{\hat n}, N_{\hat \al}$ and $N_{\hat 0}$ are all $ \mc O(r^0)$, and $N_{n} = \mc O(r^{-D-1})$ and so is highly suppressed.

    We can also write $N_{\hat n}$ compactly as
    \begin{align}
        N_{\hat n} &\simeq +\f{i l}{r \sqrt{-\bar g^{tt}}}. \label{Nhatn}
    \end{align}

    \subsection{Frame Components of Weyl Tensor}

        For $D$ even, Hamamoto \cite{Hamamoto} gives $Q_T$ for the Einstein spaces. For these the $X_\mu$ are given by \eqref{Xmu}. For these, 
    \begin{align}
        Q_T &= l^{-2} \sum_{\mu=1}^n x_\mu^2 + c_{2(n-1)} + \sum_{\mu=1}^n \f{b_\mu x_\mu}{U_\mu}.
    \end{align}
    Adapting the above to the Generalized Kerr--AdS case, the full $Q_T$ is
    \begin{align}
        Q_T &= \sum_{\mu=1}^n Q_\mu \nn 
        &= \sum_{\mu=1}^n \bar Q_\mu + (Q_n-\bar Q_n) \nn 
        &= \bar Q_T + H,
    \end{align}
    where the $\bar Q_\mu$ are the values for AdS. These have $b_\mu =0 $ and so we have
    \begin{align}
        \bar Q_T &= \f{\sum_{\mu=1}^n x_\mu^2}{l^2} + c_{2(n-1)}.
    \end{align}
    Comparing to the expression for $\bar X_\mu = Y(x_\mu)$ we have $c_{2(n-1)} = - \sum_{j=0}^{n-1} a_j^2/l^2$, so that
    \begin{align}
        \bar Q_T &= \f{\sum_{\mu=1}^n x_\mu^2 - \sum_{j=0}^{n-1} a_j^2}{l^2}.
    \end{align}

    Consequently, in even dimensions,
    \begin{align}
        \bt^2 &= Q_T = \f{\sum_{\mu=1}^n x_\mu^2 - \sum_{j=0}^{n-1} a_j^2}{l^2} + H.
    \end{align}

    In odd dimensions the argument is essentially unaltered. We have
    \begin{align}
        Q_T + S &= (\bar Q_T + S) + H.
    \end{align}
    $\bar Q_T + S$ for AdS is, again from Hamamoto \cite{Hamamoto} (using the expression for the constant-curvature space, where $\hat V = 0$),
    \begin{align}
        \bar Q_T + S &= c_{2n} \sum_{\mu=1}^n x_\mu^2 + c_{2(n-2)},
    \end{align}
    with $c_{2n} = l^{-2}$. For the constant-curvature space, 
    \begin{align}
        (-1)^{n-1} X_\mu &= \sum_{k=-1}^n c_{2k} x^{2k},
    \end{align}
    where $c_{-2} = (-1)^{n+1} c$ where $c$ is the constant appearing in the $e^{\hat 0}$. Comparing with the expression for $Y(y)$ we have $c_{2(n-1)} = - \sum_{j=0}^n a_j^2/l^2$. So we have
    \begin{align}
        \bar Q_T + S &= \f{\sum_{\mu=1}^n x_\mu^2 - \sum_{j=0}^n a_j^2}{l^2} \nn 
        \bt^2 &= \f{\sum_{\mu=1}^n x_\mu^2 - \sum_{j=0}^n a_j^2}{l^2} + H.
    \end{align}
    For odd and even dimensions,
    \begin{align}
        \bt^2 &= \f{\sum_{\mu=1}^n x_\mu^2 - \sum_{j=0}^{n-1+\ve} a_j^2}{l^2} + H \nn 
        &= \overline{\bt^2} + H.
    \end{align}

    \subsection{Weyl Frame Components} \label{WeylFrameComponents}
    
    \subsubsection{Calculation of $C_{n \hat n \al \hat \al}$} \label{Cnhatnalhatal}

    Now, $C_{n\hat n \alpha \hat \alpha} = R_{n \hat n \alpha \hat \alpha}$, given by (from \eqref{Rmn})
    \begin{align}
        C_{n \hat n \alpha \hat \alpha} &= -2 q_{n\al} \nn 
        &= -\f{1}{x_n^2-x_\al^2}\left( x_\al \f{\pa \bt^2}{\pa x_n} - x_n \f{\pa \bt^2}{\pa x_\al}\right) \nn 
        &= \bar C_{n\hat n \al \hat \al} - \f{1}{x_n^2-x_\al^2} \left( x_\al \f{\pa H}{\pa x_n} - x_n \f{\pa H}{\pa x_\al}\right),
    \end{align}
    using the $\bt^2 = \overline{\bt^2} + H$ decomposition. The Weyl tensor is zero in the background so this is simply
    \begin{align}
        C_{n \hat n \al \hat \al} &= - \f{1}{x_n^2-x_\al^2} \left( x_\al \f{\pa H}{\pa x_n} - x_n \f{\pa H}{\pa x_\al}\right) \nn 
        &= \f{1}{r^2+y_\al^2} \left( -i y_\al \f{\pa H}{\pa r} - i r \f{\pa H}{\pa y_\al}\right) \nn 
        &= - \f{i}{r^2+y_\al^2} \left( y_\al \f{\pa H}{\pa r} + r \f{\pa H}{\pa y_\al}\right).
    \end{align}

    $H = 2 \mu(r) r^{1-\ve}/U_n$, and $y_\al$ appears only via the $(r^2+y_\al^2)$ factor in $U_n$. 
    \begin{align}
        C_{n \hat n \al \hat \al} &= -\f{i}{r^2+y_\al^2} \left( y_\al \f{\pa H}{\pa r} - \f{2 r y_\al}{r^2+y_\al^2} H\right) \nn
        &= -\f{i y_\al}{r^2+y_\al^2} \left( \f{\pa H}{\pa r} - \f{2 r}{r^2+y_\al^2} H\right). \label{Cnnhatalphaalphahat1}
    \end{align}

    For Kerr--AdS, $H = 2 m r^{1-\ve}/U_n$, which for large $r$ goes as $H \simeq 2 m r^{3-D}$. We then have, for large $r$,
    \begin{align}
        C_{n \hat n \al \hat \al} &\simeq \f{i (D-1) y_\al H}{r^3} \nn 
        &\simeq -2(D-1) m i y_\al r^{-D}. \label{Cnnhatalphaalphahat2}
    \end{align}
    Changing $m \to \mu(r) = m + O(r^{-1})$ (for example, for Kerr--Newman) will not affect the asymptotic form. $dS$ goes as $r^{D-2}$, and so the combination $C_{n \hat n \al \hat \al} dS$ will go as $r^{-2}$ and so will vanish in the large $r$ limit.

    \subsubsection{Calculation of $C_{n \hat n n \hat n}$} \label{Cnnnn}

    $R_{n \hat n n \hat n}$ is
    \begin{align}
        R_{n \hat n n \hat n} &= -\f12 \f{\pa^2 \bt^2}{\pa x_n^2} \nn 
        &= + \f12 \f{\pa^2 \bt^2}{\pa r^2} \nn 
        &= -l^{-2} + \f12 \f{\pa^2 H}{\pa r^2}.
    \end{align}

    We then have
    \begin{align}
    C_{n\hat n n \hat n} &= R_{n \hat n n \hat n} - \f{2 \mc R_{nn} }{D-2} + \f{ R}{(D-1)(D-2)},
    \end{align}
    using that $\mc R_{nn} = \mc R_{\hat n \hat n}$. I'll skip straight to the Kerr--AdS case where $\mc R_{nn} = -(D-1)/l^2, R = -D(D-1)/l^2$, in which case
    \begin{align}
        C_{n \hat n n \hat n} &= \f{1}{2} \f{\pa^2 H}{\pa r^2} \nn 
        &\simeq (D-3)(D-2) m r^{1-D},
    \end{align}
    using $H \simeq 2m r^{3-D}$. 

    \subsubsection{Calculation of $C_{n \hat \al n \hat \al}$ and $C_{n \hat 0 n \hat 0}$}
    The calculation for $C_{n \hat \al n \hat \al}$ is similar to that for $C_{n \hat n n \hat n}$. Again I will go straight to the Kerr--AdS case where $\mc R_{A B} = - (D-1)l^{-2} \de_{AB}$ and $R = - D(D-1)l^{-2}$.
    \begin{align}
        C_{n \hat \al n \hat \al} &= R_{n \hat \al n \hat \al} - \f{\mc R_{nn} + \mc R_{\al \al}}{D-2} + \f{R}{(D-1)(D-2)} \nn 
        &= R_{n \hat \al n \hat \al} + l^{-2} \nn 
        &= - p_{n \al} + l^{-2} \nn
        &= -\f{1}{2(-r^2-y_\al^2) } \left( r \f{\pa \bt^2}{\pa r} - y_\al \f{\pa \bt^2}{\pa y_\al}\right)+ l^{-2} \nn 
        &= \f{1}{2(r^2+y_\al^2)} \left( r \f{\pa \bt^2}{\pa r} - y_\al \f{\pa \bt^2}{\pa y_\al}\right) + l^{-2}\nn 
        &= \f{1}{2(r^2+y_\al^2)} \left( r \f{\pa \bar \bt^2}{\pa r} - y_\al \f{\pa \bar \bt^2}{\pa y_\al}\right) + \f{1}{2(r^2+y_\al^2)} \left( r \f{\pa H}{\pa r} - y_\al \f{\pa H}{\pa y_\al}\right) + l^{-2}\nn 
        &= \bar R_{n \hat \al n \hat \al} + \f{1}{2(r^2+y_\al^2)} \left( r \f{\pa H}{\pa r} - y_\al \f{\pa H}{\pa y_\al}\right) + l^{-2}\nn 
        &=  \f{1}{2(r^2+y_\al^2)} \left( r \f{\pa H}{\pa r} - y_\al \f{\pa H}{\pa y_\al}\right).
    \end{align}
    As in \eqref{Cnnhatalphaalphahat1}, $\pa H/\pa y_\al = - 2 y_\al H/(r^2+y_\al^2)$. As stated previously, in Kerr--AdS, $H = 2m r^{1-\ve} / U_n \simeq 2 m r^{3-D}$ (large $r$). $y_\al \pa H/\pa y_\al$ is of order $H/r^2$ or $\mc O(r^{1-D})$ whereas $r \pa H/\pa r$ is of order $H$, and so will dominate, so that we have
    \begin{align}
        C_{n \hat \al n \hat \al} &\simeq \f{1}{2r} \f{\pa H}{\pa r} \nn 
        &\simeq -(D-3) m r^{1-D}.
    \end{align}

    For $C_{n \hat 0 n \hat 0}$ we have, similarly, again in Kerr--AdS (with $\mc R_{AB} = -(D-1)l^{-2} \de_{AB}, R = -D(D-1)l^{-2}$),
    \begin{align}
        C_{n \hat 0 n \hat 0} &= R_{n \hat 0 n \hat 0} - \f{\mc R_{nn} + \mc R_{\hat 0 \hat 0}}{D-2} + \f{R}{(D-1)(D-2)} \nn 
        &= R_{n \hat 0 n \hat 0} + l^{-2} \nn 
        &= \f{1}{2r} \f{\pa \bt^2}{\pa r} + l^{-2} \nn 
        &= \f{1}{2r} \f{\pa \bar \bt^2}{\pa r} + \f{1}{2r} \f{\pa H}{\pa r} + l^{-2} \nn 
        &= \bar R_{n \hat 0 n \hat 0} + \f{1}{2r} \f{\pa H}{\pa r} + l^{-2} \nn 
        &= - l^{-2} + \f{1}{2r} \f{\pa H}{\pa r} + l^{-2} \nn 
        &= \f{1}{2r} \f{\pa H}{\pa r}.
    \end{align}
    With $H \simeq 2m r^{3-D}$ for large $r$ this is
    \begin{align}
        C_{n \hat 0 n \hat 0} &\simeq -(D-3) m r^{1-D}.
    \end{align}

\section{Evaluation of \texorpdfstring{$\mc F$}{F}} \label{evaluation}

    For completeness I will evaluate $\mc F$. $C^{n \hat n n \hat n}$ for Kerr--AdS is $\simeq -(D-3) m r^{1-D}$ as calculated in Section \ref{Cnnnn}. $N_{\hat n}$ is given by \eqref{Nhatn}.
    
    We need $dS$. $dS = \sqrt{\sigma} d^{D-2} x$, where $\sigma_{IJ}$ is the metric on the constant-$t,r$ integration surface. Using the Kerr--Schild decomposition, as shown in Section \ref{AreaRevisited}, $\textrm{det} \sigma$ for the full spacetime can be related to $\textrm{det} \bar \sigma$ for the background spacetime by
    \begin{align}
        \textrm{det} \sigma = \textrm{det} \bar \sigma ( 1 + H k_J k^J)
    \end{align}
    where $k$ is the Kerr--Schild null vector and $J$ is the index for the coordinates, excluding $t$ and $r$. Since $k_ak^a = 0$ including $t$ and $r$, the multiplicative factor is
    \begin{align}
        1 - H k_t k^t - H k_r k^r &\simeq 1,
    \end{align}
    since $H(k_t k^t + k_r k^r) = \mc O(r^{1-D})$. We then have
    \begin{align}
        \sqrt{\sigma} &\simeq \sqrt{\bar \sigma} \nn 
        &= \sqrt{ \f{\bar g}{\bar g_{tt} \bar g_{rr}}}. \label{sqrtsigma}
    \end{align}
    $N_{\hat n}$, from Section \ref{dtcalculation}, is $N_{\hat n} \simeq + i l r^{-1} /\sqrt{-g^{tt}}$; $g^{tt} = \bar g^{tt} - H (k^t)^2 \simeq \bar g^{tt} = \bar g_{tt}^{-1}$, giving
    \begin{align}
        N_{\hat n} \sqrt \sigma &\simeq \f{i l}{r} \f{\sqrt{-\bar g}}{\sqrt{\bar g_{rr}}} \nn 
        &\simeq i \sqrt{-\bar g}. \label{Nnhatsqrtsigma}
    \end{align}
    We then have from \eqref{FQCbetaformula}
    \begin{align}
        \mc F &= Q_C[\bt] = \f{1}{8\pi(D-3)} \lim_{r\to\infty} \oint r C^{n \hat n n \hat n} \sqrt{-\bar g} d^{D-2} x \nn 
        &= \f{(D-2)m}{8\pi} \lim_{r\to\infty} \oint r^{2-D} \sqrt{-\bar g} d^{D-2}x.
    \end{align}
    Using \eqref{Gibbonssqrtg} and taking the leading term in $r$ we find
    \begin{align}
        \mc F &= \f{(D-2)m}{8\pi} \oint \f{\prod_{i=1}^{n-1+\ve} \mu_i}{\mu_n \prod_{j=1}^n \Xi_j} d^{D-2} x,
    \end{align}
    which, from Chapter \ref{VolumeAreaChapter}, gives
    \begin{align}
        \mc F &= \f{(D-2) \mc A_{D-2} m}{8 \pi \prod_{j=1}^n \Xi_j},
    \end{align}
    matching the results from GPP \cite{GibbonsPerry}.

\chapter{Extended Abstract}

\emph{This is a longer version of the abstract.}

Gibbons et al.~\cite{GibbonsPerry} and Cveti\v{c} et al.~\cite{Cvetic} consider the first law of black hole thermodynamics as well as Smarr relations in the Kerr--anti-de Sitter spacetime. Their papers suggest some ``open questions,'' or at least partially open questions, some of which have to do with the volume associated with black holes, which I address by adapting the work of Barnich and Comp\`ere \cite{BarnichCompere} to define a $(D-2)$-form $\bs I_\chi$ associated with a generic Killing vector $\chi$, for which the energy $\mc E$ associated with an asymptotically static Killing vector $\xi$ can be written as $\mc E = \oint \bs I_\xi$ on an arbitrary 2-surface enclosing the black hole. I use this form to address these ``open questions'' in the literature. 

Given a $D$-dimensional region $\mc R$ in a $D$-dimensional spacetime, if there exists a divergence-free vector field $v^a$ such that $v$ is tangent to the boundary of $\mc R$, I define the vector volume associated with $v,\mc R$ to be $\mc V_{v,\mc R} = \int_\Si v^a d\Sigma_a$, where $\Si$ is a hypersurface spanning $\mc R$ and where $\Si$ intersects each of the integral curves of $v^a$ lying within $\mc R$ exactly once (at least in some part of the spacetime which can be covered by coordinates adapted to the vector $v^a$). For a finite region, this volume as defined vanishes, but for a region infinite in extent it can be nonzero. I show that, with certain caveats, this is constant and independent of choice of such surface $\Si$. I explore some more of its properties and relate the quantity to black holes. This definition coincides with one by Parikh \cite{Parikh} and I also show that this volume, applied to a black hole using the canonically normalized stationarity Killing vector, is equal to the geometric volume of Cveti\v{c} et al.~\cite{Cvetic}. 

In Gibbons et al.~\cite{GibbonsPerry}, the energy for arbitrary dimensional ($D \geq 4$) Kerr--anti-de Sitter black holes $\mc E$ was found by integrating a first-law variation formula $\de \mc E = \sum_i \Om_i \de \mc J_i + T \de S$, with black hole angular momenta $\mc J_i$, angular velocity $\Om_i$, temperature $T$ and entropy $S$. The cosmological constant $\tilde \La$ satisfying $R_{ab} = \tilde \La g_{ab}$ for Ricci tensor $R_{ab}$, metric $g_{ab}$, was kept fixed. This energy was found to correspond to the Ashtekar--Magnon--Das (AMD) energy, but only when calculated in an asymptotically static frame, rather than an asymptotically rotating one. In Cveti\v{c} et al.'s~paper \cite{Cvetic}, the first law was extended by interpreting $\mc E$ as an enthalpy, and interpreting $\tilde \La$ as a thermodynamic variable proportional to a pressure, so that the first law now reads $\de \mc E = \sum_i \Om_i \de \mc J_i + T \de S + \Th \de \tilde \La$, where $\Th$ is a thermodynamic potential proportional to the so-called thermodynamic volume $V_{th} = -16\pi \Th/(D-2)$. Due to a scaling symmetry, the above variables also automatically satisfy the Smarr relation $(D-3) \mc E = (D-2) \left( \sum_i \Om_i \mc J_i + T S\right) - 2 \Th \tilde \La$. 

If instead a certain asymptotically rotating frame is used, the equivalent AMD ``energy'' (which I denote by $\mc F$), along with angular velocities $\om_i$ in the asymptotically rotating frame, satisfies the modified Smarr relation $(D-3) \mc F = (D-2) \left(\sum_i \Om_i \mc J_i + TS\right) - 2 \Th' \tilde \La$, associated with a geometric volume $V_{geo} = -16\pi \Th'/(D-2)$. This geometric volume, given by $\int \sqrt{-g} d^{D-1}x$ with metric determinant $g$ and product of coordinate differentials with the time omitted $d^{D-1}x$ and integrated below the horizon down to the singularity, is the quantity which is equal to my vector volume. In general $V_{th} \neq V_{geo}$. \cite{Cvetic} also noted the intriguing relation $V_{geo} = r_+ A/(D-1)$, where $r_+$ is the value of the usual spheroidal (Boyer--Lindquist) $r$ coordinate on the horizon. 

These papers (either implicitly or explicitly) raise the following questions: what is it that makes it necessary to base the conserved energy (or enthalpy) on an asymptotically static frame, rather than an asymptotically rotating one? Why does the volume appear more naturally in the relation with the asymptotically rotating frame rather than the asymptotically static one? Why is the volume proportional to the black hole area, with proportionality constant equal to $r_+/(D-1)$? I address each of these questions. 

Barnich and Comp\`ere \cite{BarnichCompere} argued that conserved quantities which satisfy a first law variation could be found by integrating through solution space, and that these integrals could then be used to construct a Smarr relation. I show how applying this method to Kerr--Schild spacetimes results in simplifications, leading to a $(D-2)$-form $\bs I_\chi$ associated with a Killing vector $\chi$. When integrated over a $(D-2)$ surface, this gives a conserved quantity for solutions to Einstein's equations. When $\chi$ is the asymptotically-static Killing vector, the result is $\mc E$, and when $\chi$ is an azimuthal symmetry Killing vector, the result is $-\mc J_i$. 

The exterior derivative of $\bs I_\chi$ is also shown to be related to the stress--energy tensor of the spacetime, and with certain caveats, is also shown to satisfy a horizon variation law even when the spacetime is not a solution to Einstein's equations. The integral of $\bs I_\chi$ over a $(D-2)$-surface is shown to be equal to the difference between the Komar integrals associated with $\chi$ for the full and background spacetimes, plus one non-Komar term. I show how the Komar term associated with the background metric is equal to the black hole vector volume. I use $\bs I_\chi$ to argue that the reason that the asymptotically rotating Killing vector is a poor candidate to base a conserved quantity on is less because it is rotating, but because there are no coordinates in which both it and the Kerr--Schild background metric do not vary when the rotational parameter varies. I show how the use of the asymptotically rotating vector simplifies the expression for $\bs I_\chi$ in such a way that the vector volume appears naturally in the Smarr relationship in this case. I also show how the volume--area relation follows directly from the existence of the Principal Conformal Killing--Yano tensor $\bs h$.


\begin{thebibliography}{200} 

\bibitem{AbbottDeser} L.~F.~Abbott and S.~Deser, Nucl.~Phys.~B \textbf{195}(1) 76--96 (1982).

\bibitem{abdelqaderlake} M.~Abdelqader and K.~Lake, Phys.~Rev.~D \textbf{86} 124037 (2012), [\href{https://arxiv.org/abs/1207.5496}{arXiv:1207.5496}].

\bibitem{AckayMatzner} S.~Ackay and R.~A.~Matzner, Class.~Quant.~Grav.~\textbf{28} 085012 (2011), [\href{https://arxiv.org/abs/1011.0479}{arXiv:1011.0479}].

\bibitem{Altamirano} N.~Altamirano, D.~Kubiz\v{nak}, R.~B.~Mann and Z.~Sherkatghanad, Galaxies \textbf{2} 89--159 (2014), doi:10.3390/galaxies2010089; also available as [\href{https://arxiv.org/abs/1401.2586}{arXiv:1401.2586}]. 

\bibitem{Apostol} T.~M.~Apostol, \emph{Calculus, Volume 2}, second edition (Wiley, 1969).

\bibitem{AshtekarMagnon}
A.~Ashtekar and A.~Magnon, Class.~Quant.~Grav.~\textbf{1} L39 (1984).

\bibitem{AshtekarDas} A.~Ashtekar and S.~Das, Class.~Quant.~Grav.~\textbf{17} L17 (2000), [\href{https://arxiv.org/abs/hep-th/9911230}{arXiv:hep-th/9911230}].


\bibitem{AshtekarPawlowski} A.~Ashtekar, T.~Pawlowski and C.~Van Den Broeck, Class.~Quant.~Grav.~\textbf{24} 625 (2007), [\href{https://arxiv.org/abs/gr-qc/0611049}{arXiv:gr-qc/0611049}].



\bibitem{BallikLake10} W.~Ballik and K.~Lake, (2010), [\href{https://arxiv.org/abs/1005.1116}{arXiv:gr-qc/1005.1116v3}].

\bibitem{Ballik} W.~Ballik and K.~Lake, Phys.~Rev.~D {\bf88} 104038 (2013), [\href{https://arxiv.org/abs/1310.1935}{arXiv:1310.1935}].

\bibitem{BCH} J.~M.~Bardeen, B.~Carter  and S.~W.~Hawking, Comm.~Math.~Phys.~\textbf{31} 161 (1973).

\bibitem{BarnichBrandt} G.~Barnich and F.~Brandt, Nucl.~Phys.~B \textbf{633} 3--82 (2002), [\href{https://arxiv.org/abs/hep-th/0111246}{arXiv:hep-th/0111246}].

\bibitem{BarnichCompere} G.~Barnich and G.~Comp\`ere, Phys.~Rev.~D \textbf{71} 044016 (2005), Erratum-idi.D \textbf{71}:029904 (2006), [\href{https://arxiv.org/abs/gr-qc/0412029}{arXiv:gr-qc/0412029}].


\bibitem{BelhajEtal:2012}
A.~Belhaj, M.~Chabab, H.~El~Moumni, and M.~Sedra, Chin.~Phys.~Lett.~{\bf 29} 100401 (2012), [\href{https://arxiv.org/abs/1210.4617}{arXiv:1210.4617}].

\bibitem{Blagojevic:2020edq}
M.~Blagojevi{\'c} and B.~Cvetkovi{\'c}, Phys.~Rev.~D \textbf{101}, no.8, 084023 (2020)
[\href{https://arxiv.org/pdf/2002.05029}{arXiv:2002.05029}].


\bibitem{Bordo} A.~B.~Bordo, F.~Gray, R.~A.~Hennigar and D.~Kubiz\v{n}\'ak, Phys.~Lett.~B \textbf{798} 134972 (2019), [\href{https://arxiv.org/abs/1905.06350}{arXiv:1905.06350}]. 

\bibitem{Brinkmann} H.~W.~Brinkmann, Math.~Ann.~\textbf{94} 119–-145 (1925).

\bibitem{CaldarelliCognola} M.~M.~Caldarelli, G.~Cognola and D.~Klemm, Class. Quant.~Grav.~\textbf{17} 399--420 (2000), [\href{https://arxiv.org/abs/hep-th/9908022}{arXiv:hep-th/9908022}].

\bibitem{CaldarelliCamps} M.~M.~Caldarelli, J.~Camps, B.~Goutéraux, B. and K.~Skenderis, Phys.~Rev.~D \textbf{87} 061502(R) (2013), [\href{https://arxiv.org/abs/1211.2815}{arXiv:1211.2815}]. 

\bibitem{Campos24} T.~de L.~Campos, M.C.~Baldiotti, C.~Molina,  Phys.~Rev.~D \textbf{110} 024049 (2024), [\href{https://arxiv.org/abs/2407.09610v1}{arXiv:2407.09610}].

\bibitem{Campos25} T.~de L.~Campos, M.C.~Baldiotti, C.~Molina, Universe 2025, 11, 215, [\href{https://arxiv.org/abs/2507.03751v1}{arXiv:2507.03751}].

\bibitem{Campos26} T.~de L.~Campos, M.C.~Baldiotti, C.~Molina, [\href{ 	arXiv:2605.16536}{arXiv:2605.16536}].

\bibitem{CM} J.~Carminati and R.~G.~McLenaghan, J.~Math.~Phys.~\textbf{32}(11) 3135--3140 (1991).

\bibitem{CarterMetric} B.~Carter, Comm.~Math.~Phys.~\textbf{10} 280--310 (1968).

\bibitem{CarterLesHouches} B.~Carter, ``Black Hole Equilibrium States'' (From \emph{Les Houches} 1972, ed.~by DeWitt).

\bibitem{Chen} W.~Chen, H.~L\"u, C.~N.~Pope, Class.~Quant.~Grav.~\textbf{23} 5323--5340 (2006), [\href{https://arxiv.org/abs/hep-th/0604125}{arXiv:hep-th/0604125}].

\bibitem{ChenLu08} W.~Chen, H.~L\"u, Phys.~Lett.~B \textbf{658} 158--163 (2008), [\href{https://arxiv.org/abs/0705.4471}{arXiv:0705.4471}].

\bibitem{Chernyavsky} D.~Chernyavsky and K.~Hajian, Class.~Quant.~Grav.~\textbf{35} (12) 125012 (2018), [\href{https://arxiv.org/abs/1710.07904}{arXiv:1710.07904}].

\bibitem{ChristodoulouRovelli} M.~Christodoulou and C.~Rovelli, Phys.~Rev.~D \textbf{91}, 064046 (2015), [\href{https://arxiv.org/abs/1411.2854}{arXiv:1411.2854}].



\bibitem{ChristodoulouDeLorenzo} M.~Christodoulou and T.~De Lorenzo, Phys.~Rev.~D \textbf{94}, 104002 (2016), [\href{https://arxiv.org/abs/1604.07222}{arXiv:1604.07222}].


\bibitem{Chrusciel} P.~T.~Chruściel, J.~Jezierski and J.~Kijowski, Phys.~Rev.~D \textbf{92}, 084030 (2015), [\href{https://arxiv.org/abs/1507.03868v2}{arXiv:1507.03868}]. 

\bibitem{Compere} G.~Comp\`ere, ``An introduction to the mechanics of black holes'' (2006), [\href{https://arxiv.org/abs/gr-qc/0611129}{arXiv:gr-qc/061129}].

\bibitem{Couch} J.~Couch, W.~Fischler, P.~H.~Nguyen, JHEP \textbf{1703} 119 (2017), [\href{https://arxiv.org/abs/1610.02038}{arXiv:1610.02038}].

\bibitem{Cvetic} M.~Cveti\v{c}, G.~W.~Gibbons, D.~Kubiz\v{n}\'ak, and C.~N.~Pope, Phys.~Rev.~D \textbf{84} 024037 (2011), [\href{https://arxiv.org/abs/1012.2888}{arXiv:1012.2888}].

\bibitem{DasMann} S.~Das, R.~B.~Mann, JHEP \textbf{0008} 033 (2000), [\href{https://arxiv.org/abs/hep-th/0008028}{arXiv:hep-th/0008028}].

\bibitem{Dereli} T.~Dereli and M.~G\"urses, Phys.~Lett.~B \textbf{171} 209 (1986).

\bibitem{DeruelleKatz05} N.~Deruelle and J.~Katz, Class.~Quant.~Grav.~\textbf{22} 421 (2005), [\href{https://arxiv.org/abs/gr-qc/0410135}{arXiv:gr-qc/0410135}].

\bibitem{DeruelleKatz06} N.~Deruelle and J.~Katz, Class.~Quant.~Grav.~\textbf{23} 753 (2006), [\href{https://arxiv.org/abs/gr-qc/0512077}{arXiv:gr-qc/0512077}].


\bibitem{Dolan:2010}
B.~P.~Dolan, Class. Quant.~Grav.~{\bf 28} 125020 (2011), [\href{https://arxiv.org/abs/1008.5023}{arXiv:1008.5023}].

\bibitem{Dolan:2011a}
B.~P. Dolan, Class.~Quant.~Grav.~{\bf 28} 235017 (2011), [\href{https://arxiv.org/abs/1106.6260}{arXiv:1106.6260}].

\bibitem{Dolan:2011b}
B.~P. Dolan, Phys.~Rev.~D {\bf 84} 127503 (2011),  [\href{https://arxiv.org/abs/1109.0198}{arXiv:1109.0198}].

\bibitem{Dolan:2012}
B.~P. Dolan, {\it Where is the $PdV$ term in the first law of black hole
  thermodynamics?},  in \emph{ Open Questions in Cosmology} (G.~J. Olomo, ed.),
  InTech, 2012, [\href{https://arxiv.org/abs/1209.1272}{arXiv:1209.1272}].

\bibitem{DolanKastor}B.~Dolan, D.~Kastor, D.~Kubiz\v{n}\'ak, R.~B.~Mann and J.~Traschen, Phys.~Rev.~D \textbf{87}, 104017 (2013), [\href{https://arxiv.org/abs/1301.5926}{arXiv:hep-th/1301.5926}].

\bibitem{ElvangEmparan} H.~Elvang and R.~Emparan, JHEP \textbf{0311} 035 (2003), [\href{https://arxiv.org/abs/hep-th/0310008}{arXiv:hep-th/0310008}].

\bibitem{EmparanReall} R.~Emparan and H.~S.~Reall, Phys.~Rev.~Lett.~\textbf{88} 101101 (2002), [\href{https://arxiv.org/abs/hep-th/0110260}{arXiv:hep-th/0110260}].

\bibitem{Frolov} V.~P.~Frolov and D.~Kubiz\v{n}\'ak, Class.~Quant.~Grav.~\textbf{25} 154005 (2008), [\href{https://arxiv.org/abs/0802.0322}{arXiv:0802.0322}].

\bibitem{FrolovWeaklyCharged} V.~P.~Frolov, P.~Krtouš, D.~Kubiz\v{n}\'ak, Phys.~Lett.~\textbf{B771} 254--6 (2017) [\href{https://arxiv.org/abs/1705.00943v2}{arXiv:1705.00943}].

\bibitem{FrolovReview} V.~P.~Frolov, P.~Krtouš, D.~Kubiz\v{n}\'ak, Rev.~Relativ.~\textbf{20}:6 (2017), [\href{https://arxiv.org/abs/1705.05482}{arXiv:1705.05482}].

\bibitem{Gao} Y.~Gao, Z.~Di, S.~Gao, Phys.~Scripta 99 9 095022 (2024), [\href{https://arxiv.org/abs/2304.10290}{arXiv:2304.10290}].

\bibitem{GaoWald} S.~Gao and R.~M.~Wald, Phys.~Rev.~D \textbf{64} 084020 (2001), [\href{https://arxiv.org/abs/gr-qc/0106071}{arXiv:gr-qc/0106071}]. 

\bibitem{GibbonsLu} G. W.~Gibbons, H.~L\"u, D.N.~Page, C.N.~Pope, Phys.~Rev.~Lett.~\textbf{93} 171102 (2004) [\href{https://arxiv.org/abs/hep-th/0409155}{arXiv:hep-th/0409155}.

\bibitem{GibbonsLu2} G. W.~Gibbons, H.~L\"u, D.N.~Page, C.N.~Pope, 
J.~Geom.~Phys.~\textbf{53} 49--73 (2005), [\href{https://arxiv.org/abs/hep-th/0404008v3}{arXiv:hep-th/0404008}].

\bibitem{GibbonsPerry} G.~W.~Gibbons, M.~J.~Perry and C.~N.~Pope, Class.~Quant.~Grav.~\textbf{22} 1503--1526 (2005), [\href{https://arxiv.org/abs/hep-th/0408217}{arXiv:hep-th/0408217}].

\bibitem{Gibbons:2012}
G.~W.~Gibbons, (2012), [\href{https://arxiv.org/abs/1201.2340}{arXiv:1201.2340}].

\bibitem{Golshani} M.~Golshani, M.~M.~Sheikh-Jabbari, V.~Taghiloo, M.~H.~Vahidinia, [\href{https://arxiv.org/abs/2407.15994}{arXiv:2407.15994}].

\bibitem{GrTensor} \emph{GRTensor} by P.~Musgrave, D.~Pollney and K.~Lake  is a package which runs within Maple. Some of the calculations used in this thesis used \emph{GRTensorII} and some used the newer \emph{GRTensorIII} (most recent version 2023). It is entirely distinct from packages distributed with Maple and must be obtained independently.  The \emph{GRTensorIII} software and documentation is distributed freely on the World-Wide-Web from the address \url{https://github.com/grtensor/grtensor}. 

\bibitem{GunasekaranEtal:2012}
S.~Gunasekaran, R.~B.~Mann, and D.~Kubiz\v{n}\'ak, JHEP {\bf 1211} 110 (2012), [\href{https://arxiv.org/abs/1208.6251}{arXiv:1208.6251}].

\bibitem{HajianGRG} K.~Hajian, General Relativity and Gravitation \textbf{48} 114 (2016), [\href{https://arxiv.org/pdf/1602.05575}{arXiv:1602.05575}]. 

\bibitem{HajianSheikh-Jabbari} K.~Hajian and M.~M.~Sheikh-Jabbari, Phys.~Rev.~D \textbf{93} 044074 (2016), [\href{https://arxiv.org/abs/1512.05584v2}{arXiv:1512.05584}].


\bibitem{HajianOzsahin} K.~Hajian, H.~Özşahin and B.~Tekin, Phys.~Rev.~D \textbf{104} 044024 (2021), [\href{https://arxiv.org/abs/2103.10983}{arXiv:2103.10983}].

\bibitem{Hajian:2025hxf}
K.~Hajian, B.~Tekin and O.~Ucanok, [\href{https://arxiv.org/abs/2511.22558v1}{arXiv:2511.22558}].

\bibitem{Hamamoto} N.~Hamamoto, T.~Houri, T.~Oota and Y.~Yasui, J.~Phys.~A: Math.~Theor.~\textbf{40} F177 (2006), [\href{https://arxiv.org/abs/hep-th/0611285}{arXiv:hep-th/0611285}].
    
\bibitem{Harville} D.~A.~Harville,~\emph{Matrix Algebra From a Statistician’s Perspective} (Springer-Verlag, 1997).

\bibitem{HawkingRad} S.~W.~Hawking, Nature \textbf{248}.5443 30--31 (1974).

\bibitem{HHTR} S.~W.~Hawking, C.~J.~Hunter and M.~M.~Taylor-Robinson, Phys.~Rev.~D \textbf{59} 064005 (1999), [\href{https://arxiv.org/abs/hep-th/9811056}{arXiv:hep-th/9811056}].

\bibitem{HawkingPage} S.~W.~Hawking and D.~N.~Page, Commun.~Math.~Phys.~\textbf{87} 577-588 (1983).

\bibitem{Hayward} S.~A.~Hayward, Class.~Quant.~Grav.~\textbf{15} 3147 (1998), [\href{https://arxiv.org/abs/gr-qc/9710089}{arXiv:gr-qc/9710089}].


\bibitem{Haywardcylinder} S.~A.~Hayward, Class.~Quant.~Grav.~\textbf{17} 1749 (2000), [\href{https://arxiv.org/abs/gr-qc/9909070}{arXiv:gr-qc/9909070v2}].

\bibitem{Hayward04} S.~A.~Hayward, Phys.~Rev.~Lett.~\textbf{93} 251101 (2004), [\href{https://arxiv.org/abs/gr-qc/0404077}{arXiv:gr-qc/0404077}].

\bibitem{HendiVahinidia:2012}
S.~Hendi and M.~Vahidinia, (2012), [\href{https://arxiv.org/abs/1212.6128}{arXiv:1212.6128}].

\bibitem{HenneauxTeitelboim85} M.~Henneaux and C.~Teitelboim, Commun.~Math.~Phys.~\textbf{98}, 391--424 (1985).

\bibitem{HenneauxTeitelboim92} M.~Henneaux and C.~Teitelboim, \emph{Quantization of Gauge Systems} (Princeton University Press, 1992).

\bibitem{Hollands} S.~Hollands, A.~Ishibashi and D.~Marolf, Class.~Quant.~Grav.~\textbf{22} 2881 (2005) [\href{https://arxiv.org/abs/hep-th/0503045}{arXiv:hep-th/0503045}].

\bibitem{Houri} T.~Houri, T.~Oota and Y.~Yasui, Phys.~Lett.~B \textbf{656} 214--216 (2007).

\bibitem{Houri08} T.~Houri, T.~Oota and Y.~Yasui, Phys.~Lett.~B \textbf{666} 391--394 (2008), [\href{https://arxiv.org/abs/0805.0838v3}{arXiv:0805.0838}].

\bibitem{Hyun:2017nkb}
S.~Hyun, J.~Jeong, S.~A.~Park and S.~H.~Yi, JHEP \textbf{04}, 048 (2017) [\href{https://arxiv.org/abs/1702.06629}{arXiv:1702.06629}].


\bibitem{ProofWiki} ``Inverse of Vandermonde Matrix'' from ProofWiki, [\url{https://proofwiki.org/wiki/Inverse\_of\_Vandermonde\_Matrix}].

\bibitem{Israel70} W.~Israel, Phys.~Rev.~D \textbf{2} 641 (1970).

\bibitem{Israel79} W.~Israel, \emph{Differential Forms in General Relativity} (Communications of the Dublin Institute for Advanced Studies \textbf{A} No, 26, Dublin, 1979).


\bibitem{Israel86} W.~Israel, Phys.~Rev.~Lett.~\textbf{57} 397 (1986).


\bibitem{IyerWald94} V.~Iyer and R.~M.~Wald, Phys.~Rev.~D \textbf{50} 846--64 (1994), [\href{https://arxiv.org/abs/gr-qc/9403028}{arXiv:gr-qc/9403028}].

\bibitem{IyerWald95} V.~Iyer and R.~M.~Wald, Phys.~Rev.~D \textbf{52} 4430 (1995), [\href{https://arxiv.org/abs/gr-qc/9503052}{arXiv:gr-qc/9503052}].


\bibitem{Jacobson} T.~Jacobson, Class.~Quant.~Grav.~\textbf{24} 5717 (2007), [\href{https://arxiv.org/abs/0707.3222}{arXiv:gr-qc/0707.3222}].

\bibitem{JacobsonVisser} T.~Jacobson, M.~Visser, SciPost Phys.~\textbf{7} 079 (2019), [\href{https://arxiv.org/abs/1812.01596}{arXiv:1812.01596}].

\bibitem{Jing:2017jxw}
Y.~D.~Jing and J.~J.~Peng, Chin.~Phys.~B \textbf{26}, 100401 (2017)
doi:10.1088/1674-1056/26/10/100401
[\href{https://arxiv.org/abs/1707.01605v1}{arXiv:1707.01605}].

\bibitem{Jorgensen} M.~Jorgensen, ``Volumes of n-dimensional spheres and ellipsoids'' (2014) [\url{https://www.whitman.edu/documents/Academics/Mathematics/2014/jorgenmd.pdf}].

\bibitem{KastorEtal:2009}
D.~Kastor, S.~Ray, and J.~Traschen, Class. Quant. Grav. {\bf 26} 195011 (2009), [\href{https://arxiv.org/abs/0904.2765}{arXiv:0904.2765}].

\bibitem{Katz85} J.~Katz, Class.~Quant.~Grav.~{\bf 2} 423 (1985).

\bibitem{KBL} J.~Katz, J.~Bi\v{c}\'ak and D.~Lynden-Bell, Phys.~Rev.~D.~\textbf{55} 5759 (1997), [\href{https://arxiv.org/abs/gr-qc/0504041}{arXiv:gr-qc/0504041}].

\bibitem{KrtousKubiznak} P.~Krtou\v{s}, D.~Kubiz\v{n}\'ak, D.~N.~Page and V.~P.~Frolov, JHEP \textbf{0702} 004 (2007), [\href{https://arxiv.org/abs/hep-th/0612029}{arXiv:hep-th/0612029}].

\bibitem{KrtousFrolov}  P.~Krtouš, V.~P.~Frolov, and D.~Kubizňák,
Phys.~Rev.~D \textbf{78}, 064022 (2008), [\href{https://arxiv.org/abs/0804.4705}{arXiv:0804.4705}].

\bibitem{Krtous16} P.~Krtouš, D.~Kubizňák and I. Kolá\v{r},  Phys.~Rev.~D \textbf{93}, 024057 (2016), [\href{https://arxiv.org/abs/1508.02642}{arXiv:1508.02642}].

\bibitem{KubiznakFrolov} D.~Kubiz\v{n}\'ak and V.P.~Frolov, Class.~Quant.~Grav.~\textbf{24} F1--F6 (2007), [\href{https://arxiv.org/abs/gr-qc/0610144v2}{arXiv:gr-qc/0610144}].

\bibitem{KubiznakThesis} D.~Kubiz\v{n}\'ak, \emph{Hidden Symmetries of Higher-Dimensional Rotating Black Holes}, Ph.D.~thesis, U.~Alberta (2008), [\href{https://arxiv.org/abs/0809.2452}{arXiv:0809.2452}].

\bibitem{KubiznakMann:2012}
D.~Kubiz\v{n}\'ak and R.~B. Mann, JHEP, {\bf 1207}  033 (2012), [\href{https://arxiv.org/abs/1205.0559}{arXiv:1205.0559}].

\bibitem{KubiznakMannTeo} D.~Kubizňák, R.~B.~Mann, M.~Teo, Class.~Quantum Grav.~\textbf{34} 063001 (2017), [\href{https://arxiv.org/abs/1608.06147}{arXiv:1608.06147}].

\bibitem{KubiznakSimovic} D.~Kubiz\v{n}\'ak and F.~Simovic, Class.~Quant.~Grav.~\textbf{33} 245001 (2016), [\href{https://arxiv.org/abs/1507.08630}{arXiv:1507.08630}].



\bibitem{LakeLectureNotes} K.~Lake, ``Perfect Fluids 1,'' \emph{General Relativity Lecture Notes}. 
\bibitem{LarranagaCardenas:2012}
A.~Larranaga and A.~Cardenas, J.~Korean Phys.~Soc.~{\bf 60} 987 (2012), [\href{https://arxiv.org/abs/1108.2205}{arXiv:1108.2205}].

\bibitem{LarranagaMojica:2012}
A.~Larranaga and S.~Mojica, The Abraham Zelmanov Journal {\bf 5} 68 (2012), [\href{https://arxiv.org/abs/1204.3696}{arXiv:1204.3696}].

\bibitem{Lee} J.~M.~Lee, \emph{Introduction to Smooth Manifolds}, 2nd edition (Springer, New York, NY, 2018).

\bibitem{LuEtal:2012}
H.~Lu, Y.~Pang, C.~N.~Pope, and J.~F.~Vazquez-Poritz, Phys.~Rev.~D \textbf{86} 044011 (2012), [\href{https://arxiv.org/abs/1204.1062}{arXiv:1204.1062}].

\bibitem{Magnon}
A.~Magnon, J.~Math.~Phys.~\textbf{26}(12) 3112--3117 (1985). 

\bibitem{Maldacena} J.~Maldacena, Adv.~Theor.~Math.~Phys.~\textbf{2}, 231 (1998) [Int.~J.~Theor.~Phys.~\textbf{38}, 1113
(1998)] [\href{https://arxiv.org/abs/hep-th/9711200}{arXiv:hep-th/9711200}].

\bibitem{Malek} T.~Málek and V.~Pravda, Class.~Quant.~Grav.~\textbf{28} 125011 (2011), [\href{https://arxiv.org/abs/1009.1727}{arXiv:1009.1727}].

\bibitem{Mann} R.~B.~Mann, International Journal of Modern Physics D34  2542001 (2025), [\href{https://arxiv.org/abs/2508.01830v1}{2508.01830}].

\bibitem{Miville-Deschenes} M.~A.~Miville-Deschênes et al.,~Astronomy and Astrophysics \textbf{641} A6--A6 (2020).

\bibitem{NUT} E.~Newman, L.~Tamburino and T.~Unti, J.~Math.~Phys.~\textbf{4} (7) 915--923 (1963).

\bibitem{Ortaggio} M.~Ortaggio, V.~Pravda and A.~Pravdová, Class.~Quant.~Grav.~\textbf{26}(2) 025008 (2009), [\href{https://arxiv.org/abs/0808.2165}{arXiv:0808.2165}].

\bibitem{OrtaggioBrinkmann} M.~Ortaggio, V.~Pravda and A.~Pravdová, Class.~Quant.~Grav.~\textbf{28} 105006 (2011), [\href{https://arxiv.org/abs/1011.3153v2}{arXiv:1011.3153}].

\bibitem{Parikh} M.~K.~Parikh, Phys.~Rev.~D \textbf{73} 124021 (2006), [\href{https://arxiv.org/abs/hep-th/0508108}{arXiv:hep-th/0508108}].

\bibitem{Parks} H.~Parks, ``The Volume of the Unit n-Ball,'' {Mathematical Association of America} \textbf{86}(4) 270--4 (2013).

\bibitem{Peng} J.-J.~Peng, C.-L.~Zou and H.-F.~Liu, Phys.~Scr.~\textbf{96} 125207 (2021).

\bibitem{PetrovKatz} A.~N.~Petrov and J.~Katz, Proceedings: Mathematical, Physical and Engineering Sciences, \textbf{458} (2018) 319--337 (2002).

\bibitem{Poisson} E.~Poisson, \emph{A Relativist's Toolkit: The Mathematics of Black-Hole Mechanics} (Cambridge University Press, Cambridge, 2004).

\bibitem{Rodriguez} N.~H.~Rodriguez and M.~J.~Rodriguez, JHEP \textbf{2022} 44 (2022), [\href{https://arxiv.org/abs/2112.00780}{arXiv:2112.00780}].

\bibitem{Romero} C.~Romero, R.~Tavakol, R.~Zalaletdinov, General Relativity and Gravitation \textbf{28} No.~3 (1996).

\bibitem{Rossi} L.~Rossi, ``The First Law of Black Hole Mechanics'' (2020), [\href{https://arxiv.org/abs/2012.04593}{arXiv:2012.04593}].

\bibitem{Silva} S.~Silva, Class.~Quant.~Grav.~\textbf{19} 3947 (2002), [\href{https://arxiv.org/abs/hep-th/0204179}{arXiv:hep-th/0204179}].

\bibitem{SmailagicSpallucci:2012}
A.~Smailagic and E.~Spallucci,  (2012), [\href{https://arxiv.org/abs/1212.5044}{arXiv:1212.5044}].

\bibitem{SmithVamanamurthy} D.~J.~Smith and M.~K.~Vamanamurthy, Mathematics Magazine \textbf{62}(2) 101--7 (1989).

\bibitem{Sopuerta} C.~Sopuerta, J.~Math.~Phys.~\textbf{39} 1024 (1998) [\href{https://arxiv.org/abs/gr-qc/9801103}{arXiv:gr-qc/9801103}.

\bibitem{Stephani} 
H.~Stephani, D.~Kramer, M.~MacCallum, C.~Hoenselaers and E.~Herlt, \emph{Exact solutions of Einstein's field equations} (Cambridge University Press, Cambridge, 2009).

\bibitem{Taub51} A.~H.~Taub, Ann.~Math.,~Second Series \textbf{53}(3) 472--490 (1951).

\bibitem{Taub81} A.~H.~Taub, Ann.~Phys.~\textbf{134} (2) 326--372 (1981). 

\bibitem{Turner} L.~R.~Turner, \emph{Inverse of the Vandermonde matrix with applications} (No.~NASA-TN-D-3547) (1966).

\bibitem{Wald84} R.~M.~Wald, \emph{General Relativity} (The University of Chicago Press, Chicago, 1984).

\bibitem{Wald90} R.~M.~Wald, J.~Math.~Phys.~\textbf{31} 2378--2384 (1990).

\bibitem{Wald93} R.~M.~Wald, Phys.~Rev.~D \textbf{48} 3427--3431 (1993), [\href{https://arxiv.org/abs/gr-qc/9307038}{arXiv:gr-qc/9307038}].

\bibitem{WaldZoupas} R.~M.~Wald and A.~Zoupas, Phys.~Rev.~D \textbf{61} 084027 (2000), [\href{https://arxiv.org/abs/gr-qc/9911095}{arXiv:gr-qc/9911095}].

\bibitem{Wei:2021lmo}
S.~W.~Wei and Y.~X.~Liu, Phys.~Rev.~D \textbf{104}, no.8, 084087 (2021)
[\href{https://arxiv.org/abs/2106.06704}{arXiv:2106.06704}].

\bibitem{Weisstein} E.~Weisstein, ``Oblate Spheroid." From Wolfram MathWorld [\url{http://mathworld.wolfram.com/OblateSpheroid.html}].

\bibitem{Wesson} P.S.~Wesson, B.~Mashhoon and H.~Liu,~Mod.~Phys.~Lett.~\textbf{A12}, 2309 (1997).

\bibitem{Xiao} Y.~Xiao, Y.~Tian, Y.-X.~Liu, Phys.~Rev.~Lett.~132 (2024) 2, 021401 [\href{https://arxiv.org/pdf/2308.12630v2}{arXiv:2308.12630}].

\bibitem{Xiao25}
Y.~Xiao, Y.-X.~Liu, Y.~Tian, H.~Zhang, [\href{https://arxiv.org/abs/2512.01916v2}{arXiv:2512.01916}].

\bibitem{YangWarpedEmbeddings} H.-X.~Yang, L.~Zhao, Mod.~Phys.~Lett.~\textbf{A25} 1521--1530 (2010) [\href{https://arxiv.org/abs/1002.1001v1}{arXiv:1002.1001}].

\bibitem{Yano} K.~Yano, \emph{The Theory of Lie Derivatives and Its Applications} (Courier Dover Publications, Mineola, New York, 2020; originally published North-Holland Publishing Company, Amsterdam, 1957).

\bibitem{Zhao} Z.-W.~Zhao, Y.-H.~Xiu and N.~Li, Phys.~Rev.~D \textbf{98} 124003 (2018), [\href{https://arxiv.org/abs/1805.04861}{arXiv:1805.04861}].



\end{thebibliography}
\end{document}